# Emergent descriptions at large charge: A foray into the structure of conformal field theories and beyond

Inaugural dissertation

of the Faculty of Science,
University of Bern

presented by
**Rafael Moser**
from Bern

Supervisor of the doctoral thesis:
**Prof. Dr. Susanne Reffert**
Institute for Theoretical Physics,
Faculty of Science,
University of Bern

UNIVERSITÄT BERN

2023

# Emergent descriptions at large charge: A foray into the structure of conformal field theories and beyond

Inaugural dissertation

of the Faculty of Science,
University of Bern

presented by

**Rafael Moser**

from Bern

Supervisor of the doctoral thesis:

**Prof. Dr. Susanne Reffert**

Institute for Theoretical Physics,

Faculty of Science,

University of Bern

Accepted by the Faculty of Science

Bern, November 21, 2023

The Dean,

Prof. Dr. Marco Herwegh

To Evelyn.

To Adrian and Ursula, my parents.

To my teachers that helped me get to this point in my academic career.

To a society that gave me the possibility to pursue higher education, let us try to extend this opportunity to all people.

# Abstract

## Emergent descriptions at large charge

by Rafael Moser

Conformal Field Theories (CFT)s play a central role in the study of Quantum Field Theory (QFT). They represent the fixed point of the Wilsonian Renormalization Group (RG) flow and any QFT is in principle describable as a relevant deformation of the associated nearby Conformal Field Theory (CFT). This thesis aims to explore the structure of CFTs with global internal symmetries and beyond via the Large-Charge Expansion (LCE), a semi-classical expansion applicable for states with large global quantum numbers.

In the first part of this thesis we study CFT and Spontaneous Symmetry Breaking (SSB). We discuss the symmetry-constraints imposed by conformal invariance on the quantum theory, introduce the concept of CFT data and the Operator–Product Expansion (OPE). Concerning SSB, we discuss the existence of Nambu–Goldstone (NG) modes, the general counting rule for the number of NG modes under the spontaneous breaking of global internal symmetries and a generalization of the Goldstone theorem at finite density.

In the second part of this thesis we discuss the current state-of-the-art understanding of the LCE and systematically study CFTs with a global $O(2)$ symmetry in the context of the LCE. We present the LCE in the broader context of the different methods available for accessing CFT data. Particularly, we discuss its relation to large-spin expansions in CFTs and the description of operators with both large spin and large charge. We discuss the emergence of effective condensed-matter descriptions, in particular superfluids, in correlators involving states with large global quantum numbers. Finally, we use the superfluid Effective Field Theory (EFT) description to systematically study two-, three- and four-point functions for CFTs with a global $O(2)$ symmetry. Using the EFT approach we derive universal results for the spectrum of scaling dimensions and three-point coefficients at large charge.

In the last part of this thesis we study CFTs in the double-scaling limit of large charge and large $N$. We discuss the $D = 3$ Wilson–Fisher (WF) fixed point at large $N$ and derive the leading order asymptotics at large charge $Q$ in the double scaling limit $Q/N$ fixed, where scaling dimensions can be studied analytically in the limit $Q/2N \gg 1$, where we recover the superfluid EFT structure, and $Q/2N \ll 1$, where we recover the free mean-field limit. These limits can be connected by resurgent analysis. We also study the spectrum of fluctuations to confirm EFT predictions. Next, we use a fixed-charge approach to gain access to the leading order effective potential for the $\varphi^4$ theory, which we then study for spacetime dimensions $2 < D < 6$. In $D = 3$, we reproduce and extend old results originally found by re-summing Feynman diagrams. In $D = 5$, under the assumption of unitarity the $\varphi^4$-model does not appear to be Ultra–Violet (UV) complete. Finally, we discuss the interacting fixed points of three-dimensional fermionic CFTs in the double-scaling limit of large charge and large $N$. While the Gross–Neveu (GN) model exhibits a Fermi-sphere description at large charge, whose fate at finite $N$ is yet to be determined, for the Nambu–Jona–Lasinio (NJL)-type models we find a Bose–Einstein Condensate (BEC). The large-charge sector of these models is therefore captured by the superfluid EFT approach.

# Acknowledgements

It is astonishing to me sometimes that an impulsive decision I made seven-and-a-half years ago has landed me here now. I could have done anything else, and I am glad that I did not. I have learned valuable lessons about the possibilities and limitations of human scientific endeavours and beyond that will not leave me for the rest of my live.

Human achievements almost never happen in a vacuum. We not only stand on the shoulders of our forefathers, but we are also continuously held up there by our peers. I would not be at this point of my career right now if it were not for my colleagues, friends and collaborators. They have helped me and guided me over the last three years towards earning my doctorate.

First and foremost, I want to acknowledge the tremendous support and guidance I have received from my supervisor, Susanne Reffert. I owe a lot to Susanne, starting from the fact that she gave me this opportunity in the first place. During the last three years — a period which includes a global pandemic — Susanne has been an immense help both academically and personally. Her guidance and expertise in research has been exceptionally important to me, always pushing me deeper into the subject and making me keep asking questions. On a personal level, I have greatly appreciated the work environment she creates and maintains, and the fruitful collaborations that have come out of it. I always felt a sense of personal freedom and belonging in her research group and under her guidance. For all of these reasons I am and will always be truly grateful to her.
At this point, I also wish to acknowledge Domenico Orlando, which regularly took up the role of a second supervisor. His broad insight into physics and his passion, which notably also extends to the more technical and computational aspect of the field, have been inspirational. Furthermore, he was a crucial component of the joyous and light-hearted atmosphere within the research group.

Second, I want to extend my gratitude towards my collaborators and co-workers at the institute, starting with Nicola Dondi and Ioannis Kalogerakis. I have worked on most of my projects together with Nicola and Ioannis and it has been both very enjoyable and a tremendous learning experience. A special thanks I want to extend to Nicola who was and is the more senior researcher in our little group. His effort and can-do attitude as well as his deep knowledge have been major factors in making this collaboration fruitful and our research possible.
I also want to thank Vito Pelizzani, Fabio Apruzzi, Giacomo Sberveglieri, Donald Youmans and Thiago Araujo for collaborations and enlightening discussions. Vito and Ioannis, I enjoyed the journey towards our doctoral degrees together with you. Once the pandemic receded I truly learned to appreciate the feeling of a shared experience you provided me with. Fabio, Donald, Giacomo and Thiago I am thankful for the discussions and insights. You made the anxiety that comes with feeling like a small fish in a big pond manageable.
My gratitude also extends to the rest of the young staff at the ITP, Alessandro, Anders, Gurtej, Aleks, Martina, Simona, Sebastian, Philipp, Nicolas and many more. Your presence and the presence of the people already mentioned above made for a great work environment and lots of light-hearted yet insightful coffee breaks. It has been a pleasure and a joy to share this formative period of my live with you.
Finally, I want to acknowledge the people I worked with during these years, Simeon Hellerman, Gilberto Colangelo, Urs Wenger, Uwe-Jens Wiese, Mikko Laine and Tim Schmidt. I learned a lot from all of you, and I am grateful for having worked with you.

I am also truly grateful for my friends and family which have supported me during this time, both directly and indirectly. A special thanks goes to my parents, Adrian and Ursula: although coming from a completely different background compared to what I am doing now, you two have always been very supportive of my endeavours. You have sacrificed a lot for me to be where I am and you will always be by my side. This thought makes me truly happy. In the same vein I also want to thank my siblings, Dominik and Andrina, for the support and the occasional fainting of interest in my research.
Lastly, I want to thank my best friend and partner (in crime), Evelyn. Having met you truly transformed my life in ways that are difficult to put in words. Over the last few years I was able to share this journey with you that now comes to an intermediate end with you obtaining your degree and me obtaining mine. It surely has been stressful sometimes, and I am thankful for the patience and care I feel coming from you everyday in helping me become a more thoughtful and caring human being. You continue to broaden my perspective on life and society everyday with your thoughtful insights and your loving warm affection. I am looking forward to many beautiful memories yet to come alongside you.

Bern, November 21, 2023 Sincerely, R. M.

# Foreword

The bulk of this thesis is based on the following three publications by the author:

[1] R. Moser et al., "Convexity, large charge and the large-n phase diagram of the $\varphi^4$ theory", Journal of High Energy Physics **02**, arXiv: 2110.07617 [hep-th], 10.1007/JHEP02(2022)152 (2022)

[2] N. Dondi et al., "Spinning correlators in large-charge cfts", Nuclear Physics B **983**, arXiv: 2203.12624 [hep-th], 10.1016/j.nuclphysb.2022.115928 (2022)

[3] N. Dondi et al., "Fermionic cfts at large charge and large n", Journal of High Energy Physics **08**, arXiv: 2211.15318 [hep-th], 10.1007/JHEP08(2023)180 (2023)

The material presented in this thesis is schematically organized as follows:

- Chapter 1 serves as a short introduction into the main subjects of relevance. It is an original presentation of textbook materials and existing results on Conformal Field Theories and Spontaneous Symmetry Breaking. It serves to make this thesis more self-contained and embeds the later chapters in a broader context within theoretical physics.
- Chapter 2 is separated in two sections. Section 2.1 is a review of the existing literature on the Large-Charge Expansion in generic Conformal Field Theories. Section 2.2 is based on [2].
- Chapter 3 is separated into three sections. Section 3.1 is a review of the literature on the Large-Charge expansion in the $O(2N)$ Wilson–Fisher fixed point at large $N$. Section 3.2 is based on [1]. Section 3.3 is based on [3].

Appendices contain details and further material not included in the main text.

# Contents

# List of Acronyms

**LCE** Large-Charge Expansion

**LCEs** Large-Charge Expansions

**LQNE** Large Quantum Number Expansion

**CSF** Conformal Superfluid

**QM** Quantum Mechanics

**QFT** Quantum Field Theory

**EFT** Effective Field Theory

**IR** Infra–Red

**UV** Ultra–Violet

**CFT** Conformal Field Theory

**CFT** Conformal Field Theories

**SCT** Special Conformal Transformation

**BCH** Baker–Campbell–Hausdorff

**GR** General Relativity

**SSB** Spontaneous Symmetry Breaking

**WF** Wilson–Fisher

**RG** Renormalization Group

**OPE** Operator–Product Expansion

**GN** Gross–Neveu

**NJL** Nambu–Jona–Lasinio

**HS** Hubbard–Stratonovich

**NG** Nambu–Goldstone

**GS** Ground State

**EoM** Equation(s) of Motion

**VEV** Vacuum Expectation Value

**DoF** Degree(s) of Freedom

**EFT** Effective Field Theory

**KG** Klein–Gordon

**GN** Gross–Neveu

**GNY** Gross–Neveu–Yukawa

**NJL** Nambu–Jona–Lasinio

**BCS** Bardeen–Cooper–Schrieffer

**PG** Pauli–Gursey

**LSM** Linear Sigma Model

**NLSM** Non–Linear Sigma Model

**WKB** Wentzel–Kramers–Brillouin

**CBZ** Caswell–Banks–Zaks

**SUSY** Supersymmetry

**BEC** Bose–Einstein Condensate

**SYM** Supersymmetric Yang–Mills

**CCWZ** Coleman–Callan–Wess–Zumino

**QCD** Quantum Chromodynamics

**YM** Yang–Mills

**QED** Quantum Electrodynamics

# Introduction

At the beginning of the 20th century — now a hundred years ago — two of the potentially most significant discoveries in modern physics were made. First came the development of the Theory of Relativity starting in 1905, which was followed by the advent of Quantum Mechanics (QM) in the 1920s. Almost immediately, while attempting to quantize the electromagnetic field, in the late 1920s physicists realized that in order to have a consistent description of nature they would need to incorporate special relativity, which was born out of classical electromagnetism, into the framework of quantum mechanics. The resulting theory — Quantum Electrodynamics (QED) — still is one of the most precise theories of nature. The advent of QED marked the birth of QFT as a subject, which after a few decades of maturing found its place as the single most successful concept in theoretical physics of the 20th and the early 21st century. Today, QFT permeates many areas of theoretical physics. High-energy and particle physics are probably the first and most obvious ones, with the success of the Standard Model of particle physics as a fundamental theory of nature, but it is also commonly used in condensed-matter physics, cosmology and inflation, quantum gravity, string theory and of course statistical mechanics. It also has applications outside of physics, in fields like mathematics, computer science and even finance.
From the modern point-of-view, the emergence of QFT is seen as the single logical consequence of the union between the principles of special relativity and and quantum mechanics. It naturally introduces the concept of the anti-particle and describes the phenomena of particle creation and annihilation, while QM always preserves the number of particles. However, its modern inception is unthinkable without the development of Feynman's path integral in 1948 [4]. First introduced in QM, this "sum over infinitely many paths" highlights both the fascinating similarities and the crucial differences between classical and quantum theory.
Let us consider QM for a moment. The path integral fundamentally encapsulates the relationship between classical mechanics and QM. Just as in the classical system, the path integral frames the corresponding quantum problem in terms of the principle of least action, with the action $S[q]$ being given by the same functional in both cases. However, while the principle of least action determines the physical evolution of a classical systems from points $q_1$ to $q_2$ purely in terms of the classical trajectory $q_{\text{cl}}(t)$ found by minimizing the action $S[q]$, the quantum amplitude instead is given by a weighted sum over all possible paths $q(t)$ connecting $q_1$ and $q_2$,

$$\langle q_2 | \, e^{-iH(t_2-t_1)} \, | q_1 \rangle = \int_{q_1}^{q_2} \mathscr{D}q \; e^{iS[q]} \, . \tag{1}$$

The initial and final states in the path integral specify the boundary conditions of the integration over the different paths $q(t)$. The quantum propagation of a particle and its classical trajectory are connected by the observation that different paths $q(t)$ within the path integral are weighted by their classical action $S[q]$. This means that the classical solution $q_{\mathrm{cl}}(t)$ along with the paths close to it contribute the most to the path integral, while paths far away from the classical solution are washed out by interference. If the propagation of a particle in a quantum system is well approximated by the properties of its classical trajectory, say for example we consider photons with wavelengths much smaller than the characteristic size of the surrounding space $\lambda \ll L$, then naturally we expect that the quantum amplitude is well approximated by only considering the paths close to the classical solution $q_{\mathrm{cl}}(t)$. Hence, the path integral localizes around the classical trajectory and is computable via a so-called semi-classical approximation. The action is expanded around its classical solution $q = q_{\mathrm{cl}} + \hat{q}$,

$$S[q] = S[q_{\mathrm{cl}}] + \frac{1}{2}\frac{\delta^2 S}{\delta q^2}\hat{q}^2 + \mathcal{O}(\hat{q}^3)\,, \tag{2}$$

and the path integral is now given by

$$\langle q_2|\, e^{-iH(t_2-t_1)}\,|q_1\rangle = e^{iS[q_{\mathrm{cl}}]}\int \mathscr{D}\hat{q}\, e^{i\frac{1}{2}\frac{\delta^2 S}{\delta q^2}\hat{q}^2 + \mathcal{O}(\hat{q}^3)}\,. \tag{3}$$

The boundary condition have to be adapted accordingly. The amplitude is now separated into a classical contribution given by $\exp(iS[q_{\mathrm{cl}}])$ plus quantum corrections. Note that the quadratic contribution in the quantum corrections — given by $(\delta^2 S/\delta q^2)\,\hat{q}^2$ — just represents the action of a harmonic oscillator.

The path-integral formulation of QM naturally generalizes to QFT, with the small issue of not having a sound mathematical basis due to the vastly increased amount of Degree(s) of Freedom (DoF) in QFT. Aside from this problem, the path integral in QFT readily captures all of the important features of the underlying quantum theory it describes and is often very efficient in simplifying computations compared to the more traditional formulations of QFT like canonical quantization.
Different quantum systems or even different regimes within the same system can be classified via the properties of their respective path integrals broadly into two general categories: weakly-coupled path integrals and strongly-coupled path integrals. The differentiate themselves by the fact that a weakly-coupled path integral in a certain limit of one or more of its parameters can be approximated by a leading classical trajectory and the corresponding loop expansion around it, mirroring the semi-classical analysis in QM in Eq. (3). In a strongly-coupled path integral, however, no generic semi-classical saddle-point approximation of the full theory is possible.

In QFT and its vast number of DoF, the predictive capabilities of physicists largely rely on the existence of a weakly-coupled path integral and the associated loop expansion. This is well exemplified in the Standard Model of particle physics, where for example collisions between particles at the $GeV$ scale are well studied using loop expansions, but processes involving the strong interaction — described by Quantum Chromodynamics (QCD) — at the same scale are hard to predict since the corresponding path integral is strongly coupled there. Any observable $\mathcal{O}$ computed in a weakly-coupled path integral consists of two contributions $\mathcal{O} = \mathcal{O}_{\mathrm{cl}} + \mathcal{O}_{\mathrm{qu}}$ coming from the classical configuration and the quantum corrections around it. The relative size between the two contributions is relevant, as only observables with larger classical contributions $\mathcal{O}_{\rm cl} \gg \mathcal{O}_{\rm qu}$ are readily computable in a semi-classical approximation. By definition, for a weakly-coupled path integral there exist at least some observables that are semi-classical and computable in a loop expansion, while in for a strongly-coupled path integral all observables are quantum in nature.
Generically, semi-classical observables are associated with states consisting of many quanta and are far removed from the vacuum state of the theory.[1] In this case a new emergent description appears in terms of an approximately free theory around an effective ground state far away from the vacuum of the underlying theory. Semi-classical methods represent an important tool to investigate non-perturbative phenomena in QFT like multiparticle production, vacuum decay and instantons. A particular class of observables admitting a semi-classical description are given by observables possessing at least one large quantum number under a conserved charge of the theory, like a large spin $J$ or a large global charge $Q$. Observables with large quantum numbers are naturally semi-classical irrespective of the existence of a small coupling in the underlying theory. Therefore, even in strongly-coupled theories, we expect the existence of a (potentially emergent) weakly-coupled semi-classical path integral description capturing any observable with a large quantum number independent of the nature of the underlying theory, at least as an effective description involving the light DoF. The fact that such light modes exist follows from the fact that states with large quantum number unavoidably break certain symmetries, which in turn implies the existence of light NG modes [5, 6].

Unsurprisingly, many important problems in theoretical physics are related to strongly-coupled theories. Luckily, in many instances involving strongly-coupled path integrals it is possible to find a subset of variables in a certain limit of the theory whose quantum fluctuations are small around an associated semi-classical trajectory. In that case, it is possible to integrate out the variables with large quantum fluctuations in the path integral and derive an effective weakly-coupled description for the remaining DoF. The process of integrating out variables within a theory is deeply tied to another fundamental concept in QFT, which is perhaps most responsible for its success as a theoretical framework: renormalization.
In its earlier iterations QFT suffered from a major flaw being that perturbative computations apparently produced meaningless infinite results. The cause of the phenomenon can be found in the large number of interacting quantum DoF that we have mentioned already. The solution to this problem and the correct way to extract a meaningful result out of the divergences turned out to be renormalization. From the modern perspective, renormalization is a fundamental feature of QFT. Although not the first ones to describe the phenomenon, the modern understanding of how renormalization in QFT works is mainly based on the contributions of Kadanoff and Wilson [7, 8].

Let us illustrate the ideas of Wilsonian renormalization. Consider a QFT described by the action $S[\Phi_i, g_n]$, with generic field content $\{\Phi_i\}$ and couplings $g_n$. The fundamental realization of renormalization is that the bare couplings $g_k$ and the normalizations of the fields $\{\Phi_i\}$ in the action do not correspond to the physically measured quantities as they do not take into account quantum effects. To make connection with reality, observables would have to be rewritten in terms of the renormalized quantities.

[1] Observables describing small fluctuations around the vacuum are only amenable to a weakly-coupled description if the system can be approximated by a free theory, *i.e.* if the theory exhibits small couplings $g_n$ accompanying the interaction terms in the action.

The bare parameters of the theory, as they are not measurable, can then be used to cancel the infinities arising in calculations in a process called regularization.
During regularization, the theory is modified by introducing a parameter — called a regulator — which regularizes the appearing divergences. The seemingly most natural choice here is a scale $\Lambda_0$ cutting off the momenta allowed within the theory to $|\mathbf{k}| \leq \Lambda_0$, but there are other well-established schemes like dimensional regularization or Pauli–Villars regularization. The different choices of regulator are called regularization schemes. Any choice of regularization scheme introduces a characteristic energy scale — called the cut-off scale $M$ — into the theory, which reproduces the original theory when taken to infinity. The now cut-off dependent divergences are then cancelled via so-called counterterms, which are cut-off-dependent parts of the bare parameters in the theory that cancel the appearing divergences. After cancellation of the divergences the cut-off scale $M$ can be taken to infinity while we recover a finite result.
If a theory is amenable to this procedure, it is called renormalizable. Renormalizability in QFT can be determined via dimensional analysis, at least superficially. During the cancellation of divergences, in principle, a renormalization point is chosen, given by the value of the cut-off scale $M$ at which divergences are cancelled. At finite values of the cut-off scale $M$ divergent terms in the theory are finite but cut-off dependent. Changes in the renormalization scale affect how much of a result comes from the bare parameters and how much comes from the divergent quantities to be computed. This can be exploited to calculate the variation of a physical parameter with the change in cut-off scale, which is encoded in the so-called beta functions of the parameters. The general theory of this kind of scale dependence is the RG.

Historically, the splitting of bare parameters into finite contributions plus counterterms came before the discovery of the RG, which essentially states that this splitting is unphysical and that all scales enter in a continuous and systematic way. We return to the QFT with action $S[\Phi_i, g_k]$ introduced above. In terms Fourier modes the path integral is written as[2]

$$Z(g_n) = \int \prod_{\mathbf{k}} \mathscr{D}\tilde{\Phi}_i(\mathbf{k})\, e^{-S[\Phi_i, g_n]}\,. \tag{4}$$

Naturally, the partition function suffers from the appearance of infinities and we define the regularized path integral via a momentum cut-off $\Lambda_0$,

$$Z_{\Lambda_0}(g_n^{(\Lambda_0)}; \Lambda_0) = \int \prod_{|\mathbf{k}|<\Lambda_0} \mathscr{D}\,\tilde{\Phi}_i(\mathbf{k})\; e^{-S[\Phi_i, g_n^{(\Lambda_0)}]}\,. \tag{5}$$

The theory is now explicitly regularized by only allowing momenta up to the cut-off $\lambda_0$, which is also the renormalization scale, in the action $S[\Phi_i, g_n^{(\Lambda_0)}]$. Note that the couplings $g_n^{(\Lambda_0)}$ should be considered as distinct from the couplings in the original theory since we have removed some DoF. In the Wilson–Kadanoff renormalization scheme we try to first integrate the modes with momenta $\Lambda_1 < \Lambda_0$. Naturally,

[2]Naturally, the path integral measure can also be in terms of position space configurations $\mathscr{D}\Phi_i(x)$.

the fields can be decomposed as

$$\Phi_i(x) = \int\limits_{|\mathbf{k}|<\Lambda_1} \frac{\mathrm{d}^D k}{(2\pi)^D}\, e^{i\mathbf{k}\cdot\mathbf{x}}\tilde{\Phi}_i(\mathbf{k}) + \int\limits_{\Lambda_1<|\mathbf{k}|<\Lambda_0} \frac{\mathrm{d}^D k}{(2\pi)^D}\, e^{i\mathbf{k}\cdot\mathbf{x}}\tilde{\Phi}_i(\mathbf{k}) =: \Phi_i^{(1)}(x) + \chi_i^{(1)}(x)\,, \tag{6}$$

where $\Phi_i^{(1)}(x)$ represents the low-energy part of the field $\Phi_i(x)$ and $\chi_i^{(1)}(x)$ is its high-energy part. The path-integral measure likewise factorizes and performing the integral over the high-energy modes $\chi_i^{(1)}(x)$ produces an effective action for the remaining low-energy DoF,[3]

$$S_{\mathrm{eff}}[\Phi_i^{(1)}, g_n(\Lambda_1); \Lambda_1] := -\log\left[\int\limits_{\Lambda_1<|\mathbf{k}|<\Lambda_0} \prod \mathscr{D}\Phi_i(\mathbf{k})\; e^{-S[\Phi_i^{(1)}+\chi_i^{(1)}, g_n^{(\Lambda_0)}]}\right]. \tag{7}$$

This process can be iterated by introducing yet a smaller scale $\Lambda_2 < \Lambda_1$ and integrating out further modes,

$$S_{\mathrm{eff}}[\Phi_i^{(2)}, g_n(\Lambda_2); \Lambda_2] = -\log\left[\int\limits_{\Lambda_2<|\mathbf{k}|<\Lambda_0} \prod \mathscr{D}\Phi_i(\mathbf{k})\; e^{-S[\Phi_i^{(2)}+\chi_i^{(2)}, g_n^{(\Lambda_0)}]}\right]. \tag{8}$$

The process of integrating out modes is called changing the scale and Eq. (7) is called the RG equation for the effective action. Importantly, while changing the scale the couplings in the action for the remaining DoF do now depend on the new scale. Starting from a generic initial action of the form

$$S[\Phi_i, g_n^{(\Lambda_0)}] = \int \mathrm{d}^D x \left[\Phi_i^*(x)(K)\Phi_i(x) + \sum_n g_n^{(\Lambda_0)}\mathscr{O}_n(x)\right], \tag{9}$$

where $K$ denotes the differential operator within the kinetic term, the effective action at the scale $\Lambda_1$ takes the general form

$$S_{\mathrm{eff}}[\Phi_i^{(1)}, g_n(\Lambda_1); \Lambda_1] = \int \mathrm{d}^D x \left[Z^{(\Lambda_1)}\Phi_i^{(1)*}(x)(K)\Phi_i^{(1)}(x) + \sum_n \left(Z^{(\Lambda_1)}\right)^{i_n/2} g_n(\Lambda_1)\mathscr{O}_n(x)\right]. \tag{10}$$

The form of the interaction terms $\mathscr{O}_n(x)$ remains the same and the wavefunction renormalization $Z^{(\Lambda_1)}$ accounts for the fact that the kinetic term may receive correction while changing the scale. At any given scale a renormalized field $\varphi_i^{(1)} = \sqrt{Z^{(\Lambda_1)}}\Phi_i^{(1)}$ can be defined. Evidently, the partition function $Z_{\Lambda_1}(g_n(\lambda_1); \Lambda_0)$ at any scale $\Lambda_1 < \Lambda_0$ is still equal to the original partition function $Z_{\Lambda_0}(g_n^{(\Lambda_0)}; \Lambda_0)$, where $Z_{\Lambda_1}(g_n(\lambda_1); \Lambda_0)$ is defined as

$$Z_{\Lambda_1}(g_n(\Lambda_1); \Lambda_0) := \int\limits_{|\mathbf{k}|<\Lambda_1} \prod \mathscr{D}\tilde{\Phi}_i(\mathbf{k})\; e^{-S_{\mathrm{eff}}[\Phi_i^{(1)}, g_n(\Lambda_1); \Lambda_1]}\,, \tag{11}$$

with the effective action $S_{\mathrm{eff}}$ given by Eq. (7). Infinitesimally, this invariance under changes of the scale

[3] We have $\int \mathscr{D}\Phi_i(\mathbf{k})\big|_{\Lambda_1<|\mathbf{k}|<\Lambda_0} \to \int \mathscr{D}\chi_i^{(1)}(x)$ when switching to position space.

results in the differential equation

$$\left(\Lambda_1 \frac{\partial}{\partial \Lambda_1} + \beta_n(\Lambda_1) \frac{\partial}{\partial g_n}\right) Z_{\Lambda_1}(g_n(\Lambda_1); \Lambda_0) = 0\,, \qquad \beta_n(\Lambda_1) := \Lambda_1 \frac{\partial g_n(\Lambda_1)}{\partial \Lambda_1} \tag{12}$$

called the RG equation for the partition function. It states that the couplings of the effective action vary to account for the change in the number of DoF in the path integral while changing the scale. The partition function is then in fact independent of the scale, provided it is below the initial cut-off scale $\Lambda_0$. The so-called beta function $\beta_n(\Lambda)$ capture the varying — or running — of the couplings and are hence very important in the context of the RG flow. A generalized statement similar to the RG equation for the partition function can be derived for arbitrary correlation functions and is called the Callan-Symanzik equation [9–11]. The overall significance remains the same: the couplings of the theory and the wavefunction renormalization factors vary in such a way as we lower the scale so that correlation functions remain unaltered.

As we change the scale, the RG generates a curve in coupling space — or parameter space — which is called the Wilsonian RG flow $g_n(\Lambda)$. The RG flow is usually expressed in terms of the beta function $\beta_n(\Lambda)$ defined above. The question naturally arises what happens when we are at a point where all the $\beta$-functions vanish. This is a special point of the RG where all the couplings are tuned to some critical value $g_n^{(\Lambda_0)} = g_n^*$ such that

$$\beta_n(\Lambda)\big|_{g_n = g_n^*} = 0\,, \qquad \forall \text{ scales } \Lambda\,. \tag{13}$$

Evidently, the couplings of the particular theories living at such points do not depend on the scale. They are known as critical points of the RG flow. Using the Callan–Symanzik equation it can be shown that two-point correlation functions are highly restrained in critical theories,

$$\langle \mathcal{O}(x)\mathcal{O}(y)\rangle \sim \frac{C(g_n^*)}{|x-y|^{2\Delta_{\mathcal{O}}}}\,, \tag{14}$$

where $\Delta_{\mathcal{O}}$ is the scaling dimension of the operator $\mathcal{O}$ within the critical theory. This power-law behaviour is a characteristic of scale-invariant theories, which besides Poincaré invariance also exhibit invariance under scale transformations

$$x \to \lambda x\,. \tag{15}$$

We conclude that critical theories in the RG flow are always scale-invariant. In fact, almost all scale invariant theories are invariant under the even bigger conformal group [12].

A QFT which is invariant under the full conformal group is called a Conformal Field Theory (CFT). CFTs play a crucial role in the context of renormalization, and hence in our understanding of the parameter space of QFTs, as they represent (almost all of) the critical point in the context the Wilsonian RG. The concept of the Wilsonian RG flow is the best tool we have to study the relationship between QFTs and the geometry of the parameter space spanned by them. Particularly, in order to probe the parameter space of QFTs we can choose a generic QFT and observe its RG flow in both directions of the energy scale to find its UV (high-energy) and Infra–Red (IR) (low-energy) limits. For a local Poincaré-invariant and unitary QFT it is generally expected that the RG flow in both directions approaches a critical point and hence a CFT. As many different QFTs generically flow to the same fixed points, CFTs can be understood as unifying landmarks in parameter space. In the case where this fixed point is interacting, we say that the theories which flow to it all belong to the same universality class. On the other hand, by perturbing — or deforming — a CFT and moving away from the critical point, breaking scale invariance in the process, the space of QFTs near the fixed point can be studied using the Wilsonian RG. Moreover, according to the Wilsonian point-of-view, starting from the correct UV CFT the entire RG flow of a given theory all the way down to the IR fixed point can be described by introducing the appropriate relevant deformation in the UV. In this sense, all QFT can be found and classified in terms of perturbations of CFTs. Illustrating this point, it is often stated that in the parameter space CFTs are beacons of light illuminating the landscape around them.
In addition, CFTs also play important roles in the description of second-order and quantum phase transitions, in string theory and in the context of the AdS/CFT correspondence for quantum gravity. Particularly, the general observation that many physical systems with vastly different descriptions converge and eventually coincide at the point of phase transition is a manifestation of the principle of Wilsonian universality.

By virtue of conformal invariance, all CFTs lack any dimensionful perturbative expansion parameters in the theory. Besides free-field CFTs, most of the interacting CFTs found in the space of QFTs and more broadly in physics are strongly-coupled theories. They do not admit any weakly-coupled path-integral description in the conventional sense. Fortunately, CFTs are highly constrained due to the extended spacetime symmetry present, which puts stringent constraints on the form of (local) correlators. The functional form of two- and three-point functions is completely fixed up to a small set of coefficients and normalization constants. Four- and higher-point functions can be reduced to a sum of three-point functions via the OPE, which in CFTs has an infinite radius of convergence. In other words, the local operators spectrum of CFTs can be defined completely algebraically in terms of a few coefficients appearing in the two-and three-point functions of the theory called the CFT data. The CFT data consists of the scaling dimensions $\Delta_i$ of the operators and the three point coefficients $C_{ijk}$. This algebraic structure provides many consistency constraints on the CFT data which are exploited by the conformal bootstrap program [13]. The conformal bootstrap aims to identify and solve CFTs in terms of the CFT data by restricting the space of their acceptable values, ideally down to a single point. On the other hand, as certain CFTs are also related to critical points of some lattice models in statistical mechanics, numerical methods such as Monte-Carlo simulations can also be applied to study CFTs [14–16].

Naturally, operators in a CFT can possess non-trivial quantum number such as spin or a global charge $Q$. As discussed at the beginning of this introduction, QFT and QM observables involving large quantum numbers generally admit a weakly-coupled description, even in otherwise strongly-coupled theories. This is a particularly intriguing observation in CFTs, as they do not admit a generic weakly-coupled description in terms of a dimensionful parameter. As it turns out, it is possible to describe certain observables involving operators with large quantum numbers under a global symmetry by an appropriate emergent description in a process referred to as the LCE. In fact, in a CFT with a global symmetry correlators of the form

$$\langle \mathcal{O}^{Q\dagger}(x_{\text{out}}) \mathcal{O}_k(x_k) \cdots \mathcal{O}_1(x_1) \mathcal{O}^Q(x_{\text{in}}) \rangle\,, \tag{16}$$

where $\mathcal{O}^Q$ possesses a parametrically larger charge than the intermediate operator insertions, do often admit an emergent description in terms of a superfluid EFT that is approximately scale invariant. [17, 18]. This can be exploited for the computation of CFT data from such correlators. In three-dimensional CFTs with a global $O(2)$ symmetry the scaling dimension of the operator $\mathcal{O}^Q$ has the form [17, 18]

$$\Delta(Q) = c_{3/2} Q^{3/2} + c_{1/2} Q^{1/2} - 0.0937 + \mathcal{O} Q^{-1/2}\,. \tag{17}$$

The scaling dimension can be extracted from the energy of the EFT ground state. The Wilsonian coefficients $c_i$ do depend on the specific underlying theory.

The emergence of an EFT description is a consequence of the fact that the presence of a state with large charge density — which in our case is $|\mathcal{O}^Q(x_{\text{in}})\rangle$ — unavoidably breaks certain spacetime and internal symmetries. The presence of SSB then dictates that the long-distance and low-energy dynamics are described by the associated NG bosons [5, 6]. In the $O(2)$ example there is a single NG boson present — called the conformal superfluid phonon — and it possesses a dispersion relation consistent with conformal invariance. Its fluctuations on top of the EFT ground state describe operators within the CFT that are located close to $\mathcal{O}^Q$ in terms of their conformal dimensions and share the same charge $Q$.

**Plan of the thesis.**

The LCE in many ways bridges the gap between different areas of physics by introducing concepts of condensed-matter physics into the study and understanding of high-energy physics. In doing so, it requires an understanding of both subjects, however. In Chapter 1 we start by reviewing two important prerequisites required to understand the existence of the large-charge program and the structure of its main results. In Section 1.1 we review CFTs from a high-energy point-of-view and discuss their structure. In Section 1.2 we discuss the features of SSB, in particular also at finite density, more from a condensed-matter perspective.

Chapter 2 consists of two parts: a snapshot of the large-charge program and a systematic analysis of the large-charge sector in CFTs invariant under a $O(2)$ symmetry. In Section 2.1 we present the landscape of the large-charge program, including many of its main results. We elucidate the interplay between large charge and other methods of computing CFT data like the conformal bootstrap and large spin expansions. And we discuss how emergent condensed-matter EFT descriptions arise in the study of CFT correlators with large-charge insertions. In Section 2.2 we systematically study the large-charge CFT data in theories hat exhibit a global $O(2)$ symmetry, which can also be a subset of a larger symmetry. In correlators involving the insertion of two heavy operators possessing a large charge $Q$ under the global $O(2)$ we perform the LCE via its superfluid EFT approach. We explain what CFT data involving the insertion of two large-charge operators is within the reach of the large-charge methodology and particularly the EFT approach, and we systematically compute said CFT data in terms of two-, three- and four-point functions.

The last part of this thesis — Chapter 3 — is dedicated to the study of the LCE at large $N$. We discuss how in the double-scaling limit of $Q, N \to \infty$, $Q/N$ fixed the large-charge sectors of strongly coupled theories become accessible beyond the EFT prescription detailed in Chapter 2. To do so we apply finite-temperature field theory techniques in yet another parallel to condensed-matter physics. We start by reviewing the WF fixed point of the $O(2N)$ vector model with a quartic interaction in $D = 3$ at fixed diagonal charges $Q = \sum_i Q_i$ and large $N$ in Section 3.1. Here, we independently recover all the important features of the large-charge superfluid EFT approach outlined in Section 2.2. However, at leading order in $N$ we have access to the scaling dimensions of operators at any value of the normalized charge $Q/2N$. Besides the large-$Q/2N$ limit, in which we recover the superfluid EFT description, we also have access analytically to the small-$Q/2N$ regime, in which results correspond to the free-field limit. In Section 3.2 we show how large-charge techniques in the $O(2N)$ $\varphi^4$-theory can be used to compute the effective potential of the theory to leading order $N$ also away from the fixed point. We use this technique to reproduce and expand upon an old computation involving the re-summation of infinitely many Feynman diagrams in the $D = 3$ $\varphi^4$-theory [19]. We then focus on the dimensions $4 < D < 6$ and discuss the effective potential there. In doing so, we find that there is no unitary UV completion of the $\varphi^4$-model in this range of dimensions, in accordance with earlier results [20]. In $D = 5$ we can compute explicitly results. We show how the effective potential can be interpreted as a complex function with a branch cut. Using large-charge techniques we can investigate the conjectured non-unitary/complex interacting UV CFT. In Section 3.3 we return to the fixed point and study fermionic CFTs in the double-scaling limit. We encounter two distinct behaviours. While in NJL-type models we find a BEC and the large-charge superfluid predictions apply, in the GN model we encounter no SSB at zero temperature and a Fermi-sphere ground state arises. The superconducting behaviour in the NJL models has a neat explanation in terms of Cooper pair condensation. The presence and stability of the Fermi-sphere ground state in the GN model, however, needs further investigation before definite conclusions can be drawn.

At the ends of Chapters 2 and 3 we provide a short conclusion and final remarks. The appendices contain additional materials supplementary to the main part of this thesis and provide technical details for computations and derivations of results.

# 1 Prerequisites: Conformal Field Theory and Spontaneous Symmetry Breaking

This chapter is meant as a short and pedagogical introduction to the two most important frameworks and ideas underlying the bulk of this thesis.
On one hand, the aim of this chapter is to build the foundation of the large-charge approach to CFTs and to motivate the research and publications of the author presented in later chapters. It serves to embed the materials and research presented in the broader context of modern physics, in particular the systematic study of QFT and CFT.
On the other hand, this chapter serves to make the present thesis more self-contained for readers at the graduate level and onwards. In particular, the discussion of SSB is important in this context and includes very recent results from the literature.
Most of the material discussed can be found in many textbooks, reviews and lecture notes, the most important of which will be pointed out below.

The large charge approach — or LCE — is a systematic way of studying CFTs that exhibit a continuous global symmetry.[1] In order to understand the ”raison d'être” of the LCE and its importance in the study of CFTs it is fundamental to understand the basic structure and properties of said theories. To this end, the first part of this chapter is a broad introduction to CFT.
To make the discussion more concise, we will focus on specific aspects of CFTs in $D \geq 3$. The analysis of CFTs is a vast and complex subject — especially in $D = 2$ — and for the purposes of this thesis we only require knowledge about three or more spacetime dimensions.
The bulk of Section 1.1 deals the implications conformal invariance in QFT, assuming that conformal symmetry is realized in the quantum theory and is not anomalous. The focus of the presentation lies on correlation functions, radial quantization, the state–operator map and the OPE. The section ends with a quick introduction to the conformal bootstrap, as it helps to motivate the large-charge program.
In Appendix A we discuss additional materials for the invested reader. We give a brief discussion of the Ising universality class in the context of CFTs and phase transitions in Appendix A.1. In Appendix A.3 we discuss some prerequisites from QFT. We give a detailed discussion of conformal invariance in classical field theory in Appendix A.2.
The presentation of Section 1.1 takes inspiration from an amalgamation of sources, the most important of which are the lecture notes by Simmons-Duffin [21], the book by di Francesco/ Mathieu/ Sénéchal [22] and the lecture notes by Rychkov [23]. Additional references of relevance are the lecture notes by Gillioz [24] as well as the books by Schottenloher [25] and Blumenhagen/ Plauschinn [26].[2]

[1] In this sense it is complementary to other approaches of accessing CFTs, like the conformal bootstrap [13].

The idea of the Large Quantum Number Expansion (LQNE) is to restrict ourselves to sectors of the theory consisting of all operators with the identical fixed quantum numbers under the global symmetry. In doing so, we explicitly and spontaneously break global and spacetime symmetries of the theory. Understanding the structure of the LCE is therefore also predicated on an understanding of the concept of SSB, in particular its generalization to finite density (and finite volume) systems. Hence, the second goal of this chapter is to introduce the idea of SSB and discuss the implications of Goldstone's Theorem. The second part of this chapter — Section 1.2 — discusses SSB of global internal symmetries as well as aspects of SSB at finite density, including a brief discussion on finite volume corrections (particularly for superfluids).
In the first part we discuss the definition and basic properties of SSB — in particular the existence of massless bosonic modes in the spectrum, the so-called NG modes — and present the most important historical results concerning the spontaneous breaking of global internal symmetries and the number and properties of the appearing NG modes. This discussion ends with the presentation of recent results that imply an exact counting rule for the number of NG bosons under certain assumptions [34–38].
Unfortunately, in the case of spontaneously broken spacetime symmetries there are no simple guiding principles or counting rules for now. Therefore, in the second part of Section 1.2 we restrict ourselves to focusing on investigating a particular class of physical systems that tend to exhibit spontaneously broken spacetime symmetries and prominently appear in the LCE: systems at finite density for a spontaneously broken charge, in particular Abelian superfluids. We present a modification of the Goldstone Theorem applicable to this class of theories. Particularly, we prove the existence of gapped NG modes in the spectrum, in addition to the usual massless NG modes. These NG modes have a gap of the order of the chemical potential and reappear when discussing the LCE in Chapter 2. Finally, we end on discussing finite-volume corrections to SSB relevant for the LCE.[3]
The original presentation of materials on SSB in Section 1.2 is influenced mainly by the reviews by Brauner [39] and Watanabe [34] as well as their and their collaborator's important contributions to the topic. The discussion of aspects of SSB at finite density follows mainly the works and contributions of Maris [40], Low and Manohar [41], Nicolis et al. [42–44], Watanabe et al. [45] and the thesis by Cuomo [46]. Finally, the inclusion of the excursus on SSB in finite volume is inspired by the review [47]. For a more comprehensive discussion of SSB in finite systems we refer to [48] and [49].

## 1.1 Conformal field theory

Traditionally, QFT as a framework is presented in terms of the isometries of the underlying spacetime, be it flat space or some curved background. However, there are instances where another spacetime symmetry group is chosen. In this language, a CFT is a QFT whose spacetime symmetries are all conformal transformations, which includes the spacetime isometries as well as scale transformations and special conformal transformations.
CFTs are fundamental to the study and understanding of QFT. As discussed in the Introduction, CFTs live at the fixed point of the Wilsonian RG flow (points where the beta-function vanishes). But CFTs also appear at other important cornerstones of modern physics.

[2]For a pedagogical introduction to QFT we refor to the books by Weiberg [27–29], Coleman [30], Schwartz [31], Fradkin [32] and also Derendinger [33].
[3]Strictly speaking, SSB — as it is generically defined in the literature — can only occur for systems in infinite volume.

- CFTs are fundamental to string theory. The world-sheet of a string is described by a two-dimensional CFT. Notable references for the reader more interested in the role of two-dimensional CFTs in string theory are [50–53]. Unfortunately, the LCE is less powerful in two-dimensional CFTs [54].

- CFTs are connected to our current understanding of quantum gravity via the AdS/CFT correspondence, *i.e.* the Gauge/Gravity duality. The AdS/CFT correspondence proposes duality between a CFT living on the boundary of spacetime and the associated gravitational theory in the bulk, suggesting that CFTs can be used to study gravitational phenomena. The seminal works on AdS/CFT — *i.e.* Gauge/Gravity — correspondence are [55–57], for a well-structured introduction see [58]. There is some literature on the LCE and its relation to the AdS/CFT correspondence [59–64].

- CFTs are of great importance in the description of critical phenomena in statistical mechanics and condensed-matter physics. Continuous — or second-order — phase transitions can generally be understood using the language of CFTs.
  In the language of statistical mechanics a system undergoing a continuous phase transitions is described by a continuous partition function $\log(Z)$ that is either non-analytic or exhibits an infinite slope at the critical point in configuration space. For a thermal system the critical point at temperature $T_c$ separates an ordered phase at low temperatures $T < T_c$ from a disordered phase at high temperatures $T > T_c$.
  In general there exists an order parameter $\langle\mathcal{O}\rangle$ that vanishes in the disordered phase $\langle\mathcal{O}\rangle = 0$ while it is non-vanishing in the ordered phase $\langle\mathcal{O}\rangle \neq 0$, hence allowing to distinguish the different phases.
  At the critical point there is a symmetry enhancement and the system becomes scale invariant. Correlation functions no longer obey a exponential decay but follow a power law and fluctuations appear at all possible wavelengths/length scales. Long range correlations develop and the microscopic structure of the system becomes irrelevant. In that sense CFTs are a natural candidate to describe critical points of condensed-matter and statistical mechanics systems in the continuum limit since scale invariance generically enhances to full conformal invariance [12]. However, there exist some exceptional critical points that are described by scale invariant theories that are not fully conformally invariant and there are scale invariant RG fixed points which do not get enhanced to full conformal invariance [12, 23]. An explicit example can be found in [65]. For the invested reader we discuss the Ising universality class in the context of CFTs in Appendix A.1.

For the purposes of this thesis we are content with studying CFTs in three or more spacetime dimensions and Euclidean signature. Unless otherwise specified, we are in $D$-dimensional Euclidean spacetime with metric $g^{\mu\nu} = \delta^{\mu\nu}$. There are some intricate issues arising in Minkowski space, for example with regard to causality. In the context of this thesis these details do not need to be discussed. For a presentation of CFTs in Minkowski signature see [24].

### 1.1.1 Conformal symmetry in the operator formalism of QFT

In QFT scale invariance is a naturally arising symmetry since fixed points of the RG flow by definition exhibit scale invariance (see the discussion in the Introduction). Although the relationship between scale and conformal invariance is very intricate [12], in most cases scale invariance enhances to full conformal invariance. As symmetries are expressed in terms of the conserved charges at the quantum level — which can be written in terms of the energy momentum tensor $T^{\mu\nu}(x)$ — in order for conformal invariance to be realized at the quantum level it is crucial that there exists a traceless stress-energy tensor

$$T^{\mu}{}_{\mu}(x) = 0\,. \tag{1.1}$$

This operator equation guarantees the existence of the four additional charges related to invariance under Special Conformal Transformation (SCT)s in addition to scale and Poincaré invariance and allows them to be expressed in terms of $T^{\mu\nu}$, as we will see.[4] Unless otherwise specified, we always assume the existence of such a traceless stress-energy tensor as a prerequisite for conformal invariance at the quantum level [12].

Assuming the existence of a traceless conserved and symmetric stress-energy tensor $T^{\mu\nu}$ allows for the totality of all conformal charges to be conveniently written as

$$Q^{(\epsilon)}(\Sigma) = -\int_{\Sigma} \mathrm{d}n_{\mu}\, \epsilon_{\nu}(x)\, T^{\mu\nu}(x)\,, \tag{1.2}$$

where $\Sigma \subset \mathbb{R}^D$ is an arbitrary $D-1$-dimensional hypersurface. The charge in Eq. (1.2) will be conserved as long as

$$\partial_{\mu}\epsilon_{\nu} + \partial_{\nu}\epsilon_{\mu} = \frac{2}{D}\,(\partial\cdot\epsilon)\,\delta_{\mu\nu}\,. \tag{1.3}$$

Eq. (1.3) is called the conformal Killing equation. The linearly independent solutions comprising the conformal group — translations, rotations, dilatations (*i.e.* scale transformations) and Special Conformal Transformations (SCTs) — are all encoded by the vector field $\epsilon_{\mu}(x)$,

$$\begin{aligned}
&\text{translations:} && p_{\mu} = \partial_{\mu}\,, && \epsilon_{\mu} = a_{\mu}\,,\\
&\text{rotations:} && m_{\mu\nu} = x_{\nu}\partial_{\mu} - x_{\mu}\partial_{\nu}\,, && \epsilon_{\mu} = \frac{1}{2}\omega_{\rho\sigma}\left(\delta^{\mu\rho}x^{\sigma} - \delta^{\mu\sigma}x^{\rho}\right),\\
&\text{dilatations:} && d = x_{\nu}\partial_{\nu}\,, && \epsilon_{\mu} = \lambda x_{\mu}\,,\\
&\text{SCTs:} && k_{\mu} = 2x_{\mu}(\mathbf{x}\cdot\partial) - \mathbf{x}^2\partial_{\mu}\,, && \epsilon_{\mu} = b^{\nu}\left(2x_{\mu}x_{\nu} - \mathbf{x}^2\delta_{\mu\nu}\right)
\end{aligned} \tag{1.4}$$

A solution $x^{\mu} \mapsto x'^{\mu} = x^{\mu} + \epsilon^{\mu}(x)$ of the conformal Killing equation Eq. (1.3) satisfies

$$\frac{\partial x'^{\mu}}{\partial x^{\nu}} = \left[1 + \frac{1}{D}(\partial\cdot\epsilon)\right]\left[\delta^{\mu}_{\nu} + \frac{1}{2}(\partial_{\nu}\epsilon^{\mu} - \partial^{\mu}\epsilon_{\nu})\right] \xrightarrow[\text{map}]{\exp[\cdot]} \Omega(x)\, R^{\mu}{}_{\nu}(x)\,, \qquad R^T R = \mathbb{1}\,. \tag{1.5}$$

Hence, a conformal transformation corresponds to an infinitesimal rescaling times an infinitesimal rotation and — upon exponentiating — a finite rotation times a local scale transformation. Therefore, a

[4] The relationship between tracelessness of $T^{\mu\nu}$ and invariance under SCTs can also be deduced in a classical analysis, see Appendix A.2.3.

conformal transformation rescaling the metric $g_{\mu\nu} = \delta_{\mu\nu}$ by an overall factor

$$g_{\mu\nu} \mapsto g_{\alpha\beta} \frac{\partial x'^{\alpha}}{\partial x^{\mu}} \frac{\partial x'^{\beta}}{\partial x^{\nu}} = \Omega^2(x) g_{\mu\nu} \,. \tag{1.6}$$

Dilatations, translations and rotations, as described in Eq. (1.4), are straightforwardly exponentiated. More interesting are SCTs. Their finite form reads

$$x^{\mu} \mapsto \frac{x^{\mu} - x^2 b^{\mu}}{1 - 2(b\cdot x) + x^2 b^2} \,, \qquad \text{with scale factor: } \Omega^2(x) = (1 - 2(b\cdot x) + b^2 x^2)^2 \,. \tag{1.7}$$

Any SCT leaves the origin invariant but maps $x^{\mu} = b^{\mu}/b^2$ to point at infinity.[5] Invariance under SCTs is crucial as it differentiates scale invariant theories from conformally invariant ones.
SCTs closely related to inversions, which are discrete conforma transformations of the form

$$\mathscr{I}: \qquad x^{\mu} \mapsto x'^{\mu} = \frac{x^{\mu}}{x^2} \,. \tag{1.8}$$

Inversions it lies outside of the identity component of the conformal group, except in the context of SCTs. As a matter of fact, an SCT is comprised of an an inversion, followed by a translation and then by another inversion,[6]

$$x^{\mu} \overset{\mathscr{I}}{\mapsto} \frac{x^{\mu}}{x^2} \overset{-b_{\mu}}{\mapsto} \frac{x^{\mu}}{x^2} - b_{\mu} \overset{\mathscr{I}}{\mapsto} \frac{x^{\mu} - x^2 b^{\mu}}{1 - 2(b\cdot x) + x^2 b^2} \,. \tag{1.9}$$

Inversions by itself cannot be obtained by exponentiating a Killing vector, but SCTs can.
Since inversions are not continuously connected to the identity, invariance under inversions needs to be checked on an individual basis.[7]

The conserved charges $Q^{(\epsilon)} = Q^{(\epsilon)}(\Sigma)$ defined in Eq. (1.2) satisfy the conformal algebra,[8]

$$[Q^{(\epsilon_1)}, Q^{(\epsilon_2)}] = Q^{([\epsilon_2,\epsilon_1])} \,, \tag{1.10}$$

where $[\epsilon_1,\epsilon_2]$ denotes the commutator of vector fields, $[\epsilon_1,\epsilon_2]^{\mu} = \epsilon_1^{\nu}(\partial_{\nu}\epsilon_2^{\mu}) - \epsilon_2^{\nu}(\partial_{\nu}\epsilon_1^{\mu})$. In terms of the linearly independent charges associated to translations, rotations, dilatations and SCTs the commutation relations read

$$\begin{aligned}
&[Q^{(M)}_{\mu\nu}, Q^{(M)}_{\rho\sigma}] = \delta_{\nu\rho} Q^{(M)}_{\mu\sigma} \pm \text{permutations}\,, && [Q^{(K)}_{\mu}, Q^{(P)}_{\nu}] = 2\delta_{\mu\nu} D - 2M_{\mu\nu}\,,\\
&[Q^{(M)}_{\mu\nu}, Q^{(P)}_{\sigma}] = \delta_{\nu\sigma} Q^{(P)}_{\mu} - \delta_{\mu\sigma} Q^{(P)}_{\nu}\,, && [Q^{(D)}, Q^{(P)}_{\mu}] = Q^{(P)}_{\mu}\,,\\
&[Q^{(M)}_{\mu\nu}, Q^{(K)}_{\sigma}] = \delta_{\nu\sigma} Q^{(K)}_{\mu} - \delta_{\mu\sigma} Q^{(K)}_{\nu}\,, && [Q^{(D)}, Q^{(K)}_{\mu}] = -Q^{(K)}_{\mu}\,.
\end{aligned} \tag{1.11}$$

All other commutators are vanishing. These are the commutation relations defining the conformal algebra.

[5]Flat space plus infinity $\mathbb{R} \cup \{\infty\}$ is Weyl equivalent to the unit sphere $S^D_1$. On the unit sphere $S^D_1$ all conformal transformations are finite, *i.e.* non-singular.

[6]SCTs are essentially translations with the points at $\{0\}$ and $\{\infty\}$ swapped by an inversion.

[7]There is a neat argument based on the embedding formalism [23, 66] implying that a CFT invariant under parity will be invariant under inversions and vice versa [23].

[8]This is a consequence of the comutation relation $[Q^{(\epsilon)}, T^{\mu\nu}] = \epsilon\cdot\partial T^{\mu\nu} + (\partial\cdot\epsilon)\, T^{\mu\nu} - \partial_{\rho}\epsilon^{\mu} T^{\rho\nu} + (\partial^{\nu}\epsilon_{\rho}) T^{\rho\mu}$, which holds true under the assumption that conformal invariance is unbroken [21].

In the operator formalism of the quantized theory the action of the topological charge operators on local operators $\mathcal{O}(x)$ of the theory is given by the integral version of the Ward identity,

$$\frac{\partial}{\partial x_\mu}\langle j_\mu^{(\epsilon)}(x)\mathcal{O}_1(x_1)\cdots\mathcal{O}_N(x_N)\rangle = -\sum_{k=1}^{N}\delta(x-x_k)\,\langle\mathcal{O}_1(x_1)\cdots G^{(\epsilon)}\mathcal{O}_k(x_k)\cdots\mathcal{O}_N,(x_N)\rangle\,, \tag{1.12}$$

with $G^{(\epsilon)}$ being the generator of the transformation encoded by $\epsilon_\mu(x)$ via the conserved current

$$j_\mu^{(\epsilon)}(x) = \epsilon^\nu(x)\,T_{\mu\nu}(x)\,. \tag{1.13}$$

In its integral form Eq. (1.12) amounts to a commutator equation,[9]

$$[Q^{(\epsilon)},\mathcal{O}(x)] = G^{(\epsilon)}\,\mathcal{O}(x)\,. \tag{1.14}$$

Just as the charge $Q^{(\epsilon)}$ can be decomposed into a linear combination of Poincaré transformations, dilatations and SCTs, so can the generator $G^{(\epsilon)}$,[10]

$$G^{(\epsilon)} = a^\mu P_\mu + \frac{\omega^{\mu\nu}}{2}M_{\mu\nu} + \lambda D + b^\mu K_\mu\,. \tag{1.15}$$

There is one important detail with respect to the conserved charges and their associated generators that needs mentioning here. As repeated action of the conserved charges reverses the order of the associated generators [21],

$$[Q^{(\epsilon_1)},[Q^{(\epsilon_2)},\mathcal{O}]] = G^{(\epsilon_2)}G^{(\epsilon_1)}\mathcal{O}\,, \tag{1.16}$$

the generators in Eq. (1.15) satisfy the commutation relations of the charges in Eq. (1.11) with reversed signs [21],[11]

$$\begin{aligned}
&[M_{\mu\nu},M_{\rho\sigma}] = \delta_{\nu\rho}M_{\sigma\mu} \pm \text{permutations}\,, && [K_\mu,P_\nu] = -2\delta_{\mu\nu}D + 2M_{\mu\nu}\\
&[M_{\mu\nu},P_\sigma] = \delta_{\mu\sigma}P_\nu - \delta_{\nu\sigma}P_\mu\,, && [D,P_\mu] = -P_\mu\,,\\
&[M_{\mu\nu},K_\sigma] = \delta_{\mu\sigma}K_\nu - \delta_{\nu\sigma}K_\mu\,, && [D,K_\mu] = K_\mu\,.
\end{aligned} \tag{1.17}$$

Topological charge operators act locally on local operators. The reason is that they can be continuously deformed to live in an arbitrarily small neighbourhood of the operator insertion. For the sake of simplicity, for now on we use the same notation for the conserved charge $Q^{(\epsilon)}$ and its associated generator counterpart $G^{(\epsilon)}$, unless it is contextually important to separate them. In the orthogonal basis of the conformal group in Eq. (1.4) and Eq. (1.11) we identify

$$Q_\mu^{(P)} \sim P_\mu\,, \qquad Q_{\mu\nu}^{(M)} \sim M_{\mu\nu}\,, \qquad Q^{(D)} \sim D\,, \qquad Q_\mu^{(K)} \sim K_\mu\,. \tag{1.18}$$

We only need to be careful as to which sets of commutation relations between Eq. (1.11) and Eq. (1.17) is applicable in each situation. Finally, we use commutator notation $[P_\mu,\mathcal{O}(x)]$ and shorthand notation

[9]After integration the left-hand side of Eq. (1.12) turns into a surface integral and then a commutator. For more details on Ward identities and a derivation of Eq. (1.12) using the path-integral formalism see Appendix A.3.3.

[10]Given a Lagrangian description of the theory the associated generators are the differential operators generating the transformation in the classical theory.

[11]These commutation relations are consistent with the classical analysis of the conformal generators, see Appendix A.2.2.

$P_\mu \circ \mathcal{O}(x)$ interchangeably.[12]

The conformal group in $\mathbb{R}^D$ is isomorphic to the Lorentz group $SO(D+1,1)$. This is best derived by taking the $\mathbb{R}^D$ variables $x_1,\ldots,x_D$ and adding two additional variables $X^+ = (x_0 + x_{D+1})$, $X^- = (x_0 - x_{D+1})$. The isomorphism now reads[13]

$$J_{\mu\nu} = M_{\mu\nu}\,, \qquad J_{+\mu} = P_\mu/2\,, \qquad J_{-\mu} = K_\mu/2\,, \qquad J_{+-} = D/2\,, \tag{1.19}$$

with $J_{\alpha\beta}$ — understood to be antisymmetric — being the generators of $SO(p+1,q+1)$. These generators satisfy the Lorentz algebra,

$$[J_{\alpha\beta}, J_{\gamma\delta}] = \eta_{\alpha\gamma} J_{\beta\delta} + \eta_{\beta\delta} J_{\alpha\gamma} - \eta_{\alpha\delta} M_{\beta\gamma} - \eta_{\beta\gamma} J_{\alpha\delta}\,. \tag{1.20}$$

The coordinates $X^+, X^-$ are called light-cone coordinates. The isomorphism gives rise to the embedding formalism, a convenient way of encoding the action of the conformal group in $\mathbb{R}^D$ via the action of the Lorentz group in $\mathbb{R}^{D+2}$ by appropriately embedding $\mathbb{R}^D$ in $\mathbb{R}^{D+2}$ [23, 66–68].

Having access to the conformal charges allows us to classify all operators of the theory into irreducible representations of said charges. The representation theory of the conformal group is built around the existence of so-called primary operators, which generate the irreducible representations of the conformal group.
The operator formalism of the quantum theory identifies primary operators in a CFT by the action of the symmetry operators at the origin. As $D$ and $M_{\mu\nu}$ commute, see Eq. (1.11), we can simultaneously diagonalize them on the space of operators of the theory, so that local operators in a CFT obey[14]

$$\begin{aligned} [M_{\mu\nu}, \mathcal{O}(0)] &= i S_{\mu\nu} \mathcal{O}(0)\,, \\ [D, \mathcal{O}(0)] &= \Delta \mathcal{O}(0)\,, \end{aligned} \tag{1.21}$$

where $S_{\mu\nu}$ are the spin matrices encoding the representation of the Lorentz group and hence the spin of $\mathcal{O}(x)$ and $\Delta$ is the scaling or conformal dimension of the operator $\mathcal{O}(x)$.[15] The defining feature of local primary operators in the space of operators is the action of $K_\mu$. The commutation relations in Eq. (1.11) imply that $K_\mu$ lowers the scaling dimension $\Delta$ of any given operator $\mathcal{O}(0)$ by one,

$$[D, [K_\mu, \mathcal{O}(0)]] = [K_\mu, [D, \mathcal{O}(0)]] + [[D, K_\mu], \mathcal{O}(0)] = (\Delta - 1) K_\mu \mathcal{O}(0)\,. \tag{1.22}$$

Given an operator $\mathcal{O}'(0)$ with scaling dimension $\Delta'$ we can act repeatedly with $K_\mu$ and obtain an operator with arbitrarily low scaling dimension, implying that scaling dimensions are not bounded from below. As in any physically sensible theory this cannot be the case (*e.g.* this makes it impossible for the theory to be unitary), this process has to eventually terminate, meaning there exists an operator $\mathcal{O}(0) = K_{\mu_1} \circ \cdots \circ K_{\mu_n} \circ \mathcal{O}'(0)$ such that

$$[K_\mu, \mathcal{O}(0)] = 0\,, \qquad \forall \mu\,. \tag{1.23}$$

[12] The notation $P_\mu \circ \mathcal{O}(x)$ also expresses the geometric procedure of surrounding the operator insertion $\mathcal{O}(x)$ with a topological charge operator $P_\mu \sim Q_\mu^{(P)}$ inside the path integral.

[13] In terms of $x_0, x_1, \ldots, x_D, x_{D+1}$ the generators are $J_{0\mu} = (P_\mu + K_\mu)/2$, $J_{D+1\mu} = (P_\mu - K_\mu)/2$, $J_{D+1\,0} = D$.

[14] The action of a local operator on a given function only depends on a single point $x$ and it is in principle possible to determine the output value at $x$ solely from the values of the input in an arbitrarily small neighbourhood of $x$. In radial quantization we condense these properties and simply define a local operator $\mathcal{O}(0)$ to be an eigenstate of the dilatation operator. On the other hand, the action of a non-local operator does not depend on a single point in spacetime. A large class of (linear) non-local operators is given by the integral transforms (e.g. the Fourier or Laplace transforms).

[15] As $\mathcal{O}(x)$ belongs to an irreducible representation of the Lorentz group — according to Schur's Lemma — any matrix commuting with all $S^{\mu\nu}$ has to be a multiple of $\mathbb{1}$. Additionally, the group of scale transformations is not compact, hence $\Delta$ is a real number.

Operators satisfying Eq. (1.23) are called primary. For any primary operator $\mathcal{O}$ there exists a tower of descendant operators — or a conformal multiplet — with increasing scaling dimension constructed by repeatedly acting with the momentum operators $P_\mu$,

$$\mathcal{O}(0) \longrightarrow P_{\mu_1} \circ \cdots \circ P_{\mu_N} \circ \mathcal{O}(0)\,, \qquad \Delta \longrightarrow \Delta + N\,. \tag{1.24}$$

Identifying all primary operators in a CFT is equivalent to identifying all operators in a CFT. As in any other (flat space) QFT, the momentum operator — or momentum generator — acts as a derivative on the space of local operators,

$$[P_\mu, \mathcal{O}(x)] = p_\mu \mathcal{O}(x) = \partial_\mu \mathcal{O}(x)\,. \tag{1.25}$$

Consider the operator insertion $\mathcal{O}(x)$ away from the origin. The exponential $\exp[\mathbf{x}\cdot\mathbf{P}]$ translates the operator $\mathcal{O}(0)$ away from the origin such that $\mathcal{O}(x) = \exp[\mathbf{x}\cdot\mathbf{P}] \circ \mathcal{O}(0)$. This tells us that the local operator $\mathcal{O}(x)$ is an infinite linear combination of descendant operators of $\mathcal{O}(0)$ (and hence clearly not an eigenstate of $D$).

$$\mathcal{O}(x) = \sum_n x_{\mu_1} \dots x_{\mu_n} P_{\mu_1} \circ \cdots \circ P_{\mu_n} \circ \mathcal{O}(0)\,. \tag{1.26}$$

The conditions in Eq. (1.21) and Eq. (1.23) allow us to construct a representation of the conformal algebra out of any primary $\mathcal{O}(0)$ and its descendants (via $P_\mu$).[16] Due to the similarities between conformal multiplets and $SU(2)$ representations we will sometimes refer to the momentum generators $P_\mu$ as raising operators and to the SCT generators $K_\mu$ as lowering operators.

In $D \geq 3$ — independent of the existence of a Lagrangian description of the theory — CFTs are formulated in terms of primary operators (or fields in the Lagrangian description). Primary operators are defined by their transformation behaviour under conformal transformations in Eq. (1.21) and Eq. (1.23). Equivalently to Eq. (1.21) and Eq. (1.23), a primary operator of scaling dimension $\Delta$ can be defined to be homogeneous (of degree $\Delta$) and transform under a conformal transformations as

$$\mathcal{O}(x) \mapsto \mathcal{O}'(x') = \left|\frac{\partial x'(x)}{\partial x}\right|^{\frac{\Delta}{D}} R(x)\mathcal{O}(x)\,, \tag{1.27}$$

where $|\partial x'/\partial x| = \Omega^{-D}$ is the Jacobian of the (finite) transformation and $R(x)$ encodes the representation of the field $\mathcal{O}(x)$ under the action of the rotation group. The equivalence of Eq. (1.27) and Eqs. (1.21), (1.23) has been known for a while and is presented properly in [69]. Eq. (1.27) presents the most general transformation behaviour under the conformal group that fields can exhibit in physically well-motivated theories [70–72].

[16] The action of the conformal generators will never take us out of the conformal multiplet. In fact, it can be shown that any local operator in a unitary CFT is a linear combination of primary and descendant operators [21].

The action of the conformal generators on local operator insertions $\mathcal{O}(x)$ away from the origin is encoded by the differential operators in Eq. (1.4),

$$\begin{aligned}
[M_{\mu\nu},\mathcal{O}(x)] &= (m_{\mu\nu} + iS_{\mu\nu})\mathcal{O}(x) = (x_\nu\partial_\mu - x_\mu\partial_\nu + iS_{\mu\nu})\mathcal{O}(x)\,,\\
[D,\mathcal{O}(x)] &= (d+\Delta)\mathcal{O}(x) = (x^\nu\partial_\nu + \Delta)\mathcal{O}(x)\,,\\
[K_\mu,\mathcal{O}(x)] &= (2x_\mu\Delta + 2ix^\nu S_{\nu\mu} + k_\mu)\mathcal{O}(x) = (2x_\mu\Delta + 2ix^\nu S_{\nu\mu} + 2x_\mu x^\nu\partial_\nu - \mathbf{x}^2\partial_\mu)\mathcal{O}(x)\,,
\end{aligned} \tag{1.28}$$

in accordance with the usual classical analysis [22] (see also Appendix A.2 for details).[17] Conveniently, the action of all of the conformal generators can be neatly summarized as

$$[Q^{(\epsilon)},\mathcal{O}(x)] = \left(\epsilon\cdot\partial + \frac{\Delta}{D}(\partial\cdot\epsilon) - \frac{i}{2}\left(\partial^\mu\epsilon^\nu\right)S_{\mu\nu}\right)\mathcal{O}(x)\,. \tag{1.29}$$

This is the most general form of the conformal generator in Eq. (1.15) and Eq. (1.15). In particular, this result implies that the stress-energy tensor $T^{\mu\nu}$ is primary with $\Delta = D$ [21]. Finally, we repeat that the condition in Eq. (1.29) at $x = 0$ defining primaries in the operator formalism is equivalent to the definition of a primary operator/field in Eq. (1.27).[18] A rigorous proof of this statement can be found in [69].

### 1.1.2 Conformal Ward–Takahashi identities

Ward-Takahashi identities are the quantum analogue of the classical conservation of currents associated to a continuous symmetry via Noether's theorem. Any local QFT has a conserved stress-energy tensor,

$$\partial_\mu T^{\mu\nu}(x) = 0\,. \tag{1.30}$$

This condition is satisfied as an operator identity and is modified once we have operator insertions in the path integral, *i.e.* if we consider correlation functions. In the presence of operator insertions we need to include contact terms,

$$\partial_\mu\langle T^{\mu\nu}(x)\mathcal{O}_1(x_1)\cdots\mathcal{O}_N(x_N)\rangle = -\sum_{k=1}^{N}\delta(x-x_k)\partial_k^\nu\langle\mathcal{O}_1(x_1)\cdots\mathcal{O}_N(x_N)\rangle\,. \tag{1.31}$$

This is the Ward identity for translational invariance with the associated current $j^{\mu\nu} = T^{\mu\nu}$. It can be derived as such from translational invariance of the path integral.[19]
Invariance under rotations — with the associated current $j^{\mu\nu\rho} = T^{\mu\nu}x^\rho - T^{\mu\rho}x^\nu$ — produces the Ward

[17] The action of the conformal charges on operator insertions can be deduced via the formula $G^{(\epsilon)}\big|_x = e^{\mathbf{x}\cdot\mathbf{P}}G^{(\epsilon)}\big|_{x=0}e^{-\mathbf{x}\cdot\mathbf{P}}$. In the computation it is important to keep in mind that the generators satisfy the commutation relations in Eq. (1.17) and not the ones in Eq. (1.11), see the discussion after Eq. (1.15).

[18] To illustrate the equivalence, consider the result in Eq. (1.39), which can be derived using Eq. (1.28) — *i.e.* Eq. (1.29) — by noting that the simultaneous action of $Q^{(\epsilon)}$ on all operator insertions must vanish as moving $Q^{(\epsilon)}$ to the boundary at infinity yields zero. The same result can be derived using Eq. (1.27) or Eq. (A.50) in Appendix A.2.2.

[19] Eq.(1.31) can equivalently be derived by coupling the QFT to a background metric $g$ near flat space. In this picture the stress-energy tensor is the response to a small metric perturbation $\delta g$ and we simply demand diffeomorphism invariance near flat space [21, 73].

identity[20]

$$\langle (T^{\mu\nu}(x) - T^{\nu\mu}(x))\mathcal{O}_1(x_1)\cdots\mathcal{O}_N(x_N)\rangle = -\sum_{k=1}^{N}\delta(x-x_k)\, iS_k^{\nu\mu}\,\langle\mathcal{O}_1(x_1)\cdots\mathcal{O}_N(x_N)\rangle\,. \tag{1.32}$$

This means that symmetricity of the stress-energy tensor — $T^{\mu\nu}(x) = T^{\nu\mu}(x)$ — in a Poincaré invariant QFT is satisfied, except at the positions of operator insertions, where we again need to add contact terms related to the spin of the insertions.

In a CFT — besides symmetricity and the conservation law — the stress-energy tensor is also traceless and satisfies Eq. (1.1).[21] The final Ward identity associated to conformal invariance — that modifies Eq. (1.1) in the presence of operator insertions — is derived from invariance under dilatations and reads[22]

$$\langle T^{\mu}{}_{\mu}(x)\mathcal{O}_1(x_1)\cdots\mathcal{O}_N(x_N)\rangle = -\sum_{i=1}\delta(x-x_i)\Delta_k\langle\mathcal{O}_1(x_1)\cdots\mathcal{O}_N(x_N)\rangle\,, \tag{1.33}$$

with $\Delta_k$ the scaling dimension of the operator/field $\phi_k$.

### 1.1.3 Conformal correlators

Some of the important results in CFT at the quantum level are the constraints imposed by conformal invariance on $N$-point correlation functions of primary operators. We will derive these constraints via the path-integral description while assuming the existence of a Lagrangian description, but we note that they can also be derived using the operator formalism as presented in Section 1.1.1.

Consider a QFT with field content $\{\phi\}$ and action $S[\phi]$. In the path integral formalism properly normalized $N$-point functions are given by

$$\langle\mathcal{O}_1(x_1)\cdots\mathcal{O}_N(x_N)\rangle = \frac{1}{Z}\int\mathcal{D}\phi\,\mathcal{O}_1(x_1)\cdots\mathcal{O}_N(x_N)\,e^{-S[\phi]}\,,\qquad Z=\int\mathcal{D}\phi\,e^{-S[\phi]}\,. \tag{1.34}$$

The path-integral measure $\mathcal{D}\phi$ is a formally defined quantity that can thought of as

$$\mathcal{D}\phi \propto \prod_{x\in\mathbb{R}^D}\mathrm{d}\phi(x)\,. \tag{1.35}$$

We assume conformal invariance of the action as well as the path-integral measure. It is important to point out that (conformal) invariance of the path-integral measure does not automatically follow from (conformal) invariance of the theory at the classical level. If a theory exhibits a classical symmetry but the path-integral measure fails to be invariant, then we speak of an anomaly and the symmetry is called anomalous.[23]

Assuming that both the action and the path-integral measure are invariant under a conformal transformation (assuming quantum conformal invariance) implies that $N$-point correlation functions satisfy

$$\langle\mathcal{O}_1(x_1')\cdots\mathcal{O}_N(x_N')\rangle = \langle\mathcal{O}_1'(x_1')\cdots\mathcal{O}_N'(x_N')\rangle\,, \tag{1.36}$$

[20] We apply Eq. (1.31) to get rid of derivatives in Eq. (1.32). The same procedure is applied to extract Eq. (1.33).

[21] There can be Weyl anomalies in curved space.

[22] There is no additional Ward identity associated to SCTs. The additional conserved charges $Q_\mu^{(K)}$ can be constructed from the traceless stress-energy tensor (see Appendix A.2.3) and do not lead to additional linearly independent Ward identities.

[23] For details on anomalies interesting references are *e.g.* [74, 75].

under a conformal transformation $\mathcal{O}(x) \mapsto \mathcal{O}'(x')$ (the field content transforms equivalently). For example, invariance under translations implies that

$$\langle \mathcal{O}(x_1+a)\cdots\mathcal{O}(x_N+a)\rangle = \langle \mathcal{O}(x_1)\cdots\mathcal{O}(x_N)\rangle = f(x_1-x_2, x_1-x_3,\ldots,x_{N-1}-x_N)\,, \tag{1.37}$$

implying that the correlation function only depends on the distance between insertions. The consequence of Lorentz invariance is that correlation functions only depend on the distance between insertions. Hence, full Poincaré invariance implies that

$$\langle \mathcal{O}_1(x_1)\cdots\mathcal{O}_N(x_N)\rangle = f(|x_1-x_2|,|x_1-x_3|,\ldots,|x_{N-1}-x_N|)\,. \tag{1.38}$$

Even though Poincaré invariance already constrains correlation functions to some extent, the additional invariance under dilatations and SCTs in fact completely fixes the form of all two- and three-point functions (between primaries).[24]

Scalar two-point functions are completely constrained by invariance under dilatations and SCTs to be of the form

$$\langle \mathcal{O}_1(x_1)\mathcal{O}_2(x_2)\rangle = \begin{cases} C_{\mathcal{O}_1}\,|x_1-x_2|^{-2\Delta} = |x_1-x_2|^{-2\Delta}\,, & \text{if } \Delta_1=\Delta_2=\Delta \\ 0 & \text{otherwise} \end{cases}\,. \tag{1.39}$$

We remark that conformal invariance alone does not directly imply the normalization $C_{\mathcal{O}_1}$ of the two-point function to be 1. However, due to the orthogonality, it is always possible to pick an appropriate basis of (scalar) operators such that all two-point functions are properly normalized with $C_{\mathcal{O}_i}=1$.[25]
In the same way the implications of conformal invariance almost completely specify scalar three-point functions,

$$\langle \mathcal{O}_1(x_1)\mathcal{O}_2(x_2)\mathcal{O}_3(x_3)\rangle = \frac{C_{\mathcal{O}_1\mathcal{O}_2\mathcal{O}_3}}{x_{12}^{\Delta_{12}^{(3)}}\,x_{23}^{\Delta_{23}^{(3)}}\,x_{31}^{\Delta_{31}^{(3)}}}\,, \tag{1.40}$$

$$\text{with}\qquad x_{ij}=|x_i-x_j|\,,\qquad \Delta_{12}^{(3)}=\Delta_1+\Delta_2-\Delta_3\,,\qquad \Delta_{23}^{(3)}=\Delta_2+\Delta_3-\Delta_1\,,\qquad \Delta_{31}^{(3)}=\Delta_3+\Delta_1-\Delta_2\,.$$

The three-point coefficients $C_{\mathcal{O}_i\mathcal{O}_j\mathcal{O}_k}$ — since field redefinitions have already been exhausted to normalize all two-point functions — are physical quantities and cannot be normalized away. They are a defining feature of a CFT and together with the scaling dimensions of the operators build the conformal data. The conformal data define a CFT completely via the OPE, as we will discuss later.
Four-point scalar correlation functions are no longer completely fixed by conformal invariance: Once there are four spacetime points available it is possible to construct so-called cross-ratios,

$$u=\frac{x_{12}x_{34}}{x_{31}x_{42}}\,,\qquad\qquad v=\frac{x_{12}x_{34}}{x_{23}x_{41}}\,, \tag{1.41}$$

which are invariant under all conformal transformations. As a consequence, four-point functions are only fixed up to a arbitrary function of $u$ and $v$,

$$\langle \mathcal{O}_1(x_1)\mathcal{O}_2(x_2)\mathcal{O}_3(x_3)\mathcal{O}_4(x_4)\rangle = \frac{f(u,v)}{x_{12}^{\Delta_{12}^{(4)}} x_{23}^{\Delta_{23}^{(4)}} x_{34}^{\Delta_{34}^{(4)}} x_{41}^{\Delta_{41}^{(4)}} x_{31}^{\Delta_{31}^{(4)}} x_{42}^{\Delta_{42}^{(4)}}}, \qquad \Delta_{ab}^{(4)} = \Delta_a + \Delta_b - \sum_i \frac{\Delta_i}{3}. \tag{1.42}$$

[24] Under SCTs the distance between two spacetime points only changes by an overall scale factor.

[25] This simply amounts to a redefinition $\mathcal{O}_i \to \mathcal{O}_i/\sqrt{C_{\mathcal{O}_i}}$.

For $N$-point correlation functions there are $N(N-3)/2$ cross-ratios that can appear.

Apart from scalar correlation functions, two-point functions of operators with non-zero spin are again completely fixed by conformal invariance. They are only non-zero if operators have identical scaling dimensions and spin.[26]
The two-point correlation function of spin-1/2 primary operators takes the form[27]

$$\langle \mathcal{O}^{\alpha}(x_1)\bar{\mathcal{O}}^{\beta}(x_2)\rangle = C_{\mathcal{O}} \frac{x_{12}^{\mu}\gamma_{\mu}^{\alpha\beta}}{x_{12}^{2\Delta+1}}, \tag{1.43}$$

where $C_{\mathcal{O}}$ is a normalization that generically depends on the dimension $D$ of spacetime and the choice of gamma matrices $\gamma_\mu$. Again, the normalization of the operators plus the normalization of the gamma matrices can be chosen such that the two-point function of spinors is properly normalized [66, 67, 77, 78] with $C_{\mathcal{O}} = 1$.
The two-point function of spin-1 primary operators — *i.e.* vector fields — with scaling dimension $\Delta$ is given by

$$\langle \mathcal{O}_\mu(x_1)\mathcal{O}_\nu(x_2)\rangle = C_{\mathcal{O}} \frac{I_{\mu\nu}(x_{12})}{x_{12}^{2\Delta}}, \qquad I_{\mu\nu}(x) = \eta_{\mu\nu} - 2\frac{x_\mu x_\nu}{x^2}. \tag{1.44}$$

The normalization $C_{\mathcal{O}}$ may be relevant, *e.g.* for conserved currents $\mathcal{O}_\mu = j_\mu$. The orthogonal matrix $I_{\mu\nu}(x)$ appearing in Eq. (1.44) is called the inversion tensor and also appears under inversions of spacetime.[28]
For higher-spin primaries — interestingly — no new conformally covariant tensors appear. The fundamental building block $I_{\mu\nu}(x)$ can be used to construct all higher-spin two-point functions. In particular, for spin-$\ell$ traceless symmetric tensors the two-point correlation function reads

$$\langle \mathcal{O}_{\mu_1\ldots\mu_\ell}(x_1)\mathcal{O}_{\nu_1\ldots\nu_\ell}(x_2)\rangle = C_{\mathcal{O}} \frac{\left(I_{\mu_1\nu_1}(x)\cdots I_{\mu_\ell\nu_\ell}(x) + \text{permutations} - \text{traces}\right)}{x_{12}^{2\Delta}}. \tag{1.45}$$

We need to subtract traces so that the result is separately traceless in the $\mu$ and $\nu$ indices (not necessarily under corrections). The normalization $C_{\mathcal{O}}$ often times can be removed, but for some particular operators there might be a non-trivial normalization appearing. For example, the normalization

[26] Lorentz invariance by itself already enforces the total spin to be vanishing.

[27] Once operators exhibit spin the embedding formalism provides a more practical and transparent way to derive the general form of the correlators [23, 67, 68, 76]. In the embedding space $\mathbb{R}^{D+2}$ operators with spin can be naturally defined via their representation under the Lorentz group due to the isomorphism in Eq. (1.19). For example, we can define a vector field under the Lorentz group in $\mathbb{R}^{D+2}$ that is projected down to a vector field under the conformal group in $\mathbb{R}^D$, simplifying the derivation of correlators via the same projection [23, 66].
Spinors in $\mathbb{R}^D$ can also be dealt with via spinors in the embedding space, but there are some subtleties arising from the Clifford algebra in different dimensions and the precise choice of gamma matrices [66, 67, 77, 78].

[28] Under an inversion $x^\mu \overset{\mathcal{I}}{\mapsto} x'^\mu$ we have $\partial x'_\mu/\partial x^\nu = \Omega(x) I_{\mu\nu}(x)$.

of the stress-energy tensor is fixed by the Ward identities and $C_T$ is physically relevant.

Three-point functions of operators with spin are fixed up to a finite number of coefficients. For example, correlation functions of the type scalar–scalar–(spin-$\ell$) are fixed up to a single constant,

$$\langle \mathcal{O}_1(x_1)\mathcal{O}_2(x_2)\mathcal{O}_{\mu_1\ldots\mu_\ell}(x_3)\rangle = \frac{C_{\phi_1\phi_2\mathcal{O}}\,(Z^{\mu_1}\cdots Z^{\mu_\ell} - \text{traces})}{x_{12}^{\Delta_{12}^{(3)}}\,x_{23}^{\Delta_{23}^{(3)}}\,x_{31}^{\Delta_{31}^{(3)}}}\,, \qquad Z^\mu = \frac{x_{13}^\mu}{x_{13}^2} - \frac{x_{23}^\mu}{x_{23}^2}\,. \tag{1.46}$$

This formula also applies for the case where $\mathcal{O}_{\mu\nu} = T_{\mu\nu}$ the stress-energy tensor. In that case the three-point coefficient is fixed by a Ward identity [79],

$$C_{\mathcal{O}_1\mathcal{O}_2 T} = -\frac{D\Delta_1}{D-1}\frac{\delta_{12}}{\Omega_D}\,, \tag{1.47}$$

where $\Omega_D$ is the volume of the unit sphere. A similar statement is true for any three-point correlator including a conserved current $\mathcal{O}_\mu = j_\mu$.
In general, there might be several spin structures consistent with conformal invariance. For three-point correlation functions consisting of multiple operators with spin there can be several linearly independent structures consistent with conformal invariance. For every linearly independent tensor structure there is a three-point coefficient, schematically we can include additional spin indices and write

$$\langle \mathcal{O}_i^a(x_1)\mathcal{O}_j^b(x_2)\mathcal{O}_k^c(x_3)\rangle = \sum_{\substack{n:\,\text{tensor}\\ \text{structures}}} O^{(n)\,abc}(x_1,x_2,x_3)\,\frac{C^{(n)}_{\mathcal{O}_i\mathcal{O}_j\mathcal{O}_k}}{x_{12}^{\Delta_{ij}^{(3)}}\,x_{23}^{\Delta_{jk}^{(3)}}\,x_{ki}^{\Delta_{ki}^{(3)}}}\,, \tag{1.48}$$

where the indices $a$, $b$, $c$ are (arbitrary) spin indices and $O^{(n)\,abc}(x_1,x_2,x_3)$ denote the different tensor structures that can appear.

### 1.1.4 Radial quantization

In the path-integral formulation of QFT symmetries are expressed in terms of conserved charges $Q_\alpha(\Sigma)$ that live on codimension-1 hypersurfaces. In the operator formalism these symmetries are implemented in terms of local (differential) operators. The fact that symmetries act locally is consistent since the charges $Q_\alpha(\Sigma)$ are topological operators that can be deformed to live in an arbitrary small neighbourhood of the operator insertions they act on (*i.e.* on a sphere $\partial B(x)$ around the insertion point $x$ with arbitrary small radius).
For any given QFT the spacetime symmetries also dictate the appropriate choices of foliation of spacetime and hence the quantization scheme. Relating hypersurfaces by a symmetry transformation guarantees that the associated Hilbert spaces are isomorphic. In a CFT the presence of scale invariance allows for a particularly convenient choice of quantization called radial quantization. The exponential of the dilatation operator $D$ — $(r/r_0)^{-D}$ with $r_0$ being the radius of some reference sphere $S^{D-1}_{r_0}$ (usually $r_0 = 1$) — in CFTs on $\mathbb{R}^D$ maps spheres $S^{D-1}_r \subset \mathbb{R}^D$ of different radii $r$ into each other. It is therefore natural to foliate spacetime into spherical hypersurfaces.

In the radial quantization picture states live on spheres $S_r^{D-1}$ and are evolved from smaller spheres $S_r^{D-1}$ to larger spheres $S_{r'>r}^{D-1}$ using the dilatation operator $D$. Each sphere has an associated Hilbert space and in order to act with a symmetry generator one inserts spherical surface operators $Q(S_{r+\epsilon}^{D-1})$ into the path integral. It is most convenient to work in polar coordinates $(r, \mathbf{n}_\Omega)$,

$$\mathrm{d}s^2_{\mathbb{R}^D} = \mathrm{d}r^2 + r^2 \mathrm{d}\Omega^2_{D-1} \,. \tag{1.49}$$

Correlation functions are interpreted as radially ordered products,

$$\begin{aligned} \langle \mathcal{O}_1(x_1)\cdots\mathcal{O}_N(x_N)\rangle &= \langle 0|\, R\{\mathcal{O}_1(x_1)\cdots\mathcal{O}_N(x_N)\}\,|0\rangle \\ &= \theta(r_1-r_2)\theta(r_2-r_3)\cdots\theta(r_{N-1}-r_N)\, \langle 0|\mathcal{O}_1(x_1)\cdots\mathcal{O}_N(x_N)\,|0\rangle \;+\text{permutations}\,. \end{aligned} \tag{1.50}$$

Initial and final states in radial quantization are vacuum states $|0\rangle, \langle 0|$ living at $r=0$ and $r=\infty$, respectively. The radial ordering prescription $R\{\ldots\}$ is consistent as operators at the same radius and different angles commute.[29]
Naturally, radial quantization can be performed around different points of $\mathbb{R}^D$. The same correlator $\langle \mathcal{O}_1(x_1)\cdots\mathcal{O}_N(x_N)\rangle$ will get a different ordering in radial quantization schemes around different points, but the overall outcomes are isomorphic to each other.[30]

The unitary evolution operator $U$ in radial quantization is the exponential of the dilatation operator. It is most conveniently written using the "radial time" coordinate $\tau = r_0 \log(r/r_0)$,

$$U = \left(\frac{r}{r_0}\right)^{-D} = e^{-\frac{\tau}{r_0} D} \,. \tag{1.51}$$

The conformal transformation $\tau = r_0 \log(r/r_0)$[31] relates flat space $\mathbb{R}^D$ to the cylinder $\mathbb{R}_\tau \times S_{r_0}^{D-1}$ via an additional local Weyl rescaling of the metric in the cylinder geometry by $e^{2\frac{\tau}{r_0}}$,

$$\mathrm{d}s^2_{\mathbb{R}^D} = \mathrm{d}r^2 + r^2\mathrm{d}\Omega^2_{D-1} \underbrace{=}_{r/r_0 = e^{\frac{\tau}{r_0}}} e^{\frac{2\tau}{r_0}} \overbrace{\left(\mathrm{d}\tau^2 + r_0^2 \mathrm{d}\Omega^2_{D-1}\right)}^{\mathrm{d}s^2_{\mathbb{R}_\tau\times S^{D-1}_{r_0}}} \underbrace{=}_{\mathrm{d}s'^2 = e^{\frac{2\tau}{r_0}}\mathrm{d}s^2} \mathrm{d}s'^2_{\mathbb{R}_\tau\times S^{D-1}_{r_0}} \,. \tag{1.52}$$

At this point it is important to note that not all local Weyl rescalings of the metric correspond to conformal transformations. While so-called Weyl invariance implies conformal invariance, the converse is not necessarily true without additional assumptions [80]. However, all unitary CFTs are believed to be Weyl invariant (up to the Weyl anomaly) [81]. Under a Weyl rescaling $g_{\mu\nu} \mapsto g'_{\mu\nu} = \Omega^2(x) g_{\mu\nu}$ correlation functions of local operators in a CFT transform as[32]

$$\langle \mathcal{O}_1^{(g)}(x_1)\cdots\mathcal{O}_N^{(g)}(x_N)\rangle_g = \left(\prod_i \Omega^{\Delta_i}(x_i)\right) \langle \mathcal{O}_1^{(\Omega^2 g)}(x_1)\cdots\mathcal{O}_N^{(\Omega^2 g)}(x_N)\rangle_{\Omega^2 g} \,. \tag{1.53}$$

[29] This is analogous to how space-like operators commute in Minkowski space.

[30] This is analogous to changing the frame of reference in a Lorentz invariant theory in Minkowski space.

[31] The transformation $\tau = r_0 \log(r/r_0)$ satisfies $\delta r/\delta\tau = r/r_0 = e^{\tau/r_0}$. Importantly, it does not amount to a pure dilatation with constant $\lambda$ as it has a spacetime dependency.

[32] In even dimensions the partition function $Z^{(g)}$ in the geometry $g$, as defined in Eq. (1.34), can transform with a Weyl anomaly, $Z^{(g)} = Z^{(\Omega^2 g)}\, e^{S_{\text{Weyl}}[g]}$. However, this does not affect correlation functions of local operators as they are properly normalized by a factor of $1/Z^{(g)}$ and the Weyl anomaly cancels out.

As a consequence, if the scale factors between the two geometries are carefully tracked, radial quantization in flat space is equivalent to equal-time quantization in $\tau$ on the cylinder.[33]
In this so-called cylinder interpretation states live on spheres $S_{r_0}^{D-1}$ and scale transformations $r \mapsto \lambda r$ become shifts in radial time $\tau \mapsto \tau + \log \lambda$. The Hamiltonian $H^{(\mathrm{cyl})}$ on the cylinder is given by the dilatation operator in flat space,

$$H^{(\mathrm{cyl})} = \frac{D}{r_0}, \tag{1.54}$$

and time evolution is generated by $U = e^{\tau H^{(\mathrm{cyl})}}$. While radial quantization relies only on scale invariance as an assumption, the cylinder interpretation relies on full conformal invariance of the theory as the non-trivial local Weyl rescaling cannot be compensated otherwise.
The statement that a CFT on the cylinder is equal to the same theory in flat space modulo scale factors is non-trivial. Consider the Ising model. Clearly, the Ising model on a cylinder is not equivalent to the Ising model in flat space. However, at the critical point where the Ising model becomes conformal and the stress-energy tensor traceless, the theory becomes insensitive to Weyl rescalings and the theories on different geometries become related.
The cylinder geometry is of particular interest because of the fact that the dilatation operator in flat space becomes the Hamiltonian on the cylinder. Because of this fact the energy of a state on the cylinder is the scaling dimension of the associated operator/state in flat space,[34]

$$E^{(\mathrm{cyl})} = \frac{\Delta}{r_0}. \tag{1.55}$$

The Weyl factor for the flat space to cylinder map appearing in Eq. (1.53) is $\Omega^2 = e^{2\frac{\tau}{r_0}}$. Given an operator $\mathcal{O}(x) = \mathcal{O}^{(\mathrm{flat})}(x)$ it is therefore natural to define the associated cylinder operator as

$$\mathcal{O}^{(\mathrm{cyl})}(\tau, \mathbf{n}) = e^{\frac{\tau}{r_0}\Delta} \mathcal{O}^{(\mathrm{flat})}(x), \qquad x = r_0 e^{\tau/r_0}\mathbf{n}\,,\ \mathbf{n} \in S_1^{D-1}. \tag{1.56}$$

The cylinder field is not an artificial construct, it is the same field as its flat counterpart but correlation functions are simply measured in a different geometry.

The cylinder is not the only geometry that is related to flat space via a Weyl rescaling. The procedure outlined around Eq. (1.53) can be applied to other symmetries that are conformally flat. For example, the stereographic projection gives a Weyl mapping between the $D$-sphere $S_{r_0}^{D}$ and flat space. Consider the transformation $r = r_0 \sin\varphi/(1-\cos\varphi)$

$$\mathrm{d}s^2_{\mathbb{R}^D} = \mathrm{d}r^2 + r^2 \mathrm{d}\Omega^2_{D-1} \overset{r=r_0\frac{\sin\varphi}{1-\cos\varphi}}{=} \frac{r_0^2}{(1-\cos\varphi)^2} \underbrace{\left(\mathrm{d}\varphi^2 + \sin^2\varphi\, \mathrm{d}\Omega^2_{D-1}\right)}_{=\mathrm{d}\Omega^2_D}. \tag{1.57}$$

This geometry is of particular interest in the study of classical conformal transformations as they are always non-singular in these coordinates. In the quantum theory the spherical geometry is less convenient as the associated choices of foliation are cumbersome. For example, if one chooses to foliate along $\varphi$, then the hypersurfaces $\Sigma(\varphi)$ are spheres with a $\varphi$-dependent radius, $\Sigma(\varphi) = S^{D-1}_{\sin\varphi}$. In other words, the generator of time translation is not a symmetry of the system. The same holds true for any other choice of time coordinate on the sphere.
Another geometry of interest is the so-called $N$–$S$ quantization scheme [23, 24], which is closely related to the cylinder and is very important in the study of Minkowskian CFTs [24].

[33] For the map from flat space to the cylinder in Eq. (1.52) correlation functions are related by $\langle \mathcal{O}_1^{(\mathrm{flat})}(x_1)\cdots\mathcal{O}_N^{(\mathrm{flat})}(x_N)\rangle_{(\mathrm{flat})} = \left(\prod_i e^{-\frac{\tau}{r_0}\Delta_i}\right)\langle \mathcal{O}_1^{(\mathrm{cyl})}(x_1)\cdots\mathcal{O}_N^{(\mathrm{cyl})}(x_N)\rangle_{(\mathrm{cyl})}$.

[34] The state-operator correspondence has important consequences here as it guarantees that states on the cylinder are in a 1-to-1 correspondence with operators inserted in flat space.

### 1.1.5 State ⇔ operator correspondence

Once we understand how local (primary) operators behave under the action of the conformal generators and how radial quantization — including the cylinder interpretation — is set up using conformal invariance, we are able to prepare states in radial quantization. For any QFT it is possible to define states via insertions of local operators in the path integral and that is no different for CFTs. Interestingly, under the assumption of conformal invariance the converse statement — that local operators can be defined via states in radial quantization — holds true, giving rise to the concept of the state–operator correspondence in CFT:

> ”For a CFT states in radial quantization are in a 1-to-1 correspondence with local operators.”

In radial quantization a state $|\psi\rangle$ in the Hilbert space on some sphere $S_r^{D-1}$ can be prepared by inserting operators in the interior $B_r(0) =: B_r$ of the sphere $S_r^{D-1}$ and performing the path integral over $B_r$. Adding no operator insertions inside $B_r$ and performing the path integral over $B_r$ produces the vacuum state $|0\rangle$ on $\partial B_r = S_r^{D-1}$. The vacuum state $|0\rangle$ is invariant under all symmetries as any topological operator $Q(S_r^{D-1})$ can be shrunk all the way to the point $x = 0$ inside of $B_r$ without crossing any operators,

$|0\rangle =$ 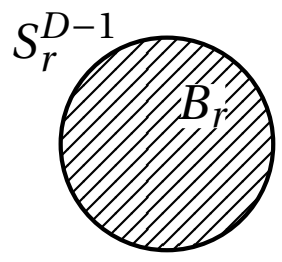

, 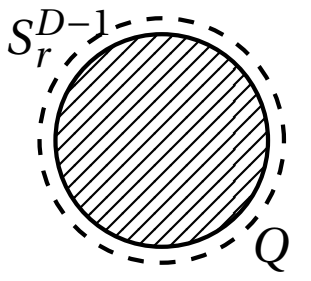

$=$ 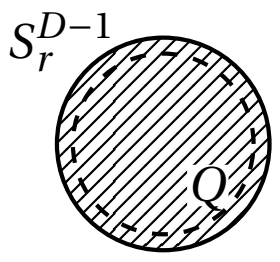

$= 0\,.$ (1.58)

Consider a CFT described by a Lagrangian $\mathscr{L}$ and a set of fields $\{\phi\}$. The Hilbert space on any given spherical hypersurface is spanned by field eigenstates $|\phi_r\rangle$, where $\phi_r(\mathbf{n})$ is a field configuration on $r\mathbf{n} \in S_r^{D-1}$. The state $|\phi_r\rangle$ and the configuration $\phi_r(\mathbf{n})$ are only defined on $S_r^{D-1}$ and not in the interior. A general state on $S_r^{D-1}$ is a linear combination of field eigenstates and can be represented by a path integral,

$$|\psi\rangle = \int \mathscr{D}\phi_r\, |\phi_r\rangle \langle\phi_r|\psi\rangle\,. \tag{1.59}$$

The coefficient $\langle\phi_r|0\rangle$ is given by the path integral over the interior $B_r$ with boundary condition $\phi_r(\mathbf{n})$,

$$\langle\phi_r|0\rangle = \int_{\substack{\phi(r\mathbf{n})=\phi_r(\mathbf{n})\\ \rho\leq r}} \mathscr{D}\phi(\rho\mathbf{n})\ e^{-S[\phi]}\,. \tag{1.60}$$

The insertion of an operator inside $x \in B_r$ defines a state $|\mathcal{O}(x)\rangle = \hat{\mathcal{O}}(x)\,|0\rangle$ on $S_r^{D-1}$, and its overlap with $|\phi_r\rangle$ is given by the path integral with an insertion of $\mathcal{O}(x)$ at $x \in B_r$,

$$|\mathcal{O}(x)\rangle = \mathcal{O}(x)\,|0\rangle =$$

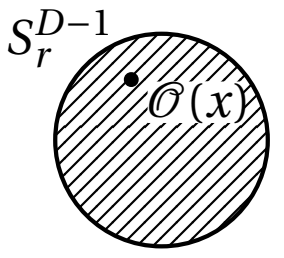

$$, \qquad \langle\phi_r|\mathcal{O}(x)\rangle = \langle\phi_r|\,\mathcal{O}(x)\,|0\rangle = \int\limits_{\substack{\phi(r\mathbf{n})=\phi_r(\mathbf{n})\\ \rho\le r}} \mathcal{D}\phi(\rho\mathbf{n})\,\mathcal{O}(x)\,e^{-S[\phi]}\,. \tag{1.61}$$

For every configuration of operator insertions inside $B_r$ there exists an associated state on the sphere $S_r^{D-1}$. These states living on the boundary $S_r^{D-1}$ can be prepared by simply inserting the associated set of operators in the path integral.

The construction of states from local operator insertions is consistent for any QFT. However, in a CFT the construction of states from local operator insertions works backwards as well, in the sense that we can construct local operators from states living on spheres with non-zero radius. Consider an eigenstate of the dilatation operator $|\mathcal{O}_i\rangle$. It satisfies

$$D\,|\mathcal{O}_i\rangle = \Delta_i\,|\mathcal{O}_i\rangle\,. \tag{1.62}$$

The states $|\mathcal{O}_i\rangle$ generate operators. To see this we cut spherical holes $B_{\rho_i}(x_i)$ with radii $\rho_i$ centred around $x_i$ out of the path integral.[35] The states $|\mathcal{O}_i\rangle$ are glued at the boundaries of the holes $B_{\rho_i}(x_i)$, respectively. The resulting quantity can be shown to behave exactly like a correlator of local operators,

$$\langle\mathcal{O}_1(x_1)\cdots\mathcal{O}_N(x_N)\rangle = \int\prod_{i=1}^{N}\mathcal{D}\phi_{ri}\,\langle\phi_{ri}|\mathcal{O}_i\rangle \int\limits_{\substack{\phi|_{\partial B_{\rho_i}}=\phi_{ri}\\ x\neq B_{\rho_i}(x_i)}} \mathcal{D}\phi(x)\;e^{-S[\phi]}\,. \tag{1.63}$$

The path integral $\mathrm{D}\phi$ is performed over the region outside of the holes $B_{\rho_i}(x_i)$ and the path integrals $\mathcal{D}\phi_{ri}$ are performed over all possible field configurations on the boundaries $\partial B_{\rho_i}(x_i)$. The expression $\phi|_{\partial B_{\rho_i}}$ denotes the restriction of the bulk field $\phi(x)$ to the $i$-th boundary $\partial B_{\rho_i}(x_i)$.
It seems crucial that the balls $B_{\rho_i}(x_i)$ do not overlap in this construction, a condition that is seemingly making this construction non-local. However, if the balls $B_{\rho_i}(x_i)$ do overlap we can apply a scale transformation under which the quantity in Eq. (1.63) satisfies

$$\langle\mathcal{O}_1(x_1)\cdots\mathcal{O}_N(x_N)\rangle = \lambda^{\sum_i\Delta_i}\,\langle\mathcal{O}_1(\lambda x_1)\cdots\mathcal{O}_N(\lambda x_N)\rangle\,. \tag{1.64}$$

For sufficiently large $\lambda$ the holes $B_{\rho_i}(x_i)$ do no longer overlap and Eq. (1.63) is well defined for any choice of radii $\rho_i$. This means that the insertion points $x_i$ can be arbitrarily close to each other for an arbitrary choice of $\rho_i$ and Eq. (1.63) defines a correlator of local quantities/operators.

The construction in Eq. (1.63) can be slightly modified and then used to define a local operator from any state $|\mathcal{O}\rangle$ on $S^{D-1}_{r>0}$ via its correlation functions with all the other operators in the theory,

$$\begin{aligned}\langle \mathcal{O}_1(x_1)\cdots\mathcal{O}_N(x_N)\mathcal{O}(0)\rangle &= \langle 0|\, R\{\mathcal{O}_1(x_1)\cdots\mathcal{O}_N(x_N)\mathcal{O}(0)\}\,|0\rangle = \langle 0|\, R\{\mathcal{O}_1(x_1)\cdots\mathcal{O}_N(x_N)\}\,|\mathcal{O}\rangle \\ &= \int \mathcal{D}\phi_r\, \langle\phi_r|\mathcal{O}\rangle \int_{\substack{\phi|_{\partial B_r}=\phi_r \\ x\neq B_r(0)}} \mathcal{D}\phi(x)\, \mathcal{O}_1(x_1)\cdots\mathcal{O}_N(x_N)\, e^{-S[\phi]}\,. \end{aligned} \tag{1.65}$$

[35] The radii $\rho_i$ should in principle be chosen such that the balls $B_{\rho_i}(x_i)$ do not overlap.

The initial state $|\mathcal{O}\rangle$ lives on the sphere $S^{D-1}_r$ and, in principle, any radius $r$ such that $r < r_i = |x_i|$, $\forall i$ can be chosen to define the correlator in the path-integral formalism.[36] The construction in Eq. (1.65) is well defined for $x_i \neq 0$, $\forall i$ and completely defines the associated local operator $\mathcal{O}(0)$ via its correlation functions with other operators.

The ability to construct unique local operators from states in radial quantization allows us define local operators in a CFT consistently as eigenstates of the dilatation operator $D$ in radial quantization.[37] Under this definition the constructions in Eq. (1.61) and Eq. (1.63) become manifestly inverse with respect to each other. A primary operator creates a primary state that transforms in a finite-dimensional representation of the rotation group plus scale transformations and gets annihilated by $K_\mu$, and vice versa,

$$\left.\begin{array}{lcl} \mathcal{O}_i(0) & \longleftrightarrow & \mathcal{O}_i(0)\,|0\rangle = |\mathcal{O}_i(0)\rangle \\ {[K_\mu,\mathcal{O}_i(0)]} = 0 & \longleftrightarrow & K_\mu\,|\mathcal{O}_i(0)\rangle = 0 \\ {[D,\mathcal{O}_i(0)]} = \Delta_i\mathcal{O}_i(0) & \longleftrightarrow & D_\mu\,|\mathcal{O}_i(0)\rangle = \Delta_i\,|\mathcal{O}_i(0)\rangle \\ {[M_{\mu\nu},\mathcal{O}_i(0)]} = iS_{\mu\nu}\mathcal{O}_i(0) & \longleftrightarrow & M_{\mu\nu}\,|\mathcal{O}_i(0)\rangle = iS_{\mu\nu}\,|\mathcal{O}_i(0)\rangle \end{array}\right. ,\qquad \forall i\,. \tag{1.66}$$

The action of the conformal group on states in radial quantization follows from the invariance of the vacuum state $|0\rangle$ under symmetry transformations. A conformal multiplet is generated by repeated action of the momentum generators $P_\mu \sim \partial_\mu$

$$\mathcal{O}(0)\,,\ \underbrace{\partial_\mu\mathcal{O}(0)\,,\ \partial_\nu\partial_\mu\mathcal{O}(0)\,,\ \ldots}_{\text{descendants}}\,, \qquad \longleftrightarrow \qquad |\mathcal{O}\rangle\,,\ \underbrace{P_\mu\,|\mathcal{O}\rangle\,,\ P_\nu P_\mu\,|\mathcal{O}\rangle\,,\ \ldots}_{\text{descendant states}}\,. \tag{1.67}$$

The conformal algebra acts exactly the same way on states $|\mathcal{O}(x)\rangle$ as on local operators $\mathcal{O}(x)$. In fact, the action of symmetry generators on local operators in any Euclidean QFT can be considered by surrounding operators with charges supported in spheres, as discussed, a procedure which in a CFT is essentially equivalent to radial quantization.

The fact that in radial quantization states are in a 1-to-1 correspondence with local operators can be understood from a geometric point of view. In an arbitrary foliation of spacetime of the form $\mathbb{R}_t \times \Sigma$ initial states live on a surface $\Sigma$ at time $t = -\infty$, while in radial quantization they live at the origin $r = 0$, *i.e.* a single point,

$$\begin{array}{lcl} \text{quantization } \mathbb{R}_t \times \Sigma & & \text{radial quantization} \\ |\mathcal{O}\rangle_{\mathbf{x}} = \mathcal{O}(-\infty,\mathbf{x})\,|0\rangle\,, \quad \mathbf{x}\in\Sigma\,, & \longleftrightarrow & |\mathcal{O}\rangle = \mathcal{O}(0)\,|0\rangle\,. \end{array} \tag{1.68}$$

[36] In practice, the same rescaling argument as in Eq. (1.64) holds such that all choices of $0 < r < r_i$ are equivalent.

[37] In unitary CFTs the dilatation operator $D$ is always diagonalizable.

This is to say that in radial quantization the entire Hilbert space can be thought of as living at a single point. The state-operator correspondence essentially states that in a CFT operators acting locally in a small neighbourhood of the origin $r = 0$ create all states in the Hilbert space of the theory. The analogous picture in a foliation of the form $\mathbb{R}_t \times \Sigma$ breaks down as you can no longer think of local operators acting at a single point in order to get the full Hilbert space back. There is always a map from operators to states via their action on the vacuum, but only for CFTs does every state correspond uniquely to a single local operator inserted at $x = 0$.

### 1.1.6 Reflection positivity on the cylinder and unitarity bounds

In radial quantization on the cylinder in and out vacuum states — which in flat space live at $r = 0, \infty$, respectively — live at cylinder time $\tau = \pm\infty$, respectively, and are related to each other via a time reflection $\tau \mapsto -\tau$. This property of the vacuum generalizes to all operators of the theory. In fact, in the cylinder interpretation for local operators on $\mathbb{R}_\tau \times S^{D_1}_{r_0}$ Hermitian conjugation is equivalent to cylinder time reflection $\tau \mapsto -\tau$.
Reflection positivity for a scalar operator in a CFT on the cylinder results in the condition

$$\mathcal{O}^{(\mathrm{cyl})\dagger}(\tau, \mathbf{n}) = \mathcal{O}^{(\mathrm{cyl})}(-\tau, \mathbf{n})\,. \tag{1.69}$$

It is important to emphasize that Hermitian conjugation in radial quantization is different to Hermitian conjugation in the usual quantization scheme with time direction $\tau = x_0$. After a Weyl rescaling, in flat space Hermitian conjugation is equivalent to an inversion of spacetime,

$$\mathcal{O}^{(\mathrm{flat})\dagger}(x) = \left(\frac{1}{x^2}\right)^{-\Delta} \mathcal{O}^{(\mathrm{flat})}\left(\frac{x}{\mathbf{x}^2}\right). \tag{1.70}$$

Conjugation properties of operators directly carry over to states in radial quantization. A (scalar) primary state $|\mathcal{O}\rangle = \mathcal{O}(0)\,|0\rangle$ behaves under Hermitian conjugation as

$$(|\mathcal{O}\rangle)^\dagger = \langle 0|\mathcal{O}(0)^\dagger = \lim_{y\to\infty} y^{2\Delta} \langle 0|\mathcal{O}(y)\,. \tag{1.71}$$

Eq. (1.70) can be generalized to fields with arbitrary spin. For example, the Hermitian conjugation of a flat space tensor operator inside the path integral in radial quantization is given by

$$\mathcal{O}^{(\mathrm{flat})\dagger}_{\mu_1\ldots\mu_N}(x) = I_{\mu_1}{}^{\nu_1}\cdots I_{\mu_N}{}^{\nu_N}\mathcal{O}^{(\mathrm{flat})}_{\nu_1\ldots\nu_N}\left(\frac{x}{\mathbf{x}^2}\right), \qquad I_\mu{}^\nu = \delta_\mu{}^\nu - 2\frac{x_\mu x^\nu}{\mathbf{x}^2}\,. \tag{1.72}$$

Applying Eq. (1.72) to the stress-energy tensor allows us to deduce the transformation property of the charge operators,

$$Q^{(\epsilon)\dagger} = -Q^{(I\epsilon I)}\,, \qquad \longleftrightarrow \qquad \begin{array}{ll} Q^{(M)\dagger}_{\mu\nu} = -Q^{(M)}_{\mu\nu} & \left(M^\dagger_{\mu\nu} = -M_{\mu\nu}\right), \\ Q^{(D)\dagger} = Q^{(D)} & \left(D^\dagger = D\right), \\ Q^{(P)\dagger}_\mu = Q^{(K)}_\mu & \left(P^\dagger_\mu = K_\mu\right). \end{array} \tag{1.73}$$

We can make use of Hermitian conjugation to compute properties of correlation functions in a purely algebraic way. For example, consider a scalar two-point function,

$$\langle \mathcal{O}(x)\mathcal{O}(y)\rangle = x^{-2\Delta}\langle \mathcal{O}|\, e^{\frac{\mathbf{y}\cdot\mathbf{K}}{\mathbf{y}^2}}\, e^{\mathbf{x}\cdot\mathbf{P}}\,|\mathcal{O}\rangle\,, \tag{1.74}$$

where $\langle \mathcal{O}|$ is given by the expression in Eq. (1.71). By expanding the exponentials using the Baker–Campbell–Hausdorff (BCH) formula and the conformal algebra it is possible to compute the two-point correlator term-by-term and recover Eq. (1.39) in an expansion in small $|y|/|x|$.[38]

Hermitian conjugation on the cylinder leads to a natural inner product on states in radial quantization. This inner product can then be used to deduce bounds that must be satisfied by a unitary (or reflection-positive) theory by demanding that all states in a conformal multiplet have positive norm. For a scalar operator this results in the following so-called unitarity bounds when considering the first few descendant states in the conformal multiplet,

$$\begin{aligned} ||\mathcal{O}\rangle|^2 &= \langle \mathcal{O}|\mathcal{O}\rangle \geq 0\,, \qquad\qquad \text{(trivial)}\\ |P_0\,|\mathcal{O}\rangle|^2 &= \langle \mathcal{O}|\,[K_0,P_0]\,|\mathcal{O}\rangle = 2\Delta\,\langle \mathcal{O}|\mathcal{O}\rangle \geq 0\,,\\ \left|P_\mu P^\mu\,|\mathcal{O}\rangle\right|^2 &= 8D\Delta\left(\Delta - \frac{D-2}{2}\right)\langle \mathcal{O}|\mathcal{O}\rangle \geq 0\,. \end{aligned} \tag{1.75}$$

The condition $\Delta = 0$ is satisfied solely by the unit operator $\mathbb{1}$. The expression $\langle \mathcal{O}|\mathcal{O}\rangle$ is just the normalization of the two-point function and can be set to 1. As it turns out, higher descendant states do not impose stronger constraints on the scaling dimension $\Delta$ [72].
Similar unitarity bounds can be computed for spinning primary operators. The most notable unitarity bounds are [21, 72, 82]

$$\text{spin } 0:\ \Delta \geq \frac{D-2}{2}\,, \qquad \text{spin } \frac{1}{2}:\ \Delta \geq \frac{D-1}{2}\,, \qquad \text{spin } l \geq 1:\ \Delta \geq D + l - 2\,. \tag{1.76}$$

These inequalities are the best you can do for spinors and traceless symmetric tensors in a generic CFT. In theories with additional symmetries — like supersymmetry — unitarity bounds can be more interesting [82].
If the scaling dimension $\Delta$ saturates one of the unitarity bounds in Eq. (1.76) the associated conformal multiplet includes a null state $\prod_i P_{\mu_i}\,|\mathcal{O}\rangle$ (state with zero norm).[39] For the unity operator $\mathbb{1}$ with scaling dimension $\Delta = 0$ this statement is trivial. For a scalar operator saturating the unitarity bound the null state is $\hat{P}^2\,|\mathcal{O}\rangle$. This means that the operator $\mathcal{O}(x)$ satisfies the condition

$$\partial^2 \mathcal{O}(x) = 0\,. \tag{1.77}$$

Hence, $\mathcal{O}(x)$ satisfies the Klein-Gordon equation for a free massless scalar field and thus decouples from the rest of the CFT. For a traceless symmetric tensor operator of spin $l$ the null state is $P_\mu\,|\mathcal{O}^{\mu\mu_2\ldots\mu_l}\rangle$ [21], hence

$$\partial_\mu \mathcal{O}^{\mu\mu_2\ldots\mu_l}(x) = 0\,. \tag{1.78}$$

[38]There is an additional constant of normalization of the form $\langle \mathcal{O}|\mathcal{O}\rangle$ appearing which can be set to 1. This is just the normalization of the two-point function.

[39]This directly follows from the derivation of the unitarity bounds from the norms of descendant states.

Therefore, any operator $\mathcal{O}^{\mu\mu_2\cdots\mu_l}$ saturating the unitarity bound is a conserved current. Notable examples here are global symmetry currents with ($l = 1$, $\Delta = D - 1$) and the stress-energy tensor with $l = 2$, $\Delta = D$.

Finally, we note that with a positive-definite inner product it is possible to prove that in a unitary CFT all operators are linear combinations of primaries and descendants of said primaries. The only other additional assumption required is that the partition function $Z(\beta) = \mathrm{Tr}\exp[-\beta D]$ of the theory on $S^1_\beta \times S^{D-1}_{r_0}$ is finite. In that case the spectral theorem tells us that $\exp[-\beta D]$ is diagonalizable, hence $D$ is diagonalizable.[40] In addition, as $D$ is Hermitian, its eigenvalues are real.
Pick any operator $\mathcal{O}$ that is an eigenvector of $D$ with eigenvalue $\Delta$.[41] Since the partition function $Z(\beta)$ is finite there is only a finite number of primaries $\mathcal{O}_i$ with a smaller scaling dimension $\Delta_i \leq \Delta$. We can now subtract the projections of $\mathcal{O}$ onto the conformal multiplets of the primaries $\mathcal{O}_i$ using the inner product. The resulting operator we call $\mathcal{O}'$.
Assume that the operator $\mathcal{O}'$ after the subtraction of all projections is non-zero $\mathcal{O}' \neq 0$. By finiteness of the partition function, repeatedly acting on $\mathcal{O}'$ with the lowering operators $K_\mu$ must eventually yield zero, hence there exists another primary with scaling dimension smaller than $\Delta$ not part of the set $\{\mathcal{O}_i\}$. We have a contradiction, hence $\mathcal{O}' = 0$ and $\mathcal{O}$ is a linear combination of the primaries $\{\mathcal{O}_i\}$ and their descendants.

### 1.1.7 The operator product expansion

In CFTs — even with a Lagrangian description — it proves useful to study operators rather than fields. Any CFT contains a distinguished set of primary operators and the conformal multiplets they produce. In a unitary CFT these are all the operators that exist, *i.e.* every operator is a linear combination of primaries and their descendents. In essence, conformal invariance constrains theories so much that a CFT can be defined purely algebraically by its primary operators and their scaling dimensions $\{\Delta\}$ as well as the three-point coefficients $C^{(n)}_{\mathcal{O}_i\mathcal{O}_j\mathcal{O}_k}$ defined in Eq. (1.40), Eq. (1.46) and Eq. (1.48). To fully understand why this is true, knowledge of the OPE and its peculiarities in CFT is required.
As discussed, a state on a given sphere $S^{D-1}_r$ is generated by inserting an operator $\mathcal{O}(x)$ inside of the sphere in the ball $B_r$, *i.e.* the insertion point satisfies $|x| < r$. Instead of a single operator insertion consider the case where there are two operator insertions $\mathcal{O}_i(x)\mathcal{O}_j(0)$ in $B_r$. Suppose we consider scalar operators for simplicity, the resulting state on the sphere $S^{D-1}_r$ is

$$\mathcal{O}_i(x)\mathcal{O}_j(0)\left|0\right\rangle . \tag{1.79}$$

However — as any state in a unitary CFT can be written as a linear combination of primaries and their descendants — this state will have an expansion in terms dilatation eigenstates. Via the state-operator correspondence this implies that the insertion of two scalar operators can be written as an expansion of single (primary and descendant) operator insertions,

$$\mathcal{O}_i(x)\mathcal{O}_j(0)\left|0\right\rangle = \sum_k F^a_{\mathcal{O}_i\mathcal{O}_j\mathcal{O}_k}(x,P)\,\mathcal{O}^a_k(y)\big|_{y=0}\left|0\right\rangle . \tag{1.80}$$

[40] In principle, boundedness of the operator $\exp[-\beta D]$ suffices for $D$ to be diagonalizable.

[41] We can assume without loss of generality that $\mathcal{O}$ is an eigenstate of $D$. If not, it is a linear combination of eigenstates.

where $\mathcal{O}_k^a$ can have non-zero spin in general, even though we started with scalar operators. However, by consistency, the operator $\mathcal{O}_k^a$ must transform in a traceless symmetric tensor representation of $SO(D)$.[42] The index $k$ runs over all primaries and the operator $F^a_{\mathcal{O}_i\mathcal{O}_j\mathcal{O}_k}(x,P)$ packages together primaries and — by acting with $P_\mu$ — their descendants.
Eq. (1.80) is exact and can be used equivalently in the path-integral formalism, provided there exists a sphere such that all other operator insertion except for two lie outside of said sphere. In the path integral formalism, schematically Eq. (1.80) amounts to[43 44]

$$\mathcal{O}_i(x_1)\mathcal{O}_j(x_2) = \sum_k F^a_{\mathcal{O}_i\mathcal{O}_j\mathcal{O}_k}(x_{12},\partial_2)\mathcal{O}_k^a(x_2)\,, \qquad \longleftrightarrow$$

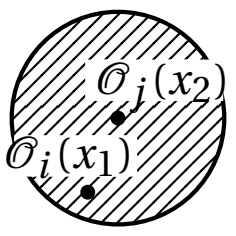

$$= \sum_k F^a_{\mathcal{O}_i\mathcal{O}_j\mathcal{O}_k}(x_{12},\partial_2)$$

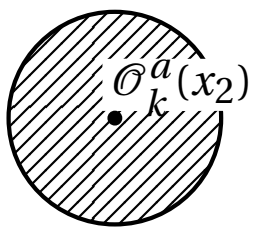

. (1.81)

Alternatively, radial quantization can be performed around a third point $x_3$, slightly modifying the OPE operators $F^a_{\mathcal{O}_i\mathcal{O}_j\mathcal{O}_k}$. And finally, operators may have non-zero spin and we need to include additional spin indices, so that the most general form of the OPE reads

$$\mathcal{O}_i^a(x_1)\mathcal{O}_j^b(x_2) = \sum_k F'^{abc}_{\mathcal{O}_i\mathcal{O}_j\mathcal{O}_k}(x_{13},x_{23},\partial_3)\mathcal{O}_k^c(x_3) \quad \overset{x_3\to x_2}{\longrightarrow} \quad \sum_k F^{abc}_{\mathcal{O}_i\mathcal{O}_j\mathcal{O}_k}(x_{12},\partial_2)\mathcal{O}_k^c(x_2)\,. \tag{1.82}$$

Conformal invariance strongly restricts the form of the OPE. For example, consider the scalar case in Eq. (1.81). Consider the OPE of two scalar operators $\mathcal{O}_i, \mathcal{O}_j$ producing a third scalar operator $\mathcal{O}_k$. Acting with both the dilatation operator $D$ and the generators of the rotation group $M_{\mu\nu}$ tells us that $F_{\mathcal{O}_i\mathcal{O}_j\mathcal{O}_k}$ has an expansion of the form

$$F_{\mathcal{O}_i\mathcal{O}_j\mathcal{O}_k}(x,\partial) \propto |x|^{\Delta_k-\Delta_i-\Delta_j}\left(1 + \#\mathbf{x}\cdot\partial + \#(\mathbf{x}\cdot\partial)^2 + \#\mathbf{x}^2\partial^2 + \ldots\right), \tag{1.83}$$

Eventually, the action of $K_\mu$ completely fixes $F_{\mathcal{O}_i\mathcal{O}_j\mathcal{O}_k}$ up to an overall constant. This procedure can of course be repeated equivalently for fields with spin.
Another way to see why the OPE is fixed by conformal invariance is by taking the correlation function with a third operator on both sides of Eq. (1.82). This relates a three-point function on the left-hand side to a sum of two-point functions on the right-hand side, both fixed by conformal invariance. For simplicity, consider the scalar case in Eq. (1.81) and a third scalar operator $\mathcal{O}_n(x_3)$ such that $x_{23} \geq x_{12}$ in order for the OPE to be valid,

$$\langle\mathcal{O}_n(x_3)\mathcal{O}_i(x_1)\mathcal{O}_j(x_2)\rangle = \sum_k F_{\mathcal{O}_i\mathcal{O}_j\mathcal{O}_k}(x_{12},\partial_2)\langle\mathcal{O}_n(x_3)\mathcal{O}_k(x_2)\rangle\,. \tag{1.84}$$

The three-point function on the left-hand side is given by Eq. (1.40). The two-point function on the right-hand side is also fixed. Assuming that the two-point functions are properly normalized, *i.e.* that we have picked and orthonormal basis of primary operators, it is given by Eq. (1.39). This yields the condition

$$\frac{C_{\mathcal{O}_n\mathcal{O}_i\mathcal{O}_j}}{x_{31}^{\Delta_n+\Delta_i-\Delta_j}\,x_{12}^{\Delta_i+\Delta_j-\Delta_n}\,x_{23}^{\Delta_j+\Delta_n-\Delta_i}} = F_{\mathcal{O}_i\mathcal{O}_j\mathcal{O}_n}(x_{12},\partial_2)\,x_{23}^{-2\Delta_n}\,. \tag{1.85}$$

[42] Tracelessness arises from restricting to irreducible representations of $SO(D)$.

[43] In Eq. (1.81) we replace the momentum operators $P_\mu$ by the appropriate derivatives in $F^a_{\mathcal{O}_i\mathcal{O}_j\mathcal{O}_k}(x,P)$.

[44] In the path integral formalism, Eq. (1.81) is valid in any correlator where the insertion points of other operators $\mathcal{O}_n(x_n)$ satisfy $|x_{2n}| \geq |x_{12}|$.

We see that the operator $F_{\mathcal{O}_i\mathcal{O}_j\mathcal{O}_n}(x,\partial)$ is proportional to the three-point coefficient $C_{\mathcal{O}_n\mathcal{O}_i\mathcal{O}_j}$. This fixes the missing factor of proportionality in Eq. (1.83). By matching the small $x_{12}/x_{23}$ expansion on both sides the result in Eq. (1.83) is recovered again. For example, in the special case of scalar operators with $\Delta_i = \Delta_j = \Delta$ the operator $F_{\mathcal{O}_i\mathcal{O}_j\mathcal{O}_k}(x,\partial)$ reads

$$F_{\mathcal{O}_i\mathcal{O}_j\mathcal{O}_k}(x,\partial) = C_{\mathcal{O}_i\mathcal{O}_j\mathcal{O}_k}|x|^{\Delta_k-2\Delta}\left(1+\frac{1}{2}\mathbf{x}\cdot\partial+\frac{\Delta_k+2}{8(\Delta_k+1)}(\mathbf{x}\cdot\partial)^2-\frac{1}{16}\frac{\Delta_k}{(\Delta_k-\frac{(D-2)}{2})(\Delta_k+1)}\mathbf{x}^2\partial^2+\dots\right). \tag{1.86}$$

Again, this procedure can be repeated for fields with arbitrary spin.

The OPE can be used to reduce any $N$-point function to a sum of $N-1$-point functions,[45]

$$\langle\mathcal{O}_1(x_1)\mathcal{O}_2(x_2)\cdots\mathcal{O}_N(x_N)\rangle = \sum_k F_{\mathcal{O}_1\mathcal{O}_2\mathcal{O}_k}(x_{12},\partial_2)\langle\mathcal{O}_k(x_2)\cdots\mathcal{O}_N(x_N)\rangle\,. \tag{1.87}$$

We can apply the OPE recursively until the point where we have reduced the $N$-point function to a sum of 1-point functions $\langle\mathcal{O}\rangle$ that either vanish if $\mathcal{O}\neq\mathbb{1}$ is not the unit operator, or are equal to one if it is. The OPE operators $F_{\mathcal{O}_i\mathcal{O}_j\mathcal{O}_k}(x_{12},\partial_2)$ appearing in the last step of this recursive procedure can be computed from the general form of a two-point function in a CFT,

$$F^{ab}_{\mathcal{O}_i\mathcal{O}_i\mathbb{1}}(x_{12},\partial_2) = f^{ab}(x_{12})\,x_{12}^{-2\Delta_i}\,, \tag{1.88}$$

where $a,b$ are arbitrary spin indices and $f^{ab}(x_{12})$ is the spin tensor structure appearing in the two-point function, with $f^{ab}(x_{12}) = 1$ for scalar fields. For the form of the spin structure see *e.g.* Eq. (1.39), Eq. (1.43), Eq. (1.44) and Eq. (1.45).

### 1.1.8 Conformal blocks

Consider the scalar four-point function in Eq. (1.42) in the special case of four identical operators with scaling dimension $\Delta$. In this special case it can be written as

$$\langle\mathcal{O}(x_1)\mathcal{O}(x_2)\mathcal{O}(x_3)\mathcal{O}(x_4)\rangle = \frac{g(\mu,\nu)}{x_{12}^{\Delta}x_{34}^{\Delta}}\,, \tag{1.89}$$

with g$(\mu,\nu)$ being an unknown function of the cross-ratios $u,v$ in Eq. (1.41). We apply to Eq. (1.89) the OPE of two scalar fields in Eq. (1.81), more conveniently written here with the three-point coefficient extracted as

$$\mathcal{O}(x_1)\mathcal{O}(x_2) = \sum_k C_{\mathcal{O}\mathcal{O}\mathcal{O}_k}F'^{a}_{\mathcal{O}\mathcal{O}\mathcal{O}_k}(x_{12},\partial_2)\mathcal{O}^a_k(x_2)\,. \tag{1.90}$$

Again, note that in general the primary $\mathcal{O}^a_k$ belongs to a traceless symmetric tensor representation with non-zero spin $\ell_k$, hence we can relate the operator $F'^{a}_{\mathcal{O}\mathcal{O}\mathcal{O}_k}(x_{12},\partial_2)$ to the three-point function in Eq. (1.46).[46] Assuming that the points $x_1,x_2,x_3,x_4$ are configured correctly, we can perform the OPE in the scalar correlator in Eq. (1.89) twice,

$$\begin{aligned}\langle \mathcal{O}(x_1)\mathcal{O}(x_2)\mathcal{O}(x_3)\mathcal{O}(x_4)\rangle &= \sum_k C^2_{\mathcal{O}\mathcal{O}\mathcal{O}_k} F'^{a}_{\mathcal{O}\mathcal{O}\mathcal{O}_k}(x_{12},\partial_2) F'^{b}_{\mathcal{O}\mathcal{O}\mathcal{O}_k}(x_{34},\partial_4)\frac{I^{ab}(x_{24})}{x_{24}^{2\Delta_k}} \\ &= x_{12}^{-\Delta}x_{34}^{-\Delta}\sum_k C^2_{\mathcal{O}\mathcal{O}\mathcal{O}_k} B_{\Delta_k\ell_k}(u,v)\,,\end{aligned} \tag{1.91}$$

where we denote the OPE contractions we have performed by square brackets above the operator insertions. In Eq. (1.91) we introduce the so-called conformal blocks $B_{\Delta_k\ell_k}(u,v)$ given in terms of OPE coefficients here by

$$B_{\Delta_k\ell_k}(u,v) = x_{12}^{\Delta}x_{34}^{\Delta} F'^{a}_{\mathcal{O}\mathcal{O}k}(x_{12},\partial_2) F'^{b}_{\mathcal{O}\mathcal{O}\mathcal{O}_k}(x_{34},\partial_4)\frac{I^{ab}(x_{24})}{x_{24}^{2\Delta_k}}\,, \tag{1.92}$$

with $\ell_k$ denoting the spin of the operator $\mathcal{O}_k^a$ (a symmetric traceless tensor). We have chosen and orthonormal basis of operators and introduced a shorthand notation for the tensor structure in Eq. (1.45),

$$\langle \mathcal{O}_k^a(x)\mathcal{O}_{k'}^b(0)\rangle = \delta_{\mathcal{O}_k\mathcal{O}_{k'}}\frac{I^{ab}(x)}{x^{2\Delta_k}}\,, \qquad I^{ab}(x) = \left(I^{\mu_1\nu_1}(x)\cdots I^{\mu_{\ell_k}\nu_{\ell_k}}(x) + \text{permutations} - \text{traces}\right). \tag{1.93}$$

[45] We suppress spin indices in Eq. (1.87).

[46] It is easily shown that if any primary $\mathcal{O}^a_k$ is absent in the OPE — *i.e.* $\langle\mathcal{O}^a_k|\hat{\mathcal{O}}(x)|\mathcal{O}\rangle = 0$ — then the whole conformal family generated by said primary is also absent.

The conformal blocks — as defined in Eq. (1.91) — are functions of the cross ratios and are related to the function $g(\mu,\nu)$ in the four-point function Eq. (1.89) by

$$g(\mu,\nu) = \sum_k C^2_{\mathcal{O}\mathcal{O}\mathcal{O}_k} B_{\Delta_k\ell_k}(u,v)\,. \tag{1.94}$$

In the case of $\mathcal{O}_k^a = \mathcal{O}_k$ being a scalar the conformal block satisfies $B_{\Delta_k 0}(u,v) = u^{\Delta_k/2}(1+\ldots)$. Note that $B_{\Delta_k\ell_k}(u,v)$ is independent of the scaling dimension $\Delta$ of the operator $\mathcal{O}(x)$ (the operator in Eq. (1.89)). This is only true because we are dealing with operators with identical scaling dimensions.

Conformal blocks are neatly understood in radial quantization. We pick an origin of radial quantization such that $\{|x_3|,|x_4|\} \geq \{|x_1|,|x_2|\}$, then

$$\langle \mathcal{O}(x_1)\mathcal{O}(x_2)\mathcal{O}(x_3)\mathcal{O}(x_4)\rangle = \langle 0| R\{\mathcal{O}(x_3)\mathcal{O}(x_4)\} R\{\mathcal{O}(x_1)\mathcal{O}(x_2)\}|0\rangle\,. \tag{1.95}$$

As conformal primaries and their descendants form a complete basis of the set of operators, we can decompose the identity operator in terms of projectors $\mathrm{Pr}(\mathcal{O}_k)$ onto the different conformal multiplets,

$$\mathbb{1} = \sum_k \mathrm{Pr}(\mathcal{O}_k)\,, \qquad \mathrm{Pr}(\mathcal{O}_k) = \sum_{i_k,j_k=\mathcal{O}_k,P\mathcal{O}_k,\ldots} |i_k\rangle\langle i_k|j_k\rangle^{-1}\langle j_k|\ . \tag{1.96}$$

Inserting this into the correlator yields

$$\begin{aligned}\langle \mathcal{O}(x_1)\mathcal{O}(x_2)\mathcal{O}(x_3)\mathcal{O}(x_4)\rangle &= \sum_k \langle 0| R\{\mathcal{O}(x_3)\mathcal{O}(x_4)\}\mathrm{Pr}(\mathcal{O}_k) R\{\mathcal{O}(x_1)\mathcal{O}(x_2)\}|0\rangle \\ &= x_{12}^{-\Delta}x_{34}^{-\Delta}\sum_k C^2_{\mathcal{O}\mathcal{O}\mathcal{O}_k} B_{\Delta_k\ell_k}(u,v)\,.\end{aligned} \tag{1.97}$$

The conformal blocks in radial quantization are therefore given by

$$B_{\Delta_k \ell_k}(u,v) = \frac{x_{12}^{\Delta} x_{34}^{\Delta}}{C^2_{\mathcal{O}\mathcal{O}\mathcal{O}_k}} \langle 0| R\{\mathcal{O}(x_3)\mathcal{O}(x_4)\} \mathrm{Pr}(\mathcal{O}_k) R\{\mathcal{O}(x_1)\mathcal{O}(x_2)\} |0\rangle \, . \tag{1.98}$$

Since the projector $\mathrm{Pr}(\mathcal{O}_k)$ commutes with all of the conformal generators, by construction, the object $\langle 0| R\{\mathcal{O}(x_3)\mathcal{O}(x_4)\} \mathrm{Pr}(\mathcal{O}_k) R\{\mathcal{O}(x_1)\mathcal{O}(x_2)\} |0\rangle$ satisfies the same Ward identities as the four-point function. Hence, the conformal blocks are clearly functions of the cross-ratios. In the path-integral formalism it is instructive to think of $\mathrm{Pr}(\mathcal{O}_k)$ as a surface operator, here inserted on a sphere separating the points $x_1, x_2$ from $x_3, x_4$.

An elegant way to compute the conformal blocks is based on the Casimir of the conformal group — $C = -\frac{1}{2} J^2$ — written in terms of the generators of $SO(D+1,1)$ in Eq. (1.19) [83]. The Casimir acts with the same eigenvalue $\lambda_{\Delta_k,\ell_k} = \Delta_k(\Delta_k - D) + \ell_k(\ell_k + D - 2)$ on every state in an irreducible representation of the conformal group, hence

$$C\,\mathrm{Pr}(\mathcal{O}_k) = \mathrm{Pr}(\mathcal{O}_k) C = \lambda_{\Delta_k,\ell_k} \mathrm{Pr}(\mathcal{O}_k) \, . \tag{1.99}$$

On the other hand, the action of the conformal Casimir on the state $\mathcal{O}(x_1)\mathcal{O}(x_2)|0\rangle$ satisfies

$$C\mathcal{O}(x_1)\mathcal{O}(x_2)|0\rangle = J^2_{1,2}\mathcal{O}(x_1)\mathcal{O}(x_2)|0\rangle \, , \tag{1.100}$$

where

$$J^2_{1,2} = -\frac{1}{2}\left(J_{\alpha\beta}\big|_{x_1} + J_{\alpha\beta}\big|_{x_2}\right)\left(J^{\alpha\beta}\big|_{x_1} + J^{\alpha\beta}\big|_{x_2}\right) . \tag{1.101}$$

The differential operators $J_{\alpha\beta}\big|_{x_i}$ give the action of the generators of $SO(D+1,1)$ on the operator $\mathcal{O}(x_i)$. We can then act with $C$ or $J^2_{1,2}$ on the conformal blocks and deduce constraints on $B_{\Delta_k \ell_k}(u,v)$. By acting with $C$ to the left on $\mathrm{Pr}(\mathcal{O}_k)$ — instead of acting with $J^2_{1,2}$ to the right on $R\{\mathcal{O}(x_1)\mathcal{O}(x_2)\}$ — we find that

$$J^2_{1,2} \langle 0| R\{\mathcal{O}(x_3)\mathcal{O}(x_4)\} \mathrm{Pr}(\mathcal{O}_k) R\{\mathcal{O}(x_1)\mathcal{O}(x_2)\} |0\rangle = \lambda_{\Delta_k,\ell_k} \langle 0| R\{\mathcal{O}(x_3)\mathcal{O}(x_4)\} \mathrm{Pr}(\mathcal{O}_k) R\{\mathcal{O}(x_1)\mathcal{O}(x_2)\} |0\rangle \, . \tag{1.102}$$

This can be rewritten in terms of the conformal blocks via Eq. (1.98) and implies that they satisfy a differential equation. To conveniently write down the differential operator we introduce simpler coordinates by using conformal transformations to move $x_4 \to \infty$, $x_1 \to 0$, $x_3 \to (1,0,\ldots,0$ and $x_2 \to (x,y,0,\ldots,0)$. In these coordinates the cross-ratios in Eq. (1.41) read

$$u(z,\bar{z}) = z\bar{z} \, , \qquad v(z,\bar{z}) = (1-z)(1-\bar{z}) \, , \qquad z = x + iy \, . \tag{1.103}$$

Written in terms of the variables $z, \bar{z}$ the conformal blocks satisfy

$$D(z,\bar{z}) B_{\Delta_k \ell_k}\big(u(z,\bar{z}), v(z,\bar{z})\big) = \lambda_{\Delta_k,\ell_k} B_{\Delta_k \ell_k}\big(u(z,\bar{z}), v(z,\bar{z})\big) \, , \tag{1.104}$$

where the differential operator $D(z,\bar{z})$ is given by

$$D(z,\bar{z}) = \sum_{\gamma=z,\bar{z}} 2\left(\gamma^2(1-\gamma)\partial_\gamma^2 - \gamma^2\partial_\gamma\right) + 2\frac{(D-2)z\bar{z}}{(z-\bar{z})}\left((1-z)\partial_z - (1-\bar{z})\partial_{\bar{z}}\right) . \tag{1.105}$$

The differential equation Eq. (1.104) together with the boundary condition $B_{\Delta_k 0}(u,v) = u^{\Delta_k/2}(1+\ldots)$ and its generalization to non-zero spin determine the conformal blocks $B_{\Delta_k \ell_k}(u,v)$. In even dimension the Casimir equation Eq. (1.104) can be solved analytically in terms of hypergeometric functions [21]. In odd dimensions there exists no closed formula in terms of elementary functions, but blocks can still be computed in a series expansion [21].

### 1.1.9 The conformal bootstrap

The existence of the OPE and its properties imply that correlation functions in a (unitary) CFT are algebraically determined by its CFT data, which consists of the scaling dimensions $\{\Delta_k\}$ of all primaries and all three-point coefficients $\{C^a_{\mathcal{O}_i\mathcal{O}_j\mathcal{O}_k}\}$ of the theory. Using the OPE any correlator can be decomposed into a sum of one-point functions with coefficients written completely in terms of the conformal data (see the discussion around Eq. (1.87)). Naturally, the question arises whether any set of random numbers $\{\Delta_k, C^a_{\mathcal{O}_i\mathcal{O}_j\mathcal{O}_k}\}$ defines a consistent CFT. The answer is not necessarily, as there are certainly constraints imposed on the conformal data by the structure of the OPE. This is where the idea of the so-called conformal bootstrap originates. The conformal bootstrap refers to a set of analytical and numerical tools motivated by the structure of the OPE that aim to probe the space of possible CFT data using these consistency constraints imposed by the OPE. The goal is to restrict the parameter space $\{\Delta_k, C^a_{\mathcal{O}_i\mathcal{O}_j\mathcal{O}_k}\}$ to the points where consistent CFTs live. In that sense the conformal bootstrap provides a fully non-perturbative approach to algebraically define CFTs without the need of a Lagrangian.

The most obvious set of constraints on the conformal data comes from the fact that the OPE has to be associative, which results in the condition

$$\begin{aligned}&\sum_{i,j} C^a_{\mathcal{O}_1\mathcal{O}_2\mathcal{O}_i} C^{ab}_{\mathcal{O}_i\mathcal{O}_3\mathcal{O}_j} F'^a_{\mathcal{O}_1\mathcal{O}_2\mathcal{O}_i}(x_{12},\partial_2) F'^{ab}_{\mathcal{O}_i\mathcal{O}_3\mathcal{O}_j}(x_{23},\partial_3)\mathcal{O}^b_j(x_3) \\ &\qquad = \sum_{i,j} C^a_{\mathcal{O}_2\mathcal{O}_3\mathcal{O}_i} C^{ab}_{\mathcal{O}_1\mathcal{O}_i\mathcal{O}_j} F'^a_{\mathcal{O}_2\mathcal{O}_3\mathcal{O}_i}(x_{23},\partial_3) F'^{ab}_{\mathcal{O}_1\mathcal{O}_i\mathcal{O}_j}(x_{13},\partial_3)\mathcal{O}^b_j(x_3)\,,\end{aligned} \tag{1.106}$$

where we have considered three scalar operators $\mathcal{O}_{1,2,3}$ for simplicity. Diagrammatically, associativity of the OPE can be represented in terms of OPE contractions the same way as in Eq. (1.91),

$$\mathcal{O}_1(x_1)\mathcal{O}_2(x_2)\mathcal{O}_3(x_2) \;=\; \mathcal{O}_1(x_1)\mathcal{O}_2(x_2)\mathcal{O}_3(x_2)\;. \tag{1.107}$$

Associativity of the OPE can be neatly encoded into constraints on four-point correlation functions, summarized in the so-called crossing-symmetry equation. At face value, crossing symmetry arises from the simple fact that in a four-point correlation function OPE contractions can be chosen in different ways and the results have to match. As it becomes cumbersome to write down constraints from OPE contractions for four-point functions — even for scalar operators, but in particular for operators with

spin — often times crossing symmetry is represented diagrammatically as

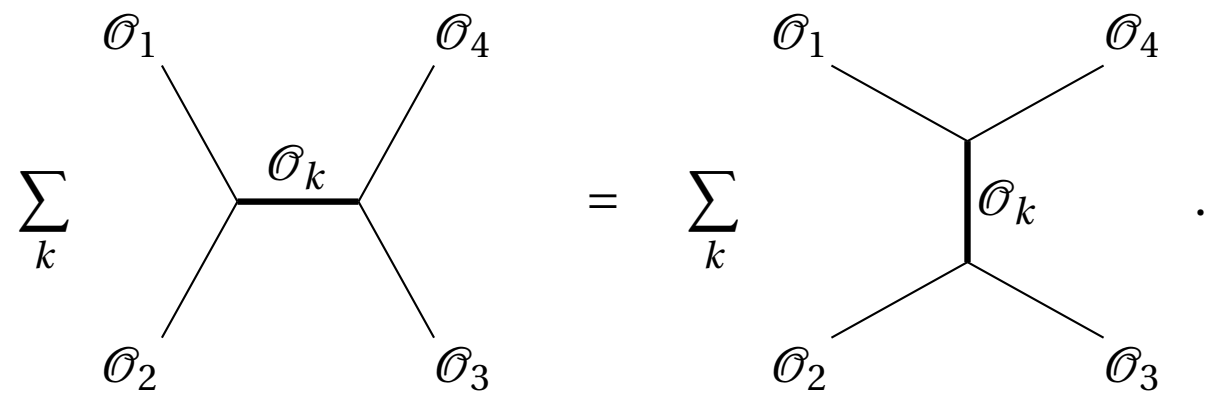

(1.108)

This is the crossing-symmetry equation for a scalar four-point function. End points represent operators in the correlator, vertices represent the OPE and internal lines represent operators that are summed over (this refers to *e.g.* the internal sum in Eq. (1.91)).[47] The diagrammatic representation in Eq. (1.108) conveniently encodes all constraints on four-point correlators coming from the OPE.

Crossing symmetry is respected by all four-point functions of local primary operators in a unitary CFT. By choosing different operators $\mathcal{O}_4$ one can recover OPE associativity from the crossing equation Eq. (1.108). The fact that all four-point functions in a unitary CFT satisfy crossing symmetry implies recursively crossing symmetry of all $N$-point functions.

For higher-point functions the diagrammatic representation in Eq. (1.108) also lends itself to cleanly represent the constraints imposed on them by crossing symmetry. For example, crossing symmetry gives the following constraint on a particular five-point function,

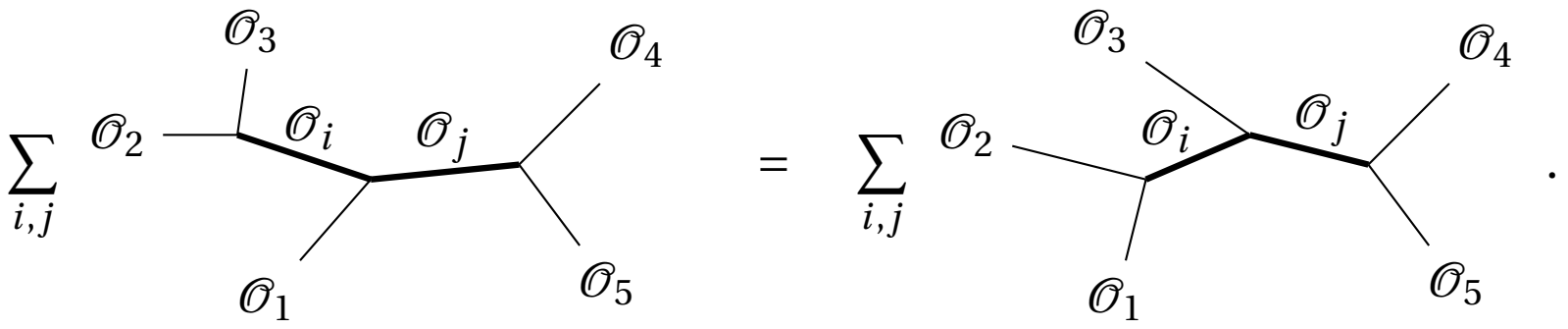

(1.109)

The existence of the crossing equation Eq. (1.108) and its implications on correlation functions of local operators is at the core of the conformal bootstrap program. The crossing symmetry equation has been known for a long time, but only in 2008 it was realized that, instead of trying to solve Eq. (1.108) exactly, the crossing-symmetry equation can be used to derive bounds on CFT data by studying its geometric properties [13].

We return to the four-point function Eq. (1.89) (*i.e.* Eq. (1.91)) of identical scalars in a unitary CFT. In this case crossing symmetry is equivalent to the condition that the four-point function is invariant under $1,2 \leftrightarrow 3,4$, or equivalently,

$$g(u,v) = \left(\frac{u}{v}\right)^{\Delta} g(v,u)\,. \tag{1.110}$$

Eq. (1.110) implies that there is an infinite number of primaries contributing to the OPE, and hence there is an infinite number of conformal blocks in the crossing equation [21]. In fact, this is a general feature — the crossing equation cannot be satisfied block-by-block.

We will use the identical scalar four-point function in Eq. (1.89) to outline the basic idea of the original conformal bootstrap first presented in [13]. We start by rewriting the crossing equation Eq. (1.108) in terms of operators for the correlator Eq. (1.110) (we suppress spin indices),[48]

$$\sum_k C^2_{\mathcal{O}\mathcal{O}\mathcal{O}_k}\left(v^\Delta B_{\Delta_k \ell_k}(u,v) - u^\Delta B_{\Delta_k \ell_k}(v,u)\right) = \sum_k C^2_{\mathcal{O}\mathcal{O}\mathcal{O}_k} A^\Delta_{\Delta_k \ell_k}(u,v) = 0\,. \tag{1.111}$$

[47] Note that in Eq. (1.108) there is no possibility to contract the operators $\mathcal{O}_1$ and $\mathcal{O}_3$ / $\mathcal{O}_2$ and $\mathcal{O}_4$ in the OPE as there is no sphere that only includes the two operator insertions we would want and no other insertion. This is a general geometrical statement satisfied by four distinct points in flat space. Hence, for a four-point function there are always only two out of the three remaining operators that can be contracted with any given operator in question.

The idea is to think of the functions $A^\Delta_{\Delta_k \ell_k}(u,v)$ as vectors in the infinite-dimensional vector space of functions of $u$ and $v$. Keep in mind that the coefficients $C^2_{\mathcal{O}\mathcal{O}\mathcal{O}_k}$ are positive. The sum runs over all scaling dimensions and spins in the $\mathcal{O}\times\mathcal{O}$ OPE. In this language Eq. (1.111) states that lot of vectors all sum to zero, which may or may not be possible depending on the vectors. The geometric way to distinguish between cases where it is possible and cases where it is not is to find a separating plane through the origin such that all vectors lie on one side of said plane.[49] If such a plane exists, the vectors $A^\Delta_{\Delta_k \ell_k}(u,v)$ cannot satisfy the crossing equation for any choice of three-point coefficients $C_{\mathcal{O}\mathcal{O}\mathcal{O}_k}$. This leads to the following algorithm for bounding operator dimensions:

- Start by making a hypothesis for which scaling dimensions $\Delta_k$ and which spins $\ell_k$ appear in the OPE.
- Search for a linear functional that is non-negative when acting on all functions $A^\Delta_{\Delta_k \ell_k}(u,v)$.
- If such a functional exists, there is a contradiction found by acting on both sides of Eq. (1.111) with said functional. In that case the hypothesis is wrong.

This algorithm can be modified so that it can be used to find bounds on the three-point coefficients $C_{\mathcal{O}\mathcal{O}\mathcal{O}_k}$ as well [84].
The algorithm outlined above can be implement by hand for simple examples — like the two-dimensional Ising model — up to a certain point of precision. However, computerized searches are the state of the art and numerical techniques can lead to very precise results. Numerical analysis requires the implementation of a cut-off for terms in the OPE, the discretization of the values of $\Delta_k$ and $\ell_k$ and the restriction of the search to a finite-dimensional subspace of the space of functions of $u, v$.

A solution to the crossing equations gives a completely non-perturbative definition for correlation functions of local operators without the need to refer to any Lagrangian. However, this is not enough to completely describe the full theory and there are additional constraints and consistency conditions coming from other considerations. We mention a few selected ones here:

- QFTs in general and CFTs in particular admit extended objects like line and surface operators, boundaries and interfaces. These objects introduce additional data into the theory. It is possible to write down OPEs and crossing relations that relate this data to itself and the more standard CFT data [85].

[48]This is best derived via the OPE contractions $\langle \mathcal{O}(x_1)\mathcal{O}(x_2)\mathcal{O}(x_3)\mathcal{O}(x_4)\rangle = \langle \mathcal{O}(x_1)\mathcal{O}(x_2)\mathcal{O}(x_3)\mathcal{O}(x_4)\rangle$, using Eq. (1.110) and of course the definition of $u$ and $v$ in Eq. (1.41).

[49]In a vector space of functions a separating plane is implemented in terms of a linear functional $\gamma$ such that $\gamma(A^\Delta_{\Delta_k \ell_k}(u,v)) \geq 0$ for all functions $A^\Delta_{\Delta_k \ell_k}(u,v)$.

- On non-trivial manifolds not conformally equivalent to flat space more conformal data is introduced, like one-point functions of local and extended operators. Theories of interest on not conformally flat geometries include *e.g.* theories at finite temperature.
- CFTs in Lorentzian signature exhibit additional interesting constraints, like bounds on energy positivity [86], causality [87] and dispersion relations [88–91].

In general, the full set of data and consistency conditions that are associated with a given CFT is not known and there are even examples of constraints on local operators beyond the scope of the OPE and crossing equations (*e.g.* modular invariance).

In rough terms the conformal bootstrap program aims to exploit the full consequences of conformal symmetry — in particular Eq. (1.108) — to strongly constrain or even solve any theory in question. As the conformal bootstrap provides a fully non-perturbative non-Lagrangian approach to CFTs, perhaps the most ambitious goal of the bootstrap program is to extend the procedure away from the fixed points of the RG flow and find fully non-perturbative algebraic description of QFT without any reference to a Lagrangian description.
Perhaps the most impressive result of the conformal bootstrap program is the numerical analysis of the three-dimensional Ising model CFT (see Appendix A.1). The three-dimensional Ising model is probably the simplest model that is, to our knowledge, not exactly solvable and experimentally relevant. As such, it is an ideal playground for the conformal bootstrap and there exists very precise data on the lowest operators coming from Monte Carlo simulations. In order to get the most precise bounds from the conformal bootstrap the fact that the Ising CFT in $D = 3$ has only two relevant scalar operators in the theory — usually denoted by $\sigma$ and $\varepsilon$ — is required as an input. Although not proven mathematically, this is an obvious experimental fact. Relevant scalars are in one-to-one correspondence with parameters that must be tuned to reach the critical point of some microscopic theory. The fact that the phase diagram of water is only two-dimensional — the two dimensions being temperature and pressure — immediately tells us that the critical point of water — the three-dimensional Ising CFT — has two relevant operators.[50] The most relevant works on bootstrapping the $D = 3$ Ising model are [92–94].
There is striking agreement between Monte Carlo simulations and the conformal bootstrap, which is strong evidence that the critical Ising model indeed flows to a conformal fixed point (as opposed to just a scale invariant one). At this point the island for allowed operator dimensions $(\Delta_\sigma, \Delta_\varepsilon)$ is far smaller than the 68% confidence region provided by the current best Monte Carlo determinations [21]. The hope is that future bootstrap studies might shrink the island $(\Delta_\sigma, \Delta_\varepsilon)$ to a single point and hence conclusively prove the IR equivalence of all theories in the Ising universality class. This would also be a strong reassurance of the principle of critical universality.

[50] For the actual Ising model the parameters in the phase diagram usually are temperature and magnetization.

## 1.2 Spontaneous breaking of global symmetries

SSB refers to the phenomenon that in certain physical systems, in particular in QFT, the ground state — or vacuum — is not invariant under all the symmetries of the theory. In other words, there are several equivalent vacua connected by the remaining symmetries. The symmetries not leaving the vacuum invariant and connecting the different ground states are referred to as broken. With the discovery of the Higgs mechanism the concept of SSB became an important cornerstone of modern particle physics. However, the first examples of SSB did appear in condensed-matter physics and the idea was then introduced into particle physics by analogy. In fact, the theory of superconductivity by Bardeen, Cooper and Schrieffer first presented 1957 [95] — or more specifically its reformulation by Nambu in 1960 [5] — provided the first paradigm for the introduction of SSB in relativistic QFT.

If nothing else is mentioned we are in $D$-dimensional Minkowski spacetime with mostly positive metric $\eta_{\mu\nu} = \text{diag}(-1, 1, \ldots, 1)$. However, we do not require theories to exhibit relativistic invariance. The results discussed generally also apply to non-relativistic theories and relativistic theories with explicitly broken spacetime symmetries. Systems without full relativistic invariance appear quite frequently in nature, most notably in condensed-matter physics [96]. In particular, in the LCE operators with large internal quantum numbers can be adequately captured by relativistic systems at finite density with respect to the associated charge $Q$. Additionally, many of the finite-density systems appearing in the LCE exhibit the spontaneous breaking of both global and spacetime symmetries [17, 18]. It is therefore important to understand the full implications of SSB at finite density including the breaking of spacetime symmetries.

### 1.2.1 Basic properties of spontaneously broken symmetries

In the study of physical systems the presence of spontaneously broken symmetries results in far-reaching consequences. In particular, the spontaneous breaking of a continuous global symmetry implies the existence of a so-called NG boson [6]. The study of SSB and the properties of the appearing NG bosons is very powerful as the low-energy dynamics of these NG bosons are governed by the symmetries and are mostly independent of the microscopic structure of the theory [97, 98]. In a standard setup with Lorentz invariance and spontaneously broken global internal symmetries — like the Higgs model — there are as many massless Goldstone bosons as there are broken generators. However, if a theory does not exhibit Poincaré invariance or if some spacetime symmetries are explicitly broken there can be a mismatch between the number of generators and the number of NG bosons [39, 99–101]. In that case there is still a general counting rule that can be derived [35–37]. Additionally, there is the possibility that not just global symmetries but also spacetime symmetries are spontaneously broken. In this case there are several complications at different levels and, unfortunately, it does not seem to be feasible to write down a general formula for the number of NG in any useful manner. It seems that they are best analysed on a case-by-case basis [34].

As first shown by Eugen Paul Wigner in 1931 any symmetry transformation in a QFT can be represented on the Hilbert space of quantum states by an operator $U$ that is either linear and unitary or anti-linear and anti-unitary [27]. The set of symmetry transformations of a QFT always forms a group, which for our purposes will be considered as continuous, since discrete groups groups do not produce any NG bosons. With the appropriate definition of the adjoint of a linear and an anti-linear operator the conditions for unitarity and anti-unitarity both take the form $U^\dagger = U^{-1}$. For a symmetry compatible with the dynamics of the system the operator $U$ commutes with the Hamiltonian $H$ of the theory. In this language SSB refers to the fact that there exist systems for which the dynamics are invariant under a given symmetry but the symmetry does not (fully) manifest itself in the spectrum of physical observables.
Consider a QFT whose action $S[\Phi_i]$ is invariant under a continuous group $G$ of symmetry transformations compatible with the dynamics of the system. Noether's theorem grants us the existence of a conserved current $j^\mu(x)$ corresponding to $G$ with the associated charge,

$$Q(t) = -\int \mathrm{d}^{D-1}x\, j^0(t,\mathbf{x})\,, \tag{1.112}$$

serving as the generator of the symmetry transformations in the quantum theory.[51] Since $Q$ commutes with $H$, the spectrum of eigenstates splits into multiplets of the symmetry and each of these multiplets corresponds to an irreducible representation of the symmetry group. The ground state $|0\rangle$ of the theory is assumed to be a discrete and non-degenerate eigenstate of $H$ with minimal energy. A spontaneously broken symmetry is a symmetry of the theory such that the state $|0\rangle$ is not an eigenstate of $Q$. Given a finite spatial domain $\Omega \subset \mathbb{R}^{D-1}$ we introduce

$$Q_\Omega(t) := -\int_\Omega \mathrm{d}^{D-1}x\, j^0(t,\mathbf{x})\,. \tag{1.113}$$

Mathematically, SSB is formulated as the statement that there exists a (not necessarily local) operator $\mathcal{O}^{(\mathrm{SB})}$ such that[52]

$$\langle\sigma_0\rangle := \lim_{\Omega\to\infty} \langle 0|\,[Q_\Omega(t),\mathcal{O}^{(\mathrm{SB})}]\,|0\rangle \neq 0\,, \tag{1.114}$$

where $|0\rangle$ is in principle any translationally invariant ground state of the theory as there are now several inequivalent ground states. The expectation value $\langle\sigma_0\rangle$ — in analogy to phase transitions — is called the order parameter.[53] Under the assumption that $|0\rangle$ were an eigenstate of $Q$ it would follow that $\langle\sigma_0\rangle$ vanishes. Hence, the condition in Eq. (1.114) is equivalent to the characterization of SSB that the ground state is not an eigenstate of $Q$. Under normal circumstances one identifies $Q = \lim_{\Omega\to\infty} Q$, however, in the case of a spontaneously broken symmetry the operator $Q$ is strictly speaking not well defined: translational invariance of the vacuum $|0\rangle$ implies translational invariance of $Q\,|0\rangle$, and therefore

$$\langle 0|\,Q(t)\,Q(t)\,|0\rangle = -\int \mathrm{d}^{D-1}x\ \langle 0|\,j^0(t,\mathbf{x})\,Q(t)\,|0\rangle = \int \mathrm{d}^{D-1}x\ \langle 0|\,j^0(t,\mathbf{0})\,Q(t)\,|0\rangle\,. \tag{1.115}$$

This diverges unless $Q\,|0\rangle = 0$, which is a contradiction to the assumption of SSB. A spontaneously broken symmetry is not realized by a unitary operator on the Hilbert space and therefore does not give rise to multiplets of symmetry in the spectrum. The operator $Q_\Omega(t)$ induces a "finite-volume symmetry

[51]The conserved Noether charges naturally satisfy the underlying group algebra [102].
[52]The limit "$\Omega\to\infty$" should be understood as blowing up $\Omega$ to the full spatial slice $\mathbb{R}^{D-1}$.
[53]We remark here that generically we have $\langle[Q_\Omega(t),\mathcal{O}^{(\mathrm{SB})}]\rangle = \langle G^{(Q)}\,\mathcal{O}^{(\mathrm{SB})}\rangle \sim \langle\mathcal{O}^{(\mathrm{SB})}\rangle$, since the charge $Q$ generates the symmetry transformation in the quantum theory. Hence, in most practical cases we would consider $\langle\sigma_0\rangle = \langle\mathcal{O}^{(\mathrm{SB})}\rangle$ as the order parameter.

transformation" $U_\Omega(\theta,t)$ that connects equivalent vacua $|\theta,t\rangle_\Omega$,

$$U_\Omega(\theta,t) = e^{-i\theta Q_\Omega(t)}\,, \qquad |\theta,t\rangle_\Omega = U_\Omega(\theta,t)\,|0\rangle\,. \tag{1.116}$$

Hence, for a system exhibiting SSB there exists a symmetry operator that does not leave the ground state intact, implying the presence of a (continuous) set of equivalent vacua and making the ground state degenerate. Like $Q$ itself, the operator $U = e^{-i\theta Q(t)}$ in the limit $\Omega \to \infty$ is not well defined and it can be shown that [39]

$$\lim_{\Omega\to\infty} \langle 0|\,|\theta,t\rangle_\Omega = 0\,. \tag{1.117}$$

Hence, in the infinite-volume limit the degenerate vacua — formally connected by a finite-volume symmetry transformation — become pairwise orthogonal. The same holds for the excited states constructed above the pairwise orthogonal vacua. As a consequence, these states cannot be accommodated in a single Hilbert space and there exists a separate Hilbert space for every equivalent ground state in the theory. All of these Hilbert spaces can equivalently be used to describe the system without any observable physical consequences.
Finite volume symmetry transformations of observables can be consistently described for theories that respect micro-causality, requiring that the commutator of any two operators separated by a spacelike distance vanishes. For any operator $A$ we define$A_\Omega(\theta,t)$ as follows:

$$A_\Omega(\theta,t) := U^\dagger_\Omega(\theta,t)\,A\,U_\Omega(\theta,t) = A + i\theta[Q_\Omega(t),A] + \mathcal{O}(\theta^2)\,, \qquad [Q_\Omega(t),A] = \int_\Omega \mathrm{d}^{D-1}x\,[j^0(t,\mathbf{x}),A]\,. \tag{1.118}$$

If we further assume that $A$ is localized in a finite domain of spacetime, then there exists a (finite) region $\Omega_A$ such that the commutator of $j^0(x)$ with $A$ vanishes outside of $\Omega_A$. This implies that the operator $A_\Omega$ is well defined as $\Omega \to \infty$. The expectation value of $A_\Omega(\theta,t)$ on the ground state $|0\rangle$ can be interpreted as the expectation value of $A$ on the transformed ground state $|\theta,t\rangle_\Omega$:

$$\langle 0|\,A_\Omega(\theta,t)\,|0\rangle = {}_\Omega\langle\theta,t|\,A\,|\theta,t\rangle_\Omega\,. \tag{1.119}$$

In Lorentz-invariant theories the micro-causality condition is automatically satisfied by relativistic causality. In non-relativistic theories interactions are often modelled by non-local instantaneous potentials, and thus micro-causality is not guaranteed. However, the above argument can still be applied provided that the range of the interaction is short enough, *i.e.* if it falls off exponentially.

### 1.2.2 Order parameter space

For any spontaneously broken symmetry there exists at least one order parameter $\langle\sigma_0\rangle$ given by the non-zero Vacuum Expectation Value (VEV) of some operator that transforms non-trivially under the symmetry group, as defined in Eq. (1.114). The set of degenerate vacua $|\theta,t\rangle_\Omega$ can be labelled by different values of the order parameter. Demonstrating that a physical system exhibits SSB requires the identification and computation of an order parameter $\langle\sigma_0\rangle$. Identifying $\langle\sigma_0\rangle$ is difficult mainly because SSB is a non-perturbative phenomenon that cannot be captured at any finite order of perturbation theory based on a symmetry-preserving ground state. Once the appropriate operator is identified, calculating the order parameter $\langle\sigma_0\rangle$ can be achieved in two main ways:

- The first approach would be to find a self-consistent non-perturbative symmetry-breaking

solution to the Equation(s) of Motion (EoM). The drawback here is that one has to introduce an Ansatz for the symmetry-breaking solution and then show that it is consistent with the EoM. Therefore, it is only possible to get solutions that are put in by hand from the beginning.

- The second approach consists of introducing an effective field $\sigma(x)$ with suitable symmetry transformation properties into the theory. The introduced field $\sigma$ will produce the NG bosons and its ground state expectation value $\langle\sigma\rangle = \langle\sigma_0\rangle$ serves as the order parameter.[54]

While the first approach starts from whatever DoF are present and is physically more satisfactory, it is often times very difficult to perform and the second, more phenomenological approach tends to be simpler in its implementation. For example, in the case of the Higgs mechanism in the standard model of particle physics an effective field is introduced by minimizing the quartic Higgs potential and expanding around its minimum $\phi_0$, which is the VEV of the effective field and also the order parameter. The introduction of an effective field for the NG bosons minimizes the potential $V$ in the Lagrangian and identifies one of its minima as the order parameter. Hence, the vacuum manifold $\mathscr{M}_V$ of the potential $V$ comprises the set of all possible values of the order parameter $\langle\sigma_0\rangle$, also called the order parameter space.
Consider a theory described by a potential $V(\phi)$ with a vacuum manifold $\mathscr{M}_V$. The potential $V(\phi)$ is invariant under the symmetry group $G$ of the theory. Hence, $G$ acts smoothly on any point $\langle\sigma_0\rangle \in \mathscr{M}_V$. The set

$$H\langle\sigma_0\rangle := \left\{g \in G : \; g\langle\sigma_0\rangle = \langle\sigma_0\rangle\right\} \tag{1.120}$$

is a subgroup of $G$ — called the little group or isotropy group of the point $\langle\sigma_0\rangle \in \mathscr{M}_V$ — and consists of all symmetry transformations left unbroken under the assumption that the order parameter takes the value $\langle\sigma_0\rangle$. The isotropy groups of two points $\langle\sigma_0\rangle, \langle\sigma_0\rangle'$ in the same group orbit $G\langle\sigma_0\rangle := \left\{g\langle\sigma_0\rangle \,|\, g \in G\right\}$ are conjugate,

$$\exists\, g \in G: \qquad H\langle\sigma_0\rangle = g^{-1} H\langle\sigma_0\rangle' g\,. \tag{1.121}$$

The converse is not always true. The set of all points with conjugate isotropy groups $H\langle\sigma_0\rangle$ is called the stratum $\mathrm{Str}(\langle\sigma_0\rangle)$ of $\langle\sigma_0\rangle$. Physically, the stratum is the set of points of the vacuum manifold that exhibit the same unbroken subgroup under SSB. As the potential $V(\phi)$ is invariant under the group action,

$$V(\phi) = V(g\phi)\,, \qquad \forall\, g \in G\,, \tag{1.122}$$

it is just a function of group orbits. It follows that the minimization of $V(\phi)$ can be reformulated as a minimization problem on the space of orbits. In the study of stationary points of $G$-invariant functions the following theorem is fundamental [103]:

**Theorem** (Michel)**.** *Let $G$ be a compact Lie group acting smoothly on a real manifold $\mathscr{M}$ and let $m \in \mathscr{M}$. Then the orbit $Gm$ is critical,* i.e. *every smooth real and $G$-invariant function on $\mathscr{M}$ is stationary on $Gm$, if and only if $Gm$ is isolated in its stratum,* i.e. *$\exists$ a neighbourhood $U_m$ of $m \in \mathscr{M}$ such that $U_m \cap \mathrm{Str}(m) = Gm$.*

[54] Again, this is consistent with Eq. (1.114) as the commutator $[Q,\sigma] = G^{(Q)}\sigma$ generates the associated symmetry transformation.

In the case that $\mathscr{M}_V$ is a linear space Theorem(Michel) simply asserts the obvious fact that $\langle\sigma_0\rangle = 0$ is a stationary point of every $G$-invariant function. However, if $\mathscr{M}_V$ is not a linear space Theorem(Michel) helps us to look for inert states, *i.e.* states for which the form of the order parameter is fixed up to a symmetry transformation (like in the Higgs case). In particular, inert states can be found among stationary points of $G$-invariant functions on manifolds of order parameters with fixed norm.[55]

### 1.2.3 The Nambu-Goldstone theorem

One of the most important consequences of SSB is the existence of soft modes in the spectrum whose energy vanishes in the IR limit — the so-called NG bosons — guaranteed by the Goldstone theorem. Consider the commutator $[\partial_\mu j^\mu, \mathcal{O}^{(\mathrm{SB})}] = 0$ of the conserved current $j^\mu$ associated to a broken charge $Q$ with the operator $\mathcal{O}^{(\mathrm{SB})}$ in Eq. (1.114) integrated over $\Omega$,

$$0 = \int_\Omega \mathrm{d}^{D-1}x\,[\partial_\mu j^\mu, \mathcal{O}^{(\mathrm{SB})}] = \partial_0\left[Q_\Omega(t), \mathcal{O}^{(\mathrm{SB})}\right] + \int_{\partial\Omega} \mathrm{d}S\left[\mathbf{n}\cdot\mathbf{j}, \mathcal{O}^{(\mathrm{SB})}\right]. \tag{1.123}$$

If the operator $\mathcal{O}^{(\mathrm{SB})}$ is localized to a finite domain of spacetime and micro-causality is satisfied,[56] then the surface integral in Eq. (1.123) vanishes as $\Omega \to \infty$ and it follows that the order parameter $\langle\sigma_0\rangle = \langle 0|\,[Q, \mathcal{O}^{(\mathrm{SB})}]\,|0\rangle$ is time-independent. By inserting the identity operator $\mathbb{1}$ in terms of all (multi-)particle eigenstates of the Hamiltonian $H$ of the theory in the correlator $\langle 0|\,[Q_\Omega, \mathcal{O}^{(\mathrm{SB})}]\,|0\rangle$ and using translational invariance we get[57]

$$\begin{aligned}\langle 0|\,[Q_\Omega(t), \mathcal{O}^{(\mathrm{SB})}]\,|0\rangle = \int \frac{\mathrm{d}^{D-1}k}{(2\pi)^{D-1}} \sum_n \Bigg[ & e^{-iE_{n_\mathbf{k}}t}\langle 0|\,j^0(t,\mathbf{0})\,|n_\mathbf{k}\rangle\langle n_\mathbf{k}|\mathcal{O}^{(\mathrm{SB})}\,|0\rangle\left(\int_\Omega \mathrm{d}^{D-1}x\,e^{i\mathbf{k}\cdot\mathbf{x}}\right) \\ & - e^{iE_{n_{-\mathbf{k}}}t}\langle 0|\mathcal{O}^{(\mathrm{SB})}\,|n_{-\mathbf{k}}\rangle\langle n_{-\mathbf{k}}|\,j^0(t,\mathbf{0})\,|0\rangle\left(\int_\Omega \mathrm{d}^{D-1}x\,e^{-i\mathbf{k}\cdot\mathbf{x}}\right)\Bigg].\end{aligned} \tag{1.124}$$

In the limit $\Omega \to \infty$ the left-hand side becomes the order parameter $\langle\sigma_0\rangle$. The function $\int_\Omega \mathrm{d}^{D-1}x\,e^{\pm i\mathbf{k}\cdot\mathbf{x}}$ on the right-hand side of Eq. (1.124) becomes sharply peaked around $\mathbf{k} = 0$ and only states with small momentum will contribute. As the left-hand side becomes time-independent this is only possible if the energy of the contributing states on the right-hand side vanishes as $\mathbf{k}$ goes to zero,

$$E_{n_\mathbf{k}} \to 0 \;\text{ as } \mathbf{k} \to 0\,, \;\text{ if }\; \langle 0|\,j^0(t,\mathbf{0})\,|n_\mathbf{k}\rangle\langle n_\mathbf{k}|\mathcal{O}^{(\mathrm{SB})}\,|0\rangle \neq 0\,. \tag{1.125}$$

If there were no contributing states in the spectrum the right-hand side of Eq. (1.124) would just be zero, a contradiction to the assumption of SSB. Hence, there must be at least one bosonic and massless state coupling to both the broken current $j^0$ and the interpolating operator $\mathcal{O}^{(\mathrm{SB})}$. This is Goldstone's theorem:

**Theorem** (Goldstone)**.** *Spontaneous breaking of a continuous global internal symmetry implies the existence of a bosonic mode* $|n^{NG}_\mathbf{k}\rangle$ *in the spectrum such that*

$$\lim_{\mathbf{k}\to 0} E_{n^{NG}_\mathbf{k}} = 0\,. \tag{1.126}$$

[55] Invariant functions can, of course, have other stationary points than those guaranteed by Theorem(Michel).

[56] In practice, the operator $\mathcal{O}^{(\mathrm{SB})}$ is often even strictly local.

[57] We assume at least discrete translational invariance in all directions here. Eigenstates of the Hamiltonian have two indices $n_\mathbf{k}, \mathbf{k}$ denoting the different multi-particle excitation branches in the spectrum as well as the continuous internal momentum eigenstates. States are normalized such that they satisfy the orthogonality condition $\langle n_\mathbf{k}|m_\mathbf{q}\rangle = (2\pi)^{D-1}\delta_{nm}\delta(\mathbf{k}-\mathbf{q})$.

In principle, some isolated states with zero momentum and energy that do not represent NG bosons — such as other degenerate vacua — can exists and give rise to very small contributions on the right-hand side of Eq. (1.124) but their contribution is suppressed in the limit $\Omega \to \infty$ [104]. The consequences of the Goldstone theorem cannot be escaped for relativistic theories. But this does not mean that the NG bosons are necessarily observable, as they may appear in the unphysical sector of the Hilbert space. In particular, this happens in the case of Lorentz-invariant gauge theories with covariant gauge fixing. On the other hand, in non-covariant gauges the Goldstone theorem does not apply. For example, in the standard model of particle physics there are no massless NG bosons present as a result of electroweak symmetry breaking.

### 1.2.4 Counting rules for the number of Nambu-Goldstone modes

While the Goldstone theorem guarantees the existence of at least one NG boson in the spectrum, it does not say anything about the exact number of NG bosons. In the case of unbroken Lorentz-invariance there exists precisely one NG boson for every broken generator $Q_a |0\rangle \neq 0$ in the Lie algebra of the symmetry group $G$.[58] However, in non-relativistic theories and theories with explicitly broken Poincaré symmetries this is not necessarily the case, as concrete examples — such as ferromagnets or three-component Fermi gases — show. In these cases there are less NG modes as there are broken generators. In models with fewer NG modes than broken symmetry generators one encounters NG modes that exhibit quadratic dispersion relations. This observation lead to the following result by Nielsen and Chadha relating the number of NG bosons to the number of broken generators [106]:

**Theorem** (Nielsen and Chadha)**.** *We assume that translational invariance is not completely broken (spontaneously) and that there are no long-range interactions. In this case the energy of a NG boson is analytic in its momentum. We denote NG modes whose energy is proportional to an odd power of the momentum as type-I and those NG modes whose energy is proportional to an even power of the momentum by type-II. In this case the number of type-I NG bosons $N_I$ plus twice the number of type-II NG bosons $N_{II}$ is greater than or equal to the number of broken symmetry generators $N_Q$,*

$$N_I + 2N_{II} \geq N_Q\,. \tag{1.127}$$

Theorem(NC) does not specify the precise power of the momentum in the dispersion relation and it only gives an inequality on the number of NG bosons. Another important result by Schäfer et al. relates the number of NG bosons to the number of pairwise commuting broken generators [39, 107].[59]

**Theorem** (Schäfer et al.)**.** *Suppose that $\langle 0| [Q_a, Q_b] |0\rangle = 0$ for all pairs of broken charges $Q_a$, $Q_b$ under SSB, then the number of NG modes $N_{NG}$ is at least equal to the number of broken generators $N_Q$*, i.e.

[58] The argument here is simple. For a linearly realized symmetry the quantum effective potential $V_{\text{eff}}$ is invariant under the same transformations as the Lagrangian, hence $\partial V_{\text{eff}}/\partial\phi T_a\phi = 0$ for all generators $T_a$ of the symmetry group. After differentiating this implies that $\partial^2 V_{\text{eff}}/\partial\phi^2 T_a\phi|_{\phi=\langle\phi\rangle} = 0$. By definition, SSB means that some of the generators do not annihilate the vacuum, $T_a\langle\phi\rangle \neq 0$. Assuming Lorentz-invariance this implies that the real vector $i T_a\langle\phi\rangle$ is a zero mode of the mass matrix corresponding to a NG boson. In the case of broken relativistic invariance the number of zero modes of the mass matrix $\partial^2 V_{\text{eff}}/\partial\phi^2$ is still equal to the number of broken generators, but no longer coincides with the number of massless modes as they no longer need to satisfy a Klein–Gordon (KG) equation [39, 105].

[59] The charges $Q_a, Q_b$ are still given by taking the limit $\Omega \to \infty$ in Eq. (1.113). They are in principle ill-defined. However, the commutator $\langle 0| [Q_a, Q_b] |0\rangle$ can be well defined and vanish — $\langle 0| [Q_a, Q_b] |0\rangle = 0$ — in the case where the structure constants $f_{abc}$ in $[Q_a, Q_b] = i f_{abc} Q_c$ are only non-zero if $Q_c$ is unbroken.

$N_{NG} \geq N_Q$.

However, the presence of a non-zero expectation value of the commutator between two charges in the theory itself is not enough to conclude that the number of NG bosons is smaller than the number of broken generators $N_{\text{NG}} < N_Q$. To conclude this additional assumptions have to be made, such as the symmetry algebra being semi-simple. Without this assumption the presence of non-trivial central charges cannot be ruled out. An example of a model where this happens would be the free non-relativistic particle [39].
Finally, there exists a relationship between non-vanishing charge densities and NG mode dispersion relations, as first shown by Leutwyler in the context of low-energy EFT [108]. An argument based just on symmetry assumptions and analyticity shows that the presence of two broken generators with $\langle 0| [Q_a, Q_b] |0\rangle \neq 0$ implies the existence of a NG boson with non-linear (*i.e.* at least quadratic) dispersion relation in the spectrum [39].

**Theorem** (Brauner)**.** *For every pair of broken generators $Q_a, Q_b$ such that $\langle 0| [Q_a, j^0_b(t,\mathbf{x})] |0\rangle \neq 0$ there is a single NG boson coupling to both charges as in Eq.* (1.125)*, and generically its dispersion relation is quadratic.*

For SSB of global internal symmetries there is an interconnectedness between the number of NG modes, their respective dispersion relations and the number and relative properties of the broken generators. Unfortunately, the results by Nielsen/Chadha, Schäfer et al. and Leutwyler only provide inequalities or bounds on the number of soft modes. This problem has been solved only very recently, in the last decade [34–38, 101]. A precise counting rule can be derived under some rather broad assumptions. Naturally, SSB is assumed, with a continuous global internal symmetry group $G$ breaking down to an unbroken subgroup $H \subset G$. The unbroken subgroup $H \subset G$ is generated by the unbroken generators $T_b$ with $Q_b |0\rangle = 0$.[60] We do not require relativistic invariance, both theories with non-relativistic spacetime symmetry groups and theories with explicitly broken Poincaré symmetries (e.g. theories at finite density) are permitted.[61]
In Poincaré-invariant theories the number of broken currents $N_Q$ is always equal to the number of NG bosons $N_{\text{NG}}$ [39], hence knowing the symmetry-breaking pattern $G \to H$ is enough to deduce the number of NG modes. In the absence or explicit breaking of relativistic invariance SSB still implies the existence of NG modes but their dispersion relations are no longer restricted by Lorentz invariance and the number of NG bosons can be fewer than the number of broken currents. The symmetry breaking pattern only gives you an upper bound on the number of NG bosons. In such cases, it is possible to derive a counting rule for the number of NG bosons by considering the commutators of (broken) generators. The following result first presented in [35, 36] relates the number of NG bosons $N_{\text{NG}}$ with the number of broken generators $N_Q$ and holds true regardless of the (unbroken) spacetime symmetries:

$$N_{\text{NG}} = N_Q - \frac{1}{2}\text{rank}(\rho)\,, \tag{1.128}$$

[60] The same holds true of course for all the degenerate vacua in Eq. (1.116). Also, to be precise, we require $H$ to be a normal subgroup, or equivalently we require left and right cosets to be equal $gH = Hg$.
[61] In order to be able to use the momentum $\mathbf{k}$ as a label we have to assume at least discrete translational invariance in all spatial directions.

where $\rho$ is the following real and anti-symmetric matrix with the commutators of (broken) charges as entries,

$$\rho_{ab} := -i \lim_{\Omega\to\infty} \frac{\langle 0|\,[Q_{\Omega\,a}, Q_{\Omega\,b}]\,|0\rangle}{|\Omega|} = \lim_{\Omega\to\infty} f_{abc} \frac{\langle 0|\,Q_c\,|0\rangle}{|\Omega|} \sim f_{abc}\,\langle 0|\, j_c^0(t,\mathbf{0})\,|0\rangle\,. \tag{1.129}$$

Here, $|\Omega|$ is the volume of the (finite) subset $\Omega \subset \mathbb{R}^{D-1}$ and $f_{abc}$ are the structure constants of the symmetry algebra. The indices $a, b$ may run over the full symmetry group or just the broken generators as the rank of the matrix remains the same in both cases. In the same vein, only broken generators contribute to the right-hand side of Eq. (1.129), *i.e.* the sum over $c$ can be considered as running only over the broken generators.[62] We schematically present the derivation of Eq. (1.128) in Appendix A.4. The real anti-symmetric matrix $\rho$ can be block-diagonalized by an orthogonal transformation $M$,

$$M^T \rho M = \begin{pmatrix} 0 & \lambda_1 & & & & & \\ -\lambda_1 & 0 & & & & & \\ & & \ddots & & & & \\ & & & 0 & \lambda_m & & \\ & & & -\lambda_m & 0 & & \\ & & & & & \ddots & \\ & & & & & & 0 \end{pmatrix}, \tag{1.130}$$

where all $\lambda_k$'s, $k = 1, \ldots, m$ are strictly non-zero and all blank entries are zero. In this form the rank of $\rho_{ab}$ is manifestly equal to $2m$ and therefore the right-hand side of Eq. (1.128) is a positive integer. We now classify the NG modes into two new categories: type A and type B.

- Type-B modes are associated with pairs of broken generators $(Q_{2l-1}, Q_{2l})$, $l = 1, \ldots, m$ in Eq. (A.112), with each pair of broken generators corresponding only to a single mode. They generically exhibit a quadratic dispersion relation, though not always as it is possible to fine-tune parameters in certain models [34].
- Type-A modes are associated to the remaining $N_Q - 2m$ broken generators and generically exhibit a linear dispersion relation. In particular, they are in a 1-to-1 correspondence with said broken generators.

Written in terms of the number of type-A modes $N_A = N_Q - \text{rank}(\rho)$ and type-B modes $N_B = \frac{1}{2}\text{rank}(\rho)$ the result in Eq. (1.128) gives the counting rule[63]

$$N_A + 2N_B = N_Q\,. \tag{1.131}$$

Theorem(NC) by Nielsen and Chadha already divides the NG modes in two classes — type-I and type-II — according to their dispersion relations, and it is worth understanding the relationship between to two classifications. In fact, the type-B NG modes in Eq. (1.131) are always type-II NG modes in

[62] It is obvious that $\rho_{ab}$ vanishes in Lorentz-invariant theories as $j_c^0$ is the temporal component of the four vector $(j_c^0, \mathbf{j}_c)$. Also, in the case where $\rho_{ab}$ does not vanish it is an acceptable choice of of order parameter characterizing the spontaneous breaking of the generator $Q_a$.

[63] In particular, the NG bosons in Lorentz-invariant theories are all of type-A.

Eq. (1.127) [35]. However, although type-I modes are type-A modes and vice versa in most generic models, they are in all generality not the same. There exists at least one (trivial non-interacting) example in which the two classifications disagree [35]. In this example, via a fine-tuning of parameters in the theory it is possible to construct a type-A mode with a quadratic dispersion relation.[64] As a consequence, this example also satisfies a strict inequality $N_{\mathrm{I}} + 2N_{\mathrm{II}} > N_Q$.

Finally, we have to to briefly touch on the spontaneous breaking of spacetime symmetries. Unfortunately, the derivation of the counting rule in Eq. (1.131) assumes that the associated symmetry transformations do not act on spacetime. It therefore fails in the case where (some of) the broken generators $Q_a$ represent a spacetime symmetry [34]. In the case of both global and spacetime symmetries spontaneously breaking the counting can only be applied to the global part or subset of the spontaneously broken charges (that do not act on spacetime).[65] Just as it is the case for non-relativistic systems or systems with explicitly broken spacetime symmetries, in systems with spontaneously broken spacetime symmetries the number of independent fluctuations can be fewer that the number of broken generators. However, naively extending the counting rule Eq. (1.131) leads to wrong predictions [34]. On top of that, even if there are additional low-energy DoF originating from the spontaneous breaking of spacetime symmetries, some of them may not actually form a propagating mode as they can become over-damped via interactions with other modes [34]. There exist attempts at generalizing the counting rule to the case with spontaneously broken spacetime symmetries [38], however, it seems that at this point in time it is not possible to write down a general counting rule for the NG bosons in a convenient way.[66]

### 1.2.5 Broken spacetime symmetries: SSB at finite density

If spacetime symmetries are allowed to be spontaneously broken the zoology of allowed configurations of NG modes does not seem to be encompassed by a set of simple principles. However, that does not mean that these classes of theories are not physically relevant. In fact, they are sometimes very important. A particular class of systems that allow for spontaneously broken spacetime symmetries are systems at finite density. They are especially relevant in the context of the LQNE [47].
We study systems at finite density for a certain spontaneously broken charge $Q$. In such systems the time evolution is governed by a Hamiltonian $H$ and the ground state of the system is the state with the lowest eigenvalue with respect to the modified Hamiltonian [96]

$$H + \mu Q\,, \tag{1.132}$$

[64] One particular example is the rather simple free theory with the Lagrangian $\mathscr{L} = (\partial_0\phi^2 - \sum_k c_k \phi(-\nabla^2)^k\phi$. The shift symmetry of this Lagrangian is always spontaneously broken and by fine-tuning $c_k = \delta_{k2}$ the type-A NG boson present in this model for arbitrary $c_k$ actually becomes a type-II NG mode.

[65] Global and spacetime symmetries do commute with each other.

[66] It is worth noting that spontaneous breaking of local/gauge symmetries occurs in physical systems as well. SSB in gauge theories works slightly differently and is also not relevant in the context of this thesis. As opposed to global and spacetime symmetries the spontaneous breaking of gauge symmetries does not lead to massless modes in the spectrum [109]. Instead, it leads to massive gauge bosons [110]. This so-called Higgs mechanism is fundamental to the standard model of particle physics [111–113].

where $\mu$ is the chemical potential associated to the charge $Q$.[67] Generically, the introduction of finite density breaks boost invariance and time translations generated by $H$ and of course the internal charge $Q$.[68] Besides the breaking of $Q$ there can be an additional set of broken global internal charges $Q_1,\ldots,Q_n$. By construction, the modified Hamiltonian in Eq. (1.132) is unbroken.
There exists a slight modification of the Goldstone Theorem applicable to (relativistic) systems at finite density [42]. Besides the existence of soft NG modes in the spectrum this theorem also implies the existence of gapped NG modes in the spectrum. Consider a theory with a global internal symmetry group $G$ — a compact $n$-dimensional Lie group generated by the charges $Q_1, Q_2,\ldots,Q_n$ — at finite density for the charge $Q := Q_1$. We denote by $|0\rangle_\mu$ the state of minimal energy for a given average charge density of $Q$ labelled by the value of the chemical potential $\mu$. The state $|0\rangle_\mu$ minimizes the modified Hamiltonian,

$$(H+\mu Q)\,|0\rangle_\mu = E_\mu^{(\min)}\,|0\rangle_\mu\,. \tag{1.133}$$

The ground state $|0\rangle_\mu$ clearly (spontaneously) breaks boost invariance. Further, in order to guarantee that rotations and spatial translations remain unbroken we assume that the system is homogeneous and isotropic. Eq. (1.133) can be satisfied in two crucially different ways:

- The first possibility corresponds to the case where $|0\rangle_\mu$ is an eigenstate of both $H$ and $Q$ separately,

  $$E_\mu^{(\min)} = E_0 + \mu Q_0\,. \tag{1.134}$$

  In this case there is no spontaneous breaking of either $H$ nor $Q$. Physically, this situation is realized in e.g. Fermi liquids [44, 117–119]. In this case the internal symmetry at finite density is a simple $U(1)$, the particle number (in the non-relativistic limit).

- The second possibility corresponds the the case where $|0\rangle_\mu$ spontaneously breaks both $H$ and $Q$ and is only an eigenstate of the linear combination $H+\mu Q$. In particular, this situation is realized in the description of zero-temperature superfluid phases [96, 115].

We are interested in the the case of broken time translation $H$ and internal charge $Q$. Additionally, there may be a number of other spontaneously broken global generators $Q_{k_2},\ldots,Q_{k_m}$, with $m<n$ the total number of global internal generators. We set $Q_{k_1} = Q_1 = Q$. In particular, all generators that do not commute with $Q$ must be spontaneously broken as well, since

$$0 \neq \langle 0|_\mu\,[Q_a, Q_1]\,|0\rangle_\mu = \langle 0|_\mu\,Q'\,|0\rangle_\mu\,, \tag{1.135}$$

with $Q' = i f_{a1j} Q_j$ depending on the structure constants $f_{abc}$ of the algebra. Additionally, some generators that commute with $Q$ may also be spontaneously broken. By the definition of SSB [39], there exists an order parameter $\mathcal{O}_I^{(\mathrm{SB})}(0)$ transforming in a non-trivial representation of the global group $G$ such that

$$\langle\sigma_0\rangle_{k_i I} := \langle 0|_\mu\,[Q_{k_i}, \mathcal{O}_I^{(\mathrm{SB})}(0)]\,|0\rangle_\mu \neq 0\,, \qquad i=1,\ldots,m. \tag{1.136}$$

[67]The charge $Q$ by itself may be thought of as the generator of a global $U(1)$, as this is the only possibility allowed by Lie theory [114]. Of course, it generically generates a certain $U(1)$ subgroup of a much larger Lie group.
[68]This is certainly true in all condensed matter states [115]. In particular, this is the symmetry-breaking pattern defining a zero-temperature superfluid phase [116].

A similar equation holds for the Hamiltonian $H$. As $H$ is spontaneously broken, it is not possible to classify states in terms of their energy eigenvalues. Instead, they are best classified in terms of their eigenvalues under the modified Hamiltonian $H+\mu Q$ and the state $|0\rangle_\mu$ represents the ground state with respect to $H+\mu Q$.[69]

We insert the unity operator $\mathbb{1}$ in terms of all (multi-)particle eigenstates of the modified Hamiltonian $H+\mu Q$ in Eq. (1.136),

$$\begin{aligned}
\langle 0|_\mu [Q_{k_i}, \mathcal{O}_I^{(\mathrm{SB})}(0)] |0\rangle_\mu &= \int \mathrm{d}^{D-1}x \left( \langle 0|_\mu j^0_{k_i}(t,\mathbf{x}) \mathcal{O}_I^{(\mathrm{SB})}(0) |0\rangle_\mu - \langle 0|_\mu \mathcal{O}_I^{(\mathrm{SB})}(0) j^0_{k_i}(t,\mathbf{x}) |0\rangle_\mu \right), \\
\langle 0|_\mu j^0_{k_i}(t,\mathbf{x}) \mathcal{O}_I^{(\mathrm{SB})}(0) |0\rangle_\mu &= \int \frac{\mathrm{d}^{D-1}k}{(2\pi)^{D-1}} \sum_n e^{-iE_{n_\mathbf{k}}t} \langle 0|_\mu e^{-it\mu Q} j^0_{k_i}(0) e^{it\mu Q} |n_\mathbf{k}\rangle \langle n_\mathbf{k}| \mathcal{O}_I^{(\mathrm{SB})}(0) |0\rangle_\mu e^{i\mathbf{k}\cdot\mathbf{x}}, \\
\langle 0|_\mu \mathcal{O}_I^{(\mathrm{SB})}(0) j^0_{k_i}(t,\mathbf{x}) |0\rangle_\mu &= \int \frac{\mathrm{d}^{D-1}k}{(2\pi)^{D-1}} \sum_n e^{iE_{n_{-\mathbf{k}}}t} \langle 0|_\mu \mathcal{O}_I^{(\mathrm{SB})}(0) |n_{-\mathbf{k}}\rangle \langle n_{-\mathbf{k}}| e^{it\mu Q} j^0_{k_i}(0) e^{-it\mu Q} |0\rangle_\mu e^{-i\mathbf{k}\cdot\mathbf{x}}.
\end{aligned} \tag{1.137}$$

Without loss of generality we have chosen $\mathcal{O}_I^{(\mathrm{SB})}$ to be Hermitian. The currents themselves do not need to be written in a real basis. It is crucial to note that $e^{i\mu Q}$ does not commute with $j^0_{k_i}$ in general. Using the algebra of $G$ and the structure constants $f_{1ac} =: (f_1)_{ac}$ we simplify the adjoint action on $j^0_{k_i}$,[70]

$$e^{-it\mu Q} j^0_{k_i}(0) e^{it\mu Q} = \left(e^{t\mu(f_1)}\right)_{k_1 c} j^0_c(0). \tag{1.138}$$

We introduce the spectral density for two operators $\mathcal{O}_{1,2}$,

$$S(\mathcal{O}_1, \mathcal{O}_2; \omega, \mathbf{k}) := 2\omega \sum_n (2\pi)\delta\left(\omega - E_{n_\mathbf{k}}\right) \langle 0|_\mu \mathcal{O}_1(0) |n_\mathbf{k}\rangle \langle n_\mathbf{k}| \mathcal{O}_2(0) |0\rangle_\mu . \tag{1.139}$$

Using the spectral density we can rewrite $\langle \sigma_0 \rangle_{k_i I}$ Eq. (1.136) as follows:[71]

$$\langle \sigma_0 \rangle_{k_i I} = \frac{1}{2\pi} \int \frac{\mathrm{d}\omega}{2\omega} e^{-i\omega t} \left(e^{t\mu(f_1)}\right)_{k_i c} \lim_{\mathbf{p}\to 0} \left[ S(j^0_c, \mathcal{O}_I^{(\mathrm{SB})}; \omega, \mathbf{p}) - S(\mathcal{O}_I^{(\mathrm{SB})}, j^0_c; -\omega, -\mathbf{p}) \right] \neq 0, \tag{1.140}$$

where it holds that $S(\mathcal{O}_I^{(\mathrm{SB})}, j^0_c; -\omega, -\mathbf{p}) = S(j^0_c, \mathcal{O}_I^{(\mathrm{SB})}; -\omega, -\mathbf{p})^\dagger$. The order parameter $\langle \sigma_0 \rangle_{k_i I}$ is non-zero due to the spontaneous breaking of $Q_{k_i}$.

Consider the broken generators $Q_{k_i}$ that commute with $Q$. Without loss of generality we take $Q_{k_i}$ Hermitian, hence $\left(e^{t\mu(f_1)}\right)_{k_i c} = \delta_{k_i c}$. As a consequence, in order to satisfy Eq. (1.140) we require that

$$\lim_{\mathbf{p}\to 0} \left[ S(j^0_{k_i}, \mathcal{O}_I^{(\mathrm{SB})}; \omega, \mathbf{p}) - S(j^0_{k_i}, \mathcal{O}_I^{(\mathrm{SB})}; -\omega, -\mathbf{p})^\dagger \right] = 4\pi\omega i m_{k_i I} \delta(\omega). \tag{1.141}$$

This result implies the existence of a Goldstone mode $|n^{\mathrm{NG}}_\mathbf{k}\rangle$ in the spectrum such that its zero-momentum limit is an exact eigenstate of $H+\mu Q$ with vanishing energy [104],

$$E_{n^{\mathrm{NG}}_\mathbf{k}}\Big|_{\mathbf{k}=0} = 0. \tag{1.142}$$

The state $|n^{\mathrm{NG}}_\mathbf{k}\rangle$ has zero matrix element with both the current and the order parameter. As $G$ is a compact Lie group, all the generators that do not commute with $Q$ (not just the broken ones) can

[69] There exist other options to classify states [43].

[70] We ignore potential t'Hooft anomalies as they are irrelevant in the limit $\mathbf{k} \to 0$.

[71] We introduce an integration over $\omega$ and perform the integration over $\mathbf{x}$ in the infinite volume limit $\Omega \to \infty$.

generically be split up into pairs of generators $Q_a^{\pm}$ such that

$$[Q, Q_a^{\pm}] = \pm q_a Q_a^{\pm}\,, \tag{1.143}$$

with the constant $c_a > 0$ depending on the underlying algebra [114]. The change of basis to $\{Q_a^{\pm}\}$ is equivalent to diagonalizing the adjoint action of $Q$. In particular, this holds true for the set of broken generators that do not commute with $Q$, which we now denote by $\{Q_{a_i}^{\pm}\} \subset \{Q_{k_i}\}$. The associated currents we denote by $j^{\mu}_{a_i^{\pm}}$ and it holds that $(j^{\mu}_{a_i^{\pm}})^{\dagger} = j^{\mu}_{a_i^{\mp}}$.
In this basis the adjoint action of $Q$ in Eq. (1.138) reads

$$\left(e^{t\mu(f_1)}\right)_{a_i^{\pm}c} j_c^{\mu} = e^{\mp t\mu q_{a_i}} j^{\mu}_{a_i^{\pm}}\,, \qquad \left(\left(e^{t\mu(f_1)}\right)_{a_i^{\pm}a_j^{\pm}} = e^{\mp t\mu q_{a_i}} \delta_{a_i a_j}\,,\ \left(e^{t\mu(f_1)}\right)_{a_i^{\pm}a_j^{\mp}} = 0\right), \tag{1.144}$$

with $q_{a_i}$ being the constant in Eq. (1.143). Now Eq. (1.140) turns into

$$\langle \sigma_0 \rangle_{a_i^{\pm} I} = \frac{1}{2\pi} \int \frac{\mathrm{d}\omega}{2\omega} e^{-i(\omega \pm \mu q_{a_i})t} \lim_{\mathbf{p}\to 0} \left[ S(j^0_{a_i^{\pm}}, \mathscr{O}_I^{(\mathrm{SB})}; \omega, \mathbf{p}) - S(j^0_{a_i^{\mp}}, \mathscr{O}_I^{(\mathrm{SB})}; -\omega, -\mathbf{p})^{\dagger} \right]. \tag{1.145}$$

Assuming definiteness $\mu q_{a_i^{\pm}} > 0$ Eq. (1.145) can only be satisfied if[72]

$$S(j^0_{a_i^-}, \mathscr{O}_I^{(\mathrm{SB})}; \omega, \mathbf{p}) = 4\pi\omega i\, m_{a_i^- I} \delta(\omega - \mu q_{a_i})\,, \qquad \lim_{\mathbf{p}\to 0} S(j^0_{a_i^+}, \mathscr{O}_I^{(\mathrm{SB})}; \omega, \mathbf{p}) = 0\,. \tag{1.146}$$

The first condition implies the existence of a gapped NG state $|n^{\mathrm{NG}}_{\mu;\mathbf{k}}\rangle$ for every pair of broken generators $Q_{a_i}^{\pm}$ [42, 43, 120, 121]. In the zero-momentum limit these gapped NG states are eigenstates of $H + \mu Q$ with energy linear in $\mu$,

$$E_{n^{\mathrm{NG}}_{\mathbf{k}}}\Big|_{\mathbf{k}=0} = \mu q_{a_i}\,. \tag{1.147}$$

Both matrix elements $\langle 0|_{\mu}\, j^0_{a_i^-}\, |n^{\mathrm{NG}}_{\mu;\mathbf{k}}\rangle$ and $\langle 0|_{\mu} \mathscr{O}_I^{(\mathrm{SB})}\, |n^{\mathrm{NG}}_{\mu;\mathbf{k}}\rangle$ are non-zero while the matrix element $\langle 0|_{\mu}\, j^0_{a_i^+}\, |n^{\mathrm{NG}}_{\mu;\mathbf{k}}\rangle$ vanishes in the zero-momentum limit.
As there are non-commuting broken charge operators present, the counting rule around Eq. (1.128) implies the existence of additional gapless NG modes — of type-B or type-II — for the spontaneously broken global symmetries not directly guaranteed by the above results. However, we do have to be careful here as there are spontaneously broken spacetime symmetries and the counting in Eq. (1.128) can only be applied to the broken global charges that do not act on spacetime.[73] In particular, explicit example of systems at finite density that do exhibit type-II modes appear in the context of the LCE in CFT [122–124].
Although maybe conceptually useful, any assumption of an underlying Lorentz invariance at zero density is not required in the proof of Eq. (1.141) and Eq. (1.146). The only requirement is unbroken rotations and spatial translations [45]. Hence, these results apply independently of the mechanism responsible for the breaking of boosts. Naturally, one might wonder whether spontaneously breaking of Lorentz invariance implies additional constraints on the spectrum. Unfortunately, this is not true in the case of the spontaneous breaking of boosts as has explicitly been shown in [44]. In general, when spacetime symmetries are broken it can happen that a single physical fluctuation may be described as the action of several different generators [41].

[72] We make use of the fact that positivity of the spectrum implies that the spectral density vanishes for negative frequencies — *i.e.* $S(j^0_{a_i^+}, \mathscr{O}_I^{(\mathrm{SB})}; \omega, \mathbf{p}) = 0$ for $\omega < 0$ — see Eq. (1.139).

[73] Technically, both $H$ and $Q$ are broken. However, the modified time-evolution operator $H + \mu Q$ replaces $H$ such that there is only one broken symmetry generator to be counted, which generically should be considered as a global symmetry. This is certainly true in all the LCE applications.

There is one comment to be made about EFT building in the presence of gapped NG modes. Generically, the low-energy dynamics of the NG modes are captured in terms of symmetries and a systematic derivative expansion [97, 98]. The gapped NG modes with a gap of order $\mu$ are needed to construct an EFT that is invariant under the full symmetry group. However, in certain examples the chemical potential itself presents the EFT cut-off (like in the LCE for certain non-Abelian symmetries [17, 18]). It can be demonstrated that a consistent EFT can be constructed by zooming in on all NG modes at small spatial momenta [46]. The gapless NG modes are soft while the gapped ones are slow and the rules are the same as in standard non-relativistic EFTs. The EFT Lagrangian formally preserves gapped NG boson number conservation, and processes that violate the number conservation are modelled inclusively by allowing complex Wilsonian coefficients, thus violating unitarity while preserving the symmetry.
If the cut-off scale $\Lambda \ll \mu$ is much smaller that the chemical potential the gapped modes can be integrated out [46]. Doing this is equivalent to specifying boundary conditions for the gapped NG modes to vanish at infinity. As a consequence, integrating out the gapped NG modes generically breaks the non-Abelian symmetry (gapped NG modes can only exist for non-Abelian symmetries) such that only the $U(1)$ symmetry associated to the chemical potential $\mu$ remains.[74]

### 1.2.6 Abelian superfluids

As an example, we investigate a particular class of finite density systems: Abelian superfluids. More precisely, we consider a relativistic superfluid with an Abelian $U(1)$ symmetry generated by the broken generator $Q$ [116]. The minimal field content of an Abelian superfluid consists of a compact Lorentz scalar that non-linearly realizes the $U(1)$ symmetry via a shift invariance,[75]

$$\chi(x) \sim \chi(x) + 2\pi\,, \qquad\qquad \chi(x) \to \chi(x) + \alpha\,. \tag{1.148}$$

The most general superfluid Lagrangian is a derivative expansions and reads

$$S_{\mathrm{SF}} = \int \mathrm{d}^D x\, \left[\mathscr{L}_1(\partial\chi) + \partial_\mu\chi\partial_\nu\chi\partial^\mu\partial^\nu\chi\mathscr{L}_2(\partial\chi) + \dots\right], \tag{1.149}$$

where $\partial\chi = \sqrt{\partial_\mu\chi\partial^\mu\chi}$ and $\mathscr{L}_{1,2}$ are arbitrary functions.[76] The action $S_{\mathrm{SF}}$ is a low-energy effective action and has to be understood as an expansion around the finite-density ground state

$$\langle\chi\rangle = \mu t + \chi_0\,, \tag{1.150}$$

[74] This exact behaviour can be observed in the LCE for the $O(2N)$ vector model [46, 47].
[75] $\chi$ is generically the phase of a complex field in the underlying theory.
[76] If the chemical potential $\mu$ itself sets the cut-off for the low-energy descriptions, then $F_1$ and $F_2$ are monomials with potentially non-integer powers: $\partial^n F_1(\mu) \sim F_1(\mu)/\mu^n$, $\partial^n F_2(\mu) \sim F_1(\mu)/\mu^{n+4}$.

with $\chi_0$ =const. and $\mu$ being the chemical potential for $Q$. More precisely, as only functions that are $2\pi$-periodic in $\chi(x)$ are well-defined operators, the ground state $|0\rangle_\mu$ satisfies

$$\langle 0|_\mu \, e^{i\chi(x)} \, |0\rangle_\mu = e^{i\mu t + i\chi_0} \,. \tag{1.151}$$

While the action in Eq. (1.149) is Poincaré invariant, the ground state spontaneously breaks boosts and time translation $H$ as well as the $U(1)$ charge $Q$. The linear combination $H + \mu Q$ remains unbroken. The energy momentum tensor and the conserved currents are given (to leading order) by

$$T_{\mu\nu} = \mathscr{L}_1'(\partial\chi)\,\partial\chi \,\frac{\partial_\mu \chi}{\partial\chi}\,\frac{\partial_\nu \chi}{\partial\chi} - \eta_{\mu\nu}\mathscr{L}_1(\partial\chi)\,, \qquad j_\mu = \mathscr{L}_1'(\partial\chi)\,\frac{\partial_\mu\chi}{\partial\chi}\,, \tag{1.152}$$

where $\partial_\mu\chi/\partial\chi$ is the superfluid four-velocity. The background solution in Eq. (1.150) produces a non-zero ground state charge density

$$j^0(\mu) := j^0\Big|_{\chi=\langle\chi\rangle} = \mathscr{L}_1'(\mu)\,. \tag{1.153}$$

The ground state energy density $e(\mu)$ and the pressure $p(\mu)$ as a function of $\mu$ are obtained by inverting the equation $T_{\mu\nu} = (e+p)(\partial_\mu\chi/\partial\chi)(\partial_\nu\chi/\partial\chi) - p\eta_{\mu\nu}$ [125]. On the ground state solution we recover the zero-temperature thermodynamic identity [126]

$$\mu\, j^0(\mu) = e(\mu) + p(\mu)\,. \tag{1.154}$$

Fluctuations $\pi(x)$ are defined by $\chi(x) = \mu t + \chi_0 + \pi(x)$. The action to leading order in derivatives and to quadratic order in the fluctuations reads

$$S_{\rm SF} \supset \left(\mathscr{L}_1''(\mu) + \ldots\right)\int \mathrm{d}^D x\, \frac{1}{2}\left[\dot\pi^2 - c_s^2(\mu)\,(\nabla\pi)^2\right], \qquad c_s^2(\mu) = \frac{\mathscr{L}_1'(\mu)}{\mu\mathscr{L}_1''(\mu)} + \ldots\,. \tag{1.155}$$

In agreement with the Goldstone Theorem and the results at finite density, the field $\pi(x)$ describes a massless excitation with linear dispersion relation $\omega = c_s|\mathbf{k}| + \ldots$. This NG mode is the so-called superfluid phonon [40].

Generically, it holds that $c_s \neq 1$ and hence $\pi(x)$ does not represent a relativistic field. However, despite the absence of any NG modes for the broken boosts, Lorentz invariance still constrains the superfluid action as it requires it to be a function of

$$\partial\chi = \sqrt{(\dot\pi + \mu)^2 - (\nabla\pi)^2}\,, \tag{1.156}$$

instead of any arbitrary combination of $\dot\pi$ and $\nabla\pi$. Finally, we remark on a simple generalization. Instead of systems with a single charge $Q$ at finite density, with minimal adaptations, it is possible to consider systems at finite density for $N$ mutually commuting broken charges $Q_k$, $k = 1, \ldots, N$. The $U(1)$ broken symmetry of the superfluid can be extended to a fully broken $U(1)^N$ symmetry. The most general form of the modified Hamiltonian becomes

$$H + \sum_k \mu_k Q_k\,. \tag{1.157}$$

The general low-energy effective action for the $U(1)^N$ superfluid is best formulated in terms of $N$ shift-invariant compact scalars $\chi_k \sim \chi_k + 2\pi$. The leading order action in the derivative expansion reads

$$S_{\rm SF} = \int \mathrm{d}^D x\, \mathscr{L}_1(\partial\chi_{ij}) + \dots\,, \qquad \partial\chi_{ij} = \sqrt{\partial_\mu \chi_i \partial^\mu \chi_j}\,. \tag{1.158}$$

The action is again to be understood as an expansion around the background solution

$$\chi_k(x) = \mu_k \tau + \chi_{k0} + \pi_k(x)\,, \qquad\qquad k = 1,\dots,N \tag{1.159}$$

Upon expanding to quadratic order it is found that the spectrum consists of $N$ massless superfluid phonons with linear dispersion relations. This is not really surprising as all the broken charges $Q_k$ are mutually commuting.

### 1.2.7 Finite-volume effects on spontaneous symmetry breaking

The definition of SSB via the existence of a broken generator in Eq. (1.114) suggests that the phenomenon can only occur in infinite systems and is absent in any theory living on a finite compact manifold.[77] However, in the context of the LCE in CFT and, in particular, in this thesis we often deal with finite-volume systems that are expected to exhibit SSB and whose low-energy dynamics are dictated by the Goldstone Theorem. In the LCE — by the state-operator correspondence — the system in flat space can be mapped to the cylinder, which is by definition a finite-volume setting. As such, the LCE requires two prerequisites that are seemingly at odds: finite volume and SSB.
For a system exhibiting SSB the algebra of observables of the theory consists of a family of inequivalent irreducible representations, each labelled by some equivalent orthogonal ground state as defined in Eq. (1.117). The broken symmetry maps representations onto each other. However, it is not a well-defined unitary operator. Hence, each representation is endowed with its own non-separable Hilbert space. The argument against SSB in finite volume systems is that all states in the theory live in a unique separable Hilbert space. As the charge operator of the (broken) symmetry is well-defined in finite volume — see Eq. (1.113) — the symmetry is realized by some well-defined unitary operator acting on the Hilbert space. This operator relates all the possible different minima in Eq. (1.117) (*i.e.* the other vacua) to the unique ground state $|0\rangle$.[78]
The tension between finite volume and SSB can be somewhat resolved in finite-density systems. As can be shown (in explicit examples), at finite density for the broken charge finite-volume effects are generically exponentially suppressed (for $D > 2$).[79] To illustrate this point consider a system with a global $U(1)$ symmetry at finite density and finite volume $V$. In infinite volume the $U(1)$ symmetry is spontaneously broken by the ground state of the theory. The fluctuations are described by the low-energy effective Lagrangian

$$\mathscr{L} = -\frac{f(\mu)}{2}\left(-(\partial_0\varphi)^2 + c_s(\mu)(\nabla\varphi)^2\right), \tag{1.160}$$

with $\varphi$ being a dimensionless real scalar field, $c_s(\mu)$ being the speed of sound and $f(\mu)$ being a dimensionful constant — $[f] = (D-2)$ — that generically depends on the VEV of the initial field as well as the chemical potential $\mu$ for the global $U(1)$ charge.[80] The $U(1)$ symmetry is non-linearly realized as a shift,

$$\varphi \mapsto \varphi + \alpha. \tag{1.161}$$

[77]The issue with finite systems hints at a much more foundational problem with our understanding of SSB. By definition, SSB appears to be an example of asymptotic emergence as no finite system should exhibit it. However, SSB is an observable effect in nature and there are numerous real systems and materials that have been shown to exhibit SSB in a finite experimental setting. This seems to put the theoretical description of SSB at odds with physical reality and contradicts the principle that no effect can be counted as a genuine physical effect if it disappears upon removing the idealizations in the theory [127]. This discrepancy between formalism and reality can be amended by exponential sensitivity to perturbations of the dynamics as the size of the system increases, causing the occurence of SSB already in finite but very large quantum systems, thus providing continuity between finite- and infinite-volume descriptions [49].

[78]We focus here on some particular aspects of finite-size effects on SSB. For a more comprehensive understanding see [48].

[79]Analogously, it can be stated that in a sector of the Hilbert space with fixed quantum number for the broken symmetry the above mentioned finite-volume effects are exponentially suppressed. Finite density systems and systems at fixed quantum number are generically related by a Legendre transform [1, 47, 128].

If we assume that in the IR the system is adequately described by an Abelian superfluid, then we have the relationship

$$f(\mu) = \frac{j^0(\mu)}{\mu c_s(\mu)}, \tag{1.162}$$

with $j^0(\mu) = Q_0/V$ being the non-zero charge density of the superfluid ground state. For more details see the short discussion of the Abelian superfluid in Section 1.2.6. The mode expansion of the field $\varphi(t,\vec{x})$ is[81]

$$\varphi(t,\mathbf{x}) = \varphi_0 + \pi_0 t + \sum_{k=1}^{\infty} \frac{1}{\sqrt{2V\omega}}\left(a(\mathbf{k})\, e^{-i\omega t + i\mathbf{k}\cdot\mathbf{x}} + a^\dagger(\mathbf{k})\, e^{i\omega t - i\mathbf{k}\cdot\mathbf{x}}\right), \qquad \omega = c_s|\mathbf{k}|, \tag{1.163}$$

The zero modes $\varphi_0$ and $\pi_0$ only appear in finite volume. By promoting the field to an operator — $\Pi(t,\mathbf{x}) = \partial_0\varphi(t,\mathbf{x})$ — and imposing equal-time canonical commutation relations,

$$[\varphi(t,\mathbf{x}), \Pi(t,\mathbf{y})] = i\delta^{(d)}(\mathbf{x}-\mathbf{y}), \tag{1.164}$$

we arrive through the standard covariant quantization procedure at the commutation relations for creation and annihilation operators [129],[82]

$$[a(\mathbf{k}), a^\dagger(\mathbf{k}')] = \delta^{D-1}_{\mathbf{k}\mathbf{k}'}, \qquad [\varphi_0, \pi_0] = \frac{i}{V}. \tag{1.165}$$

Note that $V^{1/(D-1)}$ is the scale fixed by the geometry. It is convenient to also express the zero modes $\varphi_0, \pi_0$ in terms of ladder operators $a_0, a_0^\dagger$,

$$a_0 = \frac{V^{\frac{(D-2)}{2(D-1)}}}{\sqrt{2}}\left(\varphi_0 + iV^{\frac{1}{(D-1)}}\pi_0\right), \qquad a_0^\dagger = \frac{V^{\frac{(D-2)}{2(D-1)}}}{\sqrt{2}}\left(\varphi_0 - iV^{\frac{1}{(D-1)}}\pi_0\right), \qquad [a_0, a_0^\dagger] = 1. \tag{1.166}$$

[80] In the context of a CFT at fixed charge the constant $f$ can only depend on the scale $\rho^{1/(D-1)}$ fixed by the charge density — $f^2 = c_f \rho^{(D-2)/(D-1)}$, $[c_f] = 0$ — as there is no other scale.

[81] Modes are discrete since we are in finite volume.

[82] The creation and annihilation operators are given by $\left(a(\mathbf{k}), a^\dagger(\mathbf{k})\right) = e^{\pm i\omega t}\int_V \frac{\mathrm{d}^{D-1}x}{\sqrt{2V\omega}}\, e^{\mp i\mathbf{k}\cdot\mathbf{x}}\left[\omega\varphi(t,\mathbf{x}) \pm i\Pi(t,\mathbf{x})\right]$.

As the non-zero modes integrate to zero, the fluctuation part of the $U(1)$ global charge operator is given by

$$Q = \int_V \mathrm{d}^{D-1}x \, \frac{\partial \mathscr{L}}{\partial(\partial_0 \varphi)} = \frac{i f(\mu)}{\sqrt{2}} V^{\frac{(D-2)}{2(D-1)}} \left(a_0^\dagger - a_0\right). \tag{1.167}$$

Via the charge operators $Q$ we can introduce the other equivalent vacua of the theory — in the same way it has been done in Eq. (1.116) — for arbitrary real $\theta$,

$$|\theta\rangle = e^{i\theta Q} |0\rangle = \exp\left(\theta \frac{f(\mu)}{\sqrt{2}} V^{\frac{(D-2)}{2(D-1)}} \left(a_0 - a_0^\dagger\right)\right)|0\rangle \,. \tag{1.168}$$

The overlap of the vacuum $|\theta\rangle$ with the Ground State (GS) $|0\rangle$ is computed using the BCH formula and reads

$$\langle 0|\theta\rangle = \langle 0| \exp\left(\theta \frac{f(\mu)}{\sqrt{2}} V^{\frac{(D-2)}{2(D-1)}} \left(a_0 - a_0^\dagger\right)\right)|0\rangle = e^{-\frac{\theta^2 j^0(\mu)^2}{4\mu^2 c_s(\mu)^2} V^{\frac{(D-2)}{2(D-1)}}} \,. \tag{1.169}$$

The overlap between equivalent vacua is exponentially suppressed for large charge densities $\rho = j^0(\mu)$ and large volumes $V$.[83] In particular, in the case of a conformal superfluid we can write the overlap between vacua as

$$\langle 0|\theta\rangle = e^{-\theta^2 \frac{c_f}{4} \left(\frac{\rho}{V}\right)^{\frac{(D-2)}{(D-1)}}} \,. \tag{1.170}$$

At finite volume $V < \infty$ all the vacua $|\theta\rangle$ have a finite non-zero overlap and the system will eventually relax to the true vacuum $|0\rangle$. In infinite volume, as $Q$ is not well-defined, each vacuum $|\theta\rangle$ belongs to a different Hilbert space and the system exhibits SSB. The large-charge limit $\rho \gg 1$ lies somewhat in between: different vacua will have a non-zero overlap, however, in the large-charge limit this overlap is exponentially suppressed (for $D > 2$). This means that, for all practical intents and purposes, the system can be treated as if there was SSB and we can apply the NG theorem up to exponential corrections.

[83] There is a caveat here as the equation $\rho = \rho(\mu)$ has to be inverted to deduce $\mu = \mu(\rho)$. As long as $\rho(\mu)$ is polynomial the speed of sound is generically constant in $\mu$ to leading order, see Eq. (1.153) and Eq. (1.155). Additionally, if $\rho(\mu)$ to leading order is at least linear in $\mu$ the exponent in Eq. (1.169) to leading order is at least constant in $\rho$. In that case the overlap is exponentially suppressed.

# 2 The large quantum number expansion in CFT and the $O(2)$ model

CFTs are strongly constrained by the extended spacetime symmetries they respect. However, due to scale invariance they do not allow for any dimensionful parameters — in particular coupling constants — within the theory and are generically strongly-coupled. Besides free massless theories there are very few weakly-coupled CFTs in physics, with one notable exception here being Caswell–Banks–Zaks (CBZ)-type fixed points [130–132]. Strongly coupled CFTs are also particularly important because they live at the end points of RG flows and therefore are important landmarks in the space of QFTs.

Besides CFTs there are many other interesting theories that do not possess a small coupling. And there are plenty of examples of strongly coupled theories in which a weakly coupled description emerges in some sector of the theory once an expansion parameter is artificially introduced. Notable examples are small-$\epsilon$ expansions [133] and large-$N$ expansions [134], both rather universal procedures applicable to many different theories. For CFTs in $D > 2$, following the advent of the AdS/CFT correspondence, a novel set of perturbative frameworks has been developed revolving around large quantum numbers.[1] Besides large-spin expansions for spinful operators [90, 136, 137] this includes Large-Charge Expansions (LCEs) [17, 18]. The LCE is generically applicable to CFTs that are invariant under a global internal symmetry group $G$, as theories of this type, generally speaking, possess an emergent perturbative regime described by an effective theory when studying operators with large quantum numbers under the symmetry $G$, even though they are often strongly coupled.

This chapter is separated into two parts. The first part serves as an introduction to the description of CFT operators at large charge and the LCE. It is an original presentation of materials and takes inspiration from many different places in the literature. A few sources that deserve to be highlighted are the works by Álvarez-Gaumé et al. [47], Monin et al. [18], Cuomo [46] and Kalogerakis [138].

The second part of the chapter presents an up-to-date analysis of the large-charge sector of $O(2)$-invariant CFTs and follows [2]. $O(2)$ models are the simplest working example in which the LCE can be applied. Nevertheless, this analysis is consequential for important physical theories such as the $O(2)$ WF fixed point [8, 139] which in $D = 3$ describes the superfluid transition in liquid Helium.[2] We systematically gather and study CFT data by computing two-, three- and four-functions accessible at large charge. A detailed outline of the contents will be given separately at the beginning of the section.

[1] Large-$N$ and small-$\epsilon$ expansions have been successfully applied to study CFTs as well, see *e.g.* [8, 135].

[2] Another example would be bosonic gauge theories with monopole operators [140], where the $O(2)$ symmetry is given by the associated topological charge.

All computations in this chapter are performed in Euclidean signature with the dimension $D$ of spacetime mostly kept arbitrary. As discussed, Section 2.1 represents an original discussion of the current understanding of the structure of CFT data at large charge. On the other hand, Section 2.2 presents original contributions of the author. It follows closely the material in [2] with some modifications.

## 2.1 The large charge program: Sectors of large charge in (strongly coupled) CFTs

We give a brief and general introduction into the most important aspects of the LCE and outline the current landscape of its applications and interplay with other methods of accessing CFT data. In doing so, we discuss some of the most relevant and interesting results in the academic literature.

### 2.1.1 Accessing CFT data

As discussed in detail in Section 1.1, CFTs are highly constrained and all $N$-point functions of local operators can essentially be computed from the CFT data. Worded differently, having access to all two- and three-point functions allows for the recursive computation of any $N$-point function. However, this is where issues arise, as in a (strongly-coupled) CFT it is not possible to compute correlators perturbatively in a small expansion parameter in the way it is usually done in QFT. Hence, for strongly-coupled interacting CFTs we require other ways of accessing correlators and compute the CFT data. Non-perturbative methods of accessing CFT data in strongly-coupled theories can be roughly classified into three categories.

#### Monte–Carlo methods and simulations

Monte–Carlo methods refers to a loose set of numerical algorithms that rely on recurring random sampling to attain results. Monte–Carlo methods have been successfully applied to CFT in order to compute CFT data and critical exponents [141, 142]. They are best suited for (generalized) spin lattice systems — such as the Ising model — and the corresponding CFTs that are found at the critical points of such systems.
Results from Monte–Carlo methods for the computation of observables involving light operators — *i.e.* operators with small scaling dimensions — are currently less precise than equivalent results from the conformal bootstrap, which we will discuss next. However, they are more versatile and can be applied to compute operators with larger scaling dimensions as well [143–146], although with limited accuracy, in contrast to the conformal bootstrap. Still, Monte–Carlo methods work best and are most widely used for lighter operators. Finally, it deserves to be mentioned that there are other inventive numerical techniques used to study CFTs such as fuzzy-sphere regularization [147], which are not directly related to Monte–Carlo simulations.

#### Conformal bootstrap methods

As we have extensively discussed the conformal bootstrap in Section 1.1.9, we will be brief here. In short, the conformal bootstrap exploits the structure of CFTs to analytically constrain and numerically solve for the conformal data. It is independent of any Lagrangian description and works with the CFT data as (independent) variables directly, constraining them via the crossing equation Eq. (1.108) [13, 92]. Its numerical constraints give the most accurate predictions for the lightest operators in the $D = 3$ Ising model to date [21].
Generally speaking, the conformal bootstrap is very efficient in computing conformal data of the lightest operators in the theory. Beyond, for heavier operators, it quickly becomes less viable.[3] There is clear overlap between Monte–Carlo methods and the bootstrap as they are used to analyse similar sectors of the CFT data.[4]

#### Artificial perturbative frameworks

Here, we refer to any analytic approach that aims to compute CFT data perturbatively by artificially introducing or constructing a small expansion parameter in the theory at hand. There are many different approaches to this end and it is not exactly a unique methodology. Accepting certain restrictions, the conformal data can then be analytically calculated in an asymptotic expansion in the small expansion parameter. Examples with widespread use here — not just in the study of CFTs — are *e.g.* large-$N$ expansions [134] and small-$\epsilon$ expansions [133]. In particular, large quantum number expansions, which we are interested in, belong in this class of methods.

The interest in large quantum number expansions for CFTs has its origin in the development of large-spin expansions [136]. The large-spin expansion originates from the fact that it is possible to extract information about operators with large spin from the crossing equation Eq. (1.110). Written in terms of the conformal cross ratios $z, \bar{z}$ — defined in Eq. (1.103) — in the limit $z \to 0$ with $\bar{z}$ fixed it is possible to extract information from Eq. (1.110) about operators with large values of the spin $\ell$ [90, 149–151].
Evidently, by virtue of the unitarity bounds in Eq. (1.76), the large-spin program probes operators with large scaling dimensions $\Delta(\ell) \geq \ell$. To leading order in the large-spin expansion we find a Regge-like behaviour with scaling dimension [137, 150, 152]

$$\Delta(\ell) = \ell + \dots . \tag{2.1}$$

With the seminal paper [17] and follow-up work [18] it was established that a similar simplification manifests itself for operators with large quantum numbers under a global $O(2)/U(1)$ (sub-)symmetry. It was found that scaling dimensions for large-charge operators behave as

$$\Delta(Q) = c_1 Q^{3/2} + \dots . \tag{2.2}$$

[3] The heavier the operators involved in the crossing equation Eq. (1.108) are, the more computationally involved conformal bootstrap techniques become.
[4] Monte–Carlo methods can also be applied to the conformal bootstrap [148].

The analytic methods universally applicable to any CFT — the conformal bootstrap, large spin and the LCE — generically access different areas of the conformal data. Whereas the conformal bootstrap allows for precision calculations of light operators, the asymptotic expansions associated to large-spin and large-charge methods compute heavy operator dimensions. The different areas of CFT data that are accessible by analytical means are schematically shown in Figure 2.1.

The large-spin expansion and the conformal bootstrap both reference the crossing equation Eq. (1.108). Since the LCE is mainly based on the state–operator correspondence, emerging condensed matter phases and semi-classical analysis, it has little to do with the crossing equation or the conformal bootstrap directly. As it became clear rather quickly that the LCE could also be applied to compute higher-point correlators and especially three-point coefficients [18], Jafferis et al. explored large-charge operators via the conformal bootstrap [154]. The authors found perfect agreement between the two methods, which is remarkable as it seems to imply that the emergent condensed-matter phases at large charge — in particular the conformal superfluid phase — are intrinsically encoded by the corresponding conformal bootstrap constraints.
As both large-charge and large-spin methods probe heavy operators in the theory, there is a lot of interplay between the two approaches and there are several different phases emerging along the trajectory in-between the two limits where either $J$ or $Q$ is parametrically small [152, 155–157]. We will discuss this in more details in Section 2.1.3.

### 2.1.2 Large-charge sectors in CFTs and their properties

The LCE is useful for the computation of CFT data, as it allows us to access information about heavy operators that is not accessible to the same level of precision with other methods. Consider a CFT invariant under a global internal symmetry group $G$. We choose a one-parameter subgroup $U(1)_{G_0} \subset G$ — which is equivalent to selecting a particular generator $Q_0^{(G)}$ of $G$ — and we consider the lowest-lying operator $\mathcal{O}^Q$ with charge $Q \gg 1$ under $U(1)_0$ — *i.e.* the operator minimizing the scaling dimension $\Delta = \Delta(Q)$ at fixed large $Q$ — which is generically a scalar.[5] The LCE allows us to access and compute correlation functions involving the insertion of $\mathcal{O}^Q$ or operators nearby in an expansion in $Q$.[6]
By the state–operator correspondence, a scalar operator $\mathcal{O}^Q$ with charge $Q$ under the $U(1)_0$ subgroup corresponds to a state $|Q\rangle$ on the cylinder $\mathbb{R}_\tau \times S^{D-1}_{r_0}$ with charge density $\rho \propto Q/r_0^{D-1}$. The cylinder compactification scale $1/r_0$ is parametrically smaller than the scale $\rho^{1/(D-1)} \propto Q^{1/(D-1)}/r_0$ introduced by the charge in the limit of $Q \gg 1$.[7] In the window described by the parametrically separated scales the CFT state is expected to correspond to some condensed-matter phase characterized by the spontaneous breakdown of certain global and spacetime symmetries [18, 115]. Hence, between the IR scale $r_0^{-1}$ and the UV scale $\rho^{1/(D-1)}$ the state $|Q\rangle$ and small excitations around it are accurately captured by an EFT describing this condensed-matter phase, which in the homogeneous and isotropic case generically is a

[5]In principle, the operator we consider can exhibit large charges $Q_1,\ldots,Q_N$ under all the mutually commuting Cartan generators $Q_1^{(G)},\ldots,Q_N^{(G)}$ of $G$. However, imposing homogeneity of the ground state configuration will force all other charges besides $Q_1$ to be parametrically smaller than $Q_1$ and vanish [17, 47].

[6]To be precise, due to charge conservation, we need an insertion of $\mathcal{O}^Q$ and $\mathcal{O}^{-Q}$. With nearby we mean operators with the same charge $Q$ and slightly higher scaling dimensions.

[7]The cylinder scale $1/r_0$ is artificial and the CFT physics is independent of $r_0$. In the LCE this manifests itself in the observation that the hierarchy of scales $1/r_0 \ll \rho^{1/(D-1)}$ is independent of $r_0$.

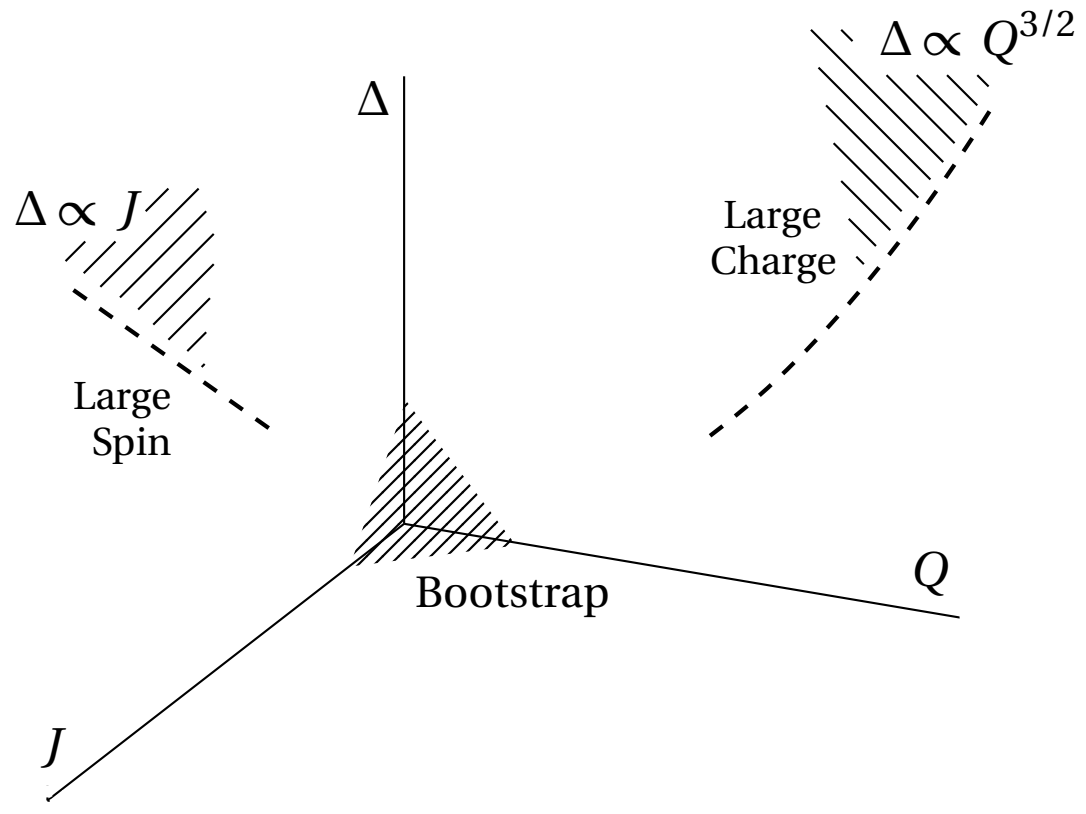

Figure 2.1: Diagrammatic representation of the space of conformal data with axes given by the scaling dimension $\Delta$, the spin $J$ and the charge $Q$. Outlined are the accessible areas for $\Delta$ as a function of $J, Q$. Figure inspired by [153].

superfluid.[8] The hierarchy of scales is

$$\frac{1}{r_0} \ll \Lambda\text{: cut-off scale of the (superfluid) EFT } \ll \frac{Q^{1/(D-1)}}{r_0}. \tag{2.3}$$

The superfluid ground state $|Q\rangle$ within the EFT computes the scaling dimension $\Delta(Q)$ of the lowest-lying operator $\mathcal{O}^Q$. Excitations on top of the EFT vacuum $|Q\rangle$ describe operators with charge $Q$ and slightly larger scaling dimensions. These are captured below the cut-off $\Lambda \ll \rho^{1/(D-1)}$ by the EFT and are described by the superfluid phonon — a NG mode coming from the spontaneous breaking of the global symmetry $U(1)_0 \subset G$ within the EFT — whose properties are largely independent of the properties of the underlying CFT. The derivative and the loop expansions within the EFT are controlled by the ratio between the UV and IR scales, which is given by the charge $1/Q^{1/(D-1)}$.

#### Path integral at fixed charge and symmetry breaking

The implementation of the LCE can be well understood from the path integral. In the following, we will consider a generic global symmetry group $G$ and not restrict our analysis to $O(2)$.
Consider a $D$-dimensional CFT described by a path integral that is invariant under a continuous global internal symmetry $G$. A primary operator within the CFT carries charges $Q_1, \ldots, Q_N$ under the Cartan generators $Q_I^{(G)}$ of $G$.[9] Working in Euclidean space, the LCE aims to investigate and extract CFT data from correlators of the form

$$\langle \mathcal{O}^{\mathbf{Q}\dagger}(x_{\text{out}}) \mathcal{O}_k(x_k) \cdots \mathcal{O}_1(x_1) \mathcal{O}^{\mathbf{Q}}(x_{\text{in}}) \rangle\,, \tag{2.4}$$

[8]The class of CFTs with a global symmetry $G$ whose large-charge operators are accurately described by a conformal superfluid — sometimes referred to as the superfluid universality class — is expected to be large. However, it excludes theories with a moduli space.

[9]Equivalently, we can ask $\mathcal{O}$ to transform within a specific representation under the action of $G$. Representations are labelled by the Cartan charges of the highest-weight element within the representation.

where the conjugate operator $\mathcal{O}^{\mathbf{Q}\dagger} \sim \mathcal{O}^{-\mathbf{Q}}$ carries the inverse charge $-\mathbf{Q} = -(Q_1, \dots, Q_N)$. In the context of the LCE $\mathbf{Q}$ will be assumed to be large and $\mathcal{O}^{\mathbf{Q}}$ to be heavy while the other operator insertions are generic and carry quantum numbers of order $\mathcal{O}(1)$. Inserted at the origin $x_{\text{in,out}} = 0, \infty$ — via the state–operator correspondence — the primary operator $\mathcal{O}^{\mathbf{Q}}(0)\,|0\rangle$ acting on the vacuum corresponds to a (primary) state $|\mathbf{Q}\rangle = |Q_1, \dots, Q_N\rangle$ of charge $\mathbf{Q}$ on the cylinder $\mathbb{R} \times S^{D-1}_{r_0}$. In- and out-states on the cylinder are defined as

$$\langle \mathbf{Q}(\tau)| := \langle \mathbf{Q}|\, e^{-H^{(\text{cyl})}\tau} = \langle \mathbf{Q}|\, e^{-\frac{\Delta(\mathbf{Q})}{r_0}\tau}, \qquad |\mathbf{Q}(\tau)\rangle = \langle \mathbf{Q}(\tau)|^{\dagger}, \tag{2.5}$$

where the cylinder Hamiltonian $H^{(\text{cyl})} = D$ is the dilatation operator in flat space and $E(\mathbf{Q}) = \Delta(\mathbf{Q})/r_0$. Hence, up to a trivial rescaling (see Eq. (1.53)), the matrix element

$$\langle \mathbf{Q}(\tau_{\text{out}})|\,\mathcal{O}_k(\tau_k, \mathbf{n}_k) \cdots \mathcal{O}_1(\tau_1, \mathbf{n}_1)\,|\mathbf{Q}(\tau_{\text{in}})\rangle \tag{2.6}$$

for the theory quantized on the cylinder is equivalent to the flat space correlator in Eq. (2.4). For large enough values of the charges $\mathbf{Q}$ we expect that the path integrals describing correlators of the form in Eq. (2.6) are dominated by a semi-classical trajectory. This trajectory specifies a definite symmetry-breaking pattern, as such a trajectory inevitably breaks the global group $G$ at least down to a subgroup $G'$. Operator insertions with small quantum numbers and higher-energy states are expected to be captured by excitations over the ground state $|\mathbf{Q}\rangle$, *i.e.* the associated classical configuration in the path integral.

In the limit of infinite separation, $\tau_{\text{in,out}} \to \pm\infty$, we can infer the symmetries respected by the leading semi-classical configuration and the ones broken by it.[10] The operator insertion $\mathcal{O}^{\mathbf{Q}}$ at 0 breaks translations $P_\mu$ while the insertion at $\infty$ breaks SCTs $K_\mu$. The fate of the rotation group $SO(D)$ is unclear, but for a scalar operator insertion it is reasonable to expect that the state $|\mathbf{Q}\rangle$ exhibits a homogeneous charge density.[11] [12] The dilatation operator — *i.e.* the cylinder Hamiltonian — may or may not be broken. However, the most natural possibility is that the semi-classical ground state at fixed charge corresponds to a superfluid phase in which both $D$ and the global internal symmetry group $G$ are spontaneously broken, leaving only the modified Hamiltonian,

$$D/r_0 + \sum_I \mu_I Q_I^{(G)}, \tag{2.7}$$

corresponding to an intact helical symmetry. Hence, the symmetry-breaking pattern we expect for the semi-classical trajectory in the presence of two large-charge operator insertions is

$$SO(D+1,1) \times G \longmapsto SO(D) \times \left(D + \sum_I \mu_I Q_I^{(G)}\right) \times G', \tag{2.8}$$

where $G' \subset G$ is the unbroken subgroup of $G$. The properties of the ground state and the fluctuations on top will be dependent on the specific symmetry-breaking pattern at hand. In the absence of any higher symmetries like Supersymmetry (SUSY) [60–63, 161–166] there is a finite gap between the NG modes coming from the symmetry breaking and any additional DoF, allowing us to compute the path integral corresponding to Eq. (2.6) using a low-energy effective action. In the case of no additional operator insertions the matrix element in Eq. (2.6) takes the form

$$\langle \mathbf{Q}| e^{-H(\tau_{\text{out}}-\tau_{\text{in}})} |\mathbf{Q}\rangle = \int \prod \mathscr{D}\chi_I \prod \mathscr{D}\pi_A \exp\left[ -S[\chi_I,\pi_A] - \frac{iQ_I}{\Omega_D r_0^{D-1}} \int_{\tau_{\text{in}}}^{\tau_{\text{out}}} \mathrm{d}\tau \int \mathrm{d}S\, \dot{\chi}_I \right], \tag{2.9}$$

where $\chi_I$ are the NG fields associated to the Cartan charges $Q_I^{(G)}$ and $\pi_A$ are additional DoF that might be present. The action $S[\chi_I,\pi_A]$ is the most general action compatible with the symmetry breaking pattern. For completeness, $\Omega_D$ refers to the volume on the unit $D-1$-sphere,

$$\Omega_D = \frac{2\pi^{D/2}}{\Gamma(D/2)}. \tag{2.10}$$

[10] This corresponds to inserting the operators $\mathcal{O}^{\mathbf{Q}}, \mathcal{O}^{\mathbf{Q}\dagger}$ at $0, \infty$, respectively.

[11] Similar arguments characterize the leading trajectories of operators carrying macroscopic spin [46].

[12] Imposing homogeneity fixes the operator $\mathcal{O}^{\mathbf{Q}}$ to be in the associated completely symmetric representation [47] (at least in the case of $O(N)$ or $U(N)$). Getting away from the completely symmetric representation appears to require relaxing the homogeneity condition [47]. Explicit inhomogeneous solutions have been constructed in the $O(4)$ model [47, 158–160].

The boundary term in the action Eq. (2.9) fixes the charge(s) of the initial and final state. In contrast, the precise value of the boundary condition for the additional modes $\pi_A$ is irrelevant in the infinite separation limit $\tau_{\text{out}} - \tau_{\text{in}} \to \infty$. For large charges $\mathbf{Q}$ the path integral in Eq. (2.9) is computable in a saddle-point approximation around the corresponding semi-classical trajectory mentioned before.

The state $|\mathbf{Q}\rangle$ itself clearly breaks neither $D$ nor the $Q_I^{(G)}$, as it is an eigenstate of all of these operators. Even though the full path integral is invariant under all symmetries, the existence of a semi-classical configuration (and the associated ground state) with the symmetry-breaking pattern Eq. (2.8) ensures that the description of the system in terms of a low-energy effective action for the associated NG modes is consistent. In particular, integration over the corresponding zero modes guarantees charge conservation on the cylinder. In contrast, in infinite volume the boundary conditions fix the value of the zero-modes as they are not normalizable.

There is the possibility that the leading trajectory does not break the dilatation operator. This situation is generically associated to the presence of a Fermi liquid phase involving fermionic excitations. The characterization of a Fermi liquid phase in terms of the NG DoF is less clear than for a superfluid phase [118, 167, 168]. We encounter this possibility in the free fermion CFT as well as in the large-$N$ GN model (at least to leading order in $N$), which we discuss later in Chapter 3.

### 2.1.3 The superfluid universality class and other emergent phases

We discuss the zoology of arising effective descriptions and emergent condensed-matter phases in the LCE. In particular, we focus on the largest subset of theories which are described by a conformal superfluid phase at large charge, also referred to as the superfluid universality class of CFTs.[13]

The superfluid EFT description has been the first emergent phase found in the large-charge sector of certain CFTs with global symmetries [17, 18]. Besides being the best-studied theories in terms of their LCE, the class of CFTs whose correlation functions with large-charge operator insertions are accurately captured by a conformal superfluid — sometimes referred to as the superfluid universality class — is also expected to be large. For a CFT in the superfluid universality class the homogeneous cylinder ground state at large charge and the fluctuations on top are captured by a conformal superfluid EFT. Theories in this class possess a global internal $O(2)$ symmetry, which may or may not be just a subgroup of a bigger global symmetry group $G \supset O(2)$.[14] Generically, conformal fixed points of scalar theories like the $O(N)$ vector model [47, 122, 124, 145, 169] belong to the superfluid universality class, with the exception of the free massless scalar. In addition, the superfluid universality class also includes models with Yukawa-type interactions [3, 170] and purely fermionic models — such as the NJL model discussed in Chapter 3 — that exhibit so-called Bardeen–Cooper–Schrieffer (BCS) superconductivity [95, 171] and allow for a Bose–Einstein Condensate (BEC). Finally, certain bosonic gauge theories with monopole operators [140, 172–174] are expected to belong to the superfluid universality class when considered at large monopole magnetic charge. In this case, the $U(1)$ global symmetry is generated by the associated topological charge.

[13] The superfluid universality class is not a universality class in the usual sense of the word. Usually, a universality class refers to the set of sometimes vastly different theories — like the the Ising model and the theoretical description of water —that flow to the same CFT at the fixed point. Here, it refers to the set of distinct conformally invariant theories that share the same effective description at large charge.

Theories with non-Abelian symmetries in the superfluid universality class are much richer than their Abelian counterparts. This is because in non-Abelian CFTs the spectrum of possible symmetry-breaking patterns and EFT descriptions is distinctly larger and there are more available DoF [122]. For example, in the $O(N)$ vector model only correlators of operators in completely symmetric representations $[Q = |\mathbf{Q}|\, 0 \dots 0]$ are described by a homogeneous superfluid ground state at large charge [46, 122].[15] When analysing the spectrum of the EFT it is found that — in addition to the usual $O(2)$ sector with the superfluid phonon and its massive partner — there are now new NG modes with quadratic dispersion relations [46, 47, 122, 124] — *i.e.* type-II or type-B NG modes — and their massive partners which are the gapped NG modes found in finite density systems (see Section 1.2.4 and Section 1.2.5). The fluctuations of these type-II modes on top of the homogeneous ground state are expected to describe operators that are in representations $[Q\, q_2 \dots q_N]$ with parametrically small asymmetries $q_2, \dots, q_N \ll Q$ [46, 47]. It is to be expected that for representations that are no longer roughly symmetric — *i.e.* if the coefficients $q_2, \dots, q_N$ become large enough — the homogeneous superfluid EFT description breaks down, just as it does for large enough spin $J \propto Q^{1/(D-1)}$ in $D$ dimensions [47, 155], and one or potentially several new inhomogeneous phases will emerge. We expect to find similar behaviour for other non-Abelian theories in the superfluid universality class besides the scalar $O(N)$ models.

These new emergent phases for operators in asymmetric representations are not well studied and there are no explicit suggestions on how they should look like generically. The only thing that is clear is these new phases should be associated with spatially inhomogeneous ground states as shown explicitly in the simplest case of the $O(4)$ vector model [47, 158, 159]. However, even the $O(4)$ model is not properly understood as recent work has shown that there is a discrepancy between theoretical predictions and lattice results for the sub-leading conformal dimensions [160]. The fact that these emerging phases describing anti-symmetric operators have to be inhomogeneous can be illustrated in the case $O(N)$ vector model with the observation that a structure which is anti-symmetric in the scalar indices requires the presence of derivatives. As a consequence, anti-symmetric operators are necessarily spinning (and potentially descendants).

[14]In most generic examples the global $O(2)$ symmetry in question corresponds to the $U(1)$ baryon symmetry counting the number of particles. The fermionic example of the NJL model is more subtle and the relationship with the particle number symmetry is only seen after a Pauli–Gursey (PG) transformation, see Section 3.3.3.

[15]To be precise, large charge in this non-Abelian setting means large dimension of the associated (completely symmetric) representation.

The picture qualitatively changes in the case of fermionic CFTs. In the presence of fermionic DoF it is certainly a possibility that the associated large-charge sector might not be captured by a bosonic superfluid EFT. And in fact, there are some fermionic CFTs that do not exhibit an emergent superfluid description at large charge and instead the fixed-charge ground state is described by a filled Fermi sphere. This is the case for the free fermion [168] and the GN model in the strict infinite-$N$ limit [3]. For interacting systems with a Fermi-sphere ground state there should be a Fermi-liquid EFT description in the spirit of [118, 167]. A proposed Fermi-liquid EFT needs to be compatible with conformal symmetry and potentially introduces new universal properties similar to the conformal superfluid EFT. Ideally, such an EFT description exhibits a BCS instability to account for fermionic models that still acquire a superfluid ground state.[16]

Both the Fermi sphere and the superfluid class of CFTs exhibit a scaling dimension that in $D$ dimensions to leading order goes like

$$\Delta(Q) \sim Q^{\frac{D}{(D-1)}} . \tag{2.11}$$

This is physically significant as it is the only scaling that allows for a non-trivial macroscopic limit. For a CFT in arbitrary $D$ on the cylinder schematically it holds that

$$Q \propto r_0^{D-1} \rho , \qquad \frac{E}{V} = \frac{\Delta(Q)}{r_0 V} \sim \frac{Q^\alpha}{r_0^D} , \tag{2.12}$$

where, to be precise, we assume that the scaling dimension $\Delta(Q)$ asymptotically scales like $Q^\alpha$ for large charge $Q$. The associated charge density $\rho$ introduces the dimensionful parameter $\rho^{1/(D-1)}$. The separation of scales $1/r_0 \ll \rho^{1/(D-1)}$ at large charge indicates the presence of a condensed-matter phase [115] naturally associated with the state $|Q\rangle$. We assume that such a condensed-matter phase exists and accurately captures the state $|Q\rangle$ and its nearby excitations. The scale $r_0$ of the cylinder $\mathbb{R} \times S^{D-1}_{r_0}$ is unphysical in any CFT and dimensional analysis dictates that

$$\frac{E}{V} = \frac{\Delta(Q)}{r_0 V} \sim \rho^{\frac{D}{(D-1)}} , \tag{2.13}$$

in order for the energy density $\Delta(Q)/(r_0 V)$ of the corresponding state $|Q\rangle$ to be well-defined in the macroscopic — or planar — limit $r_0 \to \infty$. If this condition is satisfied, the energy density on the cylinder neither blows up nor does it dilute in this limit, consistent with the fact that the separation of scales $1/r_0 \ll Q^{1/(D-1)}/r_0$ remains intact for $r_0 \to \infty$ and hence the condensed-matter description necessarily does so as well. The universality class of CFTs with a non-trivial macroscopic limit includes both the superfluid universality class and the Fermi-sphere class and is believed to be very large; essentially any theory without flat directions and higher symmetries is believed to lie in this universality class.

The exists a distinct number of CFTs for which the energy density of the large-charge ground state dilutes in the planar limit. We refer to this set of CFTs whose large-charge operators do not possess a non-trivial macroscopic limit on the cylinder as the moduli class. Theories in the moduli class do not exhibit a condensed-matter EFT description. This includes the free scalar theory [154] and $\mathcal{N} = 2$ Supersymmetric Yang–Mills (SYM) models [60–64].[17] In this case the scaling dimension goes like

$$\Delta(Q) \sim Q\,. \tag{2.14}$$

[16] Theoretically, it is feasible that in addition there exists a class of theories described by a non-Fermi liquid phase [175, 176], however, in contrast to the Fermi sphere ground state, such behaviour has not yet been observed in the context of the LCE.

However, not all supersymmetric models lie in the moduli class. In particular, the $\mathcal{N} = 2$ supersymmetric $W = \Phi^3$ model in three dimensions exhibits a scaling dimension $\Delta(Q)$ that goes like $Q^{3/2}$ at large charge $Q$ [47].

To summarize, in the literature on the LCE we essentially encounter two scenarios. In generic CFTs we expect and find the operator that minimizes the conformal dimension at fixed charge $Q \gg 1$ to correspond to a finite density state with a non-trivial macroscopic limit satisfying the scaling law $\Delta(Q) \sim Q^{D/(D-1)}$. On the other hand, in theories with a moduli space of vacua the lightest operator of charge $Q \geq 0$ generically has vanishing energy density in the macroscopic flat space limit, violating the above scaling relation. This is in particular the case in many superconformal field theories and for the free massless scalar where $\Delta(Q) \sim Q$.[18] In principle, for this to occur it suffices there to be an operator which scales like $\Delta(Q) \sim Q^{\alpha}$ with $\alpha < D/(D-1)$. However, in all known examples this relationship is linear with $\alpha = 1$, suggesting this behaviour to represent a universal feature for CFTs with a flat direction in which SSB is present also in the moduli space of vacua. In this context, there exists a result from condensed-matter physics proving that a ground state with off-diagonal long-range order — which is a fundamental feature of SSB present in the moduli space of vacua of the theory — implies a linear relationship between energy and charge [177]. It is, however, an open problem to adapt this proof from condensed-matter systems to CFTs. The observation that in the large-charge literature the conformal dimensions $\Delta(Q)$ of the lowest-lying operators at fixed charge all either satisfy the scaling $\Delta(Q) \sim Q^{D/(D-1)}$ in generic theories or the linear scaling $\Delta(Q) \sim Q$ in certain theories with higher symmetries leads to formulation of the following conjecture [178]:

**Conjecture:** In any CFT in $D > 2$, the relationship between scaling dimension $\Delta(Q)$ and charge $Q$ of the lightest operator at fixed charge $Q$ — $\Delta(Q) \sim Q^{\alpha}$ — satisfies either $\alpha = D/(D-1)$, in which case there exists an emergent condensed-matter description, or $\alpha = 1$, in which case the theory exhibits higher symmetries and a moduli space of vacua.

The above conjecture can also be formulated in a slightly weaker form by requiring only that $1 \leq \alpha \leq D/(D-1)$ [178]. As mentioned, the upper bound $\alpha = D/(D-1)$ comes from the requirement of a finite charge density in the macroscopic limit on the cylinder, a physically well-motivated assumption. The lower bound $\alpha = 1$ represent the smallest value of $\alpha$ for which the scaling dimensions $\Delta(Q)$ still is a convex function of $Q$. It has been suggested that convexity of the scaling dimension $\Delta(Q)$ (and hence

[17] In $D = 2$ it was demonstrated that the $U(1)$ large-charge sector decouples from the rest of the system and cannot control the low-energy dynamics [168]. Hence, it is not possible to write down an EFT description there as well. However, it was later shown that the LCE can be fruitful when paired with an additional controlling parameter and evaluated in the double-scaling limit [54].

[18] In these superconformal field theories the lightest operator at fixed charge is the BPS operator which always satisfies a linear relation between scaling dimension and charge.

$\alpha \geq 1$) follows from consistency of the large-charge EFT under the assumption that said EFT consists of a single NG boson [179]. It would be certainly interesting and illuminating to establish the claims of either the weaker or stronger form of this conjecture using non-perturbative techniques such as the conformal bootstrap.

### 2.1.4 Connecting large charge and large spin

Large-spin expansions in CFTs have been established well before the LCE. Once the LCE as a method for computing conformal data has been established — as both methods allow for the computation of CFT data involving heavy operators — it is natural to try to understand what lies at the intersection between the two methods. In the following we consider $U(1)$-invariant CFTs in $D = 3$.[19]

The case of large-charge operators in the $O(2)$ model additionally carrying a small non-zero spin $J$ has already been discussed in the seminal paper [17]. They can be described via the superfluid phonon excitations on top of the ground state at large charge as long as the total spin $J$ is smaller than the EFT cut-off $J \lesssim \Lambda \ll \sqrt{Q}$ (in $D = 3$). The scaling dimension of such an operator goes like

$$\Delta(Q, J) = c_1 Q^{3/2} + \frac{J}{\sqrt{2}} + \dots, \qquad J \ll \sqrt{Q}. \tag{2.15}$$

The validity of the phonon description and the phase diagram of heavy operators have been thoroughly investigated [155, 156, 180, 181]. For spins larger than the cut-off $\Lambda \ll \sqrt{Q}$ the superfluid EFT breaks down and a new effective description should emerge. On the other end of the phase diagram, for $J \gg Q^{\alpha}$, where $\alpha \geq 3/2$ is a real number larger than the power of the leading term in the LCE, the large spin EFT takes over and there is a Regge-like behaviour [137, 150, 152]. It can be explicitly shown that the Regge limit is valid for $J \gg Q^2$ and the scaling dimension is [156]

$$\Delta(Q, J) = J + \gamma Q + \dots, \qquad J \gg Q^2. \tag{2.16}$$

In between the Regge limit and the superfluid EFT at least one other phase has to exists. Interestingly, the phase diagram is richer than originally anticipated and there are three new phases between the superfluid limit at very large $Q$ and the Regge limit at very large $J$ [155, 156, 180]. The phase diagram is schematically depicted in Figure 2.2, which in the context of computing CFT data is an extension of the diagram in Figure 2.1 for the $O(2)$ model in $D = 3$.
Beyond the validity of the superfluid EFT, for $\sqrt{Q} \ll J$ the superfluid develops vortices.[20] For spins smaller that $Q$ the ground state corresponds to a vortex-antivortex pair with scaling dimension [155, 156]

$$\Delta(Q, J) = c_1 Q^{3/2} + \frac{Q}{6c_1} \log \frac{J^2}{Q} + \dots, \qquad \sqrt{Q} \ll J \leq Q. \tag{2.17}$$

For $J \geq Q$ the number of vortices increases and the ground state becomes a vortex distribution. If the spin increases further to $J \gg Q$ the number of vortices becomes very large such that it can be approximated by a smooth distribution. Hence, the ground state for $\sqrt{Q} \ll J \ll Q^{3/2}$ corresponds to a rigid body — or vortex crystal — and the corresponding scaling dimension is [155, 156]

$$\Delta(Q,J) = c_1 Q^{3/2} + \frac{1}{2c_1}\frac{J^2}{Q^{3/2}} + \dots, \qquad \sqrt{Q} \ll J \leq Q^{3/2}. \tag{2.18}$$

[19] Although it is not understood as well as the $D = 3$ case, there is some work that has been done in $D = 4$ [156, 180]. We refrain from discussing this here.

[20] This is consistent with experimental observation, as BECs in the lab have been shown to develop an increasing number of vortices as the angular momentum is increased [182–185].

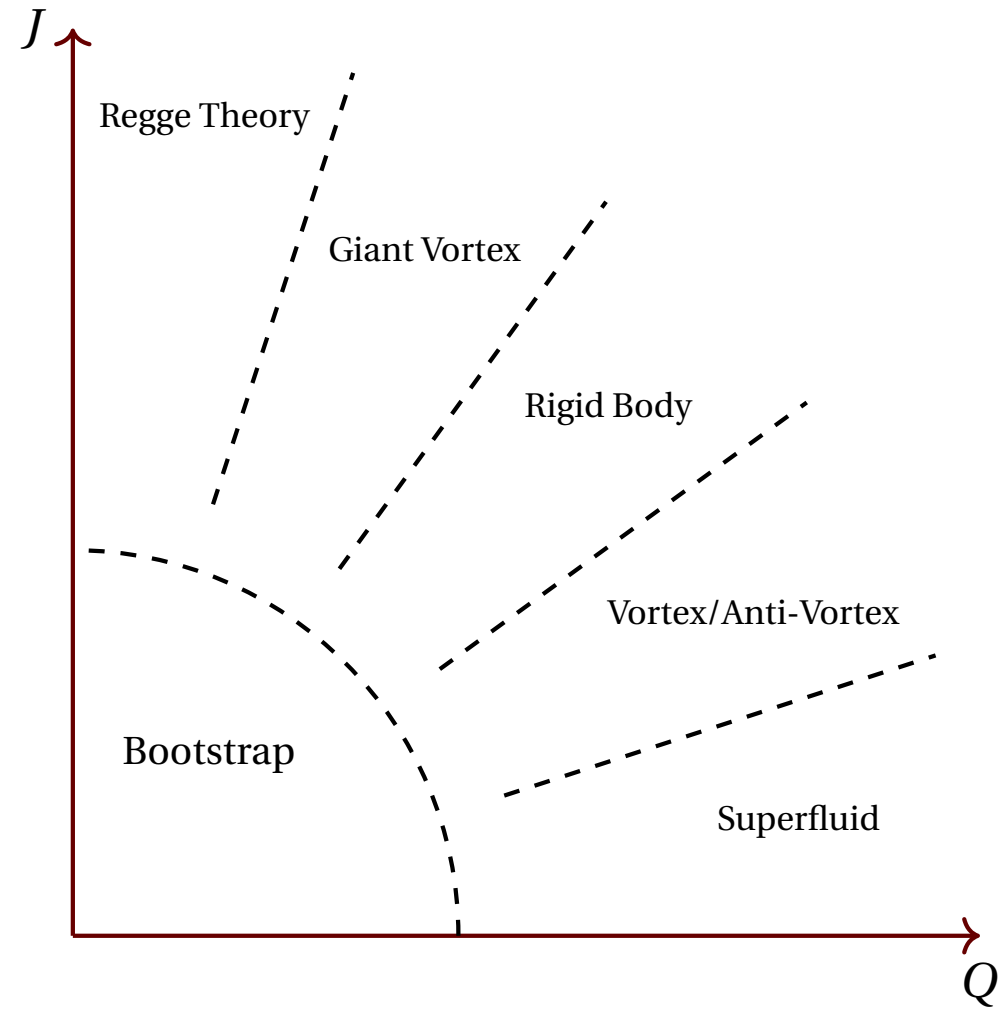

Figure 2.2: Phases for the ground state of the $O(2)$ model in $D = 3$ on the cylinder. Figure adopted from [156].

For $J \sim Q^{3/2}$ the angular velocity of the rigid body becomes relativistic, the vortex density grows and once the velocity exceeds the speed of sound presumably an instability develops. For $Q^{3/2} \ll J \ll Q^2$ a giant vortex phase emerges where the superfluid does not extent throughout $S^2_{r_0}$, but is localized on a strip around the equator. In its domain the superfluid is spinning relativistically fast and the scaling dimension becomes [156]

$$\Delta(Q,J) = J + \frac{9c_1^2}{4\pi}\frac{Q^3}{J} + \dots, \qquad Q^{3/2} \ll J \leq Q^2. \tag{2.19}$$

For $J \sim Q^2$ the domain where the superfluid is localized becomes very narrow and the EFT breaks down. For $J \gg Q^2$, in the Regge limit, the EFT describes $Q$ partons spinning around the equator [156].[21]

## 2.2 The $O(2)$ model: Computing CFT data at large charge

In the following we focus our analysis on the superfluid EFT description for the case where the global internal symmetry in question is Abelian, $G = U(1)$.[22] This Section follows closely [2] which attempts to bridge a gap in the literature by systematically computing relevant CFT data. Before the publication of [2] the primary area of interest was the computation of conformal dimensions within the LCE [17, 18, 46, 47, 122, 186–188] — with some of the results independently verified via lattice computations [143–145, 160] – and the phase diagram for large spin and large charge operators [155–157, 180]. Some other conformal data in the form of three- and four-point functions has turned up dispersed in the literature [18, 46, 47, 154, 168, 187–189] but there is no systematic collection and presentation of results anywhere. This Section that follows closely [2] bridges that gap in the literature and presents and extensive but not exhaustive list of conformal correlators and data in $U(1)$ invariant CFTs at large charge in a systematic way and in a common language.

[21] The coefficient $\gamma$ in Eq (2.16) is the scaling dimension of one such parton [156].

[22] Generically, the group $G = U(1)$ can be part of a bigger symmetry group as long as the associated $U(1)$ generator commutes with all other generators.

The plan is as follows. In Section 2.2.1 through Section 2.2.6 we review the basic ideas of the $O(2)$ sector at large charge, focusing on canonical quantization, path-integral methods, computing the basic two-point functions of the ground state $\langle Q|Q\rangle$ as well as the one-phonon state $\langle {}^{Q}_{\ell_2 m_2}|{}^{Q}_{\ell_1 m_1}\rangle$ and discussing the quantum corrections to the scaling dimension from $\langle Q|Q\rangle$. Additionally, in Appendix B.3, we discuss the two-loop correction to the scaling dimensions of the primary operator corresponding to the large-charge ground state in detail. In Section 2.2.7 we present correlators of the $O(2)$ conserved current $J$ and the stress-energy tensor $T$. We improve upon the state-of-the art and compute relevant correlators here for spinning one-phonon states $\langle {}^{Q}_{\ell_2 m_2}|\cdots|{}^{Q}_{\ell_1 m_1}\rangle$ representing excitations of the large-charge ground state. Although we do not directly use conformal symmetry, our results are consistent with the expected form of conformal correlators. Finally, in Section 2.2.8 we give an overview of correlators in which a small-charge operator is inserted between large-charge states.
In Appendix B.1 we collect important properties and identities of hyperspherical harmonics in $D$ dimensions and we make the connection to the constraints on correlators from conformal invariance at large separation. The Casimir energy of the fluctuations is computed in various dimensions in Appendix B.2. Finally, in Appendix B.5 we give some calculational details for correlators and in Appendix B.4 we give details for the loop calculations.

### 2.2.1 Effective field theory for the O(2) sector at large charge

We consider a CFT in $D$-dimensional flat space with an $O(2)$ internal symmetry which can generically be a subgroup of a larger global symmetry. The $O(2)$ symmetry is generated by its generator $Q = Q^{(O(2))}$. This set is not empty as it in particular includes the $O(2)$ WF fixed point for $2 < D < 4$ [8, 17, 47] and, in particular, this set includes the interacting CFT that describes the superfluid transition in liquid Helium.
In our presentation we generally work in the cylinder frame $\mathbb{R} \times S^{D-1}$. We consider the state $|Q\rangle$ in the cylinder generated by the scalar primary $\mathcal{O}^Q$ with $O(2)$ charge $Q$ and we are interested in correlators of such primaries at long distances, which can be expressed on the cylinder as

$$\langle Q,\infty|Q,-\infty\rangle = \lim_{\beta\longrightarrow\infty} \langle Q|e^{-\beta H^{\text{cyl}}}|Q\rangle \,. \tag{2.20}$$

By the state–operator correspondence the cylinder Hamiltonian $H^{\text{cyl}} = D/r_0$ is the dilatation operator in flat space. There is strong indication [17, 18, 47] that as the charge $Q$ becomes very large, this correlator on the cylinder admits a description in terms of a weakly coupled EFT based on the coset

model

$$\frac{SO(D+1,1)\times U(1)_Q}{SO(D)\times U(1)_{D+\mu Q}}\,, \qquad \text{valid for energy scales} \qquad \frac{1}{r_0} \ll \Lambda \ll \frac{Q^{1/(D-1)}}{r_0} \sim \mu\,, \tag{2.21}$$

where $r_0$ is the radius of the cylinder $\mathbb{R}\times S^{D-1}_{r_0}$. The parameter $\mu(Q)$ will be introduced later as part of the ground state solution and is best interpreted as the chemical potential dual to the quantum number $Q$ (the fixed control parameter). The symmetry-breaking pattern of the coset model in Eq. (2.21),

$$SO(D+1,1)\times U(1)_Q \longrightarrow SO(D)\times U(1)_{D+\mu Q}\,, \tag{2.22}$$

is known as the conformal superfluid phase.[23] We quickly repeat why we expect this symmetry breaking pattern. The leading trajectory and hence the emergent effective description must have the same symmetries as the correlator in Eq. (2.20). The insertions of $\mathcal{O}^Q$ and $\mathcal{O}^{-Q}$ at $\pm\infty$ in cylinder time effectively break translations and SCTs. We assume that $\mathcal{O}^Q$ is a scalar and rotations remain unbroken (homogeneity). As $|Q\rangle$ is both an eigenstate of $H^{\text{cyl}} = D/r_0$ and the charge operator $Q = Q^{(O(2))}$ we expect that the leading trajectory should be at least invariant under some effective time translation operator $H'$ involving $H^{\text{cyl}}$ and $Q$. Thus, the leading semi-classical solution will correspond to a homogeneous state on $S^{D-1}$ characterized by a large charge density $\rho = Q/r_0^{D-1}\Omega_D$, where $\Omega_D$ is the volume of the sphere. The simplest option respecting all of these conditions is a generalized superfluid [17, 18] with effective time translation $H^{\text{cyl}}+\mu Q$. Predictions that are derived from this assumption can be independently verified via lattice computations [143–145, 160].[24]

The most general EFT action that non-linearly realizes the symmetry-breaking pattern in Eq. (2.22) can be systematically constructed via the CCWZ approach [97, 98] in the case of broken spacetime symmetries [190]. This procedure presents a systematic way to obtain all terms in the derivative expansion up to a given order. However, invariance under Weyl rescalings of the metric $g_{\mu\nu}\mapsto\Omega^2(x)g_{\mu\nu}$ can be exploited to construct the most general effective Lagrangian in a less systematic but more convenient way.[25]
Realizing the symmetry-breaking pattern in Eq. (2.22) does not require any DoF for the spontaneously broken boosts and translations. It suffices to introduce a single shift-invariant superfluid NG field $\chi$ associated to the breaking of the $U(1)$ symmetry,

$$\chi = -i\mu t + \pi(\tau,\mathbf{n})\,, \tag{2.23}$$

where $\pi(\tau,\mathbf{n})$ are the fluctuations over the fixed-charge ground state $\chi^{\circledast} = -i\mu\tau$. The chemical potential $\mu$ will be determined eventually by the fixed charge $Q$. The leading order action of the corresponding

---

[23]The state $|Q\rangle$ itself is not a superfluid state as it is an eigenstate of the charge operator. To be precise, we just assume that the two-point function in Eq. (2.20) is dominated by a saddle corresponding to a superfluid state.

[24]The symmetry-breaking pattern in Eq. (2.22) can be more directly interpreted by taking the decompactification limit $r_0\to\infty$ of the cylinder. The conformal group on $\mathbb{R}'^D = \lim_{r\to\infty}\mathbb{R}\times S^{D-1}_{r_0}$ can mapped into the original generators on $\mathbb{R}^D$ and vice versa. In particular, the unbroken generators of $SO(D)$ are mapped into translations and rotations on $\mathbb{R}^{D-1}$ and the generator of effective time translations $H^{\text{cyl}}+\mu Q$ is mapped into effective time translation $P'_0+\mu Q$. This is exactly the symmetry of homogeneous and isotropic condensed matter [115] as conformal invariance and boosts are spontaneously broken due to the existence of a finite charge and energy density.

[25]All unitary CFTs are believed to be Weyl invariant (up to the Weyl anomaly) [81].

EFT in Euclidean spacetime[26] on the cylinder in terms of the NG field $\chi$ can be found by demanding Weyl invariance and shift invariance for $\chi$ [17, 18].[27] Equivalently, we can specialize the Abelian superfluid action in Eq. (1.149) to the cylinder and demand Weyl invariance,

$$S[\chi] = -c_1 \int_{\mathbb{R}\times S^{D-1}_{r_0}} \mathrm{d}\tau \mathrm{d}S\, \left(-\partial_\mu\chi\partial^\mu\chi\right)^{D/2}, \tag{2.24}$$

where $c_1$ is an unknown Wilsonian coefficient that depends on the UV theory (*i.e.* the starting $\mathrm{CFT}_D$) and $\mathrm{d}^D x\sqrt{g} = \mathrm{d}\tau\mathrm{d}S = r_0^{D-1}\mathrm{d}\tau\mathrm{d}\Omega$. Sub-leading higher derivative terms in the action can be constructed by noticing that the modified metric

$$g'_{\mu\nu} = \left(\partial\chi\right)^2 g_{\mu\nu}, \qquad \partial\chi = \left(-\partial_\mu\chi\partial^\mu\chi\right)^{1/2} \tag{2.25}$$

is Weyl- invariant. Now we can build invariant operators by simply replacing the metric $g_{\mu\nu}$ by $g'_{\mu\nu}$ while still respecting diffeomorphism invariance. The building blocks compatible with the modified metric $g'_{\mu\nu}$ for the EFT action are $\partial_\mu\chi, g', \nabla'_\mu, \mathscr{R}'_{\mu\nu\rho\sigma}$, where $\mathscr{R}'_{\mu\nu\rho\sigma}$ is the Riemann tensor. In this language the EFT action can straightforwardly be constructed term-by-term and we find

$$\begin{aligned} S[\chi] &= -\int_{\mathbb{R}\times S^{D-1}_{r_0}} \mathrm{d}^D x\sqrt{g'}\left[c_1 - c_2\mathscr{R}' + c_3\mathscr{R}'^{\mu\nu}\partial_\mu\chi\partial_\nu\chi + \mathscr{O}\left(\nabla'^4_\mu\right)\right] \\ &= -c_1\int_{\mathbb{R}\times S^{D-1}_{r_0}} \mathrm{d}\tau\mathrm{d}S\,\left(-\partial_\mu\chi\partial^\mu\chi\right)^{\frac{D}{2}} + c_2\int_{\mathbb{R}\times S^{D-1}_{r_0}} \mathrm{d}\tau\mathrm{d}S\,\left(-\partial_\mu\chi\partial^\mu\chi\right)^{\frac{D}{2}}\left[\frac{\mathscr{R}}{(\partial\chi)^2} + (D-1)(D-2)\frac{\nabla(\partial\chi)^2}{(\partial\chi)^4}\right] + \dots, \end{aligned} \tag{2.26}$$

where we neglected terms which vanish on the EoM $\nabla'_\mu\partial^\mu\chi = 0$ of the leading order action in Eq. (2.24). The $c_i$'s are all Wilsonian coefficients determined by the underlying CFT. The action Eq. (2.26) is to be interpreted as an action for the fluctuation $\pi(\tau,\mathbf{n})$ with cut-off $\Lambda \sim \mu$ given by the chemical potential $\dot\chi^{\circledast} = -i\mu$, so that a hierarchy is generated and controlled by the dimensionless ratio $(R\mu) \gg 1$.[28] All observables within the EFT are expressed as an expansion in inverse powers of $\mu$. In particular, the ground-state action takes the form

$$S^{\circledast} = \left(\frac{\tau_2-\tau_1}{r_0}\right)\sum_{r=0}^{\infty}\alpha_r (r_0\mu)^{D-2r}, \tag{2.27}$$

where the coefficients $\alpha_r$ depend on $c_1, c_2$ and all other Wilsonian coefficients appearing in the expansion in Eq. (2.26). Other than the scaling behaviour there is a number of universal predictions that do not depend on the Wilsonian parameters of the EFT [122, 169, 188].

Technically, we will make use of the fact that the EFT at large charge to leading order is a free theory, which allows us to perform explicit computations also for a strongly-coupled theory. The leading term in the EFT only receives corrections sub-leading in $Q$ from curvature terms. The quantum fluctuations arising from the leading term in the EFT are of order $(Q)^0$. In odd dimensions, the tree-level expressions do not have any $(Q)^0$ contribution, so all terms at this order are due to quantum fluctuations and are universal. In even dimensions, the universal $(Q)^0$ term [17, 18] is replaced by a $(Q)^0 \log Q$ term [188]. Studying the structure of the higher-order loop corrections we find in even $D$ logarithmic $l$-loop contributions of the form [2][29]

$$\Delta_l \supset \frac{1}{Q^{(l-1)D/(D-1)}} \left(\alpha_0 + \alpha_1 \log Q + \ldots + \alpha_l (\log Q)^l\right). \tag{2.28}$$

[26] Our convention for Euclidean space is $\tau = it$, so that $i\partial_\tau = \partial_t$.

[27] The shift-symmetric field $\chi$ has mass dimension $[\chi] = 0$. Due to the shift-symmetry, only the derivatives of $\chi$ are physically meaningful.

[28] The derivative expansion is controlled by the chemical potential $\mu = \langle(\partial\chi)\rangle$ as well as the $c_i$'s. The system becomes strongly coupled at energies $E \sim \mu$.

We will now review the classical and quantum treatment of the action in Eq. (2.24) and Eq. (2.26), from which we will be able to compute important CFT correlators and corrections to the scaling dimension of the primary $\mathcal{O}^Q$.

### 2.2.2 Classical treatment

For our purposes we generally neglect curvature couplings and consider only the leading order action in Eq. (2.24). The first few corrections coming from sub-leading terms in the action do in principle give corrections with positive scaling in $Q$, however, they are theory-dependent and do not give universal predictions. After expanding to quadratic order in the superfluid phonon fluctuations $\pi(\tau, \mathbf{n})$, the EFT Lagrangian in Eq. (2.24) reads

$$\mathscr{L} = -c_1 \mu^D - i c_1 \mu^{D-1} D \dot{\pi} + c_1 \mu^{D-2} \frac{D(D-1)}{2} \left(\dot{\pi}^2 + \frac{1}{D-1} (\partial_i \pi)^2\right) + \mathcal{O}\left(\mu^{D-3}\right). \tag{2.29}$$

The canonically conjugate momentum to the NG field $\pi$ is defined in the usual manner from the quadratic Lagrangian,

$$\Pi = i \frac{\delta \mathscr{L}}{\delta \dot{\pi}}\bigg|_{\text{lin}} = c_1 D \mu^{D-1} + i c_1 D (D-1) \mu^{D-2} \dot{\pi}. \tag{2.30}$$

To leading order we recover the standard canonical Poisson brackets between $\pi$ and $\Pi$ of the free theory. We start by studying the spectrum of the quadratic Lagrangian. The fields $\pi$ and $\Pi$ can be decomposed into a complete set of solutions of the EoM [122]:

$$\begin{aligned} \pi(\tau, \mathbf{n}) &= \pi_0 - \frac{i \Pi_0 \tau}{c_1 \Omega_D r_0^{D-1} D(D-1) \mu^{D-2}} + \frac{\sum_{\ell \geq 1, m} \left(\frac{a_{\ell m}}{\sqrt{2\omega_\ell}} e^{-\omega_\ell \tau} Y_{\ell m}(\mathbf{n}) + \frac{a_{\ell m}^*}{\sqrt{2\omega_\ell}} e^{\omega_\ell \tau} Y_{\ell m}^*(\mathbf{n})\right)}{\sqrt{c_1 r_0^{D-1} D(D-1) \mu^{D-2}}}, \\ \Pi(\tau, \mathbf{n}) &= c_1 D \mu^{D-1} + \frac{\Pi_0}{\Omega_D r_0^{D-1}} + i \sqrt{\frac{c_1 D(D-1) \mu^{D-2}}{r_0^{D-1}}} \sum_{\ell, m} \left(-a_{\ell m} \sqrt{\frac{\omega_\ell}{2}} e^{-\omega_\ell \tau} Y_{\ell m}(\mathbf{n}) + a_{\ell m}^* \sqrt{\frac{\omega_\ell}{2}} e^{\omega_\ell \tau} Y_{\ell m}^*(\mathbf{n})\right), \end{aligned} \tag{2.31}$$

where $\pi_0$ and $\Pi_0$ are constant zero modes of the fields that we have separated from the other modes, $\Omega_D = \frac{2\pi^{D/2}}{\Gamma(D/2)}$ is again the volume of the unit $D-1$-sphere and the $Y_{\ell m}$ are hyperspherical harmonics.[30] The dispersion relation associated to the oscillator modes is

$$r_0 \omega_\ell = \sqrt{\frac{\ell(\ell + D - 2)}{(D-1)}}. \tag{2.32}$$

[29] This result is especially relevant for applications in the context of resurgent asymptotics as in odd dimensions large-$Q$ expansions are expected to be log-free trans-series with non-perturbative corrections related to worldline instantons [191, 192].

[30] The index $m$ is a vector index with $D-2$ components. We collect our conventions and important properties of the hyperspherical harmonics in Appendix B.1.

Adding higher-curvature terms in the EFT action will add sub-leading corrections in $1/Q$ (*i.e.* in $1/\mu$) to the expression for the dispersion relation in Eq. (2.32). In contrast to the leading contribution, these corrections will depend on the Wilsonian coefficients $c_i$ and are therefore not universal [46]. The complex Fourier coefficients $a_{\ell m}$ can be extracted from the fields,

$$a_{\ell m} = \sqrt{\frac{c_1 D(D-1)\mu^{D-2}}{2\omega_\ell r_0^{D-1}}} \int_{S_{r_0}^{D-1}} \mathrm{d}S \left[\pi(\tau,\mathbf{n})\partial_\tau \left(Y^*_{\ell m}(\mathbf{n})e^{\omega_\ell \tau}\right) - (\partial_\tau \pi(\tau,\mathbf{n}))\, Y^*_{\ell m}(\mathbf{n})e^{\omega_\ell \tau}\right], \tag{2.33}$$

and the canonical Poisson bracket between $\pi$ and $\Pi$ — $[\pi,\Pi] = 1$ — corresponds to the brackets $\{a_{\ell m}, a^\dagger_{\ell' m'}\} = \delta_{\ell\ell'}\delta_{mm'}$ for the Fourier modes. The classical $O(2)$ conserved charge to leading order does not get any corrections from the Fourier modes,

$$j^\mu = \frac{\delta \mathscr{L}}{\delta \partial_\mu \chi}, \qquad Q = \int \mathrm{d}S\, j^\tau = c_1 D \Omega_D (\mu r_0)^{D-1} + \Pi_0 . \tag{2.34}$$

The leading contribution to the charge comes from the homogeneous term — the zero mode on the sphere — which corresponds to the ground state $\langle(\partial\chi)\rangle = \mu$. Finally, this relates the EFT scale $\mu$ — the chemical potential — to the ground state charge $Q_0 = Q\big|_{\Pi=0}$ as

$$\mu = \left[\frac{Q_0}{c_1 D r_0^{D-1}\Omega_D}\right]^{1/(D-1)} . \tag{2.35}$$

Evidently, the $O(2)$ charge is our controlling parameter since $\Lambda r_0 \ll \mu r_0 \sim Q_0^{1/(D-1)}$ and the validity of the EFT is controlled by $1/(\mu r_0)$. Hence, all observables within the EFT — such as the ground-state action in Eq. (2.27) — can be expressed as an expansion in $1/\sqrt{Q_0}$. To leading order in the fluctuations $\pi$, the charge $Q$ of a generic solution of the EoM depends additively on the zero mode $\Pi_0$,

$$Q = Q_0 + \Pi_0 . \tag{2.36}$$

Using the state–operator correspondence, we can now compute the classical scaling dimension of the operator $\mathscr{O}^Q$ with charge $Q \sim Q_0$ from the cylinder Hamiltonian. We find that a generic solution to the EoM corresponds to an operator with scaling dimension $\Delta(Q) = \Delta_Q$ given by

$$\Delta_Q \sim D = r_0 E_{\text{cyl}} = \Delta_0 + \frac{\partial \Delta_0}{\partial Q_0}\Pi_0 + \frac{1}{2}\frac{\partial^2 \Delta_0}{\partial Q_0 \partial Q_0}\Pi_0^2 + r_0 \sum_{\ell \geq 1, m} \omega_\ell a^*_{\ell m} a_{\ell m} , \tag{2.37}$$

where we have defined the quantities

$$\Delta_0 = c_1 (D-1)\Omega_D (\mu r_0)^D + \mathscr{O}\left((\mu r_0)^{D-2}\right), \qquad \frac{\partial \Delta_0}{\partial Q_0} = \mu r_0 , \qquad \frac{\partial^2 \Delta_0}{\partial Q_0 \partial Q_0} = \frac{1}{c_1 D(D-1)\Omega_D (\mu r_0)^{D-2}} . \tag{2.38}$$

The quantity $\Delta_0$ corresponds to the leading (classical) contribution to the action in Eq. (2.27). In terms of the charge $Q_0$ in Eq. (2.35) it reads

$$\Delta_0 = \frac{c_1 (D-1)\Omega_D}{(c_1 D \Omega_D)^{D/(D-1)}} (Q_0)^{\frac{D}{(D-1)}} + \mathscr{O}\left((Q_0)^{\frac{(D-2)}{(D-1)}}\right) . \tag{2.39}$$

If we include all curvature corrections in Eq. (2.26) that produce terms with a positive $Q$-scaling the classical contribution to the scaling dimension is of the form [47]

$$\Delta_0 = d_{\frac{D}{(D-1)}} (Q_0)^{\frac{D}{(D-1)}} + d_{\frac{(D-2)}{(D-1)}} (Q_0)^{\frac{(D-2)}{(D-1)}} + \dots + \begin{cases} d_0 + \mathcal{O}\left((Q_0)^{\frac{-2}{(D-1)}}\right), & \text{for } D \text{ even}, \\ d_{\frac{1}{(D-1)}} (Q_0)^{\frac{1}{(D-1)}} + \mathcal{O}\left((Q_0)^{\frac{-1}{(D-1)}}\right), & \text{for } D \text{ odd}. \end{cases} \tag{2.40}$$

The coefficients can in principle be computed but do depend on the Wilsonian coefficients appearing in the action Eq. (2.26). For example, in $D = 3$ the classical contribution in terms of the Wilsonian coefficients $c_1, c_2$ reads [46]

$$\Delta_0 = \frac{2c_1(4\pi)}{(3c_1(4\pi))^{3/2}} Q_0^{3/2} + \frac{2c_2(4\pi)}{(3c_1(4\pi))^{1/2}} Q_0^{1/2} + \mathcal{O}\left(Q_0^{-1/2}\right). \tag{2.41}$$

Finally, we remark that the Hamiltonian for the field $\chi$ is shifted *w.r.t.* the one for $\pi$ by $\mu Q$. The effective time evolution for the fluctuation $\pi$ is generated by $H^{(\text{cyl})} \sim H_\chi = H_\pi + \mu Q$, as expected for a superfluid phonon.

### 2.2.3 Canonical quantization

Radial quantization in flat space corresponds to canonical quantization in the cylinder frame, obtained by $\tau$-slicing and associating a Hilbert space to every fixed $\tau \in \mathbb{R}$. This poses no conceptual problems at all, as the cylinder is a direct product of the time direction and a curved manifold. The coefficients in the mode decompositions in Eq. (2.31) are promoted to field operators with non-vanishing commutators,

$$[\pi_0, \Pi_0] = i \quad \text{(zero modes)}, \qquad [a_{\ell m}, a^\dagger_{\ell' m'}] = \delta_{\ell\ell'}\delta_{mm'}. \tag{2.42}$$

The commutation relations in Eq. (2.42) are equivalent to the canonical equal-$\tau$ commutator,

$$[\pi(\tau,\mathbf{n}), \Pi(\tau,\mathbf{n}')] = i\delta_{S^{D-1}}(\mathbf{n},\mathbf{n}'), \tag{2.43}$$

where $\delta_{S_1^{D-1}}(\mathbf{n},\mathbf{n}')$ is the invariant delta function on the unit sphere $S_1^{D-1}$. To build a representation of the Heisenberg algebra on the cylinder we start with a vacuum $|Q\rangle$ that satisfies

$$a_{\ell m}|Q\rangle = \Pi_0 |Q\rangle = 0. \tag{2.44}$$

Since we are in finite volume, the $O(2)$ charge is not broken — as discussed in Section 1.2.1 and Section 1.2.7 — and is a well-defined operator acting on the Hilbert space as

$$Q = \int \mathrm{d}S\, \Pi(\tau,\mathbf{n}) = Q_0 \mathbb{1} + \Pi_0, \qquad Q|Q\rangle = Q_0|Q\rangle. \tag{2.45}$$

The non-zero charge $Q_0$ of the vacuum can be increased by acting on it with the zero mode $\pi_0$. All other Fourier modes carry zero charge under the global $O(2)$,

$$[Q, \pi_0] = -i, \qquad [Q, a_{\ell m}] = [Q, a^\dagger_{\ell m}] = 0. \tag{2.46}$$

Starting from the vacuum $|Q\rangle$ we can obtain a state with charge $Q_0+q$ and scaling dimension $\Delta_0(Q_0+q)$ that is also annihilated by all modes $a_{\ell m}$,[31]

$$|Q+q\rangle = e^{i\pi_0 q}|Q\rangle = \exp\left[\frac{iq}{\Omega_D r_0^{D-1}}\int \mathrm{d}S\,\pi(\tau,\mathbf{n})\right]|Q\rangle\,. \tag{2.47}$$

It is important to emphasize that this does not lead to a degeneracy in the spectrum. While these states are all annihilated by the ladder operators — since $[a_{\ell m},\pi_0]=0$ — they are not annihilated by the zero mode $\Pi_0$. Hence, these states are gapped and do not represent degenerate vacua,

$$\Delta_0(Q_0+q)-\Delta_0(Q_0)\sim q\,(\mu r_0)\,. \tag{2.48}$$

Since $\pi_0$ is the only operator in the theory on which the $O(2)$ charge $Q$ acts non-trivially, it has to be compact,

$$\pi_0 \sim \pi_0 + 2\pi\mathbb{1}\,. \tag{2.49}$$

As a consequence, the variable $q$ in Eq. (2.47) is quantized and a natural number,

$$q\in\mathbb{Z}\,. \tag{2.50}$$

This clearly shows that the states with charge $Q_0+q$ live at the EFT cut-off and will not be discussed any further.

The commutation relations in Eq. (2.46) allow us to identify $\mu$ as the chemical potential via a variational approach. Fixing the charge $Q$ imposes the first-class constraint

$$\langle Q|Q^{(O(2))}|Q\rangle = Q_0\,. \tag{2.51}$$

The classical system can be equivalently quantized using a variational approach by finding the state $|Q\rangle$ that minimizes $\langle Q|\,H^{(\mathrm{cyl})}\,|Q\rangle$ under the charge-fixing constraint, which is implemented via a Lagrange multiplier — the chemical potential $\mu$ — so that the ground state minimizes $\left(H^{(\mathrm{cyl})}+\mu Q\right)|Q\rangle = E(Q)$. The ground state solution $\langle Q|\dot\chi|Q\rangle = -i\mu$ fixes the chemical potential $\mu$,

$$\langle Q|\dot\chi|Q\rangle = \langle Q|[H^{(\mathrm{cyl})},\chi]|Q\rangle = \mu\,\langle Q|[Q,\chi]|Q\rangle = -i\mu\,. \tag{2.52}$$

The quantized quadratic Hamiltonian $D/r_0 = H^{(\mathrm{cyl})}$ corresponding to the classical expression of the scaling dimension in Eq. (2.37) can be written as the sum of a normal-ordered[32] operator $:H^{(\mathrm{cyl})}:$ plus a vacuum contribution,

$$H^{(\mathrm{cyl})} =\, :H^{(\mathrm{cyl})}: + \frac{\Delta_1}{r_0}\mathbb{1}, \qquad \text{where} \qquad \Delta_1 := \frac{1}{2}\sum_{\ell\geq 1,m}(r_0\omega_\ell)\,. \tag{2.53}$$

[31] $\Delta_0=\Delta_0(Q_0)$ is defined via Eq. (2.35) and Eq. (2.38).

[32] Normal ordering refers to the vacuum $|Q\rangle$ where $\langle Q|:H^{(\mathrm{cyl})}:|Q\rangle = \frac{\Delta_0}{r_0}$.

The vacuum contribution — also called the Casimir energy — needs to be regulated and has physical consequences. We compute $\Delta_1$ in various dimensions in Appendix B.2. It first appeared for $D = 3$ in [18] and for $D = 4,5,6$ in [188]. From the point of view of the LCE, the one-loop correction comes at order $\mathcal{O}\big((Q_0)^0\big)$. Hence, in principle, in the tree-level computation we also need to keep track of all the curvature terms that do appear in the action Eq. (2.26) up to this order. In arbitrary dimension $D$ there will be $\lceil (D+1)/2 \rceil$ terms with positive $Q$-scaling, see Eq. (2.40), each controlled by a Wilsonian coefficient [47]. However, there are no new universal predictions that can be extracted from curvature corrections in the action (at the classical level) and we know the $Q$-scaling of these curvature terms, so computing them in terms of the Wilsonian coefficients $c_i$ is not necessarily useful.

Within the EFT there is a spectrum of excited states on top of the ground state $|Q\rangle$. The commutators between $H^{(\text{cyl})}$ and the various Fourier modes show which ones generate excited states when acting on the vacuum:

$$\begin{aligned} &[H^{(\text{cyl})}, a_{\ell m}] = -\omega_\ell a_{\ell m}, && [H^{(\text{cyl})}, a^\dagger_{\ell m}] = \omega_\ell a^\dagger_{\ell m}, \\ &[r_0 H^{(\text{cyl})}, \pi_0] = -i\frac{\partial \Delta_0}{\partial Q_0} - i\frac{\partial^2 \Delta_0}{\partial Q_0^2}\Pi_0, && [H^{(\text{cyl})}, \Pi_0] = 0. \end{aligned} \tag{2.54}$$

The Hilbert space of the theory is given by the Fock space spanned by all the states that are generated via repeated action of the different creation operators on the vacuum,

$$a^\dagger_{\ell_1 m_1} \dots a^\dagger_{\ell_k m_k} |Q\rangle \tag{2.55}$$

A state of the form in Eq. (2.55) has charge $Q_0$ and scaling dimension

$$\Delta = \Delta_0 + \Delta_1 + \sum_{i=1}^{k} (r_0 \omega_{\ell_k}). \tag{2.56}$$

These states are also referred to as superfluid phonon states in the literature. We finish the discussion of canonical quantization with some prescient comments:

- From the perspective of the underlying $\text{CFT}_D$, the superfluid phonon states in Eq. (2.55) correspond to primary operators with generically different quantum numbers compared to $\mathcal{O}^Q$ (the scalar operator corresponding to the vacuum $|Q\rangle$) but the same charge under $O(2)$. The only exception to this rule are states including at least one $a^\dagger_{1m}$ which are descendant operators in the conformal multiplets. This can be seen from the fact that any operator $a^\dagger_{1m}$ adds energy

$$\Delta E = r_0 \omega_1 = 1. \tag{2.57}$$

- Superfluid phonon states of the form $a^\dagger_{\ell_1 m_1} \dots a^\dagger_{\ell_k m_k} |Q\rangle$ generically correspond to spinning primaries in the appropriate reducible representation. This can be seen from the action of the rotation group $SO(D)$ on the phonon states.[33] The $SO(D)$ part of the cylinder isometry group acts on the Hilbert space in terms of some unitary operator $U$. Under the action of $U$ the Fourier

[33] In the $\text{CFT}_D$ this is the group of Euclidean rotations.

mode operators transform as

$$U(R)\, a^\dagger_{\ell m} U^\dagger(R) = \sum_{m'} D^\ell_{mm'}(R^{-1})\, a^\dagger_{\ell m'}\,, \qquad R \in SO(D)\,. \tag{2.58}$$

This follows from the decomposition in Eq. (2.31) and the properties of the hyperspherical harmonics, where $D^\ell_{mm'}$ is a finite-dimensional irreducible representation of $SO(D)$ that generalizes Wigner's D-symbol to $D \geq 3$.

- It is not possible to describe all phonon states within the EFT. When the $\ell$-quantum number becomes too large, their contribution $r_0\omega_\ell$ can compete with the leading $\Delta_0$ term, breaking the large-$Q$ expansion. The same issue arises if the number of creation operators describing the phonon state becomes too large. The leading contribution to the scaling dimension goes like $\Delta_0 \sim Q^{D/(D-1)}$ — where $Q \sim Q_0$ — but higher-curvature terms in Eq. (2.26) will introduce lower order corrections up to $Q^{1/(D-1)}$. Phonon states with energies $\omega_\ell$ comparable to the lowest classical contribution to the scaling dimension should be excluded from the EFT. This sets a natural cut-off for the $\ell$-quantum number in terms of the charge,

$$\ell_{\text{cutoff}} \sim Q^{1/(D-1)}\,. \tag{2.59}$$

Operators with such high spin should be described by new coset models with more complicated breaking pattern. The newly arising phases for different (relative) values of charge $Q$ and spin $\ell$ are well understood in $D = 3$ [155–157, 180]. We discuss the corresponding phase diagram extensively in Section 2.1.4.

The structure of the spectrum and the existence of the above-mentioned charged spinning primaries described in terms of superfluid phonon states is a direct prediction of the superfluid hypothesis for a generic $O(2)$-invariant $\text{CFT}_D$. Canonical quantization is the appropriate framework for this discussion, but we will generically expect corrections to scaling dimensions and the spectrum structure coming from interactions in Eq. (2.29), which correspond to sub-leading terms in the controlling parameter $Q \sim Q_0$. The sub-leading corrections are best discussed within a path integral formulation, so that ordinary loop expansion techniques can be employed.

### 2.2.4 Path-integral methods

The path-integral as a framework is equivalent to canonical quantization but allows for the seamless incorporation of sub-leading corrections from interaction terms in the Lagrangian. To set up the path integral we note that an equivalent basis of the Hilbert space at fixed $\tau$ is given by the field and momentum eigenstates,

$$\chi\,|\chi\rangle = \chi(\mathbf{n})\,|\chi\rangle\,, \qquad \Pi\,|\Pi\rangle = \Pi(\mathbf{n})\,|\Pi\rangle\,. \tag{2.60}$$

Their overlap is fixed by the canonical commutation relations and reads

$$\langle\chi|\Pi\rangle = e^{i\int \mathrm{d}S\,[\chi\Pi]}\,. \tag{2.61}$$

Generically, the vacuum of the theory — $|Q\rangle$ — can be expressed as a superposition of momentum eigenstates with the coefficient of the $\Pi_0$-component set to zero (see Eq. (2.45)),

$$|Q\rangle = \mathcal{N}_Q \int \mathcal{D}\Pi\, \delta(\Pi_0)\Psi_Q(\Pi)\, |\Pi\rangle\,, \tag{2.62}$$

where $\mathcal{N}_Q$ is a standard normalization factor. In the limit of large separation — *i.e* in the limit $\lim_{\tau\to\infty} Q(\tau)\,|0\rangle$ — correlators cease to depend on the specific details of the vacuum wave function $\Psi_Q$, which in this limit will only affect the overall normalization. Without loss of generality, we choose $\Psi_Q = 1$ and with Eq. (2.45) the overlap of the vacuum $|Q\rangle$ with the field eigenstates is given by

$$\langle\chi|Q\rangle = \begin{cases} \mathcal{N}_Q \exp\left[\frac{iQ}{\Omega_D R^{D-1}} \int \mathrm{d}S\,\chi\right] & \text{if } \chi \text{ is constant}, \\ 0 & \text{otherwise}. \end{cases} \tag{2.63}$$

The zero mode component $\pi_0$ of any field configuration can be extracted by integrating over the sphere on a given $\tau$-slice, see Eq. (2.31),

$$\chi_0 = \int \mathrm{d}S\,\chi\,. \tag{2.64}$$

The bracket $\langle\chi|Q\rangle$ defines the correct boundary conditions for all correlators of the form $\langle Q|\ldots|Q\rangle$ computed in the path-integral representation. These boundary conditions generalize the standard open boundary conditions on a line segment in the special case of $D = 1$.
With this technology we are able to construct path integrals that compute the norm of the states in Eq. (2.55) at large cylinder-time separation $\tau_{2,1} \to \pm\infty$. By the state–operator correspondence, the norm of a state in Eq. (2.55) computes the two-point function of the corresponding primary in the $\mathrm{CFT}_D$. Unless otherwise specified, we always consider correlators in which the vacuum $|Q\rangle$ is inserted at large separation $\tau_1 \to -\infty$ on the cylinder (the same goes for $\langle Q|$ at $\tau_2 \to \infty$). In this limit the details on the boundary conditions that the vacuum imposes are largely irrelevant. However, in the following we generically write down correlators at finite separation $\tau_2 - \tau_1$ and the limit $\tau_2, \tau_1 \to \pm\infty$ is implicit. For insertions at past or future infinity in $\tau$ the insertion point $\mathbf{n}$ on the sphere is irrelevant as these $\tau$-slices correspond to the origin and the point at infinity in flat space, allowing for the shorthand notation $\mathcal{O}^Q(\tau,\mathbf{n}) \to \mathcal{O}^Q(\tau)$.

### 2.2.5 Two-point functions

We compute the norms of the vacuum $|Q\rangle$ and the one-phonon states $a^\dagger_{\ell m}|Q\rangle$ — and therefore the two-point functions of the associated operators — in the path-integral formalism.

#### $\langle Q|Q\rangle$ correlator

The path integral for the vacuum correlator with cylinder times $\tau_2 > \tau_1$ can be written using Eq. (2.63) and reads

$$\langle Q|e^{-\frac{(\tau_2-\tau_1)}{R}D}|Q\rangle = |\mathcal{N}_Q|^2 \int \mathrm{d}\chi \exp\left[-S[\chi] - \frac{iQ}{\Omega_D r_0^{D-1}} \int_{\tau_1}^{\tau_2} \mathrm{d}\tau \int \mathrm{d}S\,\dot\chi\right] := \mathcal{A}(\tau_1,\tau_2)\,, \tag{2.65}$$

where we have introduced the shorthand notation $\mathscr{A}(\tau_1,\tau_2)$ for future convenience. We can consider the path integral in Eq. (2.65) as the working definition of the $\langle Q|Q\rangle$ correlator, without referring to the canonically quantized picture any more and taking the EFT action in Eq. (2.24) as a starting point.
The integral in Eq. (2.65) can be computed in a saddle-point approximation around a field configuration $\chi^{\text{saddle}}(\tau,\mathbf{n})$ that represents a solution to the minimization problem

$$\delta S[\chi] = \int_{\tau_1}^{\tau_2} \mathrm{d}\tau \mathrm{d}S \left(-\partial_\mu \frac{\partial \mathscr{L}}{\partial(\partial_\mu \chi)}\right)\delta\chi + \int \mathrm{d}S \left(\frac{\partial \mathscr{L}}{\partial(\partial_\tau \chi)} + \frac{iQ}{\Omega_D r_0^{D-1}}\right)\delta\chi\Big|_{\tau_1}^{\tau_2}. \tag{2.66}$$

The bulk EoM requires the divergence of the (Euclidean) $O(2)$ conserved current,

$$j_\mu = \frac{\partial \mathscr{L}}{\partial(\partial^\mu \chi)} = c_1 D(-\partial_\nu\chi\partial^\nu\chi)^{D/2-1}\partial_\mu\chi, \tag{2.67}$$

to vanish. The second part of the EoM specifies the appropriate boundary conditions. The general solution compatible with the boundary conditions is the homogeneous superfluid configuration $\chi^{\text{saddle}}(\tau,\mathbf{n}) = -i\mu\tau + \pi_0$ [17], with $\pi_0$ constant and the chemical potential $\mu$ fixed by the boundary condition,

$$c_1 D\mu^{D-1} = \frac{Q}{\Omega_D r_0^{D-1}}, \tag{2.68}$$

This is the same relationship that we had already found in Eq. (2.35). The action expanded to quadratic order in the fluctuations around the ground state $\chi(\tau,\mathbf{n}) = -i\mu\tau + \pi(\tau,\mathbf{n})$ becomes

$$S = \Delta_0 \frac{\tau_2-\tau_1}{r} + c_1\mu^{D-2}\frac{D(D-1)}{2}\int_{\tau_1}^{\tau_2}\mathrm{d}\tau \int \mathrm{d}S \left(\dot{\pi}^2 + \frac{1}{(D-1)r_0^2}(\partial_i\pi)^2\right) + \mathscr{O}(\mu^{D-3}). \tag{2.69}$$

The boundary condition eliminates the linear term in Eq. (2.29) and therefore also the zero-mode terms appearing in Eq. (2.37), as expected, where the charge $Q$ here in Eq. (2.69) can be identified with $Q_0$ in Eq. (2.37). This ground state is a good starting point for setting up a loop expansion controlled by $\mu t_0$.
The normalization factor $\mathscr{N}_Q$ is chosen such that the cylinder correlator takes the form

$$\mathscr{A}(\tau_1,\tau_2) = r_0^{-2(\Delta_0+\Delta_1+\dots)}\exp\left[-\frac{(\tau_2-\tau_1)}{r_0}\underbrace{\left(\Delta_0+\Delta_1+\dots\right)}_{\Delta_Q}\right], \qquad \Delta_Q = \Delta(Q), \tag{2.70}$$

which corresponds to the properly normalized two-point function in $\mathbb{R}^D$. The first quantum correction $\Delta_1$ introduced in Eq. (2.53) is given by the Casimir energy of the fluctuation $\pi$ around the homogeneous ground state $\chi^{\text{saddle}} = -i\mu\tau$.
By the state–operator correspondence, the reference states $|Q\rangle, \langle Q|$ correspond to insertions of the associated scalar primaries $\mathscr{O}^Q, \mathscr{O}^{Q\dagger}$ with scaling dimension $\Delta(Q) = \Delta_Q$ at $\tau_{1,2} = \mp\infty$:

$$|Q\rangle = \lim_{\tau_1\to-\infty}\mathscr{O}^Q(\tau_1)|0\rangle, \qquad \langle Q| = \lim_{\tau_2\to\infty}\langle 0|\mathscr{O}^Q(\tau_2)^\dagger, \tag{2.71}$$

where the spatial dependence $\mathbf{n}_{1,2}$ is irrelevant at infinite separation and hence omitted. We recall that conjugation on the cylinder corresponds to time reversal and charge conjugation,

$$\mathscr{O}^Q(\tau,\mathbf{n})^\dagger = \mathscr{O}^{-Q}(-\tau,\mathbf{n}). \tag{2.72}$$

Finally, we repeat the Weyl map from the cylinder to flat space (see Eq. (1.53)),

$$\langle \mathcal{O}^{-Q}(x_2)\mathcal{O}^{Q}(x_1)\rangle_{(\text{flat})} = \left(\frac{|x_1|}{r_0}\right)^{-\Delta_Q}\left(\frac{|x_2|}{r_0}\right)^{-\Delta_Q}\langle \mathcal{O}^{-Q}(\tau_2,\mathbf{n}_2)\mathcal{O}^{Q}(\tau_1,\mathbf{n}_1)\rangle_{(\text{cyl})}\,. \tag{2.73}$$

### $\langle {}^{Q}_{\ell_2 m_2}|{}^{Q}_{\ell_1 m_1}\rangle$ correlator

The second class of two-point functions we study in detail are correlators of one-phonon states obtained by acting with a single creation operator $a^\dagger_{\ell m}$ on the vacuum $|Q\rangle$,

$$|{}^{Q}_{\ell m}\rangle = a^\dagger_{\ell m}|Q\rangle\,, \qquad\qquad \text{where} \qquad\qquad |{}^{Q}_{00}\rangle = |Q\rangle\,. \tag{2.74}$$

The correlator $\langle {}^{Q}_{\ell_2 m_2}|{}^{Q}_{\ell_1 m_1}\rangle$ is best computed in the canonical quantization picture, using the commutation relations of the creation and annihilation operators $a_{\ell m}, a^\dagger_{\ell m}$,

$$\langle {}^{Q}_{\ell_2 m_2}|{}^{Q}_{\ell_1 m_1}\rangle = \langle Q|a_{\ell_2 m_2}\, e^{-\frac{(\tau_2-\tau_1)}{r_0}D}\, a^\dagger_{\ell_1 m_1}|Q\rangle = \frac{\mathcal{A}(\tau_1,\tau_2)}{e^{(\tau_2-\tau_1)\omega_\ell}}\delta_{\ell_1\ell_2}\delta_{m_1 m_2} = \frac{e^{-\Delta\frac{(\tau_2-\tau_1)}{r_0}}}{r_0^{-\Delta}}\delta_{\ell_1\ell_2}\delta_{m_1 m_2}\,, \tag{2.75}$$

where $\Delta = \Delta_Q + r_0\omega_\ell$ is the conformal dimension in Eq. (2.56) for $k = 1$. The result in Eq. (2.75) is consistent with the general structure of a conformal two-point function on the cylinder at large separation $\tau_2 - \tau_1 \to \infty$ given in Eq. (B.23).

This result holds true up to quadratic order in the Hamiltonian, however, we expect that loop corrections will shift the spectrum in a complicated way which also depends on the underlying CFT$_D$. For this reason, it is convenient to formulate the correlator as a path integral by expressing $a_{\ell m}$ in terms of the fields as in Eq. (2.33),

$$\begin{aligned}\langle {}^{Q}_{\ell_2 m_2}|{}^{Q}_{\ell_1 m_1}\rangle = \frac{c_1 D(D-1)\mu^{D-2}}{2r_0^{D-1}\sqrt{\omega_{\ell_2}\omega_{\ell_1}}}\int \mathrm{d}S(\mathbf{n}_2)\int \mathrm{d}S(\mathbf{n}_1)\, Y^*_{\ell_2 m_2}(\mathbf{n}_2)\, Y_{\ell_1 m_1}(\mathbf{n}_1)\\ \times \mathcal{A}(\tau_1,\tau_2)\lim_{\substack{\tau'\to\tau_2\\ \tau\to\tau_1}}\left(\omega_{\ell_2}-\partial_{\tau'}\right)\left(\omega_{\ell_1}+\partial_\tau\right)\langle \pi(\tau',\mathbf{n}_2)\pi(\tau,\mathbf{n}_1)\rangle\,,\end{aligned} \tag{2.76}$$

where the two-point function of the superfluid NG fluctuations is given by

$$\langle \pi(\tau_2,\mathbf{n}_2)\pi(\tau_1,\mathbf{n}_1)\rangle = \left(\langle Q|e^{-\frac{(\tau_2-\tau_1)}{r_0}D}|Q\rangle\right)^{-1}\int \mathcal{D}\pi\; \pi(\tau_2,\mathbf{n}_2)\,\pi(\tau_1,\mathbf{n}_1)\,e^{-S[\pi]}\,, \tag{2.77}$$

where $S[\pi]$ is the action in Eq. (2.69). Unsurprisingly, the full information about the spectrum is all contained in the two-point function of the superfluid fluctuation $\pi$. In this formalism, using the tree-level propagator, the correlator in Eq. (2.77) on the cylinder is computed to be

$$\langle \pi(\tau_2,\mathbf{n}_2)\pi(\tau_1,\mathbf{n}_1)\rangle = \frac{1}{c_1 D(D-1)(\mu r_0)^{D-2}}\left(\sum_{\ell=1}^{\infty}\sum_m e^{-\omega_\ell|\tau_2-\tau_1|}\frac{Y_{\ell m}(\mathbf{n}_2)^*\,Y_{\ell m}(\mathbf{n}_1)}{2r_0\omega_\ell} - \frac{|\tau_2-\tau_1|}{2r_0\Omega_D}\right), \tag{2.78}$$

and the result in Eq. (2.75) is recovered. The state $|{}^{Q}_{\ell m}\rangle$ on the cylinder defines a spin-$\ell$ symmetric and traceless tensor operator of charge $Q$ inserted in the infinite past,

$$|{}^{Q}_{\ell m}\rangle = \lim_{\tau_1\to-\infty}\mathcal{O}^{Q}_{\ell m}(\tau_1)\,|0\rangle\,. \tag{2.79}$$

The procedure outlined for the computation of one-phonon correlators $\langle {}^{Q}_{\ell_2 m_2} | {}^{Q}_{\ell_1 m_1} \rangle$ at tree-level — particularly in canonical quantization — is easily generalized to states with more phonon excitations. For example, for states with two phonon excitations we find that

$$\begin{aligned}\langle {}^{Q}_{(\ell_2 m_2)\otimes(\ell_2' m_2')} | {}^{Q}_{(\ell_1 m_1)\otimes(\ell_1' m_1')} \rangle &= \langle Q | a_{\ell_2 m_2} a_{\ell_2' m_2'} e^{-\frac{(\tau_2-\tau_1)}{r_0} D} a^\dagger_{\ell_1' m_1'} a^\dagger_{\ell_1 m_1} | Q \rangle \\ &= \mathscr{A}(\tau_1,\tau_2)\, e^{-(\tau_2-\tau_1)\left(\omega_{\ell_2}+\omega_{\ell_2'}\right)} \left( \delta_{\ell_1 \ell_2} \delta_{m_1 m_2} \delta_{\ell_1' \ell_2'} \delta_{m_1' m_2'} + \delta_{\ell_1 \ell_2'} \delta_{m_1 m_2'} \delta_{\ell_1' \ell_2} \delta_{m_1' m_2} \right). \end{aligned} \tag{2.80}$$

In the computation for states with higher numbers of phonon excitations the energy is corrected accordingly and there is an ever increasing sum over all possible permutations of Kronecker deltas. As long as none of the $\ell$-quantum numbers are equal to one, these states are primary, but they will not be in irreducible representations like the one-phonon states. For example, in $D = 3$ — by virtue of the Clebsch–Gordan decomposition — we have

$$\ell \otimes \ell' = (\ell + \ell') \oplus (\ell + \ell' - 2) \oplus \cdots \oplus \left| \ell - \ell' \right| . \tag{2.81}$$

### 2.2.6 Quantum corrections

Quantum correction to the tree-level computation of $\langle Q|Q\rangle$ and the associated scaling dimension can be computed in perturbation theory. The perturbation theory for the superfluid EFT action in Eq. (2.24) is best set up by compactifying the radial time coordinate $\tau$ on the thermal circle $S^1_\beta$ and considering the theory on $S^1_\beta \times S^{D-1}_{r_0}$. The original EFT predictions are recovered in the zero-temperature limit $\beta \to \infty$. The fluctuations $\pi$ on $S^1_\beta \times S^{D-1}_{r_0}$ can be decomposed into modes as

$$\pi(\tau, \mathbf{n}) = \sqrt{\frac{\beta}{r_0}} \sum_{n\in\mathbb{Z}} \sum_{\ell \geq 1, m} Y_{\ell m}(\mathbf{n}) e^{i\omega_n \tau} \pi_{n\ell m}, \qquad \pi^*_{n\ell m} = (-1)^{m_{D-2}} \pi_{-n\ell m^*}, \tag{2.82}$$

where the appropriate notation for the $m$-type quantum numbers follows the standard-tree convention for the hyperspherical harmonics [193], see Appendix B.1. In addition, we have introduced the Matsubara frequencies $\omega_n$ given by

$$\omega_n = \frac{2\pi n}{\beta}, \tag{2.83}$$

the bosonic eigenvalues on the thermal circle.[34] On $S^1_\beta \times S^{D-1}_{r_0}$ there is a unique zero mode $\pi_0$ appearing in the mode decomposition, which we can exclude in Eq. (2.82) as it never appears in the derivative-only interactions within the EFT. In the space of Fourier modes the propagator can be computed from the quadratic part of the action in Eq. (2.29) and reads

$$\langle \pi_{n\ell m} \pi_{n'\ell' m'} \rangle = \frac{1}{c_1 D(D-1)(\mu r_0)^{D-2} \beta^2} G_{n\ell} (-1)^{|m|} \delta_{-n'n} \delta_{\ell'\ell} \delta_{-m'm}, \qquad G_{n\ell} := \frac{1}{\omega_n^2 + \omega_\ell^2}, \tag{2.84}$$

where $\omega_\ell^2$ denotes the dispersion relations in Eq. (2.32) — given by the rescaled eigenvalues of the hyperspherical harmonics $Y_{\ell m}$ — and $G_{n\ell}$ is the propagator in Fourier space. The zero mode $\pi_0$, which remains present at finite temperature, does not mix with the other Fourier modes and has

[34] Standard references for thermal field theory methods are [194, 195].

$\langle \pi_0 \pi_0 \rangle$ = constant. It cannot receive corrections at any order in perturbation theory because all vertices within the EFT contain only derivatives of the field $\pi(\tau, \mathbf{n})$.

### First quantum correction: one- loop scaling dimension $\Delta_1$

We start by reviewing the computation of the first quantum correction, the one-loop scaling dimension $\Delta_1$ (the Casimir energy) for the primary $\mathcal{O}^Q$, first defined in Eq. (2.53). In the path-integral formalism this contribution arises from the Gaussian integration over the quadratic part of the action in Eq. (2.69).[35] On $S^1_\beta \times S^{D-1}_{r_0}$ the one-loop contribution $\Delta_1$ becomes

$$\Delta_1 = -\lim_{\beta\to\infty} \frac{\partial}{\partial \beta} \log Z_0 = \frac{1}{2} \sum_{\ell>0} \mathrm{Deg}_D(\ell)\, (r_0 \omega_\ell)\,, \tag{2.85}$$

where we have performed the sum over the Matsubara frequencies $\vartheta_n$ as in Appendix B.4. The result coincides with the expression from canonical quantization found in Eq. (2.53). We indicated with $\mathrm{Deg}_D(\ell)$ the degeneracy of the $\ell$-th eigenvalue $\lambda_\ell$ of the Laplacian $\Delta_{S^{D-1}}$, see Eq. (B.4). The sum appearing in Eq. (2.85) and Eq. (2.53) above is clearly divergent and needs to be regularized. To regularize this sum the use of a momentum-dependent regulator is natural. For convenience it is best to introduce the family $\Sigma(s)$ of regularized sums defined by

$$\Sigma(s) := \lim_{\Lambda\to\infty} \sum_{\ell>0} \mathrm{Deg}_D(\ell)\, (r_0 \omega_\ell)^s e^{-\omega_\ell^2/\Lambda^2}\,. \tag{2.86}$$

In this notation, the one-loop scaling dimension is given by $\Delta_1 = \frac{1}{2}\Sigma(1)$. We compute $\Delta_1$ in detail for various spacetime dimensions in Appendix B.2. For $D$ odd there is no classical term at the same order as the Casimir energy $\Delta_1$, and hence $\Delta_1$ is fully calculable as there is no local counter-term that renormalizes it. For example, in $D = 3$ the Casimir energy $\Delta_1$ is a universal contribution and reads

$$\Delta_1 \Big|_{D=3} = -0.0937255\,. \tag{2.87}$$

For $D$ even there is a classical term at order $(Q)^0$ and $\Delta_1$ gets renormalized. However, there is a universal contribution proportional to $\log(Q)$ appearing [188].

As a last remark we note that, generally, in loop calculations the controlling parameter $Q$ — which defines the EFT UV scale via $\Lambda r_0 \sim \mu r_0 \sim Q^{1/(D-1)}$ — cuts the phonon states running in internal lines, as discussed in Eq. (2.59).

[35] As a path integral the first quantum correction is $\Delta_1 \sim \int \mathcal{D}\pi \exp\left[ -\frac{c_1 \mu^{D-2} D(D-1)}{2} \int \mathrm{d}\tau \mathrm{d}S \left( \dot{\pi}^2 + \frac{(\partial_i \pi)^2}{(D-1) r_0^2} \right) \right]$.

### Higher loop corrections

For generic spacetime dimension $D$ the EFT action includes all possible $k$-point vertices. After expanding the action im powers of $\mu$, the interaction part of the action $S_{\text{int}}$ is of the form

$$S_{\text{int}} = \sum_{k=3}^{\infty} \mu^{D-k} S^{(k)}. \tag{2.88}$$

We consider the two-loop correction $\Delta_2$ to the scaling dimension. We analyse and compute this contribution in detail in Appendix B.3 and Appendix B.4. This contribution only gets contributions from diagrams involving three-point and four-point vertices. These vertices are

$$S^{(3)} = i\frac{c_1}{6} D(D-1)(D-2) \int_0^\beta \mathrm{d}\tau \int_{S_{r_0}^{D-1}} \mathrm{d}S\, \dot{\pi} \left[ \dot{\pi}^2 + \frac{3}{(D-1)} \frac{1}{r_0^2} (\partial_i \pi)^2 \right], \tag{2.89}$$

$$S^{(4)} = -\frac{c_1}{24} D(D-1)(D-2) \int_0^\beta \mathrm{d}\tau \int_{S_{r_0}^{D-1}} \mathrm{d}S \left[ \frac{3}{r_0^4 (D-1)} (\partial_i \pi)^4 + \frac{6}{r_0^2} \left( \frac{D-3}{D-1} \right) \dot{\pi}^2 (\partial_i \pi)^2 + (D-3) \dot{\pi}^4 \right], \tag{2.90}$$

and the $\Delta_2$-correction to the partition function is

$$\Delta_2 = -\mu^{D-4} \langle S^{(4)} \rangle_c + \frac{1}{2} \mu^{2D-6} \langle S^{(3)} S^{(3)} \rangle_c, \tag{2.91}$$

where $\langle \ldots \rangle_c$ indicates connected contractions only. As propagators scale as $\sim \mu^{2-D}$, both contributions enter at order $\mu^{-D} \sim Q^{-\frac{D}{(D-1)}}$ modulo possible powers of $(Q)^0 \log Q$. Applying this counting to higher loop corrections reveals that an $l$-loop diagram produces a term in the scaling dimension of $\mathcal{O}^Q$ that scales like

$$\Delta_l \sim Q^{-\frac{(\ell-1)D}{(D-1)}}. \tag{2.92}$$

In the limit $\beta \to \infty$, after computing the Matsubara sums appearing in the expression for $\Delta_2$, the two-loop scaling dimension of the operator $\mathcal{O}^Q$ can be reduced to the expression

$$\begin{aligned} \Delta_2 = \frac{1}{16 c_1 D(D-1)\Omega_D (\mu r_0)^D} \Bigg[ & \frac{(r_0\Lambda)^2}{6\pi} (D-2) \left[ 2D - 4 + (D-5)\Sigma(0) \right] \Sigma(0) - \frac{(r_0\Lambda)}{\sqrt{\pi}} (D-2)^2 (1-\Sigma(0)) \Sigma(1) \\ & - \frac{(D-2)}{3} \left[ (D-2)(\Sigma(2) + 6\Sigma(0)\Sigma(2) + 2\Sigma(-1)\Sigma(3)) - (5D-16)\Sigma(1)^2 - 8\Sigma^{(2\ell)} \right] \Bigg], \end{aligned} \tag{2.93}$$

where $\Lambda$ is a smooth cut-off introduced during regularization, $\Sigma(s)$ is defined in Eq. (2.86) and $\Sigma^{(2\ell)}$ is a sum that cannot easily be reduced to a combination of sums of the form in Eq. (2.86),

$$\Sigma^{(2\ell)} := \sum_{\ell_a, \ell_b, \ell_c} S_{\ell_a \ell_b \ell_c} \triangle_{\ell_a \ell_b \ell_c} \frac{\omega_{\ell_a} \omega_{\ell_b} \omega_{\ell_c}}{\omega_{\ell_a} + \omega_{\ell_b} + \omega_{\ell_c}}. \tag{2.94}$$

For completeness, the symbol $\triangle_{\ell_a \ell_b \ell_c}$ guarantees that the $SO(D)$ quantum numbers $\ell_a$, $\ell_b$, $\ell_c$ satisfy a triangle inequality, which essentially corresponds to discrete momentum conservation on $S_{r_0}^{D-1}$,

$$\triangle_{\ell_a\ell_b\ell_c} := \begin{cases} 1 & \text{if } |\ell_b - \ell_a| \leq \ell_c \leq \ell_b + \ell_a \text{ and } \ell_c - \ell_a - \ell_b \text{ even}, \\ 0 & \text{otherwise}, \end{cases} \tag{2.95}$$

and the symmetric structure $S_{\ell_a\ell_b\ell_c}$ is defined as

$$\begin{aligned} \frac{r_0^{2D-2}}{(D-2)\Omega_D} S_{\ell_a\ell_b\ell_c} &:= \triangle_{\ell_a\ell_b\ell_c} \frac{R^{2D-2}}{(D-2)\Omega_D} \frac{(D+2\ell_a-2)(D+2\ell_b-2)(D+2\ell_c-2)}{2\Gamma(D-1)\Gamma\left(\frac{D}{2}-1\right)^2} \\ &\times \frac{\Gamma\left(\frac{\ell_{abc}}{2}+\frac{(D-2)}{2}\right)}{\Gamma\left(\frac{\ell_{abc}}{2}+1\right)} \frac{\Gamma\left(\frac{\ell_{cab}}{2}+\frac{(D-2)}{2}\right)}{\Gamma\left(\frac{\ell_{cab}}{2}+1\right)} \frac{\Gamma\left(\frac{\ell_{bca}}{2}+\frac{(D-2)}{2}\right)}{\Gamma\left(\frac{\ell_{bca}}{2}+1\right)} \frac{\Gamma\left(\frac{\ell_a+\ell_b+\ell_c}{2}+\frac{(2D-4)}{2}\right)}{\Gamma\left(\frac{\ell_a+\ell_b+\ell_c}{2}+\frac{D}{2}\right)}, \end{aligned} \tag{2.96}$$

where we have introduced the shorthand notation $\ell_{abc} = \ell_a + \ell_b - \ell_c$. The sum in Eq. (2.94) can be regularized and computed in theory, however, we restrict ourselves to noting that it possesses two divergent regimes. In the case $\ell_a \sim \ell_b \gg 1$ grows like $\sim \Sigma(1)$, and in the case $\ell_a \sim \ell_b \sim \ell_c \gg 1$ it grows like $\sim \Sigma(2)$.[36] In even dimensions $D$ the result in Eq. (2.93) is of the form

$$\Delta_2 \supset \frac{1}{Q^{D/(D-1)}} \left(\alpha_0 + \alpha_1 \log Q + \alpha_2 (\log Q)^2\right). \tag{2.97}$$

In particular, we have the appearance of a non-universal $(Q)^0 \log Q^2$ term. On general grounds we expect this result to generalize to arbitrary loop order so that in even $D$ the scaling dimension will have an $\ell$-loop contribution of the form

$$\Delta_l \supset \frac{1}{Q^{(l-1)D/(D-1)}} \left(\alpha_0 + \alpha_1 \log Q + \ldots + \alpha_l (\log Q)^l\right). \tag{2.98}$$

This result is of particular importance for applications in the context of resurgent asymptotics. In odd dimensions, LCEs are expected to produce log-free trans-series with non-perturbative corrections related to worldline instantons [191, 192]. This is a consequence of quasi-zero mode integration in quantum mechanics problems [196–198]. As trans-series in general contain logarithmic terms, the absence of such terms represents an important simplification. It would be interesting to better understand the appearance of $(Q)^0 \log Q$ terms for LCEs in even dimensions $D$ from the point of view of resurgence.

### 2.2.7 Correlators with current insertions

The EFT around the fixed-charge ground state in the LCE has the advantage that, while it is a free theory, it adequately captures some of the physics at the strongly-coupled fixed point. At tree level, the operator algebra in Eq. (2.42) is in perfect agreement with the predictions from the path-integral formulation and we can use Eq. (2.42) by itself to compute three- and four-point functions in the associated strongly coupled system to leading order in $Q$.

[36] The first case corresponds to an equilateral triangle described by two integers and the second case corresponds to a degenerate triangle described by three integers. These two cases are simply discrete versions of different collinear divergences in ordinary loop integrals.

In this section, we improve upon the current state-of-the-art and systematically review and compute three-and four-point functions with current insertions between spinning large-charge primaries $\mathcal{O}^Q_{\ell m}$. A few of these correlators have already appeared for the case $\ell = 0$ in the literature [18, 46, 154, 168] before the publication of [2]. Although we never directly use conformal invariance to compute correlators, there is always perfect agreement between our results and the expected structure of conformal correlators on the cylinder at large separation $\tau_2 - \tau_1$, which we have collected in Appendix B.1.2. Our results are to be understood as an expansion in $Q$ and only contain the classical contribution plus the leading order quantum correction.

#### Conserved currents and Ward identities in the EFT

The relevant classical conserved currents from the EFT action in Eq. (2.24) read

$$j_\mu = c_1 D(-\partial_\nu\chi\partial^\nu\chi)^{D/2-1}\partial_\mu\chi\,, \tag{2.99}$$

$$T_{\mu\nu} = c_1\left[D(-\partial_\alpha\chi\partial^\alpha\chi)^{D/2-1}\partial_\mu\chi\partial_\nu\chi + g_{\mu\nu}(-\partial_\alpha\chi\partial^\alpha\chi)^{D/2}\right]. \tag{2.100}$$

Their integrals over the sphere yield the conserved charges of the theory. We expand the above currents up to quadratic order in the fluctuations $\pi$ around the ground state$\chi^{\circledast} = -i\mu\tau$,

$$j_\tau = -i\frac{Q}{\Omega_D r_0^{D-1}}\left[1 + \frac{i}{\mu}(D-1)\dot\pi - \frac{(D-2)(D-1)}{2\mu^2}\left(\dot\pi^2 + \frac{(\partial_i\pi)^2}{r_0^2(D-1)}\right) + \mathcal{O}(\mu^{-3})\right], \tag{2.101a}$$

$$j_i = \frac{Q}{\Omega_D r_0^{D-1}}\left[\frac{1}{\mu r_0}\partial_i\pi + \frac{i}{\mu}\frac{(D-2)}{\mu r_0}\dot\pi\partial_i\pi + \mathcal{O}(\mu^{-3})\right], \tag{2.101b}$$

$$T_{\tau\tau} = -\frac{\Delta_0}{\Omega_D r_0^D}\left[1 + i\frac{D}{\mu}\dot\pi - \frac{D(D-1)}{2\mu^2}\left(\dot\pi^2 + \frac{(D-3)(\partial_i\pi)^2}{r_0^2(D-1)^2}\right) + \mathcal{O}(\mu^{-3})\right], \tag{2.101c}$$

$$T_{\tau i} = -i\frac{\Delta_0}{\Omega_D r_0^D}\left[\frac{1}{\mu r_0}\frac{D}{D-1}\partial_i\pi + \frac{i}{\mu}\frac{D}{\mu r_0}\dot\pi\partial_i\pi + \mathcal{O}(\mu^{-3})\right] \tag{2.101d}$$

$$T_{ij} = \frac{\Delta_0}{\Omega_D r_0^D}\left(\frac{h_{ij}}{(D-1)}\left[1 + i\frac{D}{\mu}\dot\pi - \frac{D(D-1)}{2\mu^2}\left(\dot\pi^2 + \frac{(\partial_i\pi)^2}{r_0^2(D-1)}\right)\right] + \frac{D}{(D-1)}\frac{\partial_i\pi\partial_j\pi}{(\mu r_0)^2} + \mathcal{O}(\mu^{-3})\right), \tag{2.101e}$$

where $h_{ij}$ is the metric on the sphere $S^{D-1}_{r_0}$. Homogeneity of the ground state $\chi^{\circledast}$ guarantees that $T_{\tau i} = T_{ij} = 0$ to leading order in the fluctuations. We emphasize that the above expressions have been calculated using only the leading term in the effective action Eq. (2.24). In terms of quantum corrections, this gives rise to contributions of order $(Q)^0$, while the effect of the sub-leading curvature terms on the fluctuations is suppressed at large charge.

Discussing correlators with current insertions in canonical quantization suffices to find the leading-order results. Integrating the insertions of $j_\tau$ and $T_{\tau\tau}$ over spatial slices produces the Hamiltonian $H^{(\text{cyl})} = D/r_0$ and the $O(2)$ charge $Q$, which are topological symmetry operators. When inserted at cylinder time $\tau$ these operators measure the scaling dimension dived by the radius $r_0$ and the $O(2)$ charge of any operator insertion that is contained within the half-cylinder $(\tau, -\infty) \times S^{D-1}_{r_0}$. This fact is expressed by the conformal Ward identities discussed in Section 1.1.2. We quickly repeat them here in a slightly reformulated form:

$$\langle Q(\tau)\prod_i \mathcal{O}_i(\tau_i,\mathbf{n}_i)\rangle = -i\sum_{\tau_i<\tau} Q_i\,\langle\prod_i \mathcal{O}_i(\tau_i,\mathbf{n}_i)\rangle\,, \qquad \langle D(\tau)\prod_i \mathcal{O}_i(\tau_i,\mathbf{n}_i)\rangle = -\sum_{\tau_i<\tau}\Delta_i\,\langle\prod_i \mathcal{O}_i(\tau_i,\mathbf{n}_i)\rangle\,. \tag{2.102}$$

The Ward identities in Eq. (2.102) are satisfied order by order in a loop expansion and can be used to constrain certain correlators with current insertions in Eq. (2.101).

### $\langle^Q_{\ell_2 m_2}|j|^Q_{\ell_1 m_1}\rangle$ correlators

The simplest correlators we consider are three-point functions with an insertion of the EFT current $j_\mu$ between one-phonon states $a^\dagger_{\ell m}|Q\rangle$. In the underlying CFT these matrix elements correspond to three-point functions of two large-charge spinning primaries $\mathcal{O}^Q_{\ell m}$ inserted at times $\tau_1,\tau_2$ with an insertion of the full CFT version of the $O(2)$ current $j_\mu(\tau,\mathbf{n})$ at time $\tau_1<\tau<\tau_2$.[37] To leading order we find that

$$\begin{aligned}\langle\mathcal{O}^{-Q}_{\ell_2 m_2} j_\tau(\tau,\mathbf{n})\mathcal{O}^Q_{\ell_1 m_1}\rangle &= -i\frac{Q}{\Omega_D r_0^{D-1}}\Bigg[\mathcal{A}_{\Delta_Q+r_0\omega_{\ell_1}}(\tau_1,\tau_2)\delta_{\ell_1\ell_2}\delta_{m_1 m_2}\\ &+\mathcal{A}^{\Delta_Q+r_0\omega_{\ell_2}}_{\Delta_Q+r_0\omega_{\ell_1}}(\tau_1,\tau_2|\tau)(D-1)(D-2)\Omega_D\frac{r_0\sqrt{\omega_{\ell_2}\omega_{\ell_1}}}{2D\Delta_0}\left(Y^*_{\ell_2 m_2}(\mathbf{n})Y_{\ell_1 m_1}(\mathbf{n})-\frac{\partial_i Y^*_{\ell_2 m_2}(\mathbf{n})\partial_i Y_{\ell_1 m_1}(\mathbf{n})}{r_0^2(D-1)\omega_{\ell_2}\omega_{\ell_1}}\right)\Bigg],\end{aligned} \tag{2.103}$$

$$\langle\mathcal{O}^{-Q}_{\ell_2 m_2} j_i(\tau,\mathbf{n})\mathcal{O}^Q_{\ell_1 m_1}\rangle = i\frac{Q(D-2)}{2\Delta_0 r_0^{D-1}D}\mathcal{A}^{\Delta_Q+r_0\omega_{\ell_2}}_{\Delta_Q+r_0\omega_{\ell_1}}(\tau_1,\tau_2|\tau)\left[\sqrt{\frac{\omega_{\ell_2}}{\omega_{\ell_1}}}Y^*_{\ell_2 m_2}(\mathbf{n})\partial_i Y_{\ell_1 m_1}(\mathbf{n})-(1\leftrightarrow 2)^*\right], \tag{2.104}$$

where we have introduced two new convenient shorthand notations,

$$\mathcal{A}_{\Delta_Q+r_0\omega_{\ell_1}}(\tau_1,\tau_2) := \mathcal{A}(\tau_1,\tau_2)\,e^{-(\tau_2-\tau_1)\omega_{\ell_1}}\,, \qquad \mathcal{A}^{\Delta_2}_{\Delta_1}(\tau_1,\tau_2|\tau) := e^{-\Delta_2(\tau_2-\tau)/r_0-\Delta_1(\tau-\tau_1)/r_0}\,. \tag{2.105}$$

This generalizes the expression $\mathcal{A}(\tau_1,\tau_2)=\mathcal{A}_{\Delta_Q}(\tau_1,\tau_2)$, which we have defined via the scalar two-point function $\langle Q|Q\rangle$ in Eq. (2.65) (with scaling dimension $\Delta(Q)=\Delta_Q$). The correlator $j_\tau$ in the case $\ell=0$ first appeared in [18] and later in [46].
From the above results and Eq. (2.75) we can additionally extract the following relevant OPE coefficient:

$$C_{\mathcal{O}^Q_{\ell m} j_\tau \mathcal{O}^Q_{\ell m}} = \frac{\langle\mathcal{O}^{-Q}_{\ell m} j_\tau(\tau,\mathbf{n})\mathcal{O}^Q_{\ell m}\rangle}{\langle\mathcal{O}^{-Q}_{\ell m}\mathcal{O}^Q_{\ell m}\rangle} = -i\frac{Q}{\Omega_D r_0^{D-1}}\,. \tag{2.106}$$

Integrating the insertion $j_\tau(\tau,\mathbf{n})$ over the spatial slice $\mathbf{n}\in S^{D-1}_{r_0}$ gives us a non-trivial consistency check via the associated Ward identity Eq. (2.102),

$$\int \mathrm{d}S(\mathbf{n})\,\langle\mathcal{O}^{-Q}_{\ell_2 m_2} j_\tau(\tau,\mathbf{n})\mathcal{O}^Q_{\ell_1 m_1}\rangle = -iQ\mathcal{A}_{\Delta_Q+r_0\omega_{\ell_1}}(\tau_1,\tau_2)\,\delta_{\ell_1\ell_2}\delta_{m_1 m_2}\,. \tag{2.107}$$

In the special case $\ell_{1,2}=0$ — corresponding to the (scalar) ground state $|Q\rangle$ — Ward identities guarantee that $\langle Q|J_i|Q\rangle=0$ to all orders in perturbation theory.
Examining the structure of the $j_\tau$ correlator in Eq. (2.103) we find that it contains a classical piece — which is homogeneous and time-independent — plus quantum corrections with spatial dependence.[38]
Hence, the classical contribution to the three-point function must be proportional to the two-point function,

$$\langle^Q_{\ell_2 m_2}|^Q_{\ell_1 m_1}\rangle = \mathcal{A}_{\Delta_Q+r_0\omega_{\ell_1}}(\tau_1,\tau_2)\,\delta_{\ell_1\ell_2}\delta_{m_1 m_2}\,. \tag{2.108}$$

[37] Recall that in the special case $\ell=1$ the operator $\mathcal{O}^Q_{\ell m}$ is not a primary, but a descendant of $\mathcal{O}^Q$.
[38] Time-independence is a consequence of charge conservation.

The inhomogeneous quantum piece can be decomposed into spherical harmonics and has the same tensor structure as the left-hand side. In addition, charge conservation enforces its integral to vanish. The classical piece of the spatial current $j_i$ in Eq. (2.101b) vanishes and we are left with only inhomogeneous quantum corrections. As the large-charge EFT is a weakly-coupled theory, the separation into a homogeneous classical part plus space-dependent quantum corrections applies to all physical observables.

In the computation of the correlators above we were able to disregard the linear terms in the fluctuations $\pi$ appearing in the expansions for the conserved charges in Eq. (2.101). Generically, we are always allowed to do this as these terms clearly vanish within one-phonon matrix elements once $\pi$ is expanded in Fourier modes. However, if we consider matrix elements between states with a differing number of phonon excitations, these linear terms do contribute. For example, if we consider matrix elements between the ground state and a one-phonon state we find to leading order that

$$\langle \mathscr{O}^{-Q} | j_\tau(\tau, \mathbf{n}) |^Q_{\ell m} \rangle = -\frac{Q(D-1)}{\Omega_D r_0^{D-1}} \sqrt{\frac{\Omega_D}{2D} \frac{r_0 \omega_\ell}{\Delta_0}} \mathscr{A}^{\Delta_Q}_{\Delta_Q + r_0 \omega_\ell}(\tau_1, \tau_2 | \tau)\, Y_{\ell m}(\mathbf{n}), \tag{2.109}$$

$$\langle \mathscr{O}^{-Q} | j_i(\tau, \mathbf{n}) |^Q_{\ell m} \rangle = \frac{Q}{\Omega_D r_0^{D-1}} \sqrt{\frac{r_0 \Omega_D}{2D \Delta_0 r_0 \omega_\ell}} \mathscr{A}^{\Delta_Q}_{\Delta_Q + r_0 \omega_\ell}(\tau_1, \tau_2 | \tau)\, \partial_i Y_{\ell m}(\mathbf{n}). \tag{2.110}$$

Similar results hold true for all correlators with current insertions of $T$ and $j$. In the following we will disregard this special case as it is a straightforward modification of the results and formulas presented in this section.

Generally, all of the correlators we discuss can also be computed for higher phonon states. For example, the correlator of $j_\tau$ between two-phonon states is given by

$$\begin{aligned}
\langle^Q_{(\ell_2 m_2)\otimes(\ell'_2 m'_2)} | j_\tau(\tau, \mathbf{n}) |^Q_{(\ell_1 m_1)\otimes(\ell'_1 m'_1)} \rangle &= -i \frac{Q}{\Omega_D r_0^{D-1}} \mathscr{A}_{\Delta_Q + r_0 \omega_{\ell_1} + r_0 \omega_{\ell'_1}}(\tau_1, \tau_2) \\
&\times \Bigg[ \left( \delta_{\ell_1 \ell_2} \delta_{m_1 m_2} \delta_{\ell'_1 \ell'_2} \delta_{m'_1 m'_2} + \delta_{\ell_1 \ell'_2} \delta_{m_1 m'_2} \delta_{\ell'_1 \ell_2} \delta_{m'_1 m_2} \right) \\
+ \Omega_D \frac{(D-2)(D-1)}{2D \Delta_0} \Bigg( & \frac{r_0 \sqrt{\omega_{\ell'_2} \omega_{\ell'_1}}}{e^{(\tau-\tau_1)(\omega_{\ell'_1} - \omega_{\ell'_2})}} \left[ Y_{\ell'_1 m'_1}(\mathbf{n}) Y^*_{\ell'_2 m'_2}(\mathbf{n}) - \frac{\partial_i Y_{\ell'_1 m'_1}(\mathbf{n}) \partial_i Y^*_{\ell'_2 m'_2}(\mathbf{n})}{r_0^2 (D-1) \omega_{\ell'_2} \omega_{\ell'_1}} \right] \delta_{\ell_2 \ell_1} \delta_{m_2 m_1} \\
& + \frac{r_0 \sqrt{\omega_{\ell'_2} \omega_{\ell_1}}}{e^{(\tau-\tau_1)(\omega_{\ell_1} - \omega_{\ell'_2})}} \left[ Y_{\ell_1 m_1}(\mathbf{n}) Y^*_{\ell'_2 m'_2}(\mathbf{n}) - \frac{\partial_i Y_{\ell_1 m_1}(\mathbf{n}) \partial_i Y^*_{\ell'_2 m'_2}(\mathbf{n})}{r_0^2 (D-1) \omega_{\ell'_2} \omega_{\ell_1}} \right] \delta_{\ell_2 \ell'_1} \delta_{m_2 m'_1} \\
& + \frac{r_0 \sqrt{\omega_{\ell_2} \omega_{\ell'_1}}}{e^{(\tau-\tau_1)(\omega_{\ell'_1} - \omega_{\ell_2})}} \left[ Y_{\ell'_1 m'_1}(\mathbf{n}) Y^*_{\ell_2 m_2}(\mathbf{n}) - \frac{\partial_i Y_{\ell'_1 m'_1}(\mathbf{n}) \partial_i Y^*_{\ell_2 m_2}(\mathbf{n})}{r_0^2 (D-1) \omega_{\ell_2} \omega_{\ell'_1}} \right] \delta_{\ell'_2 \ell_1} \delta_{m'_2 m_1} \\
& + \frac{r_0 \sqrt{\omega_{\ell_2} \omega_{\ell_1}}}{e^{(\tau-\tau_1)(\omega_{\ell_1} - \omega_{\ell_2})}} \left[ Y_{\ell_1 m_1}(\mathbf{n}) Y^*_{\ell_2 m_2}(\mathbf{n}) - \frac{\partial_i Y_{\ell_1 m_1}(\mathbf{n}) \partial_i Y^*_{\ell_2 m_2}(\mathbf{n})}{r_0^2 (D-1) \omega_{\ell_2} \omega_{\ell_1}} \right] \delta_{\ell'_2 \ell'_1} \delta_{m'_2 m'_1} \Bigg) \Bigg].
\end{aligned} \tag{2.111}$$

While technically difficult to compute, these higher-phonon matrix elements pose no conceptual issue and we will refrain from computing more of them for the rest of this section.

### $\langle^{Q}_{\ell_2 m_2}| j j |^{Q}_{\ell_1 m_1}\rangle$ correlators

We consider next the case of two insertions of the current $j_\tau$ at cylinder times $\tau < \tau'$ between one-phonon states generated by $\mathcal{O}^{Q}_{\ell m}$ at $\tau_2 > \tau > \tau' > \tau_1$. For two insertions of the temporal part $j_\tau$ we find that

$$\begin{aligned}\langle \mathcal{O}^{-Q}_{\ell_2 m_2} j_\tau(\tau,\mathbf{n}) j_\tau(\tau',\mathbf{n}')\mathcal{O}^{Q}_{\ell_1 m_1}\rangle &= -\mathscr{A}_{\Delta_Q + r_0\omega_{\ell_1}}(\tau_1,\tau_2)\frac{Q^2}{\Omega_D^2 r_0^{2D-2}}\delta_{\ell_1\ell_2}\delta_{m_1 m_2} \\ &\quad\times\left[1 + \frac{(D-1)^2}{2D\Delta_1}\sum_{\ell} e^{-|\tau-\tau'|\omega_\ell} r_0\omega_\ell \frac{(D+2\ell-2)}{(D-2)} C_\ell^{\frac{D}{2}-1}(\mathbf{n}\cdot\mathbf{n}')\right] \\ &+\left\{\mathscr{A}^{\Delta_Q + r_0\omega_{\ell_2}}_{\Delta_Q+r_0\omega_{\ell_1}}(\tau_1,\tau_2|\tau)\frac{Q^2(D-1)^2}{2\Omega_D r_0^{2D-2} D}\frac{r_0\sqrt{\omega_{\ell_1}\omega_{\ell_2}}}{\Delta_0}\left[-\frac{Y^*_{\ell_2 m_2}(\mathbf{n})Y_{\ell_1 m_1}(\mathbf{n}')}{e^{-(\tau-\tau')\omega_{\ell_1}}}\right.\right. \\ &\left.\left.+\frac{(D-2)}{(D-1)}\left(\frac{\partial_i Y_{\ell_1 m_1}(\mathbf{n})\partial_i Y^*_{\ell_2 m_2}(\mathbf{n})}{(D-1)\, r_0^2\omega_{\ell_1}\omega_{\ell_2}} - Y_{\ell_1 m_1}(\mathbf{n})Y^*_{\ell_2 m_2}(\mathbf{n})\right)\right] + \left[(\tau,\mathbf{n}) \leftrightarrow (\tau',\mathbf{n}')\right]\right\}, \end{aligned} \tag{2.112}$$

where for ease of notation we have introduced the so-called Gegenbauer polynomials $C_\ell^{D/2-1}$, defined in Eq. (B.15). The special case $\ell = 0$ of the above correlator has appeared first in [46].
We can again perform a consistency check with respect to the Ward identity Eq. (2.102) by integrating the result over the sphere centred around $(\tau,\mathbf{n})$, producing a conserved charge and removing the $\tau$-dependence from the correlator,

$$\int \mathrm{d}S(\mathbf{n})\,\langle \mathcal{O}^{-Q}_{\ell_2 m_2} j_\tau(\tau,\mathbf{n}) j_\tau(\tau',\mathbf{n}')\mathcal{O}^{Q}_{\ell_1 m_1}\rangle = -iQ\,\langle \mathcal{O}^{-Q}_{\ell_2 m_2} j_\tau(\tau',\mathbf{n}')\mathcal{O}^{Q}_{\ell_1 m_1}\rangle\,. \tag{2.113}$$

We compute the remaining $jj$ matrix elements,

$$\langle \mathcal{O}^{-Q}_{\ell_2 m_2} j_\tau(\tau,\mathbf{n}) j_i(\tau',\mathbf{n}')\mathcal{O}^{Q}_{\ell_1 m_1}\rangle = 0\,, \tag{2.114}$$

$$\begin{aligned}\langle \mathcal{O}^{-Q}_{\ell_2 m_2} j_i(\tau,\mathbf{n}) J_j(\tau',\mathbf{n}')\mathcal{O}^{Q}_{\ell_1 m_1}\rangle &= \mathscr{A}^{\Delta_Q + r_0\omega_{\ell_2}}_{\Delta_Q+r_0\omega_{\ell_1}}(\tau_1,\tau_2|\tau)\frac{Q^2}{2\Omega_D r_0^{2D-2}\Delta_0 D} \\ &\times\left[\partial_i\partial'_j\sum_\ell \frac{e^{-|\tau-\tau'|\omega_\ell}}{r_0\omega_\ell}\frac{(D+2\ell-2)}{(D-2)\Omega_D}C_\ell^{\frac{D}{2}-1}(\mathbf{n}\cdot\mathbf{n}')\delta_{\ell_2\ell_1}\delta_{m_2 m_1}\right. \\ &\left.+\frac{\partial_j Y^*_{\ell_2 m_2}(\mathbf{n}')\partial_i Y_{\ell_1 m_1}(\mathbf{n})}{e^{(\tau-\tau')\omega_{\ell_2}} r_0\sqrt{\omega_{\ell_1}\omega_{\ell_2}}} + \frac{\partial_i Y^*_{\ell_2 m_2}(\mathbf{n})\partial_j Y_{\ell_1 m_1}(\mathbf{n}')}{e^{-(\tau-\tau')\omega_{\ell_1}} r_0\sqrt{\omega_{\ell_1}\omega_{\ell_2}}}\right]. \end{aligned} \tag{2.115}$$

The tree-level contribution to the $j_\tau j_i$ correlator vanishes, however, there is no symmetry protection for this matrix element and we generically expect that sub-leading corrections appear.
In the scalar case $\ell_{1,2} = 0$ the $jj$ matrix elements reduce to

$$\langle \mathcal{O}^{-Q} j_\tau(\tau,\mathbf{n}) j_\tau(\tau',\mathbf{n}')\mathcal{O}^{Q}\rangle = -\frac{Q\mathscr{A}(\tau_1,\tau_2)}{(\Omega_D r_0^{D-1})^2}\left[Q + \frac{(D-1)}{2\mu}\sum_\ell \omega_\ell e^{-|\tau-\tau'|\omega_\ell}\frac{(D+2\ell-2)}{(D-2)}C_\ell^{\frac{D}{2}-1}(\mathbf{n}\cdot\mathbf{n}')\right], \tag{2.116}$$

$$\langle \mathcal{O}^{-Q} j_i(\tau,\mathbf{n}) j_j(\tau',\mathbf{n}')\mathcal{O}^{Q}\rangle = \frac{Q\mathscr{A}(\tau_1,\tau_2)}{2\mu\Omega_D(D-1)r_0^{2D-1}}\partial_i\partial'_j\sum_\ell \frac{(D+2\ell-2)\,C_\ell^{\frac{D}{2}-1}(\mathbf{n}\cdot\mathbf{n}')}{(D-2)\Omega_D\omega_\ell e^{|\tau-\tau'|\omega_\ell}}\,, \tag{2.117}$$

$$\langle \mathcal{O}^{-Q} j_\tau(\tau,\mathbf{n}) j_i(\tau',\mathbf{n}')\mathcal{O}^{Q}\rangle = 0\,. \tag{2.118}$$

In this case the correlator $j_\tau j_i$ is protected by symmetry and vanishes exactly on the homogeneous ground state associated to $\mathcal{O}^Q$ due to rotational invariance. Finally, we remark that all of these correlators again satisfy the Ward identity Eq. (2.102).

### $\langle^{Q}_{\ell_2 m_2}|T|^{Q}_{\ell_1 m_1}\rangle$ correlators

We move on to correlators with insertions of the stress-energy tensor $T$. Here, we analyse matrix elements with an insertion of $T$ at cylinder time $\tau$ between spinning operators $\mathcal{O}^Q_{\ell m}$ at $\tau_2 > \tau > \tau_1$. The matrix element of $T_{\tau\tau}$ between one-phonon states reads

$$\langle \mathcal{O}^{-Q}_{\ell_2 m_2} T_{\tau\tau}(\tau,\mathbf{n}) \mathcal{O}^{Q}_{\ell_1 m_1}\rangle = -\mathscr{A}^{\Delta_Q + r_0\omega_{\ell_2}}_{\Delta_Q + r_0\omega_{\ell_1}}(\tau_1,\tau_2|\tau)\,\frac{1}{\Omega_D r_0^D}\Bigg[(\Delta_0+\Delta_1)\delta_{\ell_2\ell_1}\delta_{m_2 m_1} + \frac{\Omega_D}{2} r_0\sqrt{\omega_{\ell_1}\omega_{\ell_2}}\Big((D-1)Y^*_{\ell_2 m_2}(\mathbf{n})Y_{\ell_1 m_1}(\mathbf{n}) - \frac{(D-3)}{(D-1)}\frac{\partial_i Y^*_{\ell_2 m_2}(\mathbf{n})\partial_i Y_{\ell_1 m_1}(\mathbf{n})}{r_0^2\omega_{\ell_1}\omega_{\ell_2}}\Big)\Bigg]. \tag{2.119}$$

We perform a consistency check with respect to the Ward identity Eq. (2.102) by integrating over the spatial slice $\mathbf{n} \in S^{D-1}_{r_0}$ at cylinder time $\tau$. The integration removes the $\tau$-dependence from the correlator, as expected,

$$\int \mathrm{d}S(\mathbf{n})\,\langle \mathcal{O}^{-Q}_{\ell_2 m_2} T_{\tau\tau}(\tau,\mathbf{n}) \mathcal{O}^{Q}_{\ell_1 m_1}\rangle = -\mathscr{A}_{\Delta_Q + r_0\omega_{\ell_1}}(\tau_1,\tau_2)\,\frac{1}{r_0}\Big(\Delta_0+\Delta_1+r_0\omega_{\ell_1}\Big)\delta_{\ell_1\ell_2}\delta_{m_1 m_2}. \tag{2.120}$$

The matrix elements involving a single insertion of the other components of $T$ read

$$\langle \mathcal{O}^{-Q}_{\ell_2 m_2} T_{\tau i}(\tau,\mathbf{n}) \mathcal{O}^{Q}_{\ell_1 m_1}\rangle = \mathscr{A}^{\Delta_Q + r_0\omega_{\ell_2}}_{\Delta_Q + r_0\omega_{\ell_1}}(\tau_1,\tau_2|\tau)\,\frac{1}{2r_0^D}\left[\sqrt{\frac{\omega_{\ell_2}}{\omega_{\ell_1}}}\,Y^*_{\ell_2 m_2}(\mathbf{n})\partial_i Y_{\ell_1 m_1}(\mathbf{n}) - (1\leftrightarrow 2)^*\right], \tag{2.121}$$

$$\langle \mathcal{O}^{-Q}_{\ell_2 m_2} T_{ij}(\tau,\mathbf{n}) \mathcal{O}^{Q}_{\ell_1 m_1}\rangle = \mathscr{A}^{\Delta_Q + r_0\omega_{\ell_2}}_{\Delta_Q + r_0\omega_{\ell_1}}(\tau_1,\tau_2|\tau)\,\frac{1}{(D-1)\Omega_D r_0^D}\Bigg[h_{ij}\Big([\Delta_0+\Delta_1]\delta_{\ell_2\ell_1}\delta_{m_2 m_1} + \frac{\Omega_D r_0\sqrt{\omega_{\ell_1}\omega_{\ell_2}}}{2}\Big[(D-1)Y^*_{\ell_2 m_2}(\mathbf{n})Y_{\ell_1 m_1}(\mathbf{n}) - \frac{\partial_i Y^*_{\ell_2 m_2}(\mathbf{n})\partial_i Y_{\ell_1 m_1}(\mathbf{n})}{r_0^2\omega_{\ell_1}\omega_{\ell_2}}\Big]\Big) + r_0\sqrt{\omega_{\ell_1}\omega_{\ell_2}}\,\Omega_D\frac{\partial_{(i} Y^*_{\ell_2 m_2}(\mathbf{n})\partial_{j)} Y_{\ell_1 m_1}(\mathbf{n})}{r_0^2\omega_{\ell_1}\omega_{\ell_2}}\Bigg]. \tag{2.122}$$

By conformal invariance, an insertion of the trace of the stress-energy tensor $T_{\tau\tau} + h^{ij}T_{ij}$ has to vanish on any phonon state. This consistency condition is satisfied here as the expression

$$\langle \mathcal{O}^{-Q}_{\ell_2 m_2} h^{ij} T_{ij}(\tau,\mathbf{n}) \mathcal{O}^{Q}_{\ell_1 m_1}\rangle = \mathscr{A}^{\Delta_Q + r_0\omega_{\ell_2}}_{\Delta_Q + r_0\omega_{\ell_1}}(\tau_1,\tau_2|\tau)\,\frac{1}{(D-1)\Omega_D r_0^D}\Bigg[(D-1)(\Delta_0+\Delta_1)\delta_{\ell_2\ell_1}\delta_{m_2 m_1} + \frac{\Omega_D r_0\sqrt{\omega_{\ell_1}\omega_{\ell_2}}}{2}\Big((D-1)^2\,Y^*_{\ell_2 m_2}(\mathbf{n})Y_{\ell_1 m_1}(\mathbf{n}) - (D-3)\frac{\partial_i Y^*_{\ell_2 m_2}(\mathbf{n})\partial_i Y_{\ell_1 m_1}(\mathbf{n})}{r_0^2\omega_{\ell_1}\omega_{\ell_2}}\Big)\Bigg] \tag{2.123}$$

sums to zero together with Eq. (2.119). On the scalar ground state $|Q\rangle$ the matrix elements reduce to

$$\langle \mathcal{O}^{-Q} T_{\tau\tau}(\tau,\mathbf{n}) \mathcal{O}^{Q}\rangle = -\mathscr{A}(\tau_1,\tau_2)\,\frac{\Delta_0+\Delta_1}{\Omega_D r_0^D}, \tag{2.124}$$

$$\langle \mathcal{O}^{-Q} T_{\tau i}(\tau,\mathbf{n}) \mathcal{O}^{Q}\rangle = 0, \tag{2.125}$$

$$\langle \mathcal{O}^{-Q} T_{ij}(\tau,\mathbf{n}) \mathcal{O}^{Q}\rangle = \frac{\mathscr{A}(\tau_1,\tau_2)}{(D-1)}\,\frac{(\Delta_0+\Delta_1)}{\Omega_D r_0^D}\,h_{ij}. \tag{2.126}$$

The three-point function of two scalar operators and $T_{\tau i}$ vanishes due to rotational invariance. Further, rotational symmetry also fixes the matrix element of $T_{ij}$ to be proportional to the sphere metric.

### $\langle^{Q}_{\ell_2 m_2}|TT|^{Q}_{\ell_1 m_1}\rangle$ correlators

Next, we compute matrix elements with two insertions of the stress-energy tensor $T$ at cylinder times $\tau > \tau'$ between one-phonon states at $\tau_2 > \tau > \tau' > \tau_1$. There is a total of six correlators to be computed. The $T_{\tau\tau}T_{\tau\tau}$ correlator reads

$$\begin{aligned}
\langle \mathcal{O}^{-Q}_{\ell_2 m_2} T_{\tau\tau}(\tau,\mathbf{n}) T_{\tau\tau}(\tau',\mathbf{n}') \mathcal{O}^{Q}_{\ell_1 m_1}\rangle &= \mathscr{A}^{\Delta_Q + r_0\omega_{\ell_2}}_{\Delta_Q + r_0\omega_{\ell_1}}(\tau_1,\tau_2|\tau)\, \frac{\Delta_0}{\Omega_D^2 r_0^{2D}} \\
&\times \Bigg[\Bigg(\Delta_0 + 2\Delta_1 + \frac{D}{2}\sum_\ell e^{-|\tau-\tau'|\omega_\ell} r_0\omega_\ell \frac{(D+2\ell-2)}{(D-2)} C_\ell^{\frac{D}{2}-1}(\mathbf{n}\cdot\mathbf{n}')\Bigg)\delta_{\ell_1\ell_2}\delta_{m_1 m_2} \\
&+ \frac{D\Omega_D}{2} r_0\sqrt{\omega_{\ell_1}\omega_{\ell_2}}\Big(Y^*_{\ell_2 m_2}(\mathbf{n}) Y_{\ell_1 m_1}(\mathbf{n}')\, e^{(\tau-\tau')\omega_{\ell_1}} + Y^*_{\ell_2 m_2}(\mathbf{n}') Y_{\ell_1 m_1}(\mathbf{n})\, e^{-(\tau-\tau')\omega_{\ell_2}}\Big)\Bigg] \\
&+ \Bigg\{\mathscr{A}^{\Delta_Q + r_0\omega_{\ell_2}}_{\Delta_Q + r_0\omega_{\ell_1}}(\tau_1,\tau_2|\tau)\, \frac{\Omega_D \Delta_0 r_0\sqrt{\omega_{\ell_1}\omega_{\ell_2}}}{2\Omega_D^2 r_0^{2D}} \Bigg[(D-1)\, Y_{\ell_1 m_1}(\mathbf{n}) Y^*_{\ell_2 m_2}(\mathbf{n}) \\
&- \frac{(D-3)}{(D-1)} \frac{\partial_i Y_{\ell_1 m_1}(\mathbf{n}) \partial_i Y^*_{\ell_2 m_2}(\mathbf{n})}{r_0^2 \omega_{\ell_1}\omega_{\ell_2}}\Bigg] + \Big[(\tau,\mathbf{n}) \leftrightarrow (\tau',\mathbf{n}')\Big]\Bigg\}. 
\end{aligned} \tag{2.127}$$

Clearly, this correlator is symmetric under $(\tau,\mathbf{n}) \leftrightarrow (\tau',\mathbf{n}')$. In the scalar case $\ell_{1,2} = 0$ the above matrix element has first appeared in [168].
The matrix elements comprised of two insertions of $T_{ij}$ and $T_{\tau i}$ are computed to be

$$\begin{aligned}
\langle \mathcal{O}^{-Q}_{\ell_2 m_2} T_{ij}(\tau,\mathbf{n}) T_{kn}(\tau',\mathbf{n}') \mathcal{O}^{Q}_{\ell_1 m_1}\rangle &= \mathscr{A}^{\Delta_Q + r_0\omega_{\ell_2}}_{\Delta_Q + r_0\omega_{\ell_1}}(\tau_1,\tau_2|\tau)\, \frac{\Delta_0}{(D-1)^2\Omega_D^2 r_0^{2D}} \\
&\times \Bigg[\Bigg(\Delta_0 + 2\Delta_1 + \frac{D}{2}\sum_\ell e^{-|\tau-\tau'|\omega_\ell} r_0\omega_\ell \frac{(D+2\ell-2)}{D-2} C_\ell^{\frac{D}{2}-1}(\mathbf{n}\cdot\mathbf{n}')\Bigg) h_{ij}h_{kn}\delta_{\ell_2\ell_1}\delta_{m_2 m_1} \\
&+ \frac{D\Omega_D}{2} r_0\sqrt{\omega_{\ell_2}\omega_{\ell_1}}\Big(Y^*_{\ell_2 m_2}(\mathbf{n}) Y_{\ell_1 m_1}(\mathbf{n}')\, e^{(\tau-\tau')\omega_{\ell_1}} + Y^*_{\ell_2 m_2}(\mathbf{n}') Y_{\ell_1 m_1}(\mathbf{n})\, e^{-(\tau-\tau')\omega_{\ell_2}}\Big) h_{ij}h_{kn}\Bigg] \\
&+ \Bigg\{\mathscr{A}^{\Delta_Q + r_0\omega_{\ell_2}}_{\Delta_Q + r_0\omega_{\ell_1}}(\tau_1,\tau_2|\tau)\, \frac{\Omega_D \Delta_0 r_0\sqrt{\omega_{\ell_1}\omega_{\ell_2}}}{2(D-1)\Omega_D^2 r_0^{2D}} \Bigg[2\frac{\partial_{(i} Y^*_{\ell_2 m_2}(\mathbf{n}) \partial_{j)} Y_{\ell_1 m_1}(\mathbf{n})}{r_0^2 (D-1)\,\omega_{\ell_1}\omega_{\ell_2}} + Y^*_{\ell_2 m_2}(\mathbf{n}) Y_{\ell_1 m_1}(\mathbf{n})\, h_{ij} \\
&- \frac{\partial_i Y^*_{\ell_2 m_2}(\mathbf{n}) \partial_i Y_{\ell_1 m_1}(\mathbf{n})}{r_0^2 (D-1)\,\omega_{\ell_1}\omega_{\ell_2}} h_{ij}\Bigg] h_{kn} + \Big[(\tau,\mathbf{n},ij) \leftrightarrow (\tau',\mathbf{n}',kn)\Big]\Bigg\}.
\end{aligned} \tag{2.128}$$

$$\begin{aligned}
\langle \mathcal{O}^{-Q}_{\ell_2 m_2} T_{\tau i}(\tau,\mathbf{n}) T_{\tau j}(\tau',\mathbf{n}') \mathcal{O}^{Q}_{\ell_1 m_1}\rangle &= -\mathscr{A}^{\Delta_Q + r_0\omega_{\ell_2}}_{\Delta_Q + r_0\omega_{\ell_1}}(\tau_1,\tau_2|\tau)\, \frac{\Delta_0 D}{2(D-1)^2\Omega_D r_0^{2D}} \\
&\times \Bigg[\partial_i\partial'_j \sum_\ell \frac{e^{-|\tau-\tau'|\omega_\ell}}{r_0\omega_\ell} \frac{(D+2\ell-2)}{(D-2)\Omega_D} C_\ell^{\frac{D}{2}-1}(\mathbf{n}\cdot\mathbf{n}')\delta_{\ell_2\ell_1}\delta_{m_2 m_1} + \frac{\partial_i Y^*_{\ell_2 m_2}(\mathbf{n}) \partial'_j Y_{\ell_1 m_1}(\mathbf{n}')}{r_0\sqrt{\omega_{\ell_2}\omega_{\ell_1}}\, e^{-(\tau-\tau')\omega_{\ell_1}}} \\
&+ \frac{\partial'_j Y^*_{\ell_2 m_2}(\mathbf{n}') \partial_i Y_{\ell_1 m_1}(\mathbf{n})}{r_0\sqrt{\omega_{\ell_2}\omega_{\ell_1}}\, e^{(\tau-\tau')\omega_{\ell_2}}}\Bigg].
\end{aligned} \tag{2.129}$$

The $T_{\tau i}T_{\tau j}$ correlator is symmetric under $(\tau,\mathbf{n},i) \leftrightarrow (\tau',\mathbf{n}',j)$.
The final three correlators are mixed. First, the matrix element of $T_{\tau\tau}T_{\tau i}$ is given by

$$\begin{aligned}
\langle \mathcal{O}^{-Q}_{\ell_2 m_2} T_{\tau i}(\tau,\mathbf{n}) T_{\tau\tau}(\tau',\mathbf{n}') \mathcal{O}^{Q}_{\ell_1 m_1}\rangle &= -\mathscr{A}^{\Delta_Q + r_0\omega_{\ell_2}}_{\Delta_Q + r_0\omega_{\ell_1}}(\tau_1,\tau_2|\tau)\, \frac{\Delta_0 D}{2\Omega_D r_0^{2D}} \frac{1}{(D-1)} \\
&\times \Bigg[\partial_i \sum_\ell e^{-|\tau-\tau'|\omega_\ell} \frac{(D+2\ell-2)}{(D-2)\Omega_D} C_\ell^{\frac{D}{2}-1}(\mathbf{n}\cdot\mathbf{n}')\delta_{\ell_2\ell_1}\delta_{m_2 m_1} + \sqrt{\frac{\omega_{\ell_2}}{\omega_{\ell_1}}} \frac{Y^*_{\ell_2 m_2}(\mathbf{n}') \partial_i Y_{\ell_1 m_1}(\mathbf{n})}{e^{(\tau-\tau')\omega_{\ell_2}}} \\
&- \sqrt{\frac{\omega_{\ell_1}}{\omega_{\ell_2}}} \frac{\partial_i Y^*_{\ell_2 m_2}(\mathbf{n}) Y_{\ell_1 m_1}(\mathbf{n}')}{e^{-(\tau-\tau')\omega_{\ell_1}}} + \frac{(D-1)}{D}\Bigg(\sqrt{\frac{\omega_{\ell_2}}{\omega_{\ell_1}}} Y^*_{\ell_2 m_2}(\mathbf{n}) \partial_i Y_{\ell_1 m_1}(\mathbf{n}) - (1 \leftrightarrow 2)^*\Bigg)\Bigg].
\end{aligned} \tag{2.130}$$

Due to the fact that $T_{\tau i}$ vanishes on the ground state, to quadratic order this correlators only receives contributions from both linear terms as well as the quadratic term of $T_{\tau i}$.
For the combination of insertions $T_{\tau i} T_{jk}$ we find

$$\begin{aligned}
\langle \mathcal{O}^{-Q}_{\ell_2 m_2} T_{\tau i}(\tau,\mathbf{n}) T_{jk}(\tau',\mathbf{n}')\mathcal{O}^{Q}_{\ell_1 m_1}\rangle &= \mathscr{A}^{\Delta_Q+r_0\omega_{\ell_2}}_{\Delta_Q+r_0\omega_{\ell_1}}(\tau_1,\tau_2|\tau)\,\frac{\Delta_0 D}{2\Omega_D r_0^{2D}}\,\frac{h_{jk}}{(D-1)^2}\\
&\times\Bigg[\partial_i \sum_\ell e^{-|\tau-\tau'|\omega_\ell}\frac{(D+2\ell-2)}{(D-2)\Omega_D}C_\ell^{\frac{D}{2}-1}(\mathbf{n}\cdot\mathbf{n}')\delta_{\ell_2\ell_1}\delta_{m_2 m_1} + \sqrt{\frac{\omega_{\ell_2}}{\omega_{\ell_1}}}\,\frac{Y^*_{\ell_2 m_2}(\mathbf{n}')\,\partial_i Y_{\ell_1 m_1}(\mathbf{n})}{e^{(\tau-\tau')\omega_{\ell_2}}}\\
&-\sqrt{\frac{\omega_{\ell_1}}{\omega_{\ell_2}}}\,\frac{Y_{\ell_1 m_1}(\mathbf{n}')\,\partial_i Y^*_{\ell_2 m_2}(\mathbf{n})}{e^{-(\tau-\tau')\omega_{\ell_1}}} + \frac{(D-1)}{D}\left(\sqrt{\frac{\omega_{\ell_2}}{\omega_{\ell_1}}}\,Y^*_{\ell_2 m_2}(\mathbf{n})\partial_i Y_{\ell_1 m_1}(\mathbf{n}) - (1\leftrightarrow 2)^*\right)\Bigg]. \qquad (2.131)
\end{aligned}$$

By conformal invariance, the correlator $\langle \mathcal{O}^{-Q}_{\ell_2 m_2} T_{\tau i}(\tau,\mathbf{n})\, h^{jk}(\mathbf{n}')\, T_{jk}(\tau',\mathbf{n}')\mathcal{O}^{Q}_{\ell_1 m_1}\rangle$ is related to the $T_{\tau i} T_{\tau\tau}$ correlator in the previous equation.
We are left with computing the $T_{\tau\tau} T_{ij}$ correlator,

$$\begin{aligned}
\langle \mathcal{O}^{-Q}_{\ell_2 m_2} T_{\tau\tau}(\tau,\mathbf{n}) T_{ij}(\tau',\mathbf{n}')\mathcal{O}^{Q}_{\ell_1 m_1}\rangle &= -\mathscr{A}^{\Delta_Q+r_0\omega_{\ell_2}}_{\Delta_Q+r_0\omega_{\ell_1}}(\tau_1,\tau_2|\tau)\,\frac{\Delta_0}{\Omega_D^2 r_0^{2D}}\,\frac{h_{ij}}{(D-1)}\\
&\times\Bigg[\Bigg(\Delta_0 + 2\Delta_1 + \frac{D\Omega_D}{2}\sum_\ell r_0\omega_\ell e^{-|\tau-\tau'|\omega_\ell}\frac{(D+2\ell-2)}{(D-2)\Omega_D}C_\ell^{\frac{D}{2}-1}(\mathbf{n}\cdot\mathbf{n}')\Bigg)\delta_{\ell_2\ell_1}\delta_{m_2 m_1}\\
&+\frac{D\Omega_D}{2}r_0\sqrt{\omega_{\ell_2}\omega_{\ell_1}}\Big(Y^*_{\ell_2 m_2}(\mathbf{n})Y_{\ell_1 m_1}(\mathbf{n}')e^{(\tau-\tau')\omega_{\ell_1}} + Y^*_{\ell_2 m_2}(\mathbf{n}')Y_{\ell_1 m_1}(\mathbf{n})e^{-(\tau-\tau')\omega_{\ell_2}}\Big)\Bigg]\\
&-\mathscr{A}^{\Delta_Q+r_0\omega_{\ell_2}}_{\Delta_Q+r_0\omega_{\ell_1}}(\tau_1,\tau_2|\tau)\,\frac{\Delta_0 r_0\sqrt{\omega_{\ell_1}\omega_{\ell_2}}}{2\Omega_D r_0^{2D}}\,h_{ij}\left[Y^*_{\ell_2 m_2}(\mathbf{n})Y_{\ell_1 m_1}(\mathbf{n}) - \frac{(D-3)}{(D-1)^2}\,\frac{\partial_i Y^*_{\ell_2 m_2}(\mathbf{n})\partial_i Y_{\ell_1 m_1}(\mathbf{n})}{r_0^2\omega_{\ell_1}\omega_{\ell_2}}\right]\\
&-\mathscr{A}^{\Delta_Q+r_0\omega_{\ell_2}}_{\Delta_Q+r_0\omega_{\ell_1}}(\tau_1,\tau_2|\tau')\,\frac{\Delta_0 r_0\sqrt{\omega_{\ell_1}\omega_{\ell_2}}}{2\Omega_D r_0^{2D}}\Bigg[h_{ij}\Big(Y^*_{\ell_2 m_2}(\mathbf{n}')Y_{\ell_1 m_1}(\mathbf{n}') - \frac{\partial_i Y^*_{\ell_2 m_2}(\mathbf{n}')\partial_i Y_{\ell_1 m_1}(\mathbf{n}')}{(D-1)\,r_0^2\omega_{\ell_1}\omega_{\ell_2}}\Big)\\
&\qquad + 2\,\frac{\partial_{(i} Y^*_{\ell_2 m_2}(\mathbf{n}')\partial_{j)} Y_{\ell_1 m_1}(\mathbf{n}')}{(D-1)\,r_0^2\omega_{\ell_1}\omega_{\ell_2}}\Bigg]. \qquad (2.132)
\end{aligned}$$

Even though this correlator is not symmetric under $(\tau,\mathbf{n}) \leftrightarrow (\tau',\mathbf{n}')$, by conformal invariance, the correlator $\langle \mathcal{O}^{-Q}_{\ell_2 m_2} T_{\tau\tau}(\tau,\mathbf{n})\, h^{ij}(\mathbf{n})\, T_{ij}(\tau',\mathbf{n}')\mathcal{O}^{Q}_{\ell_1 m_1}\rangle$ is.
All of the above correlators satisfy the Ward identity Eq. (2.102) for insertions of $T_{\tau\tau}$. For example, the matrix element with two insertions of $T_{\tau\tau}$ satisfies

$$\int \mathrm{d}S(\mathbf{n})\,\langle \mathcal{O}^{-Q}_{\ell_2 m_2} T_{\tau\tau}(\tau,\mathbf{n}) T_{\tau\tau}(\tau',\mathbf{n}')\mathcal{O}^{Q}_{\ell_1 m_1}\rangle = -\frac{(\Delta_0+\Delta_1+r_0\omega_{\ell_2})}{r_0}\,\langle \mathcal{O}^{-Q}_{\ell_2 m_2} T_{\tau\tau}(\tau',\mathbf{n}')\mathcal{O}^{Q}_{\ell_1 m_1}\rangle\,. \qquad (2.133)$$

Finally, the above $TT$ matrix elements restricted to the scalar case $\ell_{1,2} = 0$ simplify as follows:

$$\langle \mathcal{O}^{-Q} T_{\tau\tau}(\tau,\mathbf{n}) T_{\tau\tau}(\tau',\mathbf{n}')\mathcal{O}^{Q}\rangle = \frac{\mathscr{A}(\tau_1,\tau_2)\Delta_0}{\Omega_D^2 r_0^{2D}}\left[\Delta_0 + 2\Delta_1 + \frac{D}{2}\sum_\ell \frac{r_0\omega_\ell}{e^{|\tau-\tau'|\omega_\ell}}\frac{(D+2\ell-2)}{(D-2)}C_\ell^{\frac{D}{2}-1}(\mathbf{n}\cdot\mathbf{n}')\right], \tag{2.134}$$

$$\langle \mathcal{O}^{-Q} T_{ij}(\tau,\mathbf{n}) T_{kn}(\tau',\mathbf{n}')\mathcal{O}^{Q}\rangle = \frac{\mathscr{A}(\tau_1,\tau_2)\Delta_0}{\Omega_D^2 r_0^{2D}}\frac{h_{ij}h_{kn}}{(D-1)^2}\left[\Delta_0 + 2\Delta_1 + \frac{D}{2}\sum_\ell \frac{r_0\omega_\ell}{e^{|\tau-\tau'|\omega_\ell}}\frac{(D+2\ell-2)}{(D-2)}C_\ell^{\frac{D}{2}-1}(\mathbf{n}\cdot\mathbf{n}')\right], \tag{2.135}$$

$$\langle \mathcal{O}^{-Q} T_{\tau i}(\tau,\mathbf{n}) T_{\tau j}(\tau',\mathbf{n}')\mathcal{O}^{Q}\rangle = -\frac{\mathscr{A}(\tau_1,\tau_2)\Delta_0 D}{2(D-1)^2\Omega_D^2 r_0^{2D}}\partial_i\partial'_j\sum_\ell \frac{e^{-|\tau-\tau'|\omega_\ell}}{r_0\omega_\ell}\frac{(D+2\ell-2)}{(D-2)}C_\ell^{\frac{D}{2}-1}(\mathbf{n}\cdot\mathbf{n}'), \tag{2.136}$$

$$\langle \mathcal{O}^{-Q} T_{\tau i}(\tau,\mathbf{n}) T_{\tau\tau}(\tau',\mathbf{n}')\mathcal{O}^{Q}\rangle = -\frac{\mathscr{A}(\tau_1,\tau_2)\Delta_0 D}{2(D-1)\Omega_D^2 r_0^{2D}}\partial_i\sum_\ell e^{-|\tau-\tau'|\omega_\ell}\frac{(D+2\ell-2)}{(D-2)}C_\ell^{\frac{D}{2}-1}(\mathbf{n}\cdot\mathbf{n}'), \tag{2.137}$$

$$\langle \mathcal{O}^{-Q} T_{\tau i}(\tau,\mathbf{n}) T_{jk}(\tau',\mathbf{n}')\mathcal{O}^{Q}\rangle = \frac{\mathscr{A}(\tau_1,\tau_2)\Delta_0 D h_{jk}}{2(D-1)^2\Omega_D^2 r_0^{2D}}\partial_i\sum_\ell e^{-|\tau-\tau'|\omega_\ell}\frac{(D+2\ell-2)}{(D-2)}C_\ell^{\frac{D}{2}-1}(\mathbf{n}\cdot\mathbf{n}'), \tag{2.138}$$

$$\langle \mathcal{O}^{-Q} T_{\tau\tau}(\tau,\mathbf{n}) T_{ij}(\tau',\mathbf{n}')\mathcal{O}^{Q}\rangle = -\frac{\mathscr{A}(\tau_1,\tau_2)\Delta_0}{\Omega_D^2 r_0^{2D}}\frac{h_{ij}}{(D-1)}\left[\Delta_0 + 2\Delta_1 + \frac{D}{2}\sum_\ell \frac{r_0\omega_\ell}{e^{|\tau-\tau'|\omega_\ell}}\frac{(D+2\ell-2)}{(D-2)}C_\ell^{\frac{D}{2}-1}(\mathbf{n}\cdot\mathbf{n}')\right]. \tag{2.139}$$

The correlator $\langle \mathcal{O}^{-Q} T_{\tau i} T_{\tau\tau}\mathcal{O}^{Q}\rangle$ was first computed in [168] in the macroscopic limit $r_0 \to \infty$.

### $\langle^{Q}_{\ell_2 m_2}|Tj|^{Q}_{\ell_1 m_1}\rangle$ correlators

The last batch of correlators we consider contain one insertion of the stress-energy tensor $T$ and one insertion of the $O(2)$ current $j$, respectively, at times $\tau > \tau'$ inserted between spinning (primary) operators $\mathcal{O}^{Q}_{\ell m}$ at cylinder times $\tau_1, \tau_2$ such that we have the ordering $\tau_2 > \tau > \tau' > \tau_1$. There is a total of six correlators involving the various components which can be computed. We start by computing the $T_{\tau i} j_\tau$ correlator,

$$\begin{aligned}\langle \mathcal{O}^{-Q}_{\ell_2 m_2} T_{\tau i}(\tau,\mathbf{n}) j_\tau(\tau',\mathbf{n}')\mathcal{O}^{Q}_{\ell_1 m_1}\rangle &= -i\mathscr{A}^{\Delta_Q + r_0\omega_{\ell_2}}_{\Delta_Q + r_0\omega_{\ell_1}}(\tau_1,\tau_2|\tau)\frac{Q}{2\Omega_D r_0^{2D-1}} \\ &\times\left[\partial_i\sum_\ell e^{-|\tau-\tau'|\omega_\ell}\frac{(D+2\ell-2)}{(D-2)\Omega_D}C_\ell^{\frac{D}{2}-1}(\mathbf{n}\cdot\mathbf{n}')\delta_{\ell_2\ell_1}\delta_{m_2 m_1} + \sqrt{\frac{\omega_{\ell_2}}{\omega_{\ell_1}}}\frac{Y^*_{\ell_2 m_2}(\mathbf{n}')\partial_i Y_{\ell_1 m_1}(\mathbf{n})}{e^{(\tau-\tau')\omega_{\ell_2}}}\right. \\ &\left. - \sqrt{\frac{\omega_{\ell_1}}{\omega_{\ell_2}}}\frac{\partial_i Y^*_{\ell_2 m_2}(\mathbf{n})Y_{\ell_1 m_1}(\mathbf{n}')}{e^{-(\tau-\tau')\omega_{\ell_1}}} + \left(\sqrt{\frac{\omega_{\ell_2}}{\omega_{\ell_1}}}Y^*_{\ell_2 m_2}(\mathbf{n})\partial_i Y_{\ell_1 m_1}(\mathbf{n}) - (1\leftrightarrow 2)^*\right)\right].\end{aligned} \tag{2.140}$$

As $T_{\tau i}$ vanishes on the ground state, only linear terms and the quadratic term of $T_{\tau i}$ contribute up to quadratic order in the fluctuations.
The matrix element with insertions of $j_i$ and $T_{\tau\tau}$ results in

$$\begin{aligned}\langle \mathcal{O}^{-Q}_{\ell_2 m_2} j_i(\tau,\mathbf{n}) T_{\tau\tau}(\tau',\mathbf{n}')\mathcal{O}^{Q}_{\ell_1 m_1}\rangle &= -i\mathscr{A}^{\Delta_Q + r_0\omega_{\ell_2}}_{\Delta_Q + r_0\omega_{\ell_1}}(\tau_1,\tau_2|\tau)\frac{Q}{2\Omega_D r_0^{2D-1}} \\ &\times\left[\delta_{\ell_2\ell_1}\delta_{m_2 m_1}\partial_i\sum_\ell e^{-|\tau-\tau'|\omega_\ell}\frac{(D+2\ell-2)}{(D-2)\Omega_D}C_\ell^{\frac{D}{2}-1}(\mathbf{n}\cdot\mathbf{n}') + \sqrt{\frac{\omega_{\ell_2}}{\omega_{\ell_1}}}\frac{Y^*_{\ell_2 m_2}(\mathbf{n}')\partial_i Y_{\ell_1 m_1}(\mathbf{n})}{e^{(\tau-\tau')\omega_{\ell_2}}}\right. \\ &\left. - \sqrt{\frac{\omega_{\ell_1}}{\omega_{\ell_2}}}\frac{\partial_i Y^*_{\ell_2 m_2}(\mathbf{n})\,Y_{\ell_1 m_1}(\mathbf{n}')}{e^{-(\tau-\tau')\omega_{\ell_1}}} + \frac{(D-2)}{D}\left(\sqrt{\frac{\omega_{\ell_2}}{\omega_{\ell_1}}}Y^*_{\ell_2 m_2}(\mathbf{n})\partial_i Y_{\ell_1 m_1}(\mathbf{n}) - (1\leftrightarrow 2)^*\right)\right].\end{aligned} \tag{2.141}$$

Here, up to quadratic order only the linear terms and the quadratic term in $j_i$ contribute, since $j_i$ itself lacks a ground-state contribution From the form of the expansion of the currents in Eq. (2.101) it is evident that this correlator is related to the previous one.

For the $T_{\tau\tau} j_\tau$ we find that

$$
\begin{aligned}
\langle \mathscr{O}^{-Q}_{\ell_2 m_2} T_{\tau\tau}(\tau,\mathbf{n}) j_\tau(\tau',\mathbf{n}') \mathscr{O}^{Q}_{\ell_1 m_1}\rangle &= i\mathscr{A}^{\Delta_Q + r_0\omega_{\ell_2}}_{\Delta_Q + r_0\omega_{\ell_1}}(\tau_1,\tau_2|\tau)\,\frac{Q(D-1)}{2\Omega_D r_0^{2D-1}}\, r_0\sqrt{\omega_{\ell_2}\omega_{\ell_1}} \\
&\times \Bigg[\Bigg(\frac{Y^*_{\ell_2 m_2}(\mathbf{n}')Y_{\ell_1 m_1}(\mathbf{n})}{e^{(\tau-\tau')\omega_{\ell_2}}} + \frac{Y^*_{\ell_2 m_2}(\mathbf{n})Y_{\ell_1 m_1}(\mathbf{n}')}{e^{-(\tau-\tau')\omega_{\ell_1}}}\Bigg) + \sum_\ell \frac{r_0\omega_\ell\, e^{-|\tau-\tau'|\omega_\ell}}{r_0\sqrt{\omega_{\ell_2}\omega_{\ell_1}}}\,\frac{(D+2\ell-2)}{(D-2)\Omega_D}\, C_\ell^{\frac{D}{2}-1}(\mathbf{n}\cdot\mathbf{n}') \\
&+ \frac{(D-2)}{D}\Bigg(Y^*_{\ell_2 m_2}(\mathbf{n}')Y_{\ell_1 m_1}(\mathbf{n}') - \frac{\partial_i Y^*_{\ell_2 m_2}(\mathbf{n}')\partial_i Y_{\ell_1 m_1}(\mathbf{n}')}{r_0^2(D-1)\omega_{\ell_2}\omega_{\ell_1}}\Bigg) + \frac{2}{(D-1)}\Bigg(\frac{1}{\Omega_D}\Big[\frac{\Delta_0+\Delta_1}{r_0\sqrt{\omega_{\ell_1}\omega_{\ell_2}}}\Big]\delta_{\ell_2\ell_1}\delta_{m_2 m_1} \\
&+ \frac{1}{2}\Big[(D-1)Y^*_{\ell_2 m_2}(\mathbf{n})Y_{\ell_1 m_1}(\mathbf{n}) - \frac{(D-3)}{(D-1)}\,\frac{\partial_i Y^*_{\ell_2 m_2}(\mathbf{n})\partial_i Y_{\ell_1 m_1}(\mathbf{n})}{r_0^2\omega_{\ell_1}\omega_{\ell_2}}\Big]\Bigg)\Bigg].
\end{aligned}
\tag{2.142}
$$

Here, in relation to the Ward identities Eq. (2.102), we have two independent consistency conditions. If we integrate this result over the spatial slice $\mathbf{n}'$ the quadratic term in the expansion of $J_\tau$ vanishes, whereas if we integrate over the $\mathbf{n}$ the quadratic term in the expansion of $T_{\tau\tau}$ remains finite. The reason for this is that the quadratic term from $T_{\tau\tau}$ has to correct the energy by $r_0\omega_{\ell_2}$, in accordance with the Ward identities Eq. (2.102).

The matrix element for insertions of $T_{\tau i}$ and $j_j$ is given by

$$
\begin{aligned}
\langle \mathscr{O}^{-Q}_{\ell_2 m_2} T_{\tau i}(\tau,\mathbf{n}) j_j(\tau',\mathbf{n}') \mathscr{O}^{Q}_{\ell_1 m_1}\rangle &= -i\mathscr{A}^{\Delta_Q + r_0\omega_{\ell_2}}_{\Delta_Q + r_0\omega_{\ell_1}}(\tau_1,\tau_2|\tau)\,\frac{Q}{2\Omega_D r_0^{2D-1}}\,\frac{1}{(D-1)} \\
&\times \Bigg[\partial_i\partial'_j \sum_\ell \frac{e^{-|\tau-\tau'|\omega_\ell}}{r_0\omega_\ell}\,\frac{(D+2\ell-2)}{(D-2)\Omega_D}\, C_\ell^{\frac{D}{2}-1}(\mathbf{n}\cdot\mathbf{n}')\,\delta_{\ell_2\ell_1}\delta_{m_2 m_1} + \frac{\partial'_j Y^*_{\ell_2 m_2}(\mathbf{n}')\partial_i Y_{\ell_1 m_1}(\mathbf{n})}{r_0\sqrt{\omega_{\ell_1}\omega_{\ell_2}}\, e^{(\tau-\tau')\omega_{\ell_2}}} \\
&+ \frac{\partial_i Y^*_{\ell_2 m_2}(\mathbf{n})\partial'_j Y_{\ell_1 m_1}(\mathbf{n}')}{r_0\sqrt{\omega_{\ell_1}\omega_{\ell_2}}\, e^{-(\tau-\tau')\omega_{\ell_1}}}\Bigg].
\end{aligned}
\tag{2.143}
$$

As both currents $T_{\tau i}$ and $j_i$ have vanishing ground-state contributions, the only contribution to quadratic order in the fluctuations comes from the two linear terms in the expansions of the currents. The correlator with insertions of $T_{ij}$ and $j_\tau$ is computed to be

$$
\begin{aligned}
\langle \mathscr{O}^{-Q}_{\ell_2 m_2} T_{ij}(\tau,\mathbf{n}) j_\tau(\tau',\mathbf{n}') \mathscr{O}^{Q}_{\ell_1 m_1}\rangle &= -i\mathscr{A}^{\Delta_Q + r_0\omega_{\ell_2}}_{\Delta_Q + r_0\omega_{\ell_1}}(\tau_1,\tau_2|\tau)\,\frac{Q}{\Omega_D r_0^{2D-1}} \\
&\times \Bigg[h_{ij}\Bigg(\frac{(\Delta_0+\Delta_1)}{\Omega_D(D-1)}\delta_{\ell_2\ell_1}\delta_{m_2 m_1} + \frac{1}{2}\sum_\ell e^{-|\tau-\tau'|\omega_\ell}\, r_0\omega_\ell\,\frac{(D+2\ell-2)}{(D-2)\Omega_D}\, C_\ell^{\frac{D}{2}-1}(\mathbf{n}\cdot\mathbf{n}')\,\delta_{\ell_2\ell_1}\delta_{m_2 m_1} \\
&+ \frac{r_0\sqrt{\omega_{\ell_1}\omega_{\ell_2}}}{2}\Bigg[\frac{Y^*_{\ell_2 m_2}(\mathbf{n}')Y_{\ell_1 m_1}(\mathbf{n})}{e^{(\tau-\tau')\omega_{\ell_2}}} + \frac{Y^*_{\ell_2 m_2}(\mathbf{n})Y_{\ell_1 m_1}(\mathbf{n}')}{e^{-(\tau-\tau')\omega_{\ell_1}}} + \Big(1+\frac{(D-2)}{D}\Big)Y^*_{\ell_2 m_2}(\mathbf{n})Y_{\ell_1 m_1}(\mathbf{n}) \\
&- \Big(1+\frac{(D-2)}{D}\Big)\frac{\partial_i Y^*_{\ell_2 m_2}(\mathbf{n})\partial_i Y_{\ell_1 m_1}(\mathbf{n})}{(D-1)\, r_0^2\omega_{\ell_1}\omega_{\ell_2}}\Bigg]\Bigg) + \frac{r_0\sqrt{\omega_{\ell_1}\omega_{\ell_2}}}{(D-1)}\,\frac{\partial_{(i} Y^*_{\ell_2 m_2}(\mathbf{n})\partial_{j)} Y_{\ell_1 m_1}(\mathbf{n})}{r_0^2\omega_{\ell_1}\omega_{\ell_2}}\Bigg].
\end{aligned}
\tag{2.144}
$$

When contracted with $h^{ij}$, by virtue of conformal invariance, this matrix element is related to the correlator in Eq. (2.142).

The last matrix element includes insertions of $j_i$ and $T_{jk}$ and reads

$$
\begin{aligned}
\langle \mathscr{O}^{-Q}_{\ell_2 m_2} j_i(\tau,\mathbf{n}) T_{jk}(\tau',\mathbf{n}') \mathscr{O}^{Q}_{\ell_1 m_1}\rangle &= i\mathscr{A}^{\Delta_Q + r_0\omega_{\ell_2}}_{\Delta_Q + r_0\omega_{\ell_1}}(\tau_1,\tau_2|\tau)\,\frac{Q}{2\Omega_D r_0^{2D-1}}\,\frac{h_{jk}}{(D-1)} \\
&\times \Bigg[\partial_i \sum_\ell e^{-|\tau-\tau'|\omega_\ell}\,\frac{(D+2\ell-2)}{(D-2)\Omega_D}\, C_\ell^{\frac{D}{2}-1}(\mathbf{n}\cdot\mathbf{n}')\,\delta_{\ell_2\ell_1}\delta_{m_2 m_1} + \sqrt{\frac{\omega_{\ell_2}}{\omega_{\ell_1}}}\,\frac{Y^*_{\ell_2 m_2}(\mathbf{n}')\partial_i Y_{\ell_1 m_1}(\mathbf{n})}{e^{(\tau-\tau')\omega_{\ell_2}}} \\
&- \sqrt{\frac{\omega_{\ell_1}}{\omega_{\ell_2}}}\,\frac{Y_{\ell_1 m_1}(\mathbf{n}')\partial_i Y^*_{\ell_2 m_2}(\mathbf{n})}{e^{-(\tau-\tau')\omega_{\ell_1}}} + \frac{(D-2)}{D}\Bigg(\sqrt{\frac{\omega_{\ell_2}}{\omega_{\ell_1}}}\, Y^*_{\ell_2 m_2}(\mathbf{n})\partial_i Y_{\ell_1 m_1}(\mathbf{n}) - (1\leftrightarrow 2)^*\Bigg)\Bigg].
\end{aligned}
\tag{2.145}
$$

As the quadratic term in the expansion of $T_{jk}$ only appears to cubic order in the expansion of the above correlator, it is fully proportional to $h_{jk}$ up to quadratic order. Higher order correction, however, will make the tensor structure more complicated.
There is significant simplification in the expressions for the $Tj$ matrix elements when restricted to the scalar ground state $\ell_{1,2}=0$,

$$\langle\mathcal{O}^{-Q}T_{\tau i}(\tau,\mathbf{n})j_\tau(\tau',\mathbf{n}')\mathcal{O}^{Q}\rangle=-i\frac{\mathscr{A}(\tau_1,\tau_2)\,Q}{2\Omega_D^2 r_0^{2D-1}}\partial_i\sum_\ell e^{-(\tau-\tau')\omega_\ell}\frac{(D+2\ell-2)}{(D-2)}C_\ell^{\frac{D}{2}-1}(\mathbf{n}\cdot\mathbf{n}'),\tag{2.146}$$

$$\langle\mathcal{O}^{-Q}j_i(\tau,\mathbf{n})T_{\tau\tau}(\tau',\mathbf{n}')\mathcal{O}^{Q}\rangle=-i\frac{\mathscr{A}(\tau_1,\tau_2)\,Q}{2\Omega_D^2 r_0^{2D-1}}\partial_i\sum_\ell e^{-(\tau-\tau')\omega_\ell}\frac{(D+2\ell-2)}{(D-2)}C_\ell^{\frac{D}{2}-1}(\mathbf{n}\cdot\mathbf{n}'),\tag{2.147}$$

$$\langle\mathcal{O}^{-Q}T_{\tau\tau}(\tau,\mathbf{n})j_\tau(\tau',\mathbf{n}')\mathcal{O}^{Q}\rangle=i\frac{\mathscr{A}(\tau_1,\tau_2)\,Q(D-1)}{2\Omega_D r_0^{2D-1}}\left[\sum_\ell\frac{r_0\omega_\ell}{e^{(\tau-\tau')\omega_\ell}}\frac{(D+2\ell-2)}{(D-2)}C_\ell^{\frac{D}{2}-1}(\mathbf{n}\cdot\mathbf{n}')+\frac{2(\Delta_0+\Delta_1)}{(D-1)\Omega_D}\right],\tag{2.148}$$

$$\langle\mathcal{O}^{-Q}T_{\tau i}(\tau,\mathbf{n})j_j(\tau',\mathbf{n}')\mathcal{O}^{Q}\rangle=-i\frac{Q\mathscr{A}(\tau_1,\tau_2)}{2(D-1)\Omega_D^2 r_0^{2D-1}}\partial_i\partial'_j\sum_\ell\frac{(D+2\ell-2)}{(D-2)}\frac{C_\ell^{\frac{D}{2}-1}(\mathbf{n}\cdot\mathbf{n}')}{e^{(\tau-\tau')\omega_\ell}r_0\omega_\ell},\tag{2.149}$$

$$\langle\mathcal{O}^{-Q}T_{ij}(\tau,\mathbf{n})j_\tau(\tau',\mathbf{n}')\mathcal{O}^{Q}\rangle=-i\frac{Q\mathscr{A}(\tau_1,\tau_2)}{\Omega_D^2 r_0^{2D-1}}h_{ij}\left[\frac{(\Delta_0+\Delta_1)}{(D-1)}+\frac{1}{2}\sum_\ell\frac{r_0\omega_\ell}{e^{(\tau-\tau')\omega_\ell}}\frac{(D+2\ell-2)}{(D-2)}C_\ell^{\frac{D}{2}-1}(\mathbf{n}\cdot\mathbf{n}')\right],\tag{2.150}$$

$$\langle\mathcal{O}^{-Q}j_i(\tau,\mathbf{n})T_{jk}(\tau',\mathbf{n}')\mathcal{O}^{Q}\rangle=i\frac{Q\mathscr{A}(\tau_1,\tau_2)}{2\Omega_D^2 r_0^{2D-1}}\frac{h_{jk}}{(D-1)}\partial_i\sum_\ell\frac{(D+2\ell-2)}{(D-2)}\frac{C_\ell^{\frac{D}{2}-1}(\mathbf{n}\cdot\mathbf{n}')}{e^{(\tau-\tau')\omega_\ell}}.\tag{2.151}$$

The two correlators with insertions of $j_i$, $T_{\tau\tau}$ and $T_{\tau i}$, $j_\tau$ have first appeared in [168] in the macroscopic limit $r_0\to\infty$.

### 2.2.8 Heavy–light–heavy correlators

Within the validity of the large-charge EFT it is possible to compute correlators involving the insertions of (spinning) primaries $\mathcal{O}^q$ with small $O(2)$ charges $q\ll Q$ (and small enough spin). These small-charge insertions act as probes around the semi-classical saddle-point configurations and do not alter the EFT description. The computation of these matrix elements has been carried out in [18, 46, 154]. For the purpose of completeness, we review the computation for some of these correlators here.

Within the EFT, every operator has to be constructed in terms of the NG field $\chi$ (or the ground state $\chi^{\circledast}=-i\mu\tau$ plus the fluctuation $\pi$, alternatively). Starting from this observation, we try to match the quantum numbers of an operator $\mathcal{O}^{q;\Delta}_{\ell m}$ with charge $q\ll Q$, scaling dimension $\Delta\ll\Delta_Q$ and which transforms in a representation of spin $\ell\sim\mathcal{O}(1)$ in terms of the NG boson alone. We find that at leading order in $Q$ the operator $\mathcal{O}^{q;\Delta}_{\ell m}$ must take the form[39]

$$\mathcal{O}^{q;\Delta}_{\ell m}=k^{(1)}_{\Delta,\ell,q}\mathrm{U}^{\nu_1\dots\nu_\ell}_{\ell m}\partial_{\nu_1}\chi\dots\partial_{\nu_\ell}\chi\left(\partial\chi\right)^{\Delta-\ell}e^{iq\chi}+\dots,\tag{2.152}$$

where $k^{(1)}_{\Delta,\ell,q}$ is a Wilsonian coefficient that is not determinable within the EFT and is therefore theory-dependent. The operator $\mathrm{U}^{\nu_1\dots\nu_\ell}_{\ell m}$ gives the change of basis from Euclidean to spherical tensors. It is explicitly given in Eq. (B.19). For simplicity, for scalar operators we will use the shorthand notation

$$\mathcal{O}^{q;\Delta}:=\mathcal{O}^{q;\Delta}_{00},\qquad\qquad k^{(1)}_{\Delta,q}:=k^{(1)}_{\Delta,0,q}.\tag{2.153}$$

[39]The sub-leading term in the LCE is also fixed and must take the form $k^{(2)}_{\Delta,q}(\partial\chi)^{\Delta-2}(\mathscr{R}+\dots)e^{iq\chi}$. Here, we have neglected higher-order corrections that are necessary to obtain a Weyl-invariant quantity.

**$\langle \mathcal{O}^{-Q-q}_{\ell_2 m_2} \mathcal{O}^{q;\Delta} \mathcal{O}^{Q}_{\ell_1 m_1} \rangle$ correlators**

We start with the insertion of a scalar operator with small charge $q \ll Q$,

$$\langle \mathcal{O}^{-Q-q}_{\ell_2 m_2}(\tau_2) \mathcal{O}^{q;\Delta}(\tau_c, \mathbf{n}_c) \mathcal{O}^{Q}_{\ell_1 m_1}(\tau_1) \rangle \,, \qquad \text{where } \left[\mathcal{O}^{q;\Delta}(\tau,\mathbf{n})\right]^\dagger = \mathcal{O}^{-q;\Delta}(-\tau,\mathbf{n}) \,. \tag{2.154}$$

The form of the classical homogeneous contribution to the above three-point function can be inferred from general symmetry considerations (see Appendix B.1.2),

$$\langle \mathcal{O}^{-Q-q}_{\ell_2 m_2}(\tau_2) \mathcal{O}^{q;\Delta}(\tau_c, \mathbf{n}_c) \mathcal{O}^{Q}_{\ell_1 m_1}(\tau_1) \rangle = C^{\Delta}_{Q+q,q,Q} \delta_{\ell_1 \ell_2} \delta_{m_1 m_2} e^{-\omega_{\ell_2}(\tau_2-\tau_1)} e^{-\Delta_{Q+q}\frac{(\tau_2-\tau_c)}{r_0}} e^{-\Delta_Q \frac{(\tau_c-\tau_1)}{r_0}} \,. \tag{2.155}$$

The above expression is consistent with the general form of a three-point function in a CFT at large separation on the cylinder given in Eq. (B.25). Dimensional analysis tells us that the OPE coefficient $\mathcal{C}^{\Delta}_{Q+q,q,Q}$ includes a factor of $r_0^{\Delta}$ coming from the insertion of $\mathcal{O}^{q;\Delta}$, which can be extracted so that

$$r_0^{-\Delta}\, \tilde{C}^{\Delta}_{Q+q,q,Q} = C^{\Delta}_{Q+q,q,Q} \,. \tag{2.156}$$

From the point-of-view of the path integral, the difference between the standard large-charge semi-classical trajectory governing the usual path integral in Eq. (2.65) and the correlator with an insertion of $\mathcal{O}^{q;\Delta}$ in Eq. (2.154) is that the insertion of an operator of charge $q$ introduces a source term to the action in Eq. (2.65) that changes the boundary conditions as well as the bulk EoM slightly. They now read

$$\nabla_\mu j^\mu = \nabla_\mu \frac{\delta S}{\delta(\partial_\mu \chi)} = \frac{iq\delta(\tau-\tau_c)\delta(\mathbf{n}-\mathbf{n}_c)}{\sqrt{g}} \,, \qquad \begin{cases} j^\mu(\tau_1 = -\infty, \mathbf{n}_1) = \dfrac{\delta^\mu_0 Q}{r_0^{D-1}\Omega_D} \,, \\[2ex] j^\mu(\tau_2 = +\infty, \mathbf{n}_2) = \dfrac{\delta^\mu_0 (Q+q)}{r_0^{D-1}\Omega_D} \,, \end{cases} \tag{2.157}$$

where $g_{\mu\nu}$ is the metric on the cylinder. Ignoring the boundary condition for a moment, the partial differential equation in Eq. (2.157) is linear and hence, on general grounds, the solution to it can be written as a sum of the homogeneous solution $\chi = -i\mu\tau + \pi(\tau,\mathbf{n})$, solving the EoM for $q = 0$, and a particular solution $q\pi_p(\tau,\mathbf{n})$. The particular solution $\pi_p$ solves the equation Eq. (2.157) for a delta-function source term and is therefore structurally identical to the Green's function, *i.e.* the propagator. However, considering the the full system of equations, we find that the new boundary conditions change the form of the constant term in the mode expansion (the $\ell = 0$ term within the expansion in Eq. (2.31)). Fortunately, to leading order the solution to Eq. (2.157) is still given by the sum of the homogeneous solution $\chi$ and the particular solution $\pi_p$.[40]

As discussed, the additional source term only appears in and alters the field normalization while the mode expansion remains equal to the one in Eq. (2.31). Hence, we can still apply the standard canonical

[40]Expanded in Fourier modes, the particular solution $\pi_p$ reads (recall that $\mu \propto Q^{\frac{1}{(D-1)}}$, see Eq. (2.35))

$$\pi_p(\tau,\mathbf{n}) = -\frac{i}{c_1 D(D-1)(\mu r_0)^{D-2}} \left[ -\frac{|\tau-\tau_c|}{2r_0\Omega_D}\theta(\tau-\tau_c) + \sum_{m,l} \frac{Y^m_l(\mathbf{n}_c)^\dagger Y^m_l(\mathbf{n}) e^{-\omega_l|\tau-\tau_c|}}{2r_0\omega_l} \right],$$

where the theta-function $\theta(\tau-\tau_c)$ is required to satisfy the different boundary conditions at $\tau_{2,1} = \pm\infty$.

quantization picture around Eq. (2.42) and in the limit of large separation $\tau_{2,1} \to \pm\infty$ identify [41]

$$\lim_{\tau_2\to\infty} \langle 0|\mathcal{O}^{-Q-q}_{\ell_2 m_2}(\tau_2) = \langle^{Q+q}_{\ell_2 m_2}| = \langle Q+q|\, a_{\ell_2 m_2}\,, \qquad \lim_{\tau_1\to-\infty} \mathcal{O}^{Q}_{\ell_1 m_1}(\tau_1)\,|0\rangle = |^{Q}_{\ell_1 m_1}\rangle = a^\dagger_{\ell_1 m_1}|Q\rangle\,. \tag{2.158}$$

To leading order the operator $\mathcal{O}^{q;\Delta}$ within the EFT is given by (see Eq. (2.152))

$$\mathcal{O}^{q;\Delta}(\tau_c,\mathbf{n}_c) = k^{(1)}_{\Delta,q}\mu^\Delta e^{\mu q \tau_c} e^{iq\pi(\tau_c,\mathbf{n}_c)+iq^2\pi_p(\tau_c,\mathbf{n}_c)} + \dots\,, \tag{2.159}$$

and the correlator in Eq. (2.154) is computed to be[42]

$$\begin{aligned}\langle \mathcal{O}^{-Q-q}_{\ell_2 m_2}\mathcal{O}^{q;\Delta}(\tau_c,\mathbf{n}_c)\mathcal{O}^{Q}_{\ell_1 m_1}\rangle &= k^{(1)}_{\Delta,q}\mu^\Delta e^{\mu q\tau_c}\langle Q+q|\, a_{\ell_2 m_2}(\tau_2)\, e^{iq\pi(\tau_c,\mathbf{n}_c)+iq^2\pi_p(\tau_c,\mathbf{n}_c)}\, a^\dagger_{\ell_1 m_1}(\tau_1)\,|Q\rangle + \dots \\ &= (\mu r_0)^\Delta \frac{k^{(1)}_{\Delta,q}}{r_0^\Delta}\delta_{\ell_2\ell_1}\delta_{m_2 m_1}\mathcal{A}^{\Delta_{Q+q}+r_0\omega_{\ell_1}}_{\Delta_Q+r_0\omega_{\ell_1}}(\tau_1,\tau_2|\tau_c) + \dots\,.\end{aligned} \tag{2.160}$$

We can generalize this result to the case where we insert an operator $\mathcal{O}^{q;\Delta}_{\ell m}$ in a representation of spin $\ell \sim \mathcal{O}(1)$. In this case we find the matrix element

$$\langle Q+q|a_{\ell_2 m_2}\mathcal{O}^{q;\Delta}_{\ell m}(\tau_c,\mathbf{n}_c)\, a^\dagger_{\ell_1 m_1}|Q\rangle = \frac{(\mu r_0)^\Delta k^{(1)}_{\Delta,\ell,q}}{r_0^\Delta}\langle \ell_2 m_2;\ell\, m|\ell_1 m_1\rangle\, \mathcal{A}^{\Delta_{Q+q}+r_0\omega_{\ell_2}}_{\Delta_Q+r_0\omega_{\ell_1}}(\tau_1,\tau_2|\tau_c) + \dots\,. \tag{2.161}$$

where $\langle \ell_2 m_2;\ell\, m|\ell_1 m_1\rangle$ represents the appropriate Clebsch–Gordan coefficient. The above expression gets corrected by quantum contributions, some of which have been computed in [154].
If we limit ourselves to the scalar ground state for $\ell_{1,2} = 0$, by rotational invariance, only scalar small-charge insertions can have non-vanishing matrix elements. In this case the correlator is given by

$$\langle Q+q|\mathcal{O}^{q;\Delta}_{\ell m}(\tau_c,\mathbf{n}_c)\,|Q\rangle \propto \delta_{\ell 0}\,\mu^\Delta \mathcal{A}^{\Delta_{Q+q}}_{\Delta_Q}(\tau_1,\tau_2|\tau_c)\,. \tag{2.162}$$

The result for this particular matrix element with scalar large-charge insertions $\mathcal{O}^Q$ has been presented originally in [18, 188].
The first sub-leading correction to the three-point coefficient $\tilde{C}^\Delta_{Q+q,q,Q}$ includes a contribution from the particular solution $\pi_p$ evaluated at $(\tau_c,\mathbf{n}_c)$ which is formally divergent. Including said first sub-leading corrections the OPE coefficients reads

$$\tilde{C}^\Delta_{Q+q,q,Q} = k^{(1)}_{\Delta,q}(\mu r_0)^\Delta\left[1 - \frac{\frac{q^2}{2}\sum_{\ell,m}\frac{Y^*_{\ell m}(\mathbf{n}_c)Y_{\ell m}(\mathbf{n}_c)}{r_0\omega_\ell}}{c_1 D(D-1)(\mu r_0)^{D-2}} + \dots\right] + \dots\,. \tag{2.163}$$

Its divergent sum can be related to the family of infinite sums in Eq. (2.86),

$$\sum_{\ell,m}\frac{Y^*_{\ell m}(\mathbf{n}_c)Y_{\ell m}(\mathbf{n}_c)}{\omega_\ell} = \frac{r_0}{\Omega_D}\sum_\ell \frac{\mathrm{Deg}_D(\ell)}{r_0\omega_\ell} = \frac{r_0}{\Omega_D}\Sigma(-1)\,. \tag{2.164}$$

[41] Note that the state $|Q+q\rangle$ from the point of view of the EFT lives at the cut-off, see the discussion around and after Eq. (2.47). The insertion of $\mathcal{O}^{q;\Delta}$ connects the two states within the validity of the EFT.

[42] We expand the leading term further in powers of $q \ll Q$.

The sum $\Sigma(-1)$ can be regularized and computed, *e.g* via dimensional regularization or zeta-function regularization.[43]
In $D=3$ the first correction presented in Eq. (2.163) is suppressed by a simple factor of $(\mu r_0) \sim \sqrt{Q}$ and is therefore dominant compared to the sub-leading correction in Eq. (2.152), which is suppressed by $(\mu r_0)^2$. Further corrections to the OPE coefficient arise from the sub-leading terms in the exponential $e^{iq\chi}$ as well as from the term $\left(\partial\chi\right)^{\Delta}$. A careful power counting reveals that the expansion in Eq. (2.163) is valid in the regime $\Delta^2 \ll Q^{3/2}$ [46]. Specializing to the case $D=3$ to compute $\Sigma(-1)$, the final result reads [188]

$$\tilde{C}^{\Delta}_{Q+q,q,Q} \propto (Q)^{\frac{\Delta}{2}} \left[1 + 0.0164523 \times \frac{q^2\sqrt{12\pi}}{\sqrt{c_1 Q}} + \dots\right] + \dots . \tag{2.165}$$

In $D=4$ the leading correction presented in Eq. (2.163) competes with the first correction to Eq. (2.152). Consequentially, the correction in Eq. (2.163) gets renormalized by a Wilsonian coefficent and there is a universal logarithmic contribution appearing [46].

### $\langle \mathcal{O}^{-Q}_{\ell_2 m_2} \mathcal{O}^{-q;\Delta} \mathcal{O}^{q;\Delta} \mathcal{O}^{Q}_{\ell_1 m_1} \rangle$ correlators

As a minor adaptation and generalization of Eq. (2.154) we are computing the four-point function

$$\langle \mathcal{O}^{-Q-q_d-q_c}_{\ell_2 m_2}(\tau_2)\, \mathcal{O}^{q_d;\Delta_d}(\tau_d, \mathbf{n}_d)\, \mathcal{O}^{q_c;\Delta_c}(\tau_c, \mathbf{n}_c)\, \mathcal{O}^{Q}_{\ell_1 m_1}(\tau_1) \rangle\,, \tag{2.166}$$

where $q_d \sim q_c \ll Q$. Computations of related four-point functions have first appeared in [154]. Similarly to Eq. (2.157), the EoM to Eq. (2.166) are given by

$$\nabla_\mu j^\mu = \frac{i q_d \delta(\tau-\tau_d)\delta(\mathbf{n}-\mathbf{n}_d)}{\sqrt{g}} + \frac{i q_c \delta(\tau-\tau_c)\delta(\mathbf{n}-\mathbf{n}_c)}{\sqrt{g}}\,, \qquad \begin{cases} j^\mu(\tau_1 = -\infty, \mathbf{n}_1) = \dfrac{\delta^\mu_0 (Q+q_d+q_c)}{r_0^{D-1}\Omega_D}\,, \\[2ex] j^\mu(\tau_2 = +\infty, \mathbf{n}_2) = \dfrac{\delta^\mu_0 Q}{r_0^{D-1}\Omega_D}\,, \end{cases} \tag{2.167}$$

The corresponding matrix element, computed to leading order reads

$$\begin{aligned} \langle \mathcal{O}^{-Q-q_d-q_c}_{\ell_2 m_2} \mathcal{O}^{q_d;\Delta_d} \mathcal{O}^{q_c;\Delta_c} \mathcal{O}^{Q}_{\ell_1 m_1} \rangle = (\mu r_0)^{\Delta_d+\Delta_c} &\frac{k^{(1)}_{\Delta_d,q_d} k^{(1)}_{\Delta_c,q_c}}{r_0^{\Delta_d+\Delta_c}} \delta_{\ell_2\ell_1}\delta_{m_2 m_1} e^{-(\tau_2-\tau_1)\omega_{\ell_2}} \\ &\times e^{-\Delta_{Q+q_d+q_c}\frac{(\tau_2-\tau_c)}{r_0} - \Delta_{Q+q_c}\frac{(\tau_d-\tau_c)}{r_0} - \Delta_Q \frac{(\tau_c-\tau_1)}{r_0}} + \dots . \end{aligned} \tag{2.168}$$

Restricting ourselves to the special case $q = q_c = -q_d$ we find

$$\langle \mathcal{O}^{-Q}_{\ell_2 m_2} \mathcal{O}^{-q;\Delta} \mathcal{O}^{q;\Delta} \mathcal{O}^{Q}_{\ell_1 m_1} \rangle = (\mu r_0)^{2\Delta} \frac{|k^{(1)}_{\Delta,q}|^2}{r_0^{2\Delta}} e^{-\Delta_Q \frac{(\tau_2-\tau_1)}{r_0} - q\frac{\partial\Delta_Q}{\partial Q}\frac{(\tau_d-\tau_c)}{r_0} - (\tau_2-\tau_1)\omega_{\ell_2}} \delta_{\ell_2\ell_1}\delta_{m_2 m_1} + \dots, \tag{2.169}$$

where the correction in $q$ is given by Eq. (2.38)

$$\frac{\partial\Delta_Q}{\partial Q} \sim \frac{\partial\Delta_0}{\partial Q} = \mu r_0\,. \tag{2.170}$$

The first sub-leading correction to the above result has appeared in [154].

[43] In zeta-function regularization the sum becomes $\Sigma(-1) = \sqrt{D-1}\,\zeta(1/2|S^{D-1})$, see Appendix B.1.

## 2.3 Conclusions and final remarks

As the state–operator correspondence maps the insertion of an operator $\mathscr{O}^Q(0)$ of charge $Q$ to a finite density state $|Q\rangle$, it connects vacuum correlators in flat space to finite density correlators on the cylinder, and for a large class of theories even to finite-density correlators in flat space again, in the macroscopic limit. Generically, we can expect operators with large quantum numbers to be composite operators that naturally produce states of many quanta, which are far removed from the vacuum. In that sense, it is not surprising that within the LCE there are emergent condensed matter phases which encode information about charged correlators in high-energy physics.

The simplest case is where the LCE maps charged operators to a conformal superfluid phase. The conformal superfluid EFT can then be used to study correlators in a $1/Q$ expansion which leads to universal predictions for the structure of conformal data in large-charge sectors of CFTs that realize the superfluid phase. As a consequence of SSB, the EFT spectrum consists of non-relativistic hydrodynamic NG modes. It becomes particularly simple in the case of a global $O(2)$ symmetry under which the operators are charged. Importantly, in this case there is a single superfluid phonon in the spectrum.
In the second part of this chapter we have computed an array of correlators and CFT data for the $O(2)$ model in $D$ dimensions. We have systematically collected three- and four-point functions using the LCE. Compared to earlier works, we have extended the analysis to correlators of spinning primaries, which are described by matrix elements with single- and multi-phonon states. For the sake of convenience, we have mostly computed matrix elements of one-phonon states but the analysis easily generalizes to multi-phonon states.
We have collected results for correlators of the form H–L–…–H with insertions of light operators or currents with parametrically smaller charges in-between two heavy large-charge operators. The insertions of such operators can be thought of as small perturbations around the leading trajectory coming from the heavy in- and out-states. In principle, the EFT might break down very close to the insertions of the light operators, but it accurately captures the overall behaviour of such correlators. While it can be easily argued that such correlators lie within the validity of the EFT, this argument breaks down once there are more than two insertions of heavy operators. The inclusion of another heavy operator between the in-and-out states can no longer be considered as a small perturbation and, in the cylinder frame, the EFT description seems no longer justified. Each heavy operator of charge $Q \gg 1$ creates a superfluid state on a sphere around its insertion point and a correlation function involving three large-charge operators therefore should describe the transition between three different superfluids. Hence, said correlator has to be associated to a new classical profile in the path integral. The key observation is, however, that also here in this case the radial field is locally gapped around all superfluid states and its corresponding modes are therefore still not excited by the new classical trajectory [189]. As a consequence, it is possible to compute these correlation functions using superfluid EFT methods. The result for the H–H–H three-point function was obtained via a numerical solution in [189]. It might be interesting to study such correlators involving additional phonon excitations.

These results generically extend to the homogeneous $O(2)$ sector of CFTs with a larger symmetry group. Unfortunately, capturing the full non-Abelian symmetry is more complicated. The first observation we make is the fact that for non-Abelian symmetry groups only representations are fixable and not charges themselves [47]. The homogeneous $O(2)$ sector in this language corresponds to operators in completely symmetric representations. More precisely, the emergent description for correlators with insertions of operators in large completely symmetric representations is that of a non-Abelian superfluid which — compared to the Abelian superfluid — includes new massless NG modes with quadratic dispersion relations and their gapped NG counterparts. Excitations of these new type-II modes on top of the superfluid ground state correspond to operators in slightly asymmetric representations [46, 47]. The Abelian sector remains present and suffices to accurately describe matrix elements without excitations of the new modes (at least at tree level, *i.e.* to leading order in $1/Q$).
In the case of a non-Abelian symmetry the landscape of EFT structures is richer as the theory includes more DoF and allows for more diverse symmetry-breaking patterns and thus more potentially allowed EFT descriptions. Thus far, the superfluid hypothesis is only capable of capturing large-charge operators in or around the completely symmetric representation — *i.e.* the homogeneous $O(2)$ sector — and a new emergent description is needed to describe other operators [46, 47]. Far away from the completely symmetric representation the corresponding ground state will be inhomogeneous. Heuristically, this can be seen from the observation that (large) asymmetric composite operators can only be constructed out of scalar building blocks by using derivatives. To date, there has no satisfactory description been proposed to describe correlators of operators in asymmetric representations. A first step towards the goal of developing an understanding for the full the non-Abelian structure of the problem is certainly to perform a similar analysis to the one for the Abelian superfluid EFT presented in Section 2.2 instead for the EFT predictions within the non-Abelian superfluid including the additional type-II NG modes and their fluctuations. In doing so, we can use an EFT construction of the form presented *e.g.* in [46, 47].

However convincing the arguments are that the superfluid hypothesis is applicable to a large class of CFTs, given that the superfluid is such a natural option, there is always the possibility that there are no theories that actually realize it.[44] Fortunately, there is strong evidence from lattice computations showing agreement with results from the LCE [143–146, 160]. In this context, since the universal scaling behaviour of $Q^{3/2}$ is only indicative of a condensed-matter phase with a non-trivial macroscopic limit and not specifically a conformal superfluid phase, it is in particular important to explicitly verify the universal predictions from the superfluid EFT hypothesis. Consequently, important universal EFT predictions, like the Casimir energy in Eq. (2.85) appearing at order $(Q)^0$ in $\Delta(Q)$ and certain specific three-point coefficients such as the one for the H–L–H correlator in Eq. (2.163) and the one for the H–H–H correlator computed in [189], have been independently verified in lattice computations to high precision [144–146].
Strikingly, in lattice computations the large-charge predictions seem to match even for operators with small charge [143, 144]. One possible explanation for this comes from Resurgence [199], as it has been shown in the $O(N)$ model at large $N$ that the asymptotic expansion at large charge exhibits a double factorial growth [191], which is stronger than the standard factorial growth observed in regular QFT. Essentially, if this is a universal feature of the LCE, results can be extrapolated to smaller values for LCEs than for other asymptotic series in QFT.

For interacting CFTs that are weakly coupled and amenable to perturbative treatments it is sometimes possible to explicitly identify the effective condensed-matter description at large charge in a double-scaling limit. Examples of theories in this category can be found in the $\epsilon$-expansion [186, 200, 201]. Other interesting cases, where a condensed matter description can be directly identified, are given by theories at large number of flavours $N$ [1, 3, 124, 128, 191]. For both the bosonic $O(N)$ vector model [124] and the fermionic NJL model [3] at large $N$ it is possible to match the predictions from the superfluid EFT and compute the Wilsonian coefficients $c_i$ in the double-scaling limit. We will discuss the LCE in the large-$N$ limit extensively in Chapter 3. Interestingly, in the case of the GN model at large $N$ a Fermi sphere instead of a superfluid ground state can be identified (at least in the strict large-$N$ limit), just like in the free-fermion CFT [3, 168].
Finally, we want to mention that the LCE has been successfully applied in the context of non-relativistic CFTs [202–206]. In this context, the prediction coming from the LCE and the description in terms of condensed matter phases might even be experimentally testable, in particular for the unitary Fermi gas in a harmonic trap [207].

[44] The question as to what possibilities there are and how prevalent the superfluid hypothesis is in the description of large-charge operators within CFTs has also been framed in the context of the conformal bootstrap [154].

# 3 The large charge expansion in the large $N$ limit of QFT

Many non-perturbative features in realistic QFTs are barely accessible and beyond perturbation theory there is a distinct lack of calculable dynamics. Moreover, a lot of important physical questions are intrinsically non-perturbative in nature. It is therefore commonplace to look for simpler models or simplifying frameworks where certain questions can be answered and some intuition for the inner workings of more complex models can be gained. One popular tactic is to introduce an additional expansion parameter, like *e.g.* a small non-integer correction to the spacetime dimension in the $\epsilon$-expansion [133], and compute observables by expanding in said parameter. In some cases, these expansions rearrange the terms in the asymptotic loop expansion from standard perturbation theory and capture certain non-perturbative features.
Large-$N$ techniques and large-$N$ field theories present one such simplifying framework allowing for the calculation of certain observables and the exploration of theoretical ideas. Historically invented to study QCD and asymptotically free Yang–Mills (YM) gauge theories by t'Hooft [208], the large-$N$ expansion nowadays has wide-ranging applications and is used to investigate many problems in physics. Since the large-$N$ limit of QFT can be interpreted as a thermodynamic limit [209], it is particularly natural in the field of condensed-matter physics. But it is also still widely used in high-energy physics. For our purposes, it is a useful framework to test the predictions of the LCE using a more direct approach than the EFT construction.

We are mainly interested in theories with quartic interactions — like $\varphi^4$-theories or GN-type models — and their interactive fixed points at large $N$. The corresponding interacting CFTs often lend themselves to large-charge analysis and it is possible to explicitly recover and identify the condensed-matter EFT descriptions. Before we introduce large-charge analysis, first, it is worth understanding why quartic interactions become simpler to deal with at large $N$. Let us consider explicitly the case of a scalar model with a quartic interaction $(\varphi_i^* \varphi_i)^2$. If this interaction could somehow be replaced by $\langle(\varphi_i^* \varphi_i)\rangle\,(\varphi_i^* \varphi_i)$, for example to leading order in an expansion, the theory would be reduced to a free one and hence would be solvable. There is a variety of schemes like mean-field theory or variational methods that exist and try to achieve this simplification. However, since these methods often do not possess or introduce a small expansion parameter, it is usually unclear how to improve upon results, and the domain of validity of the underlying assumption is generally unknown.

The approximation $(\varphi_i^*\varphi_i)^2 \approx \langle(\varphi_i^*\varphi_i)\rangle\,(\varphi_i^*\varphi_i)$ would be justified, if the fluctuations of the composite field $(\varphi_i^*\varphi_i)$ are much smaller that the fluctuations of the fields $\varphi_i$ themselves. Large-$N$ techniques solve this problem via the central limit theorem. For a large number of fields $\varphi_i$, $i = 1,\ldots,N$ the composite field $(\varphi_i^*\varphi_i)$ is the sum of many terms and, if said terms are sufficiently uncorrelated, it has small fluctuations. To put it in other terms, collective excitations generically tend to be classical. Additionally, the large-$N$ expansion comes with a very natural small expansion parameter given by $1/N$, which organizes well-defined perturbative expansion [134].
The goal of the large-$N$ expansion is to construct an EFT for the identified collective DoF by integrating out the original fields in the action, a set of fields whose cardinality scales like $N$, and hence not only solve the QFT in the (strict) large-$N$ limit but also have access to a systematic $1/N$ expansion. Large-$N$ results tend to complement results from the conventional perturbative RG [134, 198]. The pertubative RG allows to study the theory around the Gaussian — or free — fixed point, whereas large $N$ can sometimes capture properties along the full RG flow up to the interacting fixed point. As with most techniques used to access QFT, the $1/N$ expansion has its limitations which are also often unpredictable. Hence, we should to be careful to discuss it in isolation.[1]

The plan of this chapter is as follows. In Section 3.1 we investigate the large-$N$ $\varphi^4$-theory and the $O(2N)$ WF fixed point. We derive the leading-order asymptotics of the critical theory at large charge and large $N$ in $D = 3$ using thermodynamical reasoning, explain how we recover the superfluid EFT predictions — including a computation of the leading-$N$ Wilsonian parameters — and discuss the fluctuation at sub-leading order to extract the universal contribution from the superfluid phonon discussed in Eq. (2.87). We also quickly comment on the resurgent analysis of the leading-order asymptotics.
In Section 3.2 we explain how the LCE at large $N$ allows us to study the $\varphi^4$-theory away from the conformal fixed point. We deduce and discuss the leading-$N$ effective potential along the full RG flow in $D = 3$, generalizing a result first derived along the critical massless trajectory by re-summing an infinite number of Feynman diagrams in [19]. We extend the analysis to $2 < D < 6$ and in particular to $D = 5$. In $D = 5$ we find tentative results indicating that the $\varphi^4$-model at large $N$ appears to not be UV complete and we uncover a non-unitary — or complex — CFT at the interacting fixed point.
In Section 3.3 we discuss three-dimensional fermionic theories at large $N$ and their large-charge sectors. We start by discussing the free fermion which already exhibits some interesting behaviour. In particular, the ground state at large charge exhibits a non-trivial macroscopic limit on the cylinder in contrast to the free boson. However, there is no SSB and said ground state no longer describes a superfluid but a Fermi sphere instead. We then move on to analysing the interacting GN model and NJL-type models at large charge and large $N$. While the GN model in the strict large-$N$ limit only exhibits a Fermi-sphere ground state, the NJL-type models allow for a Bose–Einstein condensate and admit a superfluid ground state in the description of certain large-charge operators.

All of the analysis and the computations in this chapter are done in Euclidean signature, mainly either in $D = 3$ or $D = 5$. While Section 3.1 represents mostly a review of the recent and established literature on the subject, Sections 3.2 and 3.3 present the original work and the contributions of the author, with the small exception of the free-fermion CFT at large charge. Section 3.2 is based on [1] (plus its companion paper [128]) and Section 3.3 is based on [3].

[1] For example, it is absolutely necessary to make sure that the $1/N$ expansion is both renormalizable and IR finite. If renormalizability and IR finiteness are not met, large-$N$ results are potentially unstable and the $1/N$ expansion might not actually exist.

## 3.1 The Wilson-Fisher fixed point at large charge and large $N$

The $O(N')$ WF fixed point is the interacting fixed point for the $\varphi^4$-theory in $D = 3$ accessible at large $N'$. In our notation the action reads

$$S[\phi_i] = \int \mathrm{d}^D x \left[\frac{1}{2}\partial_\mu\phi_i\partial^\mu\phi_i + \frac{r}{2}\phi_i\phi_i + \frac{g}{2N'}(\phi_i\phi_i)^2\right]. \tag{3.1}$$

We divide the interaction term by a factor of $N'$ so that all terms in the action are of order $N'$. In flat space, the IR interacting fixed point is reached in the massless limit $r = 0$. If $N' = 2N$ is even, the action is more conveniently written in terms of complex fields (the index $i$ runs from 1 to $N$),[2]

$$S[\phi_i] = \int \mathrm{d}^D x \left[\partial_\mu\phi_i^*\partial^\mu\phi_i + r\phi_i^*\phi_i + \frac{g}{N}(\phi_i^*\phi_i)^2\right]. \tag{3.2}$$

We will generally use this form of the action.[3]

### 3.1.1 The Wilson-Fisher fixed point and large $N$

In $2 < D < 4$ the coupling $g$ of the $\varphi^4$-interaction is relevant and the theory described by the action in Eq. (3.2) is free at short distances in the UV and strongly coupled at long distances in the IR where it can be assumed to flow to an interacting and generically strongly-coupled fixed point called the WF fixed point. The WF CFT living there is of particular importance in condensed-matter physics and the description of second-order phase transitions in statistical physics. It describes the critical points of systems such as superfluid Helium, vapour liquid, ferromagnetic transitions and binary mixtures [134].

#### Weakly-coupled WF fixed point in small $\epsilon$

There is strong evidence for the existence of the WF fixed point between $2 < D < 4$ — in particular also for the physical dimension $D = 3$ — and for arbitrary $N$ [11, 135, 221]. Its existence is generally agreed upon and has first been proven perturbatively in the small-$\epsilon$ expansion around four dimensions [8, 133], where significant simplifications occur. In fact, in dimensions $D = 4 - \epsilon$, for $\epsilon \ll 1$ the IR WF fixed point of the $\varphi^4$-theory Eq. (3.2) becomes weakly-coupled and accessible via perturbation theory. The one-loop beta function for the quartic coupling $g$ around the IR fixed point in $D = 4 - \epsilon$ for the $O(2N)$

[2] Clearly, although not manifestly, this action remains $O(2N)$ invariant instead of just $U(N) \subset O(2N)$.

[3] Additionally, we can include a $|\phi_i|^6$-term in the action Eq. (3.2), which by dimensional analysis is marginal in $D = 3$. However, this term is marginally irrelevant in the presence of the $|\phi_i|^4$-term [19, 135] and hence we omit it. The theory without a $|\phi_i|^4$-term and only including $|\phi_i|^6$-term is itself interesting and worth studying [210–220]. We leave the large-charge analysis of the $|\phi_i|^6$-model for future research.

invariant theory then reads [222]

$$\beta_g = -\frac{g}{N}\epsilon + (N+4)\frac{g^2}{4\pi^2 N^2} + \dots, \tag{3.3}$$

and the WF fixed point lies at the critical value of the coupling[4]

$$\frac{g_*}{N} = \frac{4\pi^2}{(N+4)}\epsilon. \tag{3.4}$$

The anomalous dimension as well as the scaling dimension of $\phi_i$ and the composite field $\phi^2 = \phi_i\phi_i$ are

$$\Delta_\phi = 1 - \frac{\epsilon}{2} + \dots, \qquad \Delta_{\phi^2} = 2 - \frac{3}{(N+4)}\epsilon + \dots. \tag{3.5}$$

The fact that the IR stable fixed point occurs for $g$ of order $\epsilon$ allows for the development of a formal expansion in $\epsilon$. Higher corrections in $\epsilon$ to these results come from higher-loop calculations and are known up to order $\epsilon^5$ [223, 224]. The coefficients of the first few terms in $\epsilon$ fall off rather rapidly and setting $\epsilon = 1$ provides an approximation for the critical theory in $D = 3$ that is in surprisingly good agreement with experimental and numerical results [133]. In $D = 4$ the $\varphi^4$-theory does not have a continuum limit and — as the only critical point reachable is the Gaussian free massless theory at large distances — it is thus known as a trivial theory there.

#### Weakly-coupled WF fixed point from large $N$

An approach complementary and similar in spirit to the small-$\epsilon$ expansion is the large-$N$ expansion. To set up the large-$N$ expansion for the $O(2N)$ vector model we consider a theory of $2N$ real scalar fields packaged into $N$ complex fields and a $O(2N)$ symmetric action of the form

$$S[\phi_i] = \int \mathrm{d}^D x \left[\partial_\mu \phi_i^* \partial^\mu \phi_i + N\Upsilon\left(\frac{\phi_i^*\phi_i}{N}\right)\right], \tag{3.6}$$

where the potential $\Upsilon(x)$ is a generic polynomial.[5] For large $N$ it can be expected that $O(2N)$ invariant quantities self-average and hence have small fluctuations.[6] This observation suggests that it is a good idea to consider $\phi_i^*\phi_i$ as a dynamical variable instead of $\phi_i$. This can be achieved via a so-called Hubbard–Stratonovich (HS) transformation [134, 225, 226]. The HS transformation introduces two additional fields $\lambda, \sigma$ while imposing the constraint $\sigma(x) = \phi_i^*(x)\phi_i(x)/N$ for every point $x$ in space by a functional integral over $\lambda(x)$,

$$\mathbb{1} = N\int \mathscr{D}\lambda\, \delta\left(\phi_i^*\phi_i - N\lambda\right) = \frac{N}{2\pi i}\int \mathscr{D}\sigma \mathscr{D}\lambda\, e^{\sigma\left(\phi_i^*\phi_i - N\lambda\right)}. \tag{3.7}$$

[4] Higher-order corrections in $\epsilon$ change the critical value $g_*$ but not its existence.

[5] We choose the notation $\Upsilon$ instead of $V$ as we write the potential as a function of $\phi^2$ instead of $\phi$.

[6] By the central limit theorem this assumption relies on the components $\phi_i$ to be weakly correlated.

Inserting the representation of the identity in Eq. (3.7) into the path integral for the action Eq. (3.6) results in the following partition function:

$$Z = \int \mathcal{D}\phi_i \mathcal{D}\phi_i^* \mathcal{D}\sigma \mathcal{D}\lambda\, e^{-S[\phi_i,\sigma,\lambda]}, \qquad S[\phi_i,\sigma,\lambda] = \int \mathrm{d}^D x \left[\partial_\mu \phi_i^* \partial^\mu \phi_i + N\Upsilon(\lambda) + \sigma\big(\phi_i^*\phi_i - N\lambda\big)\right]. \tag{3.8}$$

The functional integral over the fields $\phi_i$ in the above representation of the partition function is now Gaussian and can be performed, effectively removing the dependence in the original fields and making the dependence on $N$ of the partition function explicit.[7] The path integral now reads[8]

$$Z = \int \mathcal{D}\sigma \mathcal{D}\lambda\, e^{-S[\sigma,\lambda]}, \qquad S[\sigma,\lambda] = N\int \mathrm{d}^D x \left[\Upsilon(\lambda) - \sigma\lambda\right] + N\mathrm{Tr}\log\big(-\nabla^2 + \sigma\big), \tag{3.9}$$

where we have used the identity $\log\det A = \mathrm{Tr}\log A$. The action Eq. (3.2) corresponds to the potential

$$\Upsilon(x) = rx + gx^2. \tag{3.10}$$

In that case the modified action is quadratic in $\lambda$,

$$S[\sigma,\lambda] = N\int \mathrm{d}^D x \left[r\lambda + g\lambda^2 - \sigma\lambda\right] + N\mathrm{Tr}\log\big(-\nabla^2 + \sigma\big), \tag{3.11}$$

and the integral over $\lambda$ can be performed, setting

$$4g\lambda + r = \sigma. \tag{3.12}$$

This results in a term of the form $r^2$ appearing in the action, which is a constant term and therefore part of the normalization. The new representation of the partition function for the $\varphi^4$-theory in Eq. (3.2) then reads

$$Z = \int \mathcal{D}\sigma\, e^{-S[\sigma]}, \qquad S[\sigma] = N\int \mathrm{d}^D x \frac{1}{4g}\left[2\sigma r - \sigma^2\right] + N\mathrm{Tr}\log\big(-\nabla^2 + \sigma\big), \tag{3.13}$$

Integrating out the $\sigma$ field and rewriting the trace-log term as a path integral gets us back to the original theory in Eq. (3.2), hence the two representations are clearly equivalent. To analyse the WF fixed point we consider the critical trajectory at $r = 0$ (in flat space).

Tuning the massless theory $r = 0$ to the conformal point at the level of the action in Eq. (3.13) is equivalent to dropping the $\sigma^2$-term.[9] In order to properly organize and perform the $1/N$ expansion we rescale the field $\sigma$ by a factor of $1/\sqrt{N}$ [227, 228],

$$\sigma = \frac{\hat{\sigma}}{\sqrt{N}}. \tag{3.14}$$

[7] Often it is more convenient to integrate only over $N-1$ components $\phi_i$ and leaving $\phi_1$ in the action. In some cases, it is even worth to leave two original fields in the action [134]. This does not affect the leading-$N$ result but only the $1/N$ corrections to the leading result.

[8] The HS formalism can be generalized slightly in order to be applicable to the most general $O(N)$-invariant action, which is found by adding the two terms $A(|\phi|^2/N)|\partial_\mu\phi|^2$ and $B(|\phi|^2/N)|\partial_\mu\phi^*\cdot\phi|^2/N$ to the action in Eq. (3.6), where $|\phi|^2 = \phi_i^*\phi_i$ and $A, B$ are two arbitrary functions. After performing the HS transform in Eq. (3.7) these terms become $A(\sigma)|\partial_\mu\phi|^2$ and $NB(\sigma)(\partial_\mu\sigma)^2$. The $\phi_i$-integrals remain Gaussian and can again be performed.

[9] This applies to both the UV free and the IR interacting fixed point [222].

The trace-log term generates a non-local effective kinetic term for the composite field $\hat{\sigma}$.

$$N\mathrm{Tr}\log\left(-\nabla^2+\hat{\sigma}/\sqrt{N}\right)=N\mathrm{Tr}\log\left(-\nabla^2\right)+\sum_{k=2}^{\infty}\frac{(-1)^{k+1}}{kN^{k/2-1}}\mathrm{Tr}\left(G\hat{\sigma}\right)^k, \tag{3.15}$$

so that to quadratic order in the field $\sigma$ the partition function is given by

$$Z=\int\mathscr{D}\hat{\sigma}\,\exp\left[\frac{1}{2}\mathrm{Tr}\left(G\hat{\sigma}\right)^2+\mathscr{O}(\hat{\sigma}^3)\right]. \tag{3.16}$$

where $G=G(x-y)$ is the propagator of the free theory, $-\nabla^2G(x-y)=\delta(x-y)$,

$$G(x-y)=\int\frac{\mathrm{d}^Dp}{(2\pi)^D}\frac{e^{ip\cdot(x-y)}}{p^2}. \tag{3.17}$$

The quadratic action in $\sigma$ is best expressed in terms of a double integral

$$S[\hat{\sigma}]=-\frac{1}{2}\int\mathrm{d}^Dx\,\mathrm{d}^Dy\,\hat{\sigma}(x)\hat{\sigma}(y)\,G(x-y)^2+\mathscr{O}(\hat{\sigma}^3). \tag{3.18}$$

The square of the $\phi$ propagator can be evaluated in momentum space,

$$G(x-y)^2=\int\frac{\mathrm{d}^Dp}{(2\pi)^D}e^{ip\cdot(x-y)}\int\frac{\mathrm{d}^Dq}{(2\pi)^D}\frac{1}{q^2(p-q)^2}=-\int\frac{\mathrm{d}^Dp}{(2\pi)^D}\frac{e^{ip\cdot(x-y)}\left(p^2\right)^{\frac{D}{2}-2}}{2^D(4\pi)^{\frac{D-3}{2}}\Gamma\left(\frac{D-1}{2}\right)\sin\left(\frac{\pi D}{2}\right)}, \tag{3.19}$$

and can be used to derive the propagator for $\sigma$,

$$\langle\tilde{\hat{\sigma}}(-p)\tilde{\hat{\sigma}}(p)\rangle=\frac{(4\pi)^{\frac{D-3}{2}}2^{D+1}\Gamma\left(\frac{D-1}{2}\right)\sin\left(\frac{\pi D}{2}\right)}{\left(p^2\right)^{\frac{D}{2}-2}},\qquad\langle\hat{\sigma}(x)\hat{\sigma}(y)\rangle=\frac{2^{D+2}\Gamma\left(\frac{D-1}{2}\right)\sin\left(\frac{\pi D}{2}\right)}{\pi^{\frac{3}{2}}\Gamma\left(\frac{D}{2}-2\right)|x-y|^4}=:\frac{C_{\hat{\sigma}}}{|x-y|^4}. \tag{3.20}$$

This is precisely the conformally invariant two-point function of a scalar operator with conformal dimension $\Delta_{\hat{\sigma}}=2$, consistent with the small-$\epsilon$ expansion.[10]
The simplest way to develop the large-$N$ perturbation theory from here is via the canonical propagator $G(x-y)$ for $\phi_i$, the propagator for $\hat{\sigma}$ in Eq. (3.20) and the action written in the form[11]

$$S[\phi_i,\hat{\sigma}]=\int\mathrm{d}^Dx\left[\partial_\mu\phi_i^*\partial^\mu\phi_i+\frac{1}{\sqrt{2N}}\hat{\sigma}\phi_i^*\phi_i\right], \tag{3.21}$$

which is recovered by rewriting the trace-log term in Eq. (3.13) as a path integral (and rescaling the field $\sigma$ accordingly). For example, the first correction in $1/N$ to the scaling dimension of $\sigma$ takes the form [222, 229–231]

$$\Delta_{\hat{\sigma}}=2+2^{D-1}(D-1)(D-2)\frac{\Gamma\left(\frac{D-1}{2}\right)\sin\left(\frac{\pi D}{2}\right)}{\pi^{\frac{3}{2}}\Gamma\left(\frac{D}{2}+1\right)}\frac{1}{2N}+\mathscr{O}(N^{-2}). \tag{3.22}$$

[10] The coefficient $C_{\hat{\sigma}}$ is positive in the range $2<D<6$.
[11] We choose to rescale by $1/\sqrt{2N}$ instead of $1/\sqrt{N}$ purely out of convenience so that our results are consistent with [222]. There is no physical difference in choosing one over the other.

After setting $D = 4 - \epsilon$ and expanding to first order in $\epsilon$, this result agrees perfectly with Eq. (3.5). While the existence of the $O(2N)$ WF fixed point in $D = 3$ from the $\epsilon$-expansion is not clear, from the large-$N$ perspective it exists everywhere for $2 < D < 4$ and survives in the limit $N \to \infty$.[12]

### 3.1.2 The large-$N$ $\varphi^4$-model at fixed charge

Large-$N$ analysis provides strong evidence that the $O(2N)$ vector model flows to the interacting WF fixed point in the IR for $2 < D < 4$. The RG flow including the interacting WF CFT survive the large-$N$ limit and the fixed point becomes a weakly-coupled theory around $N \to \infty$. The CFT living at this fixed point is an ideal playground to test the LCE in a setting that is under perturbative control.

We leave the dimension $D$ arbitrary for now. While fixing the charge in an Abelian model like the $O(2)$ WF CFT is straightforward, in non-Abelian theories the global internal symmetry still acts on the charge matrix $Q$, so that only the spectrum of $Q$ and hence its representation remain invariant. Importantly, it has been noticed that the homogeneous finite-density superfluid configuration and the corresponding symmetry-breaking pattern discussed in Chapter 2 only capture correlators of charged operators in completely symmetric representations [46, 47, 122].

In the critical $O(2N)$ vector model at large $N$ we choose to fix the value $Q_i$ of the $N$ Cartan charges/generators $Q_i^{(O(2N))}$, $i = 1, \ldots, N$ rotating the field $\phi_i$ individually, which equals the number of eigenvalues of the charge matrix $Q$ and hence represents the maximum amount of entries of $Q$ that we are allowed to fix under the group action. Without using an EFT description this is best achieved by restricting the path integral to the sector of operators $\mathcal{O}^Q$ with fixed eigenvalues under the Cartan charges, $Q_i^{(O(2N))}\mathcal{O}^Q = Q_i \mathcal{O}^Q$. This procedure is best discussed in finite-temperature field theory as fixing the value of the Cartan charges has a very natural interpretation at finite temperature where it corresponds to fixing the number of particles, *i.e.* considering the canonical ensemble.[13]

We start with the $O(2N)$-invariant $\varphi^4$ vector model with action Eq. (3.2) at the WF fixed point on the cylinder $S^1_\beta \times S^{D-1}_{r_0}$, where we have compactified the cylinder-time coordinate.[14] The length of the circle $\beta = 1/T$ is the inverse temperature and we recover the theory on the regular cylinder $\mathbb{R} \times S^{D-1}_{r_0}$ in the limit $\beta \to \infty$. In the following we always implicitly assume the zero-temperature limit to recover the flat space WF CFT at the fixed point of the RG flow,

$$S^1_\beta \times S^{D-1}_{r_0} \sim \lim_{\beta \to \infty} S^1_\beta \times S^{D-1}_{r_0} = \mathbb{R} \times S^{D-1}_{r_0} . \tag{3.23}$$

The diagonal Cartan charges of the Lagrangian in Eq. (3.2) rotating the fields $\phi_i$ individually are given

[12] Besides using small-$\epsilon$ and large-$N$ expansions the WF CFTs of three-dimensional $O(N)$ models have also been studied using Monte-Carlo simulations [14, 16, 143, 144, 160] and the conformal bootstrap [232, 233]. Detailed information on the spectrum of light operators is known and there is strong agreement between all of the different methods. As a consequence, the existence of these conformal theories is widely accepted.

[13] Naturally, it is also possible to perform all computations in flat space and at zero temperature, see [20].

[14] In curved space the mass $r$ now also includes a coupling to the Ricci curvature $\mathcal{R}$. Since we consider the theory on the cylinder $S^1_\beta \times S^{D-1}_{r_0}$ the curvature is constant, $\mathcal{R} \propto 1/r_0^2$.

by ($\tau = x_0$ in Euclidean signature)

$$Q_i^{(O(2N))} = -\int_{S_{r_0}^{D-1}} \mathrm{d}S \left[\partial_\tau \phi_i^* \phi_i - \phi_i^* \partial_\tau \phi_i\right]. \tag{3.24}$$

The partition function on the thermal circle $S^1_\beta$ can be conveniently written as the trace of the exponential of the Hamiltonian. Bosonic DoF obey periodic boundary conditions on the thermal circle (while fermionic DoF obey anti-periodic ones), therefore

$$Z = \mathrm{Tr}_{S^1_\beta \times S_{r_0}^{D-1}} \left[e^{-\beta H^{(\mathrm{cyl})}}\right] = \int_{\text{periodic}} \mathscr{D}\phi_i \mathscr{D}\phi_i^* \; e^{-S[\phi_i]}, \tag{3.25}$$

where $H^{(\mathrm{cyl})}$ at zero temperature $\beta \to \infty$ and at the fixed point corresponds to the dilatation operator $D$. We can restrict the trace over the thermal circle in Eq. (3.25) to operators of fixed diagonal charges $Q_i$ under the $O(2N)$ symmetry. For later convenience we introduce the charge densities $\rho_i = Q_i/(r_0^{D-1}\Omega_D)$,

$$Z_c(\rho_1, \dots, \rho_N) := \mathrm{Tr}_{S^1_\beta \times S_{r_0}^{D-1}} \left[e^{-\beta H^{(\mathrm{cyl})}} \prod_{i=1}^N \delta\big(Q_i^{(O(2N))} - Q_i\big)\right]. \tag{3.26}$$

In finite-temperature field theory language this is the canonical partition function at fixed particle number (as particles carry the charge in this language). Using a Fourier transform we can rewrite the path integral in Eq. (3.26),

$$\begin{aligned} Z_c(\rho_1, \dots, \rho_N) &= \mathrm{Tr}_{S^1_\beta \times S_{r_0}^{D-1}} \left[e^{-\beta H^{(\mathrm{cyl})}} \prod_{i=1}^N \int \frac{\mathrm{d}\theta_i}{2\pi} e^{i\theta_i\left(Q_i^{(O(2N))} - Q_i\right)}\right] \\ &= \left[\prod_{i=1}^N \int \frac{\mathrm{d}\theta_i}{2\pi} e^{-i\theta_i Q_i}\right] \mathrm{Tr}_{S^1_\beta \times S_{r_0}^{D-1}} \left[e^{-\beta\left(H^{(\mathrm{cyl})} - i\beta^{-1}\sum_i \theta_i Q_i^{(O(2N))}\right)}\right], \end{aligned} \tag{3.27}$$

where we can identify the grand-canonical partition function $Z_{gc}$ with imaginary chemical potentials $\mu_i = i\theta_i/\beta$ for the fields $\phi_i$ by

$$Z_{\mathrm{gc}}(\mu_1, \dots, \mu_N) := \mathrm{Tr}_{S^1_\beta \times S_{r_0}^{D-1}} \left[e^{-\beta\left(H^{(\mathrm{cyl})} - \sum_i \mu_i Q_i^{(O(2N))}\right)}\right] \overset{N\to\infty}{=:} e^{-(2N)\beta V \Omega(\mu_1, \dots, \mu_N)}, \quad V_{S_{r_0}^{D-1}} = r_0^{D-1}\Omega_D. \tag{3.28}$$

In thermodynamical language, the quantity $\Omega(\mu_1, \dots, \mu_N)$ — given as the logarithm of $Z_{gc}$ in the thermodynamical limit $N \to \infty$ — is called the grand potential. Note that Eq. (3.28) describes the partition function for a theory at finite density with imaginary chemical potentials $\mu_i = i\theta_i/\beta$. Equivalently to the (conformal) superfluid EFT in Chapter 2, the effective time translation is generated by the modified Hamiltonian $H - \sum_i \mu_i Q_i^{(O(2N))}$.
Analogously, the canonical partition function $Z_c$ defines the free energy density $f_c(\rho)$ per DoF by

$$Z_c(\rho_1, \dots, \rho_N) \overset{N\to\infty}{=:} e^{-(2N)\beta V f_c(\rho_1, \dots, \rho_N)}, \qquad V_{S_{r_0}^{D-1}} = r_0^{D-1}\Omega_D, \tag{3.29}$$

which is the ground-state energy at fixed charge. The free energy at zero temperature and at the conformal fixed point is related to the scaling dimension of the lowest-lying operator of fixed charge,

$$\Delta(Q_1, \dots) = -\lim_{\beta\to\infty} \frac{r_0}{\beta} \log Z_c(\rho_1, \dots)\Big|_{\mathrm{FP}} = (2N) \lim_{\beta\to\infty} V r_0 f_c(\rho_1, \dots)\Big|_{\mathrm{FP}} + \mathscr{O}(N^{-1}), \qquad Q_i = \rho_i/V. \tag{3.30}$$

From here on we will no longer differentiate notation-wise between the charge operator $Q_i^{(O(2N))}$ and the charge $Q_i$,

$$\text{shorthand notation:} \quad Q_i^{(O(2N))} \sim Q_i\,. \tag{3.31}$$

The grand-canonical partition function $Z_{\text{gc}}(\mu)$ has a path-integral representation, which will be the starting point for our analysis. Since the current $j^\mu$ of the $O(2N)$ symmetry depends on the canonically conjugate momenta of the fields $\phi_i$, the sum over momenta in the partition function and the switch from Hamiltonian to Lagrangian formalism are non-trivial. The resulting action $S_\mu$ can be understood as introducing constant background fields for the gauged $U(1)^N$ diagonal symmetry in the original action Eq. (3.2) while keeping periodic boundary conditions within the path integral,

$$Z_{\text{gc}}(\mu) = \int_{\text{periodic}} \mathscr{D}\phi_i \mathscr{D}\phi_i^*\, e^{-S_\mu[\phi_i]}\,, \quad S_\mu[\phi_i] = \int_{S^1_\beta \times S^{D-1}_{r_0}} \mathrm{d}^D x\, \left[(\partial_\tau - \mu_i)\phi_i^*(\partial_\tau + \mu_i)\phi_i + |\nabla\phi_i|^2 + r|\phi_i|^2 + \frac{g}{N}|\phi_i|^4\right]. \tag{3.32}$$

We discuss the detailed derivation of the finite-density action $S_\mu$ for a generic $O(2N)$ vector model in Appendix C.1. Since we are on a compact spatial manifold, we expect $\Omega(\mu)$, and hence $f_{gc}(\rho)$, to be smooth and its derivatives to be well-defined. Additionally, CP-invariance implies that the grand potential $\Omega$ is an even function of all the $\theta_i$-variables.[15]

In the thermodynamic limit $N \to \infty$ the quantities in Eq. (3.28) and Eq. (3.27) are computable in a saddle-proint approximation around the leading semi-classical trajectory.[16] The saddle-point equation for the partition function in Eq. (3.27) reads

$$i\frac{\rho_i}{(2N)} + \beta\frac{\partial}{\partial\theta_i}\Omega(i\theta_1/\beta,\ldots,i\theta_N/\beta) = 0\,. \tag{3.33}$$

Importantly, at the saddle point Eq. (3.33) of the canonical partition function in Eq. (3.27) the chemical potentials $\theta_i$ are imaginary, due to the fact that the derivative of the grand potential $\Omega$ with respect to $\theta_i$ is an odd real function.[17] Equivalently, the chemical potentials $\mu_i$ are real. We will therefore change variables from $\theta_i$ to $\mu_i$ in the canonical partition function Eq. (3.27),

$$Z_c(\rho_1,\ldots,\rho_N) \sim \int [\mathrm{d}\mu_i]\, e^{-\beta\left[\sum_i \mu_i Q_i - \frac{1}{\beta}\log Z_{gc}(\mu_1,\ldots,\mu_n)\right]} \overset{N\to\infty}{\to} \int [\mathrm{d}\mu_i]\, e^{-(2N)\beta\left[\sum_i \mu_i \frac{Q_i}{2N} + V\Omega(\mu_1,\ldots,\mu_N)\right]}\,, \tag{3.34}$$

As $\mu_i \to -\mu_i$ is a symmetry, we can assume without loss of generality that $\mu_i \geq 0$.

[15]Under time reversal T the chemical potential transforms as $\mu_i \to -\mu_i$. Due to CPT symmetry, under the assumption of CP invariance, the transformation $\mu_i \to -\mu_i$ now has to be a symmetry of the theory.

[16]Charge quantization implies that the integrand in Eq. (3.27) is a periodic function in each of the $\theta_i$-variables. Hence, as they cancel each other due to periodicity, we do not have to worry about contributions from the end points of the integrals to the asymptotic expansion around the saddle points.

[17]Taking the complex conjugate of the saddle-point equation Eq. (3.33) we deduce that at the minimum $\theta_a^* = -\theta_a$, as $\partial\Omega/\partial\theta_i$ is an odd function.

At the large-$N$ saddle point,[18]

$$Z_c \overset{N\to\infty}{\simeq} e^{-(2N)\beta V f_c(\rho)}, \qquad Z_{gc} \overset{N\to\infty}{\simeq} e^{-(2N)\beta V \Omega(\mu)}, \qquad \frac{\rho_i}{2N} + \frac{\partial}{\partial \mu_i}\Omega(\mu) = 0. \tag{3.35}$$

the free energy and the grand potential are related by a Legendre transform,

$$f_c(\rho_1, \dots) = \sup_{\mu_i}\Big(\sum_i \mu_i \frac{\rho_i}{2N} + \Omega(\mu_1, \dots)\Big) = \sum_i \mu_i \frac{\rho_i}{2N} + \Omega(\mu_1, \dots)\Big|_{\frac{\rho_i}{2N} = -\frac{\partial}{\partial \mu_i}\Omega}. \tag{3.36}$$

This is a non-trivial consistency check of our construction: at the saddle point we reproduce the usual Legendre transform that relates the two thermodynamic potentials.

The idea is to compute the grand potential $\Omega(\mu)$ for the $O(2N)$ vector model in an expansion in large $N$ and then use the saddle-point condition to relate $\Omega(\mu)$ to the free energy $f_c$, with the end goal being to extract the scaling dimension from $f_c$. We consider the action $S_\mu = S_\mu[\phi_i]$ in Eq. (3.32) and perform a HS transform of the form in Eq. (3.7). The action becomes

$$S_\mu[\phi_i, \sigma, \lambda] = \int\limits_{S^1_\beta \times S^{D-1}_{r_0}} \mathrm{d}^D x \, \left[(\partial_\tau - \mu_i)\phi_i^* (\partial_\tau + \mu_i)\phi_i + |\nabla \phi_i|^2 + r|\phi_i|^2 + \sigma\big(|\phi_i|^2 - N\lambda\big) + Ng\lambda^2\right], \tag{3.37}$$

where we have consciously left the mass and curvature term $r|\phi_i|^2$ untouched by the HS transform. Integrating out the field $\lambda$ — as it was done in Eq. (3.13) — yields the action

$$S_\mu[\phi_i, \sigma] = \int\limits_{S^1_\beta \times S^{D-1}_{r_0}} \mathrm{d}^D x \, \left[(\partial_\tau - \mu_i)\phi_i^* (\partial_\tau + \mu_i)\phi_i + |\nabla \phi_i|^2 + (r + \sigma)|\phi_i|^2) - \frac{N}{4g}\sigma^2\right]. \tag{3.38}$$

The WF conformal point on the cylinder is reached for $|g| \to \infty$ [222] and $r$ of order $\mathcal{O}(1)$ [124], where the action becomes

$$S_\mu[\phi_i, \sigma] = \int\limits_{S^1_\beta \times S^{D-1}_{r_0}} \mathrm{d}^D x \, \left[(\partial_\tau - \mu_i)\phi_i^* (\partial_\tau + \mu_i)\phi_i + |\nabla \phi_i|^2 + (r + \sigma)|\phi_i|^2\right]. \tag{3.39}$$

### 3.1.3 Leading order asymptotics at large charge

We compute the free energy of the $O(2N)$ model restricted to the completely symmetric representation (the Abelian sector) — which corresponds to the energy of the homogeneous ground state in the sector of total fixed charge $Q$ [122, 234] — in the double-scaling limit of large $N$ and large $Q$.
The original fields $\phi_i$ only appear quadratically in the critical action in Eq. (3.39) and hence the

[18]The function $\exp[-(2N)\beta V \Omega(\mu)]$ is the leading-order result for the grand-canonical partition function $Z_{gc}$ in a $1/N$ expansion,

$$Z_{gc}(\mu_1, \dots) e^{\beta V (2N) \Omega(\mu)(\mu_1, \dots) + \mathcal{O}(N^0)}.$$

This expansion is well-behaved for the usual reasons that large-$N$ expansions with a scalar collective field are. A similar statement holds true for $Z_c$.

functional integral can be performed. Afterwards, the grand-canonical partition function in terms of the collective field $\sigma$ only reads

$$Z_{gc} = \int_{\text{periodic}} \mathscr{D}\sigma \; e^{-S_\mu[\sigma]}, \qquad S_\mu[\sigma] = \sum_{i=1}^{N} \operatorname{Tr}\log\left( -(\partial_\tau + \mu_i)^2 - \Delta_{S_{r_0}^{D-1}} + r + \sigma \right). \tag{3.40}$$

In Section 3.1.1, in particular Eq. (3.20), we have discussed and presented evidence that at zero chemical potential and zero temperature this partition function corresponds to the WF CFT at large $N$. At $N \to \infty$ the conformal dimension $\Delta_\sigma$ of the collective field $\sigma$ is equal to 2.

As we are in finite volume, the mode decomposition for the fields $\phi_i$ admits a zero-mode, which can in principle be extracted from the functional determinant [124]. We instead choose to consider said zero-mode while evaluating the functional determinant in Eq. (3.40). We further assume that $\sigma$ admits a constant VEV,[19]

$$\sigma = \langle\sigma\rangle + \hat{\sigma}/\sqrt{N}, \qquad \langle\sigma\rangle = \text{const.}, \tag{3.41}$$

whose value will be determined by the saddle-point equations. The action can then be expanded in powers of the fluctuation $\hat{\sigma}$,

$$\begin{aligned} S_\mu[\sigma] &= \sum_{i=1}^{N} \operatorname{Tr}\log\left( -(\partial_\tau + \mu_i)^2 - \Delta_{S_{r_0}^{D-1}} + r + \langle\sigma\rangle + \frac{\hat{\sigma}}{\sqrt{N}} \right) \\ &= \sum_{i=1}^{N} \operatorname{Tr}\log\left( -(\partial_\tau + \mu_i)^2 - \Delta_{S_{r_0}^{D-1}} + r + \langle\sigma\rangle \right) + \sum_{i=1}^{N} \sum_{k=2}^{\infty} \frac{(-1)^{k+1}}{k N^{k/2}} \operatorname{Tr}\left(G_i \hat{\sigma}\right)^k, \end{aligned} \tag{3.42}$$

where $G_i(x-y)$ is the propagator associated to the field $\phi_i$,

$$\left( -(\partial_\tau + \mu_i)^2 - \Delta_{S_{r_0}^{D-1}} + r + \langle\sigma\rangle \right) G_i(x-y) = \frac{1}{\sqrt{g}} \delta(x-y). \tag{3.43}$$

For completeness, the action for the path-integral representation of the canonical partition function reads

$$S_Q = \beta \sum_i \mu_i Q_i + S_\mu = \beta \sum_i \mu_i Q_i + \sum_i \operatorname{Tr}\log\left( -(\partial_\tau + \mu_i)^2 - \Delta_{S_{r_0}^{D-1}} + r + \langle\sigma\rangle \right) + \sum_i \sum_{k=2}^{\infty} \frac{(-1)^{k+1}}{k N^{k/2}} \operatorname{Tr}\left(G_i \hat{\sigma}\right)^k. \tag{3.44}$$

The large-$N$ saddle point is found by minimizing the ground state action in Eq. (3.44) for $\hat{\sigma} = 0$. The functional determinant in the action can be evaluated, which we do in Appendix C.2. The grand

[19] If the field $\sigma$ admits a homogeneous/constant VEV, then it will correspond to the lowest-energy ground state as any inhomogeneous solution in comparison will correspond to a (ground) state with higher energy.

potential $\Omega(\mu_1,\ldots)$ becomes[20]

$$\begin{aligned}\Omega(\mu_1,\ldots) &= \frac{1}{2N}\sum_i \zeta_i^2\left(r+\langle\sigma\rangle-\mu_i^2\right)+\frac{1}{(2N)V}\sum_i\sum_\ell \mathrm{Deg}_D(\ell)\left[\left[\omega_\ell+\frac{1}{\beta}\log(1-e^{-\beta(\omega_\ell+\mu_i)})(1-e^{-\beta(\omega_\ell-\mu_i)})\right]\right],\\ \omega_\ell^2 &= \frac{\ell(\ell+D-2)}{r_0^2}+r+\langle\sigma\rangle.\end{aligned} \tag{3.45}$$

This result is convergent as long as

$$|\mu_i|\le\sqrt{r+\langle\sigma\rangle}, \qquad \forall i=1,\ldots,N. \tag{3.46}$$

The saddle-point equations are obtained by deriving the ground-state action *w.r.t.* the variables $\langle\sigma\rangle$, $\zeta_i$ and $\mu_i$, where the BEC parameters $\zeta_i$ come from the zero modes of the fields $\phi_i$. Since the parameters $\zeta_i$ are a priori note determined, they should be treated as variational parameters related to the charge carried by the condensed particles, *i.e.* the ground state.

At zero temperature $\beta\to\infty$ we reach the WF fixed point in flat space,

$$\begin{aligned}\Omega(\mu_1,\ldots) &= \frac{1}{(2N)}\sum_i\zeta_i^2\left(r+\langle\sigma\rangle-\mu_i^2\right)+\frac{1}{2V}\sum_\ell \mathrm{Deg}_D(\ell)\omega_\ell,\\ f_c(\rho_1,\ldots) &= \sum_i \mu_i\frac{\rho_i}{(2N)}+\Omega(\mu_1,\ldots), \qquad \omega_\ell^2=\frac{\ell(\ell+D-2)}{r_0^2}+r+\langle\sigma\rangle.\end{aligned} \tag{3.47}$$

The infinite sum in the grand potential is the zeta function $\zeta(-1/2\,|\,S_{r_0}^{D-1}, r+\langle\sigma\rangle)$ on the two-sphere,[21]

$$\zeta(s\,|\,S_{r_0}^{D-1},r) := \sum_\ell \mathrm{Deg}_D(\ell)\left(\frac{\ell(\ell+D-2)}{r_0^2}+r\right)^{-s}. \tag{3.48}$$

The saddle-point equations in $\langle\sigma\rangle$, $\zeta_i$ and $\mu_i$ read

$$\begin{aligned}\zeta_i &: \quad \zeta_i\left(r+\langle\sigma\rangle-\mu_i^2\right)=0, \quad i=1,\ldots,N,\\ \langle\sigma\rangle &: \quad \frac{1}{(2N)}\sum_i\zeta_i^2+\frac{1}{2V}\frac{1}{2}\zeta(1/2\,|\,S_{r_0}^{D-1},r+\langle\sigma\rangle)=0,\\ \mu_i &: \quad \frac{\rho_i}{(2N)}-\frac{2\mu_i}{(2N)}\zeta_i^2=0.\end{aligned} \tag{3.49}$$

The first equation either fixes the chemical potentials $\mu_i$ to be equal to the VEV of the collective field $\sigma$ or sets $\zeta_i$ to zero (the case of no condensate). As all solutions except for the trivial one $\zeta_i=0$, $\forall i$ can be shown to be equivalent, we choose all the $\mu_i$ to be equal,

$$\mu_i^2=r+\langle\sigma\rangle:=\mu^2, \qquad i=1,\ldots,N. \tag{3.50}$$

[20] This result can be generalized from the sphere $S_{r_0}^{D-1}$ to any other spatial manifold — like the torus — by simply replacing the eigenvalues $\omega_\ell\to\omega(\mathbf{p})=\sqrt{E(\mathbf{p})^2+r+\langle\sigma\rangle}$ and the corresponding multiplicities $\mathrm{Deg}_D(\ell)\to\mathrm{Deg}_D(\mathbf{p})$.

[21] This definition can be of course be extended to arbitrary manifolds by replacing the eigenvalues and degeneracies.

The third equation equation relates the charge densities $\rho_i$ to the BEC parameter $\zeta_i$,

$$2\mu_i\zeta_i = \rho_i\,, \qquad\qquad i = 1,\dots,N. \tag{3.51}$$

From a statistical mechanics perspective, this equation simply represents the fact that at zero temperature all Bose particles and hence all the charge reside in the ground state. We define the total charge and charge density,

$$\rho := \sum_i \rho_i = 2\sum_i \mu_i\zeta_i\,, \qquad\qquad Q := V\rho = \sum_i Q_i\,. \tag{3.52}$$

After implementing the first two saddle-point equations, the remaining equation gives the condition for the Legendre transform defining the relationship between the free energy density and the grand potential,

$$\frac{\rho}{2N} = -\frac{\mu}{2V}\zeta(1/2\,|\,S^{D-1}_{r_0},\mu^2) = -\frac{1}{2V}\frac{\partial}{\partial\mu}\zeta(-1/2\,|\,S^{D-1}_{r_0},\mu^2) = -\frac{\partial}{\partial\mu}\Omega(\mu)\,. \tag{3.53}$$

The final result for the leading-$N$ free energy (density) per DoF at fixed charge $Q = V\rho$ for the $O(2N)$ WF fixed point can conveniently be written as [22]

$$V f_c(\rho) = f_c(Q) = \mathrm{LT}[-V\Omega](Q/2N) = \mu\frac{Q}{2N} + \frac{1}{2}\zeta(-1/2\,|\,S^{D-1}_{r_0},\mu^2)\Big|_{\frac{Q}{2N}=-\frac{\mu}{2}\zeta(1/2\,|\,S^{D-1}_{r_0},\mu^2)}\,. \tag{3.54}$$

The operator $\mathrm{LT}[h](x)$ denotes the Legendre transform of the convex function $h$, which for a differentiable function can be written in terms of the derivative of $h$,

$$\mathrm{LT}[h](x) = \sup_y\big(xy - h(y)\big) = xy - h(y)\Big|_{x=h'(y)}\,. \tag{3.55}$$

If the function $h$ is not convex, then the Legendre transform does not exists. If the function $h$ is convex, then $\mathrm{LT}[h]$ is convex too and the Legendre transform is an involution $\mathrm{LT}[\mathrm{LT}[h]](y) = h(y)$. As a consequence, the convexity properties of the grand potential $\Omega(\mu)$ are crucial for the theory to be well-defined (at least in the large-$N$ limit).
The natural parameter is $Q/2N$ which we hold fixed. The system can be studied both at $Q/2N \gg 1$ and $Q/2N \ll 1$.

The large-$N$ result in Eq. (3.54) is a one-loop result capturing the leading quantum contributions. It allows us to construct a one-loop effective action using thermodynamical reasoning [235]. Given the physical description of a system in terms of a grand potential $\Omega(\mu)$ (always at finite density) we can write down an effective action capturing the microscopic details of the theory in terms of the NG mode

[22] The contribution from the zero modes to the energy density vanishes at the saddle-point. In terms of $\rho$ the free energy density reads $f_c(\rho) = \mathrm{LT}[-\Omega](\rho)$.

$\chi$ coming from the breaking of the global symmetry, the superfluid phonon,

$$S = \int \mathrm{d}^D x\, (2N)\, \Omega\left(\left|\partial_\mu \chi \partial^\mu \chi\right|^{1/2}\right). \tag{3.56}$$

The action in Eq. (3.56) has to always be understood as an expansion around the superfluid ground state $\chi^{\circledast} = -i\mu\tau$. By construction, on the ground state $\chi^{\circledast}$ the EFT Hamiltonian associated to the action in Eq. (3.56) produces the (large-$N$) free energy in Eq. (3.54). This observation also connects the large-$N$ fixed-charge analysis of the Abelian sector in the $O(2N)$ vector model in the limit $Q/2N \gg 1$ to the EFT approach outlined in Chapter 2.

### 3.1.4 Q/2N ≫ 1 and Q/2N ≪ 1

We analyse the leading-order result in Eq. (3.54) for the energy at fixed charge in the $O(2N)$ WF CFT in the limits $Q/2N \gg 1$ and $Q/2N \ll 1$ using analytic methods. We are doing computations explicitly in $D = 3$ in this section and the next two sections.

#### Q/2N ≫ 1

To compute the free energy in an asymptotic expansion around $Q/2N \gg 1$ we use the Mellin representation of the sphere zeta function,

$$\zeta(s\,|\,S^{D-1}_{r_0}, \mu^2) = \frac{1}{\Gamma(s)} \int_0^\infty \frac{\mathrm{d}t}{t}\, t^s e^{-\mu^2 t} \mathrm{Tr}\left[e^{\Delta_{S^{D-1}_{r_0}} t}\right], \tag{3.57}$$

where $\Gamma(s)$ is the gamma function and $\mathrm{Tr}[e^{\Delta_{S^{D-1}_{r_0}} t}]$ is the (trace of the) heat kernel for the Laplacian on the sphere. Roughly, in the limit $Q/2N \gg 1$ we expect $\mu \gg 1$ to be parametrically large as well and hence the $t$-integral localizes around $t = 0$, where the heat kernel admits an asymptotic expansion. We use Weyl's asymptotic formula expressed in terms of heat kernel coefficients for an arbitrary manifold $\mathcal{M}$ [236–238],

$$\mathrm{Tr}\left[e^{\Delta_{\mathcal{M}} t}\right] = \sum_{n=0}^{\infty} K_n t^{\frac{n}{2}-1}. \tag{3.58}$$

Heat kernel coefficients in the asymptotic expansion are computable via geometric invariants [239, 240]. For a manifold without boundary all odd coefficients vanish. In this case the two leading coefficients are given in terms of the volume $V$ and the scalar curvature $\mathcal{R}$ of the manifold $\mathcal{M}$,

$$K_0 = \frac{V}{4\pi}, \qquad K_2 = \frac{V\mathcal{R}}{24\pi}. \tag{3.59}$$

For every consecutive order in the LCE we need another term in the heat kernel expansion.
For the two-sphere in $D = 3$ there exists an explicit integral representation of the heat kernel [241],

$$\mathrm{Tr}\left[e^{\Delta_{S^2_{r_0}} t}\right] = r_0^3 \frac{2e^{\frac{t}{4r_0^2}}}{\sqrt{\pi}\, t^{3/2}} \int_0^1 \mathrm{d}u\, e^{-r_0^2 \frac{u^2}{t}} \frac{u}{\sin(u)}, \tag{3.60}$$

which can be used to find an integral representation of the zeta function Eq. (3.57) admitting a natural asymptotic expansion in terms of a conformally coupled scalar with mass term $r - 1/4r_0^2$ [124],[23]

$$\zeta(s\,|\,S^2_{r_0},\mu^2) = \frac{r_0^2}{(s-1)}\left(\mu^2 - \frac{1}{4r_0^2}\right)^{1-s} + \frac{1}{12}\left(\mu^2 - \frac{1}{4r_0^2}\right)^{-s} + \frac{7s}{480r_0^2}\left(\mu^2 - \frac{1}{4r_0^2}\right)^{-1-s} + \dots . \tag{3.61}$$

This result allows us to express $\mu$ and $f_c$ — and hence the scaling dimension $\Delta(Q)$ — as a function of $Q/2N$,

$$\begin{aligned} r_0\,\mu(Q) &= \left(\frac{Q}{2N}\right)^{\frac{1}{2}} + \frac{1}{12}\left(\frac{Q}{2N}\right)^{-\frac{1}{2}} + \mathcal{O}(Q^{-3/2})\,, \\ \frac{\Delta(Q)}{2N} &= r_0 f_c(Q) = \frac{2}{3}\left(\frac{Q}{2N}\right)^{\frac{3}{2}} + \frac{1}{6}\left(\frac{Q}{2N}\right)^{\frac{1}{2}} + \mathcal{O}(Q^{-1/2})\,. \end{aligned} \tag{3.62}$$

In the spirit of Eq. (3.56) we observe that from the conformal dimension $\Delta(Q)$ at large $N$ we can extract explicitly the form of the Wilsonian coefficients $c_1, c_2, \dots$ (to leading order in $N$) which are outside of the scope of the EFT approach outlined in Chapter 2 (in particular Section 2.2.2). Additionally, they can be compared to numerical results from the lattice [143, 144]. For the $O(4)$ model the discrepancy is already of the order of just 10% [124].

**Q/2N ≪ 1**

As the saddle-point Eq. (3.54) is valid for any value of the charge — as opposed to the EFT construction discussed in Chapter 2 which is only valid at large charge — we can also expand the grand potential $\Omega(\mu)$ and the free energy $f_c(Q)$ in the opposite limit $Q/2N \ll 1$. Contrary to what we would naively expect, the small-charge limit does not correspond to an expansion in the chemical potential around $\mu \sim 0$, but instead around $\mu \sim 1/2r_0$ [124]. The first hint of this fact comes from the form of the expansion of the zeta function around large values of $\mu$ in Eq. (3.61), where the natural expansion parameter appears to be $\mu^2 - 1/4r_0^2$.

The origin of the small-charge expansion around $\mu \sim 1/2r_0$ is very natural from the point-of-view of conformal invariance. Weyl invariance enforces the curvature coupling of a scalar field on the cylinder to be non-zero and hence produces an effective mass term $r|\phi|^2 = (1/4r_0^2)|\phi|^2$. The saddle-point equation for $\zeta_i$ in Eq. (3.49) fixes $\mu^2 = r + \langle\sigma\rangle$, and therefore it is natural to expand $\mu^2$ at the conformal point around the conformal mass $r = (1/4r_0^2)$.

In the small-$\mu$ limit the zeta function in Eq. (3.48) can be expanded in a binomial expansion for $-1/4 < 0 < \mu^2 < 3/4$,

$$\begin{aligned} \zeta(s\,|\,S^2_{r_0},\mu^2) = \sum_{\ell=0}^{\infty} 2(\ell+\frac{1}{2})\left[\ell(\ell+1){r_0}^{-2} + \mu^2\right]^{-s} &= 2r_0^{2s}\sum_{\ell=0}^{\infty}(\ell+\frac{1}{2})^{1-2s}\left[1 + \frac{(\mu r_0)^2 - \frac{1}{4}}{(\ell+\frac{1}{2})^2}\right]^{-s} \\ &= 2r_0^{2s}\sum_{k=0}^{\infty}\zeta(2s+2k-1;1/2)\left((\mu r_0)^2 - \frac{1}{4}\right)^k\,, \end{aligned} \tag{3.63}$$

[23]The term $-1/4r_0^2$ is produced from the conformally coupled scalar curvature term $\xi\mathscr{R}|\phi|^2$ which on the two-sphere $S^2_{r_0}$ becomes $-1/4r_0^2$.

where $\zeta(s;a)$ denotes the Hurwitz zeta function, formally given by

$$\zeta(s;a) = \sum_{\ell=0}^{\infty} (\ell + a)^{-s} . \tag{3.64}$$

At next-to-leading order $\mu$ receives a correction of order $Q/N$ to the conformal value,

$$r_0\,\mu(Q) = \frac{1}{2} + \frac{8}{\pi^2}\left(\frac{Q}{2N}\right) + \mathcal{O}(Q^2)\,, \tag{3.65}$$

and the conformal dimension reproduces to leading order the result of the (free) mean-field theory,[24]

$$\Delta(Q) = r_0 f_c(Q) = \frac{Q}{2} + \frac{8N}{\pi^2}\left(\frac{Q}{2N}\right)^2 + \mathcal{O}(Q^3)\,. \tag{3.66}$$

In the small-charge limit $Q/2N \ll 1$ we can therefore identify the lowest-lying operator associated to $\Delta(Q)$ with $\varphi^Q$, which has engineering dimension $[\varphi^Q] = Q/2$.

### 3.1.5 Resurgent analysis at $N \to \infty$

It has been observed in lattice results that the predictions from the LCE, at least for the scaling dimension $\Delta(Q)$, remain accurate for small values of the charge [143, 144]. In fact, only a minimal number of terms in the effective action Eq. (2.26) and hence the asymptotic expansion of $\Delta(Q)$ suffices to reproduce lattice results with high precision. This behaviour cannot be explained from the point-of-view of the EFT but rather in terms of the non-perturbative physics that specifies the optimal truncation of the asymptotic series, which can be predicted to be at $N^* = \mathcal{O}(\sqrt{Q})$ with an error of order $\mathcal{O}(e^{-\pi\sqrt{Q}})$ explaining the small error and the small amount of terms necessary to reproduce the scaling dimension up to $Q = \mathcal{O}(10)$ [191].

As it was first pointed out by Dyson [242], perturbative expansions in QM and QFT are usually asymptotic series and not convergent ones. The LCE is no different in this regard, and at large $N$ the asymptotic nature of the expansion in $Q \gg 1$ is manifest due to its connection to the manifestly asymptotic expansion of the zeta function in Eq. (3.61). The asymptotic nature of perturbative expansions implies the existence of non-perturbative contributions and phenomena in the underlying theory. The modern approach to analytically study non-perturbative corrections is resurgence [243, 244]. Within the EFT construction we do not have access to enough information about the non-perturbative sector, however, at $N \gg 1$ we have access to the full theory and an explicit form of the energy $r_0\Delta(Q)$ valid for all values of $Q$. As first shown in [191], in the double-scaling limit $Q \to \infty$, $N \to \infty$, $Q/N$ fixed there is enough structure to perform a complete resurgent analysis, which also produces the results for the optimal truncation discussed above.

[24]To leading order in large $N$ the WF fixed point is equivalent to a mean-field theory — or generalized free-field theory — with an action of the form $\int \left(|\partial\phi_i|^2 + \langle|\phi_i|^2\rangle\,|\phi_i|^2\right)$. In this theory an operator of charge $Q$ is simply of the form $\phi_{i_1}\ldots\phi_{i_Q} \sim \phi^Q$. Importantly, in contrast to the small-$\epsilon$ expansion around $D = 4$, the large $N$ expansion of the WF fixed point is not an expansion around the free — or Gaussian — conformal fixed point.

The resurgent analysis of the LCE in the Abelian sector of the $O(2N)$ model at large $N$ can be performed as a result of the Seeley–DeWitt expansion [245–248] of the heat kernel trace in Eq. (3.57). The trace of the conformal Laplacian can be written as [191]

$$\mathrm{Tr}\left[e^{\left(\Delta_{S^2_{r_0}}-\frac{1}{4r_0^2}\right)t}\right]=\frac{r_0^2}{t}+\sum_{k\in\mathbb{Z}}(-1)^k\left[\frac{r_0^2}{t}-2\frac{\pi r_0^3|k|}{t^{3/2}}F\Big(\frac{\pi r_0|k|}{\sqrt{t}}\Big)\right], \qquad F(z)=e^{-z^2}\int_0^z \mathrm{d}u\, e^{-u^2}. \tag{3.67}$$

Dawson's function $F(z)$ appearing in the above representation of the heat kernel admits an asymptotic expansion so that after some formal manipulations the heat kernel can be written as

$$\mathrm{Tr}\left[e^{\left(\Delta_{S^2_{r_0}}-\frac{1}{4r_0^2}\right)t}\right]\sim\frac{r_0^2}{t}\sum_{n=0}^{\infty}a_n\left(\frac{t}{r_0^2}\right)^n, \qquad a_n:=\frac{(1-2^{1-2n})B_{2n}}{(-1)^{n+1}n!}\sim\frac{2}{\sqrt{\pi n}}\frac{n!}{\pi^{2n}}, \tag{3.68}$$

where the divergence is a consequence of the asymptotic expansion of Dawson's function. As is standard for asymptotic expansions, this expansion is only formally valid and needs a summing prescription. Assuming that the series can be completed into a resurgent trans-series implies that the above expansion gets supplemented by non-perturbative terms of the form

$$\mathrm{Tr}\left[e^{\left(\Delta_{S^2_{r_0}}-\frac{1}{4r_0^2}\right)t}\right]\supset 2i\left(\pi\frac{r_0^2}{t}\right)^{\frac{3}{2}}(-1)^{k+1}|k|e^{-\frac{(\pi r_0 k)^2}{t}}. \tag{3.69}$$

The relative coefficients cannot be fixed from the resurgent analysis of the asymptotic expansion, which is a reflection of the fact that any choice of coefficients results in the same asymptotic expansion.
The resurgent analysis of the heat kernel trace directly identifies the structure of the trans-series of the grand potential

$$V\Omega(\mu)=\frac{1}{2}\zeta(-1/2\,|\,S^2_{r_0},\mu^2)=\frac{1}{2\Gamma(s)}\int_0^\infty\frac{\mathrm{d}t}{t}t^s e^{-\mu^2 t}\mathrm{Tr}\left[e^{\Delta_{S^2_{r_0}}t}\right]\bigg|_{s=-1/2}. \tag{3.70}$$

As a result, the grand potential $\Omega$ and its Legendre dual — the free energy $f_c$ — include non-perturbative corrections of the form

$$V\Omega(\mu)\supset\sqrt{\mu^3 r_0}\frac{(-1)^k e^{-2\pi\mu r_0|k|}}{(2\pi|k|)^{3/2}}\sum_{i=0}^{\infty}\left(\frac{\gamma_i}{\gamma_0}\right)(2\pi\mu r_0|k|)^{-i}, \qquad f_c(Q)\supset\frac{1}{r_0}\left(\frac{Q}{2N}\right)^{\frac{3}{4}}\frac{(-1)^k e^{-2\pi|k|\sqrt{q}}}{(2\pi|k|)^{3/2}}+\dots. \tag{3.71}$$

where the coefficients $\gamma_i$ can be computed recursively [191]. This result corresponds to a $(2n)!$ divergence of the perturbative series of $f_c(Q)$.
Importantly, from the EFT point-of-view the $(2n)!$ divergence is a tree-level effect. If we identify the grand potential with a superfluid effective action as in Eq. (3.56) the Wilsonian coefficients form a divergent series, whose $(2n)!$ divergence is to be contrasted with the $n!$ divergence that we generically expect in QFT from the proliferation of Feynman diagrams. Here, the classical divergence is more important than the quantum one.

The trace of the heat kernel can be Borel re-summed, leading to a closed form expression that includes the non-pertubative corrections. In general, the Borel re-summation is a prescription from resurgence that gives meaning to factorially divergent asymptotic series by systematically incorporating the associated non-perturbative terms. There will be ambiguities left associated to the coefficients of the non-perturbative corrections, which in this case can be fixed by physical inputs (*e.g* reality of the heat trace) [191]. This in turn fixes the ambiguities for $\Omega(\mu)$ and $f_c(\mu)$. After Borel re-summation, the heat kernel trace becomes

$$\mathrm{Tr}\left[e^{\left(\Delta_{S^2_{r_0}}-\frac{1}{4r_0^2}\right)t}\right]=\frac{2}{\sqrt{\pi}}\left(\frac{r_0^2}{t}\right)^{\frac{3}{2}}\mathrm{P.V.}\int_{\mathscr{C}^\pm}\mathrm{d}\zeta\,\frac{\zeta\, e^{-\zeta^2\frac{r_0^2}{t}}}{\sin(\zeta)}\,, \tag{3.72}$$

where $\mathscr{C}^\pm$ is the contour along the positive real axis avoiding the simple poles $\zeta = k\pi$, $k \in \mathbb{N}^+$ of $\sin(\zeta)$ in the Borel transform of the asymptotic expansion of the heat kernel either above ($\mathscr{C}^+$) or below ($\mathscr{C}^-$). The prefix P.V. tells us to take the principal value of the $\zeta$-integral. Despite its appearance, the Borel re-summation prescription in Eq. (3.72) is unambiguous, a property that is mirrored in various systems involving ordinary differential equations [249].

The results for the scaling dimension derived from the Borel re-summed heat kernel trace in Eq. (3.72) match the result from the small-charge regime $Q/2N \ll 1$ to high precision. For a more detailed discussion on the resurgent analysis of the large-$N$ and large-$Q$ double-scaling limit in the $O(2N)$ vector model we refer to [191].

### 3.1.6 Subleading corrections: the NG modes

We end this Section with a discussion of the sub-leading correction in the large-$N$ analysis of the $O(2N)$ WF fixed point in $D = 3$. We consider the $O(2N)$ WF CFT (at $|g| \to \infty$ in our conventions) in terms of the action in Eq. (3.39) including the original field $\phi_i$ and separate the collective field into its VEV plus fluctuations,

$$S_\mu[\phi_i,\hat{\sigma}]=\int\limits_{S^1_\beta\times S^{D-1}_{r_0}}\mathrm{d}^D x\left[(\partial_\tau-\mu_i)\phi_i^*(\partial_\tau+\mu_i)\phi_i+|\nabla\phi_i|^2+(r+\langle\sigma\rangle)|\phi_i|^2+\frac{\hat{\sigma}}{\sqrt{N}}|\phi_i|^2\right], \tag{3.73}$$

where we have rescaled the fluctuation resulting in a self-consistent $1/N$ expansion following the standard procedure [227, 228]. This introduces a hierarchy among the terms as in Eq. (3.42), which is important once we go beyond leading order. We are in a regime with two large numbers $N, Q$, where $N \gg Q/N$ controls the splitting between tree-level and quantum corrections. Additionally, the ratio $Q/N$ controls an expansion of physical observables at every fixed order in $N$.

Most importantly, at sub- leading order we expect to encounter the universal $Q^0$-term in $D = 3$ discussed within the EFT approach in Section 2.2 — in particular around Eq. (2.53) and Eq. (2.87) — coming from the superfluid phonon. To do so it is more convenient to choose a slightly different approach to the one presented in Section 3.1.3 and Appendix C.2. We choose to incorporate the ground state for $\phi_i$ — the BEC parameter $\zeta_i$ — within the action before integrating out the fluctuations $\hat{\phi}_i = \phi_i - \zeta_i e^{i\varphi_i}$ of $\phi_i$.

The action becomes (up to a total derivative)

$$S_\mu[\hat\phi_i,\hat\sigma] = \int\limits_{S^1_\beta\times S^{D-1}_{r_0}} \mathrm{d}^D x \left[\zeta_i^2(r+\lambda-\mu^2) + \hat\phi_i^*(-(\partial_\tau+\mu_i)^2 + r + \langle\sigma\rangle + \hat\sigma/\sqrt{N})\hat\phi_i + \frac{\hat\sigma}{\sqrt{N}}\left(\zeta_i e^{-i\varphi_i}\hat\phi_i + \hat\phi_i^*\zeta_i e^{i\varphi_i}\right)\right]. \tag{3.74}$$

To analyse the fluctuation we integrate out all but one of the fields $\phi_i$. As discussed in Section 3.1.3 around the saddle-point equations in Eq (3.49), there are several equivalent choices of the ground state that can be rotated into each other. We choose to rotate the BEC parameters so that all of the charge lies in the condensate of the field $\phi_N$ [124],

$$\rho = \sum_i \rho_i = 2\sum_i \mu_i\zeta_i = 2\mu\zeta_N, \qquad \zeta_1 = \cdots = \zeta_{N-1} = 0. \tag{3.75}$$

There are now $N-1$ fields with equal dispersion relations and one field that is different. The action becomes (we also choose $e^{i\varphi_N}=1$)

$$\begin{aligned} S_\mu[\hat\phi_i,\hat\sigma] = {}& \beta V(r+\langle\sigma\rangle-\mu^2)\zeta_N^2 + \sum_{i=1}^{N-1}\int\limits_{S^1_\beta\times S^{D-1}_{r_0}} \mathrm{d}^D x\left[\hat\phi_i^*(-(\partial_\tau+\mu)^2 + r + \langle\sigma\rangle + \hat\sigma/\sqrt{N})\hat\phi_i\right] \\ & + \int\limits_{S^1_\beta\times S^{D-1}_{r_0}} \mathrm{d}^D x\left[\hat\phi_N^*(-(\partial_\tau+\mu)^2 + r + \langle\sigma\rangle + \hat\sigma/\sqrt{N})\hat\phi_N + \frac{\hat\sigma}{\sqrt{N}}\zeta_N(\hat\phi_N^* + \hat\phi_N)\right]. \end{aligned} \tag{3.76}$$

We first discuss the dispersion relations of the fields $\hat\phi_1,\dots,\hat\phi_{N-1}$ before we integrate them out later. Their dispersion relations are all the same and correspond each to a pair of modes — one gapless and one gapped NG mode — with inverse propagator and dispersion relations (in flat space)

$$\tilde G^{-1}_{(0)}(\omega,\mathbf{p}) = \frac{1}{2}\begin{pmatrix} 0 & \omega^2+\mathbf{p}^2-2\mu\omega \\ \omega^2+\mathbf{p}^2+2\mu\omega & 0\end{pmatrix}, \qquad \begin{cases} \omega^2 = -\frac{\mathbf{p}^4}{4\mu^2}+\dots, \\ \omega^2 = -2\mathbf{p}^2 - 4\mu^2 + \dots . \end{cases} \tag{3.77}$$

The first mode is massless and corresponds to a particle with quadratic dispersion relation. These modes are the $N-1$ non-relativistic NG modes also found in [46, 47, 122], which are naturally arranged into the fundamental representation of the $U(N-1)$ subgroup which remains unbroken by the large-$N$ large-$Q$ saddle-point. Their massive counterparts are the gapped NG modes first encountered in Section 1.2.5.

We can expand the action in Eq. (3.76) in the fluctuations $\hat\sigma/\sqrt{N}$ and integrate out the fields $\phi_1,\dots,\phi_{N-1}$,

$$\begin{aligned} S_\mu[\hat\phi_N,\hat\sigma] = {}& \beta V(r+\langle\sigma\rangle-\mu^2)\zeta_N^2 + N\sum_\ell \mathrm{Deg}_D(\ell)\left[\left[\omega_\ell + \frac{1}{\beta}\log(1-e^{-\beta(\omega_\ell+\mu_i)})(1-e^{-\beta(\omega_\ell-\mu_i)})\right]\right. \\ & + \sum_{k=2}^{\infty}\frac{(N-1)\mathrm{Tr}(G_{(0)}\hat\sigma)^k}{(-1)^{k+1}k(N)^{k/2}} + \int\limits_{S^1_\beta\times S^{D-1}_{r_0}} \mathrm{d}^D x\left[\hat\phi_N^*(-(\partial_\tau+\mu)^2 + r+\langle\sigma\rangle+\hat\sigma/\sqrt{N})\hat\phi_N + \frac{\hat\sigma}{\sqrt{N}}\zeta_N(\hat\phi_N^*+\hat\phi_N)\right]. \end{aligned} \tag{3.78}$$

As we are interested in the fluctuations and the first quantum corrections, we only care about the quadratic part of the action and drop other contributions.[25] In addition, we can express $\zeta_N$ and $r+\langle\sigma\rangle$ via the saddle-point equations in Eq. (3.49),[26]

$$S^{(2)}[\hat{\phi}_N,\hat{\sigma}] = -\frac{1}{2}\underbrace{\frac{(N-1)}{N}}_{=1+\mathcal{O}(N^{-1})}\mathrm{Tr}(G_{(0)}\hat{\sigma})^2 + \int_{S^1_\beta\times S^{D-1}_{r_0}} \mathrm{d}^D x\,\left[\hat{\phi}^*_N\big(-(\partial_\tau+\mu)^2+\mu^2\big)\hat{\phi}_N + \sqrt{\frac{-\zeta(1/2\,|\,S^{D-1}_{r_0},\mu^2)}{2V}}\,\hat{\sigma}(\hat{\phi}^*_N+\hat{\phi}_N)\right]. \tag{3.79}$$

We of course consider the above quadratic action in the limit $\beta\to\infty$ where $S^1_\beta\to\mathbb{R}$. The term $\mathrm{Tr}(G_{(0)}\hat{\sigma})^2$ is non-local and computable in the low-energy limit [124],

$$\mathrm{Tr}(G_{(0)}\hat{\sigma})^2 = \int_{S^1_\beta\times S^{D-1}_{r_0}}\mathrm{d}^D x_1\int_{S^1_\beta\times S^{D-1}_{r_0}}\mathrm{d}^D x_2\,\left[\hat{\sigma}(x_1)\,\hat{\sigma}(x_2)G_{(0)}(x_2-x_1)^2\right] \sim \int_{S^1_\beta\times S^{D-1}_{r_0}}\mathrm{d}^D x\,\hat{\sigma}^2\frac{1}{4V}\zeta(3/2\,|\,S^{D-1}_{r_0},\mu)\,, \tag{3.80}$$

so that the quadratic action becomes [124]

$$S^{(2)}[\hat{\phi}_N] = \int_{S^1_\beta\times S^{D-1}_{r_0}}\mathrm{d}^D x\,\left[\hat{\phi}^*_N\big(-(\partial_\tau+\mu)^2+\mu^2\big)\hat{\phi}_N + \sqrt{\frac{-\zeta(1/2\,|\,S^{D-1}_{r_0},\mu^2)}{2V}}\,\hat{\sigma}(\hat{\phi}^*_N+\hat{\phi}_N) - \frac{1}{8V}\zeta(3/2\,|\,S^{D-1}_{r_0},\mu)\,\hat{\sigma}^2\right]. \tag{3.81}$$

We now specialize to the case $D=3$. There is no kinetic term for $\hat{\sigma}$ and we can integrate it out. Additionally, we can use the asymptotic expansion of the heat kernel in Eq. (3.58) to extract via Eq. (3.57) the leading-order results from the zeta functions [124]. The action in $D=3$ becomes

$$S^{(2)}[\hat{\phi}_N] = \int_{S^1_\beta\times S^2_{r_0}}\mathrm{d}^2 x\left[\hat{\phi}^*_N\big(-(\partial_\tau+\mu)^2+\mu^2\big)\hat{\phi}_N + \mu^2(\hat{\phi}^*_N+\hat{\phi}_N)^2\right], \tag{3.82}$$

and the inverse propagator for the two remaining modes reads

$$2\tilde{G}^{-1} = \begin{pmatrix}\omega^2-\mu^2+p^2+\mu^2+4\mu^2 & 2\mu\omega\\ -2\mu\omega & \omega^2-\mu^2+p^2+\mu^2\end{pmatrix} = \begin{pmatrix}\omega^2+p^2+4\mu^2 & 2\mu\omega\\ -2\mu\omega & \omega^2+p^2\end{pmatrix}. \tag{3.83}$$

From there we get the dispersion relations of the fluctuations (in flat space),

$$\omega^2 = -\mathbf{p}^2-4\mu^2\pm\sqrt{16\mu^4+4\mathbf{p}^2\mu^2} = \begin{cases}-\frac{1}{2}\mathbf{p}^2+\dots\\ -\frac{3}{2}\mathbf{p}^2-8\mu^2+\dots\end{cases}. \tag{3.84}$$

We recover the conformal superfluid NG mode and hence can also identify the universal contribution to the conformal dimension $\Delta(Q)$ in Eq. (2.87) first presented in Section 2.2. In the large-$N$ expansion this contribution appears at sub-leading order $\mathcal{O}(N^0)$ in $N$.

[25] The constant part of the action corresponds to the leading classical trajectory and was already thoroughly discussed in Section 3.1.3.

[26] $G_{(0)}$ satisfies $(-(\partial_\tau+\mu)^2+r+\langle\sigma\rangle)G_{(0)}(x-y)=\delta(x-y)$.

## 3.2 Convexity and the effective potential for $\varphi^4$ from large charge

Working in the double-scaling limit for the $O(2N)$ vector model introduces a lot of structure that can be used to perform calculations. We have discussed extensively how this structure can be used to access to strongly-coupled (for every finite $N$) fixed point in Section 3.1.1, in particular for the case of the WF fixed point in $D = 3$. Perhaps surprisingly, large-charge methods also allow us to study the theory away from the fixed point purely at the level of the grand potential $\Omega(\mu)$ without resorting to Feynman-diagram techniques [1, 128]. It is possible for us to derive a relationship between the one-loop effective action and the grand potential (which is also a one-loop quantity) and to describe in a compact way the phase diagram of the theory with the potential $V(\phi) = r|\phi|^2 + (g/N)|\phi|^4$, generalizing and expanding earlier works [19, 135, 250, 251].

Beyond analysing the $\varphi^4$-theory in $2 < D < 4$ and $D = 3$, where it flows to the interacting WF fixed point, we are able to analyse the model in $4 < D < 6$ and $D = 5$, where we would expect an interacting fixed point in the UV [222, 252]. We find that the $\varphi^4$-model is not UV complete if we require unitarity, which is consistent with recent works [20, 253, 254]. Nevertheless, we can compute a possible completion in terms of a complexified effective potential in $D = 5$ describing the flow to a non-unitary CFT, with results in agreement with the existing literature [20, 253–255]. Importantly, this complex CFT remains unstable for both small- and large-charge operators even though in the small-charge regime scaling dimensions are still purely real [20, 254–257].

In this Section we are mostly content with working in flat space or on the torus. In terms of Weyl's asymptotic expansion in Eq. (3.59) and Eq. (3.58) the flat space result will generically capture the leading order result on the sphere as well, but we cannot access to the sub-leading corrections in flat space.

### 3.2.1 The effective action in (scalar) QFT and its relation to the grand potential

It is important to understand the underlying reason for the relationship between the grand potential $\Omega(\mu)$ used in Section 3.1.1 to set up the LCE for the WF fixed point at large $N$ and the quantum effective potential. For this reason we repeat some important QFT prerequisites about effective actions before we present the relationship between the two potentials.

#### The effective potential in QFT and its convexity properties

In this part we mainly review some standard arguments for the properties of effective potentials in unitary QFTs [105, 258–260]. For simplicity we restrict ourselves to scalar $O(2)$ models, but the discussion trivially generalizes to $O(2N)$ vector models (and generally other unitary QFTs).

Consider a theory described by a complex scalar field $\phi$ that is invariant under an $O(2)$ symmetry acting linearly on $\phi$,

$$\phi \to e^{i\alpha}\phi. \tag{3.85}$$

Via the standard procedure we add a linear source term for the $O(2)$ symmetry in the path integral to write the generating functional $Z[j]$,

$$Z[j] = \langle 0 \rangle_j = \frac{\int \mathcal{D}\phi\mathcal{D}\phi^* \, e^{-S[\phi]-\int \mathrm{d}^D x\,[j^*\phi + j\phi^*]}}{\int \mathcal{D}\phi\mathcal{D}\phi^* \, e^{-S[\phi]}} \,. \tag{3.86}$$

The logarithm of $Z[J]$ is the so-called connected generating functional, the generating functional for all the connected correlation functions of the theory,

$$W[j] := \log Z[j] \,. \tag{3.87}$$

In analogy to simple probability theory, we can perform a functional Legendre transformation LT$[\cdot]$ of $W[j]$ to change the dependent variables from the sources $j, j^*$ to field variables $\phi^{(c)}, \phi^{(c)*}$. The so-called classical fields $\phi_c, \phi_c^*$ are defined via the quantum field equations as

$$\phi^{(c)} = \frac{\delta W}{\delta j^*} \,, \qquad\qquad \phi^{(c)*} = \frac{\delta W}{\delta j} \,. \tag{3.88}$$

The Legendre dual of $W$ is called the effective action $\Gamma[\phi^{(c)}]$,

$$\Gamma[\phi^{(c)}] := \mathrm{LT}[W]\,[\phi^{(c)}] = j^*\phi^{(c)} + j\,\phi^{(c)*} - W[j]\Big|_{j=j(\phi^{(c)}),\, j^*=j^*(\phi^{(c)})} \,. \tag{3.89}$$

By expanding around the value $\phi_c = \text{const.}$, the effective action $\Gamma[\phi_c]$ becomes

$$\Gamma[\phi^{(c)}] = \int \mathrm{d}^D x \left[ V_{\text{eff}}(|\phi^{(c)}|) + Z[\phi^{(c)}]\,\partial_\mu\phi^{(c)*}\partial^\mu\phi^{(c)} + \ldots \right] , \tag{3.90}$$

where the function $V_{\text{eff}}(|\phi^{(c)}|)$ appearing in the expansion of $\Gamma[\phi^{(c)}]$, which includes all terms in $\Gamma[\phi^{(c)}]$ without derivatives, is the so-called effective potential encoding all quantum corrections to the classical potential $V(|\phi^{(c)}|)$ in the original action $S[\phi]$. We note that the $O(2)$ symmetry, which we assume to be realized at the quantum level, requires all functions appearing in $\Gamma[\phi^{(c)}]$ to depend only on the absolute value $|\phi^{(c)}|$, in particular the effective potential.

If we consider a unitary theory with a positive definite path-integral measure $\int \mathcal{D}\phi\mathcal{D}\phi^*$, the corresponding generating functional $W[J]$ is a convex function of the sources.[27] The argument proving this fact about unitary QFT is based on Hölder's inequality. Given a positive measure $\mathrm{d}\mu$ and two positive functions $f$, $g$ the following inequality holds:

$$\int \mathrm{d}\mu\, f^\lambda g^{1-\lambda} \le \left[\int \mathrm{d}\mu\, f\right]^\lambda \left[\int \mathrm{d}\mu\, g\right]^{1-\lambda} , \qquad\qquad \forall 0 \le \lambda \le 1 \,. \tag{3.91}$$

In the case of our scalar field theory we can define the normalized measure

$$\mathrm{d}\mu := \frac{\mathcal{D}\phi\mathcal{D}\phi^*\, e^{-S[\phi]}}{\int \mathcal{D}\phi\mathcal{D}\phi^*\, e^{-S[\phi]}} \,. \tag{3.92}$$

[27] In fact, this is more generally true for any parameter that enters linearly in the action $S[\phi]$.

We consider the expectation value

$$\langle\alpha\rangle := \frac{1}{Z[0]}\int \mathscr{D}\phi \int \mathscr{D}\phi^*\, e^{-S[\phi]} e^{\int \mathrm{d}^D x \sum_i \alpha_i f_i(\phi)} = \int \mathrm{d}\mu\, e^{\int \mathrm{d}^D x \sum_i \alpha_i f_i(\phi)}\,, \tag{3.93}$$

with the coefficients $\alpha_i$ being functions of spacetime. By virtue of Hölder's inequality, for $0 < \lambda < 1$ the expectation value

$$\langle \lambda\alpha + (1-\lambda)\rangle = \int \mathrm{d}\mu\, e^{\int \mathrm{d}^D x \sum_i [\lambda\alpha_i + (1-\lambda)\beta_i] f_i(\phi)} \tag{3.94}$$

satisfies

$$\langle \lambda\alpha + (1-\lambda)\beta\rangle \le \langle\alpha\rangle^{\lambda}\langle\beta\rangle^{1-\lambda}\,. \tag{3.95}$$

As a consequence, the logarithm of the expectation value, which is related to the generating functional, satisfies the convexity property,

$$\log\langle \lambda\alpha + (1-\lambda)\beta\rangle \le \lambda\log\langle\alpha\rangle + (1-\lambda)\log\langle\beta\rangle\,. \tag{3.96}$$

In particular, this property holds for the connected generating functional $W[j]$, making it a complex function of the sources $j, j*$.
As the Legendre transform is an involution on convex functions — meaning that the Legendre transform of a convex function is convex — hence, the effective action $\Gamma[\phi^{(c)}]$ is a convex function of the classical fields $\phi^{(c)}, \phi^{(c)}$. This property has to persist for constant values of $\phi^{(c)}$, so that the effective potential $V(|\phi^{(c)}|)$ itself has to be convex, provided the theory is unitary.[28]

### The grand potential in a scalar theory and its relationship to the effective potential

Under the assumption that in some simplifying limit the effective action $\Gamma[\phi^{(c)}]$ has a canonical kinetic term,[29]

$$\Gamma[\phi^{(c)}] = \int \mathrm{d}^D x\,\left[\partial_\mu\phi^{(c)*}\partial_\mu\phi^{(c)} + V_{\text{eff}}(|\phi^{(c)}|)\right]. \tag{3.98}$$

We can relate the grand potential $\Omega(\mu)$ to the effective potential. The conserved $O(2)$ charge is of the form ($\tau = x_0$ in Euclidean signature)

$$Q = -\int \mathrm{d}^{D-1}x\,\left[\partial_\tau\phi^{(c)*}\phi^{(c)} - \phi^{(c)*}\partial_\tau\phi^{(c)}\right]. \tag{3.99}$$

To study the system at fixed charge, we make the finite-density ansatz[30]

$$|\phi^{(c)}| = \zeta\,, \qquad\qquad \arg(\phi^{(c)}) = -i\mu\tau\,, \tag{3.100}$$

[28]This property of the effective potential persists in finite volume. For strongly-coupled systems that are generally studied numerically, it is more convenient to introduce another, closely related quantity, the so-called constraint effective potential $U$, roughly defined as

$$e^{-U(\varphi)} = \int \mathscr{D}\phi\mathscr{D}\phi^*\,\delta(\phi-\varphi)\,e^{-S[\phi]}\,. \tag{3.97}$$

The constraint effective potential $U(\varphi)$ is related to the effective potential $V(|\phi^{(c)}|)$ by another Legendre transform. As a quantity, $U(\phi)$ is in general not convex but coincides with the effective potential in the decompactification limit [261].

[29]Generically, this is the case if higher-derivative terms in the effective action are suppressed and $Z(\phi^{(c)})$ in Eq. (3.90) can be written as $Z(\phi^{(c)}) = 1+$ sub-leading terms. This is expected to be true for a wide array of scalar theories.

[30]The imaginary unit in the argument is always due to Wick rotation.

which corresponds to a ground state with a fixed chemical potential $\mu$ (*e.g.* a superfluid). Notice that $\zeta^2$ is related to the charge of the ground state as

$$Q\big|_{\phi^{(c)}=\zeta e^{\mu\tau}} = 2\mu V \zeta^2, \tag{3.101}$$

where $V$ is the spatial volume. The modulus $\zeta$ of $\phi^{(c)}$ — the radial mode — of the field is removed from the action via its EoM,

$$\frac{\mathrm{d}}{\mathrm{d}(\zeta^2)}\left[\partial_\mu\phi^{(c)*}\partial_\mu\phi^{(c)} + V_{\text{eff}}(|\phi^{(c)}|)\right]_{\phi^{(c)}=\zeta e^{\mu\tau}} = \frac{\mathrm{d}}{\mathrm{d}(\zeta^2)}\left[-\mu^2\zeta^2 + V_{\text{eff}}(\zeta)\right] = -\mu^2 + \frac{\mathrm{d}V_{\text{eff}}(\zeta)}{\mathrm{d}(\zeta^2)} = 0, \tag{3.102}$$

Now we can write the VEV of the action and the Lagrangian as a function of the chemical potential $\mu$ alone,

$$-\Omega(\mu) = \zeta^2\mu^2 - V_{\text{eff}}(\zeta)\Big|_{\zeta=\zeta(\mu^2)}. \tag{3.103}$$

The corresponding fixed-charge energy density $f_c(\rho)$ for the charge density $\rho = Q/V$ is computed using the momentum associated to $\mu$, which is given by said charge density. Given the solution to $\zeta = \zeta(\mu)$ in Eq. (3.103), the energy density reads

$$\rho = -\frac{\delta\Omega}{\delta\mu} = 2\mu\zeta^2(\mu), \qquad f_c(\rho) = \left[\rho\mu + \Omega(\mu)\right]_{\mu=\mu(\rho)}. \tag{3.104}$$

Both Eq. (3.103) and Eq. (3.104) describe a Legendre transform relating the effective potential $V_{\text{eff}}(\zeta)$ to the grand potential $\Omega(\mu)$ and the grand potential to the free energy $f_c(\rho)$. We introduce the following notation

$$\Upsilon(|\phi^{(c)}|^2) := V_{\text{eff}}(|\phi^{(c)}|), \qquad \varpi(\mu^2) := -\Omega(\mu), \tag{3.105}$$

In this language the chain of Legendre transformations becomes

$$V_{\text{eff}}(|\phi^{(c)}|) = \Upsilon(|\phi^{(c)}|^2) \overset{(\leftrightarrow)}{\to} \text{LT}[\Upsilon](\mu^2) = \varpi(\mu^2) = -\Omega(\mu) \overset{(\leftrightarrow)}{\to} \text{LT}[-\Omega](\rho) = f_c(\rho). \tag{3.106}$$

For convex functions $V_{\text{eff}}$, $\Upsilon$, $\varpi$, $\Omega$ and $f_c$ the Legendre transform is an involution and the arrows can be reversed. In this case we can compute *e.g.* the effective potential from and vice versa.
If the convexity condition is not met in some areas of the parameter space, the minimization condition via the derivative can still be applied and admits in general complex solutions. There are generally speaking two possible stances we can take with regards to this issue:

Either we define the relationship between different quantities strictly via the Legendre transform and its supremum definition,

$$\text{LT}[f](y) = \sup_x(xy - f(x)). \tag{3.107}$$

In this case $\text{LT}[f](y)$ always takes values on the extended real line $\mathbb{R}\cup\{\pm\infty\}$ and in intervals where $f(x)$ ceases to be convex the Legendre transform returns a constant value.

Or, on the other hand, we can simply use the naive definition of the minimization condition via the derivative, as seen above in Eq. (3.103) and Eq. (3.104). In general, this leads to a multi-valued complex function with branch points and cuts. If the function $f(x)$ is convex and differentiable, both definitions agree, the Legendre transform is an involution and differentiable. The supremum definition also always results in a convex function.

While it is clear that the supremum definition is the correct one to use in the context of classical thermodynamics [259], in the case of quantum systems the situation is less clear-cut. In this case also complex saddles of the path integral have a meaning, a fact which is of particular importance in the field of resurgence [262].

We stress that of all functions that we have discussed and introduced around Eq. (3.106), only the effective potential $V_{\text{eff}}$ is always convex in a unitary theory. In this Section we will repeatedly make use of this property as a necessary condition for unitarity of the theory. Particularly, we note that the convexity of $V_{\text{eff}}$ and $\Upsilon$ are non-trivially related,

$$\frac{\mathrm{d}^2\Upsilon\big(|\phi^{(c)}|^2\big)}{\mathrm{d}\big(|\phi^{(c)}|^2\big)^2} = \frac{1}{4|\phi^{(c)}|^2}\left[\frac{\mathrm{d}^2 V_{\text{eff}}\big(|\phi^{(c)}|\big)}{\mathrm{d}|\phi^{(c)}|^2} - \frac{1}{|\phi^{(c)}|}\frac{\mathrm{d}V_{\text{eff}}\big(|\phi^{(c)}|\big)}{\mathrm{d}|\phi^{(c)}|}\right]. \tag{3.108}$$

**Effective action and grand potential at large N**

As discussed around Eq. (3.56), using thermodynamical reasoning [235], the grand potential $\Omega(\mu)$ for the $O(2N)$ vector model at large $N$ describes a one-loop effective action. Requiring that the physics of the grand potential $\Omega(|\partial_\mu\chi\partial^\mu\chi|^{1/2})$ is equivalently captured by the one-loop (or leading-$N$) effective action $\Gamma^{(0)}[\phi_i^{(c)}]$, which is manifestly $O(2N)$ invariant, leads to the same relationship between the leading-$N$ effective potential $V_{\text{eff}}^{(0)}$ and the grand potential $\Omega(\mu)$ in terms of the Legendre transform described in Eq. (3.103) and Eq. (3.106) [128].[31] The only difference in this case is that $\Omega(\mu)$ at large $N$, as defined in Eq. (3.28), is divided by the number of DoF of the theory when compared to the definition of $\Omega(\mu)$ in Eq. (3.103),

$$-\Omega(\mu) = \frac{|\phi_i^{(c)}|^2}{(2N)}\mu^2 - \frac{V_{\text{eff}}^{(0)}(|\phi_i^{(c)}|)}{(2N)}\bigg|_{|\phi_i^{(c)}|=|\phi_i^{(c)}|(\mu^2)}, \qquad |\phi_i^{(c)}| = \sqrt{\sum_i \phi_i^{(c)*}\phi_i^{(c)}}. \tag{3.109}$$

We therefore expect to be able to study and potentially derive the effective potential $V_{\text{eff}}^{(0)} = V_{\text{eff}}^{(0)} + \mathscr{O}(N^0)$ at leading order in $N$ (which is a one-loop observable) via our knowledge of the grand potential $\Omega(\mu)$. The rescaling of $\Omega$ at large $N$ compared to $\Omega$ in Eq. (3.103) poses no conceptual issue as the Legendre transform preserves homogeneity in the variables,

$$f(x) = ag(x) \;\to\; \mathrm{LT}[f](y) = a\,\mathrm{LT}[g](y/a), \qquad f(x) = g(ax) \;\to\; \mathrm{LT}[f](y) = \mathrm{LT}[g](y/a). \tag{3.110}$$

Finally, at large $N$, based on the structure of the large-$N$ expansion, there is very good reason to expect that the kinetic term in the leading-order effective action $\Gamma^{(0)}[\phi_i^{(c)}]$ is in fact canonical, *i.e.* $Z[\phi_i^{(c)}] = 1 + \mathcal{O}(N^{-1})$.

[31] The line of argument here is essentially that the effective action $\Gamma^{(0)}[\phi_i^{(c)}]$ defines a Linear Sigma Model (LSM) and the grand potential defines a Non–Linear Sigma Model (NLSM), both of them describing the same system (at fixed chemical potential). Hence, on the ground state they have to agree.

### 3.2.2 $\varphi^4$-theory away from the fixed point in $2 < D < 6$

We want to apply the general considerations presented in the previous Section 3.2.1 to the $O(2N)$ invariant $\phi^4$-theory between two and six dimensions. As we will show, it is possible to compute the effective potential $\Omega(\mu)$ along the flow for any value of the quartic coupling $g$ and the mass $r$, and use it to derive the one-loop effective potential of the theory.

Without setting the coupling to the fixed point value $|g| \to \infty$ we start our discussion with the action in Eq. (3.38) in terms of a collective DoF for the $\varphi^4$-theory along the flow from the free to the interacting fixed point,

$$S_\mu[\phi_i, \sigma] = \int_{S^1_\beta \times T_L^{D-1}} \mathrm{d}^D x \left[ (\partial_\tau - \mu_i)\phi_i^* (\partial_\tau + \mu_i)\phi_i + |\nabla \phi_i|^2 + (r + \sigma)|\phi_i|^2) - \frac{N}{4g}\sigma^2 \right], \tag{3.111}$$

Instead of the cylinder, for our purposes it suffices to consider consider the theory in flat space $S^1_\beta \times T_L^{D-1}$, where $T_L^{D-1}$ is the torus of length $L$ and volume $V = L^{D-1}$, as there is a priori no reason to put the theory on the cylinder away from the fixed point.[32]

Depending on the dimension $D$ of spacetime, the quartic interaction — now replaced by a quadratic interaction in the collective field $\sigma$ — from the point-of-view of the free theory is either relevant or irrelevant,

- Between spacetime dimensions $2 < D < 4$, and in particular in $D = 3$, the quartic operator is relevant. After fine-tuning the mass $r$ to the conformal mass the model flows from a free Gaussian fixed point in the UV ($g = 0$) to a strongly-coupled interacting WF fixed point in the IR at $g \to \infty$.
- In $4 < D < 6$ the quartic operator is irrelevant and, after fine-tuning $r$ to the conformal mass, the expectation is that the RG flow connects a IR free Gaussian CFT ($g = 0$) to a strongly-coupled interacting fixed point in the UV ($g \to \infty$).

First we remark that the expression for the grand potential at the WF fixed point that we have discussed in the context of large-charge asymptotics in Section 3.1.3 is obtained from the grand potential $\Omega(\mu)$ along the RG flow in the limit $|g| \to \infty$ with $r$ being fine-tuned to the conformal mass $r_{\text{conf}}$ (zero in flat space and $1/4r_0^2$ on the cylinder), $\Omega(\mu)\big|_{\text{WFFP}} = \lim_{g\to\infty} \lim_{r \to r_{\text{conf}}} \Omega(\mu)$.

We derive the general form of the grand potential. At leading order in $N$ the collective field $\sigma$ does not fluctuate and we can replace it in the action by its VEV,

$$S_\mu[\phi_i, \langle\sigma\rangle] = \int_{S^1_\beta \times T^{D-1}_L} \mathrm{d}^D x \left[ (\partial_\tau - \mu_i)\phi_i^* (\partial_\tau + \mu_i)\phi_i + |\nabla\phi_i|^2 + (r + \langle\sigma\rangle)|\phi_i|^2) - \frac{N}{4g}\langle\sigma\rangle^2 \right]. \tag{3.112}$$

[32] Alternatively, we could also move to infinite space $S^1_\beta \times \mathbb{R}^{D-1}$ without any conceptual hurdles. However, it is also convenient to explicitly see factors of the volume $V$ in our computations and results.

The appearance of the running term $-\frac{N}{4g}\langle\sigma\rangle^2$ poses no obstruction to the analysis outlined in Section 3.1.3 and Appendix C.2. At zero temperature we find that the grand potential is given by

$$\Omega(\mu_1, \dots) = \frac{1}{(2N)} \sum_i \zeta_i^2 \left( r + \langle\sigma\rangle - \mu_i^2 \right) + \frac{1}{2V} \zeta(-1/2 \,|\, T_L^{D-1}, r + \langle\sigma\rangle) - \frac{1}{8g}\langle\sigma\rangle^2 . \tag{3.113}$$

where $V = L^{D-1}$ is the volume of the torus. We note that the zeta function on the sphere in Eq. (3.47) is replaced here by the same zeta function evaluated on the torus. The saddle-point equations in Eq. (3.49) are modified slightly to incorporate the running term,

$$\begin{aligned} \zeta_i &: \quad \zeta_i \left( r + \langle\sigma\rangle - \mu_i^2 \right) = 0, \quad i = 1, \dots, N, \\ \langle\sigma\rangle &: \quad \frac{1}{(2N)} \sum_i \zeta_i^2 + \frac{1}{2V}\frac{1}{2}\zeta(1/2 \,|\, T_L^{D-1}, r + \langle\sigma\rangle) - \frac{1}{4g}\langle\sigma\rangle = 0, \\ \mu_i &: \quad \frac{\rho_i}{(2N)} - \frac{2\mu_i}{(2N)}\zeta_i^2 = 0. \end{aligned} \tag{3.114}$$

Despite the appearance of the flow term, the saddle point equations are solved equivalently to the ones in Eq. (3.49) and the grand potential and free energy are given by

$$\Omega(\mu) = \frac{1}{2V}\zeta(-1/2 \,|\, T_L^{D-1}, r + \langle\sigma\rangle) - \frac{1}{8g}\langle\sigma\rangle^2, \qquad f_c(\rho) = \mathrm{LT}[-\Omega](\rho), \tag{3.115}$$

with $\mu = \mu_1 = \dots = \mu_N$ and $\rho = (Q_1 + \dots + Q_N)/V$. It is convenient to use the Mellin representation of the zeta function in Eq. (3.57) and apply Weyl's asymptotic expansion in Eq. (3.58) on the torus,

$$\mathrm{Tr}\left[ e^{t\Delta_{T_L^{D-1}}} \right] = \frac{V}{(4\pi t)^{\frac{D-1}{2}}} + \mathcal{O}\left(e^{-\frac{L^2}{4t}}\right), \qquad \zeta(s \,|\, T_L^{D-1}, \mu^2) = \frac{V\,\Gamma(s - \frac{D-1}{2})}{(4\pi)^{\frac{D-1}{2}}\,\Gamma(s)} \mu^{D-1-2s}, \tag{3.116}$$

so that the grand potential can be written as

$$\Omega(\mu) = -\left[ \frac{\Gamma(-\frac{D}{2})}{2(4\pi)^{D/2}} \mu^D + \frac{(\mu^2 - r)^2}{8g} \right]. \tag{3.117}$$

If we fine-tune $r$ to the conformal mass, $r = 0$ in flat space, we expect this function to describe the flow of the theory from the Gaussian to the strongly-coupled fixed point at leading order in $N$. The only non-trivial dependence on the spacetime is in the function $\Gamma(-D/2)$ in the first term in Eq. (3.117). A plot of $\Gamma(-D/2)$ for relevant spacetime dimensions can be seen in Figure 3.1. It is positive for $4n - 2 < D < 4n$ and negative for $4n < D < 4n + 2$, while it diverges at $D = 2n$, $n \in \mathbb{N}$, consistent with the fact that there exists no WF fixed point in even dimensions $D = 2, 4$. We can now apply the technology developed in the previous Section to derive an expression for the leading-order effective potential $V_{\text{eff}}^{(0)}$ in terms of $2N$ real classical fields packaged into $N$ complex classical fields $\phi_i^{(c)}$.[33] $O(2N)$ invariance implies that the effective potential only depends on the invariant combination $\phi_i^{(c)*}\phi_i^{(c)}$, which we will indicate as $|\phi_i^{(c)}|^2$.

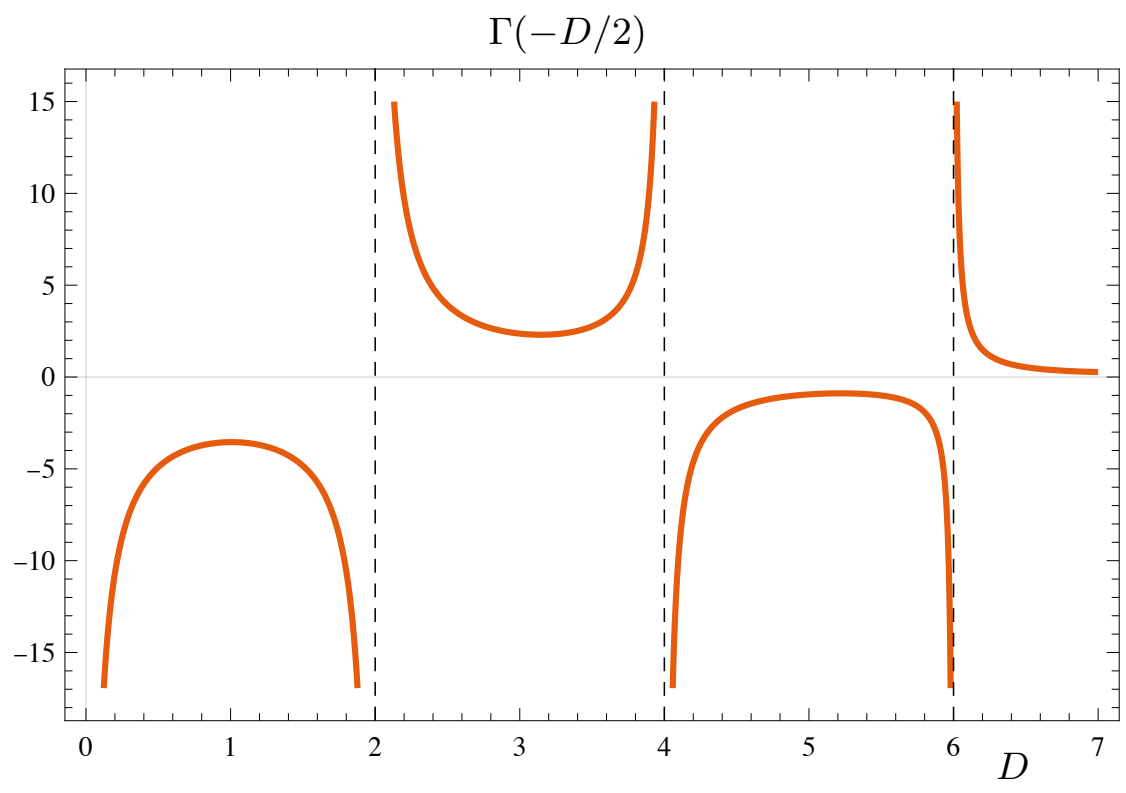

Figure 3.1: The function $\Gamma(-D/2)$ from the leading term in the grand potential on the torus for $0 < D < 7$. The function is positive for $2 < D < 4$, $6 < D < 8$ and negative for $4 < D < 6$.

$$|\phi_i^{(c)}|^2 := \sum_i \phi_i^{(c)*}\phi_i^{(c)}. \tag{3.118}$$

Lastly, following the chain of transformations in Eq. (3.106), we can turn the convexity of the effective potential into a consistency condition on $\varpi(\mu^2) = -\Omega(\mu)$ by identifying any possible inflection points of $\varpi(\mu^2)$,

$$\varpi''(\mu^2) = \frac{D(D-2)}{8(4\pi)^{D/2}}\Gamma(-D/2)\left(\mu^2\right)^{\frac{D-4}{2}} + \frac{1}{4g}. \tag{3.119}$$

This can only be zero for $\Gamma(-D/2) < 0$, which is satisfied in $4 < D < 6$ but not in $2 < D < 4$, see Figure 3.1. In the following we will focus on these particular intervals — $2 < D < 4$ and $4 < D < 6$ — as they are generally of most interest in physics.

### 3.2.3 2<D<4

In spacetime dimensions $2 < D < 4$ the gamma function in Eq. (3.117) is positive and hence $\varpi(\mu^2)$ is always convex for all values of $\mu^2$.[34] The theory fulfils our necessary condition for unitarity, and in $D = 3$ the effective potential can be computed explicitly [128]. In $D = 3$ the grand potential reads

$$\Omega(\mu) = -\left[\frac{\mu^3}{12\pi} + \frac{(\mu^2 - r)^2}{8g}\right]. \tag{3.120}$$

[33]We emphasize again that $V_{\text{eff}}^{(0)}$ is determined by the condition that at fixed charge it must reproduce the physics described by $\Omega(\mu)$ in the double-scaling limit [128].

[34]The same statement is not necessarily true for the grand potential $\Omega(\mu)$ itself, as $-\Omega''(\mu) = \frac{D(D-1)\Gamma(-D/2)}{2(4\pi)^{D/2}}\mu^{D-2} + \frac{(12\mu^2-4r)}{8g}$. In $D = 3$ for $r > 0$ this has the positive solution $\mu = (\sqrt{g^2 + 12\pi^2 r} - g)/6\pi$. If we take a look at the inverse Legendre transform in $\mu$, then for $r > 0$ there are regions that cannot be reached when starting from a fixed-charge description. This is unsurprising, as we expect that the fixed-charge regime is generally different from the fixed-chemical-potential regime [47, 128].

We evaluate the constraint of the Legendre transform between $\Upsilon(|\phi_i^{(c)}|^2)$ and $\varphi(\mu^2) = -\Omega(\mu)$,

$$\frac{|\phi_i^{(c)}|^2}{2N} - \varpi'(\mu^2) = \frac{|\phi_i^{(c)}|^2}{2N} - \left[\frac{\mu}{8\pi} + \frac{\mu^2 - r}{4g}\right] = 0 . \tag{3.121}$$

Depending on the value of $r$ there are two regimes here:

- For $r \geq 0$ there is a unique solution given by

$$\mu^2\Big(\frac{|\phi_i^{(c)}|^2}{2N}\Big) = \left(\frac{g}{4\pi}\right)^2 \left[1 + \sqrt{1+\eta}\right]^2 , \qquad \eta = \left(\frac{16\pi}{g}\right)^2 \left[\frac{g|\phi_i^{(c)}|^2}{2N} + \frac{r}{4}\right]. \tag{3.122}$$

  Therefore, the leading-$N$ effective potential is given by

$$\frac{V_{\text{eff}}^{(0)}(|\phi_i^{(c)}|)}{2N} = \frac{g^3}{3\times 2^8 \pi^4}\left[1 + \frac{3}{2}\eta + \frac{3}{8}\eta^2 - (1+\eta)^{3/2}\right] - \frac{r^2}{8g} . \tag{3.123}$$

- For $r < 0$ there exists a solution only for $\frac{|\phi_i^{(c)}|^2}{2N} > -\frac{r}{4g} > 0$. For smaller values of the classical fields the supremum is obtained for the value of $\mu^2$ that minimizes $\varpi(\mu^2)$, which is $\mu^2 = 0$. For small values of the classical fields the effective potential becomes constant. For big values we recover the same form as above,

$$\frac{V_{\text{eff}}^{(0)}(|\phi_i^{(c)}|)}{2N} = \begin{cases} -\frac{r^2}{8g} & \text{for } 0 < \frac{|\phi_i^{(c)}|}{\sqrt{2N}} < \sqrt{-\frac{r}{4g}} , \\ \frac{g^3}{3\times 2^8\pi^4}\left[1 + \frac{3}{2}\eta + \frac{3}{8}\eta^2 - (1+\eta)^{3/2}\right] - \frac{r^2}{8g} & \text{for } \frac{|\phi_i^{(c)}|}{\sqrt{2N}} > \sqrt{-\frac{r}{4g}} . \end{cases} \tag{3.124}$$

The regions $r \lesseqgtr 0$ correspond to the broken and unbroken phases of the tree-level potential

$$V(|\phi_i|) = r|\phi_i|^2 + \frac{g}{N}|\phi_i|^4 . \tag{3.125}$$

Particularly interesting is the broken phase here, where the tree-level potential here has the form of the Mexican hat potential with minima,

$$\frac{|\phi_i|^2}{2N} = -\frac{r}{4g} . \tag{3.126}$$

The double-well shaped tree-level potential has a flex and ceases to be convex in-between its minima. The quantum corrections in the effective potential have to correct this, as it is well-understood that the effective potential is always convex, even in finite volume [258, 259]. We can confirm this here based on the Legendre transform preserving convexity, as the effective potential becomes constant between minima and washes out the concave region of the tree-level potential.[35] [36]

[35] This is very analogous to the classical Maxwell rule for coexisting phases in thermodynamics [128, 260]. As a consequence, the effective potential has a cusp in its second derivative at the location of the tree-level minimum.

[36] It is natural to wonder now what the acceptable vacua of the theory in the context of SSB are, as it is no longer true that all of the minima (and hence the vacua) of the effective potential are connected by a symmetry. A strong argument for the actual vacua of the theory to still be at the location of the minima of the tree-level potential can be put forward using cluster decomposition [128, 260].

For the critical trajectory at $r = 0$ connecting the UV Gaussian fixed point with the IR WF fixed point this result for the effective potential at leading order in $N$ had originally been found in [19] by re-summing an infinite number of Feynman diagrams that we managed to avoid.
The expression for the effective potential in Eq. (3.123) at $r = 0$ becomes more transparent, if we expand it in the limits $g \to 0, \infty$:

- In the limit $g \to 0$ Eq. (3.123) reproduces the standard loop expansion at large $N$ around the Gaussian fixed point [19],

$$V_{\text{eff}}^{(0)}(|\phi_i^{(c)}|) = \frac{g}{N}|\phi_i^{(c)}|^4 \left[1 - \frac{1}{3\pi}\frac{\sqrt{2Ng}}{|\phi_i^{(c)}|} + \mathcal{O}(g)\right]. \tag{3.127}$$

- For large values of the coupling around $g \to \infty$ Eq. (3.123) produces a perturbative expansion in $1/g$ around the strongly-coupled WF point,

$$V_{\text{eff}}^{(0)}(|\phi_i^{(c)}|) = \frac{16\pi^2}{3N^2}|\phi_i^{(c)}|^6 \left[1 - 24\pi^2\frac{|\phi_i^{(c)}|^2}{2Ng} + \mathcal{O}\left(g^{-2}\right)\right]. \tag{3.128}$$

As a last remark, we emphasize that the condition of convexity is only a necessary condition for unitarity. For example, event though convexity is always fulfilled for $2 < D < 4$, it appears that the theory in $D = 4 - \epsilon$ is not unitary [263], despite the fact that the effective potential is convex there.

### 3.2.4 4<D<6

In spacetime dimensions $4 < D < 6$ the gamma function in Eq. (3.117) is negative and the grand potential $\Omega(\mu) = \varpi(\mu^2)$ has a flex for positive values of $\mu^2$,

$$\varpi''(\mu^2_{\text{flex}}) = 0, \qquad \text{for} \qquad (\mu^2_{\text{flex}})^{(D-4)/2} = \frac{2(4\pi)^{D/2}}{gD(D-2)\,|\Gamma(-D/2)|}. \tag{3.129}$$

This flex separates a convex region of the grand potential at small values of $g$ around the IR free Gaussian fixed point from a concave region around the conjectured strongly-coupled UV fixed point, see Figure 3.2. We can express the location of the flex in terms of the classical fields $|\phi_i^{(c)}|^2$,

$$\frac{|\phi_i^{(c)}|^2_{\text{flex}}}{2N} = \frac{(D-4)(4\pi)^{\frac{D}{D-4}}}{4\left(\frac{D}{2}\,|\Gamma(-\frac{D}{2})|\right)^{\frac{2}{D-4}}(D-2)^{\frac{D-2}{D-4}}}\, g^{-\frac{D-2}{D-4}} - \frac{r}{4}g^{-1}. \tag{3.130}$$

Here, the working definition of the Legendre transform in terms of the derivative leads to a multi-variable complex function. We can avoid this only by using the supremum definition of the Legendre transform in Eq. (3.107). For $|\phi_i^{(c)}|^2 < |\phi_i^{(c)}|^2_{\text{flex}}$ the supremum can still be obtained via the working definition consisting of differentiating the argument $|\phi_i^{(c)}|^2\mu^2/2N - \varpi(\mu^2)$ at fixed $|\phi_i^{(c)}|^2$ and expressing $\mu^2$ as a function of $|\phi_i^{(c)}|^2$.
For $|\phi_i^{(c)}|^2 > |\phi_i^{(c)}|^2_{\text{flex}}$, however, there is no real value of $\mu^2$ such that $|\phi_i^{(c)}|^2 = \varpi'(\mu^2)$, see Figure 3.2. The supremum in this range is obtained by minimizing $\varpi(\mu^2)$. However, as the function $\varpi(\mu^2)$ is not

bounded from below, the solution is $\{+\infty\}$.[37]

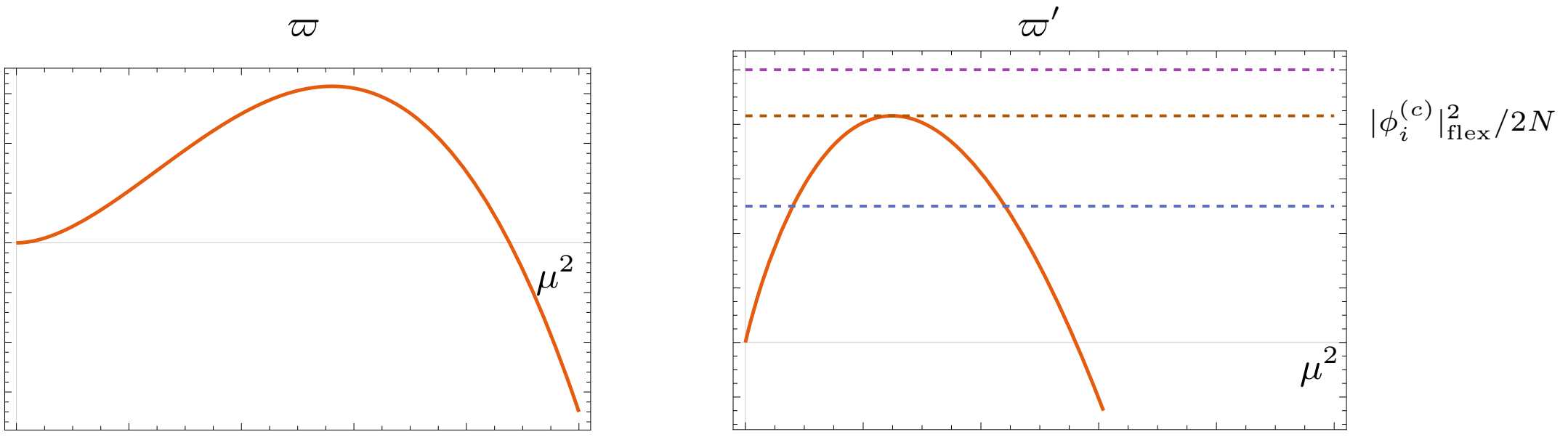

Figure 3.2: The function $\varpi(\mu^2)$ (left) and its first derivative $\varpi'(\mu^2)$ (right) in flat space (on the torus) for $D = 5$. The grand potential $\Omega(\mu) = \varpi(\mu^2)$ has a flex at $\mu^2 = \mu^2_{\text{flex}}$, which separates a convex region $\{\mu^2 < \mu^2_{\text{flex}}\}$ from a concave region $\{\mu^2 > \mu^2_{\text{flex}}\}$. The maximization condition $|\phi_i^{(c)}|^2_{\text{flex}}/2N = \varpi'(\mu^2)$ admits two solutions in the convex region $|\phi_i^{(c)}|^2/2N < |\phi_i^{(c)}|^2_{\text{flex}}/2N = \varpi'(\mu^2_{\text{flex}})$ — the physical branch corresponds to the first intersection (right) — and no real solutions for $|\phi_i^{(c)}|^2/2N > |\phi_i^{(c)}|^2_{\text{flex}}/2N$.

Further, for $r < 0$ — the broken phase — the tree-level potential again takes the form of a Mexican hat potential. Via the supremum definition of the Legendre transform, the non-convex region of the tree-level potential is washed out by the leading-order effective potential $V^{(0)}_{\text{eff}}$, which takes a constant value there. All in all, we have

$$\Upsilon\Big|_{r\geq 0} = \begin{cases} \frac{|\phi_i^{(c)}|^2}{2N}\mu^2 - \varpi(\mu^2) & \text{for } \frac{|\phi_i^{(c)}|^2}{2N} < \frac{|\phi_i^{(c)}|^2_{\text{flex}}}{2N}, \\ +\infty & \text{for } \frac{|\phi_i^{(c)}|^2}{2N} \geq \frac{|\phi_i^{(c)}|^2_{\text{flex}}}{2N}, \end{cases} \qquad \Upsilon\Big|_{r<0} = \begin{cases} -\frac{r^2}{8g} & \text{for } 0 < \frac{|\phi_i^{(c)}|^2}{2N} < -\frac{r}{4g}, \\ \frac{|\phi_i^{(c)}|^2}{2N}\mu^2 - \varpi(\mu^2) & \text{for } -\frac{r}{4g} < \frac{|\phi_i^{(c)}|^2}{2N} < \frac{|\phi_i^{(c)}|^2_{\text{flex}}}{2N}, \\ +\infty & \text{for } \frac{|\phi_i^{(c)}|^2}{2N} \geq \frac{|\phi_i^{(c)}|^2_{\text{flex}}}{2N}, \end{cases} \tag{3.131}$$

Concretely, we consider the case $D = 5$,

$$\Omega(\mu) = \varpi(\mu^2) = \frac{\mu^5}{120\pi^2} - \frac{(\mu^2 - r)^2}{8g}, \tag{3.132}$$

where the flex of $\varpi(\mu^2)$ is given by

$$\mu^2_{\text{flex}} = \frac{(4\pi)^4}{4g^2}, \qquad \frac{|\phi_i^{(c)}|^2_{\text{flex}}}{2N} = \frac{(4\pi)^4}{48g^3} - \frac{r}{4g}. \tag{3.133}$$

As a consequence, the large-$N$ effective potential $V^{(0)}_{\text{eff}}$ describes a box with infinitely high walls and width

$$2|\phi_i^{(c)}|_{\text{flex}}\Big/\sqrt{2N} = 2\frac{(4\pi)^2}{4\sqrt{3}\,g^{3/2}}\sqrt{1 - \frac{12rg^2}{(4\pi)^4}}. \tag{3.134}$$

If we follow the RG flow along the critical trajectory $r = 0$ in reverse order from the IR free theory towards the UV, we can observe the walls closing in on each other. Hence, the UV interacting CFT cannot be reached. This signals a breakdown of the effective theory along the RG flow $g \to \infty$; the theory is incomplete and requires a UV completion.[38] The effective potential $V_{\text{eff}}^{(0)}$ in $D = 5$ is schematically presented in Figure 3.3.

[37] Technically speaking, we are using the notion of convex conjugate, which is defined on the extended real line $\mathbb{R} \cup \{\pm\infty\}$).

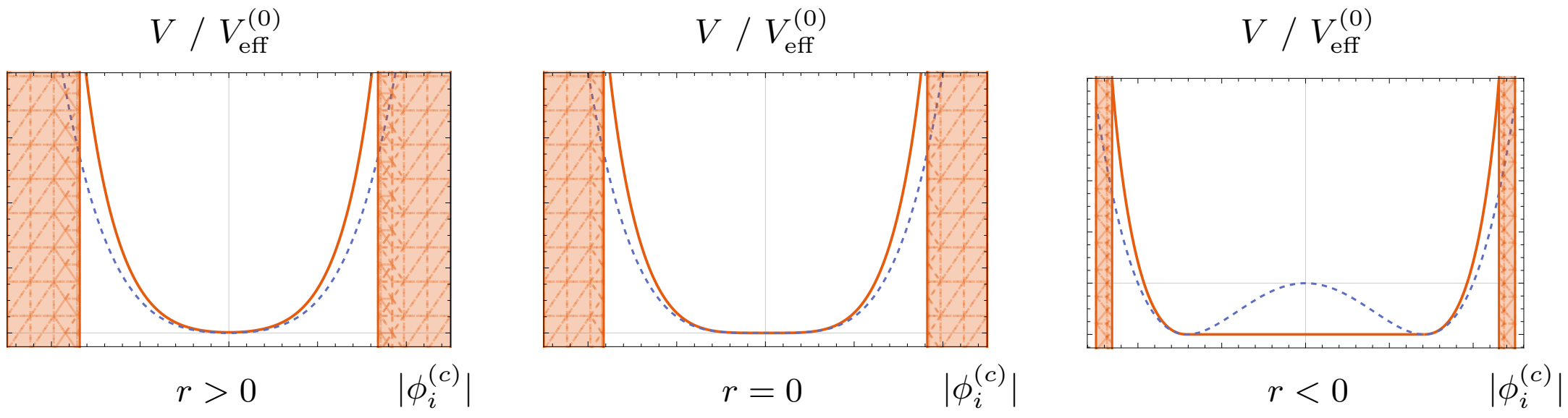

Figure 3.3: Effective potential $V_{\text{eff}}^{(0)}$ in $D = 5$: unbroken phase ($r > 0$), critical trajectory ($r = 0$) and broken phase ($r < 0$). The red lines represent the effective potential $V_{\text{eff}}^{(0)}$, which is always convex, while the dotted lines represent the tree-level potential $V$. Inside the shaded region the effective potential $V_{\text{eff}}^{(0)}$ is literally infinite, which signals the breakdown of the theory and the lack of a UV completion. Between the shaded regions the behaviour of the effective potential is qualitatively the same as in $D = 3$, discussed in Section 3.2.3.

Inside the box, the effective potential $V_{\text{eff}}^{(0)}$ can be computed and expressed in terms of trigonometric functions,

$$\begin{aligned}\frac{V_{\text{eff}}^{(0)}(|\phi_i^{(c)}|)}{2N} = \frac{2^{10}\pi^8}{5u^5}\Big(4\sin(\frac{10\theta+\pi}{6}) + 20\sin(\frac{4\theta+\pi}{6}) + 10\cos(\frac{4\theta+\pi}{3}) + 20\cos(\frac{\theta+\pi}{3}) \\ - 20\cos(\theta) - 7\Big) - \frac{r^2}{8g},\end{aligned} \tag{3.135}$$

where

$$\cos(\theta) = 1 - 2\frac{|\phi_i^{(c)}|^2}{|\phi_i^{(c)}|^2_{\text{flex}}} - \frac{24 r g^2}{(4\pi)^4}\left(1 - \frac{|\phi_i^{(c)}|^2}{|\phi_i^{(c)}|^2_{\text{flex}}}\right). \tag{3.136}$$

If we expand around the IR free Gaussian fixed point at $g \to 0$ and $r \to 0$, this result becomes more transparent and accurately captures all the (infinitely many) leading-$N$ corrections to the quartic tree-level potential,

$$\begin{aligned}V_{\text{eff}}^{(0)}(|\phi_i^{(c)}|) = \left(|\phi_i^{(c)}|^2 r + \frac{N r^{5/2}}{60\pi^2}\dots\right) + \frac{g}{N}\left(|\phi_i^{(c)}|^4 + \frac{|\phi_i^{(c)}|^2 r^{3/2}}{12N\pi^2} + \frac{r^3}{576\pi^4}\dots\right) \\ + g^2\left(\frac{|\phi_i^{(c)}|^4\sqrt{r}}{8N\pi^2} + \frac{|\phi_i^{(c)}|^2 r^2}{96\pi^4}\dots\right) + \dots .\end{aligned} \tag{3.137}$$

Along the critical trajectory $r = 0$ this result reduces to an expansion in the only dimensionless

[38] We note that away from criticality along a massive trajectory $r > 0$ the inflection point $|\phi_i^{(c)}|^2_{\text{flex}} 2N$ will eventually become imaginary.

combination of quantities within the theory given by $g^{3/2}|\phi_i^{(c)}|$,

$$V(|\phi_i^{(c)}|) = \frac{g}{N}|\phi_i^{(c)}|^4\left[1+\frac{4}{5}\left(\frac{g^{3/2}|\phi_i^{(c)}|}{6\pi^2\sqrt{2N}}\right)+\left(\frac{g^{3/2}|\phi_i^{(c)}|}{6\pi^2\sqrt{2N}}\right)^2+\frac{3}{2}\left(\frac{g^{3/2}|\phi_i^{(c)}|}{6\pi^2\sqrt{2N}}\right)^3+\frac{5}{2}\left(\frac{g^{3/2}|\phi_i^{(c)}|}{6\pi^2\sqrt{2N}}\right)^4+\dots\right]. \tag{3.138}$$

This expansion also has an interpretation in terms of Feynman diagrams computed around the IR free fixed point.

Under the assumption that the supremum definition of the Legendre transform is the correct one, the effective potential in the far UV at $g\to\infty$ becomes infinite everywhere and the theory is in need of a UV completion. Alternatively, it is possible that the UV theory violates unitarity and hence the effective potential does not need to be convex, allowing for observables to have non-zero imaginary parts. If we instead decide to extend the definition of the Legendre transform relating the effective potential to the grand potential in order to allow for a complex solution of the maximization condition

$$|\phi_i^{(c)}|^2/2N = \varpi'(\mu^2), \tag{3.139}$$

we find a branch point at $|\phi_i^{(c)}|^2 = |\phi_i^{(c)}|^2_{\text{flex}}$. The maximization equation describes a Riemann surface and the branch point at $|\phi_i^{(c)}|^2_{\text{flex}}$ is joined to infinity by a branch cut. We can choose one branch and expand the complex effective potential in the UV around the location of the conjectured strongly-coupled UV CFT at $g\to\infty$,

$$V_{\text{eff}}^{(0)}(|\phi_i^{(c)}|) = \frac{12}{5}\left(\frac{3\pi^2}{N}\right)^{\frac{2}{3}} e^{\frac{2\pi i}{3}}|\phi_i^{(c)}|^{\frac{10}{3}}\left[1+\left(\frac{N\pi^4}{3}\right)^{\frac{1}{3}}\frac{5e^{\frac{\pi i}{3}}}{|\phi_i^{(c)}|^{\frac{2}{3}}g}+\left(\frac{N\pi^4}{3}\right)^{\frac{2}{3}}\frac{20e^{\frac{2\pi i}{3}}}{|\phi_i^{(c)}|^{\frac{4}{3}}g^2}-\frac{200N\pi^4}{9|\phi_i^{(c)}|^2g^3}+\mathcal{O}\left(g^{-4}\right)\right]. \tag{3.140}$$

The effective potential is a multi-valued complex function with three branches. Each of the different branches corresponds to a different choice of phase in the coefficient of the leading $|\phi_i^{(c)}|^{10/3}$-term in Eq. (3.140).[39]

### 3.2.5 The strongly-coupled fixed point in D = 5

We investigate the UV interacting fixed point of the $\varphi^4$-theory in $D=5$ via the LCE. Based on the results of the previous section, we expect the theory to behave distinctly different to the $O(2N)$ WF fixed point in $D=3$.

As discussed extensively in both Section 3.1.2 and Section 3.2.1, the free energy (density) — the energy of the system at fixed charge — is obtained via a Legendre transform

$$f_c(Q) = \sup_{\mu>0}\left(\frac{Q}{2N}\mu + V\Omega(\mu)\right), \qquad \left(f_c(\rho) = \sup_{\mu>0}\left(\frac{\rho}{2N}\mu+\Omega(\mu)\right)\right). \tag{3.141}$$

[39]The relative phases in the coefficients of the sub-leading terms will change as well. Also note that one of the three branches corresponds to a trivial phase of $e^{2\pi i}=1$ in the coefficient of the leading term. This still represents a complex solution, since on this branch sub-leading terms in the coupling $g$ will again come with non-trivial phases.

Clearly, as the Legendre transform preserves homogeneity (see Eq. (3.110)), the free energy density as a function of the charge density $f_c(\rho)$ has the same functional form as the free energy as a function of the charge $f_c(Q)$.

Via the state–operator correspondence, at the fixed points the free energy on the sphere can be identified with and computes the scaling dimension of the associated operator in flat space,[40]

$$\Delta(Q) = r_0 f_c(Q)\,. \tag{3.142}$$

We fine-tune the $\varphi^4$-theory to flow to the interacting fixed point by setting $r$ to the conformal mass (on the cylinder we have $r_{\text{conf}} = 1/4r_0^2$) and $g \to \infty$. The running term in Eq. (3.115) vanishes at the interacting fixed point and the grand potential on the cylinder is given in terms of a zeta function for the Laplacian on the sphere,

$$\Omega(\mu) = \frac{1}{2V}\zeta(-1/2\,|\,S^{D-1}_{r_0}, \mu^2)\,, \tag{3.143}$$

where $V$ is the volume of the sphere. In odd spacetime dimensions $D$ there exists both a convergent series expansion around the conformal mass $r_{\text{conf}}$ for small values of $\mu$ and an asymptotic expansion for large values of $\mu$.

Convexity of $\Omega(\mu)$ and convexity of $\varpi(\mu^2)$ — in terms of their second derivatives – are non-trivially related. Importantly, the existence of non-convex regions of $\Omega(\mu)$ does not signal a breakdown of the theory, in contrast to $\varpi(\mu^2)$. There always exists a proper expression for the free energy $f_c(Q) = \text{LT}[-V\Omega](Q)$ that is convex as a function of the charge $Q$, with $Q$ assumed to be positive. At worst, the Legendre transform just tells you that it is infinite.

Regions in which $\Omega(\mu)$ is not convex cannot be reached from the fixed-charge regime (see *e.g.* [47]). Such concave regions of $\Omega(\mu)$ can exist also in well-defined theories [128]. Even if $\Omega(\mu)$ is convex everywhere, the existence of regions that cannot be accessed at fixed charge cannot be ruled out; the charge is positive and $Q = -V\Omega'(\mu)$ has no solutions for values of $\mu$ where $-\Omega(\mu)$ is decreasing. This is *e.g.* the case in the WF CFT, where the boundary of the accessible region — with $-\Omega'(\mu) > 0$ — is at the conformal mass $r = 1/4r_0^2$.

**Interlude: The WF CFT in D = 3**

We first review the strongly-coupled fixed point in $D = 3$, which we have extensively discussed in Section 3.1. In full, the zeta function in $D = 3$ cannot be expressed in elementary functions. At large $\mu$ it can be written as a Mellin transform of the heat kernel and expanded in terms the asymptotic

[40]Alternatively, we can derive the conformal dimensions directly from the effective potential via the Callan–Symanzik equations, see *e.g.* [264].

expansion of said heat kernel [124, 191],

$$\zeta(s\,|\,S^2_{r_0},\mu^2) = \frac{1}{\Gamma(s)}\int_0^\infty \frac{\mathrm{d}t}{t^{1-s}}\, e^{-m^2 t}\mathrm{Tr}\Big[e^{t\Delta_{S^2_{r_0}}}\Big], \qquad \mathrm{Tr}\Big[e^{t\left(\Delta_{S^2_{r_0}}-\frac{1}{4r_0^2}\right)}\Big] \sim \frac{r_0^2}{t} - \sum_{n=1}^{\infty}\frac{(1-2^{1-2n})B_{2n}}{(-1)^n n!}\left(\frac{t}{r_0^2}\right)^{n-1}. \tag{3.144}$$

The large-$\mu$ expansion for the grand potential then reads [191]

$$\Omega(\mu) = -\frac{1}{r_0 V}\left(\mu^2 r_0^2 - \frac{1}{4}\right)^{3/2}\sum_{n=0}^{\infty}\frac{\Omega_n}{\left(\mu^2 r_0^2 - \frac{1}{4}\right)^n}, \qquad \Omega_n = \frac{1}{4\pi}\sum_{k\neq 0}\frac{(-1)^{k+1}}{(k\pi)^{2n}}\Gamma\left(n+\frac{1}{2}\right)\Gamma\left(n-\frac{3}{2}\right). \tag{3.145}$$

In the opposite limit of small $\mu$ a convergent expansion can be derived from the definition of the zeta function in terms on the eigenvalues of the Laplacian, see Eq. (3.63),

$$\Omega(\mu) = \frac{r_0^{2s}}{V}\sum_{k=0}^{\infty}\binom{-s}{k}\zeta(2s+2k-1;1/2)\left(m^2 r_0^2 - \frac{1}{4}\right)^k\Bigg|_{s=-1/2}, \tag{3.146}$$

where $\zeta(s;a)$ denotes the Hurwitz zeta function. From the explicit form of the convergent expansion we can deduce that $-\Omega(\mu)$ is always convex and has a minimum at the value of the conformal mass, where it also vanishes itself,

$$\Omega(\mu)\Big|_{\mu=\frac{1}{2r_0}} = 0, \qquad\qquad \Omega'(\mu)\Big|_{\mu=\frac{1}{2r_0}} = 0. \tag{3.147}$$

The Legendre transform can be performed order-by-order, as done in Section 3.1.4. In doing so, we find a solution to the maximization condition $Q = -V\Omega'(\mu)$ for positive values of $Q$ in the region $\mu^2 > 1/4r_0^2$, which is precisely the conformal mass (see Figure 3.4). We quickly repeat the results from Section 3.1.4,

$$\frac{\Delta(Q)}{2N} = r_0 f_c(Q) = \frac{2}{3}\left(\frac{Q}{2N}\right)^{\frac{3}{2}} + \frac{1}{6}\left(\frac{Q}{2N}\right)^{\frac{1}{2}} + \ldots, \qquad \frac{\Delta(Q)}{2N} = r_0 f_c(Q) = \frac{1}{2}\left(\frac{Q}{2N}\right) + \frac{4}{\pi^2}\left(\frac{Q}{2N}\right)^2 + \ldots. \tag{3.148}$$

As discussed in Section 3.1.5, a resurgent analysis can be performed, interpolating between the two regions [191].

#### The interacting fixed point in D = 5: the zeta function on the four sphere

As we have seen in the previous section, in $D = 5$ the effective potential $V^{(0)}_{\text{eff}}$ obtained via a Legendre transform from the grand potential $\varpi(\mu^2)$ exhibits an inaccessible region that hides the expected strongly-coupled UV fixed point. Based on this observation we expect that — if it exists — the conjectured UV CFT at the very least violates unitarity, if not worse. We can learn more about the spectrum of scaling dimensions in this five-dimensional non-unitary CFT using the LCE technology developed in Section 3.1 originally for the WF fixed point in $D = 3$, which we have outlined again above.

We start by analysing the trace of the heat kernel on the four-sphere $S^4_{r_0}$. The heat kernel traces on even-dimensional spheres are all related via a recursion relation to derivatives of the heat kernel on the

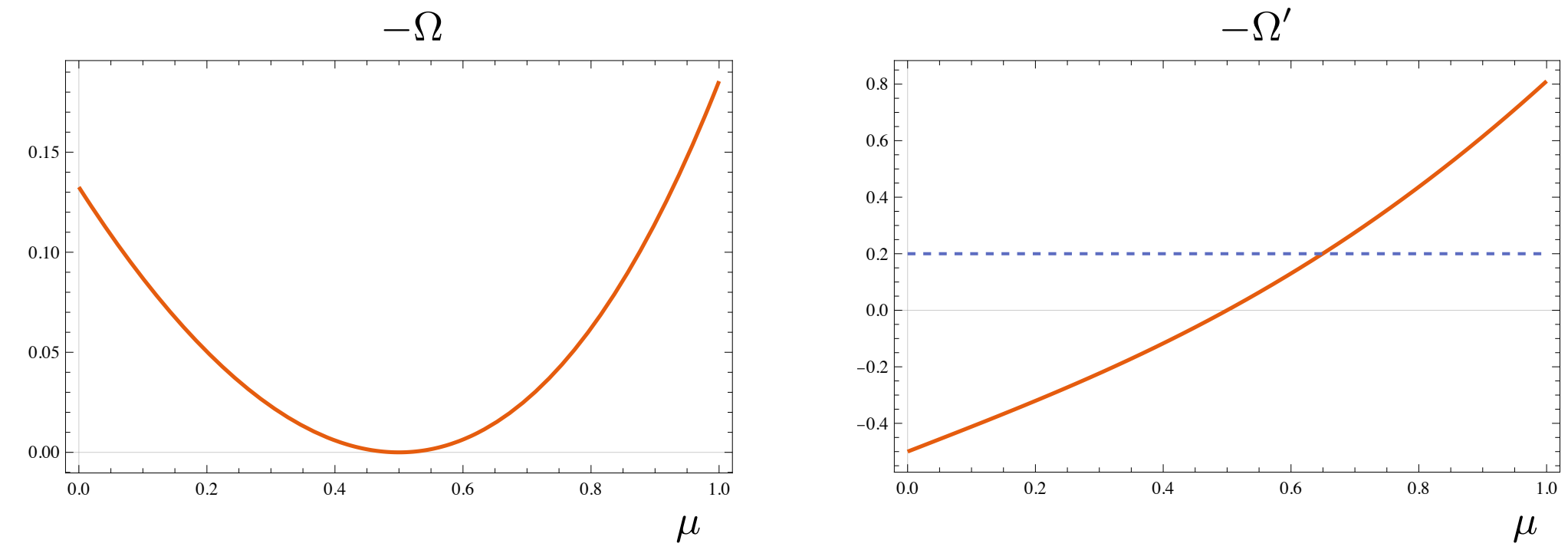

Figure 3.4: The grand potential $-\Omega(\mu)$ (left) and its first derivative $-\Omega'(\mu)$ (right) in $D=3$ computed on the two-sphere. Evidently, the function $-\Omega(\mu)$ is convex for all values of $\mu$. It has a minimum at $\mu = 1/2r_0$, where it vanishes. The maximization condition has a positive solution for $Q$ only in the region $\{\mu \geq 1/2r_0\}$.

two-sphere [265],[41]

$$\begin{aligned}\mathrm{Tr}[e^{\Delta_{S^{2n}_{r_0}} t}](t) &= \sum_{l\geq 0}\left[\frac{2l+2n-1}{2n-1}\prod_{k=1}^{2n-2}\frac{l+k}{k}\right] e^{-\frac{l(l+2n-1)t}{r_0^2}} \\ &= \frac{e^{(n-\frac{1}{2})^2\frac{t}{r_0^2}}}{(2n-1)!}\sum_{j=0}^{n-1}\beta_{j;n}(-1)^j r_0^{2j}\frac{\mathrm{d}^j g(t)}{\mathrm{d}t^j}\end{aligned}, \quad \text{where } g(t) = \mathrm{Tr}\left[e^{t(\Delta_{S^2_{r_0}}-\frac{1}{4r_0^2})}\right]. \tag{3.149}$$

The coefficients $\beta_{j;n}$ are defined via the relationship

$$\frac{2s}{(2n-1)!}\prod_{j=\frac{1}{2},\frac{3}{2},\dots}^{n-3/2}(s^2-j^2) = \frac{2s}{(2n-1)!}\sum_{j=0}^{n-1}\beta_{j;n}s^{2j}. \tag{3.150}$$

The relationship in Eq. (3.149) can be used to derive a similar formula for the zeta function on a generic even-dimensional sphere, relating the higher-dimensional zeta function to a sum of zeta functions on the two-sphere,[42]

$$\zeta(s\,|\,S^{2n}_{r_0}, m) = \frac{r_0^{2s}}{(2n-1)!}\sum_{j=0}^{n-1}\beta_{j;n}\sum_{k=0}^{j}\binom{j}{k}(-1)^k\left[\mu^2 r_0^2 - \left(n-\frac{1}{2}\right)^2\right]^{j-k} r_0^{2k}\zeta(s-k\,|\,S^2_{r_0}, m^2-\left(n^2-n\right)r_0^{-2}). \tag{3.151}$$

For large values of $\mu$, we can again expand the heat kernel trace in an asymptotic expansion for small values of $t$, as the Mellin integral localizes around $t=0$. For the four-sphere the relation in Eq. (3.149) reduces to

$$\mathrm{Tr}[e^{\Delta_{S^4_{r_0}} t}] = \frac{e^{\frac{9}{4r_0^2}t}}{6}\left[\beta_{0;2} - \beta_{1;2}r_0^2\frac{\mathrm{d}}{\mathrm{d}t}\right]g(t), \qquad \begin{aligned}\beta_{0;2} &= -\tfrac{1}{4} \\ \beta_{1;2} &= 1\end{aligned}. \tag{3.152}$$

[41] Similarly, heat kernel traces on odd-dimensional spheres can be related to the trace of the heat kernel on the circle [265].

[42] To derive this expression we need to perform an integration by parts in the Mellin integral. Hence, it needs to be suitably analytically continued for $\mu^2 r_0^2 < \left(n-\frac{1}{2}\right)^2$. The value $\left(n-\frac{1}{2}\right)^2$ is related to the conformal mass in $2n$ dimensions.

The term $9/4r_0^2$ is just the conformal mass on the cylinder in $D = 5$ (the four-sphere). For the asymptotic expansion on the four-sphere we then find that

$$\zeta(s\,|\,S^4_{r_0},\mu^2) = \frac{r_0^{2s}\left(\mu^2 r_0^2 - \frac{9}{4}\right)^{2-s}}{6(s-1)(s-2)} + \frac{r_0^{2s}}{24}\sum_{k\geq 0}\frac{(-1)^k\left(2^{2k+1}-1\right)B_{2k}}{2^{2k-1}\left(\mu^2 r_0^2-\frac{9}{4}\right)^{k+s-1}}\frac{\Gamma(s+k-1)}{\Gamma(s)k!}\left[\frac{\left(2^{2k-1}-1\right)}{\left(2^{2k+1}-1\right)} - \frac{kB_{2k+2}}{(k+1)B_{2k}}\right]. \tag{3.153}$$

As this expansion is asymptotic, it can be studied using resurgent techniques, just like it has been done for the two-sphere in [191] (see Appendix C.3).

For small values of $\mu$ the zeta function in $D = 5$ on the four-sphere can be expanded in a convergent series valid in the regime $0 < \mu r_0 < 3/\sqrt{2}$ using a binomial expansion, similarly to the case $D = 3$ treated in Section 3.1.4,

$$\zeta(s\,|\,S^4_{r_0},\mu^2) = \sum_{l\geq 0}\frac{\left[\frac{2l+3}{3}\prod_{k=1}^{2}\frac{l+k}{k}\right]}{\left(\frac{l(l+3)}{r_0^2}+\mu^2\right)^s} = \frac{r_0^{2s}}{3}\sum_{k\geq 0}\binom{-s}{k}\left[\zeta\left(2s+2k-3;\frac{3}{2}\right)-\frac{1}{4}\zeta\left(2s+2k-1;\frac{3}{2}\right)\right]\left(\mu^2 r_0^2-\frac{9}{4}\right)^k. \tag{3.154}$$

The small-$\mu$ and large-$\mu$ limits can be connected via a resurgent analysis, as we show in Appendix C.3. The overall result can be written in terms of the principal value of an integral involving Bessel functions. The grand potential takes the form

$$\Omega(\mu) = \frac{\mu^4 r_0^3}{24\pi V}\,\text{P.V.}\int_0^\infty \frac{\mathrm{d}y}{y\sin(y)}\left[2K_4(2\mu r_0 y) + \left(2+\frac{1}{\mu^2 r_0^2}\right)K_2(2\mu r_0 y)\right]. \tag{3.155}$$

In the regime at small $\mu$, where the convergent expansion is valid and can be trusted, the zeta function and hence the grand potential on the four-sphere in Figure 3.5 have a maximum as well as a minimum and hence also an inflection point, as opposed to the zeta function on the two-sphere in Figure 3.4, which only possesses a minimum. As a consequence the grand potential $-\Omega(\mu)$ in $D=5$ has a convex region around its minimum and a concave region around its maximum separated by a flex,[43]

$$r_0\mu_{\text{min}} \approx 0.9927\dots, \qquad r_0\mu_{\text{fl}} \approx 1.266\dots, \qquad r_0\mu_{\text{max}} = \frac{3}{2}. \tag{3.156}$$

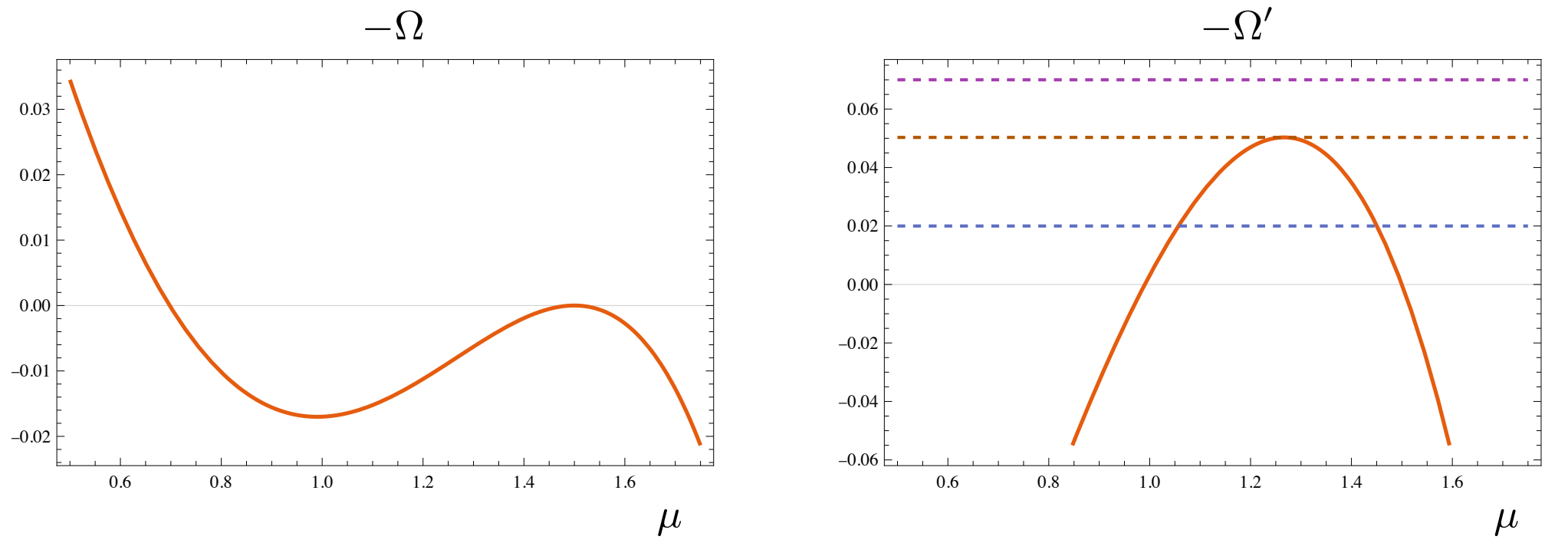

Figure 3.5: The grand potential $-\Omega(\mu)$ (left) and its first derivative $-\Omega'(\mu)$ (right) in $D = 5$ computed on the four-sphere. The function $-\Omega(\mu)$ is only convex for $\mu \leq \mu_{\text{fl}} \approx 1.266\ldots$ and concave otherwise. It has a minimum at $\mu r_0 = \mu_{\min} r_0 \approx 0.9927$ and a maximum at $\mu r_0 = \mu_{\max} r_0 = 3/2$. The maximization condition admits two positive solution for $0 \leq Q/2N < Q_{\text{fl}}/2N = -V\Omega'(\mu_{\text{fl}}) \approx 0.5029\ldots$ and no real solution for $Q > Q_{\text{fl}}$. The leftmost solutions for $0 \leq Q < Q_{\text{fl}}$ results in a convex free energy, but the expression does not satisfy the consistency condition for the scaling dimension of the unit operator as $r_0 f_c(0) = V\Omega(\mu_{\min}) \approx 0.017\cdots > 0$.

In terms of its supremum definition the Legendre transform is non-trivial (not infinite) only for values of the charge $Q$ smaller than

$$\frac{Q_{\text{fl}}}{2N} = -V\Omega'(\mu_{\text{fl}}) \approx 0.05029\dots. \tag{3.157}$$

For $Q > Q_{\text{fl}}$ the free energy is infinite as $-\Omega(\mu)$ is not bounded from below,

$$\frac{\Delta(Q)}{2Nr_0} = f_c(Q) = \begin{cases} \frac{Q}{2N}\mu + V\Omega(\mu)\Big|_{\mu=\mu(Q)} & \text{for } 0 < Q/2N < Q_{\text{fl}}/2N \approx 0.05029\dots, \\ +\infty & \text{for } Q/2N > Q_{\text{fl}}/2N. \end{cases} \tag{3.158}$$

The function $f_c(Q)$ obtained from this procedure cannot possibly describe the conformal dimension of a fixed-charge operator within a CFT. By virtue of the state–operator correspondence $r_0 f_c(Q)$ at $Q=0$ has to correspond to the identity operator, which enforces the consistency condition $f_c(0)=0$. We find instead that

$$r_0 f_c(0) = V\Omega(\mu_{\text{min}}) \approx 0.01699\dots \neq 0, \tag{3.159}$$

implying that we are not in a CFT and hence not at the critical point. After expanding in $Q$, the free energy is given by

$$r_0 f_c(Q) \approx 0.01699\dots + 0.9903\dots\left(\frac{Q}{2N}\right) + 1.516\dots\left(\frac{Q}{2N}\right)^2 + \dots. \tag{3.160}$$

We see that the leading term in $Q$ is linear, implying that we find ourselves in a massive phase here.

Alternatively, we can drop the assumption of unitarity of the theory and decide to ignore convexity while using the working definition of the Legendre transform in terms of the derivative. Instead of the minimum, we can also choose to expand around the maximum $\mu_{\text{max}}^2 = 9/(4r_0^2)$ situated at the value of the conformal mass on the four-sphere. In this case we find that the free energy is given by

$$r_0 f_c(Q) = \frac{3}{2}\left(\frac{Q}{2N}\right) - \frac{32}{3\pi^2}\left(\frac{Q}{2N}\right)^2 + \mathcal{O}\left(Q^3\right). \tag{3.161}$$

This result has previously appeared in the literature in [20], where the value of $Q_{\text{fl}}$ had been identified as the critical value above which the scaling dimension $\Delta(Q)$ becomes complex. This result is consistent with the conformal dimension of an operator given by the $Q$-th power of a field of dimension $3/2$ (see also [255]).

It is important to better understand the meaning of this expression. On one hand, we are discussing a potential UV continuation via a hard-to-justify effective potential that is complex. On the other hand, we are also expanding around a maximum in the context of the Legendre transform for the free energy $f_c(Q)$, in a concave region of the grand potential $-\Omega(\mu)$ that is not expected to contribute to said Legendre transform.

[43] The value of the flex can be found with quite good precision using an optimal truncation of the large-$\mu$ expansion. However, the asymptotic expansion cannot reproduce the whole structure of $-\Omega(\mu)$, in particular it cannot reproduce the maximum at $r_0\mu_{\text{max}} = 3/2$.

Finally, we can choose to continue the free energy $f_c(Q)$ beyond the critical value $Q_{\text{fl}}$ to a multi-valued complex function. The point $Q = Q_{\text{fl}}$ becomes a branch point and the free energy develops four branches. In doing so, we can study the large-$Q$ behaviour using the large-$\mu$ expansion of the zeta function in Eq. (3.153) for each of the four branches of $f_c(Q)$. Using said expansion we find that the scaling dimension is given by

$$\frac{\Delta(Q)}{2N} = r_0 f_c(Q) = \left[ f_1 \frac{4\sqrt{3}}{5} \left(\frac{Q}{2N}\right)^{\frac{5}{4}} - \frac{f_2}{\sqrt{3}} \left(\frac{Q}{2N}\right)^{\frac{3}{4}} + \dots \right], \tag{3.162}$$

where the complex phases $f_1$ and $f_2$ can take different values depending on the choice of branch,

$$\begin{array}{lllll} & \text{branch 1} & \text{branch 2} & \text{branch 3} & \text{branch 4} \\ f_1 & e^{i\pi/4} & e^{-i\pi/4} & e^{i3\pi/4} & e^{-i3\pi/4} \\ f_2 & e^{i3\pi/4} & e^{-i3\pi/4} & e^{i\pi/4} & e^{-i\pi/4} \end{array} . \tag{3.163}$$

These results agree perfectly with earlier computations found within the scientific literature in [20]. Note that we can reproduce the leading term in Eq. (3.162) from the effective potential in Eq. (3.140). This can be seen from the fact that the effective potential in Eq. (3.140) at $g \to \infty$ (and $r = 0$) directly computes the fixed-charge vacuum energy $E_{T_L^4}$ on the torus, which by virtue of Weyl's asymptotic expansion of the heat kernel in Eq. (3.58) and Eq. (3.59) is equal to the leading coefficient of the vacuum energy on the sphere (modulo a factor coming from the differing volumes of sphere and torus),

$$V_{\text{eff}}^{(0)}(|\phi_i^{(c)}|) = \kappa |\phi_i^{(c)}|^{10/3}, \qquad E_{T_L^4} = \frac{8N\sqrt{3}}{4L} \kappa^{3/8} \left(\frac{Q}{2N}\right)^{5/4} . \tag{3.164}$$

Differently said, since the ground state is homogeneous, the leading term in the large charge expansion of the energy on the four sphere is given by

$$E_{S^4_{r_0}}\Big|_{Q^{5/4}} = \frac{L}{V_{S^4_{r_0}}^{1/4}} E_{T_L^4} = \frac{\Delta(Q)}{r_0}\Big|_{Q^{5/4}} . \tag{3.165}$$

**A phase transition on the cylinder away from the fixed point**

The presence of a complex effective potential can be interpreted in terms of unstable states (see for example [266]). While following a given state within the theory at hand, it is possible that while changing some parameter in the theory the state becomes unstable and develops an imaginary part in its energy, which in term has the interpretation of a decay rate.
If we move away from the conformal fixed point in the critical limit $g \to \infty$ on the cylinder (along the critical trajectory with $r$ fine-tuned to the conformal mass $r = 9/(4r_0^2)$, we observe that something similar happens in the $O(2N)$ model in $D = 5$. In the IR at small $g$ the states in the theory are perfectly fine and then become unstable as we move towards the UV. Since we have access to an explicit convergent expansion for the zeta function on the sphere at small $\mu$, we can study this phenomenon for states with small charges $Q$ and analyse the implications for the free energy on the sphere.

Generally, if a function $f(x)$ possesses a critical point at $x = x_c$, then its Legendre transform $\mathrm{LT}[f](y)$ around $y = 0$ can be expanded and expressed in terms of the local properties of $f(x)$,

$$\mathrm{LT}[f](y) = \left[ -f(x_c) + x_c y + \frac{1}{2f''(x_c)} y^2 + \dots \right]. \tag{3.166}$$

For the free energy $f_c(Q)$ on the four-sphere we therefore find that

$$f_c(Q) = \left[ V\Omega(\mu_c) + \mu_c \left(\frac{Q}{2N}\right) - \frac{1}{2V\Omega''(\mu_c)} \left(\frac{Q}{2N}\right)^2 + \dots \right]_{\mu_c = 3/(2r_0)}. \tag{3.167}$$

If we keep only the first two terms, then Eq. (3.167) corresponds to the free energy of a fixed-charge state describing a massive particle with mass equal to $\mu_c$.

The grand potential on the cylinder, fine-tuned to the critical trajectory $r = r_{\text{conf}} = 9/4r_0^2$, is of the form

$$\Omega(\mu) = \frac{1}{2V} \zeta(-1/2 \,|\, S^4_{r_0}, \mu^2) - \frac{(\mu^2 - \frac{9}{4r_0^2})^2}{8g}. \tag{3.168}$$

The zeta function and its first derivative vanish at $\mu^2 = 9/4r_0$, and the same evidently holds true for the RG flow term. The grand potential $\Omega(\mu)$ expanded around the value of the conformal mass $\mu^2 = 9/4r_0^2$ is given by

$$\Omega(\mu) = -\frac{1}{r_0 V} \left[ \frac{\pi^2}{384} \left( \frac{128 r_0}{g} - 1 \right) \left( \mu^2 r_0^2 - \frac{9}{4} \right)^2 + \frac{(\pi^2 - 12)\pi^2}{2^8 3^2} \left( \mu^2 r_0^2 - \frac{9}{4} \right)^3 + \dots \right]. \tag{3.169}$$

Importantly, we observe that the nature of the critical point at $\mu_c = 3/2r_0$ changes depending on the value of the coupling $g$. For $g < 128 r_0$ the critical point $\mu_c$ is a local minimum, while for large values of the coupling along the flow $g \to \infty$ it turns into a local maximum and a new local minimum appears at $\mu_c' < \mu_c = 3/2r_0$, see Figure 3.6.

The Legendre transform, in terms of its supremum definition, tracks the position of the minimum. Hence, for small values of $g$ the free energy is determined via the behaviour of the grand potential $\Omega(\mu)$ around $\mu_c = 3/2r_0$

$$r_0 f_c(Q) = \frac{3}{2} \left(\frac{Q}{2N}\right) + \frac{48g}{(2\pi)^2 (128 r_0 - g)} \left(\frac{Q}{2N}\right)^2 + \dots, \tag{3.170}$$

as long as the coefficient of the quadratic term in $\mu$ is positive. At $g = 128 r_0$ this expression exhibits a pole and is no longer valid. For larger values of $g$ the free energy $f_c(Q)$ then depends on the local properties of the grand potential $\Omega(\mu)$ around the new minimum at $\mu_c' < 3/2r_0$, which as we have seen in Eq. (3.160), describes a massive phase in the limit $g \to \infty$. However, it is important to point out that this new minimum is only a local minimum and metastable. For large enough charge $Q$, the free energy $f_c(Q)$ obtained via the supremum definition of the Legendre transform becomes infinite because the grand potential $-\Omega(\mu)$ is no longer bounded from below for $g > 128r_0$.

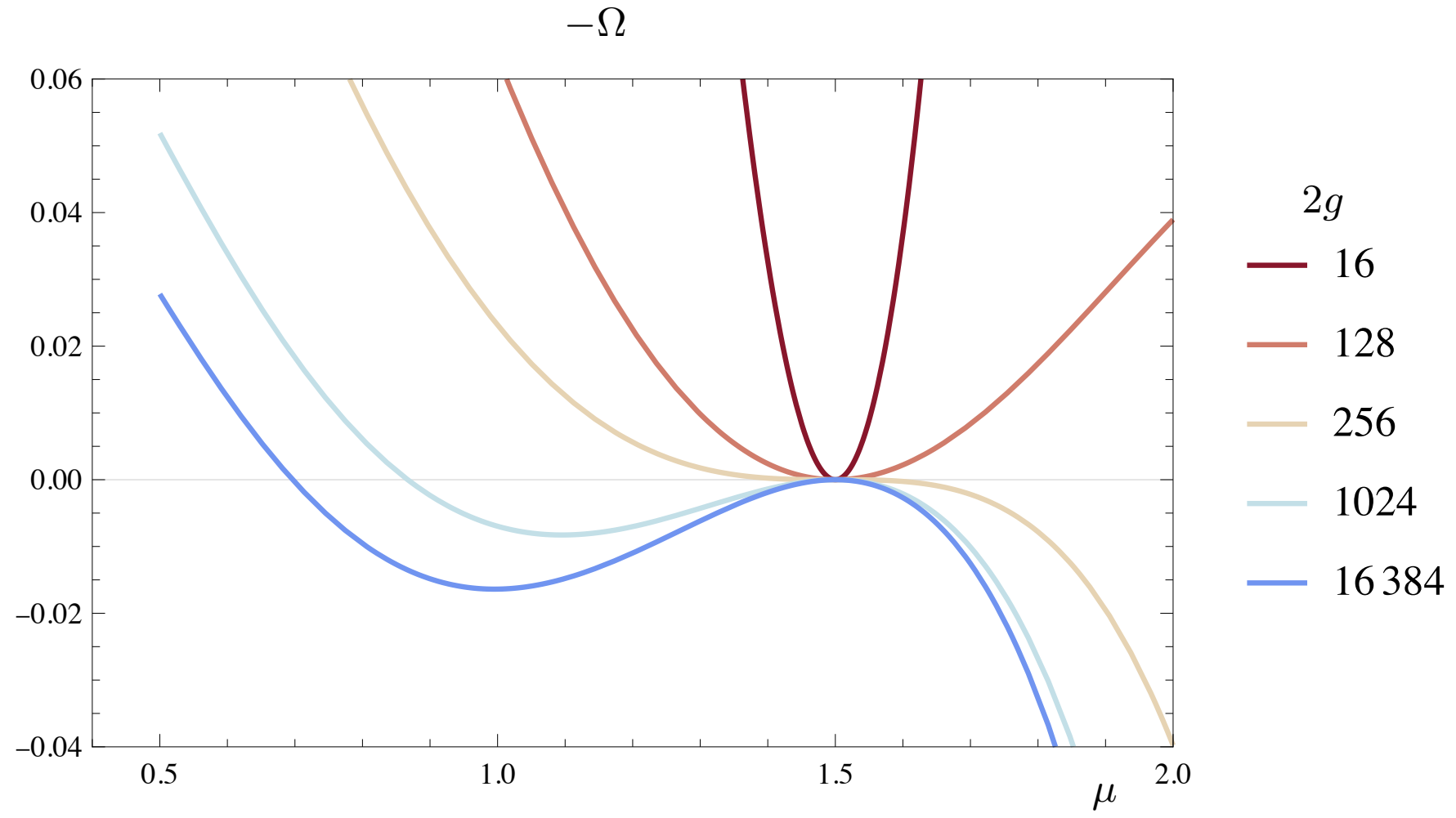

Figure 3.6: The grand potential $-\Omega(\mu)$ on the cylinder $\mathbb{R}\times S^4_{r_0}$ for different values of the coupling $g$ along the critical trajectory $r = 9/4r_0^2$ in $D = 5$. For $g < 128r_0$, the critical point $\mu_c = 3/2r_0$ represents a local minimum. As we increase the value of the coupling $g$ along the flow to the UV, the point $\mu_c$ becomes a maximum and a new minimum appears at $\mu_c' < 3/2r_0$.

On the other hand, for $g > 128r_0$, the naive definition of the Legendre transform allows for the expansion around the maximum $\mu_c = 3/2r_0$ which now corresponds to a concave and unstable region of the grand potential $-\Omega(\mu)$, which we have done in Eq. (3.161) in the limit $g \to \infty$. The free energy $f_c(Q)$ obtained in this way is still real for small enough values of $Q$, but the coefficient of the of the quadratic term in $Q$ is negative which is a sign of instability. Beyond small charge $Q$, the free energy then develops an imaginary part, see the discussion around Eq. (3.162), clearly violating unitarity.
Generically, when expanded around a maximum, the free energy $f_c(Q)$ as a function of the charge is concave. It appears as though the absence of convexity is a general feature of any expansion around an unstable state.

#### The free energy in $\mathbf{D = 4+\epsilon}$ and $\mathbf{D = 6-\epsilon}$

In Section 3.2.4 we have extensively discussed that the conjectured strongly-coupled UV fixed point cannot be reached for $4 < D < 6$, unless we analytically continue the effective potential to a complex function with a branch cut describing a Riemann surface. We can, however, compute the free energy $f_c(Q)$ on the sphere at the conformal point, which we naturally identify with the scaling dimension of the lowest operator of a given charge in flat space. In the large-charge limit the zeta function can be computed perturbatively using Weyl's asymptotic expansion of the heat kernel in Eq. (3.58), as discussed extensively in Section 3.1.4 and also in this section, *i.e.* Section 3.2.5. In arbitrary real dimension $D$ we have (in terms of the Legendre transform $\mathrm{LT}[\,\cdot\,]$)

$$\frac{\Delta(Q)}{2N} = r_0\,\mathrm{LT}[-V\Omega](Q/2N), \quad V\Omega(\mu) = \frac{1}{2}\zeta(-\frac{1}{2}\,|\,S^{D-1}_{r_0},\mu^2) = \frac{\Gamma\left(s-\frac{D-1}{2}\right)}{\Gamma(s)\,\mu^{2s+1-D}}K_0\big|_{S^{D-1}_{r_0}}\Bigg|_{s=-\frac{1}{2}} + \dots, \tag{3.171}$$

where the heat kernel coefficient $K_0$ evaluated on the $D-1$-sphere is given in Eq. (3.59). For $s=\pm$, since the maximization condition includes $\zeta(1/2\,|\,S^{D-1}_{r_0},\mu^2)$, the Gamma function $\Gamma(s)$ in the numerator diverges. We can, however, resort to analytic continuation and study the behaviour around $D=4$ and $D=6$. The geometric invariants have to be analytically continued as well and are given by

$$m^{3\pm\epsilon}K_0^{S^{3\pm\epsilon}} \tag{3.172}$$

$$\mu^{3\pm\epsilon}K_0\Big|_{S^{3\pm\epsilon}_{r_0}} = \frac{2^{-2\mp\epsilon}\sqrt{\pi}}{\Gamma(2\pm\frac{\epsilon}{2})}(r_0\mu)^{3\pm\epsilon} = \frac{\sqrt{\pi}}{4}\left[1\pm\left(\frac{\gamma}{2}-\log(2)-\frac{1}{2}+\log(r_0\mu)\right)\epsilon\right](r_0\mu)^3, \tag{3.173}$$

$$K_0^{S^{5\pm\epsilon}}m^{5\pm\epsilon} \tag{3.174}$$

$$\mu^{5\pm\epsilon}K_0\Big|_{S^{5\pm\epsilon}_{r_0}} = \frac{2^{-4\mp\epsilon}\sqrt{\pi}}{\Gamma(3\pm\frac{\epsilon}{2})}(r_0\mu)^{5\pm\epsilon} = \frac{\sqrt{\pi}}{32}\left[1\mp\left(\log(2)+\frac{3}{4}-\frac{\gamma}{2}-\log(r_0\mu)\right)\epsilon\right](r_0\mu)^5, \tag{3.175}$$

where $\gamma$ is the Euler-Mascheroni constant.Using the expansion of the Gamma function,

$$\Gamma(\epsilon-n) = \frac{(-1)^n}{n!}\epsilon^{-1}+\mathcal{O}\left(\epsilon^0\right), \tag{3.176}$$

the leading behaviour of the zeta functions in $D=4+\pm\epsilon$ and $D=6\pm\epsilon$ are given by

$$\zeta(1/2\,|\,S^{3\pm\epsilon}_{r_0},\mu^2) = \pm\epsilon^{-1}\frac{r_0^3\mu^2}{2}+\dots, \qquad \zeta(-1/2\,|\,S^{3\pm\epsilon}_{r_0},\mu^2) = \pm\epsilon^{-1}\frac{r_0^3\mu^4}{8}+\dots, \tag{3.177}$$

$$\zeta(1/2\,|\,S^{5\pm\epsilon}_{r_0},\mu^2) = \mp\epsilon^{-1}\frac{r_0^5\mu^4}{32}+\dots, \qquad \zeta(-1/2\,|\,S^{5\pm\epsilon}_{r_0},\mu^2) = \mp\epsilon^{-1}\frac{r_0^5\mu^6}{192}+\dots. \tag{3.178}$$

Around the physical dimension $D=4$ the maximization condition reads $Q/N=\mp(r_0m)^3/(2\epsilon)$ and, as expected, there is a real solution for positive charge $Q$ only in $D=4-\epsilon$. The free energies and conformal dimensions are found to be

$$\begin{cases} \frac{\Delta_{-\epsilon}(Q)}{2N} = \frac{3}{2^{4/3}}\epsilon^{1/3}\left(\frac{Q}{2N}\right)^{4/3}+\dots & \text{for } D=4-\epsilon, \\ \frac{\Delta_{+\epsilon}(Q)}{2N} = \frac{3}{2^{4/3}}e^{i\pi(2k+1)/3}\epsilon^{1/3}\left(\frac{Q}{2N}\right)^{4/3}+\dots & \text{for } D=4+\epsilon, \end{cases} \tag{3.179}$$

where the integer $k=0,1,2$ depends on the choice of branch for the analytically continued free energy $f_c(Q)$. In a similar fashion, around $D=6$ the maximization condition reads $Q/N=\pm(r_0m)^5/(32\epsilon)$, with a real solution for positive charge only in $D=6+\epsilon$. The associated conformal dimensions are given by

$$\begin{cases} \frac{\Delta_{-\epsilon}(Q)}{2N} = \frac{52^{1/5}}{3}e^{i\pi(2k+1)/5}\epsilon^{1/5}\left(\frac{Q}{2N}\right)^{6/5}+\dots & \text{for } D=6-\epsilon, \\ \frac{\Delta_{+\epsilon}(Q)}{2N} = \frac{52^{1/5}}{3}\epsilon^{1/5}\left(\frac{Q}{2N}\right)^{6/5}+\dots & \text{for } D=6+\epsilon, \end{cases} \tag{3.180}$$

where the integer $k=0,1,2,4,5$ again depends on the choice of branch for the analytically continued free energy $f_c(Q)$. These results perfectly agree with earlier results from the literature in [20, 234].
As expected, in the range $4<D<6$ the Legendre transform requires an analytic continuation and results in conformal dimensions $\delta(Q)$ with non-zero imaginary parts corresponding to complex free energies $f_c(Q)$ exhibiting a branch cut. The theory in $D = 4+\epsilon$ has been scrutinized more closely in [267], confirming its instability for operators of any value of the charge $Q$.

## 3.3 From Fermi spheres to superfluids: Fermionic CFTs at large charge and large $N$

When studying bosonic theories at large charge, the superfluid hypothesis underlying the EFT approach is a very natural one, although there are important outliers not captured by the EFT, like the free boson [168]. However, once we consider fermionic theories the superfluid description can still be justified sometimes, but it is no longer the only game in town as options like a Fermi-sphere ground state are reasonable as well. In particular, the superfluid paradigm does not apply to the free fermion, where sectors of fixed charge are described by Fermi surfaces [168]. The Fermi sphere, as we will see later, also has a non-trivial macroscopic limit on the cylinder [168]. Unfortunately, the characterization of the corresponding EFT description is more complicated and less clear than for a superfluid phase [118, 167, 168].

Beyond the study of bosonic theories, there is a distinct lack of literature on the LCE approach applied to fermionic CFTs. Only a handful of forays have been made into the domain of fermions [168, 170, 268]. In particular, it is important to better understand the landscape of emergent condensed-matter phases and possible EFT description in fermionic models. As we have seen in Section 3.1, for the $O(2N)$ model at large $N$ we recover from general principles the EFT description used to study the $O(2)$ model. In analogy, it appears to be a good idea to study fermionic models in simplifying limits — like large $N$ or small $\epsilon$, where the interacting fixed points become perturbative — to potentially gain access to the EFT descriptions of the large-charge sectors in these models.

This is exactly what we are attempting to do in this section, where we systematically study several different theories with four-fermion interactions in $D = 3$ and Euclidean signature, in the limit where the number $N$ of flavours of fermions becomes large. In this section we review the free fermion CFT and study the GN model as well as the chiral GN or NJL model and its $SU(2) \times SU(2)$ generalization. Additionally, we discuss the Cooper model, which is related to the NJL model by a Pauli–Gursey (PG) transformation [269–272].
We find two qualitatively different behaviours. While in the free fermion and the large-$N$ GN model there is no SSB occurring in sectors of large charge — *i.e.* large baryon number — and the physics is that of an (approximate) Fermi surface, the four-fermion interactions in NJL-type models allow for SSB to occur in sectors of large "chiral" charge. In this case the large-charge sector is described by the usual superfluid description and we can verify the prediction from the large-charge EFT discussed in Chapter 2. For the GN model we could not yet determine whether the Fermi surface persists for finite $N$ once we include sub-leading corrections.
Besides the discussion of the free fermion CFT, which represents a review of existing material, the rest of the material covered in this section, which appeared in [2], represents original work by the author.

We apply essentially the same technology which we used to analyse the bosonic theories at the beginning of this chapter (see Section 3.1). For convenience, we repeat the most important points again here. Independently of the presence of SSB — by the state–operator correspondence — finite-density ground states in the cylinder in critical theories compute the scaling dimensions of the associated CFT primary operators in flat space. If $\mathcal{O}^Q$ is the lightest primary operator of charge $Q$ under a certain global symmetry $Q$, then the corresponding state on the cylinder is understood to minimize the modified Hamiltonian $H^{\text{cyl}} - \mu Q$ on the cylinder, with $H^{\text{cyl}}$ being the dilatation operator/cylinder Hamiltonian and $Q$ the charge operator. Qualitatively, this picture is expected to be independent of the underlying bosonic or fermionic nature of the theory. In particular, we expect the lowest-lying operator $\mathcal{O}^Q$ to still be a scalar in a fermionic theory.

The most convenient way of selecting the correct state on the cylinder that computes the right scaling dimension $\Delta(Q)$ is to consider the thermal CFT on $S^1_\beta \times S^{D-1}_{r_0}$ and study the theory of interest in the zero-temperature-limit via its grand-canonical partition function $Z_{gc}(\mu)$,

$$Z_{gc}(\mu) = \text{Tr}\left[e^{-\beta(H^{(\text{cyl})}-\mu Q)}\right] \xrightarrow{\beta\to\infty} \langle\mathcal{O}_Q|\mathcal{O}_Q\rangle\, e^{-\beta(E(Q)-\mu Q)}\,, \tag{3.181}$$

where $\langle\mathcal{O}_Q|\mathcal{O}_Q\rangle$ is a normalization factor and the energy $E(Q)$ in the thermodynamic limit $N\to\infty$ is equal to the zero-temperature free energy $f_c(Q)$ at charge $Q$ of the system first introduced in Section 3.1.2. In doing so, we proceed in analogy to the discussion around Eq. (3.27) presented in the context of the $O(2N)$ vector model.[44] All of the fermionic theories investigated in the present section possess a path-integral representation of their respective canonical partition function $Z_{gc}(\mu)$, which can be computed order-by-order at large values of the number of flavours $N$. In the thermodynamic limit $N\to\infty$, as usual, the grand canonical partition function is given by the grand potential $\Omega$,

$$Z_{gc}(\mu) \xrightarrow{N\to\infty} e^{-(2N)\beta V\Omega(\mu)}\,, \tag{3.182}$$

which we care for mainly in the zero-temperature limit. When comparing to Eq. (3.181),[45] we conclude that for the CFT living at $\beta\to\infty$ we have

$$r_0\Delta(Q) = E(Q) = (2N)f_c(Q) + \mathcal{O}(N^0) = (2N)V\Omega(\mu) + \mu Q\Big|_{\mu=\mu(Q)} + \mathcal{O}(N^0)\,, \qquad \frac{Q}{(2N)V} = -\frac{\partial\Omega}{\partial\mu}\,, \tag{3.183}$$

without necessarily needing to reference the canonical partition function (see Section 3.1.2). As usual, $f_c(Q)$ denotes the free energy. For certain purposes we are content with performing computations on $S^1_\beta \times T^{D-1}_L$. In these instances we will use the same notation as on the sphere and simply replace the volume $V$.[46]

With regards to the large-$N$ methodology, for the interacting theories considered (GN and NJL-type models) we perform a HS transformation, introducing a (complex) scalar collective field $\sigma$ ($\Phi$ if complex) and reducing the four-fermion interactions to Yukawa-type interactions. The field $\sigma(\Phi)$ can both be kept dynamical and non-dynamical. If $\sigma(\Phi)$ is kept non-dynamical, the interacting fixed point in $D=3$ is found in the UV and is only treatable within the context of the large-$N$ expansion. If instead the scalar field $\sigma(\Phi)$ is made dynamical, then we obtain the UV completion of these fermionic models, and the same interacting fixed point arises in the IR (in $D = 3$). The large-charge primary $\mathcal{O}_Q$ is part of the CFT spectrum independently of the specific realization chosen.

[44] In contrast to the $O(2N)$ model in Section 3.1.2 and 3.1.3 we do, however, start directly by only fixing the charge associated to a single $U(1)$ subgroup of the global symmetry, in contrast to the $U(1)^N$-diagonal subgroup fixed for the bosonic vector model. Note that the homogeneous ground state in the bosonic model also corresponds to the case of a single $U(1)$ subgroup at finite density, the $U(1)$ baryon charge, see Section 3.1.3.

[45] The normalization of the state $|\mathcal{O}_Q\rangle$ is irrelevant here for our purposes and can be ignored as a constant contribution.

[46] We also do not distinguish between a torus of large volume and infinite flat space.

### 3.3.1 Fermi spheres in the large charge expansion: The free fermion

Although the class of CFTs whose large-charge sector is described by a conformal superfluid encompasses many different theories, there are a few important examples that lie outside of it. The simplest and most obvious example is the free complex scalar theory, which does not lead to a state of finite density in the macroscopic limit on the cylinder [168]. In the free scalar case the lowest-lying operator of charge $Q$ is simply

$$\mathcal{O}^Q \propto \phi^Q, \qquad \Delta(Q) = r_0 E(Q) = Q. \tag{3.184}$$

In the macroscopic limit $r_0 \to \infty$ on the sphere the charge density $\rho = Q/V$ remains finite, and hence the energy density behaves like

$$\frac{E(Q)}{V} = \frac{\Delta(Q)}{\Omega_D r_0^D} \to 0, \tag{3.185}$$

where $\Omega_D$ is the volume of the unit $(D-1)$-sphere. As a consequence, in the large-charge sector of the free scalar there is no SSB at large charge and also no emergent condensed-matter description. The same is not true, however, for the free fermion CFT, as we will see.

#### Comments about notation

Our discussion of the free fermion CFT — besides introducing the Fermi-sphere ground state and hence a new condensed-matter description at large charge — serves to set up our discussion of the interacting fermionic models of GN-type, which admit a second-order critical phase transition and are computable in the large-$N$ limit, in the later parts of Section 3.3. More precisely, the models we will consider are the GN model and the NJL model along with its minimal $SU(2) \times SU(2)$ generalization. They can all be obtained by deforming the free-fermion CFT with a four-fermion interaction and a coupling $g$. As the quartic interaction in the fermionic fields is irrelevant for $D > 2$, and hence the theory is not renormalizable in standard perturbation theory, these models are typically studied either in $D = 4 - \epsilon$ or $D = 2 + \epsilon$ at small $\epsilon$ and fixed values of $N$, or in the large-$N$ limit for $2 < D < 4$.[47]
It has been shown that results from small-$\epsilon$ expansions in $4 - \epsilon$ and $2 + \epsilon$ at fixed values of $N$ are completely consistent with the large-$N$ expansion for general dimension $2 < D < 4$, which in particular also includes $D = 3$ [273]. Moreover, the conformal phases found in the simplifying limits of large $N$ and small $\epsilon$ are known (or strongly believed, *e.g.* from lattice studies [274]) to exist in the physical dimension $D = 3$ also at finite values of $N$.

Importantly, in particular with regards to the NJL-type models, we rely on some notion of chirality within the theory. Of course, in $D = 3$ there exists no natural notion of chirality. This issue can be amended by dimensionally reducing the corresponding four-dimensional model in $D = 4, (3+1)$ down to a model in three dimensions, $D = 3$, and using the four-component fermions from four dimensions. In $D = 3$ then, a four-component Dirac fermion $\Psi$ sits in a reducible representation consisting of two two-component Dirac fermions $\psi_{a,b}$. Additionally, a Dirac fermion in $D = 3+1$ can be decomposed into two Majorana fermions $\Psi = \lambda_1 + i\lambda_2$, so that, so that starting from $N$ Dirac fermion in $D = 4$ we obtain a theory of $4N$ Majorana fermions in $D = 3$ [2],[48]

$$\Psi\big|_{D=4} = \lambda_1 + i\lambda_2 \overset{\text{dim. red.}}{\longrightarrow} \Psi\big|_{D=3} = \begin{pmatrix} \psi_a \\ \psi_b \end{pmatrix} = \begin{pmatrix} \lambda_{a1} + i\lambda_{a2} \\ \lambda_{b1} + i\lambda_{b2} \end{pmatrix}. \tag{3.186}$$

[47]These four-fermion models can be understood via their UV completions in terms of Yukawa-type theories, which flow to the same interacting fixed point now located in the IR (in $D = 3$). The associated Yukawa-type theories are in principle renormalizable.

In addition, in $D = 3$ there exist two inequivalent two-dimensional representations of the Clifford algebra. They are related by a parity transformation and can be identified in terms of a differing sign in the gamma matrices $\gamma_\mu$. Therefore, the two representations can be denoted as $\pm\gamma_\mu$. As a consequence, in principle, there are $2N+1$ possible inequivalent choices to arrange the $4N$ Majorana fermions. In practice, there are only two interesting situations: either all Majorana fermions are in the same representation $\pm\gamma_\mu$, or half of the fermionic fields are in one representation $\pm\gamma_\mu$ while the others sit in the other representation $\mp\gamma_\mu$.[49]
Starting from a $U(N)$ invariant theory in $D = 4$, in the first case the system in $D = 3$ is $O(4N)$ invariant while in the second case the system exhibits a $O(2N) \times O(2N) \times \mathbb{Z}_2$ symmetry. In the case where all the Majorana fermions sit in the same representation there is no parity invariance — which is present in the four-dimensional theory — as all spinors share the same eigenvalue under the operator $\gamma_0\gamma_1\gamma_2$. For this reason, since we want to retain the notion of parity because we are interested in the dimensionally reduced four-dimensional model, we will make the latter parity-invariant choice. Concretely, for the presentation of our computations and results this means that we will arrange our two-dimensional (Dirac) spinors $\psi_i$, $i = 1, \ldots, 2N$ in $D = 3$ into four-dimensional reducible spinors $\Psi_i = (\psi_i, \psi_{i+N})$, $i = 1, \ldots, N$ and use four dimensional reducible gamma matrices of the form

$$\Gamma_\mu = \sigma_3 \otimes \gamma_\mu = \begin{pmatrix} \gamma_\mu & 0 \\ 0 & -\gamma_\mu \end{pmatrix}. \tag{3.187}$$

In this way we can actually define a $\Gamma_5$-matrix in $D = 3$ and a notion of chirality. Importantly, in contrast to the four-dimensional theory, after dimensional reduction this chiral symmetry in $D = 3$ is actually a flavour symmetry, *i.e.* it is part of the global symmetry group. For more details about notation we refer to Appendix C.4.3.

[48] A Dirac fermion $\psi_a$ in three-dimensional spacetime can be decomposed into two (symplectic) Majorana fermions $\lambda_{a1} + i\lambda_{a2}$ [275–277].

[49] Somewhat confusingly, in the presence of a standard four-fermion interaction of the form $(\bar{\psi}\psi)^2$ both of these situations are called GN model in the literature [278].

#### Asymptotics for the free fermion at large charge

We discuss the free fermion CFT in $D = 3$ (Euclidean) with $2N$ flavours of Dirac fermions organized into $N$ four-dimensional reducible spinor representations, as outlined above,

$$S = \sum_i \int_{S^1_\beta \times S^2_{r_0}} \mathrm{d}^3x\, \bar{\Psi}_i \Gamma_\mu \partial^\mu \Psi_i = \sum_i \int_{S^1_\beta \times S^2_{r_0}} \mathrm{d}^3x\, \Psi_i^\dagger \Gamma_3 \Gamma_\mu \partial^\mu \Psi_i\,. \tag{3.188}$$

Note that this action is secretly $O(4N)$ invariant and not just $O(2N) \times O(2N) \times \mathbb{Z}_2$, as discussed above, since there is a product of two gamma matrices appearing in the kinetic term which neutralizes the difference in sign. Nevertheless, we will stick with this notation in order to be consistent with the rest of the section.

The large-charge sector of the free fermion has been first studied in [168], where the authors deduce the general form of the large-charge operator for a single free fermion via the the Pauli exclusion principle and then perform a combinatorial computation to deduce that the scaling dimension of the lightest operator of large charge $Q$ has a scaling dimension that behaves like

$$\Delta_{\text{FS}}(Q) = \frac{2}{3} Q^{3/2} + \frac{1}{12} Q^{1/2} + \dots\,, \tag{3.189}$$

where the subscript $_{\text{FS}}$ denotes that the ground state is a filled Fermi sphere and not a superfluid. We will reproduce the same result using thermodynamical reasoning.

Similarly to the treatment $O(2N)$ vector model in Section 3.1.2, the action in the grand-canonical partition function $Z_{gc}(\mu)$ is found by introducing the appropriate chemical potential inside the action in Eq. (3.188) as a (constant) background field in the time direction,

$$S_\mu[\Psi_i] = \int_{S^1_\beta \times S^2_{r_0}} \mathrm{d}^3x\, \bar{\Psi}_i \left(\Gamma_\nu \partial^\nu - \mu \Gamma_3\right) \Psi_i\,. \tag{3.190}$$

In analogy with the results for the saddle-point equations of the $O(2N)$ vector model, here we introduce a single chemical potential $\mu$ for the global $U(1)_B$ baryon number symmetry, which we formally expect to correspond to the homogeneous sector or the completely symmetric representation like in the case of the $O(2N)$ vector model (again see Section 3.1.2 for details). Worded differently, we are only interested in CFT data pertaining to operators charged under the $U(1)_B$ symmetry.

As we are dealing with a free theory, the path integral can be explicitly performed using standard techniques. The cylinder computation at finite temperature via the summation over Matsubara frequencies is outlined in Appendix C.6. The resulting grand potential is given by

$$\Omega(\mu) = \frac{S_\mu\big|_{N\to\infty}}{(2N)\beta V} = -\frac{1}{\beta V} \sum_{j=\frac{1}{2},\frac{3}{2},\dots} 2(2j+1) \sum_{n\geq 0} \left[\beta\omega_\ell + \log\left(1 + e^{-\beta(\omega_\ell - \mu)}\right) + \log\left(1 + e^{-\beta(\omega_\ell + \mu)}\right)\right], \tag{3.191}$$

where $\omega_j^2 = (j+1/2)^2/r_0^2$. In the zero temperature limit $\beta \to \infty$ the grand potential reduces to

$$\Omega(\mu) \overset{\beta\to\infty}{\longrightarrow} -\frac{2}{V}\sum_j (2j+1)\omega_j - \frac{2}{V}\sum_{\omega_j<\mu}(2j+1)\left(\mu-\omega_j\right) = -\frac{2}{V}\Big[\sum_{\omega_j>\mu}(2j+1)\omega_j + \mu\sum_{\omega_j<\mu}(2j+1)\Big], \tag{3.192}$$

and now unsurprisingly describes a filled Fermi-sphere ground state of massless fermionic particles (massless due to conformal invariance). The associated Fermi momentum is given by $\mu r_0$. Using Eq. (3.183) the charge and energy density are found in terms of the Fermi momentum

$$\frac{Q}{2N} = \lfloor \mu r_0 \rfloor (\lfloor \mu r_0 \rfloor + 1), \qquad \frac{f_c(Q)}{2N} = \frac{1}{3r_0}\lfloor \mu r_0 \rfloor (\lfloor \mu r_0 \rfloor + 1)(2\lfloor \mu r_0 \rfloor + 1). \tag{3.193}$$

The floor function in the above expressions implements the fact that on the cylinder the energy levels $\omega_j$ are discretized. Hence, any chemical potential value $\mu r_0$ that lies in between two energy levels corresponds to the same filled Fermi sphere.

Using the form of the $U(1)_B$-charge in Eq. (3.193) the scaling dimension of the corresponding scalar CFT operator $\mathcal{O}_{\text{FS}}^Q$ in flat space can be expressed in closed form as

$$\frac{\Delta_{\text{FS}}(Q)}{2N} = \frac{1}{3}\lfloor \mu r_0 \rfloor (\lfloor \mu r_0 \rfloor + 1)(2\lfloor \mu r_0 \rfloor + 1) = \frac{1}{3}\left(\frac{Q}{2N}\right)\sqrt{4\left(\frac{Q}{2N}\right)+1}, \tag{3.194}$$

where we again denote the fact that this operator corresponds to a filled Fermi sphere ground state on the cylinder — which has a non-trivial macroscopic limit and describes a condensed-matter phase — by the subscript $_{\text{FS}}$. Further, we have normalized the charge $Q$ and scaling dimension $\Delta_{\text{FS}}(Q)$ by $2N$, which is the total number of three-dimensional Dirac-fermion flavours present. The scaling dimension $\Delta_{\text{FS}}(Q)$ is shown later on in Section 3.3.3 in Figure 3.8 as a function of $Q$. Notably, since the Fermi shell always has to be filled, only discrete values of the charge $Q$ — specifically $Q/2N = 0, 2, 6, \ldots$ — are allowed.

We note that the Fermi-sphere ground state is an eigenstate of both the Hamiltonian $H^{(\text{cyl})}$ and the $U(1)_B$ charge $Q$. As a consequence, there is no SSB on the ground state — in accordance with Section 1.2.5 — and we do not have any NG fluctuations on top of the ground state as in the superfluid case. The ground state is an eigenstate of both the Hamiltonian $H^{(\text{cyl})}$ and the $U(1)_B$ charge $Q$. There are no phonon states describing new primaries with a gap of order $\mathcal{O}(1)$ from the Fermi-sphere primary $\mathcal{O}_{\text{FS}}^Q$ in the CFT context. Instead, there are particle-hole excitations that exhibit the same charge as the Fermi-sphere ground state and have a gap of order $\mathcal{O}(1)$. It would be interesting to see whether there can be a description in terms of a Fermi liquid for the EFT, in particular with regards to the GN model which we will discuss in Section 3.3.2.

The $Q \to \infty$ asymptotics of the expression in Eq. (3.194) for the scaling dimension of the Fermi-sphere operator $\mathscr{O}_{\text{FS}}^{Q}$ can be systematically computed; the first few orders are

$$\frac{\Delta_{\text{FS}}(Q)}{2N} = \frac{2}{3}\left(\frac{Q}{2N}\right)^{3/2} + \frac{1}{12}\left(\frac{Q}{2N}\right)^{1/2} - \frac{1}{192}\left(\frac{Q}{2N}\right)^{-1/2} + \mathscr{O}\left(Q^{-3/2}\right), \qquad \frac{Q}{2N} \to \infty. \tag{3.195}$$

After setting the number of flavours $N$ to $N = 1/2$ — which in our convention of using the reducible representation described in Eq. (3.187) and Appendix C.4.3 corresponds to a single flavour of fermions in $D = 3$ — the finite-density ground state and the associated free energy $f_c(Q)$ in Eq. (3.193) correspond exactly to the parity-even scalar primary operator $\mathscr{O}_{\text{FS}}^{Q}$ first constructed explicitly in [168] within the single-flavour free fermion. [50]

### 3.3.2 The Gross-Neveu model at large $N$: Interacting fermionic CFTs at large charge

Primarily, we focus in this section on fermionic models in three-dimensional Euclidean space admitting a second-order phase transition, which are compatible with the large-$N$ limit allowing us to take advantage of the simplifications in said limit. We start by discussing the GN model.
In the four-dimensional reducible representation we employed — see the beginning of Section 3.3.1 and Appendix C.4.3 — the Gross–Neveu model [279] is described by the action

$$S = \int \mathrm{d}^3 x \left[ \sum_{i=1}^{N} \bar{\Psi}_i \Gamma_\mu \partial^\mu \Psi_i - \frac{g}{N}\Big( \sum_{i=1}^{N} \bar{\Psi}_i \Psi_i \Big)^2 \right]. \tag{3.196}$$

In this form it has a global $O(2N) \times O(2N) \times \mathbb{Z}_2$ symmetry. Out of the global symmetry group we will denote the diagonal $U(1)$ subgroup — the baryon number symmetry — by $U(1)_B$,

$$U(1)_B : \Psi_i \to e^{i\alpha} \Psi_i . \tag{3.197}$$

As discussed extensively in Section 3.1.1, the standard large-$N$ analysis is carried out via a HS transformation introducing an auxiliary (real) scalar field $\sigma$ replacing the quartic interaction.[51] The resulting action for the GN model is given by

$$S = \int \mathrm{d}^3 x \left[ \bar{\Psi}_i \left( \Gamma_\mu \partial^\mu + \sigma \right) \Psi_i + \frac{N}{4g} \sigma^2 \right]. \tag{3.198}$$

The UV critical limit for the GN model is reached by neglecting the $\sigma^2$ term. It corresponds to a second-order phase transition which separates the broken and unbroken phases for the breaking of the $\mathbb{Z}_2$ chiral symmetry acting as[52]

$$\Psi \to -\Gamma_5 \Psi . \tag{3.199}$$

For finite values of $N$, the theory in Eq. (3.196) is not renormalizable and the appropriate UV completion is a Gross–Neveu–Yukawa (GNY) model that is obtained by promoting the collective field $\sigma$ to a

[50] As we will see, at large $N$ this operator is the lowest primary of $U(1)_B$ charge $Q$ even if the CFT is an interacting one (in both GN and NJL models), due to the fact that the auxiliary field $\sigma$ does not condense.

[51] The HS transformation can always be applied independently of the nature of the field content of the theory to replace quartic (self)-interactions. The procedure always looks the same [134].

[52] See Appendix C.4.3 for the conventions around the $\Gamma_5$-matrix. The use of four-dimensional language here should always be seen in the context outlined at the beginning of Section 3.3.1.

dynamical field [280], resulting in the action

$$S = \int \mathrm{d}^3 x \left[ \bar{\Psi}_i \left( \Gamma_\mu \partial^\mu + \sigma \right) \Psi_i + \frac{1}{2 g_Y} \partial_\mu \sigma \partial^\mu \sigma \right]. \tag{3.200}$$

In the IR limit, the Yukawa coupling $g_Y$ is relevant and grows large, formally reproducing the critical action of the GN model as the kinetic term drops out. The critical point of the GNY model becomes weakly-coupled in $4-\epsilon$, allowing for a study of the GN critical phase at finite values of $N$. The theory in Eq. (3.196) on the other hand is only accessible in a $1/N$ expansion. As we will see, to leading order in $N$ there is no SSB for the GN model in sectors of large baryon number and the physics is approximately that of a Fermi surface, in the strict infinite-$N$ limit at least. We have not yet determined whether the Fermi surface persists once we consider sub-leading corrections in $N$.

Generally speaking, both the GN model as well as the NJL-type models we discuss later in Section 3.3.3 are obtained from the free fermion CFT in Eq. (3.188) via an irrelevant coupling $g$ and a four-Fermi interaction. Assuming that the model in question has a fundamental UV scale $\Lambda$ (given by a specific underlying lattice discretization), at zero temperature and zero density the irrelevant coupling possesses a critical value $1/g_c \sim \Lambda$ at which we find a scale-invariant theory separating two distinct phases characterized by some spontaneously broken symmetries.[53] In flat space specifically, if we wish to take a formal continuum limit $\Lambda \to \infty$ for these kinds of models with irrelevant couplings, we run into the problem that they are not renormalizable in the $1/N$ expansion [134]; the associated RG flow only joins the free fermion CFT in the far IR at $g = 0$ to the interacting conformal phase at $g = g_c$ in the UV. Our goal is to study these theories — GN and NJL — at the critical limit of the phase-transition point at finite charge density.
There is evidence — *e.g.* on the lattice [274] — that these conformal phases at $g = g_c$ for the GN and NJL models survive at finite values of $N$, however a proper RG flow analysis requires a UV completion, as these theories are not renormalizable away from any simplifying limit like small $\epsilon$ or large $N$. As discussed, working UV completions are found in terms of Yukawa-type models [280] with the distinct advantage that the Yukawa coupling $g_Y$ is relevant in $D < 4$. Yukawa-type models in $2 < D < 4$ are UV free and strongly coupled in the IR, where we recover the interacting CFT of the associated four-Fermi model (GN or NJL here). The Yukawa-CFT becomes again weakly coupled in $D = 4-\epsilon$ allowing for the perturbative computation of conformal data, *e.g.* at large charge [170]. As we are working in the large-$N$ limit, it is sufficient for us to consider the minimal models including only fermionic matter, hence we only briefly comment on the UV completions.

#### Symmetry-breaking at large $N$: Leading-order action and gap equation

The presence of a condensate in the large-charge sector is a crucial ingredient determining whether or not the LCE leads to simplifications, as we have seen in the case of the free boson or the free fermion in Section 3.3.1. Without the presence of a condensate there is no SSB and no superfluid description. To see, if this is the case, we start by discussing the symmetry breaking for the GN model.

[53] The exact value of $g_c$ may depend on the geometry of spacetime and the regularization procedure, making certain geometries and regularization procedures preferable over others. Generically, in zeta-function regularization (*e.g.* on the sphere) the critical value of the coupling is $1/g_c \to 0$, essentially obstructing the broken phase. For example, in the three-dimensional $O(2N)$ vector model discussed in Section 3.1 the value of the critical coupling was $g_c = \infty$ (we have used zeta-function regularization there).

We consider the GN model for $N$ reducible four-dimensional Dirac fermions in the representation outlined in Appendix C.4.3 at finite $U(1)_B$-chemical potential $\mu$ and finite temperature. To analyse the symmetry-breaking it suffices to consider the theory in flat space $\mathbb{R}^2$. After introducing the auxiliary field $\sigma$ the action is

$$S = \int_{S^1_\beta \times \mathbb{R}^2} \mathrm{d}^3 x \left[ \bar{\Psi}_i (\Gamma_\nu \partial^\nu - \mu \Gamma_3 + \sigma) \Psi_i + \frac{N}{4g} \sigma^2 \right] . \tag{3.201}$$

The model needs to be equipped with a cut-off scale $\lambda$ in momentum space in order to be consistent. At generic values of the parameters of the theory $g, \beta, \mu$ the auxiliary field $\sigma$ is allowed to acquire a VEV, which spontaneously breaks the discrete $\mathbb{Z}_2$ parity symmetry $\Psi \to -\Gamma_5 \Psi$. Generically, we can assume that the $\sigma$-VEV is given by a homogeneous configuration $\langle\sigma\rangle = \text{const}$. Under the standard procedure, the large-$N$ effective action computable in a self-consistent $1/N$ expansion can be obtained by expanding around the saddle $\sigma = \langle\sigma\rangle + \hat{\sigma}/\sqrt{N}$ [134],

$$S_\mu[\hat{\sigma}] = N \left\{ \beta V \frac{\langle\sigma\rangle^2}{4g} - \operatorname{Tr}\log\left(D^{(\mu)}\right)^{-1} \right\} + \frac{1}{2} \left[ \operatorname{Tr}(D^{(\mu)} \hat{\sigma} D^{(\mu)} \hat{\sigma}) + \frac{1}{4g} \int_{S^1_\beta \times \mathbb{R}^2} \hat{\sigma}^2 \right] + \mathcal{O}\left(N^{-1}\right) , \tag{3.202}$$

where we have introduced the fermion propagator $D^{(\mu)}(x-y)$ at finite chemical potential,[54]

$$D^{(\mu)}(x-y) = \langle x | (\Gamma_\nu \partial^\nu - \mu \Gamma_3 + \langle\sigma\rangle)^{-1} | x \rangle . \tag{3.203}$$

We find the value of the condensate $\langle\sigma\rangle$ by minimizing the leading-$N$ part of the action in Eq. (3.202), which is equivalently found by setting the fluctuation $\hat{\sigma}$ to zero. Using the summation over the fermionic Matsubara frequencies discussed in Appendix C.7 and the momentum-space representation $\tilde{D}^{(\mu)}$ of the propagator in Eq. (3.203) we find the grand potential $\Omega(\mu)$ as defined in Eq. (3.182),

$$\Omega(\mu) = \frac{\langle\sigma\rangle^2}{8g} - \int^\Lambda \frac{\mathrm{d}^2 p}{(2\pi)^2} \left[ \omega_p + \frac{1}{\beta} \log\left(1 + e^{-\beta(\omega_p + \mu)}\right) + \frac{1}{\beta} \log\left(1 + e^{-\beta(\omega_p - \mu)}\right) \right] , \tag{3.204}$$

where we have introduced the flat space energies $\omega_p^2 = \mathbf{p}^2 + \langle\sigma\rangle^2$. For consistency, we need to assume that $\langle\sigma\rangle, \mu \ll \Lambda$. The regularized integral over the momenta can be performed and the gap equation $0 = \partial\Omega/\partial\langle\sigma\rangle$ becomes

$$0 = \langle\sigma\rangle \left[ \frac{1}{2g} - \frac{\Lambda}{\pi} - \frac{1}{\pi} \left( \langle\sigma\rangle - \frac{1}{\beta} \log\left(1 + e^{\beta(\langle\sigma\rangle + \mu)}\right) - \frac{1}{\beta} \log\left(1 + e^{\beta(\langle\sigma\rangle - \mu)}\right) \right) \right] . \tag{3.205}$$

We can introduce the critical coupling

$$g_c^{-1} = 2\Lambda/\pi . \tag{3.206}$$

The trivial solution $\langle\sigma\rangle = 0$ to the gap equation corresponds to the free fermion CFT. In order to find a non-trivial solution to the gap equation we need to solve

$$0 = \frac{1}{2} \left( \frac{1}{g} - \frac{1}{g_c} \right) - \frac{1}{\pi} \left( \langle\sigma\rangle - \frac{1}{\beta} \log\left(1 + e^{\beta(\langle\sigma\rangle + \mu)}\right) - \frac{1}{\beta} \log\left(1 + e^{\beta(\langle\sigma\rangle - \mu)}\right) \right) . \tag{3.207}$$

[54] We use the notation $x = (\tau, \mathbf{x})$ for points on the spacetime $S^1_\beta \times \mathbb{R}^2$.

The solution $\langle\sigma\rangle$ to the above equation is given in closed form by

$$e^{\beta\langle\sigma\rangle} = \frac{1}{2}\left[e^{\beta\frac{\pi}{2}\left(\frac{1}{g_c}-\frac{1}{g}\right)} - 2\cosh(\beta\mu) + \sqrt{\left(e^{\beta\frac{\pi}{2}\left(\frac{1}{g_c}-\frac{1}{g}\right)} - 2\cosh(\beta\mu)\right)^2 - 4}\right]. \tag{3.208}$$

At zero density $\mu = 0$ and zero temperature $\beta \to \infty$ there exists a non-trivial solution to the gap equation only for values of the coupling $g$ that lie in the broken phase $g > g_c$. The solution there reads

$$\langle\sigma\rangle\Big|_{\mu,\beta^{-1}=0} = \frac{\pi}{2}\left(\frac{1}{g_c} - \frac{1}{g}\right). \tag{3.209}$$

This corresponds to the well-known and well-understood second-order quantum phase transition (since it occurs at zero temperature) of the large-$N$ GN model at the critical value of the coupling $g = g_c$, which separates the parity-unbroken from the parity-broken phase.
At zero temperature and finite density, this solution survives outside of criticality only as long as

$$\mu < \mu_c = \langle\sigma\rangle\big|_{\mu,\beta^{-1}=0}. \tag{3.210}$$

For values of the chemical potential $\mu$ greater that $\mu_c$ parity is restored again. Further, at the quantum critical point $g = g_c$ there exists no non-vanishing solution for any value of $\mu$. This is consistent with the fact that within the CFT the chemical potential $\mu$ is sourcing a parity even primary operator $\mathcal{O}^Q_{\text{FS}}$, as we will discuss later.[55]

In the critical limit, the zero-temperature ground state is a filled Fermi sphere of massless fermions with $\langle\sigma\rangle = 0$. The grand potential there reads

$$\Omega(\mu) = -\int_{\mu<|\mathbf{p}|<\Lambda} \frac{\mathrm{d}^2 p}{(2\pi)^2}\omega_p - \mu\int_{|\mathbf{p}|<\mu} \frac{\mathrm{d}^2 p}{(2\pi)^2} = -\frac{\Lambda^3}{6\pi} - \frac{\mu^3}{12\pi}. \tag{3.211}$$

Using Eq. (3.183) the $U(1)_B$ charge density $\rho = Q/V$ and the renormalized (free) energy $f_c(Q)$ of the Fermi-sphere ground state reads

$$\frac{\rho}{2N} = \frac{\mu^2}{4\pi}, \qquad \frac{f_c(Q)}{V} = \frac{1}{6\pi}\left(4\pi\frac{\rho}{2N}\right)^{3/2}. \tag{3.212}$$

This expression computes the leading-order result in the LCE for the scaling dimension $\Delta_{\text{FS}}(Q)$ of the Fermi-sphere operator $\mathcal{O}^Q_{\text{FS}}$ [168], which is the lightest primary of charge $Q$ within the theory. In fact, there is absolutely no difference with respect to the free-fermion CFT result in Eq. (3.194) at this order.

At this point, we want to stress that it is a priori not clear what the implications of the absence of SSB at leading order in the large-$N$ expansion are. At this order interactions between fermions are suppressed by the coupling $g_{\text{eff}} = \mathcal{O}(1/N)$ going to zero as $N \to \infty$. Therefore, at infinite $N$ the ground state is described by an exactly free Fermi surface. At finite values of $N$, however, corrections from interactions between fermions can in principle be studied in the correct framework of the Fermi surface EFT [118].

[55] It is maybe worth remarking that the spontaneously broken chiral symmetry of the GN model is discrete, and hence the Goldstone theorem does not predict the existence of a NG boson in this case.

In doing so, the corrections — in addition to their large-$N$ suppression — need to be properly organized in a large-$Q$ expansion corresponding to the appropriate low-energy expansion in powers of $1/(\mu r_0)^2$.

One distinct possibility is that the physics at the lowest energies remains the one of a weakly-interacting Fermi surface. But this is not the only possibility, since at finite $N$ we know that the Fermi surface is never exactly free, and interactions in the Fermi surface are not automatically suppressed at low enough energies, unlike in the superfluid EFT case [17, 122]. It is well-known that the effect of the four-Fermi interaction $(\bar{\Psi}\Psi)^2$ runs logarithmically with the excitations above the Fermi surface [118], hence running to strong coupling as $g_{\text{eff}} \propto \log(\mu/E_{\text{IR}})$ given the existence of any attractive four-Fermi channel. In particular, this suggests that the Fermi surface may always develop a condensate of Cooper pairs at an IR scale of $L^{-1} = E_{\text{IR}} \sim \exp\left(-(\text{constant})/g_{\text{UV}}\right)\mu = \exp\left(-(\text{constant}) \times N\right)\mu$, which leads to a gap of order $E_{\text{gap}} \sim \exp\left(-(\text{ constant}) \times N\right)\sqrt{Q}/L$ in the fermionic sector. This scenario seems to be somewhat generic, and it would predict that the lowest-lying operator at large charge $Q$ is described by a purely bosonic superfluid EFT. However, we note that this would be the case only for ultra-large values of the charge $Q$ exponentially large in $N$ so that the enhancement of the gap by the factor $\sqrt{Q}$ is able to overcome the exponential suppression in $N$.

Since we have not done the analysis of the four-Fermi interaction about the Fermi-sphere ground state, we do not know which of these two possibilities is actually realized.

#### Spectrum of fluctuations for the Gross–Neveu model

Now that we have identified the large-charge ground state, we can study the spectrum of fluctuations on top of it, while still working in flat space for convenience. As the ground state of the GN model is a filled Fermi sphere and no NG bosons arise, the supefluid EFT predictions from Section 2.2 do not apply here.

In principle, the fluctuations around the Fermi sphere ground state of the GN model can be both fermionic and bosonic in nature. The fermionic fluctuations are then clearly of the particle-hole-type, just like it happens in the free fermion critical theory (*i.e.* in the massless free fermion). Potentially occurring bosonic fluctuations on the other hand are due to the collective or composite field $\sigma$, more precisely its fluctuations.

In order to understand the effect of the $\sigma$-fluctuations on the ground state we have to compute the effective propagator of the $\hat{\sigma}$-field, which is equal to the full $\sigma$-field as the VEV of $\sigma = \hat{\sigma}/\sqrt{N}$ vanishes, $\langle\sigma\rangle = 0$. The effective propagator can be read off of the next-to-leading order quadratic effective action for the $\hat{\sigma} \sim \sigma$-field in Eq. (3.202). Most conveniently, this action is computed in momentum space[56]

$$\operatorname{Tr}\left(D^{(\mu)}\sigma D^{(\mu)}\sigma\right) = -\sum\!\!\!\!\!\!\int \mathrm{d}^2 p\, \tilde{\sigma}(-P)\tilde{\sigma}(P) \sum\!\!\!\!\!\!\int \frac{\mathrm{d}^2 k}{\beta(2\pi)^2} \operatorname{Tr}\left[\tilde{D}^{(\mu)}(K)\tilde{D}^{(-\mu)}(P-K)\right], \tag{3.213}$$

[56]Here, we use the notation $P = (\omega_n, \mathbf{p})$ for the momenta on $S^1_\beta \times \mathbb{R}^2$, where $\omega_n$ still are the fermionic Matsubara frequencies.

where we have made use of the symmetry property

$$D^{(\mu)}(x-y) = -D^{(-\mu)}(y-x) \tag{3.214}$$

satisfied the fermion propagator $D^{(\mu)}(x-y)$. The expression for the momentum-space finite-density fermionic propagator is given by

$$\tilde{D}^{(\mu)}(P) = i\frac{\Gamma_\nu\left(P^{(\mu)}\right)^\nu}{\left(P^{(\mu)}\right)^2}, \qquad P^{(\mu)} = (\omega_n - i\mu, \mathbf{p}). \tag{3.215}$$

The one-loop integral in Eq. (3.213) after some simple algebraic transformations can be reduced to

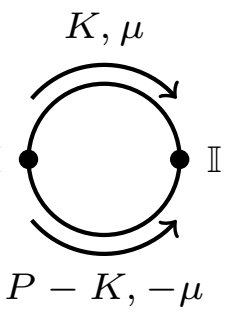

$$= 2P^2 I_2 - 4I_1, \tag{3.216}$$

where $I_1, I_2$ denote two scalar master integrals that are computed in detail in Appendix C.7. This result suffices to be able to obtain the action of the quadratic fluctuations $S^{(2)}_{\text{eff}}$ in a mass-independent regularization scheme in which the critical coupling $g_c$ is given by $g_c^{-1} = 0$. At zero temperature we find that

$$S^{(2)}_{\text{eff}} = \frac{1}{2}\text{Tr}(\sigma D^{(\mu)}\sigma D^{(\mu)}) = \frac{1}{2}\sum\!\!\!\!\!\!\int \mathrm{d}^2 p\, \tilde{\sigma}(-P)\tilde{\sigma}(P)\left[\frac{\sqrt{P^2}}{4} + \frac{\mu}{\pi}\right]. \tag{3.217}$$

We note that the non-local action above in fact does not describe stable bosonic fluctuations lying on top of the Fermi-sphere ground state. In particular, the momentum-independent part cannot properly be interpreted as actual mass term. Instead, it should rather be interpreted as a decay constant which is of order $\sim \mu$, similar to how it is the case in the unbroken phase of the model [281]. This is consistent with the fact that the $\sigma$-field describes fluctuations of a $\bar{\psi}\psi$ bound state on top of the Fermi sphere, which is only stable in the broken phase of the theory. As a consequence, it cannot generate any new conformal primaries within the local CFT spectrum at large charge. On the other hand, the fermionic particle-hole excitations are stable. The local CFT spectrum therefore only consists of operators associated with the stable particle-hole excitations.

#### Conformal dimensions and local CFT spectrum for the GN model

Up to this point we were content with treating the critical GN model in flat space by studying its finite-density properties in a large spatial box or torus of volume $V$. This was sufficient for the results we cared about. However, if we want to compute the scaling dimension of the lightest operator of charge $Q$, in order to take advantage of the state–operator correspondence, we have to put the theory on the cylinder $\mathbb{R}\times S^2_{r_0}$. This way the energy of the finite-density ground state — a Fermi sphere in the GN case — corresponds to the scaling dimension of the corresponding local primary operator within the CFT living at the critical point.

Fermionic theories — like their bosonic counterparts — can be mapped from flat space $\mathbb{R}^3$ to the cylinder $\mathbb{R} \times S^2_{r_0}$ using a Weyl transformation, as outlined in Appendix C.4.2, after which the finite-density kinetic term for the fermionic fields is given by

$$S = \int_{\mathbb{R}\times S^2} \mathrm{d}^3x \left[\bar{\Psi}(\Gamma_\mu D^\mu - \mu\Gamma_\tau + \sigma)\Psi\right], \qquad \text{with} \qquad Q = \int_{S^2} \bar{\Psi}\Gamma_\tau\Psi, \tag{3.218}$$

where $\Gamma_\mu D^\mu$ denotes the Dirac operator on the cylinder $\mathbb{R} \times S^2_{r_0}$.

Analogously to the results for the grand potential $\Omega(\mu)$ of the GN model in flat space in Eq. (3.204) and of the free fermion model on the cylinder in Eq. (3.191), the thermodynamic grand potential at finite temperature on the sphere $S^1_\beta \times S^2_{r_0}$ for the GN model is found to be

$$\Omega(\mu) = \frac{\langle\sigma\rangle^2}{8g} - \frac{1}{V}\sum_{j=\frac{1}{2}} (2j+1)\left[\sqrt{\omega_j^2 + \sigma_0^2} + \text{thermal contributions}\right], \tag{3.219}$$

where $V = 4\pi r_0^2$ is the volume of the sphere and $\omega_j = (j+1/2)/r_0$ are the eigenvalues of the Dirac operator on $S^2_{r_0}$. Clearly, the gap equation still does not admit any non-trivial solution $\langle\sigma\rangle \neq 0$ at zero temperature and at criticality for any value of $\mu > 0$, just as it did in flat space.[57] The zero temperature grand potential now simply reduces to the free-fermion result in Eq. (3.192)

$$\Omega(\mu) = -\frac{1}{4\pi r_0^2}\left[\sum_{\omega_j>\mu}(2j+1)\omega_j + \mu\sum_{\omega_j<\mu}(2j+1)\right], \tag{3.220}$$

where we have explicitly introduced the sphere volume $V = 4\pi r_0^2$. Hence, it describes the Fermi-sphere ground state for massless fermions, which we have analysed and computed in Section 3.3.1 already. We briefly repeat the results for the $U(1)_B$ charge and the scaling dimension here,

$$\frac{Q}{2N} = \lfloor\mu r_0\rfloor(\lfloor\mu r_0\rfloor+1), \qquad \frac{\Delta_{\mathrm{FS}(Q)}}{2N} = \frac{1}{3}\lfloor\mu r_0\rfloor(\lfloor\mu r_0\rfloor+1)(2\lfloor\mu r_0\rfloor+1) = \frac{1}{3}\left(\frac{Q}{2N}\right)\sqrt{4\left(\frac{Q}{2N}\right)+1}. \tag{3.221}$$

Only discrete values of the charge permissible — $Q/2N = 0, 2, 6, \ldots$ — since the Fermi shell needs to always be filled. The floor function implements the discretized cylinder energy levels and the formal macroscopic limit $r_0 \to \infty$ reproduces the flat space results derived in Eq. (3.212). This limit is of course analogous to the large-chemical-potential limit $\mu \to \infty$. The ground state corresponds again to the scalar primary operator $\mathcal{O}_{\mathrm{FS}}$ first constructed explicitly in [168] in the free fermion CFT. The lowest primary $\mathcal{O}_{\mathrm{FS}}$ in the GN model is exactly the same operator as in the free fermion CFT due to the fact that there is no SSB and the auxiliary HS field $\sigma$ does not condense, at least to leading order in $N$.

We reiterate that for the large-charge sector of the GN model there are no NG fluctuations which describe new primaries with gap of order order $\mathcal{O}(1)$ from the Fermi-sphere primary $\mathcal{O}_{\mathrm{FS}}$. Instead, new generally spinful and fermionic primaries with the same charge $Q$ and a gap of order $\mathcal{O}(1)$ relative to $\mathcal{O}_{\mathrm{FS}}$ are created via particle-hole excitations. As we have discussed before, the fluctuations of the collective field $\sigma$ cannot describe new primaries within some hypothetical EFT description around the Fermi-sphere ground state. In principle, this is a large-$N$ result and it is important to see if these effects persist for finite values of $N$, and if there exists some sort of EFT description in terms of a Fermi liquid in the spirit of [118, 167].

[57] The sums in the above expression are best regularized in a way as to preserve diff-invariance, *e.g.* by a smooth cut-off function like $e^{-\omega_j/\Lambda}$. Alternatively, zeta-function regularization, as done for the free fermion in Appendix C.6, works as well.

Finally, we want to stress again that the lack of SSB within the large-charge sector of the GN model to leading order in $N$ can in principle have two distinct explanations:

- Either, the four-Fermi interaction around the leading-$N$ Fermi sphere truly is repulsive in every single channel and the large-$N$ ground state in the GN model truly is a Fermi sphere. This would actually be the first non-trivial (non-free) example of such behaviour at large charge;

- Or the four-Fermi has at least one attractive channel, therefore runs logarithmically to strong coupling, exhibits a BCS condensate and also a gap in the fermionic sector of the theory of order $\exp(-1/g_{\text{eff}})$, where $g_{\text{eff}} \propto 1/N$ is the effective coupling at the cut-off. This would also mean that the condensate and the gap are exponentially small and suppressed to all orders in $1/N$ perturbation theory.

### 3.3.3 The NJL and the Cooper model at large $N$: Superfluid phases in fermionic CFTs

The Nambu–Jona–Lasinio (NJL) model — or chiral GN model — is a well-established and time-honoured four-Fermi interaction model exhibiting a continuous chiral symmetry in four spacetime dimensions. By dimensional reduction, in the language outlined in the beginning of Section 3.3.1 and Appendix C.4.3, the action in terms of four-dimensional reducible Dirac spinors in $D = 3$ reads

$$S = \int \mathrm{d}^3 x \left[ \sum_{i=1}^{N} \bar{\Psi}_i \Gamma_\mu \partial^\mu \Psi_i - \frac{g}{N} \left[ \Big( \sum_{i=1}^{N} \bar{\Psi}_i \Psi_i \Big)^2 - \Big( \sum_{i=1}^{N} \bar{\Psi}_i \Gamma_5 \Psi_i \Big)^2 \right] \right]. \tag{3.222}$$

In $D = 3$ the $\bar{\Psi}\Gamma_5\Psi$ bilinear is invariant under an $Sp(2N)$ symmetry and hence the total global internal symmetry group of the model is reduced with respect to the GN model in Eq. (3.196) down to

$$[O(2N) \times O(2N)] \cap Sp(2N) = U(N) = SU(N) \times U(1)_B \,. \tag{3.223}$$

In addition, on top of the standard $U(1)_B$ baryon symmetry, there now exists an extra $U(1)_A$ symmetry that extends the $\mathbb{Z}_2$ chiral symmetry of the GN model, bringing the total symmetry up to $SU(N) \times U(1)_B \times U(1)_A$. We mainly focus our attention on the Abelian $U(1)_B \times U(1)_A$ symmetry and take advantage of the unbroken global $SU(N)$ group factor still present to consider the theory in the large-$N$ limit.

The continuous $U(1)_A$ chiral symmetry of the model only arises when the two quartic interaction in Eq. (3.222) are linearly combined with precisely the same coupling $g$ and opposite signs. It acts linearly on the spinors as

$$\Psi_i \to e^{i\alpha\Gamma_5} \Psi_i \,. \tag{3.224}$$

The $U(1)$-NJL model therefore exhibits a different phase transition compared to the GN model, where the critical phase of the theory separates two phases in which the continuous $U(1)_A$ chiral symmetry is either broken or unbroken (compared to the discrete $\mathbb{Z}_2$ symmetry defining the different phases for the GN model). In contrast to the GN model, where the spontaneously broken chiral symmetry is discrete, in the broken phase of the NJL model a NG boson now arises. As we will see, this broken phase persists also at finite density.

Later, in Section 3.3.4 we will introduce a $SU(2)$-generalization of the NJL model given by the action in Eq. (3.222). For this reason, in order to avoid any confusion, we will refer to the NJL model as the $U(1)$-NJL model and to its generalization as the $SU(2)$-NJL model.

Just as the GN model, the $U(1)$-NJL model represents a deformation of the free-fermion CFT in Eq. (3.188) via an irrelevant four-Fermi interaction and is therefore not renormalizable in standard perturbation theory. At finite values of $N$, the $U(1)$-NJL model is tractable via its known Yukawa-type UV completion in a small-$\epsilon$ expansion around $D = 2+\epsilon$ and $D = 4-\epsilon$. This $U(1)$-NJL–Yukawa UV completion of the four-Fermi $U(1)$-NJL model in $2 < D < 4$ encompasses a complex scalar DoF $\Phi$ and is given by[58]

$$S = \int \mathrm{d}^D x \left[ \bar{\Psi}_i \left( \Gamma_\mu \partial^\mu + \Phi \left( \frac{(1+\Gamma_5)}{2} \right) + \Phi^* \left( \frac{(1-\Gamma_5)}{2} \right) \right) \Psi_i + \frac{1}{g_Y} \partial_\mu \Phi^* \partial^\mu \Phi \right]. \tag{3.225}$$

In this action, the $U(1)_A$ chiral symmetry is realized as

$$\Psi_i \to e^{i\alpha\Gamma_5} \Psi_i, \qquad \Phi \to e^{-2i\alpha} \Phi. \tag{3.226}$$

The above action can be deduced by performing the appropriate HS transform in the $U(1)$-NJL action in Eq. (3.222) replacing the four-Fermi interaction with a scalar collective field $\Phi$,

$$S = \int \mathrm{d}^3 x \left[ \bar{\Psi}_i \left( \Gamma_\mu \partial^\mu + \Phi \left( \frac{(1+\Gamma_5)}{2} \right) + \Phi^* \left( \frac{(1-\Gamma_5)}{2} \right) \right) \Psi_i + \frac{N}{4g} |\Phi|^2 \right], \tag{3.227}$$

tuning the the theory to the critical UV limit by neglecting the $|\Phi|^2$ term and introducing a kinetic term for the field $\Phi$ along with the Yukawa coupling $g_Y$. The collective field $\Phi$ replaces the complex bilinear $\bar{\Psi}_i \Psi_i + \bar{\Psi}_i \Gamma_5 \Psi_i$, while its complex conjugate $\Phi^*$ replaces $\bar{\Psi}_i \Psi_i - \bar{\Psi}_i \Gamma_5 \Psi_i$.

The UV interacting fixed point at $g = g_c$ of the $U(1)$-NJL model is recovered as the IR interacting fixed point is reached at $g_Y \to \infty$ of the associated $U(1)$-NJL-Yukawa theory. As in the GN model, in the IR limit the no longer dynamical field $\Phi$ in the Yukawa-type model is identified with the HS collective field $\Phi$.

[58] While the GNY includes a free scalar DoF, the $U(1)$-NJL–Yukawa comes with a complex one. This not surprising given the fact that the chiral symmetry in the GN model is $\mathbb{Z}_2$ while the chiral symmetry for the $U(1)$-NJL model is a continuous $U(1)_A$.

#### Cooper model and the PG transformation

As we will see later, the NJL model supports a condensate and a superfluid ground state at large $U(1)_A$ charge, and just like the GN model no condensate at finite or large $U(1)_B$ baryon charge. At first sight, there seems to be no intuitive way to understand as to why there is condensate at finite density for the axial symmetry. However, as it turns out, there is a way to understand the presence of the axial condensate in rather intuitive physical terms. This interpretation comes about from the fact that there exists a transformation — usually referred to as a Pauli–Gursey (PG) transformation [270, 271] — which relates the $U(1)$-NJL model to a model with a Cooper-type di-fermion interaction term [269, 272].

Generically, both GN-type and NJL-type models exhibit a fermion-anti-fermion interaction. Often, in particular in the context of condensed-matter physics, physical systems are instead described by a di-fermion interaction. For example, models with di-fermion interactions are important in the study of superconductivity (arising via the process of Cooper pairing). Generally speaking, large-$N$ fermionic models intended to study and describe superconductivity at finite $U(1)_B$ baryon charge include di-fermion interaction terms of the form

$$(4f)_{\mathrm{CP}} = \frac{g}{N}\sum_{i,j} \bar{\Psi}_i C \bar{\Psi}_i^T \, \Psi_j^T C \Psi_j \,. \tag{3.228}$$

Additionally, there may also be more standard GN-type or NJL-type interactions present in the action [269, 282]. For our purposes, it suffices to consider the model with just the above Cooper-pair interaction term $(4f)_{\mathrm{CP}}$. The full action reads

$$S = \int \mathrm{d}^3 x \left[ \bar{\Psi}_i \Gamma_\mu \partial^\mu \Psi_i + \frac{g}{N} \bar{\Psi}_i C_4 \bar{\Psi}_i^T \, \Psi_j^T C_4 \Psi_j \right]. \tag{3.229}$$

In our notation for the four-dimensional reducible representation of the gamma matrices we have $C_4 = \Gamma_2$, for more details see Appendix C.4.3. At criticality — in contrast to the situation in the GN model discussed in Section 3.3.2 — the Cooper model with the action above exhibits a non-trivial solution to the gap equation at zero temperature also at finite $U(1)_B$ baryon chemical potential giving rise to a superconducting phase and a finite-density quantum phase transition.

Perhaps surprisingly, the Cooper model at finite $U(1)_B$ chemical potential turns out to be dual to the $U(1)$-NJL model at finite $U(1)_A$ chemical potential in the sense of [269], shedding light on the physical nature of the axial condensate appearing in the NJL model. This duality is found by applying the Pauli–Gursey (PG) transformation [270–272] given by

$$\Psi_i \mapsto \frac{1}{2}\left[(1-\Gamma_5)\Psi_i - (1+\Gamma_5) C_4 \bar{\Psi}_i^T\right], \qquad \bar{\Psi}_i \mapsto \frac{1}{2}\left[\bar{\Psi}_i (1+\Gamma_5) - \Psi_i^T C_4 (1-\Gamma_5)\right]. \tag{3.230}$$

For a more detailed discussion of the above PG transformation we refer to the Appendix C.5 outlining the properties of said transformation.

In the following we perform all relevant computations within the $U(1)$-NJL model at finite $U(1)_A$ density.

Naturally, all computations can be repeated in the Cooper model with action in Eq. (3.229) leading to the exact same results. The main advantage of the Cooper model is that the physical nature of the elusive axial condensate present in the $U(1)$-NJL model at large $N$ becomes much more transparent: it is in fact a condensate of Cooper pairs, describing a system exhibiting superconductivity. The attractive di-fermion interaction $(4f)_{\text{CP}}$ gives rise to a Cooper instability within the theory, effectively leading to a system of condensing bosons at large charge. This nicely explains the similarity of our results in the $U(1)$-NJL model to those derived for the $O(2N)$ scalar model in Section 3.1.

We end this segment with a few relevant remarks:

- In fact, both the Cooper model and $U(1)$-NJL model exhibit a $U(1)_A \times U(1)_B$ symmetry. Importantly, under the PG transformation in Eq. (3.230) the $U(1)_B$ chemical potential term of the Cooper model gets mapped into (minus) the $U(1)_A$-chemical potential term of the $U(1)$-NJL model, and vice versa. Hence, any results obtained in the $U(1)$-NJL model at finite $U(1)_A$ axial chemical potential necessarily also applies to the Cooper model at finite $U(1)_B$ baryon chemical potential.
- At the critical point, relevant quantities computed in both the Cooper model and the $U(1)$-NJL model agree up to a PG transformation. We are able to explicitly check this matching in both models at leading order in $N$ on the ground-state energy at finite density for the respective $U(1)$ symmetries.
- Finally, the PG transformation in Eq. (3.230) corresponds to a linear involution at the level of the path integral in the fermionic variables $\Psi_i^\dagger, \Psi_i$. As a consequence, it can only affect the path-integral measure $\mathscr{D}\Psi_i^\dagger \mathscr{D}\Psi_i$ up to a trivial rescaling.

**Symmetry-breaking at large $N$: Leading-order action and gap equation**

As discussed quickly in Section 3.3.2, the presence of a condensate is very important for the efficacy of the large-charge approach. We investigate this question for the NJL model by discussing the symmetry-breaking for the NJL model at large $N$.

We are interested in the $U(1)$-NJL model, invariant under the above discussed $U(1)_A \times U(1)_B$ symmetry, at finite $U(1)_A$ axial chemical potential $\mu$. After having introduced the appropriate complex HS field $\Phi$ the action can be written as

$$S = \int_{S^1_\beta \times \mathbb{R}^2} \left[ \bar{\Psi}_i \left( \Gamma_\nu \partial^\nu - \mu \Gamma_3 \Gamma_5 + \Phi P_+ + \Phi^* P_- \right) \Psi + \frac{N}{4g} |\Phi|^2 \right], \tag{3.231}$$

where $P_\pm = (1 \pm \Gamma_5)/2$ denotes the standard four-dimensional chiral projectors. The axial chemical potential $\mu$ is sourcing a finite charge density for the chiral symmetry acting as

$$\Psi_i \to e^{i\alpha\Gamma_5} \Psi_i, \qquad \Phi_i \to e^{-2i\alpha} \Phi_i. \tag{3.232}$$

In the standard large-$N$ methodology the collective field $\Phi$ at large $N$ exhibits small fluctuations $\hat{\varphi}$ and is decomposed in terms of a constant (*i.e.* homogeneous) VEV $\langle\Phi\rangle$ plus fluctuations,

$$\Phi = \langle\Phi\rangle + \hat{\Phi}/\sqrt{N}, \qquad \Phi^* = \langle\Phi\rangle^* + \hat{\Phi}^*/\sqrt{N}, \tag{3.233}$$

generating a consistent perturbative $1/N$ expansion [134]. The thermodynamic potential $\Omega(\mu)$ is found as the leading-$N$ contribution to the action in Eq. (3.231) by setting the fluctuations of the collective field $\Phi$ to zero, $\hat{\Phi} = \hat{\Phi}^* = 0$,

$$\Omega(\mu) = \frac{|\langle\Phi\rangle|^2}{8g} - \frac{1}{2}\int^{\Lambda} \frac{\mathrm{d}^2 p}{(2\pi)^2}\left[\omega_+ + \omega_- + \frac{2}{\beta}\log\left(1 + e^{-\beta\omega_+}\right) + \frac{2}{\beta}\log\left(1 + e^{-\beta\omega_-}\right)\right], \tag{3.234}$$

where we have introduced a momentum cut-off $\Lambda$ and the one-particle on-shell energies

$$\omega_\pm^2 := |\langle\Phi\rangle|^2 + \left(|\mathbf{p}| \pm \mu\right)^2. \tag{3.235}$$

We see immediately, that $\langle\Phi\rangle$ can be assumed to be real. Compared to the GN model in Section 3.3.2, the novelty in the $U(1)$-NJL model is that — since $\omega_\pm \geq 0$ — no Fermi sphere solution can arise as long as $\langle\Phi\rangle \neq 0$. Under the assumption that this is the case we can neglect the logarithmic thermal contributions in the zero-temperature limit. Hence, the grand potential at zero temperature for $\langle\Phi\rangle \neq 0$ reads

$$\lim_{\beta\to\infty} \Omega(\mu) = \frac{\langle\Phi\rangle^2}{8g} - \frac{1}{2}\int^{\Lambda} \frac{\mathrm{d}^2 p}{(2\pi)^2}\left[\sqrt{(p+\mu)^2 + \langle\Phi\rangle^2} + \sqrt{(p-\mu)^2 + \langle\Phi\rangle^2}\right]. \tag{3.236}$$

After introducing the critical coupling $g_c^{-1} = 2\Lambda/\pi$, the gap equation for the collective field $\Phi$ at zero temperature reads

$$0 = 4\frac{\partial^2\Omega}{\partial\langle\Phi\rangle^2} = \frac{1}{2}\left(\frac{1}{g} - \frac{1}{g_c}\right) + \frac{1}{\pi}\left[\sqrt{\langle\Phi\rangle^2 + \mu^2} - \mu\cdot\operatorname{arctanh}\left(\frac{\mu}{\sqrt{\langle\Phi\rangle^2 + \mu^2}}\right)\right]. \tag{3.237}$$

We look for a real and positive solution $\langle\Phi\rangle > 0$. In the critical limit $g = g_c$ we find such a non-trivial solution to the gap equation and finite value of the axial chemical potential,

$$\langle\Phi\rangle = \mu\sqrt{\kappa_0^2 - 1}, \tag{3.238}$$

where the number $\kappa_0$ is found to be the solution to the equation

$$\kappa_0 \tanh(\kappa_0) = 1, \tag{3.239}$$

and numerically we have

$$\langle\Phi\rangle/\mu = 0.6627\ldots. \tag{3.240}$$

The existence of this solution implies that the finite-axial-density ground state spontaneously breaks the $U(1)_A$ symmetry by giving a non-trivial VEV $\langle\phi\rangle$ to the collective field $\Phi$ — which also plays the role of the order parameter associated to the spontaneously broken $U(1)_A$ axial symmetry. Symmetry-restoration can only occur at zero density $\mu = 0$, since conformal invariance prohibits the existence of another scale separating the broken phase and the unbroken phase.

The renormalized grand potential $\Omega(\mu)$ can be computed for a generic constant and complex configuration of the collective field $\Phi = \Phi_c$ =const. without minimizing, making the situation more explicit. As the divergent part of $\Omega(\mu)$ is $\mu$-independent, we can compute its minimal subtraction and find

$$\begin{aligned}\Omega(\mu) - \Omega(0) &= -\frac{1}{2}\int \frac{d^2 p}{(2\pi)^2}\left[\omega_+\big|_{|\Phi_c|} + \omega_-\big|_{|\Phi_c|} - 2\sqrt{\mathbf{p}^2 + |\Phi_c|^2}\right] \\ &= -\frac{1}{12\pi}\left[3|\Phi_c|^2\mu \operatorname{arctanh}\left(\frac{\mu}{\sqrt{|\Phi_c|^2+\mu^2}}\right) + (\mu^2 - 2|\Phi_c|^2)\sqrt{|\Phi_c|^2+\mu^2} + 2|\Phi_c|^3\right].\end{aligned} \tag{3.241}$$

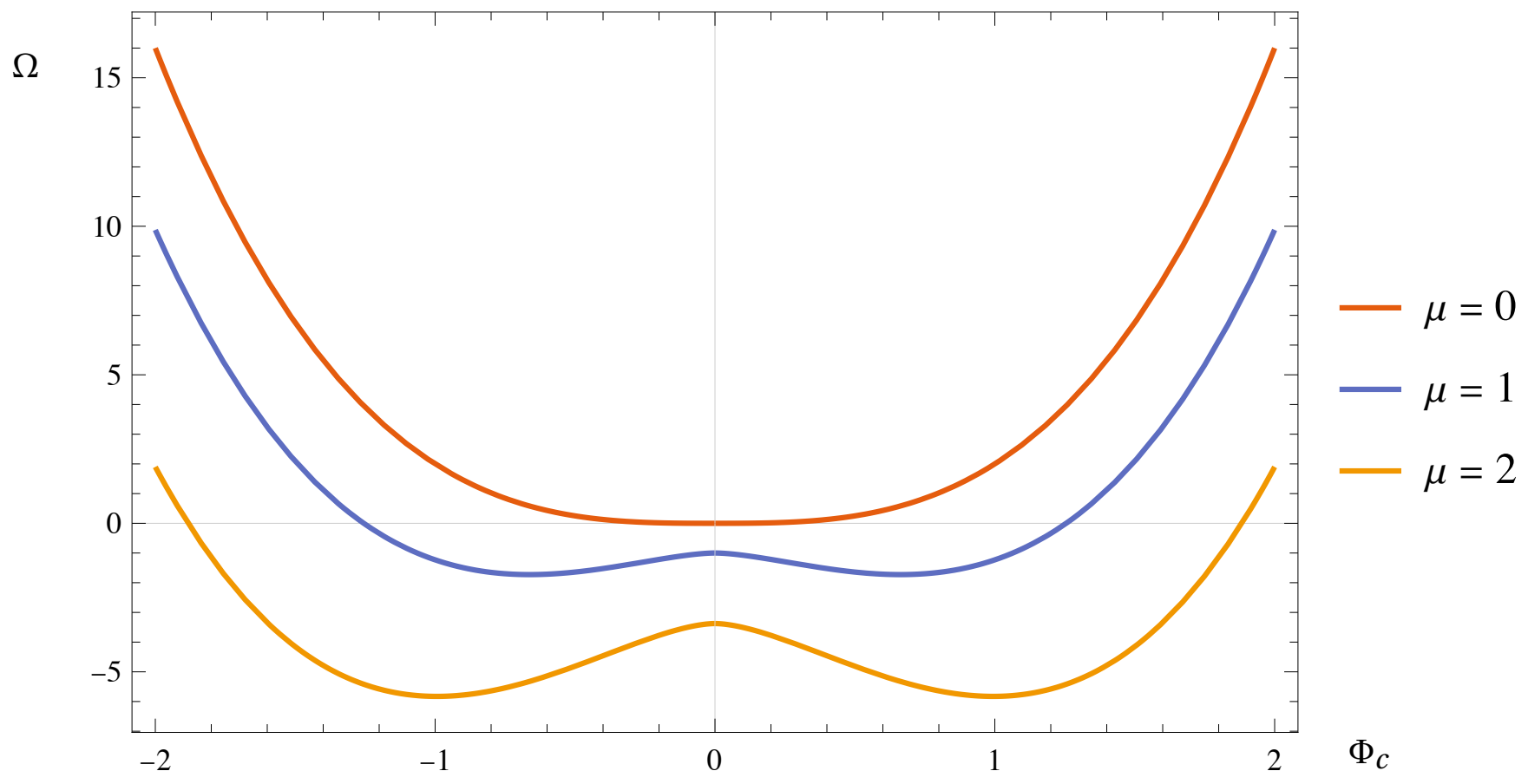

Figure 3.7: The thermodynamic potential $\Omega(\mu)$ — which represents the leading-$N$ result — for the $U(1)$-NJL model as a function of $\Phi = \Phi_c$ =const. for different values of the chemical potential $\mu$. The two minima appearing at $\Phi_c \neq 0$ for $\mu > 0$ signal the spontaneous breaking of the $U(1)_A$ symmetry.

Afterwards, we add again the renormalized value of $\Omega(0)$, which turns out to be exactly the same integral already solved for the GN model in Section 3.3.2,

$$\Omega(0) = \frac{|\Phi_c|^3}{6\pi}. \tag{3.242}$$

In the end, the renormalized grand potential $\Omega(\mu)$ reads

$$\Omega(\mu) = -\frac{1}{12\pi}\left[3|\Phi_c|^2\mu \operatorname{arctanh}\left(\frac{\mu}{\sqrt{|\Phi_c|^2+\mu^2}}\right) + (\mu^2 - 2|\Phi_c|^2)\sqrt{|\Phi_c|^2+\mu^2}\right]. \tag{3.243}$$

As expected, this grand potential is invariant under the $U(1)_A$ symmetry, but has a $U(1)$s worth of equivalent vacua at $|\Phi_c| = \mu(\kappa_0^2 - 1)^{1/2}$. In Figure 3.7 we plot the grand potential $\Omega(\mu)$ as a function of $\Phi_c$ for $\text{Im}(\Phi_c) = 0$ and for different values of the chemical potential $\mu$.

The obtained ground-state solution for the $U(1)$-NJL model corresponds to a superfluid of $U(1)_A$ charge density $\rho = Q/V$ and energy density $f_c(Q)/V$, respectively, given by

$$\frac{\rho}{2N} = \frac{\kappa_0^3\mu^2}{4\pi}, \qquad \frac{f_c(Q)}{V} = \frac{1}{6\pi\kappa_0^{3/2}}\left(4\pi\frac{\rho}{2N}\right)^{3/2}. \tag{3.244}$$

This ground-state energy computes the leading-order result for the conformal dimension $\Delta_{\text{SF}}(Q)$ of the associated operator $\mathscr{O}^Q_{\text{SF}}$ corresponding to the superfluid ground state within the LCE. We use the subscript $_{\text{SF}}$ to denote that the operator corresponds to a superfluid ground state and to distinguish it from the Fermi-sphere operator $\mathscr{O}^Q_{\text{FS}}$ encountered both for the free-fermion CFT in Section 3.3.1 and the GN model in Section 3.3.2. Importantly, no Fermi-sphere ground state arises in the finite axial-chemical potential and zero baryon-chemical potential sector of the $U(1)$-NJL model, as all charge is contained within the superfluid condensate.

It is important to investigate whether a superfluid ground state also arises at finite $U(1)_B$ baryon charge. Unfortunately, at zero $U(1)_A$ charge density with chemical potential $\mu_A = 0$ and finite $U(1)_B$ charge density with chemical potential $\mu_B \neq 0$ we recover precisely the same eigenvalues already appearing in the GN model at finite $U(1)_B$ charge density, see Section 3.3.2,

$$\omega_\pm = \sqrt{\mathbf{p}^2 + \langle\Phi\rangle^2} \pm \mu_B \,. \tag{3.245}$$

This again describes a Fermi-sphere ground state with charge and energy density given as in Eq. (3.212), hence corresponding to the same Fermi-sphere operator $\mathscr{O}^Q_{\text{FS}}$ within the CFT.

There exists the possibility that a transition between a superfluid ground state and a Fermi-sphere ground state occurs somewhere in the $(\mu_A, \mu_B)$ phase diagram, where the $U(1)_A$ symmetry is restored even for finite values of $\mu_A$. The question as to whether or not this transition exists is certainly interesting and worth investigating. It will be particularly relevant to describe large-charge operators within the CFT given by the critical $U(1)$-NJL theory that are charged under both of the $U(1)$ symmetries. As we are focussing on operators which are only charged under a single $U(1)$, in the context of this chapter the existence or non-existence of this transition is irrelevant.

#### Spectrum of fluctuations for the Nambu–Jona–Lasinio model

After having identified the large-charge ground state at finite $U(1)_A$ chemical potential as a superfluid, we can study the spectrum of fluctuations on top if it. In particular, we can expect to find the superfluid phonon — equipped with the dispersion relation $\omega = -|\mathbf{p}|/2 + \dots$ and paired with a massive mode of mass of order $\mu$ — which is well-known from the discussion of the large-charge EFT description in Chapter 2 and also found in the $O(2N)$ vector model at large $N$ in Section 3.1.6. To do so, we again work in flat space.

As we have shown, fixing the axial charge in the $U(1)$-NJL model leads to a non-trivial VEV $\langle\Phi\rangle \neq 0$ for the field $\Phi$. From the form the of the functional determinant in Eq. (3.236) we can deduce that all fermions in the theory acquire a mass of order $\mu$,

$$m_F^2 = \mu^2 + \langle\Phi\rangle^2 = \kappa_0^2\mu^2 \,, \tag{3.246}$$

implying that the $U(N)$ flavour symmetry remains unbroken, while the $U(1)_A$ axial symmetry is spontaneously broken.

The large-charge EFT predictions outlined in Section 2.2 can be verified in the present model explicitly by computing the sub-leading in the $1/N$ expansion of the functional determinant, which is quadratic in the fluctuations $\hat{\Phi}, \hat{\Phi}^*$ of the collective field and gives the propagator of said fluctuations over the vacuum $\langle\Phi\rangle$. For our purposes, it is most convenient to decompose the fluctuations into real and imaginary parts,

$$\hat{\Phi} = \hat{\sigma} + i\hat{\pi}\,. \tag{3.247}$$

In order to identify the spectrum of fluctuations over the condensate $\langle\Phi\rangle$, we compute the inverse propagator of the fields $\hat{\sigma}, \hat{\pi}$ at the one-fermion-loop order. The large-$N$ action around the vacuum $\langle\Phi\rangle$, which we assume to be real, is given by

$$S_{\text{eff}} = \int \mathrm{d}^3x \left[\bar{\Psi}_i\Big(\Gamma_\nu\partial^\nu + \langle\Phi\rangle - \mu\Gamma_3\Gamma_5\Big)\Psi_i + \frac{1}{\sqrt{N}}(\hat{\sigma}\bar{\Psi}_i\Psi_i + i\hat{\pi}\bar{\Psi}_i\Gamma_5\Psi_i)\right], \qquad \langle\Phi\rangle = \mu\sqrt{\kappa_0^2 - 1}\,. \tag{3.248}$$

The momentum-space representation of the fermionic propagator reads

$$\tilde{D}^{(\mu,\langle\Phi\rangle)}(P) = \big(-i\Gamma_\nu P^\nu + \langle\Phi\rangle - \mu\Gamma_3\Gamma_5\big)^{-1} = \frac{\big(\omega^2 + k^2 + \langle\Phi\rangle^2 - \mu^2 + 2\mu(i\omega\Gamma_3 + \Phi_0)\Gamma_3\Gamma_5\big)}{\big(\omega^2 + \langle\Phi\rangle + (\mu+k)^2\big)\big(\omega^2 + \langle\Phi\rangle^2 + (\mu-k)^2\big)}\big(i\Gamma_\nu P^\nu + \langle\Phi\rangle - \mu\Gamma_3\Gamma_5\big)\,, \tag{3.249}$$

where $P = (\omega, \mathbf{p})$ denotes the momentum in flat space at zero temperature. Due to the absence of any Fermi-sphere contribution we can without obstruction directly work at zero temperature $\beta \to \infty$ and omit all of the temperature-dependent contributions at the start of the computation. Hence, we can also consider the Matsubara frequencies as continuous $\omega_n \to \omega$. Also note that the fermion propagator is properly anti-symmetric,

$$\tilde{D}^{(\mu,\langle\Phi\rangle)}(-P) = -\tilde{D}^{(-\mu,-\langle\Phi\rangle)}(P)\,. \tag{3.250}$$

Via the fermionic propagator $D^{(\mu,\langle\Phi\rangle)}(P)$ we can derive the inverse propagator for the scalar fluctuations $\hat{\sigma}, \hat{\pi}$ in terms of the following momentum-space loop integrals:

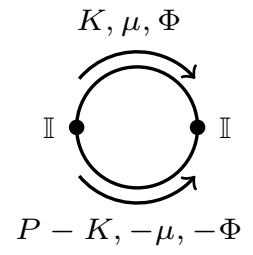

$$= \tilde{G}^{-1}_{\sigma\sigma}(P) = -\int \frac{\mathrm{d}^3k}{(2\pi)^3}\mathrm{Tr}\big[\tilde{D}^{(\mu,\langle\Phi\rangle)}(K)\,\tilde{D}^{(-\mu,-\langle\Phi\rangle)}(P-K)\big]\,, \tag{3.251}$$

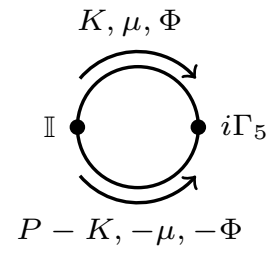

$$= \tilde{G}^{-1}_{\sigma\pi}(P) = -i\int \frac{\mathrm{d}^3k}{(2\pi)^3}\mathrm{Tr}\big[\tilde{D}^{(\mu,\langle\Phi\rangle)}(K)\,\Gamma_5\,\tilde{D}^{(-\mu,-\langle\Phi\rangle)}(P-K)\big]\,, \tag{3.252}$$

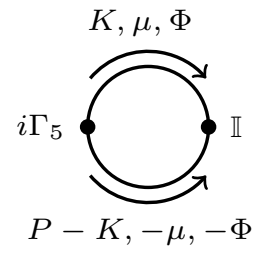

$$= \tilde{G}^{-1}_{\pi\sigma}(P) = -i\int \frac{\mathrm{d}^3k}{(2\pi)^3}\mathrm{Tr}\big[\Gamma_5\,\tilde{D}^{(\mu,\langle\Phi\rangle)}(K)\,\tilde{D}^{(-\mu,-\langle\Phi\rangle)}(P-K)\big]\,, \tag{3.253}$$

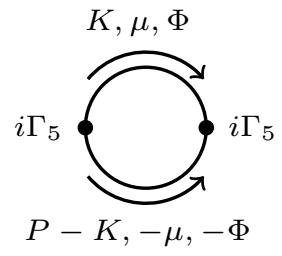

$$= \tilde{G}^{-1}_{\pi\pi}(P) = \int \frac{\mathrm{d}^3k}{(2\pi)^3}\mathrm{Tr}\big[\Gamma_5\,\tilde{D}^{(\mu,\langle\Phi\rangle)}(K)\,\Gamma_5\,\tilde{D}^{(-\mu,-\langle\Phi\rangle)}(P-K)\big]\,. \tag{3.254}$$

All of the above loop integrals can be conveniently expanded in the regime $(P/\mu) \ll 1$, which we care about while extracting the dispersion relations. Note that the zeroth order corresponds to the $P = 0$ result and needs to be regularized by subtracting the corresponding expression at $\mu = 0$, since the divergence is always $\mu$-independent. For details on the computations see Appendix C.7.3. The result for the zeroth order in $P$ of the above loop integrals reads

$$\tilde{G}^{-1}(P)\Big|_{\mathcal{O}(0)} = \begin{pmatrix} \tilde{G}^{-1}_{\sigma\sigma}(0) & \tilde{G}^{-1}_{\sigma\pi}(0) \\ \tilde{G}^{-1}_{\pi\sigma}(0) & \tilde{G}^{-1}_{\pi\pi}(0) \end{pmatrix} = \frac{\kappa_0 \mu}{\pi} \begin{pmatrix} 1 & 0 \\ 0 & 0 \end{pmatrix}, \tag{3.255}$$

where the coefficient $\kappa_0$ is again given by the solution to $\kappa_0 \tanh(\kappa_0) = 1$ and we have $\langle \Phi \rangle = \mu\sqrt{\kappa_0^2 - 1}$. Expectedly, beyond the zeroth order no regularization is required any more. The linear order in $P$ reads

$$\tilde{G}^{-1}(P)\Big|_{\mathcal{O}(P/\mu)} = \omega \frac{\kappa_0}{2\pi} \begin{pmatrix} 0 & -1 \\ 1 & 0 \end{pmatrix}, \tag{3.256}$$

and the quadratic order is given by

$$\tilde{G}^{-1}(P)\Big|_{\mathcal{O}(P^2/\mu^2)} = \begin{pmatrix} \frac{(2\kappa_0^2-1)\omega^2}{12\pi\kappa_0(\kappa_0^2-1)\mu} + \frac{(3\kappa_0^6-2\kappa_0^4-2\kappa_0^2+2)\mathbf{p}^2}{24\pi\kappa_0^3(\kappa^2-1)\mu} & 0 \\ 0 & \frac{\kappa_0\omega^2}{4\pi(\kappa_0^2-1)\mu} + \frac{\kappa_0^3\mathbf{p}^2}{8\pi(\kappa_0^2-1)\mu} \end{pmatrix}. \tag{3.257}$$

The dispersion relations of the fluctuation modes on top of the superfluid ground state are extracted from the zeroes of the inverse propagator, which is given by the matrix

$$\tilde{G}^{-1}(P) = \begin{pmatrix} \frac{\kappa_0\mu}{\pi} + \frac{2\kappa_0^2(2\kappa_0^2-1)\omega^2 + (3\kappa_0^6-2\kappa_0^4-2\kappa_0^2+2)\mathbf{p}^2}{24\pi\kappa_0^3(\kappa_0^2-1)\mu} & -\frac{\kappa_0}{2\pi}\omega \\ \frac{\kappa_0}{2\pi}\omega & \frac{2\kappa_0\omega^2+\kappa_0^3\mathbf{p}^2}{8\pi(\kappa_0^2-1)\mu} \end{pmatrix} + \mathcal{O}\left(P^3/\mu^3\right). \tag{3.258}$$

They are computed to be

$$\omega_1^2 = -\frac{1}{2}\mathbf{p}^2 + \dots, \tag{3.259}$$

$$\omega_2^2 = -12\frac{\left(\kappa_0^2-1\right)\kappa_0^4}{\left(2\kappa_0^2-1\right)}\mu^2 - \frac{\left(5\kappa_0^6 - 5\kappa_0^4 - \kappa_0^2 + 2\right)}{2\kappa_0^2(2\kappa_0^2-1)}\mathbf{p}^2 + \dots. \tag{3.260}$$

We directly recognize the expected massless conformal superfluid NG mode with speed of sound $c_s^2 = 1/2$. Additionally, and also as expected from the EFT approach, we find a gapped NG mode of mass of order $\mathcal{O}(\mu)$.

The result in Eq. (3.259) can be considered as one of the main results of this section, constituting one of the main prediction from the EFT approach outlined in Section 2.2 and hence providing strong evidence for the idea that the large-charge superfluid EFT approach can be applied to the large-axial-charge sector of the $U(1)$-NJL model.

**Conformal dimensions and local CFT spectrum for the NJL model**

To compute the scaling dimension $\Delta_{\text{SF}}(Q)$ of the lowest-lying operator $\mathscr{O}^{Q}_{\text{SF}}$ of fixed $U(1)_A$ charge $Q$ (and zero $U(1)_B$ charge) within the CFT living at the critical point of the $U(1)$-NJL model we now leave flat space and put the theory on the cylinder $\mathbb{R}\times S^2_{r_0}$. On the cylinder we can take advantage of the state–operator correspondence to relate the superfluid ground-state energy $f_c(Q)$ at fixed $U(1)_A$ charge to the scaling dimension $\Delta_{\text{SF}}(Q)$ of the associated operator.

As discussed, in the $U(1)$-NJL model the ground state at finite $U(1)_A$ charge is a superfluid, while at finite $U(1)_B$ charge it is a Fermi sphere. We focus on the superfluid case, as the Fermi-sphere case works equivalently to the GN model at finite $U(1)_B$ charge discussed in Section 3.3.2 and also produces exactly the same results.
At criticality and in the zero-temperature limit the grand potential $\Omega(\mu)$ at finite $U(1)_A$ density reads

$$\Omega(\mu) = -\frac{1}{2V}\sum_{j=\frac{1}{2}}^{\infty}(2j+1)\left[\omega_+ + \omega_-\right], \qquad \omega_\pm^2 = \langle\Phi\rangle^2 + \left(\omega_j \pm \mu\right)^2, \tag{3.261}$$

where $\omega_j = (j+1/2)/r_0$ are the eigenvalues of the Dirac operator on the sphere and $V = \Omega_3 r_0^2 = 4\pi r_0^2$ is the volume of the two-sphere. Following the standard procedure outlined in Eq. (3.183), we determine the axial charge $Q$ and the scaling dimension $\Delta_{\text{SF}}(Q)$ of the superfluid large-charge operator $\mathscr{O}^{Q}_{\text{SF}}$,

$$\frac{Q}{2N} = \frac{1}{2}\sum_{j=\frac{1}{2}}^{\infty}(2j+1)\left[\frac{\omega_j+\mu}{\omega_+} - \frac{\omega_j-\mu}{\omega_-}\right], \tag{3.262}$$

$$\frac{\Delta_{\text{SF}}(Q)}{2N} = f_c(Q) = -\frac{r_0}{2}\sum_{j=\frac{1}{2}}^{\infty}(2j+1)\left[\omega_+ + \omega_-\right] + (\mu r_0)\frac{Q}{2N}. \tag{3.263}$$

The assumed to be real VEV of the auxiliary field $\langle\Phi\rangle$ needs to be evaluated on the solution of the gap equation $g_{\text{gap}}(\langle\Phi\rangle) = 0$ on the sphere, (we note that the flat space solution in Eq. (3.238) will only reproduce the leading-order result for the VEV on the sphere)

$$g_{\text{gap}}(\langle\Phi\rangle) := \frac{1}{2r_0}\sum_{j=\frac{1}{2}}^{\infty}(2j+1)\left[\frac{1}{\omega_+} + \frac{1}{\omega_-}\right]\Bigg|_{\Phi=\Phi_0} = 0. \tag{3.264}$$

While the charge $Q$ is given by a finite expression, the expressions for the scaling dimension $\Delta_{\text{SF}}(Q)$ and the gap equation $g_{\text{gap}}(\langle\Phi\rangle)$ are formally divergent and need to be properly regularized. We choose here to regularize all expressions by removing the leading divergence in the sums and adding them back later in a zeta-function regulated version. Following this procedure we find that

$$g_{\text{gap}}(\langle\Phi\rangle)\Big|^{(\text{reg})} = \frac{1}{2}\sum_{j=\frac{1}{2}}^{\infty}\left[(2j+1)\frac{1}{r_0}\left[\frac{1}{\omega_+} + \frac{1}{\omega_-}\right] - 4\right] + 2\zeta(0), \tag{3.265}$$

$$\frac{\Delta_{\text{SF}}(Q)}{2N}\Big|^{(\text{reg})} = -\frac{1}{2}\sum_{j=\frac{1}{2}}^{\infty}\left[(2j+1)r_0\left[\omega_+ + \omega_-\right] - 4r_0^2\omega_j^2 - 2r_0^2\langle\Phi\rangle^2\right] - r_0^2\langle\Phi\rangle^2\zeta(0) + (\mu r_0)\frac{Q}{2N}. \tag{3.266}$$

Since all of the infinite sums are convergent now, the regulated expressions are amenable to numerical analysis for arbitrary values of the charge $Q$. Additionally, the first few terms in the expansions of $\Delta_{\mathrm{SF}}(Q)$ at $Q/2N \gg 1$ and $Q/2N \ll 1$ can be computed analytically.

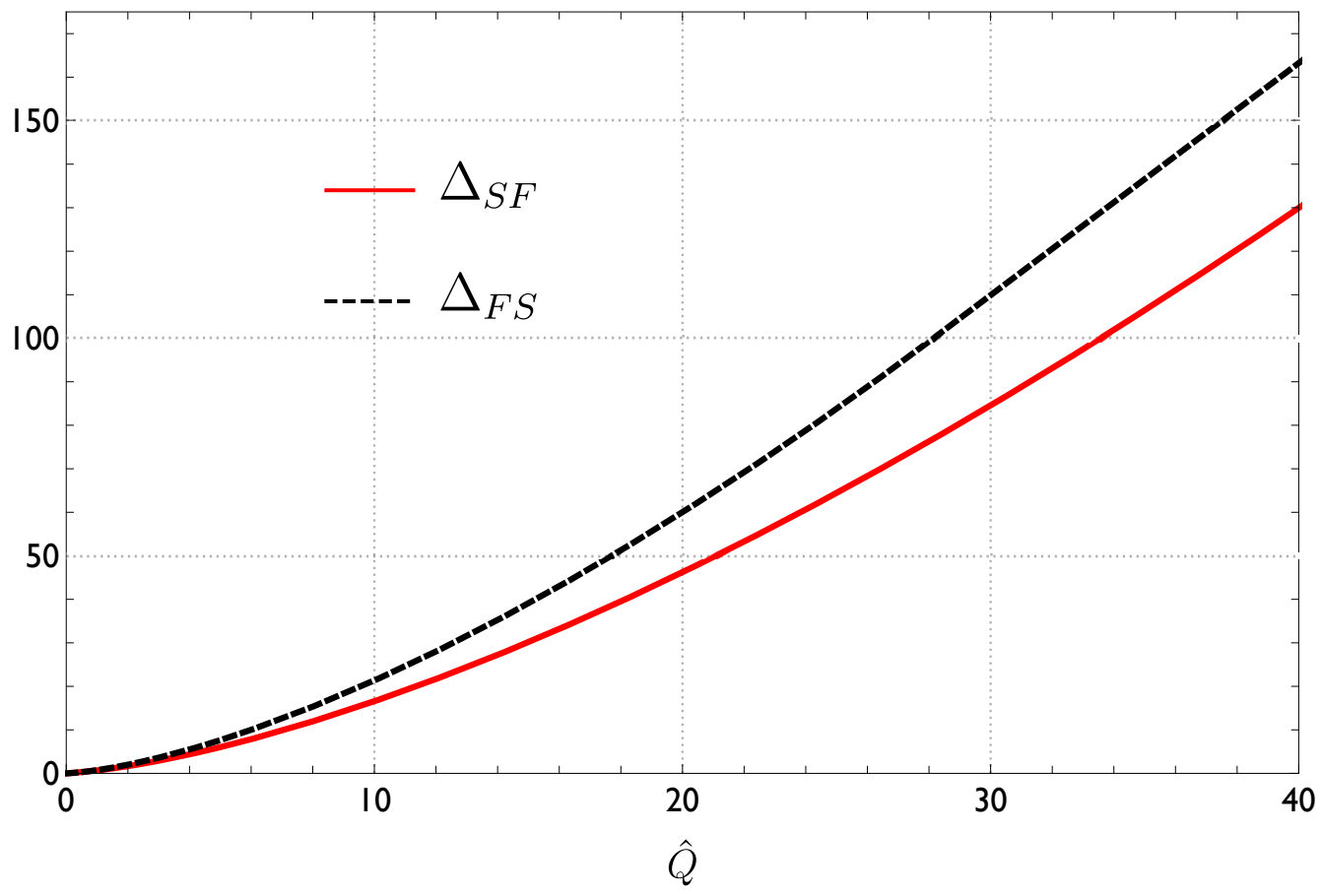

Figure 3.8: Scaling dimensions $\Delta_{\mathrm{FS}}$ and $\Delta_{\mathrm{SF}}$ of the Fermi-sphere primary operator $\mathcal{O}^{Q}_{\mathrm{FS}}$, appearing both in the GN and NJL models as well as the free fermion CFT at finite $U(1)_B$ charge $Q$, and the superfluid primary operator $\mathcal{O}^{Q}_{\mathrm{SF}}$, appearing in the NJL model at finite $U(1)_A$ charge $Q$ (and the Cooper model at finite $U(1)_B$ charge $Q$). We have divided all expressions by a factor of $2N$. Importantly, the Fermi-sphere operator $\mathcal{O}_{\mathrm{FS}}$ is only defined for values of the charge $Q$ corresponding to a filled Fermi sphere, see Eq. (3.221).

We plot the numerical result for the scaling dimension $\Delta_{\mathrm{SF}}(Q)$ at finite $U(1)_A$ charge $Q$ in Figure 3.8 and compare it to the scaling dimension $\Delta_{\mathrm{FS}}(Q)$ of the Fermi-sphere operator $\mathcal{O}^{Q}_{\mathrm{FS}}$ with finite $U(1)_B$ charge $Q$, which appears both in the NJL and GN model at finite $U(1)_B$ and zero $U(1)_A$ chemical potential.

**Q/2N ≫ 1**

There are three dimensionful parameters appearing in Eq. (3.262) and Eq. (3.264) at finite $U(1)_A$ charge — given by $\mu, r_0, \langle\Phi\rangle$ — but only two dimensionless ratios,

$$\mu r_0\,, \qquad\qquad r_0\langle\Phi\rangle\,. \tag{3.267}$$

In the large-charge limit $Q/2N \gg 1$ the chemical potential $\mu$ represents the dominant scale, so that

$$\mu r_0 \gg 1\,. \tag{3.268}$$

In the limit of large chemical potential the solution $\langle\Phi\rangle$ to the gap equation $g_{\mathrm{gap}}(\langle\Phi\rangle) = 0$ in Eq. (3.264) can be expanded into an expression of the form

$$r_0\langle\Phi\rangle = \sqrt{\kappa_0^2 - 1}\left(\mu r_0 + \frac{\kappa_1}{\mu r_0} + \frac{\kappa_2}{(\mu r_0)^3} + \dots\right). \tag{3.269}$$

Expectedly, this represents an expansion in $1/r_0^2$ since the radius $r_0$ of the two-sphere $S^2_{r_0}$ only enters via the conformal mass involving the scalar curvature $\mathscr{R}$ (which has engineering dimension $[\mathscr{R}] = 2$). In order to derive the precise values of the coefficients $\kappa_i$ we have to solve the gap equation,

$$g_{\text{gap}}(\langle\Phi\rangle) = g_{\text{gap}}(\langle\Phi\rangle)\big|^{(\text{reg})} + g_{\text{gap}}(\langle\Phi\rangle)\big|^{(\text{div})} = 0\,, \tag{3.270}$$

where we have conveniently separated the expression for the gap equation into its regular and divergent parts,

$$g_{\text{gap}}(\langle\Phi\rangle)\big|^{(\text{reg})} = \sum_{\ell=1} \ell \left[ \frac{1}{\sqrt{(\ell+\mu r_0)^2 + (r_0\langle\Phi\rangle)^2}} + \frac{1}{\sqrt{(\ell-\mu r_0)^2 + (r_0\langle\Phi\rangle)^2}} - 2\frac{1}{\sqrt{\ell^2 + (r_0\langle\Phi\rangle)^2}} \right], \tag{3.271}$$

$$g_{\text{gap}}(\langle\Phi\rangle)\big|^{(\text{div})} = 2\sum_{\ell=1} \ell \frac{1}{\sqrt{\ell^2 + (r_0\langle\Phi\rangle)^2}}\,. \tag{3.272}$$

We note that in order to write down these expressions we have also shifted the summation convention in Eq. (3.264) to $j = \ell - 1/2$. The regular part of the gap equation can now be computed in an asymptotic expansion around $\mu r_0 \gg 1$ using the standard Euler–Maclaurin formula and expressed in terms of the coefficients $\kappa_i$ in Eq. (3.269),

$$g_{\text{gap}}(\langle\Phi\rangle)\big|^{(\text{reg})} = 2\mu r_0 \left[ -\kappa_0 + \sqrt{\kappa_0^2 - 1} + \operatorname{arccoth}(\kappa_0) \right] + \frac{1}{6\mu r_0} \left[ -\frac{1}{\kappa_0} - 12\kappa_0\kappa_1 + \frac{1 + 12(\kappa_0^2 - 1)\kappa_1}{\sqrt{\kappa_0^2 - 1}} \right] + \dots\,. \tag{3.273}$$

The divergent part needs regularization, which we perform using the zeta-function procedure. This results in[59]

$$g_{\text{gap}}(\langle\Phi\rangle)\big|^{(\text{div})} = 2\sum_{\ell=1}^{\infty} \ell(\ell^2 + r_0^2\langle\Phi\rangle^2)^{-s}\bigg|_{s=1/2} = \frac{2}{\Gamma(s)} \int_0^\infty \frac{\mathrm{d}t}{t}\, t^s \sum_{\ell=1}^{\infty} \ell e^{-(\ell^2 + r_0^2\langle\Phi\rangle^2)t}\bigg|_{s=1/2}\,. \tag{3.274}$$

Evidently, as can be seen from the flat space result in Eq. (3.238) that coincides with the leading expression on the sphere, in the limit $\mu r_0 \gg 1$ we also necessarily have $r_0\langle\Phi\rangle \gg 1$. Hence, the zeta-function regulated expression for the divergent part $g_{\text{gap}}|^{(\text{div})}$ can be obtained in an expansion around $\langle\Phi\rangle r_0 \gg 1$. In this limit the zeta-function integral in the above expression localizes around $t = 0$. The appropriate expression for the asymptotic expansion is therefore found by expanding the integral around $t = 0$ using the standard $t$-expansion formula for the sum over $\ell$ given by

$$\sum_{\ell=1}^{\infty} \ell e^{-\ell^2 t} = \frac{e^{-t}}{12}\left(2t + 5 + \frac{6}{t} + \dots\right)\,. \tag{3.275}$$

The resulting expression for the divergent part reads

$$g_{\text{gap}}(\langle\Phi\rangle)\big|^{(\text{div})} = -2r_0\langle\Phi\rangle - \frac{1}{6r_0\langle\Phi\rangle} - \frac{1}{120(r_0\langle\Phi\rangle)^3} + \dots\,. \tag{3.276}$$

After inserting the Ansatz in Eq. (3.269) into $g_{\text{gap}}|^{(\text{div})}$ we can now solve the full gap equation $g_{\text{gap}}|^{(\text{reg})} + g_{\text{gap}}|^{(\text{div})} = 0$ order by order in $\mu r_0$ and find the precise values of the coefficients $\kappa_i$.

[59] It can be easily checked that these expansions coincide with the cut-off-independent part in a smooth cut-off regularization scheme.

As expected, the first coefficient $\kappa_0$ satisfies the irrational equation $\kappa_0 \tanh(\kappa_0) = 1$. The other coefficients $\kappa_{i>0}$ can be written in terms of the irrational number $\kappa_0 \approx 1.19968$,

$$\kappa_0 \tanh \kappa_0 = 1 , \qquad \kappa_1 = -\frac{1}{12\kappa_0^2} , \qquad \kappa_2 = \frac{33 - 16\kappa_0^2}{1'440\kappa_0^6} , \qquad \dots . \tag{3.277}$$

We can then apply the same strategy to the computation of the divergent sum appearing in the scaling dimension in Eq (3.262), which comes from the thermodynamic potential $\Omega(\mu)$. We therefore divide $\Omega(\mu)$ into two contributions, divergent and regular,

$$V\Omega(\mu)\big|^{(\text{reg})} = -\sum_{\ell=1} \ell \left[ \sqrt{(\ell + \mu r_0)^2 + (r_0 \langle \Phi \rangle)^2} + \sqrt{(\ell - \mu r_0)^2 + (r_0 \langle \Phi \rangle)^2} - 2\sqrt{\ell^2 + (r_0 \langle \Phi \rangle)^2} \right] , \tag{3.278}$$

$$V\Omega(\mu)\big|^{(\text{div})} = -2 \sum_{\ell=1}^{\infty} \ell \sqrt{\ell^2 + (r_0 \langle \Phi \rangle)^2} . \tag{3.279}$$

The regular part can be evaluated using the Euler–Maclaurin formula and to leading order reads

$$V\Omega(\mu)\big|^{(\text{reg})} = -\frac{(r_0\mu)^3}{3} \left( 3(\kappa_0^2 - 1) \operatorname{arccoth}(\kappa_0) + 3\kappa_0 - 2\kappa_0^3 + 2(\kappa_0^2 - 1)^{\frac{3}{2}} \right) + \dots . \tag{3.280}$$

The divergent part is again treated in zeta-function regularization. After performing the regularization it reads

$$V\Omega(\mu)\big|^{(\text{div})} = \frac{1}{2} \left[ \frac{4(r_0 \langle \Phi \rangle)^3}{3} + \frac{r_0 \langle \Phi \rangle}{3} - \frac{1}{60 r_0 \langle \Phi \rangle} + \dots \right] , \tag{3.281}$$

which can be written as a function of $r_0\mu$ using Eq. (3.269) and Eq. (3.277).

After inverting the expression for the charge $Q$ in Eq. (3.262) we obtain the relation $\mu = \mu(Q)$. Since the sum in the expression for $Q = Q(\mu)$ is already convergent, we can simply apply the Euler–Maclaurin formula to invert it. After combining the relation $\mu = \mu(Q)$ with the result of the gap equation $g_{\text{gap}}$ in Eq. (3.269) and the coefficients $\kappa_i$ in Eq. (3.277) we can finally compute the asymptotic expansion of the scaling dimension $\Delta_{\text{SF}}(Q)$ for the superfluid primary operator $\mathscr{O}_{\text{SF}}^{Q}$ in the limit $Q/2N \gg 1$,

$$\frac{\Delta_{\text{SF}}(Q)}{2N} = \frac{2}{3} \left( \frac{Q}{2N\kappa_0} \right)^{3/2} + \frac{1}{6} \left( \frac{Q}{2N\kappa_0} \right)^{1/2} + \frac{11 - 6\kappa_0^2}{720\kappa_0^2} \left( \frac{Q}{2N\kappa_0} \right)^{-1/2} + \dots . \tag{3.282}$$

We note that the leading term is consistent with the expression for the ground state free energy density $f_c(Q)/V$ found in flat space in Eq. (3.244). As usual, we can think of the sub-leading terms in the LCE as an expansion in the scalar curvature, with the leading term depending only on the volume $V = 4\pi r_0^2$ of the two-sphere.

The form of the scaling dimension $\Delta_{\text{SF}}(Q)$ is again consistent with the predictions from the large-charge EFT in Section 2.2: the asymptotic expansion of $\Delta_{\text{SF}}(Q)$ is consistent with the general form of the classical contribution to the scaling dimension expected from the structure of the EFT Lagriangian. In the spirit of the discussion around Eq. (3.56), the unknown Wilsonian coefficients $c_i$ of the corresponding superfluid EFT describing the large-axial-charge sector of the $U(1)$-NJL model can be computed to leading order in $N$ and depend on the irrational number $\kappa_0$.
The conformal NG mode, whose existence we have demonstrated in Eq. (3.259), is present for any value of $N$ and gives rise to the same universal contribution to the scaling dimension at order $\mathcal{O}(Q^0)$ discussed around Eq. (2.87) within the EFT approach. In the large-$N$ expansion this contribution appears at order $\mathcal{O}(N^0)$, consistent with what we find for the $O(2N)$ vector model in Section 3.1. Consequently, the excitations of the conformal NG mode generate the spectrum on top of the superfluid ground state, *i.e.* in terms of the flat space CFT on top of the primary operators $\mathcal{O}^Q_{\text{SF}}$.

**Q/2N ≪ 1**

Unsurprisingly, the scaling dimension $\Delta_{\text{SF}}(Q)$ is analytically accessible in the limit $Q/2N \ll 1$ as well. Due to conformal invariance, at $Q = 0$ the free energy has to vanish as this point corresponds to the conformal dimension associated with the identity operator, which is the unique operator in the CFT with vanishing conformal dimension. Within the description of the theory in terms of the collective scalar field $\Phi$ we have to consider the conformal mass of said scalar field coming from the conformal coupling to the curved background of the cylinder. This corresponds to a mass $m = 1/2r_0$ for the field $\Phi$ on the three-dimensional cylinder. In our description of the theory there is no explicit mass term present and hence the conformal mass of $\Phi$ has to come from the chemical potential $\mu$ (compare this to the $Q/2N \ll 1$ regime for the $O(2N)$ model discussed in Section 3.1.4). In fact, a direct computation shows that for the chemical potential equal to the conformal mass — $\mu = 1/2r_0$ — the free energy and the charge both vanish and the gap equation $g_{\text{gap}} = 0$ is solved by a vanishing VEV $\langle\Phi\rangle = 0$.
As a consequence, the small-charge expansion has to be performed around the point $\mu = 1/(2r_0)$ and it is convenient for us to write $\mu$ as

$$\mu = \frac{1}{2r_0} + \hat{\mu}. \tag{3.283}$$

We can then expand all expressions in Eq.(3.262) around $\hat{\mu} \ll 1$ to find the correct small-charge expansion of the scaling dimension. At the point $\hat{\mu} = 0$ there is full symmetry restoration as the auxiliary field $\Phi$ does not acquire a VEV. This observation allows us to write an Ansatz for the VEV of the field $\Phi$ of the form

$$\hat{\mu} = \mu_2 \langle\Phi\rangle^2 r_0 + \mu_4 \langle\Phi\rangle^4 r_0^3 + \dots. \tag{3.284}$$

Evidently, in the limit $\hat{\mu} \ll 1$ we have $r_0\langle\Phi\rangle \ll 1$, so that the corresponding charge $Q$ can be expanded in $r_0\langle\Phi\rangle \ll 1$ and becomes

$$\frac{Q}{2N} = \frac{\pi^2}{4}(r_0\langle\Phi\rangle)^2 - \frac{\pi^2}{16}(\pi^2 - 16\mu_2)(r_0\langle\Phi\rangle)^4 + \frac{\pi^2}{48}\left(\pi^4 + 12\pi^2(\mu_2^2 - 2\mu_2) + 48\mu_4\right)(r_0\langle\Phi\rangle)^6 + \dots. \tag{3.285}$$

We note that this expression manifestly vanishes for $\langle\Phi\rangle = 0$. Solving the gap equation order-by-order in small $\hat{\mu}$ allows us to determine the coefficients $\mu_i$ appearing in Eq. (3.284). For this purpose, once again, the gap equation is separated into a divergent and a convergent part. Both of them admit a well-defined expansion in the limit $r_0\langle\Phi\rangle \ll 1$. However, in contrast to the computation in the large-$Q$ regime, we subtract the $\hat{\mu} = 0$ contribution instead of the $\mu = 0$ contribution. The convergent part becomes

$$g_{\text{gap}}(\langle\Phi\rangle)\Big|^{(\text{reg})} = \frac{\pi^2}{2}\mu_2(r_0\langle\Phi\rangle)^2 - \frac{\pi^2}{4}(\pi^2\mu_2 - 4\mu_2^2 - 2\mu_4)(r_0\langle\Phi\rangle)^4 + \dots. \tag{3.286}$$

The divergent part is regularized order-by-order using zeta-function regularization and reads as follows:

$$
\begin{aligned}
g_{\text{gap}}(\langle\Phi\rangle)\Big|^{(\text{div})} &= \sum_{\ell=0}^{\infty}(2\ell+1)\frac{1}{\sqrt{\left(\ell+\frac{1}{2}\right)^2+(r_0\langle\Phi\rangle)^2}} \\
&= 2\sum_{\ell=0}\left(\ell+\frac{1}{2}\right)\sum_{k=0}\binom{-1/2}{k}(r_0\langle\Phi\rangle)^{2k}\left(\ell+\frac{1}{2}\right)^{2(-\frac{1}{2}-k)} \\
&= 2\sum_{k=0}\binom{-1/2}{k}(r_0\langle\Phi\rangle)^{2k}\zeta\left(2k;\frac{1}{2}\right) = -\frac{\pi^2(r_0\langle\Phi\rangle)^2}{2}+\frac{\pi^4(r_0\langle\Phi\rangle)^4}{8}+\dots, \qquad (3.287)
\end{aligned}
$$

where $\zeta(s;a)$ denotes the Hurwitz zeta function. Putting together $g_{\text{gap}}|^{(\text{reg})}$ and $g_{\text{gap}}|^{(\text{div})}$ we can now solve for the coefficients $\mu_i$. The first two coefficients $\mu_1$, $\mu_2$ are given by

$$
\mu_2 = 1, \qquad \mu_4 = \frac{\pi^2-8}{4}, \qquad \dots. \qquad (3.288)
$$

We then apply the same procedure to compute the divergent sums appearing in the scaling dimension $\Delta_{\text{SF}}(Q)$ in Eq. (3.262), *i.e.* the grand potential $\Omega(\mu)$ since the charge $Q$ is convergent. Again, the grand potential is divided into two quantities $\Omega\Big|^{(\text{reg})}+\Omega\Big|^{(\text{div})}$ given by

$$
\begin{aligned}
V\Omega(\mu)\Big|^{(\text{reg})} = -\sum_{\ell=1}\ell\Bigg[&\sqrt{\left(\ell+\frac{1}{2}+\hat{\mu}r_0\right)^2+(r_0\langle\Phi\rangle)^2}+\sqrt{\left(\ell-\frac{1}{2}-\hat{\mu}r_0\right)^2+(r_0\langle\Phi\rangle)^2} \\
&-\sqrt{\left(\ell+\frac{1}{2}\right)^2+(r_0\langle\Phi\rangle)^2}+\sqrt{\left(\ell-\frac{1}{2}\right)^2+(r_0\langle\Phi\rangle)^2}\Bigg], \qquad (3.289)
\end{aligned}
$$

$$
V\Omega(\mu)\Big|^{(\text{div})} = -2\sum_{\ell=1}^{\infty}\left(\ell+\frac{1}{2}\right)\sqrt{\left(\ell+\frac{1}{2}\right)^2+(r_0\langle\Phi\rangle)^2}. \qquad (3.290)
$$

The regular part $\Omega\Big|^{(\text{reg})}$ is dealt with by expanding the expression inside the sum in the limit $(r_0\langle\Phi\rangle)\ll 1$. After performing the expansion, at leading order it reads

$$
V\Omega(\mu)\Big|^{(\text{reg})} = \frac{1}{2}\left[\frac{\pi^2}{2}\mu_2(r_0\langle\Phi\rangle)^4+\dots\right]. \qquad (3.291)
$$

The divergent part of the grand potential instead is regularized order-by-order using zeta-function regularization. It can be written in terms of Hurwitz zeta functions $\zeta(s;a)$ and to leading order it reads

$$
\begin{aligned}
V\Omega(\mu)\Big|^{(\text{div})} &= -2\sum_{\ell=0}^{\infty}\left(\ell+\frac{1}{2}\right)\sum_{k=0}\binom{1/2}{k}(r_0\langle\Phi\rangle)^{2k}\left(\ell+\frac{1}{2}\right)^{2\left(\frac{1}{2}-k\right)} \\
&= -2\sum_{k=0}\binom{1/2}{k}(r_0\langle\Phi\rangle)^{2k}\zeta\left(2k-2;\frac{1}{2}\right) = \frac{\pi^2}{8}(r_0\langle\Phi\rangle)^4+\dots. \qquad (3.292)
\end{aligned}
$$

We are left with inverting the condition $Q = Q(\langle\Phi\rangle)$,

$$
r_0\langle\Phi\rangle(Q) = \frac{2}{\pi}\left(\frac{Q}{2N}\right)^{1/2}+\frac{\pi^2-16}{\pi^3}\left(\frac{Q}{2N}\right)^{3/2}+\dots, \qquad (3.293)
$$

and the conformal dimension $\Delta_{\mathrm{SF}}(Q)$ in the limit of small charge is then given by

$$\frac{\Delta_{\mathrm{SF}}(Q)}{2N} = \frac{1}{2}\left(\frac{Q}{2N}\right) + \frac{2}{\pi^2}\left(\frac{Q}{2N}\right)^2 + \dots . \tag{3.294}$$

As expected, since the collective field $\Phi$ has charge two, the leading order result at small charge — $\Delta_{\mathrm{SF}}(Q) = Q/2$ — gives the correct relation for the scaling dimension of the operator $\Phi^{Q/2}$ in terms of its charge in the free mean-field limit. This also consistent with the associated result found for the $O(2N)$ vector model in Section 3.1.4.

### 3.3.4 SU(2): A generalization of the NJL model

A direct generalization of the $U(1)$-NJL model is found by replacing the $U(1)_A$ axial symmetry to a $SU(2)_L \times SU(2)_R$ symmetry [283, 284]. We refer to this model as the $SU(2)$-NJL model, but it is also known as the isoNJL model. It can be constructed by doubling number of fermions present in the $U(1)$-NJL model — which is done in terms of introducing two-flavour fermions $\Psi_{i,f}$, $f = 1,2$ — and arranging them in an action of the form

$$S[\Psi_{i,f}] = \int \mathrm{d}^3x \left[ \sum_{i=1}^{N}\sum_{f=1}^{2} \bar{\Psi}_{i,f}\Gamma^\mu\partial_\mu\Psi_{i,f} - \frac{g}{N}\left[ \left( \sum_{i=1}^{N}\sum_{f=1}^{2} \bar{\Psi}_{i,f}\Psi_{i,f} \right)^2 - \sum_{a=1}^{3}\left( \sum_{i=1}^{N}\sum_{f=1}^{2} \bar{\Psi}_{i,f}\Gamma_5\sigma^a_{fg}\Psi_{i,g} \right)^2 \right]\right], \tag{3.295}$$

where the $(2\times2)$-matrices $\sigma^a$, $a = 1,2,3$ are the three Pauli-matrices, the generators of the $SU(2)$ Lie algebra. The fermionic fields $\Psi_{i,f}$ in $D = 3$ used to write down the action of the model are still given by spinors in the usual four-dimensional reducible representation, see the beginning of Section 3.3.1 and Appendix C.4.3 for details. In contrast to the $SU(N) \times U(1)_B \times U(1)_A$ symmetry exhibited by the $U(1)$-NJL model from the previous Section 3.3.3, the $SU(2)$-NJL model boasts a $SU(N) \times U(1)_B \times SU(2)_L \times SU(2)_R$ symmetry. In particular, the $U(1)_B \times SU(2)_L \times SU(2)_R$ subgroup will be of relevance to us, with the $SU(N)$ part being mostly there to allow for a large-$N$ analysis, just as in the $U(1)$ case.

The global (sub)group $SU(2)_L \times SU(2)_R$ acts linearly on the two-flavour four-dimensional fermions $\Psi_{i,f}$ as

$$\Psi_{i,f} \to e^{i\frac{1+\Gamma_5}{2}\omega^a_L\sigma^a_{fg}}\Psi_{i,g}, \qquad \text{and} \qquad \Psi_{i,f} \to e^{i\frac{1-\Gamma_5}{2}\omega^a_R\sigma^a_{fg}}\Psi_{i,g}. \tag{3.296}$$

We note that this symmetry can only be present due to the pseudo-real character of the non-Abelian Lie group $SU(2)$, as for $SU(2)$ the totally symmetric symbol $d^{abc}$ vanishes.

Analogously to its $U(1)$ counterpart, the $SU(2)$-NJL model is tractable in $D = 2+\epsilon$, $D = 4-\epsilon$ via its Yukawa-type UV completion. The $SU(2)$-Yukawa-NJL model is found by introducing a set of four real scalar fields $\sigma, \pi^{a=1,2,3}$ resulting in the action

$$S[\Psi_i, \sigma, \pi^a] = \int \mathrm{d}^3x \left[ \bar{\Psi}_i\left(\Gamma^\mu\partial_\mu + \sigma + i\pi^a\sigma^a\Gamma_5\right)\Psi_i + \frac{1}{2g_Y}\left(\partial_\mu\sigma\partial^\mu\sigma + \partial_\mu\pi^a\partial^\mu\pi^a\right)\right]. \tag{3.297}$$

For $2 < D < 4$ the Yukawa-type model described by this action flows in the IR to the same critical point as the $SU(2)$-NJL model in the UV. Naturally, in the far IR at $g_Y \to \infty$ of the $SU(2)$-Yukawa-NJL model in Eq. (3.297) the fields $\sigma$, $\pi^a$ can be understood as non-dynamical scalar collective fields introduced by a HS transform in the $SU(2)$-NJL model in Eq. (3.295). On the action in Eq. (3.297) the $SU(2)_L \times SU(2)_R$ symmetry acts infinitesimally as

$$\delta_{L,R}\Psi = i\left(\frac{(1 \pm \Gamma_5)}{2}\right)\omega^a \sigma^a \Psi, \qquad \begin{cases} \delta_{L,R}\sigma = \pm\omega^a\pi^a, \\ \delta_{L,R}\pi^a = \mp\omega^a\sigma + \epsilon^{abc}\pi^b\omega^c. \end{cases} \tag{3.298}$$

The four scalar fields can be conveniently combined into a quaternionic field $\Phi = \sigma + \pi^a\sigma^a$ on which the $SU(2)_L \times SU(2)_R$ symmetry in terms of finite transformations acts linearly,

$$\begin{cases} \Psi_i \to e^{i\frac{(1+\Gamma_5)}{2}\omega^a\sigma^a}\Psi_i, \\ \Phi \to \Phi e^{-i\omega^a\sigma^a}, \end{cases} \qquad \begin{cases} \Psi_i \to e^{i\frac{(1-\Gamma_5)}{2}\omega'^a\sigma^a}\Psi_i, \\ \Phi \to e^{i\omega'^a\sigma^a}\Phi. \end{cases} \tag{3.299}$$

### Symmetry-breaking at large $N$: Leading-order action and gap equation

The symmetry-breaking for the $SU(2)_L \times SU(2)_R$ at first glance can be expected to be more complex than its $U(1)$ counterpart. We have three linearly independent charge densities — $U(1)_B$, $SU(2)_L$, $SU(2)_R$ — that we can source. At finite $U(1)_B$ charge there is no symmetry breaking and the ground state is again a Fermi sphere, just as in the GN or the $U(1)$-NJL model, with the difference that the number of DoF is now doubled (see the discussion of the GN model in Section 3.3.2 for details). Hence, we focus in the model at finite chemical potential for the $\sigma^3$ and $\Gamma_5\sigma^3$ Cartan generators of the $SU(2)$ symmetries. The critical action in flat space at finite chemical potential for either of symmetry generators reads

$$S[\Psi_i, \sigma, \pi^a] = \int_{S^1_\beta\times\mathbb{R}^2} \mathrm{d}^3x \left[\bar{\Psi}_i\left(\Gamma_\nu\partial^\nu + \sigma + i\pi^a\sigma^a\Gamma_5 - \begin{Bmatrix} \mu_V\Gamma_3\sigma^3 \\ \mu_A\Gamma_3\Gamma_5\sigma^3 \end{Bmatrix}\right)\Psi\right], \tag{3.300}$$

where $\sigma^a$, $a = 1,2,3$ again denotes the Pauli matrices and $\sigma$ itself is one of the auxiliary HS fields. We assume that the auxiliary fields all admit a constant VEV and can be decomposed into said VEV plus fluctuations sub-leading in $N$. The grand potential $\Omega^{V,A}(\mu_{V,A})$ for the two different choices of chemical potential is given by

$$\Omega^{V,A}(\mu_{V,A}) = -\int^\Lambda \frac{\mathrm{d}^2p}{(2\pi)^2}\left[\omega_+^{V,A} + \omega_-^{V,A} + \frac{2}{\beta}\log\left(1 + e^{-\beta\omega_+^{V,A}}\right) + \frac{2}{\beta}\log\left(1 + e^{-\beta\omega_+^{V,A}}\right)\right], \tag{3.301}$$

where we have introduced the one-particle on-shell energies given by

$$\omega_\pm^V = \sqrt{|\langle\Phi_2\rangle|^2 + \left(\sqrt{|\mathbf{p}| + |\langle\Phi_1\rangle|} \pm \mu_V\right)^2}, \qquad \Omega_\pm^A = \sqrt{|\langle\Phi_1\rangle|^2 + \left(\sqrt{|\mathbf{p}| + |\langle\Phi_2\rangle|} \pm \mu_A\right)^2}, \tag{3.302}$$

with $|\langle\Phi_{1,2}\rangle|^2$ given by certain linear combinations of the VEVs of the fields $\sigma, \pi^a$,

$$|\langle\Phi_1\rangle|^2 = \langle\sigma\rangle^2 + \langle\pi_3\rangle^2, \qquad |\langle\Phi_2\rangle|^2 = \langle\pi_1\rangle^2 + \langle\pi_2\rangle^2. \tag{3.303}$$

In both cases there exists no non-trivial solution for the zero temperature gap equations in which both linear field combinations $\Phi_{1,2}$ acquire a VEV. At finite $\mu_V$ chemical potential one finds the solution $|\langle\Phi_1\rangle| = 0$, $|\langle\Phi_2\rangle| = \mu_V\sqrt{\kappa_0^2 - 1}$, where $\kappa_0$ is precisely the same irrational number found for the $U(1)$-NJL model in Eq. (3.238), while at finite $\mu_A$ chemical potentials the values of the two VEVs $\langle\Phi_{1,2}\rangle$ are reversed. As discussed, a finite $U(1)_B$ chemical potential does not lead to SSB at finite density and zero temperature.

In both cases $\mu_{V,A} \neq 0$ the one-particle on-shell energies are positive for any value of the momentum $|\mathbf{p}|$, hence no Fermi sphere arises. There is a condensate and we find the same superfluid regime as for the $U(1)$-NJL model in Eq. (3.238) leading to the same results for the charge and energy density, see Section 3.3.3 for more details.[60] The only difference is an overall factor of two coming from the fact that there are now twice as many DoF.

Finally, we can compute the ground state solutions for finite left/right chemical potentials $\mu_{L,R}$, which are sourcing the charge densities $\bar{\Psi}_i(1 \pm \Gamma_5)\Gamma_3\sigma^3\Psi_i$, respectively. Interestingly, the result in both cases for $\mu_{L,R} \neq 0$ is that the ground state is a filled Fermi sphere, and hence no SSB arises.

#### Spectrum of fluctuations for the $SU(2)$-NJL model: Symmetry breaking patterns and NG modes

Analysing and computing the spectrum of fluctuations on top of the superfluid ground state in the $SU(2)$-NJL model can be performed analogously to how it is done in the case of the $U(1)$-NJL model for both cases $\mu_{V,A} \neq 0$. The main difference is that there are now four real scalar fields instead of two entering the calculations for the inverse propagator in Eq. (3.251). Fortunately, the inverse propagator is block-diagonal, consisting of two non-trivial 2×2 blocks. These blocks get exchanged in the computation for the two cases $\mu_{V,A} \neq 0$. One of the 2×2 blocks produces the same spectrum of fluctuations found for the $U(1)$-NJL model, the superfluid NG mode and its gapped counterpart. The other produces two extra degenerate gapped modes with mass of order $\mu_{V,A}$ and dispersion relation

$$\omega^2 = -4\kappa_0^2\mu^2 - \frac{\left(\kappa_0^2 - 1\right)\mathbf{p}^2}{\kappa_0^2} + \dots. \tag{3.304}$$

Based on these results we can formally conclude that the same conformal superfluid EFT — in the spirit of Section 2.2.1 — describes the fluctuations on top of the large-charge ground state of both the $U(1)$-NJL model and the $SU(2)$-NJL model.

It appears that the large-charge sectors of the $U(1)$-NJL and $SU(2)$-NJL models are identical. No new NG boson with potentially quadratic dispersion relation is found in the spectrum of fluctuations of the $SU(2)$-NJL model, with the two newly-appearing gapped modes exhibiting an unbroken $U(1)$ invariance. As we will discuss now, this observation is in fact consistent with the NG boson counting rules outlined in Section 1.2.

[60] It seems reasonable to expect that there is again a very natural explanation for the presence of the condensate in terms of a dual model, just like for the $U(1)$-NJL and the Cooper model. It seems certainly feasible to find a transformation similar to the PG transformation in Eq. (3.230) for the $SU(2)$-NJL model.

We consider the formal macroscopic limit $r_0 \to \infty$ for the theory situated on the cylinder $\mathbb{R} \times S^2_{r_0}$. In this limit we can label the spacetime generators on the cylinder using a flat-space notation, see [18]. The actions for the $U(1)$- and $SU(2)$-NJL models in Eq. (3.231) and Eq. (3.300) at zero chemical potential and at criticality possess the total symmetries

$$SO(4,1)_{\text{conf}} \times SU(N) \times U(1)_B \times \begin{cases} U(1)_A & (U(1)\text{-NJL}), \\ SU(2)_L \times SU(2)_R & (SU(2)\text{-NJL}), \end{cases} \tag{3.305}$$

including both global and spacetime symmetries. To study the models at finite density we introduce the respective axial chemical potential terms, where we denote the chemical potentials as $\mu$ for $U(1)$-NJL and as $\mu_A$ for $SU(1)$-NJL (since there are two distinct chemical potentials $\mu_{V,A}$). The addition of these terms to the respective actions explicitly breaks some of the original symmetries and reduces the total symmetries of the two models to

$$\mathbb{R}_\tau \times SO(3)_{\text{rot}} \times SU(N) \times U(1)_B \times \begin{cases} U(1)_A & (U(1)\text{-NJL}), \\ U(1)^{(3)}_A \times U(1)^{(3)}_V & (SU(2)\text{-NJL}), \end{cases} \tag{3.306}$$

where $\mathbb{R}_\tau$ denotes cylinder-time translation symmetry and $SO(3)_{\text{rot}}$ are the isometries of the unit sphere $S^2_1$, and $U(1)^{(3)}_{A,V}$ are the global Abelian symmetries generated by the Cartan generators $\Gamma_5\sigma^3$ and $\sigma^3$ of the two global $SU(2)$ symmetries in the $SU(2)$-NJL model (the Cartans of $SU(2)_L \times SU(2)_R$). In both cases the large-$N$ ground state spontaneously breaks time translation $\mathbb{R}_\tau$ and the $U(1)_A$ global symmetry, while keeping the linear combination corresponding to the helical generators $H + \mu Q$ intact. This is the typical setting in which SSB of a global internal symmetry group occurs at finite density in a not necessarily Lorentz-invariant theory.[61]
As discussed in Section 1.2.5, in this setting the counting of NG bosons is non-trivial, unlike in the Lorentz-invariant case [34, 36, 37, 106, 285]. As in both the $U(1)$-NJL model and the $SU(2)$-NJL model we find a single broken generator, there is only one NG mode exhibiting a linear dispersion relation that can appear. Here, this is the well-known conformal superfluid NG boson.[62]

The only difference between the two NJL models is found in the additional gapped sector present in the $SU(2)$-NJL model. The fluctuations on top of the ground state in the $SU(2)$-NJL additionally contain two degenerate gapped NG modes which combine into a complex scalar that is charged under the unbroken $U(1)^{(3)}_V$ symmetry. In the case of the $SU(2)$-NJL model at finite vector chemical potential this analysis remains the same, however, the $U(1)^{(3)}_A$ and $U(1)^{(3)}_V$ symmetries switch roles.

[61] Strictly speaking, Eq. (3.306) describes the (Euclidean) cylinder equivalent of a theory that is not invariant under Lorentz-boost.

[62] The above argument hides the fact that the underlying CFT is Lorentz invariant. Alternatively, we might think of boosts and time translation to be spontaneously broken at finite density. In this picture, the counting of NG modes remains consistent as the spontaneously broken spacetime symmetries (boosts and time translation) do not result in additional NG modes at finite density.

**Conformal dimensions and local CFT spectrum for the $SU(2)$-NJL model**

This is a very short paragraph as all the leading-$N$ computations here are completely equivalent to the ones for the $U(1)$-NJL model, hence we do not repeat them here and refer to Section 3.3.3 for the analysis. In particular, we refer the large-$Q$ and small-$Q$ results for the scaling dimension of the associated flat-space operator in Eq. (3.282) and Eq. (3.294). The only diffrence we need to be careful of is the fact that the number of DoF given by $2N$ appearing all over in these equations has to be doubled $2N \to 4N$ in the case of the $SU(2)$-NJL model as we are now dealing with two-flavour fermions and hence twice as may DoF. Beyond this small adaptation, the results match perfectly between the models at leading order in $N$.

## 3.4 Conclusions and final remarks

Strongly-coupled systems can often be accessed by introducing an additional control parameter, with the goal of of organizing analytic computations of observables perturbatively in the appropriate limit. Large $N$ represents one such class of techniques, in which a large number of DoF is introduced in order to take advantage of the central limit theorem and compute observables in a $1/N$ expansion around a semi-classical description at leading order in $N$. Remarkably, the leading-order semi-classical trajectory captures many qualitative features of the quantum system.
Introducing an additional control parameter leads to even further simplifications and allows to extract more physical information from the semi-classical large-$N$ expansion. In the context of the LCE in CFTs there is a very natural candidate for such an additional parameter, which is the global charge $Q$. From the point of view of large $N$ this introduces more structure into the theory, allowing for the computation of observables in a double-scaling limit

$$Q, N \to \infty, \qquad Q/N \text{ fixed}. \tag{3.307}$$

From the point of view of the LCE, the additional control parameter at large $N$ allows us to access and study large-charge sectors of CFTs without resorting to the appropriate physically motivated emergent condensed-matter EFT description. Importantly, only the superfluid EFT description is well understood, excluding any CFTs not belonging to the superfluid universality from being studied. At large $N$ we can in principle study the large-charge sectors of theories outside of the superfluid universality class — which is what we have done to some extent in Section 3.3 — and potentially even start to develop other EFT descriptions, like for example a Fermi-liquid EFT for models with a Fermi-sphere ground state [118, 167] compatible with conformal symmetry. Particularly intriguing is also the possibility of a potential non-Fermi liquid phase [175, 176, 286], if such a phase can in fact be shown to exist in large-charge sectors of CFTs.[63]
The most obvious candidate in which to study possible large-charge Fermi-liquid EFT descriptions is the GN model, which at large $N$ can be distinguished from the free fermion without interactions by the presence of a scalar mode that does not condense on the ground state of the theory but still appears in and enriches the local CFT spectrum. However, before the issue of constructing the proper fermionic EFT description can be addressed, it is first and foremost important to deduce the fate of the large-$N$

[63] The EFT-like description of strange-metal phases remains an open problem in condensed-matter physics. However, the question whether they can arise in large-charge CFTs and are compatible with the constraints of conformal invariance can certainly be investigated in a setup like large $N$.

Fermi-sphere ground state of the GN model. As pointed out, the absence of SSB in the large-$N$ limit of the GN model either indicates that the physics remains that of a Fermi sphere at all energies and that there is a transition happening at finite $N$. Or it holds true that SSB is simply exponentially suppressed in $1/N$, hence making it invisible to all orders in perturbation theory. Which of the two possibilities is realized can be investigated by analysing the Four-Fermi interaction on top of the Fermi sphere and observing whether there exists and attractive four-Fermi channel or not. This question needs to be answered at some point in the near future.
Generally speaking, properly extending the large-charge analysis to next-to-leading order in $1/N$ is certainly important for all theories discussed in this chapter, as in addition to the GN model the sub-leading corrections in $1/N$ have not yet been computed neither for the $O(N)$ vector nor the NJL model (besides extracting the spectrum of fluctuations).

The observation that large-charge methods in the double-scaling limit at large $N$ allow us to access the full RG flow in the $O(2N)$ vector model represents an interesting and seemingly accidental development within the large-charge approach. The quartic interaction term in the action responsible for triggering the RG flow becomes a constant on the large-$N$ trajectory — by virtue of the central limit theorem — and hence the addition of said term in the computation of the grand potential $\Omega(\mu)$ is trivial. After relating the grand potential $\Omega(\mu)$ to the leading-$N$ effective action $V_{\text{eff}}^{(0)}$ in the bosonic theory, the addition of the flow term becomes very much non-trivial.[64] This allows us to reproduce a very complicated computation involving the re-summation of infinitely many Feynman diagrams with a computation of just a few lines.
This construction also allows us to study the $\varphi^4$-model away from spacetime dimensions $2 < D < 4$ where it flows to the WF fixed point. Particularly, the range $4 < D < 6$ deserves scrutiny. There we can show that under the assumption of unitarity, the $\varphi^4$-model, now boasting an irrelevant coupling $g$, is not UV complete in flat space, consistent with earlier results [20, 253, 254]. In fact, although we have access to the full RG flow, it is impossible to follow in reverse order along the RG flow back into the far UV, as the effective potential $V_{\text{eff}}^{(0)}$ becomes infinite everywhere. We can study the conjectured non-unitary and complex CFT in the UV and outline its pathologies for both small- and large-charge operators. By analysing the RG flow on the cylinder we find that there is a second-order phase transition along the flow separating the well-behaved IR phase from the pathological UV phase of the theory. The new metastable massive phases that arises in the UV might be explored in terms of instantons on the cylinder, extending the treatment in [253].
It is feasible to extend the spacetime dimensions beyond $2 < D < 6$. The range of dimensions $0 < D < 2$ is of questionable interest and rigour, but in $6 < D < 8$ the $\varphi^4$-theory can be expected to be convex based on basic observations from the large-charge approach to the effective potential. However, in this range the collective field introduced by the HS transformation violates the CFT unitary bounds.
While it might not necessarily be useful to push the spacetime dimensions even further for the $\varphi^4$-theory, this construction could be used to analyse $O(2N)$ vector models with different interaction terms in the spacetime dimension $2 < D < 6$. Particularly interesting could be the $\varphi^6$-theory in $2 < D < 6$ [210–220]. It is known that the sextic interaction is marginally irrelevant in the presence of the quartic term and hence does not affect the RG flow. Without the quartic interaction the $\varphi^6$-model in $2 < D < 4$ is conjectured to flow from a IR free CFT to a UV interacting CFT. There is a debate on whether the

[64] This relationship between $\Omega(\mu)$ and $V_{\text{eff}}^{(0)}$ is not reproduced in a fermionic theory like the GN model, unfortunately.

$\varphi^6$-model is even mathematically sound and perhaps an analysis of the effective potential using large-charge techniques could shed some further light on this issue.

Finally, the resurgent analysis deserves mentioning here as well. It can definitely be performed for the large-$N$ fermionic models discussed in Section 3.3 by analogy with the $O(N)$ vector model, particularly for the NJL model, although unfortunately numerical techniques will most certainly be required. Nevertheless, a resurgent analysis of the fermionic models could shed some further light on the for QFT atypical $(2n)!$ divergence found in the $O(2N)$ vector model.[65] Further, in the bosonic theory there exists an interpretation of the results coming from the standard resurgence computations in terms of a worldline path integral, a quantum mechanical path integral of a spin zero particle moving on the sphere [191]. It would be interesting to see whether a similar interpretation is found for one or several of the fermionic theories discussed.

[65]The results obtained from resurgence are important as they will enable us to extrapolate the large-charge expansion also to operators with small charge, which in turn would allow for making contact with results from Monte–Carlo simulations and the numerical bootstrap. In this context, extending the resurgence calculus to much smaller values of $N$ — say the $O(2)$ model — would be particularly desirable. However, beyond computing sub-leading corrections in $1/N$, it seems much more ambitious to perform a resurgent analysis in the $O(2)$ model than extending the analysis to other models in the large-$N$ limit, particularly to those that also exhibit a superfluid ground state.

# A Appendices to Chapter 1

## A.1 The Ising universality class

This appendix discusses the Ising model and its relationship to the study of CFTs. Our aim is to motivate the study of CFTs from a statistical mechanics perspective in the context of the second-order phase transitions present in the description of different Ising-type models.

To illustrate the role of CFTs in the description of critical phenomena let us consider the Ising model from statistical mechanics. The Ising model describes the interactions of classical spins on a lattice and is described by the Hamiltonian

$$E[\sigma] = -J \sum_{\langle ij \rangle} \sigma_i \sigma_j - H \sum_i \sigma_i \,, \tag{A.1}$$

where $i$ runs over all lattices sites and $\langle ij \rangle$ runs over all nearest-neighbour interactions. The Ising model is the simplest working theory describing classical magnetism and capturing many of the important properties of ferromagnetic materials. In particular, in $D > 1$ it exhibits the ferromagnetic-to-paramagnetic second-order phase transition.

Instead of solving the Ising model for a particular lattice completely, sometimes there is a much simpler way of finding the critical temperature $T_c$ where the second-order phase transition occurs. As it turns out, there is a set of duality relations that relate the low-temperature ordered regime of Ising-type models to the high-temperature unordered regime of dual models and vice versa. This structure within the theory may be abused to derive the critical temperature where the two regimes meet. These so-called Kramers–Wannier-type dualities are a general feature of any abelian theory [287, 288]. They are closely related to the Heisenberg principle and identify a variable $\mu_i$ that is dual to the spin $\sigma_i$ in Eq. (A.1).[1] In this regard, they are important in the understanding of the Ising model for $D \geq 2$, where they might not tell you much about the critical temperature. For example, for the Ising model in $D = 3$ on the cubic lattice the dual model is a gauge theory [287].

[1] The variable $\mu_i$ is part of the description of the dual model and lives on a dual lattice.

It is important to understand how characteristic quantities behave for a system undergoing a phase transition. For a spin system the correlation functions $\Gamma_{ij} = \langle(\sigma_i - \langle\sigma_i\rangle)(\sigma_j - \langle\sigma_j\rangle)\rangle$ — which describe how different spins on the lattice are correlated with each other — are of particular interest. From explicitly or numerically solvable models and experimental observations we know that for spin systems the correlation function behaves like

$$\Gamma(r) \overset{T=T_c}{\sim} r^{2-D-\eta} \tag{A.2}$$

at the critical point, where $\eta$ is the so-called critical exponent associated to the correlation function. There is a whole range of critical exponents describing the behaviour of thermodynamic quantities at and around the fixed point $T_c$. Critical exponents appear to be universal and do not depend on the microscopic structure of the system. Instead, the properties of the system at the critical point only depend on fundamental parameters like the dimension $D$ or the symmetries of the order parameter. This observation leads to the concept of universality: Many vastly different physical models at their critical points are accurately described by a small set of scale invariant theories.
As a consequence, systems originating from completely separate corners of physics can sometimes be described by the same theory at criticality. Because of the fact that CFTs describe critical points of continuous phase transitions, the concept of universality is intricately linked to the concept of the Renormalization Group (RG) flow in Quantum Field Theory (QFT). The fact that CFTs are sparse in theory space and many QFTs flow to the same CFT (IR equivalence) mirrors the idea of universality in condensed-matter physics and statistical mechanics.

At criticality, due to scale invariance the Ising model (and more generally lattice models) become independent of the lattice spacing $a$, which is a microscopic feature. It is then very natural to take the continuum limit $a \to 0$, a process that is also necessary to be able to describe the system in terms of a CFT. To illustrate this point further, we consider the case of the two-dimensional Ising model, which is solvable. The exact solution to the two-dimensional Ising model has first been provided in 1944 by Onsager [289]. By now all critical exponents are exactly known and they are rational numbers [290], in particular we have

$$\eta = \frac{1}{4}. \tag{A.3}$$

Having access to the exact solution provides a strong basis for finding and substantiating the correct CFT description of its critical point. In the continuum limit the spin operator $\sigma_i$ of the Ising model is not adequate to describe the system. Instead, in $D = 2$ we can introduce operators that are constructed non-locally out of the spin operators. This is achieved via the so-called transfer matrix method [291] — a clever reformulation of the partition function as a trace over matrix products — and a Jordan-Wigner transformation [291, 292]. The resulting fermionic operators $\bar{\psi}$, $\psi$ obey the proper commutation relations. The final result of this procedure is the observation that the two-dimensional Ising model is equivalently written as a system of lattice fermions. After careful manipulation of the partition function, it can be shown that in the continuum limit around the critical point it is described by a free real (Majorana) fermion with the action [22]

$$S = \frac{1}{2\pi}\int d^2z \left(\bar{\psi}\partial\bar{\psi} + \psi\bar{\partial}\psi + m\bar{\psi}\psi\right), \qquad m \propto (\beta - \beta_c). \tag{A.4}$$

The mass term spoils conformal invariance away from the critical point. The resulting two-dimensional

CFT has central charge $c = 1/2$ while the fermions have conformal weights $(1/2, 0)$ and $(0, 1/2)$, respectively. In addition, it is possible to identify the spin operator $\sigma(z,\bar{z})$ within the CFT as a twist operator with conformal weight $(1/16, 1/16)$ [22, 293]. Consistently, its correlation function is of the form

$$\langle \sigma(0,0)\sigma(z,\bar{z})\rangle = \frac{1}{|z|^{1/4}}\,. \tag{A.5}$$

Further, it is also possible to identify a disorder operator $\mu(z,\bar{z})$ dual to the spin $\sigma(z,\bar{z})$. The operator $\mu(z,\bar{z})$ can be related to the spin operator $\sigma(z,\bar{z})$ it via its OPE with the fermion operator $\psi(z,\bar{z})$. Essentially, the disorder operator $\mu(z,\bar{z})$ is the dual variable $\mu_i$ identified by the Kramers–Wannier duality and its conformal weights are equal to the conformal weights of $\sigma(z,\bar{z})$.

The Ising CFT in $D = 2$ can also be expressed in terms of bosonic DoF instead of the fermionic fields that we have discussed up to now. There are several different ways in which this can be achieved, some of which include a superposition of two Ising models on the same lattice [22]. Independent of the procedure applied, the end result is that the critical Ising model CFT is equivalent the $\mathscr{M}_3$ minimal model, sometimes also referred to as the $(4,3)$ or $m = 3$ minimal model [22, 293, 294].[2] For completeness, a minimal model is a CFT whose spectrum is constructed from finitely many irreducible representations of the Virasoro algebra. The discrete unitary series of minimal models $\mathscr{M}_p$ have central charge

$$c = 1 - \frac{6}{p(p+1)}\,. \tag{A.6}$$

The $\mathscr{M}_3$ minimal model is made up of only three operators (at least for its holomorphic part): the identity $\mathbb{1}$, the spin $\sigma$ and the energy $\epsilon$, which can be readily identified with the fermion mass term $\bar{\psi}\psi$. The fact that $\mathscr{M}_3$ is equivalent to the theory of a free Majorana fermion lies at the heart of Onsager's exact solution to the Ising model in two dimensions.

Beyond $D = 2$, the Ising model can be related to another field-theoretic model. There is a succinct argument which shows that the long-range physics of the Ising model in arbitrary dimensions is captured by the $\phi^4$-model [295]. This argument, which we will outline here now, relies on a mathematical trick called a Hubbard–Stratonovich (HS) transformation. Consider the Ising-model partition function with the Hamiltonian in Eq.(A.1). We rewrite the interaction term,

$$\beta J \sum_{\langle ij\rangle} \sigma_i \sigma_j = \beta \sum_{ij} \sigma_i J_{ij} \sigma_j\,, \tag{A.7}$$

where the matrix $J_{ij} = J_{ji}$ encodes the nearest neighbour interactions. Next, we introduce a factor of unity in the partition function as follows:

$$1 = \int \mathscr{D}\lambda_i\, e^{-\frac{1}{4\beta}\sum\limits_{ij}\lambda_i (J^{-1})_{ij}\lambda_j} = \int \mathscr{D}\lambda_i\, e^{-\frac{1}{4\beta}\sum\limits_{ij}\lambda_i (J^{-1})_{ij}\lambda_j - \beta\sum\limits_{ij}\sigma_i J_{ij}\sigma_j + \sum\limits_{i}\sigma_i\lambda_i}\,, \tag{A.8}$$

where the path-integral measure $\prod_i \mathscr{D}\lambda_i$ is properly normalized. In the last step we have applied the transformation $\lambda_i \mapsto \lambda_i - 2\sum_k \beta J \lambda_k$. Introducing this expression into the partition function of the Ising

[2] More precisely, it can be stated that the two-dimensional critical Ising model's spin and energy correlation functions are described by $\mathscr{M}_3$.

model yields

$$Z = \sum_{\{\sigma_i\}} e^{\beta \sum_{ij} \sigma_i J_{ij} \sigma_j + \beta H \sum_i \sigma_i} = \int \mathscr{D}\lambda_i \sum_{\{\sigma_i\}} e^{-\frac{1}{4\beta}\sum_{ij} \lambda_i (J^{-1})_{ij} \lambda_j + \sum_i \sigma_i (\beta H + \lambda_i)} . \tag{A.9}$$

The spin interactions are removed at the expense of introducing a set of continuous fields $\lambda_i$. This is the philosophy of the HS transformation: the interaction of one field in the action is removed at the expense of introducing a new field using a generalized version of the Gaussian integral. The sum over spins can now be performed explicitly,[3]

$$Z = \int \mathscr{D}\lambda_i \, e^{-\frac{1}{4\beta}\sum_{ij} \lambda_i (J^{-1})_{ij} \lambda_j + \sum_i \log\cosh(\lambda_i + \beta H)} = \int \mathscr{D}\phi_i \, e^{-\beta \sum_{ij} \phi_i J_{ij} \phi_j + \beta H \sum_i \phi_i + \sum_i \log\cosh(2\beta \sum_i J_{ij} \phi_j)} , \tag{A.10}$$

where we have performed a field redefinition $\phi_i = \sum_j (J^{-1})_{ij}(\lambda_j + \beta H)/(2\beta)$ in the last step. In order to actually get to a proper $\phi^4$-action from here we need to consider the limiting behaviour $|\phi_i| \ll 1$ of only small fluctuations.[4] By switching to a Fourier representation and expanding the action we find that[5]

$$\begin{aligned} S[\tilde{\phi}] = \sum_{\mathbf{k}} \tilde{\phi}(\mathbf{k}) \left( \beta \tilde{J}(\mathbf{0})(1 - 2\beta \tilde{J}(\mathbf{0})) + \frac{1}{2} \beta \tilde{J}''(\mathbf{0})(1 - 4\beta \tilde{J}(\mathbf{0}))\mathbf{k}^2 \right) \tilde{\phi}(-\mathbf{k}) + \beta H \sum_{\mathbf{k}} \tilde{\phi}(\mathbf{k}) \delta_{\mathbf{k},\mathbf{0}} \\ + \frac{4\beta^4 \tilde{J}(\mathbf{0})^4}{3N} \sum_{\mathbf{k}_1,\ldots} \tilde{\phi}(\mathbf{k}_1)\tilde{\phi}(\mathbf{k}_2)\tilde{\phi}(\mathbf{k}_3)\tilde{\phi}(\mathbf{k}_4) \delta_{\mathbf{k}_1 + \cdots + \mathbf{k}_4, \mathbf{0}} + \mathscr{O}(\mathbf{k}^4, \beta^2 H^2, \tilde{\phi}^6) . \end{aligned} \tag{A.11}$$

After switching back to real space and taking the continuum limit $\sum_i \to \int \mathrm{d}^d x$ the action now reads[6]

$$S[\phi] = \int \mathrm{d}^d x \left[ \frac{1}{2}(\partial\phi)^2 + \frac{\tilde{J}(\mathbf{0})(1 - 2\beta \tilde{J}(\mathbf{0}))}{\beta \tilde{J}''(\mathbf{0})(1 - 4\beta \tilde{J}(\mathbf{0}))} \phi^2 + \frac{4\beta^4 \tilde{J}(\mathbf{0})^4}{3\beta \tilde{J}''(\mathbf{0})(1 - 4\beta \tilde{J}(\mathbf{0}))} \phi^4 + \beta H \phi \right] . \tag{A.12}$$

This relates the behaviour of the Ising model at low temperatures to the $\phi^4$-theory, which is a continuum QFT. The microscopic derivation that we have presented here even relates the coupling constants of $\phi^4$ to the coefficients and the temperature of the Ising model. In particular, the coefficient of the mass term relates the phase transition of the Ising model to the occurrence of SSB in the $\phi^4$-model.

In $D = 2$, the fact that the Ising model can be related to the $\phi^4$-model begs the question how the $\phi^4$-theory is related to the minimal model $\mathscr{M}_3$. This relationship can be made rigorous and it has been successfully argued that $\mathscr{M}_p$ minimal models describe the critical points of interacting scalar field theories with interaction terms $\phi^{2(p-1)}$ [296, 297]. In particular, the $\mathscr{M}_3$ model describes the interacting fixed point of the scalar field theory with a $\phi^4$ interaction. Despite the vast body of knowledge already assembled on the subject, there is research ongoing in the field of two-dimensional Ising models to this day, see for example [298–300].

In $D = 3$, the interacting fixed point of the $\phi^4$- model appears in the description of a range of continuous phase transitions besides the Ising model.[7] For example, in the theory of liquid states the Ising-model CFT describes the continuous phase transition at the critical point in the pressure–temperature plane, which is located at the end of the first-order liquid-gas transition line. The order parameter for this phase transition is the density $\rho - \rho_c$ instead of the spin $\sigma$.[8] It is an astounding fact that the second-order phase transition of water is described by the same CFT as the three-dimensional Ising model [301]. For a discussion of the mapping between fluid and magnetic Hamiltonians see [302]. For a general review of critical phenomena in liquid states see [303].

[3] Factors of 2 associated to the hyperbolic cosines and the factor of $\exp[-(\beta h^2/4)\sum_{ij}(J^{-1})_{ij}]$ are absorbed into the integral measure.

[4] This is tantamount to working at low enough temperatures such that the exponential weight $\beta J$ inhibits large fluctuations.

[5] We are still on a lattice, hence the Fourier representations are $\phi_i = \sum_{\mathbf{k}} e^{-i\mathbf{k}\cdot\mathbf{r}_i} \tilde{\phi}(\mathbf{k})/\sqrt{N}$, $J_{ij} = \sum_{\mathbf{k}} e^{-i\mathbf{k}\cdot(\mathbf{r}_i - \mathbf{r}_j)} \tilde{J}(\mathbf{k})/N$.

[6] There is a trivial rescaling $\phi \to \phi/\sqrt{\beta \tilde{J}''(\mathbf{0})(1 - 4\beta \tilde{J}(\mathbf{0}))}$ that needs to be performed here as well.

The Ising model and its critical point are a pure manifestation of the Wilsonian universality principle, the fact that a sparse number of CFTs lies at the fixed points of the RG flows of completely dissimilar physical and mathematical systems. The Wilsonian understanding of universality directly translates to the concept of universality in condensed-matter physics and statistical mechanics. It is reflected in the fact that at the critical point of a continuous phase transition microscopic properties become irrelevant and there are many different physical systems converging to an identical description at their respective critical points. The ubiquity of these equivalences is the phenomenon of critical universality. There are also many more examples of theories belonging to the Ising universality class, some of which are important in high-energy physics like the finite transition in the theory of electroweak interactions. For a non-exhaustive list of examples and references thereof see [304].

Although we only discussed thermal phase transition, there are phase transition that occur at zero temperature. These quantum phase transitions are not driven by thermal fluctuations and instead by quantum fluctuations dictated by the Heisenberg uncertainty principle. For a reference to quantum phase transition see [305].
Further, the examples presented above are all Euclidean CFTs. However, Lorentzian CFTs also appear in physics, for example thin-film superconductors can be described by the Lorentzian $O(2)$ model [306, 307]. The associated Wick-rotated theory — the Euclidean $O(2)$ model — describes the superfluid transition of $^4$He [308]. This particular example is beautiful as the critical exponents of the two theories agree even experimentally, making this an instance of Wick rotation working in nature.

## A.2 Conformal invariance in classical field theory

We discuss the consequences of conformal invariance in classical field theory in $D \geq 3$. We will focus on theories with a Lagrangian description $\mathscr{L}$.

### A.2.1 Conformal transformations

We analyse the constraints of conformal invariance in a general context and introduce to conformal group. The conformal group is comprised of all spacetime transformations $x \mapsto x'$ for which the pull-back metric,

$$g'_{\mu\nu} = \frac{\partial x^\mu}{\partial x'^\rho}\frac{\partial x^\nu}{\partial x'^\sigma} g_{\rho\sigma} = \Omega^2(x) g_{\mu\nu}\,, \tag{A.13}$$

only differs by a (local) scale factor $\Omega^2(x)$.[9] It is a simple exercise to verify that the conformal mappings form a group on any given (curved) spacetime.[10] For our purposes it is enough to work in flat space $\mathbb{R}^d$ with a flat indefinite metric $g_{\mu\nu} = \eta_{\mu\nu}$ of signature $(p,q)$. In flat space, it is manifestly clear that the Poincaré group (isometries) is a subgroup of the conformal group, the case $\Omega^2(x) \equiv 1$ corresponds to transformations of the Poincaré group.

[7]The interacting fixed point of the $\phi^4$-model in $2 < D < 4$ is also referred to as the WF fixed point.

[8]However, the liquid-gas transition does not exhibit the $\mathbb{Z}_2$ symmetry present in magnetic systems. This leads to the observation of $\mathbb{Z}_2$-non-invariant corrections to observables and, in particular, to the critical exponents.

To determine the conformal generators we study the properties of infinitesimal transformations $x^\mu \mapsto x^\mu + \epsilon^\mu(x) + \mathcal{O}(\epsilon^2)$. Conformal invariance implies that

$$\partial_\mu \epsilon_\nu(x) + \partial_\nu \epsilon_\mu(x) = \frac{2}{D}\partial_\sigma \epsilon^\sigma \eta_{\mu\nu}\,, \qquad \Omega^2(x) = 1 + \frac{2}{D}\partial_\sigma\epsilon^\sigma + \mathcal{O}(\epsilon^2)\,. \tag{A.14}$$

Eq. (A.14) is the conformal Killing equation. For elements of the Poincaré group the factor of proportionality — given by $\partial\cdot\epsilon$ — is zero. By deriving both sides of the equation and permuting the indices we can derive the identities

$$\frac{(D-2)}{D}\partial_\sigma\partial_\mu\epsilon^\mu = \Box\epsilon_\sigma\,, \qquad \partial_\mu\partial_\nu\epsilon_\rho = \frac{1}{D}\left(\eta_{\rho\nu}\partial_\mu + \eta_{\rho\mu}\partial_\nu - \eta_{\mu\nu}\partial_\rho\right)\partial_\sigma\epsilon^\sigma\,. \tag{A.15}$$

Applying another derivative results in the condition

$$\left[\eta_{\mu\nu} + (D-2)\partial_\mu\partial_\nu\right]\partial_\sigma\epsilon^\sigma = 0\,, \tag{A.16}$$

implying that in $D > 2$ the third derivative of $\epsilon_\nu$ has to vanish.[11] There are two special cases to be quickly commented on:

- In the case $D = 1$ there is no constraint imposed on $\epsilon$, *i.e.* $\partial\epsilon$. Basically, any smooth transformation is conformal in $D = 1$.[12]

- For $D = 2$ the first condition in Eq. (A.15) becomes $0 = \Box\epsilon_\sigma$. In Euclidean space $\eta_{\mu\nu} = \delta_{\mu\nu}$ Eq. (A.14) turns into the Cauchy-Riemann equations,

  $$\partial_0\epsilon_1 = -\partial_1\epsilon_0\,, \qquad \partial_0\epsilon_0 = \partial_1\epsilon_1\,. \tag{A.17}$$

  For this reason, conformal invariance in $D = 2$ is special, as the set of local conformal transformations is the set of all (anti-)holomorphic functions. Hence, conformal invariance in $D = 2$ is a vastly more powerful tool than in higher dimensions.[13] For more details on CFTs in $D = 2$ see [22, 293]

---

[9] As a consequence, the conformal group preserves angles between vector fields, which are given by $\cos\theta_{XY} = \frac{X(x)\cdot Y(x)}{\sqrt{X(x)^2\,Y(x)^2}}$.

[10] Conformal invariance may be discussed on arbitrary semi-Riemannian smooth manifolds equipped with a metric tensor $g_{\mu\nu}(x)$ (see [25]).

[11] By contracting Eq. (A.16) with the metric the condition $\Box\partial_\sigma\epsilon^\sigma = 0$ is derived.

[12] In $D = 1$ there is no notion of angles, *i.e.* we have $\frac{X(x)\cdot Y(x)}{\sqrt{X(x)^2\,Y(x)^2}} = X(x)\,Y(x)$.

[13] A small caveat here is that not all conformal transformations are globally invertible, only the Möbius transformations are.

or any other resource on the subject.

As mentioned, in $D \geq 3$ the third derivative of $\epsilon^\mu$ has to vanish, hence the infinitesimal transformation is at most quadratic,

$$\epsilon^\mu(x) = a^\mu + \lambda x^\mu + \omega^{\mu\nu} x_\nu + c^{\mu\nu\rho} x_\nu x_\rho \,. \tag{A.18}$$

We derive the conformal Lie algebra by acting with the infinitesimal transformations in Eq. (A.18) on smooth functions of spacetime and from there deducing the form of the spacetime generators.

The constant term $\epsilon^\mu = a^\mu$ corresponds to spacetime translations and trivially satisfies Eq. (A.14). The corresponding spacetime generators are

$$x^\mu \mapsto x^\mu + a^\mu = x^\mu + a^\nu P_\nu x^\mu \,, \qquad P_\mu = \partial_\mu \,. \tag{A.19}$$

The linear term in Eq. (A.18) separates into symmetric and anti-symmetric parts by virtue of Eq. (A.14). The anti-symmetric part has to obey

$$\omega^{\mu\nu} + \omega^{\nu\mu} = 0 \,. \tag{A.20}$$

The infinitesimal scale factor in Eq. (A.14) disappears for $x^\mu \mapsto x^\mu + \omega^{\mu\nu} x_\nu$. These transformations are the infinitesimal (proper orthochronous) Lorentz transformations $\Lambda^\mu{}_\nu = \delta^\mu{}_\nu + \omega^\mu{}_\nu + \mathcal{O}(\omega^2)$ in Minkowski space,[14] and in Euclidean spacetime they correspond to infinitesimal rotations. The associated spacetime generators $L_{\mu\nu}$ are given by

$$x^\mu \mapsto x^\mu + \frac{\omega^{\nu\rho}}{2} L_{\nu\rho} x^\mu \,, \qquad L_{\nu\rho} = x_\rho \partial_\nu - x_\nu \partial_\rho \,. \tag{A.21}$$

An infinitesimal rotation can be integrated to deduce the associated finite transformation,

$$x^\mu \mapsto \Lambda^{\mu\nu} x_\nu \,, \qquad \Lambda^T \eta \Lambda = \eta \,. \tag{A.22}$$

The symmetric part of the term linear in $x$ in Eq. (A.18) — given by $x^\mu \mapsto (1+\lambda) x^\mu$ — in the form that we have written it trivially satisfies Eq. (A.14).[15] These transformations are scale transformations, also called dilatations. The generator of scale transformations is given by

$$x^\mu \mapsto x^\mu + \lambda D_0 x^\mu \,, \qquad D_0 = x^\nu \partial_\nu \,. \tag{A.23}$$

The finite scale transformation associated to the (infinitesimal) parameter $\lambda$ simply is

$$x^\mu \mapsto e^\lambda x^\mu \,, \qquad e^\lambda = 1 + \lambda + \mathcal{O}(\lambda^2) \,. \tag{A.24}$$

The quadratic term $x^\mu \mapsto x^\mu + c^{\mu\nu\rho} x_\nu x_\rho$ is best treated by considering the implication of the second condition in Eq. (A.15). After defining $b_\mu := c^\nu{}_{\nu\mu}/D$ we can bring the transformation into the following

[14] The proper orthochronous Lorentz transformations form the identity component of the Lorentz group, they satisfy the conditions $\det(\Lambda) = 1$, $\Lambda^0{}_0 \geq 1$.

[15] The condition in Eq. (A.14) automatically imposes the symmetric part to be proportional to the metric.

form:

$$c^{\mu\nu\rho}x_\nu x_\rho \overset{\text{Eq. (A.15)}}{=} 2(x\cdot b)x^\mu - x^2 b^\mu\,. \tag{A.25}$$

A transformation of this form is called a Special Conformal Transformation (SCT). The spacetime generators of SCTs are given by

$$x^\mu \mapsto x^\mu + b_\nu K_0^\nu x^\mu\,, \qquad\qquad K_0^\mu = 2x^\mu x^\nu \partial_\nu - x^2 \partial^\mu\,. \tag{A.26}$$

The associated finite transformation reads

$$x^\mu \mapsto \frac{x^\mu - x^2 b^\mu}{1-2(b\cdot x)+x^2 b^2} = x^\mu + 2(b\cdot x)x^\mu - x^2 b^\mu\,. \tag{A.27}$$

Any SCT leaves the origin invariant but is singular at the point $x^\mu = b^\mu/b^2$, which is mapped to infinity. For this and other reasons, in CFT spacetime is generally best thought of as the sphere $\mathbb{R}^D \cup \{\infty\}$ with the point at infinity $\{\infty\}$ included.[16] The conformal group does not generally leave $\{\infty\}$ invariant, as opposed to the Poincaré group.

The scale factor associated with a SCT can be computed and reads

$$\Omega^2(x) = (1-2(b\cdot x)+b^2x^2)^2\,. \tag{A.28}$$

Invariance under SCTs is crucial as it differentiates scale invariant theories from conformally invariant ones. In fact, the generators of Poincaré together with the dilatation operator $D$ form a subalgebra of the conformal algebra allowing for the possibility of a purely scale-invariant theory.

Besides the conformal transformations captured by the Lie algebra there is an important set of discrete transformations that should be discussed. Consider the following discrete transformation:

$$\mathscr{I}: \qquad x^\mu \mapsto x'^\mu = \frac{x^\mu}{x^2}\,. \tag{A.29}$$

This transformation is called an inversion and it lies outside of the identity component of the conformal group. In fact, a SCT can be reproduced by an inversion, followed by a translation and then by another inversion,

$$x^\mu \overset{\mathscr{I}}{\mapsto} \frac{x^\mu}{x^2} \mapsto \frac{x^\mu}{x^2} - b_\mu \overset{\mathscr{I}}{\mapsto} \frac{x^\mu - x^2 b^\mu}{1-2(b\cdot x)+x^2b^2}\,. \tag{A.30}$$

In that sense, SCTs can be thought of as translations that move the point at infinity while leaving the origin invariant (as opposed to regular translations that leave infinity invariant and move the origin).[17] Since inversions are not continuously connected to the identity, CFTs may not be invariant under inversions. Only in the context of SCTs is invariance guaranteed, as these transformations are continuously connected to the identity and hence captured by the Lie algebra. Invariance under inversions needs to be checked on an individual basis.[18]

[16] Flat space $\mathbb{R}\cup\{\infty\}$ is Weyl equivalent to the unit sphere $S_1^D$. On the unit sphere $S_1^D$ all conformal transformations are finite/non-singular.

[17] Translations and SCTs are analogous to each other, with the points at $\{0\}$ and $\{\infty\}$ swapped by an inversion.

[18] There is a neat argument based on the embedding formalism implying that a CFT invariant under parity will be invariant

### A.2.2 Conformal algebra and the representation theory of fields

As the spacetime generators associated to the conformal transformations have to form a representation of the conformal algebra, we can use their explicit form to deduce all relevant commutation relations defining said algebra,

$$\begin{aligned}
&[M_{\mu\nu}, P_\rho] = \eta_{\mu\rho} P_\nu - \eta_{\nu\rho} P_\mu\,,\\
&[M_{\mu\nu}, M_{\rho\sigma}] = \eta_{\mu\rho} M_{\nu\sigma} + \eta_{\nu\sigma} M_{\mu\rho} - \eta_{\mu\sigma} M_{\nu\rho} - \eta_{\nu\rho} M_{\mu\sigma}\,,\\
&[D, P_\mu] = -P_\mu\,,\\
&[D, K_\mu] = K_\mu\,,\\
&[M_{\mu\nu}, K_\rho] = \eta_{\mu\rho} K_\nu - \eta_{\nu\rho} K_\mu\big)\,,\\
&[P_\mu, K_\nu] = 2\big(\eta_{\mu\nu} D + M_{\mu\nu}\big)\,.
\end{aligned} \tag{A.31}$$

The other correlators are vanishing. Any representation of the conformal algebra necessarily satisfies Eq. (A.31). For completeness and continuation, we compile all the spacetime generators quickly,

$$P_\mu = \partial_\mu\,, \qquad K_\mu = 2x_\mu x^\nu \partial_\nu - x^2 \partial_\mu\,, \qquad M_{\mu\nu} \sim L_{\mu\nu} = x_\nu \partial_\mu - x_\mu \partial_\nu\,, \qquad D = x^\nu P_\nu\,. \tag{A.32}$$

In $D$ dimension the number of generators of the conformal group is $(D+2)(D+1)/2$. Importantly, the conformal group in $\mathbb{R}^D$ with signature $(p,q)$ is isomorphic to the Lorentz group $SO(p+1,q+1)$ in $\mathbb{R}^{D+2}$ equipped with signature $(p+1,q+1)$. To see this, we take the $\mathbb{R}^D$ variables $x_1,\ldots,x_D$ and add two additional variables $X^+ = (x_0 + x_{D+1})$, $X^- = (x_0 - x_{D+1})$. In this language the isomorphism reads[19]

$$J_{\mu\nu} = M_{\mu\nu}\,, \qquad J_{+\mu} = P_\mu/2\,, \qquad J^{(c)}_{-\mu} = K_\mu/2\,, \qquad J^{(c)}_{+-} = D/2\,, \tag{A.33}$$

with $J_{\alpha\beta}$ — understood to be antisymmetric — being the generators of $SO(p+1,q+1)$. These generators satisfy the Lorentz algebra,

$$[J_{\alpha\beta}, J_{\gamma\delta}] = \eta_{\alpha\gamma} J_{\beta\delta} + \eta_{\beta\delta} J_{\alpha\gamma} - \eta_{\alpha\delta} M_{\beta\gamma} - \eta_{\beta\gamma} J_{\alpha\delta}\,. \tag{A.34}$$

The coordinates $X^+, X^-$ are called light-cone coordinates. The isomorphism gives rise to the embedding formalism, a convenient way of encoding the action of the conformal group in $\mathbb{R}^D$ via the action of the Lorentz group in $\mathbb{R}^{D+2}$ by appropriately embedding $\mathbb{R}^D$ in $\mathbb{R}^{D+2}$ [23, 66–68].

Having an understanding of the structure of the conformal group and algebra, we can discuss the field content of a CFT at the level of the classical Lagrangian. Consider a theory with field content $\{\phi^i\}$ described by the action

$$S = \int \mathrm{d}x^D\, \mathscr{L}(\phi^i, \partial_\mu \phi^i, x)\,. \tag{A.35}$$

For simplicity, we suppress the field indices with the shorthand notation $\phi = (\phi_1, \phi_2, \ldots, \phi_k)$. Just like the Lorentz group in Poincaré-invariant QFT, conformal transformations in general act non-trivially on the fields of the theory. Given an infinitesimal conformal transformation as in Eq. (A.18) we are looking under inversions and vice versa [23].

[19] In terms of $x_0, x_1, \ldots, x_D, x_{D+1}$ the generators read $J_{0\mu} = (P_\mu + K_\mu)/2$, $J_{D+1\mu} = (P_\mu - K_\mu)/2$, $J_{D+10} = D$.

for the most general form of the representation $T_\alpha$ of the conformal algebra on the space of fields, such that the field content transforms as

$$\phi'(x) = (\mathbb{1} - \epsilon^\alpha T_\alpha)\phi(x)\,. \tag{A.36}$$

To restrict the form of $T_\alpha$ we use representation theory and physical intuition. The end result is that, besides the notion of spin introduced by the representation theory of the Lorentz algebra, invariance under dilatations introduces the scaling dimension $\Delta$, which is defined as the degree of homogeneity of a given field.

Unsurprisingly, the representation theory of the Poincaré subalgebra is analogous to regular Poincaré-invariant field theory. The generators of translations — the momentum operators — purely act on spacetime,

$$P_\mu\phi(x) = \partial_\mu\phi(x)\,. \tag{A.37}$$

Every field is a scalar quantity with respect to translations. The physical intuition here is that under a translation of spacetime $x \mapsto x' = x + a$ it is most natural to identify the value of the field $\phi'(x')$ at $x'$ the field value $\phi(x)$ at $x$, hence

$$\phi'(x') = \phi(x)\,. \tag{A.38}$$

Having access to the momentum operators $P_\mu = \partial_\mu$ allows us to study the rest of the algebra from their action on fields at the point $x = 0$.[20] The momentum operators generate spacetime translations and define the value of the field at $x \neq 0$ via its value at $x = 0$ by imposing that fields transform in the adjoint representation of the group of spacetime translations,

$$\phi(x) = e^{x\cdot P}\phi(0)e^{-x\cdot P}\,, \tag{A.39}$$

with $P = (P_1, \ldots, P_D)$. Further, every operator $\mathcal{O}(x)$ in the theory should transform in the same way as the fields, since it necessarily holds true that

$$\mathcal{O}(x)\phi(x) = \mathcal{O}(x)e^{x\cdot P}\phi(0)e^{-x\cdot P} \stackrel{!}{=} e^{x\cdot P}\mathcal{O}(0)\phi(0)e^{-x\cdot P}\,. \tag{A.40}$$

Generators of symmetry transformations are no exception to this rule and satisfy Eq. (A.39). As a consequence, we are able to study the representation theory of the conformal group by simply studying the subgroup that leaves $x = 0$ invariant, called the stability subgroup or little group. In the case of the conformal group the stability subgroup is spanned by rotations, dilatations and SCTs. Hence, we impose

$$K_\mu\phi(0) = k_\mu\phi(0)\,, \qquad D\phi(0) = \Delta\phi(0)\,, \qquad M_{\mu\nu}\phi(0) = iS_{\mu\nu}\phi(0)\,, \tag{A.41}$$

where the matrix-valued quantities $\{k_\mu, \Delta, S_{\mu\nu}\}$ satisfy the reduced conformal algebra,

$$[k_\mu, \Delta] = -k_\mu\,, \qquad [k_\rho, iS_{\mu\nu}] = -\left(\eta_{\rho\mu}k_\nu - \eta_{\rho\nu}k_\mu\right)\,, \tag{A.42}$$

with all other commutators vanishing. The little group in Eq. (A.42) is the starting point to determine the complete action of the generators of the conformal group on any field in accordance with the standard theory of induced representations [70].

[20]Translations correspond to the constant term in Eq. (A.18). All other conformal transformations vanish at $x = 0$.

First, we notice that letting $k_\mu \neq 0$ turns out to be inconsistent, both at the classical and quantum level. Schur's Lemma enforces $\Delta$ to be a multiple of the identity,[21] hence Eq. (A.42) implies that[22]

$$[\Delta, k_\mu] = 0 = -k_\mu\,. \tag{A.43}$$

Furthermore, the group of dilatations is non-compact. As a consequence, the scaling dimension $\Delta$ is a real number, as any finite-dimensional representation of a non-compact Lie algebra is non-unitary.

The conformal generators are translated to non-zero values via the adjoint action of the exponential of $P$ in Eq. (A.39) and the commutation relations in Eq. (A.31),

$$\begin{aligned} D &= D\big|_{x=0} + x^\mu \left[P_\mu, D\right]_{x=0} = \Delta + x\cdot P\,,\\ M_{\mu\nu} &= iS_{\mu\nu} + x_\nu P_\mu - x_\mu P_\nu\,,\\ K_\mu &= 2x_\mu\Delta + 2ix^\nu S_{\nu\mu} + 2x_\mu \mathbf{x}\cdot P - x^2 P_\mu\,. \end{aligned} \tag{A.44}$$

These generators satisfy Eq. (A.31) since the matrix-valued quantities $\Delta$ and $S_{\mu\nu}$ commute with the spacetime-dependent parts. The spin operators $S^{\mu\nu}$ encode the (irreducible) representation of the Lorentz group on the field content $\{\phi_i\}$ and hence define the spin of $\{\phi_i\}$.[23] Physically speaking, the choice of representation determines the spin of the field. The scaling dimension $\Delta$ of a field is a real number and corresponds to the degree of homogeneity of said field,

$$\phi(e^\lambda x) = e^{-\lambda\Delta}\phi(x)\,. \tag{A.45}$$

As an example, consider the action of a massless free scalar field in flat spacetime,

$$S = \int \mathrm{d}^D x \partial_\mu\phi\partial^\mu\phi\,. \tag{A.46}$$

Under a scale transformation $x \mapsto e^\lambda x$ the action transforms as

$$S \mapsto S' = \int \mathrm{d}^D(e^\lambda x)\frac{\partial\phi(e^\lambda x)}{\partial(e^\lambda x^\mu)}\frac{\partial\phi(e^\lambda x)}{\partial(e^\lambda x_\mu)} = e^{-\lambda(\Delta-\frac{D}{2}+1)}S\,. \tag{A.47}$$

The action is therefore invariant under dilatations if and only if the scaling dimension $\Delta$ is given by

$$\Delta = \frac{D-2}{2}\,. \tag{A.48}$$

[21] Demanding that $\phi(x)$ belong to an irreducible representation of the Lorentz group implies that any matrix commuting with all of the spin matrices $S^{\mu\nu}$ has to be a multiple of $\mathbb{1}$, according to Schur's Lemma.

[22] Requiring that the spectrum of operators be bounded from below enforces $k_\mu = 0$ in the quantum theory [69].

[23] The spin is given by the square of the Pauli-Lubanski vector, $W_\mu = \frac{1}{2}\epsilon_{\mu\nu\rho\sigma}P^\nu\left(L^{\mu\nu} + S^{\mu\nu}\right)$, which is the second Casimir of the Lorentz group [33].

A mass term of the form $m^2\phi^2$ is forbidden by scale invariance. A term of the form $\phi^n$ is allowed if and only if $(D-2)n = 2D$. For $D = 4$ this means that a $\phi^4$ interaction is allowed, for $D = 3$ it is a $\phi^6$ interaction. Performing the same exercise for a free spinor field in $D$ spacetime dimensions yields

$$\Delta = \frac{D-1}{2}\,. \tag{A.49}$$

In $D \geq 3$ — independent of the existence of a Lagrangian description of the theory — CFTs are formulated in terms of primary operators (or fields in the Lagrangian description). Primary operators/fields are defined by their transformation behaviour under conformal transformations. A primary operator is homogeneous (of degree $\Delta$) and transforms under conformal transformations $\eta_{\mu\nu} \mapsto |\partial x'/\partial x|^{-2/D}\eta_{\mu\nu}$ as

$$\phi(x) \mapsto \phi'(x') = \left|\frac{\partial x'(x)}{\partial x}\right|^{\Delta/D} R\big[\Lambda(x)\big]\,\phi(x)\,, \tag{A.50}$$

where $|\partial x'/\partial x| = \Omega^{-D}$ is the Jacobian of the transformation and $R[\Lambda(x)]$ encodes the representation of the field $\phi(x)$ under the action of the Lorentz/rotation group. Note that the transformation behaviour Eq. (A.50) is equivalent to the general form of the conformal generators in Eq. (A.44) [69]. This presents the most general transformation behaviour under the conformal group that fields can exhibit in physically well-motivated theories [70–72]. [24]

### A.2.3 Conformal stress-energy tensor

For a Poincaré-invariant field theory (with Lagrangian description $\mathscr{L}$) there always exists a symmetric stress-energy tensor $T^{(B)}_{\mu\nu}$, which is derived from translational invariance and supplemented via rotational invariance and the Belinfante procedure to be symmetric.[25] By definition, the conserved Noether charges corresponding to translations and rotations can be written in terms of said stress-energy tensor,

$$Q^{(P)}_{\mu} = -\int \mathrm{d}^{D-1}x\, T^{(B)}_{0\mu}\,, \qquad Q^{(M)}_{\mu\nu} = -\int \mathrm{d}^{D-1}x\left(x_\nu T^{(B)}_{0\mu} - x_\mu T^{(B)}_{0\nu}\right). \tag{A.51}$$

The conserved charges are closely related to the spacetime generators in Eq. (A.44). In fact, classically, the Noether charges are the generators of said transformations in the sense that

$$\{Q^{(P)}_{\mu}, \phi\} \sim P_\mu\phi = \partial_\mu\phi\,, \qquad \{Q^{(M)}_{\mu\nu}, \phi\} \sim M_{\mu\nu}\phi = \left(x_\nu\partial_\mu - x_\mu\partial_\nu + iS^{\mu\nu}\right)\phi\,, \tag{A.52}$$

where $\{\cdot,\cdot\}$ denotes the (equal-time) functional Poisson bracket in the Hamiltonian formalism.[26] Given scale invariance on top of Lorentz invariance, under very basic assumptions, it is possible to define a symmetric and traceless energy momentum tensor $\overset{(C)}{T}_{\mu\nu}$. To do so one defines the so-called virial,

$$V^\mu = \frac{\partial \mathscr{L}}{\partial(\partial_\nu \phi)} \left(\eta^{\mu\nu}\Delta + i S^{\mu\nu}\right)\phi. \tag{A.53}$$

[24] Primary operators can be defined in two equivalent ways, either by Eq. (A.50) or via the action of $D$ and $K_\mu$ on the operators at the origin [69–72].

[25] There is a manifestly symmetric definition of the energy momentum tensor from General Relativity (GR) based on variation of the metric in the action [73]. It coincides with the Belinfante stress-energy tensor.

[26] In Euclidean space the Poisson bracket can be defined on arbitrary codimension-1 surfaces instead of equal-time hypersurfaces.

Under the assumption that the virial can be written as a total derivative — $V^\mu = \partial_\nu N^{\nu\mu}$ — we can define the tensor[27]

$$\chi^{\sigma\rho\mu\nu} = \frac{2\left[(D-1)\left(\eta^{\sigma\rho}N^{(\mu\nu)} - \eta^{\sigma\mu}N^{(\rho\nu)} - \eta^{\sigma\nu}N^{(\rho\mu)} + \eta^{\mu\nu}N^{(\sigma\rho)}\right) - (\eta^{\sigma\rho}\eta^{\mu\nu} - \eta^{\sigma\mu}\eta^{\rho\nu})N^{\alpha}{}_{\alpha}\right]}{(D-1)(D-2)}. \tag{A.54}$$

The modified traceless stress-energy tensor is then given by

$$\overset{(C)}{T}_{\mu\nu} = \overset{(B)}{T}_{\mu\nu} + \frac{1}{2}\partial_\sigma\partial_\rho\chi^{\sigma\rho\mu\nu}, \qquad \overset{(C)}{T}_{\mu}{}^{\mu} = 0. \tag{A.55}$$

The additional term does neither spoil the conservation law or the symmetricity of $\overset{(B)}{T}_{\mu\nu}$, nor does it change the conserved charges $Q^{(P)}_\mu, Q^{(M)}_{\mu\nu}$ in Eq.(A.51). It is traceless because of the conservation of the current associated to scale transformations $j^\mu_D$,

$$\overset{(C)}{T}_{\mu}{}^{\mu} = \partial_\mu j^\mu_D = 0, \qquad j^\mu_D = T^{\mu\nu}x_\nu + \Delta\frac{\partial\mathscr{L}}{\partial(\partial_\mu\phi)}\phi. \tag{A.56}$$

Given the existence of a traceless stress-energy tensor as described above, the conserved charge associated to Poincaré and scale transformations can be written as

$$Q^{(D)} = -\int \mathrm{d}^{D-1}x\, \overset{(C)}{T}_{0\nu}x^\nu, \qquad Q^{(P)}_\mu\Big|_{\overset{(B)}{T}_{0\nu}\to\overset{(C)}{T}_{0\nu}}, \qquad Q^{(M)}_{\mu\nu}\Big|_{\overset{(B)}{T}_{0\nu}\to\overset{(C)}{T}_{0\nu}}. \tag{A.57}$$

with $Q^{(P)}$, $Q^{(M)}$ given in Eq. (A.51). Additionally, tracelessness of the stress-energy tensor allows for the construction of four additional conserved currents,

$$j^{\mu\nu}_K = x^2\,\overset{(C)}{T}{}^{\mu\nu} - 2x^\mu x_\sigma \overset{(C)}{T}{}^{\sigma\nu}, \qquad Q^{(K)}_\mu = \int \mathrm{d}^{D-1}x\left(x^2\,\overset{(C)}{T}_{0\mu} - 2x_\mu x^\sigma\,\overset{(C)}{T}_{0\sigma}\right). \tag{A.58}$$

The charges $Q^{(K)}_\mu$ generate SCTs. Hence, full conformal invariance is obtained directly from tracelessness of the stress-energy tensor in a scale-invariant theory (at the classical level).

Note that the generators of Poincaré together with the generator of scale transformations form a subalgebra of the conformal group. In fact, conformal invariance implies scale invariance from the closure of the algebra in Eq. (A.31). The converse is not necessarily true, as discussed.

To guarantee conformal invariance at the quantum level, it is necessary that tracelessness of the stress-energy tensor also extends to the quantum theory. This is, however, only a necessary condition. To what extent scale invariance implies conformal invariance remains an open question for $D > 2$.[28] Under certain assumptions the conformal group is the maximally enhanced bosonic symmetry of space-time for massless particles [12].

[27] The notation $^{(\mu\nu)}$ denotes symmetrization, *i.e.* $A^{(\mu\nu)} = (A^{\mu\nu} + A^{\nu\mu})/2$.

[28] For CFTs in $D = 2$ there is a rigorous argument that scale invariance enhances to full conformal invariance at the quantum level under some basic assumptions such as unitarity, causality, a discrete spectrum of scaling dimensions, existence of a scale

Generically, we will denote the traceless stress-energy tensor without the index — $T_{\mu\nu} \sim T^{(C)}_{\mu\nu}$ — and assume it exists — both classically and quantum — unless otherwise specified.

## A.3 Some QFT prerequisites

In this appendix we collect some important QFT prerequisites necessary to discuss CFTs.

### A.3.1 Generators and Infinitesimal Transformations

Consider a classical field theory that is invariant under a certain (spacetime) symmetry group. A generic symmetry transformation acts on the spacetime coordinates,

$$x^\mu \mapsto x'^\mu = x^\mu + \epsilon^\alpha \frac{\delta x^\mu}{\delta \epsilon^\alpha} + \mathcal{O}(\epsilon^2)\,. \tag{A.59}$$

In addition, the fields $\{\phi_i(x)\}$ comprising the theory are allowed to transform in non-trivial representations of the symmetry group. As a consequence, symmetry transformations can non-trivially act on the fields $\phi_i(x)$ of the theory,

$$\phi_i(x) \mapsto \phi'_i(x') = \phi_i(x) + \epsilon^\alpha \frac{\delta F_{ij}}{\delta \epsilon^\alpha} \phi_j(x) + \mathcal{O}(\epsilon^2)\,, \tag{A.60}$$

where $F(x,\epsilon)_{ij}$ relates the old field $\phi_i(x)$ at $x$ to the new field $\phi'_i(x') = F_{ij}\phi_j(x)$ at $x'$ and hence encodes the representation of $\phi_i$. The most obvious examples of fields with non-trivial transformation behaviour and hence non-trivial representations are fields with non-zero spin.
It is convenient to rewrite the transformation of fields by introducing a set of generators $\{G_\alpha\}$ encoding the transformation as in

$$\Phi'_i(x) - \Phi_i(x) = -\epsilon^\alpha G_{\alpha\, ij} \Phi_j(x) + \mathcal{O}(\epsilon^2)\,, \qquad G_{\alpha\, ij} = \frac{\delta x^\mu}{\delta \epsilon^\alpha} \partial_\mu \delta_{ij} - \frac{\delta F_{ij}}{\delta \epsilon^\alpha}\,, \tag{A.61}$$

or equivalently

$$\Phi'_i(x') - \Phi_i(x) = \epsilon^\alpha \left( \frac{\delta x^\mu}{\delta \epsilon^\alpha} \partial_\mu \delta_{ij} - G_{\alpha\, ij} \right) \Phi_j(x) + \mathcal{O}(\epsilon^2)\,. \tag{A.62}$$

The differential operators $\{G_\alpha\}$ are the generator of the symmetry transformation in the classical theory. Evidently, the set of generators $\{G_\alpha\}$ form a representation of the Lie algebra on the space of fields associated to the symmetry group in question. For smooth functions and fields that transform in the trivial representation of the symmetry group — so-called scalar quantities — the generators $\{G_{\alpha\, ij}\}$ current and unbroken scale invariance [12]. The conjecture is that this result extends to $D > 2$ [12].

reduce to

$$G_{\alpha ij} = \frac{\delta x^\mu}{\delta \epsilon^\alpha} \partial_\mu \delta_{ij} \,. \tag{A.63}$$

### A.3.2 Stress-energy tensor in field theory

Consider a theory with field content $\{\phi_i\}$ and Lagrangian description $\mathscr{L}(\phi_i, \partial_\mu \phi_i)$. Under a symmetry transformation $x^\mu \mapsto x^\mu + \epsilon^\alpha \frac{\delta x^\mu}{\delta \epsilon^\alpha}$, $\phi_i(x) \mapsto \phi_i(x) + \epsilon^\alpha \frac{\delta F_{ij}}{\delta \epsilon^\alpha} \phi_j(x)$ the action transforms as

$$\delta S = \int \mathrm{d}^D x \epsilon^\alpha \partial_\mu j^\mu_\alpha \,, \qquad j^\mu_\alpha = \left[ \frac{\partial \mathscr{L}}{\partial(\partial_\mu \phi_i)} \partial_\nu \phi_i - \delta^\mu{}_\nu \mathscr{L} \right] \frac{\delta x^\nu}{\delta \epsilon^\alpha} - \frac{\partial \mathscr{L}}{\partial(\partial_\mu \phi_i)} \frac{\delta F_{ij}}{\delta \epsilon^\alpha} \phi_j \,. \tag{A.64}$$

This is just Noether's theorem: Any continuous symmetry that leaves the action invariant implies the existence of a conserved current $j^\mu_\alpha$. The associated conserved charge also derived from Noether's theorem,

$$Q_\alpha = \int \mathrm{d}^{D-1} x \, j^0_\alpha \,, \tag{A.65}$$

generates the symmetry transformation at the classical level via its Poisson bracket and is therefore related to the generator of the symmetry $G_\alpha$ in Eq. (A.61). We denote the Poisson bracken in the Hamiltonian formalism by $\{\cdot,\cdot\}$. The Poisson bracket is defined via the conjugate momenta $\{\pi\}$ and functional derivatives,

$$\{F, G\} = \int \mathrm{d}^{D-1} x \left[ \frac{\delta F}{\delta \pi_i} \frac{\delta G}{\delta \phi_i} - \frac{\delta F}{\delta \phi_i} \frac{\delta G}{\delta \pi_i} \right] \,, \qquad \pi_i = \frac{\partial \mathscr{L}}{\partial(\partial_0 \phi_i)} \,. \tag{A.66}$$

We note that there is an implicit sum over the field content of the theory in the Poisson bracket. The fundamental property of the Poisson bracket is $\{\pi_i(x), \phi_j(y)\} = \delta_{ij} \delta^{d-1}(x-y)$. In terms if its Poisson bracket the conserved Noether charge satisfies

$$\phi'_i(x) - \phi_i(x) = -\epsilon^\alpha G_{\alpha ij} \phi_j(x) = \epsilon^\alpha \{Q_\alpha, \phi_i(x)\} \,. \tag{A.67}$$

The relationship between conserved charges and symmetry transformations does carry over to the quantum theory. The stress-energy (or energy-momentum) tensor is the conserved current associated to translations $\epsilon^\nu = a^\nu$ of spacetime,

$$j^{\mu\nu}_P = T^{\mu\nu} = \left[ \frac{\partial \mathscr{L}}{\partial(\partial^\mu \phi_i)} \partial^\nu \phi_i - \eta^{\mu\nu} \mathscr{L} \right] \,. \tag{A.68}$$

The stress-energy tensor is conserved, $\partial^\mu T_{\mu\nu} = 0$. In a theory that is not only scale but also rotationally/Lorentz invariant is possible to find a Belinfante tensor $B_{\alpha\mu\nu} = -B_{\mu\alpha\nu}$ such that the Belinfante stress-energy tensor, defined as

$$T^{(B)}_{\mu\nu} = T_{\mu\nu} + \partial^\alpha B_{\alpha\mu\nu} \,, \tag{A.69}$$

is symmetric, still conserved and the Noether charge for translations remains unaffected,

$$\int \mathrm{d}^D x T^{(B)}_{0\nu} = \int \mathrm{d}^D x T_{0\nu} \,. \tag{A.70}$$

The Belinfante tensor $B_{\alpha\mu\nu}$ is derived from the conserved current associated to Lorentz transformations $j_M^{\alpha\mu\nu}$,

$$j_M^{\alpha\mu\nu} = \frac{1}{2}\Big(T^{\alpha\mu}x^\nu - T^{\alpha\nu}x^\mu + i\frac{\partial\mathscr{L}}{\partial(\partial_\alpha\phi_i)}S^{\mu\nu}\phi_i\Big), \qquad B_{\alpha\mu\nu} = \frac{i}{2}\left(\frac{\partial\mathscr{L}}{\partial(\partial_\alpha\phi_i)}S^{\nu\mu} + \frac{\partial\mathscr{L}}{\partial(\partial_\mu\phi_i)}S^{\alpha\nu} + \frac{\partial\mathscr{L}}{\partial(\partial_\nu\phi_i)}S^{\alpha\mu}\right)\phi_i\,. \tag{A.71}$$

Rotational invariance and the conservation of the Lorentz current $\partial_\alpha j_M^{\alpha\mu\nu} = 0$ imply that

$$\partial_\alpha \frac{\partial\mathscr{L}}{\partial(\partial_\alpha\phi_i)} iS^{\mu\nu}\phi_i = \partial^\alpha B_{\alpha\mu\nu} + \partial^\alpha B_{\alpha\nu\mu} = T^{\mu\nu} - T^{\nu\mu}\,. \tag{A.72}$$

Hence, the Belinfante stress-energy tensor is manifestly symmetric.

### A.3.3 Ward identities

We consider a QFT — described by the path integral $Z = \int \mathscr{D}\phi \exp[-S[\phi]]$ — that is invariant under a symmetry $\phi(x) \mapsto \phi'(x')$, so that correlation functions satisfy

$$\langle \mathscr{O}_1(x_1')\cdots\mathscr{O}_N(x_N')\rangle = \langle \mathscr{O}_1'(x_1')\cdots\mathscr{O}_N'(x_N')\rangle\,, \tag{A.73}$$

Under an infinitesimal transformation $\phi'(x) = \phi(x) - \epsilon^\alpha G_\alpha(x)$ encoded by the generator $G_\alpha$ given in Eq. (A.61) the action transforms as

$$\delta S = \int \mathrm{d}^D x\, \epsilon^\alpha \partial_\mu J_\alpha^\mu(x)\,, \tag{A.74}$$

see Eq. (A.64). Under the assumption that the path-integral measure remains invariant — *i.e.* that the symmetry is not anomalous — a correlation function $\langle X\rangle$, with $X = \phi_1(x_1)\cdots\phi_N(x_N)$, satisfies $\delta\langle X\rangle = 0$. This is just the infinitesimal form of Eq. (A.73). After expanding to linear order in the small parameter $\epsilon^\alpha$, we find the condition

$$\langle \delta X\rangle = \langle \delta S\rangle\,, \qquad \langle \delta S\rangle = \int \mathrm{d}^D x\, \langle \epsilon^\alpha \partial_\mu j_\alpha^\mu X\rangle\,. \tag{A.75}$$

The variation of $X = \phi_1(x_1)\cdots\phi_N(x_N)$ can be rewritten in terms of an integral to match the right-hand side,

$$\delta X = -\int \mathrm{d}^D x\, \epsilon^\alpha \sum_{k=1}^{N} \delta(x - x_k)\, \phi_1(x_1)\cdots G_\alpha^{(c)}\phi_k(x_k)\cdots\phi_N(x_N)\,. \tag{A.76}$$

The integrands on the left- and right-hand sides of Eq. (A.75) have to agree, which implies the equality

$$\frac{\partial}{\partial x^\mu}\langle j_\alpha^\mu(x)\phi_1(x_1)\cdots\phi_N(x_N)\rangle = -\sum_{k=1}^{N} \delta(x-x_k)\,\langle \phi_1(x_1)\cdots G_a\phi_k(x_k)\cdots\phi_N,(x_N)\rangle\,, \tag{A.77}$$

for any symmetry $G_\alpha$ realized at the quantum level. This is the so-called Ward identity for the symmetry $G_\alpha$ associated to the current $j_\alpha^\mu$. The Ward identity essentially tells us how symmetry transformations act on correlation functions in the quantum theory and how insertions of the associated conserved current can be evaluated. In addition, the Ward identity also identifies the associated charge,

$$Q_\alpha = -\int \mathrm{d}^{D-1}x\, j^0_\alpha\,, \tag{A.78}$$

as the generator of the symmetry transformation in the quantum theory: suppose that the spacetime coordinate $x^0_1$ is different from all other $x^0_i$, $i = 2, \ldots, N$. In that case we integrate the Ward identity in Eq. (A.77) over a thin box $\{x \in \mathrm{R}^D :\ x^0 \in [t_-, t_+]\}$ such that $x^0_1 \in [t_-, t_+]$, $x^0_{i\neq 1} \notin [t_-, t_+]$. Using the divergence theorem, the left-hand side turns into a surface integral,[29]

$$\langle \big(Q_\alpha(t+)\phi_1(x_1) - \phi_1(x_1)Q_\alpha(t_-)\big)\phi_2(x_2)\cdots\phi_N(x_N)\rangle = \langle \big(G_\alpha\phi_1(x_1)\big)\phi_2(x_2)\cdots\phi_N(x_N)\rangle\,. \tag{A.79}$$

In the limit $t_\pm \to x^0_1$ we conclude that

$$[Q_\alpha(x^0_1), \phi_1(x_1)] = G_\alpha\phi_1(x_1)\,. \tag{A.80}$$

The fact that the conserved current $j^\mu_\alpha$ satisfies the Ward identity in Eq. (A.77) turns the conserved charge $Q_\alpha$ into a topological operator, meaning that it can be freely moved around between different times $x^0$ in the path integral as long as it does not cross any operator insertions in the process.

In Euclidean field theory there is no preferred time direction and charges can equivalently be defined via an arbitrary closed codimension-1 surface $\Sigma$. For example, the integral of $T^{\mu\nu}$ over $\Sigma$ returns the generator of translations,

$$Q^{(P)}_\mu(\Sigma) = -\int_\Sigma \mathrm{d}n^\nu\, T_{\mu\nu}(x)\,. \tag{A.81}$$

Consider the particular choice $\Sigma = \partial B_\epsilon(x_1)$ of a small ball around $x_1$ excluding any other operator insertion. We have

$$\langle Q^{(P)}_\nu\big(\partial B_\epsilon(x_1)\big)\phi_1(x_1)\cdots\phi_N\rangle = \partial_\nu\langle\phi_1(x_1)\cdots\phi_N\rangle\,. \tag{A.82}$$

We can deduce that in the Euclidean path integral surrounding the operator insertion $\phi_1(x_1)$ with the topological surface operator $Q^{(P)}_\nu(\Sigma)$ is equivalent to taking a derivative, *i.e.* acting with the associated symmetry generator. The action of the conserved charge on $\phi_1(x_1)$ and the other operator insertions within the path integral remains unchanged for any arbitrary $\Sigma$ separating the operator insertions (hence it is called topological). Similar statements will hold for any other symmetry of the theory. The important take-away here is that the existence of a codimension-one topological operator in Euclidean QFT is equivalent to the presence of a symmetry.[30]

### A.3.4 Quantization

The process of quantization in QFT divides spacetime $\mathbb{R}^D$ up into hypersurfaces $\Sigma$ plus an additional "time" variable $t$ — schematically $\mathbb{R}^D \sim \Sigma\times[t_i, t_f]$ — with a symmetry transformation mapping hypersurfaces $\Sigma = \Sigma(t)$ at different times $t$ into each other. In order to make sense of the quantization procedure, the theory should be invariant under the spacetime symmetry relating different hypersurfaces. If this is the case, it is then possible define a Hilbert space on any arbitrary hyperplane and relate the different Hilbert spaces to each other.

[29]Correlation functions are vacuum expectation values of time-ordered products in the operator formalism.

[30]There are topological operators with support on manifolds with different codimensions and they correspond to so-called generalized symmetries [102, 309, 310].

Each hypersurface $\Sigma$ is equipped with a Hilbert space and the symmetry generator $G_t$ of translations in $t$ describes the evolution of a system or state between Hilbert spaces. In-states $|\Psi_{\text{in}}\rangle$ at $t$ are created by inserting operators in the "past" $t_0 < t$ while out-states $\langle\Psi_{\text{out}}|$ are created by inserting operators in the "future" $t_1 > t$. The (unique) vacuum state of the Hilbert space is denoted by $|0\rangle$. Initial and final states are generically given by vacuum states in the infinite past and the infinite future, respectively, and they satisfy $\langle 0|0\rangle = 1$.[31] The overlap of in- and out-states at time $t$ is equivalent to the correlation function of the inserted operators.[32] The time-evolution operator $U(t_1 - t_0) = e^{-G_t(t_1-t_0)}$ connects in-states at $t_0$ and out-states at $t_1$, so that the associated correlation function — the overlap of $|\Psi_{\text{in}}\rangle$ and $\langle\Psi_{\text{out}}|$ — reads $\langle\Psi_{\text{out}}|\,U(t_1 - t_0)\,|\Psi_{\text{in}}\rangle$.

Consider a QFT described by an action $S$ and a path integral $Z$. The path-integral description of the theory can be interpreted in terms of different time-evolutions into different Hilbert spaces dictated by the spacetime symmetries of the theory. Every possible foliation of spacetime corresponds to a quantization of the theory. For example, in a Euclidean theory with rotational invariance on $\mathbb{R}^D$ any spatial direction $\mathbf{r}_0$ is an acceptable choice of time direction. Here, the symmetry transformation described by the evolution operator $G_t$ discussed above is translation in the $\mathbf{r}_0$-direction and Hilbert spaces live on the hyperplanes orthogonal to the chosen direction.
After quantization, correlation functions — described by operator insertions in the path integral — get interpreted as time-ordered expectation values,[33]

$$\langle\mathcal{O}_1(x_1)\cdots\mathcal{O}_N(x_N)\rangle = \langle 0|\,T\{\mathcal{O}_1(x_1)\cdots\mathcal{O}_N(x_N)\}\,|0\rangle\ . \tag{A.83}$$

The time-ordering is performed with respect to the specific foliation chosen during quantization. A different choice of quantization results in different Hilbert spaces and different quantum operators, however, results in terms of correlation functions remain equivalent.[34]

Naturally, the question arises whether the choice of quantization has any consequence on the realization of the symmetries in the quantum theory. Consider an operator insertion $\mathcal{O}(x)$ at time $t$, two hypersurfaces $\Sigma_{1,2}$ at times $t_1 < t < t_2$ and a symmetry operator $Q_\alpha(\Sigma_{1,2})$ associated to a symmetry transformation respected by the quantum theory. By construction, $Q_\alpha$ is a conserved charge and topological. We are free to move it around as long as it does not cross any other insertions.[35] We are interested in correlators with insertions of $Q_\alpha(\Sigma_2) - Q_\alpha(\Sigma_1)$, $\mathcal{O}(x)$ and no other insertions in between $t_1$ and $t_2$. After quantization the difference $Q_\alpha(\Sigma_2) - Q_\alpha(\Sigma_1)$ inside any correlation function becomes a commutator with the operator insertion $\mathcal{O}(x)$ as we collide the hypersurfaces $\Sigma_{1,2}$ at time $t$. On the other hand, as $Q_\alpha$ is topological, we can also deform the total surface $\Sigma_2 - \Sigma_1$ to a sphere $\partial B(x)$ surrounding $\mathcal{O}(x)$,

[31] Other choices of initial and final states correspond to non-trivial boundary conditions in the path integral.
[32] For particle scatterings $\langle\Psi_{\text{in}}|\Psi_{\text{out}}\rangle$ is equivalent the $S$-matrix.
[33] Times are ordered with the largest time on the left and the smallest time on the right in the commutator.
[34] This can be demonstrated explicitly in specific examples, see e.g. [21].
[35] The fact that $Q_\alpha$ is topological is the quantum equivalent to its conservation classically.

implying that the following expressions are all equivalent:

$$\langle \big(Q_\alpha(\Sigma_2) - Q_\alpha(\Sigma_1)\big)\mathcal{O}(x)\dots\rangle = \langle Q_\alpha(\Sigma = \partial B(x))\,\mathcal{O}(x)\dots\rangle = \langle 0|\,T\{[Q_\alpha,\mathcal{O}(x)]\,\dots\}\,|0\rangle = \langle 0|\,T\{\big(G_\alpha\mathcal{O}(x)\big)\dots\}\,|0\rangle\,, \tag{A.84}$$

where $G_\alpha$ is the generator of the symmetry transformation acting locally on the space of operators in the quantized theory (see Eq. (A.77)). It is evident that this result is independent of the choice of quantization/foliation. The commutator $[Q_\alpha,\mathcal{O}(x)] = G_\alpha\mathcal{O}(x)$ is local while the charge $Q_\alpha(\Sigma)$ itself is non-local. This is consistent because $Q_\alpha$ is topological and — independent of the foliation — can be deformed such that it is supported on an arbitrary small sphere around the insertion point $x$. We conclude that it is justified to replace the charge $Q_\alpha$ in the path integral by the local operator $G_\alpha$ in the operator formalism without specifying a quantization scheme. The identifications between the operator formalism in the quantized theory and the path-integral formalism schematically look as follows:

$$G_\alpha\mathcal{O}(x) = [Q_\alpha,\mathcal{O}(x)] \sim Q_\alpha \circ \mathcal{O}(x) \sim$$

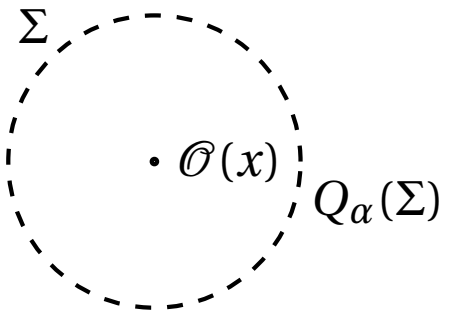

. (A.85)

The shorthand notation $Q_\alpha \circ \mathcal{O}(x)$ signifies that we surround $\mathcal{O}(x)$ with the surface operator $Q_\alpha$. It can be interpreted equivalently to a commutator in the operator formalism.[36] In terms of the generators $G_\alpha$, the notation $G_\alpha\mathcal{O}(x) = [G_\alpha,\mathcal{O}(x)]$ is valid due to the Jacobi identity (this is the adjoint action). Alternatively, we can think of states $|\mathcal{O}(x)\rangle = \mathcal{O}(x)\,|0\rangle$ on the Hilbert space. The vacuum $|0\rangle$ is invariant under all symmetry operators $Q_\alpha \sim G_\alpha$ of the theory and, with regards to notation, it is true that $G_\alpha\mathcal{O}(x)\,|0\rangle = [\hat{G}_\alpha,\mathcal{O}(x)]\,|0\rangle$.

Finally, the expression $Q_N \circ \dots \circ Q_1 \circ \mathcal{O}(x)$ means surrounding the insertion $\mathcal{O}(x)$ in the path integral with topological surface operators of increasing size,

$$Q_N \circ \dots \circ Q_1 \circ \mathcal{O}(x) \sim$$

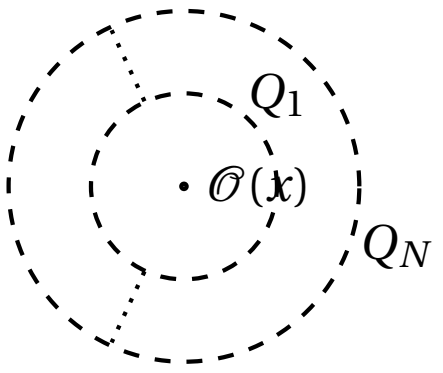

. (A.86)

Suppose we want to change the order of the inserted charges/surface operators. In that case the commutation relations obeyed by the conserved charges tell us how to properly reorder insertions of said surface operators, as this is equivalent to commuting symmetry generators in the operator formalism.[37]

[36] The way local operators transform under symmetries is always insensitive to IR details like SSB or compactification in time (finite temperature), as commutators can always be computed at short distances.

[37] For the conformal charges these commutation relations are given by Eq. (1.11).

### A.3.5 Hermitian conjugation in Euclidean spacetime and reflection positivity

Unitarity is a fundamental property that a QFT in Lorentzian signature can exhibit. In many physical situations we require unitarity of the theory in question as it is equivalent to the preservation of probability [27–29].[38] In unitary theories conserved charges, including the Hamiltonian, are Hermitian operators that generate unitary transformations at the quantum level. In Euclidean signature, however, unitarity for Lorentzian QFTs is mapped into a property called reflection positivity.

Consider a Lorentzian QFT with Hermitian stress-energy generators $(H, P_i^{(L)})$ and a local Hermitian operator $\mathcal{O}^{(L)}(0) = \mathcal{O}^{(L)\dagger}(0)$. It holds that

$$\mathcal{O}^{(L)}(t,\mathbf{x}) = e^{iHt+i\mathbf{x}\cdot\mathbf{P}^{(L)}}\mathcal{O}^{(L)}(0)\, e^{-iHt-i\mathbf{x}\cdot\mathbf{P}^{(L)}}\,, \tag{A.87}$$

implying that $\mathcal{O}^{(L)}(x)$ is Hermitian as well. After performing a Wick rotation $t = -i\tau$, in Euclidean signature the above relationship turns into

$$\mathcal{O}^{(E)}(\tau,\mathbf{x}) = e^{H\tau+i\mathbf{x}\cdot\mathbf{P}^{(L)}}\mathcal{O}^{(L)}(0)\, e^{-H\tau-i\mathbf{x}\cdot\mathbf{P}^{(L)}} \qquad \mathcal{O}^{(E)}(\tau,\mathbf{x}) := \mathcal{O}^{(L)}(-i\tau,\mathbf{x})\,. \tag{A.88}$$

Therefore, the corresponding Euclidean operator $\mathcal{O}^{(E)}$ satisfies

$$\mathcal{O}^{(E)\dagger}(\tau,\mathbf{x}) = \mathcal{O}^{(E)}(-\tau,\mathbf{x})\,. \tag{A.89}$$

This property can be generalized to operators with arbitrary spin. For tensor fields, for example, reflection positivity reads

$$\mathcal{O}^{(E)\dagger}_{\mu_1\ldots\mu_N}(\tau,\mathbf{x}) = \Theta_{\mu_1}{}^{\nu_1}\cdots\Theta_{\mu_N}{}^{\nu_N}\mathcal{O}^{(E)}_{\nu_1\ldots\nu_N}(-\tau,\mathbf{x})\,, \qquad \Theta_\mu{}^\nu = \delta_\mu{}^\nu - 2\delta_\mu{}^0\delta_0{}^\nu\,. \tag{A.90}$$

Operators that satisfy the condition in Eq (A.90) — and are therefore Hermitian in Lorentzian signature — are sometimes called real operators and we will refer to them as such here.

As can be seen from Eq. (A.90), Hermitian conjugation in Euclidean space becomes a reflection in Euclidean time. This means that the choice of time direction is important for the definition of Hermitian conjugation. As a consequence, the conjugation properties of operators depend on how we choose to quantize the theory. This is fundamentally different to the situation in Lorentzian signature, where the conjugation properties of local operators are independent of the frame of reference. For example, we can consider the associated charges to the momentum generators in Euclidean signature and their conjugation properties in a foliation along the $x^0 = \tau$ direction,

$$Q_\mu^{(P)} = -\int \mathrm{d}^{D-1}x\, T_{\mu 0}(0,\mathbf{x})\,, \qquad \longrightarrow \qquad Q_0^{(P)\dagger} = Q_0^{(P)}\,,\ Q_i^{(P)\dagger} = -Q_i^{(P)}\,. \tag{A.91}$$

Evidently, we can relate the generators $Q_\mu^{(P)}$ to their Lorentzian counterparts discussed above as follows:

$$Q_0^{(P)} \sim P_0 = \hat{H}\,, \qquad Q_i^{(P)} \sim P_i = iP_i^{(L)}\,. \tag{A.92}$$

[38] In Lorentzian QFT unitarity is often formulated as the requirement that the $S$-matrix is unitary.

If we had quantized using a different time direction, we would have concluded that the momentum generator in the new time direction is Hermitian while all the other generators are anti-Hermitian. We see that Hermitian conjugation is fundamentally dependent on the choice of foliation.

A natural question to ask is what conditions does a Euclidean QFT have to satisfy if it computes a Wick-rotated unitary Lorentzian theory. First and foremost, in such a theory it is necessarily the case that the norms of states in the theory are all positive. Given a state of the form

$$|\psi\rangle = \mathcal{O}_1(-\tau_1, \mathbf{x}_1)\cdots\mathcal{O}_N(-\tau_N, \mathbf{x}_N)\,|0\rangle\,, \tag{A.93}$$

with operator insertions at negative Euclidean times, the condition of positive norms implies that

$$\langle\psi|\psi\rangle = \langle 0|\mathcal{O}_1(\tau_1, \mathbf{x}_1)\cdots\mathcal{O}_N(\tau_N, \mathbf{x}_N)\mathcal{O}_1(-\tau_1, \mathbf{x}_1)\cdots\mathcal{O}_N(-\tau_N, \mathbf{x}_N)\,|0\rangle \geq 0\,. \tag{A.94}$$

In terms of correlators, this statement is the requirement that time-reflection symmetric correlation functions have a positive path-integral description. This property is called reflection positivity and it is a necessary condition for a Euclidean theory to be given by a Wick-rotated unitary Lorentzian theory. In particular, in the context of CFTs it is very natural to define a theory in Euclidean spacetime, and in that case reflection positivity always has to be checked.

Beyond the scope of this discussion, there are important distinctions between Lorentzian and Euclidean signature. For example, out-of-time-order correlators are not very meaningful in Euclidean signature as the Hamiltonian is bounded only from below and not from above. Hence, out-of-time-order correlation functions are formally infinite. In Lorentzian QFT, out-of-time-order correlators — so called Wightman functions [24] — are meaningful and interesting observables because of the additional imaginary unit in the definition of the evolution operator. In this context, however, it deserves to be mentioned that both time-ordered correlation functions and Wightman functions in Lorentzian QFT can be obtained from time-ordered correlators in Euclidean QFT by carefully performing the Wick rotation [21].

.

.

## A.4 A schematic derivation of the Nambu-Goldstone mode counting rule

In this Appendix we review the derivation of the general counting rule for NG modes in non-relativistic systems discussed in Section 1.2.4. The derivation presented here is based on [36]. For more details we refer to the above publication and also [34, 35, 47, 311, 312]. We only give a rough sketch of proof and ignore certain more involved details.

The proof and derivation of the general counting rule follows from a more fundamental effective Lagrangian approach. The effective Lagrangian $\mathscr{L}_{\text{eff}}$ captures adequately the low-energy long-distance fluctuations and the macroscopic order of the theory. The main input for the construction of the effective Lagrangian is the symmetry-breaking pattern

$$G \to H, \tag{A.95}$$

where $G$ is the full global symmetry group of the theory and $H$ is the unbroken subgroup at low energies. As usual in effective field theory, the most general Lagrangian capturing the dynamics of the system in question is constructed by including all terms — or operators — that respect the symmetries of said system. Naturally, the low-energy long-distance behaviour is described by fluctuations in the direction of the coset of $G$ and $H$,

$$G/H = \{gh \,|\, g \in G,\ h \in H\}. \tag{A.96}$$

These fluctuations correspond to the massless NG bosons present because of SSB. We explicitly assume here that we are so low in energies such that there are no other long-distance DoFs present. In our construction, we introduce one field $\pi_a$ for each broken generator $T_a^{(\mathrm{br})}$. For now we refer to all of these fields as NG modes, even though in general some of the corresponding modes are not necessarily independent of each other in the absence Lorentz invariance, as we will see. The NG fields $\pi_a$ naturally map spacetime into the manifold $G/H$. In the following, we write down the effective Lagrangian for the NG modes in a systematic expansion in powers of derivatives, which is consistent since higher-derivative terms become less important at longer distances.

Naturally, the NG fields $\pi_a$ transform under the action of $G$. However, since they form a parametrization of the coset $G/H$, the fields $\pi_a$ cannot linearly realize the full symmetry group $G$. Instead, they form a non-linear realization of $G$. In general, under the action of $G$ the NG modes transform as

$$\delta^{(\epsilon)}\pi_a = \epsilon_b h_{ba}(\pi), \tag{A.97}$$

where the generators $h_{ba}(\pi)$ can be viewed as vector fields on $G/H$,

$$h_b(\pi) = h_{ba}(\pi)\partial_a, \qquad \partial_a := \frac{\partial}{\partial \pi_a}. \tag{A.98}$$

The generators $h_b$ satisfy the commutation relations associated to $G$,

$$[h_b, h_c] = f_{bcd} h_d. \tag{A.99}$$

where $b, c, d$ refer to generators of $G$. Crucially for us later, on the level of the action a symmetry transformation can always change the Lagrangian density by a total derivative.

In the continuum limit (at sufficiently long distances) we assume spatial translational invariance as well as rotational invariance. If the system exhibits Lorentz invariance, the form of the effective Lagrangian $\mathscr{L}_{\mathrm{eff}}$ is highly constrained,

$$\mathscr{L}_{\mathrm{eff}} = \frac{1}{2} g_{ab}(\pi)\partial_\mu \pi_a \partial^\mu \pi_b + \mathscr{O}(\partial^4). \tag{A.100}$$

Invariance under $G$ at the level of the Lagrangian requires that $g_{ab}(\pi)$ be a $G$-invariant metric on $G/H$. If we drop Lorentz invariance the general form of the effective Lagrangian exhibits substantially more freedom,

$$\mathscr{L}_{\text{eff}} = c_a(\pi)\dot{\pi}_a + \frac{1}{2}g^{(1)}_{ab}(\pi)\dot{\pi}_a\dot{\pi}_b + \frac{1}{2}g^{(2)}_{ab}(\pi)\partial_i\pi_a\partial_i\pi_b + \mathscr{O}(\partial_\tau^3, \partial_\tau\partial_i^2, \partial_i^4)\,. \tag{A.101}$$

Importantly, spatial isotropy only allows for quadratic terms in the spatial derivatives within the Lagrangian but permits a linear term in $\dot{\pi}_a$. The coefficients $c_a(\pi)$, $g^{(1)}_{ab}(\pi)$ and $g^{(2)}_{ab}(\pi)$ are dependent on the symmetry-breaking pattern $G \to H$. Under the action of the symmetry group $G$ the Lagrangian density in Eq. (A.101) can only change by a total derivative. This is the case if and only if under the infinitesimal transformation $\delta^{(\epsilon)}\pi_a = \epsilon_b h_{ba}(\pi)$ we have

$$\big(\partial_b c_a(\pi) - \partial_a c_b(\pi)\big)h_{cb}(\pi) = \partial_a e_c(\pi)\,. \tag{A.102}$$

Importantly, The functions $e_c(\pi)$ are related to the charge densities of the system. Using the fact that the variation of the Lagrangian in Eq. (A.101) is given by the surface term

$$\delta\mathscr{L}_{\text{eff}} = \epsilon_b\partial_t\big(c_a(\pi)h_{ba}(\pi) + e_b(\pi)\big)\,, \tag{A.103}$$

we can derive the form of the Noether current for the global symmetry, which is given by

$$j^0_c = e_c(\pi) - g^{(1)}_{ab}(\pi)h_{ca}(\pi)\dot{\pi}_b\,. \tag{A.104}$$

Since the ground state $|0\rangle$ is time-independent, we see that

$$\langle 0|j^0_c(x)|0\rangle = e_c(\pi)\big|_{\pi=0}\,. \tag{A.105}$$

In particular, the charge densities must vanish in the Lorentz-invariant case, which is obviously true as $e_c$ is zero if all $c_a$ vanish. The matrix $\rho_{ij}$ defined in Eq. (1.129) is now given by

$$\rho_{ab} := -i\lim_{\Omega\to\infty}\langle 0|[Q_{\Omega a}, j^0_b(x)]|0\rangle = -i\,\langle 0|[Q_a, j^0_b(x)]|0\rangle = h_{ac}\partial_c e_b\big|_{\pi=0}\,. \tag{A.106}$$

For unbroken generators this vanishes by definition. For the broken generators we can further deduce that

$$h_{ca}h_{db}\big(\partial_b c_a - \partial_a c_b\big)\big|_{\pi=0} = \rho_{cd} \tag{A.107}$$

This differential equation can be solved around the origin. The Taylor expansion of $c_a(\pi)$ is given by

$$c_a(\pi) = c_a(0) + (S_{ab} + A_{ab})\pi_b + \mathscr{O}(\pi^2)\,, \tag{A.108}$$

where $S_{ab}$ and $A_{ab}$ stand for the symmetric and anti-symmetric parts of the derivative $\partial_b c_a|_{\pi=0}$, respectively. As $c_a(0)$ and $S_{ab}$ only lead to total derivative terms in the effective Lagrangian $\mathscr{L}_{\text{eff}}$, they can be ignored or dropped. In total we have

$$c_a(\pi)\dot{\pi}_a = A_{ab}\dot{\pi}_a\pi_b + \text{total derivative terms} + \mathscr{O}(\pi^3)\,. \tag{A.109}$$

The equation for the anti-symmetric part,

$$2h_{ca}h_{db}\big|_{\pi=0}A_{ab} = \rho_{cd}\,, \tag{A.110}$$

has a unique solution which reads

$$c_a(\pi)\dot{\pi}_a = \frac{1}{2}\rho_{ab}\dot{\pi}'_a\pi'_b + \mathcal{O}(\pi'^3)\,, \qquad\qquad \pi'_a := \pi_b\big(h^{-1}(0)\big)_{ba}\,. \tag{A.111}$$

As the matrix $\rho_{ab}$ is real and anti-symmetric, it can always be brought into a suitable block-diagonal form,

$$M^T\rho M = \begin{pmatrix} 0 & \lambda_1 & & & & & \\ -\lambda_1 & 0 & & & & & \\ & & \ddots & & & & \\ & & & 0 & \lambda_m & & \\ & & & -\lambda_m & 0 & & \\ & & & & & \ddots & \\ & & & & & & 0 \end{pmatrix}, \tag{A.112}$$

by an orthogonal transformation $M$. Here, we have $\lambda_\alpha \neq 0$ for $\alpha = 1, \dots, m$ with $2m = \text{rank}(\rho)$. The explicit expression of the effective Lagrangian in Eq. (A.101) now includes

$$c_a(\pi)\dot{\pi}_a = \frac{1}{2}\sum_{\alpha=1}^{m}\lambda_\alpha\left(\dot{\pi}'_{2\alpha-1}\pi'_{2\alpha} - \dot{\pi}'_{2\alpha}\pi'_{2\alpha-1}\right). \tag{A.113}$$

When compared to the standard and familiar form of any Lagrangian on the phase space,[39]

$$\int \mathrm{d}^{D-1}x\,\mathscr{L} = p_i\dot{q}_i - H\,, \tag{A.114}$$

this implies that $\pi'_{2\alpha-1}$ and $\pi'_{2\alpha}$ are canonically conjugate variables, which together represent a single Degree(s) of Freedom (DoF) rather than two. This proves the counting rule.

We end this appendix with two remarks. First, we note that the definition of Degree(s) of Freedom (DoF) used here is the conventional one in physics. In other words, we need to specify both the initial value of any given DoF itself as well as its time derivative as initial conditions. And second, the Lagrangian formalism is mandatory in our discussion. If we would want to move to the Hamiltonian formalism, we then would run into the issue that the presence of the first-order derivative terms affect the definition of the canonical momenta.

[39] Here, Darboux's theorem states that we can choose a local coordinate system in such a way that higher-order terms $\mathcal{O}(\pi'^3)$ vanish.

# B Appendices to Chapter 2

## B.1 Hyperspherical harmonics and conformal symmetry

In this appendix we collect useful formulas concerning hyperspherical harmonics in $D$-dimensional space. We mostly follow [193]. In the second part we discuss the constraints imposed by conformal symmetry on two-, three- and four-point functions introduced in Section 1.1 in the limit of infinite separations and in the language of hyper-spherical harmonics on the cylinder.

### B.1.1 Hyperspherical harmonics and their properties

Hyperspherical harmonics — denoted by $Y_{\ell m}$ — are the eigenfunction of the Laplacian on $S_{r_0}^{D-1}$, $D \geq 3$,

$$-\Delta_{S_{r_0}^{D-1}} Y_{\ell m}(\mathbf{n}) = \lambda_\ell Y_{\ell m}(\mathbf{n}), \qquad \lambda_\ell := \ell(\ell + D - 2), \tag{B.1}$$

where $\ell = 0, 1, \ldots$ is a positive integer and $m$ in $D > 3$ is a vector consisting of $D-2$ components which in the standard-tree convention satisfy

$$l \geq m_1 \geq m_2 \geq \cdots \geq m_{D-3} \geq |m_{D-2}|. \tag{B.2}$$

We note that the lowest component $m_{D-2}$ can be associated to the standard $SO(3)$ quantum number appearing in the spherical harmonics in $D = 3$. This is also the only component which can be negative. In our convention we denote the vector with the sign of $m_{D-2}$ switched as the conjugate $m^*$. This is reasonable, as it appears in the conjugation property of the hyperspherical harmonics,

$$Y_{\ell m}^* = (-1)^{m_{D-2}} Y_{\ell m^*}. \tag{B.3}$$

As can be seen above, the eigenvalues $\lambda_\ell$ only depend on $\ell$ and not on $m$. The multiplicity $\mathrm{Deg}_D(\ell)$ of the eigenvalue $\lambda_\ell$ is given by the all the different choices of $m$ and reads

$$\mathrm{Deg}_D(\ell) = \frac{(D + 2\ell - 2)\Gamma(D + \ell - 2)}{\Gamma(\ell + 1)\Gamma(D - 1)}. \tag{B.4}$$

Due to the fact that the Laplacian is self-adjoint, the hyperspherical harmonics $Y_{\ell m}$ form an orthonormal basis of $L^2(S_{r_0}^{D-1})$,

$$(Y_{\ell m}, Y_{\ell' m'}) = \int_{S_{r_0}^{D-1}} \mathrm{d}S\, Y_{\ell m}(\mathbf{n}) Y^*_{\ell' m'}(\mathbf{n}) = \delta_{\ell\ell'}\delta_{mm'}\,, \tag{B.5}$$

where we have defined the rescaled volume element on the sphere as

$$\mathrm{d}S = r_0^{D-1}\mathrm{d}\Omega\,. \tag{B.6}$$

Certain useful identities can be obtained by summing spherical harmonics and their derivatives over the indices $m$

$$\sum_m Y_{\ell m}(\mathbf{n}) Y^*_{\ell m}(\mathbf{n}) = \frac{\mathrm{Deg}_D(\ell)}{\Omega_D}\,, \tag{B.7}$$

$$\sum_m Y_{\ell m}(\mathbf{n})\partial_i Y^*_{\ell m}(\mathbf{n}) = 0\,, \tag{B.8}$$

$$\sum_m \partial_i Y_{\ell m}(\mathbf{n})\partial_j Y^*_{\ell m}(\mathbf{n}) = \frac{\mathrm{Deg}_D(\ell)}{\Omega_D}(r_0^2\lambda_\ell) h_{ij}(\mathbf{n})\,, \tag{B.9}$$

where $\Omega_D = \frac{2\pi^{D/2}}{\Gamma(D/2)}$ volume of the $D-1$-dimensional unit sphere. Sums over the eigenvalues $\lambda_\ell$ — *i.e.* the dispersion relations $\omega_\ell$ in Eq. (2.32) — in terms of Eq. (2.86),

$$\sum_{\ell,m} \omega_\ell^s Y_{\ell m}(\mathbf{n}) Y^*_{\ell m}(\mathbf{n}) = \frac{\Sigma(s)}{\Omega_D r_0^s} = \frac{\zeta(-s/2|S_{r_0}^{D-1})}{(D-1)^{s/2} r_0^s \Omega_D}\,, \qquad \omega_\ell^2 = \frac{\ell(\ell+D-2)}{r_0^2(D-1)} = \frac{\lambda_\ell}{r_0^2(D-1)}\,. \tag{B.10}$$

The cut-off independent part of $\Sigma(s)$ ($\Lambda$-independent) is related to the zeta function on the sphere [313],

$$\zeta(s|S_{r_0}^{D-1}) = \mathrm{Tr}\left[\left(-\Delta_{S_{r_0}^{D-1}}\right)^s\right]\,. \tag{B.11}$$

For $s=1$ we recover the expression for the Casimir energy of the superfluid phonon discussed and computed in Appendix B.2,

$$\sum_{\ell,m} \omega_\ell Y_{\ell m}(\mathbf{n}) Y^*_{\ell m}(\mathbf{n}) = \frac{\Sigma(1)}{r_0\Omega_D} = \frac{2\Delta_1}{r_0\Omega_D}\,. \tag{B.12}$$

Sums with open derivative indices can be computed analogously,

$$\sum_{\ell,m} \omega_\ell^s \partial_i Y_{\ell m}(\mathbf{n})\partial_j Y^*_{\ell m}(\mathbf{n}) = \frac{\Sigma(s+2)}{r_0^s\Omega_D} h_{ij} = \frac{\zeta(-s/2-1|S_{r_0}^{D-1})}{(D-1)^{s/2+1} r_0^s \Omega_D} h_{ij}\,, \tag{B.13}$$

which for $s=1$ becomes

$$\sum_{\ell,m} \frac{1}{\omega_\ell}\partial_i Y_{\ell m}(\mathbf{n})\partial_j Y^*_{\ell m}(\mathbf{n}) = \frac{r_0\Sigma(1)}{\Omega_D} h_{ij} = \frac{2r_0\Delta_1}{\Omega_D} h_{ij}\,. \tag{B.14}$$

Another useful tool are the Gegenbauer polynomials, which are defined from the hyperspherical harmonics on the unit sphere $S^{D-1}\, r_0 = 1$ as follows:

$$C_\ell^{D/2-1}(\mathbf{n}\cdot\mathbf{n}') = \frac{(D-2)\Omega_D}{(D+2\ell-2)} \sum_m Y^*_{\ell m}(\mathbf{n}) Y_{\ell m}(\mathbf{n}')\,. \tag{B.15}$$

Monomials on the unit sphere can be decomposed in terms of these Gegenbauer polynomials,

$$(\mathbf{n}\cdot\mathbf{n}')^\ell = \frac{\ell!}{2^\ell} \sum_{s=0}^{\lfloor \frac{\ell}{2} \rfloor} \frac{\left(\frac{D}{2}-1+\ell-2s\right)\Gamma\left(\frac{D}{2}-1\right)}{s!\,\Gamma\left(\frac{D}{2}+\ell-s\right)} C_{\ell-2s}^{D/2-1}(\mathbf{n}\cdot\mathbf{n}')\,. \tag{B.16}$$

The Gegenbauer polynomials also satisfy an addition property of the form

$$C_{\ell_a}^{D/2-1}(\mathbf{n}\cdot\mathbf{n}') C_{\ell_b}^{D/2-1}(\mathbf{n}\cdot\mathbf{n}') = \sum_{k=0}^{\min(\ell_a,\ell_b)} \langle k|\ell_a \ell_b\rangle\, C_{\ell_a+\ell_b-2k}^{D/2-1}(\mathbf{n}\cdot\mathbf{n}')\,, \tag{B.17}$$

with the coefficients $\langle k|\ell_a\ell_b\rangle$ given by

$$\begin{aligned}\langle k|\ell_a\ell_b\rangle = &\left(\frac{D}{2}-1-2k+\ell_a+\ell_b\right) \frac{\Gamma(\ell_a+\ell_b+1-2k)}{\Gamma\left(\frac{D}{2}-1\right)^2 \Gamma(\ell_a+\ell_b-2k+D-2)} \\ &\times \frac{\Gamma\left(\frac{D}{2}+k-1\right)\Gamma(\ell_a+\ell_b-k+D-2)\Gamma\left(\ell_a-k+\frac{D}{2}-1\right)\Gamma\left(\ell_b-k+\frac{D}{2}-1\right)}{\Gamma(k+1)\Gamma\left(\ell_a+\ell_b-k+\frac{D}{2}\right)\Gamma(\ell_a-k+1)\Gamma(\ell_b-k+1)}\,. \end{aligned} \tag{B.18}$$

We note that this is simply a generalization of angular-momentum addition in $D = 3$.

### B.1.2 Constraints from conformal symmetry

As discussed in detail in Section 1.1, conformal invariance strongly constrains the form of correlation functions. Here, we care about correlators where the outermost insertion are at infinite separation which — by the state–operators correspondence — corresponds to matrix elements of states inserted at $\tau_{2,1} = \pm\infty$. Furthermore, for spinful operators it is most convenient to work in the spherical tensor basis and hence in terms hyperspherical harmonics.

In the standard basis in Cartesian coordinates an object which transforms in an irreducible representation of the rotation group $SO(D)$ is represented by a completely symmetric and traceless tensor $T_{\nu_1\ldots\nu_\ell}$. Naturally, the same object can be consistently represented in the spherical basis as a tensor $T_{\ell,m}$ labelled by $\ell, m$. The operators $\mathrm{U}_{\ell m}^{\nu_1\ldots\nu_\ell}$ which allows us to pass from one basis to the other is represented by the following integral on the unit sphere:

$$\mathrm{U}_{\ell m}^{\nu_1\ldots\nu_\ell} = k_{D,\ell} \int_{S_1^{D-1}} \mathrm{d}\Omega\, n^{\nu_1}\ldots n^{\nu_\ell}\, Y^*_{\ell m}(\mathbf{n})\,, \tag{B.19}$$

where the normalization $k_{D,\ell}$ is fixed by the condition that the operator U squares to one,

$$|\mathrm{U}_{\ell m}|^2 = \delta_{\mu_1\nu_1}\ldots\delta_{\mu_\ell\nu_\ell} \left(\mathrm{U}_{\ell m}^{\nu_1\ldots\nu_\ell}\right)^* \mathrm{U}_{\ell m}^{\mu_1\ldots\mu_\ell} = 1\,, \tag{B.20}$$

and reads

$$k_{D,\ell} = \sqrt{\frac{2^\ell}{\Omega_D} \frac{\Gamma\left(\frac{D}{2}+\ell\right)}{\ell!\,\Gamma\left(\frac{D}{2}\right)}}\,. \tag{B.21}$$

For example, we can consider the simplest non-trivial case given by a vector $V_\mu$ in $D = 3$, which is mapped to the object $V_{1m}$ with components

$$\begin{pmatrix} V_{1,-1} \\ V_{1,0} \\ V_{1,1} \end{pmatrix} = \begin{pmatrix} -\frac{1}{\sqrt{2}}(V_1 + iV_2) \\ V_3 \\ \frac{1}{\sqrt{2}}(V_1 - iV_2) \end{pmatrix}. \tag{B.22}$$

We want to discuss the structure of conformal correlators in the limit of large separation. The two-point function of two primary operators is non-vanishing only if they live in conjugate representations of the conformal group and hence exhibit the same scaling dimension, see Section 1.1. On the cylinder at large separation $\tau_2 - \tau_1 \gg 1$ the two-point function of the operator $\mathscr{O}^{q;\Delta}_{\ell\bar{m}}(\tau_1, \mathbf{n}_1)$ simplifies to

$$\langle \mathscr{O}^{-q;\Delta}_{\ell\bar{m}}(\tau_2, \mathbf{n}_2)\, \mathscr{O}^{q;\Delta}_{\ell\bar{m}}(\tau_1, \mathbf{n}_1)\rangle = e^{-(\tau_2-\tau_1)\Delta/r_0}\, I^\ell_{m\bar{m}}(\mathbf{n}_2) := \mathscr{A}^\Delta(\tau_1,\tau_2)\, I^\ell_{m\bar{m}}(\mathbf{n}_2)\,, \tag{B.23}$$

where we have made use of the fact that in the limit of large separation the unit vector in the direction of the separation between the two insertions is given by

$$\mathbf{n} = \frac{x-y}{|x-y|} = \frac{e^{\tau_2/r_0}\mathbf{n}_2 - e^{\tau_1/r_0}\mathbf{n}_1}{\left|e^{\tau_2/r_0}\mathbf{n}_2 - e^{\tau_1/r_0}\mathbf{n}_1\right|} \xrightarrow{\tau_{2,1}\to\pm\infty} \mathbf{n}_2\,. \tag{B.24}$$

The tensor structure $I^\ell_{m\bar{m}}$ appearing in the two-point function is the intertwiner between conjugate representations and is equal to the appropriate structure constructed from the inversion tensor from Eq. (1.44) and written in spherical coordinates (see Section 1.1 for details).

Similarly, the three-point function of scalar primaries — given in Eq. (1.40) — is fixed up to a constant. In the limit of large separation $\tau_{2,1} \to \pm\infty$ on the cylinder the general form of the three-point function simplifies considerably. Importantly, the dependence on the scaling dimension of the operator insertion in the middle drops out,

$$\langle \mathscr{O}^2(x_2)\, \mathscr{O}^c(x)\, \mathscr{O}^1(x_1)\rangle \longrightarrow C_{1c2}\, e^{-\Delta_2(\tau_2-\tau)/r_0}\, e^{-\Delta_1(\tau-\tau_1)/r_0} = \mathscr{A}^{\Delta_2}_{\Delta_1}(\tau_1,\tau_2|\tau)\, C_{1c2}\,. \tag{B.25}$$

In the special case where $\Delta_1 = \Delta_2 = \Delta$ the dependence on the insertion point $\tau/x$ drops out completely,

$$\langle \mathscr{O}^2(x_2)\, \mathscr{O}^c(x)\, \mathscr{O}^1(x_1)\rangle \longrightarrow C_{1c2}\, e^{-\Delta(\tau_2-\tau_1)/r_0} = \mathscr{A}^\Delta(\tau_1,\tau_2)\, C_{1c2}\,. \tag{B.26}$$

The scalar four-point function, given in Eq. (1.42), additionally depends on the conformal cross-ratios as defined in Eq. (1.41). In the limit of large separation the dependence on the cross-ratios simplifies and can be collected in a function $f_c(\tau' - \tau, \mathbf{n}\cdot\mathbf{n}')$ that depends only on the intermediate insertions $\mathscr{O}^c(x)\mathscr{O}^d(x')$. Furthermore, the large-separation limit of the scalar four-point function can also be expressed in terms of the function $\mathscr{A}(\tau_1, \tau_2)$. The overall result then reads

$$\langle \mathcal{O}^2(x_2)\mathcal{O}^c(x)\mathcal{O}^d(x')\mathcal{O}^1(x_1)\rangle = e^{-\Delta_2\frac{(\tau_2-\tau)}{r_0}-\Delta_1\frac{(\tau-\tau_1)}{r_0}} f_c(\tau'-\tau, \mathbf{n}\cdot\mathbf{n}') = \mathscr{A}^{\Delta_2}_{\Delta_1}(\tau_1,\tau_2|\tau)\, f_c(\tau'-\tau,\mathbf{n}\cdot\mathbf{n}'). \quad \text{(B.27)}$$

For spinful operators the above result have to be supplemented by the appropriate tensor structure [76, 314], see also Section 1.1 for details. For example, for the scalar–scalar–(spin-$\ell$) correlator the result in Eq. (B.25) has to be multiplied by the tensor structure $(V^{(ijk)}\cdot t)^\ell$, where $t$ is an auxiliary (complex) vector that squares to zero to ensure tracelessness and

$$V^{(ijk)} = \frac{|x_{ki}||x_{kj}|}{|x_{ij}|}\left(\frac{x_{ki}}{|x_{ki}|^2} - \frac{x_{kj}}{|x_{kj}|^2}\right), \quad \text{(B.28)}$$

In the spherical basis this object is particularly simple. We note that, by construction, $\mathrm{U}^{\nu_1\dots\nu_\ell}_{\ell m}$ is manifestly traceless and anti-symmetric. Hence, there is no need to worry about traces and

$$\begin{aligned} V^{(ijk)}_{\ell m} &= \mathrm{U}^{\mu_1\dots\mu_\ell}_{\ell m} V^{(ijk)}_{\mu_1}\dots V^{(ijk)}_{\mu_\ell} = k_{\ell,D}\int \mathrm{d}\Omega\, Y^*_{\ell m}(\mathbf{n})\left(\mathbf{n}\cdot V^{(ijk)}\right)^\ell \\ &= \frac{1}{k_{\ell,D}}\frac{\left||x_{kj}|^2x_{ki}-|x_{ki}|^2x_{kj}\right|^\ell}{|x_{ij}|^\ell|x_{ki}|^\ell|x_{kj}|^\ell} Y^*_{\ell m}\left(\frac{|x_{kj}|^2x_{ki}-|x_{ki}|^2x_{kj}}{\left||x_{kj}|^2x_{ki}-|x_{ki}|^2x_{kj}\right|}\right). \end{aligned} \quad \text{(B.29)}$$

On the cylinder — $x_i = r_0 e^{\tau_2/r_0}\mathbf{n}_2$, $x_j = r_0e^{\tau_1/r_0}\mathbf{n}_1$, $x_k = r_0 e^{\tau/r_0}\mathbf{n}$ — in the limit of large separation $\tau_{2,1} = \pm\infty$, in agreement with representation theory, this expression simplifies to

$$V^{(ijk)}_{\ell m} = \frac{1}{k_{\ell,D}} Y^*_{\ell m}(\mathbf{n})\left(1+\mathcal{O}\left(e^{-(\tau_2-\tau)/r_0}\right)\right). \quad \text{(B.30)}$$

## B.2 Casimir energy in various dimensions

The first quantum correction to the scaling dimension of the scalar primary $\mathcal{O}^Q$ is given by the Casimir energy of the superfluid NG mode within the EFT on $\mathbb{R}\times S^{D-1}_{r_0}$. It reads

$$\Delta_1 = \frac{1}{2\sqrt{D-1}}\sum_{\ell=1}^{\infty}\mathrm{Deg}_D(\ell)\sqrt{\ell(\ell+D-2)}. \quad \text{(B.31)}$$

In terms of the family of regulated sums $\Sigma(s)$ defined in Eq. (2.86) the Casimir energy reads,

$$\Sigma(1) = \sum_{\ell=1}^{\infty}\mathrm{Deg}_D(\ell)\frac{\sqrt{\ell(\ell+D-2)}}{\sqrt{(D-1)}}\, e^{-\ell(\ell+D-2)/\Lambda^2}. \quad \text{(B.32)}$$

We note that a cut-off regularization is very natural for the Casimir energy in Eq. (B.31) as the EFT is only able to describe phonon states $a^\dagger_\ell|Q\rangle$ with spin smaller than the EFT cut-off $\ell \ll \Lambda$, which is given by the charge $\Lambda\sim r_0\mu\sim Q^{\frac{1}{D-1}}$. Here, we choose to regulate the Casimir energy using a smooth cut-off [313] for reasons of convenience.
Once regulated, the sum in Eq. (B.31) can be computed in an asymptotic series around $\Lambda\to\infty$. To do so we note that for large values of $\ell$ the summand of the original unregularized can be expanded in powers of $1/\ell$ with the first term given by

$$\mathrm{Deg}_D(\ell)\sqrt{\ell(\ell+D-2)} \xrightarrow{\ell\to\infty} \sum_{k=1} a_k(D)\ell^{D-k} . \tag{B.33}$$

With this in consideration, the regulated sum can be split into three terms,

$$\sum_{\ell=0}^{\infty} \mathrm{Deg}_D(\ell)\sqrt{\ell(\ell+D-2)}\, e^{-\ell(\ell+D-2)/\Lambda^2} = \Sigma_{\text{div.}} + \Sigma_{\log} + \Sigma_{\text{conv.}} , \tag{B.34}$$

where

$$\begin{aligned} \Sigma_{\text{conv.}} &= \sum_{\ell=0}^{\infty}\left(\mathrm{Deg}_D(\ell)\sqrt{\ell(\ell+D-2)} - \sum_{k=1}^{D+1} a_k(D)\ell^{D-k}\right) e^{-\ell(\ell+D-2)/\Lambda^2} , \\ \Sigma_{\log} &= a_{D+1}(D)\sum_{\ell=0}^{\infty}\frac{1}{\ell} e^{-\ell(\ell+D-2)/\Lambda^2} , \\ \Sigma_{\text{div.}} &= \sum_{\ell=0}^{\infty}\left(\sum_{k=1}^{D} a_k(D)\ell^{D-k}\right) e^{-\ell(\ell+D-2)/\Lambda^2} . \end{aligned} \tag{B.35}$$

We discuss all three contributions separately. By construction, the first part $\Sigma_{\text{conv.}}$ represents a convergent series,

$$\mathrm{Deg}_D(\ell)\sqrt{\ell(\ell+D-2)} - \sum_{k=1}^{D+1} a_k(D)\ell^{D-k} \sim \mathcal{O}\left(\frac{1}{\ell^2}\right) \qquad \text{as} \qquad \ell\to\infty . \tag{B.36}$$

No further regulation is need as we have $\Sigma_{\text{conv.}} = \text{const.} + \mathcal{O}(1/\Lambda)$. Next, the term $\Sigma_{log}$ is only non-trivial in even dimensions $D$, which can be seen from the fact that the coefficients $a_k(D)$ satisfy

$$a_{2k}(D) \propto (D-2k+1) , \qquad \forall k \in \mathbb{N} , \tag{B.37}$$

On the other hand, the coefficients $a_{2k+1}(D)$ do not possess zeroes for any integer dimension $D > 2$ for all $k \in \mathbb{N}$ This term leads to a contribution of the form $\log\Lambda$ first found in [188] in this context. The correct form of the asymptotic expansion for $\Sigma_{log}$ around $\Lambda\to\infty$ can be found via the standard Euler-Maclaurin formula,

$$\begin{aligned} \sum_{\ell=1}^{\infty}\frac{1}{\ell} e^{-\ell(\ell+D-2)/\Lambda^2} &\sim \int_0^{\infty}\frac{\mathrm{d}x}{x} e^{-x(x+D-2)/\Lambda^2} + \frac{1}{2} e^{-(D-1)/\Lambda^2} - \sum_{k=1}^{\infty}\frac{B_{2k}}{(2k)!}\left(\frac{\partial}{\partial x}\right)^{2k-1}\frac{1}{x} e^{-x(x+D-2)/\Lambda^2}\bigg|_{x=1} \\ &\sim \frac{1}{2}\left(\gamma + \log\Lambda^2\right) + \mathcal{O}\left(\Lambda^{-1}\right) , \end{aligned} \tag{B.38}$$

where $\gamma$ is the Euler–Mascheroni constant. Hence, the $\Sigma_{\log}$ term is responsible for the existence for a term of the form

$$\Delta_1\big|_{D=\text{even}} \supset \frac{a_{D+1}(D)}{2(D-1)\sqrt{D-1}} \log Q . \tag{B.39}$$

in the one-loop scaling dimension $\Delta_1$ in even spacetime dimensions $D$. This contribution cannot be corrected by any classical contributions and therefore represents a universal prediction independent of the details of the underlying CFT besides its global symmetry group.
In odd spacetime dimensions $D$ the universal contribution is instead simply of the order $(Q)^0$. It is computed from the other two terms as in $(\Sigma_{\text{div.}} + \Sigma_{\text{conv.}})|_{\text{const.}}$. We note that the term $\Sigma_{\text{div.}}$ also contains terms scaling with positive powers of the cut-off $\Lambda$. These terms can be absorbed into the Wilsonian coefficients $c_i$ within the EFT. The fact that we have performed the regularization using a symmetry-preserving regulator guarantees this to be the case. We can estimate which powers of $\Lambda$ are going to appear using the fact that

$$\sum_{\ell=0}^{\infty} \ell^{\alpha} e^{-\ell(\ell+D-2)/\Lambda^2} \sim \frac{1}{2}\Gamma\left(\frac{\alpha+1}{2}\right)\Lambda^{\alpha+1} + \sum_{k=1}^{\infty} a_k(\alpha)\Lambda^{\alpha-k}, \tag{B.40}$$

which is valid for any $\alpha \in \mathbb{N}$, with the coefficients $a_k(\alpha)$ being computable order-by-order. For the lowest few integer spacetime dimensions we find that

$$\Sigma_{\text{div.}}\Big|_{D=3} = \frac{\sqrt{\pi}}{2}\Lambda^3 - \frac{1}{4} + \mathcal{O}\left(\Lambda^{-1}\right), \tag{B.41}$$

$$\Sigma_{\text{div.}}\Big|_{D=4} = \frac{1}{2}\Lambda^4 + \frac{1}{4}\Lambda^2 - \frac{9}{20} + \mathcal{O}\left(\Lambda^{-1}\right), \tag{B.42}$$

$$\Sigma_{\text{div.}}\Big|_{D=5} = \frac{\sqrt{\pi}}{8}\Lambda^5 + \frac{\sqrt{\pi}}{6}\Lambda^3 - \frac{21}{64} + \mathcal{O}\left(\Lambda^{-1}\right), \tag{B.43}$$

$$\Sigma_{\text{div.}}\Big|_{D=6} = \frac{1}{12}\Lambda^6 + \frac{5}{24}\Lambda^4 + \frac{1}{6}\Lambda^2 - \frac{18'553}{30'240} + \mathcal{O}\left(\Lambda^{-1}\right). \tag{B.44}$$

We summarize the relevant results in the computation of the universal contributions to the one-loop scaling dimension $\Delta_1$ for different integer dimensions $D$ in Table B.1.

| $D$ | $\Sigma_{\text{div.}}\big\|_{\text{const.}}$ | $\Sigma_{\log}$ | $\Sigma_{\text{conv.}}$ | $\Delta_{1,\text{univ.}}$ |
|---|---|---|---|---|
| 3 | $-\frac{1}{4}$ | 0 | -0.01509 | $-0.09372 \times (Q)^0$ |
| 4 | $-\frac{9}{20}$ | $-\frac{1}{8}\left(\frac{\gamma}{2}+\log\Lambda\right)$ | 0.1106 | $-\frac{1}{48\sqrt{3}}\log Q$ |
| 5 | $-\frac{21}{64}$ | 0 | -0.1035 | $-0.1079 \times (Q)^0$ |
| 6 | $-\frac{18'553}{30'240}$ | $-\frac{1}{6}\left(\frac{\gamma}{2}+\log\Lambda\right)$ | 0.1990 | $-\frac{1}{60\sqrt{5}}\log Q$ |
| 7 | $-\frac{4'735}{12'288}$ | 0 | -0.1684 | $-0.1130 \times (Q)^0$ |
| 8 | $-\frac{534'983}{725'760}$ | $-\frac{981}{5'120}\left(\frac{\gamma}{2}+\log\Lambda\right)$ | 0.2655 | $-\frac{981}{71'680\sqrt{7}}\log Q$ |
| 9 | $-\frac{1'273'741}{2'949'120}$ | 0 | -0.2203 | $-0.1153 \times (Q)^0$ |
| 10 | $-\frac{10'420'037}{12'418'560}$ | $-\frac{22}{105}\left(\frac{\gamma}{2}+\log\Lambda\right)$ | 0.3192 | $-\frac{11}{2'835}\log Q$ |
| 11 | $-\frac{277'116'003}{587'202'560}$ | 0 | -0.2641 | $-0.1163 \times (Q)^0$ |

Table B.1: Relevant values from the computation of the sums in Eq. (B.35) for different integer spacetime dimensions $D$. We denote the final result for the universal one-loop contribution to the scaling dimension of the scalar operator $\mathcal{O}^Q$ by $\Delta_{1,\text{univ.}}$.

## B.3 More details on the two-loop corrections in the $O(2)$ model

In this Appendix we provide more details for the loop computations in Section 2.2.6. As discussed, loop corrections can be set up at finite temperature on the thermal circle $S^1_\beta \times S^{D-1}_{r_0}$. The CFT predictions are obtained in the zero temperature limit $\beta \to \infty$. The fluctuations are decomposed into modes as

$$\pi(\tau,\mathbf{n}) = \sqrt{\frac{\beta}{r_0}} \sum_{n\in\mathbb{Z}} \sum_{\ell\geq 1,m} Y_{\ell m}(\mathbf{n}) e^{i\omega_n\tau} \pi_{n\ell m}, \qquad \pi^*_{n\ell m} = (-1)^{m_{D-2}} \pi_{-n,\ell,m^*}\,. \tag{B.45}$$

The notation for the $m$-type quantum numbers here is as outlined in Appendix B.1. The Matsubara frequencies are $\omega_n = 2\pi n/\beta$. The unique zero-mode on the thermal circle can be excluded as it never appears in the derivative-only interactions within the EFT. The propagator in Fourier space is computed from the quadratic part of the action,

$$\langle \pi_{n\ell m}\pi_{n'\ell' m'}\rangle = \frac{1}{c_1 D(D-1)(\mu r_0)^{D-2}} \frac{1}{\beta^2} \underbrace{\frac{1}{\omega_n^2+\omega_\ell^2}}_{:=G_{n\ell}} \delta_{n,-n'}\delta_{\ell\ell'}(-1)^{|m|}\delta_{m,-m'}\,, \tag{B.46}$$

where the dispersion relations $\omega_\ell^2$ are given in Eq. (2.32). Further, $G_{n\ell} = (\omega_n^2+\omega_\ell^2)^{-1}$ denotes the propagator in Fourier space, with $\omega_n$ being the Matsubara frequencies. The zero mode has constant norm $\langle\pi_0\pi_0\rangle = \text{const.}$, does not mix with the other modes and does not get corrected at any order in perturbation theory since all vertices contain derivatives.

In arbitrary dimension $D$ of spacetime the EFT action includes all possible $k$-point vertices so that the interaction is of the form

$$S_{\text{int}} = \sum_{k=3}^{\infty} \mu^{D-k} S^{(k)}\,. \tag{B.47}$$

In particular, the two-loop correction $\Delta_2$ to the scaling dimension of the primary operator $\mathscr{O}^Q$ only gets contributions from diagrams involving three- and four-point vertices,

$$S^{(3)} = \frac{i}{6} c_1 D(D-1)(D-2) \int_0^\beta \mathrm{d}\tau \int_{S^{D-1}_{r_0}} \mathrm{d}S\, \dot\pi \left[\dot\pi^2 + \frac{3}{(D-1)}\frac{1}{r_0^2}(\partial_i\pi)^2\right], \tag{B.48}$$

$$S^{(4)} = -\frac{1}{24} c_1 D(D-1)(D-2) \int_0^\beta \mathrm{d}\tau \int_{S^{D-1}_{r_0}} \mathrm{d}S \left[\frac{3}{r_0^4(D-1)}(\partial_i\pi)^4 + \frac{6}{r_0^2}\left(\frac{D-3}{D-1}\right)\dot\pi^2(\partial_i\pi)^2 + (D-3)\dot\pi^4\right]. \tag{B.49}$$

The overall two-loop correction is computed as

$$\log Z = \log Z_0 - \mu^{D-4}\langle S^{(4)}\rangle_c + \frac{1}{2}\mu^{2D-6}\langle S^{(3)}S^{(3)}\rangle_c\,, \tag{B.50}$$

where we have indicated with the notation $\langle\ldots\rangle_c$ that only connected contractions contribute. As propagators scale like $\mu^{2-D}$ it is evident that both contributions in Eq. (B.50) enter at order $\mu^{-D} \sim Q^{-\frac{D}{D-1}}$ modulo powers of $(Q)^0 \log Q$.[1]

[1]Following the same power counting we deduce that an $\ell$-loop diagram contributes at order $Q^{-\frac{(\ell-1)D}{D-1}}$.

### B.3.1 Tadpole (sub)diagrams are vanishing

We start by verifying that all tadpole-type diagrams in the two-loop contribution $\Delta_2$ are identically vanishing.[2] In particular, this guarantees that the ground state introduced in three different contexts in Sections 2.2.1, 2.2.2 and 2.2.5 is stable under quantum corrections. This can be understood as a consequence of the $SO(D) \times O(2)_{\text{shift}}$ symmetry.
For a semi-diagrammatic proof of this statement we note that any tadpole sub-diagram is generated via contractions from a vertex with an odd number of legs, *i.e.* $\langle S^{(2k-1)}\rangle_c$ for some $k \in \mathbb{N}$. Due to the $SO(D)$ invariance the quantum-corrected propagator $\langle \pi_{n\ell m}\pi_{n'\ell' m'}\rangle$ remains proportional to $\delta_{m m'}$. Generically, every term of this kind will contain $k$ pairs of paired hyperspherical harmonics $Y^* Y$ and one unpaired harmonic $Y$,

$$\langle S^{(2k-1)}\rangle \supset \sum_{m,m_1\dots,m_k} \int_{S^{D-1}_{r_0}} \mathrm{d}S\, Y_{\ell m} Y_{\ell_1 m_1} Y^*_{\ell_1 m_1} \dots Y_{\ell_k m_k} Y^*_{\ell_k m_k}\,. \tag{B.51}$$

The sum $\sum_m Y_{\ell m} Y^*_{\ell m}$ is constant (see Appendix B.1), and hence

$$\sum_{m,m_1\dots,m_k} \int_{S^{D-1}} \mathrm{d}S\, Y_{\ell m} Y_{\ell_1 m_1} Y^*_{\ell_1 m_1} \dots Y_{\ell_k m_k} Y^*_{\ell_k m_k} \propto \sum_m \int_{S^{D-1}} \mathrm{d}S\, Y_{\ell m} = 0\,. \tag{B.52}$$

This is a consequence of the fact that the $O(2)$ shift symmetry guarantees that the zero mode $\ell = 0$ cannot appear in perturbation theory. The same argument applies if derivatives of hyperspherical harmonics appear, in this case we can use the identities in Eq. (B.7).

### B.3.2 Two-loop scaling dimension $\Delta_2$

In the computation of the two-loop scaling dimension we require the family of regulated sums defined in Eq. (2.86),

$$\Sigma(s) := \lim_{\Lambda\to\infty} \sum_{\ell>0} \mathrm{Deg}_D(\ell)(r_0\omega_\ell)^s e^{-\omega_\ell^2/\Lambda^2}\,. \tag{B.53}$$

The use of a momentum-dependent regulator is natural as the EFT can only capture phonon state with small enough values of the $\ell$ quantum number. This fact also cuts the allowed phonon states running in internal limes, as discussed in Eq. (2.59).
For two-loop computations and beyond the regularization procedure needs to take into account the Matsubara frequencies $\omega_n$ as well since vertices with derivative couplings generically lead to sums of the from

$$\sum_{n\in\mathbb{Z}} \sum_{\ell>0} \frac{\omega_n^2 \omega_\ell^2}{\omega_n^2 + \omega_\ell^2}\,, \tag{B.54}$$

where also the Matsubara sum is now divergent. A linear regularization procedure is a convenient choice here [315],

$$\mathrm{Reg}\left[\sum_{n\in\mathbb{Z}} \sum_{\ell>0} \big(\alpha f(n,\ell) + \beta g(n,\ell)\big)\right] = \alpha \mathrm{Reg}\left[\sum_{n\in\mathbb{Z}} \sum_{\ell>0} f(n,\ell)\right] + \beta \mathrm{Reg}\left[\sum_{n\in\mathbb{Z}} \sum_{\ell>0} g(n,\ell)\right]. \tag{B.55}$$

[2]More generally, no vertex appearing in Eq. (B.47) generates non-vanishing tad-pole sub-diagrams.

In addition, it should be symmetric under $n \leftrightarrow -n$ and independent of the $m$ quantum numbers. We again choose a smooth cut-off regularization scheme [313, 316] which now involves the thermal circle as well,

$$\text{Reg}\left[\sum_{n\in\mathbb{Z}}\sum_{\ell>0} f(n,\ell)\right] = \sum_{n\in\mathbb{Z}}\sum_{\ell>0} f(n,\ell)\, e^{-(\omega_n^2+\omega_\ell^2)/\Lambda^2}\,. \tag{B.56}$$

Both, the sum over the dispersion relations $\omega_\ell$ in Eq. (2.32) and the sum over the Matsubara frequencies $\omega_n$ are separately regularized. After performing the proper regularization, we take the zero-temperature limit $\beta \to \infty$ while keeping $r_0\Lambda$ fixed. In the end we are left with regulated sums of the form in Eq. (B.53). Finally, standard Feynman diagram techniques can be used in finite volume on $S^1_\beta \times S^{D-1}_{r_0}$.

First, we consider contributions coming from the quartic vertex in the action. The overall contribution at finite temperature is found to be

$$\langle S^{(4)}\rangle = -\frac{c_1 D(D-2)}{24}\left[\frac{3}{r_0^4}\,[\text{diagram 1}] + \frac{6(D-3)}{r_0^2}\,[\text{diagram 2}] + (D-1)(D-3)\,[\text{diagram 3}]\right]. \tag{B.57}$$

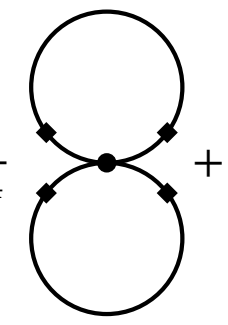
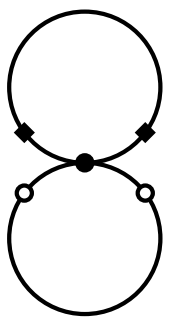
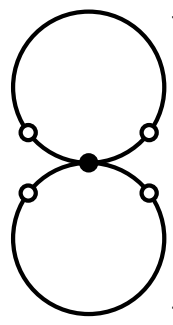

In our notation vertices are black dots, square black dots indicate spatial derivatives $\partial_i$ acting on the corresponding legs and white dots indicate temporal derivatives $\partial_\tau$. In our notation we have also suppressed the permutation of the derivatives on the legs, which need to be taken into account as independent Wick rotations. We provide detailed computations for each contribution in Appendix B.4. We find that

$$\mu^{D-4}\langle S^{(4)}\rangle_c = -\frac{(D-2)}{8c_1 D(D-1)\Omega_D(\mu r_0)^D}\left(\frac{r_0}{\beta}\right)\left[(D-3)\left[\sum_{n,\ell}\text{Deg}_D(\ell)\right]^2 - 4\left[\sum_{n\ell}\text{Deg}_D(\ell)G_{n\ell}\omega_\ell^2\right]^2\right], \tag{B.58}$$

where $G_{n\ell}$ is the propagator in Fourier space and $\deg_D(\ell)$ denote the multiplicities of the eigenvalues $\ell$. Each sum is regulated as in Eq. (B.56). After performing the sum over Matsubara frequencies we take the limit $\beta \to \infty$ and find the contribution to the scaling dimension from the quartic vertices,

$$\Delta_2^{(4)} = -\frac{(D-2)}{8c_1 D(D-1)\Omega_D(\mu r_0)^D}\left[(D-3)\frac{(\Lambda r_0)^2}{4\pi}\Sigma(0)^2 - \Sigma(1)^2\right]. \tag{B.59}$$

In contrast to the quartic term in the action — at the order we are working at — all contributions coming from Feynman diagrams involving cubic vertices are appearing at two-loop order. The overall contribution is given by

$$\langle S^{(3)}S^{(3)}\rangle_c = -\frac{c_1^2}{36}D^2(D-1)^2(D-2)^2\left[[\text{diagram 4}] + \frac{6}{r_0^2(D-1)}[\text{diagram 5}] + \frac{9}{r_0^4(D-1)^2}[\text{diagram 6}]\right], \tag{B.60}$$

where we again omit permutations of the different derivatives on the internal legs. We note that, as discussed in Appendix B.3.1, there are no contributions involving tadpole diagrams. We present computations for each Feynman grap separately in Appendix B.4. The final result can be expressed in a completely symmetrized form as follows:

$$\mu^{2D-6}\langle S^{(3)}S^{(3)}\rangle_c = -\frac{(D-2)}{12c_1 D(D-1)\Omega_D(\mu r_0)^D}\left(\frac{r_0}{\beta}\right)\sum_{\substack{\ell_a,\ell_b,\ell_c\\ n_a+n_b+n_c=0}}\sum G_{n_a\ell_a}G_{n_b\ell_b}G_{n_c\ell_c}S_{\ell_a\ell_b\ell_c}$$
$$\times\Bigg[2\omega_{n_a}^2\omega_{n_b}^2\omega_{n_c}^2 - \omega_{n_a}\omega_{n_b}\omega_{n_c}\left[\omega_{n_a}(\omega_{\ell_b}+\omega_{\ell_c})^2 + \text{(cyclic perm.)}\right]$$
$$+\frac{1}{2}\omega_{n_a}^2(\omega_{\ell_b}^2+\omega_{\ell_c}^2-\omega_{\ell_a}^2)^2 + \text{(cyclic perm.)}$$
$$-\omega_{n_a}\omega_{n_b}\left[\omega_{\ell_c}^4 - (\omega_{\ell_a}^2-\omega_{\ell_b}^2)^2\right] + \text{(cyclic perm.)}\Bigg], \quad \text{(B.61)}$$

where $\omega_{n_i}$ denotes the Matsubara frequencies and $\omega_{\ell_i}$ the dispersion relations given in Eq. (2.32). The symbol $\triangle_{\ell_a\ell_b\ell_c}$ essentially corresponds to (discrete) momentum conservation on $S_{r_0}^{D-1}$ and enforces the $SO(D)$ quantum numbers $\ell_{a,b,c}$ to satisfy a triangle inequality,

$$\triangle_{\ell_a\ell_b\ell_c} := \begin{cases} 1 & \text{if } |\ell_b-\ell_a| \le \ell_c \le \ell_b+\ell_a \text{ and } \ell_c-\ell_a-\ell_b \text{ even}, \\ 0 & \text{otherwise}. \end{cases} \quad \text{(B.62)}$$

The symmetric structure $S_{\ell_a\ell_b\ell_c}$ appearing in this expression involves the symbol $\triangle_{\ell_a\ell_b\ell_c}$ as well and is defined in Section 2.2.6 and also in Appendix B.4 in Eq. (B.87). We recover the corresponding contribution to the scaling dimension of the primary operator $\mathcal{O}^Q$ is obtained in the zero-temperature limit $\beta\to\infty$ while computing the Matsubara sums along the same line as discussed before. The overall end result including both cubic and quartic contributions to the two-loop scaling dimension reads

$$\Delta_2 = \frac{1}{16c_1 D(D-1)\Omega_D(\mu r_0)^D}\Bigg[\frac{(r_0\Lambda)^2}{6\pi}(D-2)\left[2D-4+(D-5)\Sigma(0)\right]\Sigma(0) - \frac{(r_0\Lambda)}{\sqrt{\pi}}(D-2)^2(1-\Sigma(0))\Sigma(1)$$
$$-\frac{(D-2)}{3}\left[(D-2)(\Sigma(2)+6\Sigma(0)\Sigma(2)+2\Sigma(-1)\Sigma(3)) - (5D-16)\Sigma(1)^2 - 8\Sigma^{(2\ell)}\right]\Bigg], \quad \text{(B.63)}$$

where we have denoted by $\Sigma^{(2\ell)}$ as sum that unfortunately cannot readily be reduced to a combination of elementary sums of the form $\Sigma(s)$,

$$\Sigma^{(2\ell)} := \sum_{\ell_a,\ell_b,\ell_c} S_{\ell_a\ell_b\ell_c}\triangle_{\ell_a\ell_b\ell_c}\frac{\omega_{\ell_a}\omega_{\ell_b}\omega_{\ell_c}}{\omega_{\ell_a}+\omega_{\ell_b}+\omega_{\ell_c}}. \quad \text{(B.64)}$$

As discussed in Section 2.2.6, this sum can in principle be regulated and computed, but we limit ourselves to observing its two divergent regimes. For $\ell_a \sim \ell_b \gg 1$ it grows as $\Sigma(1)$ and for $\ell_a \sim \ell_b \sim \ell_c \gg 1$ it grows as $\Sigma(2)$. These limit just represent discrete version of different collinear divergences found in ordinary loop integrals.

For even dimensions the result in Eq. (B.63) includes a (non-universal) $(Q)^0 \log Q^2$ term,

$$\Delta_2 \supset \frac{1}{Q^{D/(D-1)}}\left(\alpha_0 + \alpha_1\log Q + \alpha_2(\log Q)^2\right). \quad \text{(B.65)}$$

We expect this result to generalize to any loop order $k$, so that in even dimensions there is a $k$-loop contribution of the form

$$\Delta_k \supset \frac{1}{Q^{(k-1)D/(D-1)}}\left(\alpha_0 + \alpha_1 \log Q + \ldots + \alpha_k (\log Q)^k\right). \tag{B.66}$$

In odd dimensions, based on resurgent analysis, we do not expect the appearance of any log terms [2, 191].

## B.4 Details of the loop computations

In this Appendix we summarize relevant technology required to perform loop computations within the EFT.

### B.4.1 Matsubara sums

Perturbation theory is set up on the thermal circle $S^1_\beta \times S^{D-1}_{r_0}$. Summations over the Matsubara frequencies $\omega_n = 2\pi n/\beta$ on the thermal circle can be performed using the identity

$$\sum_{n\in\mathbb{Z}} f\left(\frac{2\pi i n}{\beta}\right) = \beta \int \frac{\mathrm{d}k}{2\pi}\left(\frac{f(ik)+f(-ik)}{2}\right) + \mathcal{O}\left(e^{-\beta}\right), \tag{B.67}$$

where the exponential corrections can be neglected in the zero-temperature $\beta \to \infty$ limit. In the computation of the different two-loop contributions there are generally speaking three relevant Matsubara sums that can appear

$$\sum_{n_a} G_{n_a \ell_a} = \frac{\beta}{2\omega_{\ell_a}}, \tag{B.68}$$

$$\sum_{n_a+n_b=\ell} G_{n_a\ell_a} G_{n_b\ell_b} = \frac{\beta}{2}\left[\frac{1}{\omega_{\ell_a}} + \frac{1}{\omega_{\ell_b}}\right] \frac{1}{\omega_\ell^2 + (\omega_{\ell_a}+\omega_{\ell_b})^2}, \tag{B.69}$$

$$\sum_n G_{n\ell} \sum_{n_a+n_b=n} G_{n_a\ell_a} G_{n_b\ell_b} = \frac{\beta^2}{4} \frac{1}{\omega_\ell \omega_{\ell_a}\omega_{\ell_b}} \frac{1}{\omega_\ell + \omega_{\ell_a} + \omega_{\ell_b}}, \tag{B.70}$$

where $G_{n\ell} = (\omega_n^2 + \omega_\ell^2)^{-1}$ denotes the propagator in momentum space, see Eq. (B.46). Other sums with powers of the Matsubara frequencies$\omega_{n_a}$, $\omega_{n_b}$ in the numerator can be expressed via the sums above using the linearity property in Eq. (B.55) satisfied by the smooth cut-off regularization procedure. This reduction procedure works analogously to the reduction of integrals to scalar master integrals in multi-loop computations [317, 318]. However, because of the derivative interactions here this reduction procedure may also produce divergent Matsubara sums. Such divergent contributions in our smooth cut-off regularization are computed in the zero-temperature limit $\beta \to \infty$ as follows:

$$\sum_{n_a} 1 \quad \longrightarrow \quad \sum_{n_a} e^{-\frac{(2\pi n_a)^2}{\beta^2\Lambda^2}} = \frac{\beta\Lambda}{2\sqrt{\pi}} + \mathcal{O}\left(\beta^{-1}\right) + \ldots. \tag{B.71}$$

### B.4.2 Kinematic vertex factors

In the context of perturbation theory o the sphere $S_{r_0}^{D-1}$ we have to compute vertex factors which are proportional to multiple integrals of hyperspherical harmonics $Y_{\ell m}$ over the sphere. In the diagrams in Eq. (B.57) that arise from the four-point vertex in the action we find have

$$T^{0\partial}(1,2,3,4) := \int_{S_{r_0}^{D-1}} Y_{\ell_1 m_1} Y_{\ell_2 m_2} Y_{\ell_3 m_3} Y_{\ell_4 m_4}\,, \tag{B.72}$$

$$T^{2\partial}(1,2,3,4) := \int_{S_{r_0}^{D-1}} Y_{\ell_1 m_1} Y_{\ell_2 m_2} \partial_j Y_{\ell_3 m_3} \partial_j Y_{\ell_4 m_4}\,, \tag{B.73}$$

$$T^{4\partial}(1,2,3,4) := \int_{S_{r_0}^{D-1}} \partial_i Y_{\ell_1 m_1} \partial_i Y_{\ell_2 m_2} \partial_j Y_{\ell_3 m_3} \partial_j Y_{\ell_4 m_4}\,. \tag{B.74}$$

Although these integrals are highly non-trivial in general, in loop computations it is sufficient for us to compute their contraction with respect to the $m$-type quantum numbers using Eq. (B.7),

$$\sum_{m_a,m_b} T^{0\partial}(a,-a,b,-b) = \frac{r_0^{D-1}}{\Omega_D}\mathrm{Deg}_D(\ell_a)\mathrm{Deg}_D(\ell_b)\,, \tag{B.75}$$

$$\sum_{m_a,m_b} T^{2\partial}(a,-a,b,-b) = \frac{r_0^{D-1}}{\Omega_D}\mathrm{Deg}_D(\ell_a)\mathrm{Deg}_D(\ell_b)\frac{\lambda_{\ell_b}}{r_0^2}\,, \tag{B.76}$$

$$\sum_{m_a,m_b} T^{2\partial}(a,b,-a,-b) = 0\,, \tag{B.77}$$

$$\sum_{m_a,m_b} T^{4\partial}(a,-a,b,-b) = \frac{r_0^{D-1}}{\Omega_D}\mathrm{Deg}_D(\ell_a)\mathrm{Deg}_D(\ell_b)\frac{\lambda_{\ell_a}}{r_0^2}\frac{\lambda_{\ell_b}}{r_0^2}\,, \tag{B.78}$$

$$\sum_{m_a,m_b} T^{4\partial}(a,b,-a,-b) = \frac{r_0^{D-1}}{\Omega_D}\frac{1}{(D-1)}\mathrm{Deg}_D(\ell_a)\mathrm{Deg}_D(\ell_b)\frac{\lambda_{\ell_a}}{r_0^2}\frac{\lambda_{\ell_b}}{r_0^2}\,, \tag{B.79}$$

where $\lambda_\ell$ are the eigenvalues of the Laplacian$-\Delta_{S_{r_0}^{D-1}}$, see Eq. (B.1). The pure two-loop topologies in the computation of $\Delta_2$ appear in the diagrams in Eq. (B.60) involving three-point vertices. In these diagrams similar structures to the ones discussed for four-point vertices above appear, with some important differences,

$$T^{0\partial}(1,2,3|4,5,6) := \int_{S_{r_0}^{D-1}} Y_{\ell_1 m_1} Y_{\ell_2 m_2} Y_{\ell_3 m_3} \int_{S_{r_0}^{D-1}} Y_{\ell_4 m_4} Y_{\ell_5 m_5} Y_{\ell_6 m_6}\,, \tag{B.80}$$

$$T^{2\partial}(1,2,3|4,5,6) := \int_{S_{r_0}^{D-1}} Y_{\ell_1 m_1} Y_{\ell_2 m_2} Y_{\ell_3 m_3} \int_{S_{r_0}^{D-1}} Y_{\ell_4 m_4} \partial_i Y_{\ell_5 m_5} \partial_i Y_{\ell_6 m_6}\,, \tag{B.81}$$

$$T^{4\partial}(1,2,3|4,5,6) := \int_{S_{r_0}^{D-1}} Y_{\ell_1 m_1} \partial_j Y_{\ell_2 m_2} \partial_j Y_{\ell_3 m_3} \int_{S_{r_0}^{D-1}} Y_{\ell_4 m_4} \partial_i Y_{\ell_5 m_5} \partial_i Y_{\ell_6 m_6}\,, \tag{B.82}$$

Via Integration-by-parts the last two structures $T^{2\partial}$ and $T^{4\partial}$ can be expressed in terms of the first structure $T^{0\partial}$,

$$T^{4\partial}(1,2,3|4,5,6) = \frac{1}{4}\left[\lambda_{\ell_3}+\lambda_{\ell_2}-\lambda_{\ell_1}\right]\left[\lambda_{\ell_6}+\lambda_{\ell_5}-\lambda_{\ell_4}\right] T^{0\partial}(1,2,3|4,5,6)\,, \tag{B.83}$$

$$T^{2\partial}(1,2,3|4,5,6) = \frac{1}{2}\left[\lambda_{\ell_6}+\lambda_{\ell_5}-\lambda_{\ell_4}\right] T^{0\partial}(1,2,3|4,5,6)\,. \tag{B.84}$$

Unfortunately, the expression for $T^{0\partial}$ is not as simple as its flat-space counterparts. This is a consequence of the fact that momentum conservation in flat space is replaced on the sphere $S^{D-1}_{r_0}$ by $SO(D)$ angular-momentum addition. Using the properties of the Gegenbauer polynomials introduced in Appendix B.1 we can compute the contraction of $T^{0\partial}$ in all of its $m$ indices,

$$\sum_{m_a,m_b,m_c} T^{0\partial}(a,b,c|a,b,c) = \frac{1}{3}\frac{r_0^{2D-2}}{(D-2)\Omega_D}\frac{(D+2\ell_a-2)(D+2\ell_b-2)}{(D+2\ell_c+2)}\mathrm{Deg}_D(\ell_c) \times \sum_{k=0}^{\min(\ell_a\ell_b)} \langle k|\ell_a\ell_b\rangle\,\delta_{\ell_c-\ell_a-\ell_b+2k} + (2\text{ perm. in }\ell_a\ell_b\ell_c)\,, \tag{B.85}$$

where the coefficients $\langle k|\ell_a\ell_b\rangle$ are given in Eq. (B.18) and we have included the permutations in $\ell_a, \ell_b, \ell_c$ to make the permutation symmetry of the expression manifest. The permutations correspond to the different choices of applying the Gegenbauer addition formula in Eq. (B.17). The summation in the above expression can be computed explicitly,

$$\sum_{k=0}^{\min(\ell_a\ell_b)} \langle k|\ell_a\ell_b\rangle\,\delta_{\ell_c-\ell_a-\ell_b+2k} = \triangle_{\ell_a\ell_b\ell_c}\frac{(D+2\ell_c-2)\Gamma(\ell_c+1)}{2\Gamma\left(\frac{D}{2}-1\right)^2\Gamma(\ell_c+D-2)} \times \frac{\Gamma\left(\frac{\ell_{abc}}{2}+\frac{(D-2)}{2}\right)}{\Gamma\left(\frac{\ell_{abc}}{2}+1\right)}\frac{\Gamma\left(\frac{\ell_{cab}}{2}+\frac{(D-2)}{2}\right)}{\Gamma\left(\frac{\ell_{cab}}{2}+1\right)}\frac{\Gamma\left(\frac{\ell_{bca}}{2}+\frac{(D-2)}{2}\right)}{\Gamma\left(\frac{\ell_{bca}}{2}+1\right)}\frac{\Gamma\left(\frac{\ell_a+\ell_b+\ell_c}{2}+\frac{(2D-4)}{2}\right)}{\Gamma\left(\frac{\ell_a+\ell_b+\ell_c}{2}+\frac{D}{2}\right)}\,. \tag{B.86}$$

Here, we have introduced the notation $\ell_{abc} = \ell_a+\ell_b-\ell_c$ and the symbol $\triangle_{\ell_a\ell_b\ell_c}$ defined in Eq. (B.62) which imposes the triangle inequality. Combining these results and using the fact that $\triangle_{\ell_a\ell_b\ell_c}$ is fully symmetric we find that

$$\begin{aligned}\sum_{m_am_bm_c} T^{0\partial}(a,b,c|a,b,c) &= \triangle_{\ell_a\ell_b\ell_c}\frac{r_0^{2D-2}}{(D-2)\Omega_D}\frac{(D+2\ell_a-2)(D+2\ell_b-2)(D+2\ell_c-2)}{2\Gamma(D-1)\Gamma\left(\frac{D}{2}-1\right)^2}\\ &\times \frac{\Gamma\left(\frac{\ell_{abc}}{2}+\frac{(D-2)}{2}\right)}{\Gamma\left(\frac{\ell_{abc}}{2}+1\right)}\frac{\Gamma\left(\frac{\ell_{cab}}{2}+\frac{(D-2)}{2}\right)}{\Gamma\left(\frac{\ell_{cab}}{2}+1\right)}\frac{\Gamma\left(\frac{\ell_{bca}}{2}+\frac{(D-2)}{2}\right)}{\Gamma\left(\frac{\ell_{bca}}{2}+1\right)}\frac{\Gamma\left(\frac{\ell_a+\ell_b+\ell_c}{2}+\frac{(2D-4)}{2}\right)}{\Gamma\left(\frac{\ell_a+\ell_b+\ell_c}{2}+\frac{D}{2}\right)}\\ &=: \frac{r_0^{2D-2}}{(D-2)\Omega_D}S_{\ell_a\ell_b\ell_c}\,,\end{aligned} \tag{B.87}$$

where have now also defined the fully symmetric structure appearing in Eq. (B.61). Sums that involve $\triangle_{\ell_a\ell_b\ell_c}$ can be computed as follows,

$$\sum_{\ell_a,\ell_b,\ell_c=1}^{\infty} \triangle_{\ell_a\ell_b\ell_c} f(\ell_a,\ell_b,\ell_c) = \sum_{\ell_a,\ell_b=1}^{\infty}\sum_{k=0}^{\min(\ell_a\ell_b)} f(\ell_a,\ell_b,\ell_a+\ell_b-2k) - \sum_{\ell_a\ell_b=1}^{\infty}\delta_{\ell_a\ell_b} f(\ell_a,\ell_a,0)\,, \tag{B.88}$$

where the second term appears as a consequence of the exclusion of the $\ell_c = 0$ term. As a consequence, certain sums involving the structure $S_{\ell_a \ell_b \ell_c}$ are explicitly computable,

$$\sum_{\ell_a,\ell_b,\ell_c=1}^{\infty} \triangle_{\ell_a\ell_b\ell_c} S_{\ell_a\ell_b\ell_c} = (D-2)\left[\sum_{\ell_a,\ell_b} \mathrm{Deg}_D(\ell_a)\mathrm{Deg}_D(\ell_b) - \sum_{\ell_a} \mathrm{Deg}_D(\ell_a)\right], \tag{B.89}$$

$$\sum_{\ell_a,\ell_b,\ell_c=1}^{\infty} \triangle_{\ell_a\ell_b\ell_c} S_{\ell_a\ell_b\ell_c}\omega_{\ell_c} = (D-2)\left[\sum_{\ell_a,\ell_b} \mathrm{Deg}_D(\ell_a)\mathrm{Deg}_D(\ell_b)\omega_{\ell_b} - \sum_{\ell_a} \mathrm{Deg}_D(\ell_a)\omega_{\ell_a}\right], \tag{B.90}$$

$$\sum_{\ell_a,\ell_b,\ell_c=1}^{\infty} \triangle_{\ell_a\ell_b\ell_c} S_{\ell_a\ell_b\ell_c}\omega_{\ell_c}^2 = 2(D-2)\sum_{\ell_a,\ell_b} \mathrm{Deg}_D(\ell_a)\mathrm{Deg}_D(\ell_b)\omega_{\ell_b}^2 . \tag{B.91}$$

### B.4.3 Graphs for $\Delta_2$

Using the notation and machinery developed in the previous section we can evaluate the four-point vertex contributions to $\Delta_2$ represented graphically in Eq. (B.57),

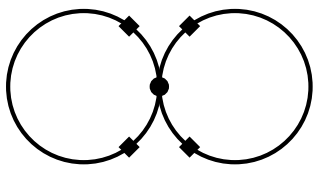

$$= \beta\left(\frac{\beta}{R}\right)^2 \sum_{\{n_i,\ell_i,m_i\}} T^{4\partial}(1,2,3,4) \times (3\text{ contractions}) \propto \left[\sum_{n,\ell} G_{n\ell}\lambda_\ell \mathrm{Deg}_D(\ell)\right]^2 , \tag{B.92}$$

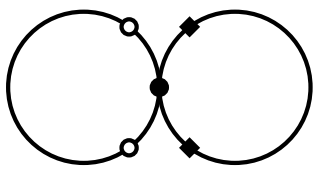

$$= -\beta\left(\frac{\beta}{R}\right)^2 \sum_{\{n_i,\ell_i,m_i\}} \omega_{n_1}\omega_{n_2} T^{2\partial}(1,2,3,4) \times (3\text{ contractions}) \propto \left[\sum_{n,\ell} G_{n\ell}\lambda_\ell \mathrm{Deg}_D(\ell)\right]\left[\sum_{n,\ell} G_{n\ell}\omega_n^2\right], \tag{B.93}$$

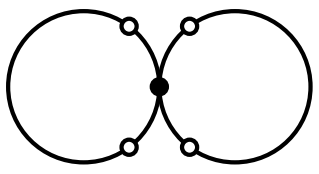

$$= \beta\left(\frac{\beta}{R}\right)^2 \sum_{\{n_i,\ell_i,m_i\}} \delta_{0\sum_i n_i}\left[\prod_{i=1}^{4}\omega_{n_i}\right] T^{0\partial}(1,2,3,4) \times (3\text{ contractions}) \propto \left[\sum_{n,\ell} G_{n\ell}\omega_n^2 \mathrm{Deg}_D(\ell)\right]^2 , \tag{B.94}$$

where $\omega_{n_i}$ denotes the Matsubara frequencies and $\omega_{\ell_i}$ are the dispersion relations given in Eq. (2.32).

In similar fashion we are able to evaluate the graphs involving three-point vertices in Eq. (B.60)

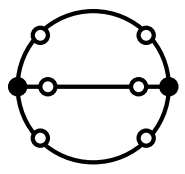

$$= -\beta^2\left(\frac{\beta}{R}\right)^3 \sum_{\{n_i,\ell_i,m_i\}}^{(\sum_i n_i=0)} \sum_{\{n_j,\ell_j,m_j\}}^{(\sum_j n_j=0)} \left[\prod_{i=1}^{6}\omega_{n_i}\right] T^{0\partial}(1,2,3|4,5,6) \times (6\text{ contractions})$$
$$\propto \sum_{n_a+n_b+n_c=0} \omega_{n_a}^2\omega_{n_b}^2\omega_{n_c}^2 \sum_{\ell_a,\ell_b,\ell_c} G_{n_a\ell_a}G_{n_b\ell_b}G_{n_c\ell_c}S_{\ell_a\ell_b\ell_c}\,, \tag{B.95}$$

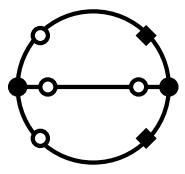

$$= \beta^2\left(\frac{\beta}{R}\right)^3 \sum_{\{n_i,\ell_i,m_i\}}^{(\sum_i n_i=0)} \sum_{\{n_j,\ell_j,m_j\}}^{(\sum_j n_j=0)} \omega_{n_1}\omega_{n_2}\omega_{n_3}\omega_{n_4} T^{2\partial}(1,2,3|4,5,6) \times (6\text{ contractions})$$
$$\propto \sum_{n_a+n_b+n_c=0} \omega_{n_a}^2\omega_{n_b}\omega_{n_c} \sum_{\ell_a,\ell_b,\ell_c} G_{n_a\ell_a}G_{n_b\ell_b}G_{n_c\ell_c}(\omega_{\ell_c}^2+\omega_{\ell_b}^2-\omega_{\ell_a}^2)S_{\ell_a\ell_b\ell_c}\,, \tag{B.96}$$

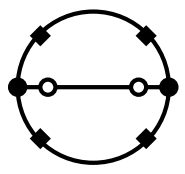

$$= -\beta^2\left(\frac{\beta}{R}\right)^3 \sum_{\{n_i,\ell_i,m_i\}}^{(\sum_i n_i=0)} \sum_{\{n_j,\ell_j,m_j\}}^{(\sum_j n_j=0)} \omega_{n_1}\omega_{n_2} T^{4\partial}(1,2,3|4,5,6) \times (6\text{ contractions})$$
$$\propto \sum_{n_a+n_b+n_c=0}\sum_{\ell_a,\ell_b,\ell_c} G_{n_a\ell_a}G_{n_b\ell_b}G_{n_c\ell_c}S_{\ell_a\ell_b\ell_c}$$
$$\times\left[2\omega_{n_a}^2(\omega_{\ell_c}^2+\omega_{\ell_b}^2-\omega_{\ell_a}^2)-4\omega_{n_a}\omega_{n_b}(\omega_{\ell_c}^2+\omega_{\ell_a}^2-\omega_{\ell_b}^2)\right](\omega_{\ell_c}^2+\omega_{\ell_b}^2-\omega_{\ell_a}^2)\,. \tag{B.97}$$

## B.5 Methods and details for the computations in Section 2.2.7

In this section we give details on the computation of three- and four-point functions in Section (2.2.7). We do this by discussing two explicit examples.

### B.5.1 Computing the $\langle\mathcal{O}^{-Q}T_{\tau\tau}\mathcal{O}^{Q}\rangle$ correlator

Using the field decomposition in terms of creation and annihilation operators in Eq. (**??**)(2.31), the expansion of $T$ and $Q$ in Eq. (2.101) and the properties of $|Q\rangle$ as a vacuum in Eq. (2.44) we can already compute the the tree-level results for the correlators in Section (2.2.7). To illustrate the computation of the three-point functions we demonstrate the computation of the correlator $\langle\mathcal{O}^{-Q}T_{\tau\tau}\mathcal{O}^{Q}\rangle$,

$$\langle\mathcal{O}^{-Q}T_{\tau\tau}(\tau,\mathbf{n})\mathcal{O}^{Q}\rangle = -\frac{\Delta_0}{r_0^D\Omega_D}\langle Q|1+i\frac{D}{\mu}\dot\pi-\frac{D}{2\mu^2}\left((D-1)\dot\pi^2-\frac{(D-3)}{r_0^2(D-1)}\pi\Delta\pi+\frac{(D-3)}{r_0^2(D-1)}\partial^i(\pi\partial_i\pi)\right)|Q\rangle\,. \tag{B.98}$$

For now, we ignore the total derivative term in this expression as we will show later that its expectation value vanishes. Setting $\mathcal{N} := c_1 \Omega_D r_0^{D-1} D(D-1)\mu^{D-2}$ we find up to quadratic order in the fields that

$$
\begin{aligned}
\langle \mathcal{O}^{-Q} T_{\tau\tau}(\tau,\mathbf{n}) \mathcal{O}^{Q} \rangle &= -\frac{\Delta_0}{r_0^D \Omega_D} \langle Q | \left[ 1 + i\frac{D}{\mu}\dot{\pi} - \frac{D}{2\mu^2}\left[ (D-1)\dot{\pi}^2 - \frac{(D-3)}{r_0^2 (D-1)} \pi \Delta \pi \right] \right]_{(\tau,\mathbf{n})} | Q \rangle \\
&= -\frac{\Delta_0}{r_0^D \Omega_D} \left\langle Q \middle| e^{-\frac{(\tau_2-\tau)}{r_0} D} \left[ 1 + i\frac{D}{\mu}\left[ -\frac{i\Pi_0}{\mathcal{N}} + \sqrt{\frac{\Omega_D}{\mathcal{N}}} \sum_{m,l} \sqrt{\frac{\omega_l}{2}} \left( a_{lm}^\dagger Y_{lm}^*(\mathbf{n}) - a_{lm} Y_{lm}(\mathbf{n}) \right) \right] \right. \right. \\
&\quad + \frac{D(D-1)}{2\mu^2} \frac{\Omega_D}{2\mathcal{N}} \sum_{\substack{m,l \\ m',l'}} \sqrt{\omega_l \omega_{l'}} \left[ a_{lm}^\dagger a_{l'm'} Y_{lm}^*(\mathbf{n}) Y_{l'm'}(\mathbf{n}) + a_{lm} a_{l'm'}^\dagger Y_{lm}(\mathbf{n}) Y_{l'm'}^*(\mathbf{n}) \right] \\
&\quad - \frac{\Pi_0^2}{\mathcal{N}^2} - 2\frac{i\Pi_0}{\mathcal{N}} \sqrt{\frac{\Omega_D}{\mathcal{N}}} \sum_{m,l} \sqrt{\frac{\omega_l}{2}} \left[ -a_{lm} Y_{lm}(\mathbf{n}) + a_{lm}^\dagger Y_{lm}^*(\mathbf{n}) \right] \\
&\quad - \frac{D}{\mu^2} \frac{\Omega_D}{2\mathcal{N}} \frac{(D-3)}{2(D-1)} \sum_{\substack{m,l \\ m',l'}} \frac{(D-1)\omega_{l'}^2}{\sqrt{\omega_l \omega_{l'}}} \left[ a_{lm}^\dagger a_{l'm'} Y_{lm}^*(\mathbf{n}) Y_{l'm'}(\mathbf{n}) + a_{lm} a_{l'm'}^\dagger Y_{lm}(\mathbf{n}) Y_{l'm'}^*(\mathbf{n}) \right] \\
&\quad \left. \left. + \frac{(D-3)}{(D-1)} \pi_0 \sqrt{\frac{\Omega_D}{\mathcal{N}}} \sum_{m,l} \frac{l(l+D-2)}{r_0^2 \sqrt{2\omega_l}} \left[ a_{lm} Y_{lm}(\mathbf{n}) + a_{lm}^\dagger Y_{lm}^*(\mathbf{n}) \right] \right] e^{-\frac{(\tau-\tau_1)}{r_0} D} \middle| Q \right\rangle, \qquad \text{(B.99)}
\end{aligned}
$$

where $D$ is the dilatation operator which on the cylinder is the Hamiltonian, $D = H^{(\mathrm{cyl})}$. Both terms linear in $a_{lm}$ and terms involving $\Pi_0$ vanish identically within the expectation value. Hence,

$$
\begin{aligned}
\langle \mathcal{O}^{-Q} T_{\tau\tau}(\tau,\mathbf{n}) \mathcal{O}^{Q} \rangle &= -\frac{\Delta_0}{r_0^D \Omega_D} \langle Q | e^{-\frac{(\tau_2-\tau)}{r_0} D} \left[ 1 + \frac{D(D-1)\Omega_D}{4\mu^2 \mathcal{N}} \sum_{\substack{m,l \\ m',l'}} \sqrt{\omega_l \omega_{l'}} [a_{lm}, a_{l'm'}^\dagger] Y_{lm}(\mathbf{n}) Y_{l'm'}^*(\mathbf{n}) \right. \\
&\quad \left. - \frac{(D-3)}{(D-1)} \frac{D}{2\mu^2} \frac{\Omega_D}{2\mathcal{N}} \sum_{\substack{m,l \\ m',l'}} \frac{(D-1)\omega_{l'}^2}{\sqrt{\omega_l \omega_{l'}}} [a_{lm}, a_{l'm'}^\dagger] Y_{lm}(\mathbf{n}) Y_{l'm'}^*(\mathbf{n}) \right] e^{-\frac{(\tau-\tau_1)}{r_0} D} | Q \rangle \\
&= -\mathcal{A}(\tau_1,\tau_2) \left[ \frac{\Delta_0}{r_0^D \Omega_D} + \frac{c_1 (D-1)\mu^D D \Omega_D \left[ (D-1) - (D-3) \right]}{4\mu^2 c_1 \Omega_D r_0^{D-1} D(D-1)\mu^{D-2}} \sum_{m,l} \omega_l Y_{lm}(\mathbf{n}) Y_{l'm'}^*(\mathbf{n}) \right] \\
&= -\mathcal{A}(\tau_1,\tau_2) \left[ \frac{\Delta_0}{r_0^D \Omega_D} + \frac{1}{2} \frac{1}{r_0^{D-1}} \sum_{m,l} \omega_l Y_{lm}(\mathbf{n}) Y_{l'm'}^*(\mathbf{n}) \right]. \qquad \text{(B.100)}
\end{aligned}
$$

The evaluation of the sum over hyperspherical harmonics can be found and is evaluated in Appendix B.4.

We are left with showing that the total derivative term vanishes,

$$
\begin{aligned}
\frac{\Delta_0}{r_0^D \Omega_D} \frac{D}{2\mu^2} \langle Q | \partial^i \left( \pi(\tau,\mathbf{n}) \partial_i \pi(\tau,\mathbf{n}) \right) | Q \rangle &= -\frac{\mathcal{A}(\tau_1,\tau_2)}{4 r_0^{D-1}} \sum_{m,l} \frac{\partial^i \left( Y_{lm}(\mathbf{n}) \partial_i Y_{lm}^*(\mathbf{n}) \right)}{\omega_l} \\
&= -\frac{\mathcal{A}(\tau_1,\tau_2)}{4 r_0^{D-1}} \sum_{l} \frac{1}{\omega_l} \frac{1}{2} \frac{\partial^i \partial_i}{2\omega_l} \frac{\mathrm{Deg}_D(\ell)}{\Omega_D} = 0.
\end{aligned} \qquad \text{(B.101)}
$$

Therefore, the final result is

$$\langle \mathscr{O}^{-Q} T_{\tau\tau}(\tau,\mathbf{n}) \mathscr{O}^{Q} \rangle = -\mathscr{A}(\tau_1,\tau_2) \frac{\Delta_0 + \Delta_1}{r_0^D \Omega_D} + \mathscr{O}\left(\mu^{D-3}\right). \tag{B.102}$$

### B.5.2 Computing the $\langle \mathscr{O}^{-Q}_{\ell_2 m_2} T_{\tau\tau} T_{\tau\tau} \mathscr{O}^{Q}_{\ell_1 m_1} \rangle$ correlator

We now illustrate the computation of the four-point functions in Section (2.2.7) by< computing the correlator $\langle \mathscr{O}^{-Q}_{\ell_2 m_2}(\tau_2)\, T_{\tau\tau}(\tau,x)\, T_{\tau\tau}(\tau',x') \mathscr{O}^{Q}_{\ell_1 m_1}(\tau_1) \rangle$. Expanded up to second order in the fields we have

$$\begin{aligned} T_{\tau\tau}(\tau,\mathbf{n})\, T_{\tau\tau}(\tau',\mathbf{n}') = \frac{\Delta_0^2}{r_0^{2D}\Omega_D^2} - \frac{\Delta_0^2}{r_0^{2D}\Omega_D^2} \frac{D^2}{\mu^2} \dot{\pi}(\tau,\mathbf{n}) \dot{\pi}(\tau',\mathbf{n}') - \frac{\Delta_0^2 (D-1)}{r_0^{2D}\Omega_D^2} \frac{D}{2\mu^2} \left[ \dot{\pi}^2 + \frac{(D-3)(\partial_i \pi)^2}{r_0^2 (D-1)^2} \right]_{(\tau,\mathbf{n})} \\ - \frac{\Delta_0^2 (D-1)}{r_0^{2D}\Omega_D^2} \frac{D}{2\mu^2} \left[ \dot{\pi}^2 + \frac{(D-3)(\partial_i \pi)^2}{r_0^2 (D-1)^2} \right]_{(\tau',\mathbf{n}')} + \mathscr{O}\left(\mu^{2D-3}\right). \end{aligned} \tag{B.103}$$

Using this result the four-point correlator becomes

$$\begin{aligned} \langle Q |\, a_{\ell_2 m_2}\, T_{\tau\tau}(\tau,\mathbf{n})\, T_{\tau\tau}(\tau',\mathbf{n}')\, a^\dagger_{\ell_1 m_1} | Q \rangle = \frac{\Delta_0^2}{r_0^{2D}\Omega_D^2} e^{-\omega_\ell(\tau_2-\tau_1)} \mathscr{A}(\tau_1,\tau_2)\, \delta_{m_1 m_2} \delta_{\ell_1 \ell_2} \\ - \frac{\Delta_0^2}{r_0^{2D}\Omega_D^2} \frac{D^2}{\mu^2} \langle Q |\, a_{\ell_2 m_2} \dot{\pi}(\tau,\mathbf{n}) \dot{\pi}(\tau',\mathbf{n}') a^\dagger_{\ell_1 m_1} | Q \rangle \\ - \frac{\Delta_0^2}{r_0^{2D}\Omega_D^2} \frac{D}{2\mu^2} \langle Q |\, a_{\ell_2 m_2} \left[ (D-1) \dot{\pi}^2(\tau,\mathbf{n}) + \frac{(D-3)}{r_0^2 (D-1)} \left(\partial_i \pi(\tau,\mathbf{n})\right)^2 \right] a^\dagger_{\ell_1 m_1} | Q \rangle \\ - \frac{\Delta_0^2}{r_0^{2D}\Omega_D^2} \frac{D}{2\mu^2} \langle Q |\, a_{\ell_2 m_2} \left[ (D-1) \dot{\pi}^2(\tau',\mathbf{n}') + \frac{(D-3)}{r_0^2 (D-1)} \left(\partial_i \pi(\tau',\mathbf{n}')\right)^2 \right] a^\dagger_{\ell_1 m_1} | Q \rangle. \end{aligned} \tag{B.104}$$

The last two terms are the same and have already been computed essentially in the derivation of the three-point function $\langle \mathscr{O}^{-Q}_{\ell_2 m_2} T_{\tau\tau}(\tau,\mathbf{n}) \mathscr{O}^{Q}_{\ell_1 m_1} \rangle$. Using $\mathscr{N} := c_1 \Omega_D r_0^{D-1} D(D-1)\mu^{D-2}$ we find that

$$\begin{aligned} - \frac{\Delta_0 (D-1)}{r_0^D \Omega_D} \frac{D}{2\mu^2} \langle Q |\, a_{\ell_2 m_2} \left[ \frac{(D-3)\left(\partial_i \pi\right)^2}{r_0^2 (D-1)^2} + \dot{\pi}^2 \right]_{(\tau,\mathbf{n})} a^\dagger_{\ell_1 m_1} | Q \rangle \\ = -\frac{\Delta_0 D}{2\mu^2 r_0^D \Omega_D} \langle Q |\, a_{\ell_2 m_2} e^{-\frac{(\tau_2-\tau)}{r_0} D} \left[ \frac{(D-3)}{r_0^2 (D-1)} \left(\partial_i \pi\right)^2 + (D-1) \dot{\pi}^2 \right] e^{-\frac{(\tau-\tau_1)}{r_0} D} a^\dagger_{\ell_1 m_1} | Q \rangle \\ = \frac{\Omega_D}{2 r_0^D \Omega_D} \mathscr{A}_{\Delta_Q + r_0 \omega_{\ell_2}}(\tau_1,\tau_2) \Bigg( \sum_{m,l} r_0 \omega_l Y_{lm}(\mathbf{n}) Y^*_{lm}(\mathbf{n}) \delta_{\ell_2 \ell_1} \delta_{m_2 m_1} \\ + (D-1) r_0 \sqrt{\omega_{\ell_1} \omega_{\ell_2}} \frac{Y^*_{\ell_2 m_2}(\mathbf{n}) Y_{\ell_1 m_1}(\mathbf{n})}{e^{-(\tau-\tau_1)(\omega_{\ell_1}-\omega_{\ell_2})}} - \frac{1}{\sqrt{\omega_{\ell_1}\omega_{\ell_2}}} \frac{(D-3)}{r_0 (D-1)} \frac{\partial_i Y^*_{\ell_2 m_2}(\mathbf{n}) \partial_i Y_{\ell_1 m_1}(\mathbf{n})}{e^{-(\tau-\tau_1)(\omega_{\ell_1}-\omega_{\ell_2})}} \Bigg), \end{aligned} \tag{B.105}$$

where we have used the property in Eq. (B.1) written in terms of the dispersion relation in Eq. (2.32),

$$-\Delta_{S^{D-1}_{r_0}} Y^*_{lm}(\mathbf{n}) = r_0^2 (D-1) \omega_l^2 Y^*_{lm}(\mathbf{n}). \tag{B.106}$$

A single terms is now left to compute,

$$\frac{\Delta_0^2}{r_0^{2D}\Omega_D^2}\frac{D^2}{\mu^2}\langle Q|\, a_{\ell_2 m_2}\dot{\pi}(\tau,\mathbf{n})\dot{\pi}(\tau',\mathbf{n}')a^\dagger_{\ell_1 m_1}|Q\rangle = -\frac{\Delta_0\Omega_D D}{2r_0^{2D}\Omega_D^2}\sum_{\substack{m',l'\\ m,l}} r_0\sqrt{\omega_{l'}\omega_l}\,\langle Q|\, a_{\ell_2 m_2}e^{-\frac{(\tau_2-\tau)}{r_0}D}\dots$$

$$\dots\Big(a^\dagger_{lm}Y^*_{lm}(\mathbf{n}) - a_{lm}Y_{lm}(\mathbf{n})\Big)e^{-\frac{(\tau-\tau')}{r_0}D}\Big(a^\dagger_{l'm'}Y^*_{l'm'}(\mathbf{n}') - a_{l'm'}Y_{l'm'}(\mathbf{n}')\Big)e^{-\frac{(\tau'-\tau_1)}{r_0}D}a^\dagger_{\ell_1 m_1}|Q\rangle$$

$$= -\frac{\Delta_0\Omega_D D\mathscr{A}(\tau_1,\tau_2)}{2r_0^{2D}\Omega_D^2 e^{\omega_{\ell_2}(\tau_2-\tau_1)}}\sum_{\substack{m',l'\\ m,l}} r_0\sqrt{\omega_{l'}\omega_l}\Bigg(\frac{Y_{l'm'}(\mathbf{n}')Y^*_{lm}(\mathbf{n})}{e^{-(\tau-\tau_1)\omega_l+(\tau'-\tau_1)\omega_{l'}}}\delta_{\ell_2 l}\delta_{\ell_1 l'}\delta_{m_2 m}\delta_{m_1 m'}$$

$$+ Y_{lm}(\mathbf{n})Y^*_{l'm'}(\mathbf{n}')e^{-(\tau-\tau')\omega_l}\Big[e^{-(\tau'-\tau_1)(\omega_l-\omega_{l'})}\delta_{\ell_1 l}\delta_{\ell_2 l'}\delta_{m_1 m}\delta_{m_2 m'} + \delta_{l'l}\delta_{m'm}\delta_{\ell_2\ell_1}\delta_{m_2 m_1}\Big]\Bigg)$$

$$= -\frac{\Delta_0\Omega_D D}{r_0^{2D}\Omega_D^2}\mathscr{A}_{\Delta_Q+r_0\omega_{\ell_2}}(\tau_1,\tau_2)\Bigg(\delta_{\ell_2\ell_1}\delta_{m_2 m_1}\sum_{m,l} r_0\omega_l Y_{lm}(\mathbf{n})Y^*_{lm}(\mathbf{n}')e^{-(\tau-\tau')\omega_l}$$

$$+ \frac{r_0\sqrt{\omega_{\ell_1}\omega_{\ell_2}}}{e^{(\tau-\tau_1)(\omega_{\ell_1}-\omega_{\ell_2})}}\Big[Y^*_{\ell_2 m_2}(\mathbf{n})Y_{\ell_1 m_1}(\mathbf{n}')e^{(\tau-\tau')\omega_{\ell_1}} + Y^*_{\ell_2 m_2}(\mathbf{n}')Y_{\ell_1 m_1}(\mathbf{n})e^{-(\tau-\tau')\omega_{\ell_2}}\Big]\Bigg). \tag{B.107}$$

After evaluating the sums appearing in the above expression using the machinery introduced in Section B.1, the overall correlator is given by

$$\langle \mathscr{O}^{-Q}_{\ell_2 m_2} T_{\tau\tau}(\tau,\mathbf{n})T_{\tau\tau}(\tau',\mathbf{n}')\mathscr{O}^{Q}_{\ell_1 m_1}\rangle = \mathscr{A}^{\Delta_Q+r_0\omega_{\ell_2}}_{\Delta_Q+r_0\omega_{\ell_1}}(\tau_1,\tau_2|\tau)\,\frac{\Delta_0}{\Omega_D^2 r_0^{2D}}$$

$$\times\Bigg[\Bigg(\Delta_0+2\Delta_1+\frac{D}{2}\sum_\ell e^{-|\tau-\tau'|\omega_\ell}r_0\omega_\ell\frac{(D+2\ell-2)}{(D-2)}C_\ell^{\frac{D}{2}-1}(\mathbf{n}\cdot\mathbf{n}')\Bigg)\delta_{\ell_1\ell_2}\delta_{m_1 m_2}$$

$$+\frac{D\Omega_D}{2}r_0\sqrt{\omega_{\ell_1}\omega_{\ell_2}}\Big(Y^*_{\ell_2 m_2}(\mathbf{n})Y_{\ell_1 m_1}(\mathbf{n}')\,e^{(\tau-\tau')\omega_{\ell_1}} + Y^*_{\ell_2 m_2}(\mathbf{n}')Y_{\ell_1 m_1}(\mathbf{n})\,e^{-(\tau-\tau')\omega_{\ell_2}}\Big)\Bigg]$$

$$+\Bigg\{\mathscr{A}^{\Delta_Q+r_0\omega_{\ell_2}}_{\Delta_Q+r_0\omega_{\ell_1}}(\tau_1,\tau_2|\tau)\,\frac{\Omega_D\Delta_0 r_0\sqrt{\omega_{\ell_1}\omega_{\ell_2}}}{2\Omega_D^2 r_0^{2D}}\Bigg[(D-1)\,Y_{\ell_1 m_1}(\mathbf{n})Y^*_{\ell_2 m_2}(\mathbf{n})$$

$$-\frac{(D-3)}{(D-1)}\frac{\partial_i Y_{\ell_1 m_1}(\mathbf{n})\partial_i Y^*_{\ell_2 m_2}(\mathbf{n})}{r_0^2\omega_{\ell_1}\omega_{\ell_2}}\Bigg] + \Big[(\tau,\mathbf{n})\leftrightarrow(\tau',\mathbf{n}')\Big]\Bigg\}. \tag{B.108}$$

# C Appendices to Chapter 3

## C.1 The action for the grand-canonical partition function

We derive the associated action in the path integral,

$$Z_{gc}(\mu_1,\dots) = \mathrm{Tr}_{S^1_\beta\times\mathcal{M}}\left[e^{-\beta(H-\mu_i Q_i)}\right] = \int \mathcal{D}\phi_i \mathcal{D}\phi_i^*\, e^{-S_\mu[\phi_i]}\,, \qquad Q_i = -\int_{\mathcal{M}} \mathrm{d}S\,\left[\partial_\tau\phi_i^*\,\phi_i - \phi_i^*\partial_\tau\phi_i\right]\,. \tag{C.1}$$

where $\mathcal{M}$ is some arbitrary spatial manifold. The underlying theory is given by

$$Z = \mathrm{Tr}_{S^1_\beta\times\mathcal{M}}\left[e^{-\beta H}\right] = \int \mathcal{D}\phi_i \mathcal{D}\phi_i^*\, e^{-S[\phi_i]}\,, \qquad S[\phi_i] = \int_{S^1_\beta\times\mathcal{M}} \mathrm{d}^D x\;\left[|\partial_\tau\phi_i|^2 + |\nabla\phi_i|^2 + V\big(|\phi_i|^2\big)\right]\,. \tag{C.2}$$

The obvious guess would be $S_\mu \overset{?}{=} S + \mu_i Q_i$, which is wrong. To see this we go to Minkowski space $i\,t = \tau$, where

$$S_{\mathrm{Mink}} = \int \mathrm{d}^D x\,\left[|\partial_t\phi_i|^2 - |\nabla\phi_i|^2 - V\big(|\phi_i|^2\big)\right]\,. \tag{C.3}$$

The corresponding Hamiltonian is

$$\pi_i = \frac{\delta}{\delta\phi_i} S_{\mathrm{Mink}} = \partial_t\phi_i^*\,, \quad \pi_i^* = \frac{\delta}{\delta\phi_i^*} S_{\mathrm{Mink}} = \partial_t\phi_i\,, \qquad H = \int \mathrm{d}S\left[|\pi_i|^2 + |\nabla\phi_i|^2 + V\big(|\phi_i|^2\big)\right]. \tag{C.4}$$

Functionally, the Hamiltonian has the same form as the Euclidean action. In terms of the canonically conjugate momenta the Cartan charges are

$$Q_i = i\int \mathrm{d}S\,\left[\pi_i\phi_i - \phi_i^*\pi_i^*\right]\,, \tag{C.5}$$

and the modified Hamiltonian reads

$$H_\mu := H - \mu_i Q_i = \int \mathrm{d}S\left[|\pi_i|^2 + |\nabla\phi_i|^2 + V\big(|\phi_i|^2\big) - i\mu_i\left(\pi_i\phi_i - \pi_i^*\phi_i^*\right)\right] \tag{C.6}$$

From this expression we return to the Lagrangian formalism,

$$\partial_t\phi_i = \frac{\delta}{\delta\pi_i} H_\mu = \pi_i^* - i\mu_i\phi_i\,, \qquad \partial_t\phi_a^* = \frac{\delta}{\delta\pi_i^*} H_\mu = \pi_i + i\mu_i\phi_i^*\,, \tag{C.7}$$

and the action becomes

$$\begin{aligned}(S_\mu)_{\text{Mink}} &= \int \mathrm{d}^D x\, \left[\partial_t\phi_i\pi_i + \partial_t\phi_i^*\pi_i^*\right] - H_\mu \\ &= \int \mathrm{d}^D x\, \left[\partial_t\phi_i\pi_i + \partial_t\phi_i^*\pi_i^* - \pi_i\pi_i^* - \nabla\phi_i^*\nabla\phi_i - V\left(\phi_i^*\phi_i\right) + i\mu_i\left(\pi_i\phi_i - \pi_i^*\phi_i^*\right)\right] \\ &= \int \mathrm{d}^D x\left[\left(\partial_t\phi_i^* - i\mu_i\phi_i^*\right)\left(\partial_t\phi_i + i\mu_i\phi_i\right) - \nabla\phi_i^*\nabla\phi_i - V\left(\phi_i^*\phi_i\right)\right].\end{aligned} \tag{C.8}$$

After a Wick roation $t = -i\tau$ to Euclidean spacetime the action becomes

$$S_\mu = \int \mathrm{d}^D x\, \left[\left(\partial_\tau\phi_i^* - \mu_i\phi_i^*\right)\left(\partial_\tau\phi_i + \mu_i\phi_i\right) + \nabla\phi_i^*\nabla\phi_i + V\left(\phi_i^*\phi_i\right)\right]. \tag{C.9}$$

Therefore, there is an extra term $\mu_i^2\phi_i^*\phi_i$ appearing in the action, which is due to the momentum-dependence of the current.

## C.2 Functional determinant for the $O(N)$ vector model

We evaluate the functional determinant Eq. (3.42) appearing in Section 3.1.3,

$$Z_{gc}(\mu) \overset{N=\infty}{\simeq} \int \mathscr{D}\phi_i \mathscr{D}\phi_i^* \,\exp\left[-\int_{S^1_\beta\times S^{D-1}_{r_0}} \mathrm{d}^D x\, \left[(\partial_\tau - \mu_i)\phi_i^*(\partial_\tau+\mu_i)\phi_i + |\nabla\phi_i|^2 + \left(r + \langle\sigma\rangle\right)|\phi_i|^2\right]\right]. \tag{C.10}$$

The functional determinant can be performed and regularized in terms of the eigenvalues on $S^1_\beta \times S^2_{r_0}$. The mode decomposition for the fields $\phi_i$ is of the form [194][1]

$$\phi_i = \zeta_i e^{i\varphi_i} + \sqrt{\frac{\beta}{V}} \sum_{n\in\mathbb{Z}}\sum_{\ell,m} e^{i\omega_n\tau}\, Y_{\ell m}(\mathbf{n})\,\tilde\phi_i(n;\ell,m)\,, \qquad \left(\sqrt{2}\,\tilde\phi_i(n;\ell,m) = \tilde\phi_i^{(\text{Re})}(n;\ell,m) + i\tilde\phi_i^{(\text{Im})}(n;\ell,m)\right), \tag{C.11}$$

where $Y_{\ell m}(\mathbf{n})$ are the hyperspherical harmonics on the unit $D-1$-sphere $S^{D-1}_{r_0}$ (with $V$ being the volume of the sphere), see Appendix B.1, and $\omega_n$ are the Matsubara frequencies,

$$\omega_n = \frac{2\pi n}{\beta}\,. \tag{C.12}$$

Here, the zero modes $\zeta_i$ and $\varphi_i$ are spacetime-independent and determine the full infrared behaviour of $\phi_i$, *i.e.* $\tilde\phi_i(0;0,0) = 0$. The variable $\zeta_i$ in particular allows for the possibility of a Bose–Einstein

[1] For an arbitrary manifold $S^1_\beta \times \mathscr{M}$ the hyperspherical harmonics are simply replaced by the appropriate eigenfunctions on $\mathscr{M}$. For example in flat space we replace $Y_{\ell m}(\mathbf{n})$ by $e^{i\mathbf{p}\cdot\mathbf{x}}$.

Condensate (BEC), the condensation of the bosons that reside in the ground state of the system. In terms of the eigenfunctions in Eq. (C.11) the leading order partition function reads

$$Z_{gc}(\mu) \overset{N\to\infty}{\simeq} \int [d\tilde{\phi}_i(n;\ell,m)]\,[d\tilde{\phi}_i^*(n;\ell,m)]\,\exp\left[-\beta V\sum_i \zeta_i^2\left(r+\langle\sigma\rangle-\mu_i^2\right)-\sum_i\sum_{n;\ell,m}\tilde{\phi}_i^*(-n;\ell,m^*)\,\tilde{G}_i\,\tilde{\phi}_i(n;\ell,m)\right], \tag{C.13}$$

where $\tilde{G}_i$ in terms of the real variables $\tilde{\phi}_i^{(\mathrm{Re})}(n;\ell,m)$, $\tilde{\phi}_i^{(\mathrm{Im})}(n;\ell,m)$ is given by[2]

$$\tilde{G}_i = \beta^2\begin{pmatrix}\omega_n^2+\omega_\ell^2-\mu_i^2 & -2\mu_i\omega_n\\ 2\mu_i\omega_n & \omega_n^2+\omega_\ell^2-\mu_i^2\end{pmatrix}, \qquad \omega_\ell^2=\frac{\ell(\ell+D-2}{r_0^2}+r+\langle\sigma\rangle\,. \tag{C.14}$$

After carrying out the integration we have

$$\begin{aligned}\Omega(\mu_1,\dots) &= -\frac{1}{(2N)\beta V}\log(Z_{gc})_{N\to\infty}=\frac{1}{(2N)\beta V}\sum_{i=1}^{N}\mathrm{Tr}\log\left(-(\partial_\tau-\mu_i)^2-\Delta_{S^{D-1}}+r+\langle\sigma\rangle\right)\\ &= \frac{1}{2N}\sum_i\zeta_i^2\left(r+\langle\sigma\rangle-\mu_i^2\right)+\frac{1}{(2N)\beta V}\sum_i\frac{1}{2}\log(\det\tilde{G}_i)\,,\end{aligned} \tag{C.15}$$

The log-det term is evaluated as follows:

$$\begin{aligned}\log(\det\tilde{G}_i) &= \log\det\prod_{n\in Z}\prod_{\ell,m}\beta^2\begin{pmatrix}\omega_n^2+\omega_\ell^2-\mu_i^2 & -2\mu_i\omega_n\\ 2\mu_i\omega_n & \omega_n^2+\omega_\ell^2-\mu_i^2\end{pmatrix}\\ &= \sum_{\{n,\ell\}}\mathrm{Deg}_D(\ell)\log\left\{\left(\left(\frac{2\pi}{\beta}n\right)^2+\omega_\ell^2-\mu^2\right)^2+4\mu^2\left(\frac{2\pi}{\beta}n\right)^2\right\}\\ &= \sum_{\{n,\ell\}}\mathrm{Deg}_D(\ell)\left\{\log\beta^2\left(\omega_n^2+(\omega_\ell-\mu)^2\right)+\log\beta^2\left(\omega_n^2+(\omega_\ell+\mu)^2\right)\right\},\end{aligned} \tag{C.16}$$

There are two important identities that apply here [194],

$$\log\left\{\frac{(2\pi n)^2+\beta^2x^2}{(2\pi n)^2+1}\right\}=\int_1^{\beta^2x^2}\frac{d\theta^2}{\theta^2+(2\pi n)^2}\,,\qquad \sum_{n\in Z}\frac{1}{(2\pi n)^2+\theta^2}=\frac{1}{2\theta}\left(1+\frac{2}{e^\theta-1}\right). \tag{C.17}$$

The term $\sum_{n\in Z}\log\left\{(2\pi n)^2+1\right\}$ is constant and temperature-independent and formally divergent. As it is independent of all variables, it will become unimportant after regularization and can be ignored. We compute,

$$
\begin{aligned}
\log(\det\tilde{G}_i) &= \sum_\ell \mathrm{Deg}_D(\ell)\left[\int_1^{\beta^2(\omega_\ell+\mu_i)^2} \mathrm{d}\theta^2\ \frac{1}{2\theta}\left(1+\frac{2}{e^\theta-1}\right) + \sum_n \log\left((2\pi n)^2+1\right) + (\mu_i \to -\mu_i)\right] \\
&= 2\sum_\ell \mathrm{Deg}_D(\ell)\Bigg[\beta\omega_\ell + \log(1-e^{-\beta(\omega_\ell+\mu_i)}) + \log(1-e^{-\beta(\omega_\ell-\mu_i)}) \\
&\qquad + \left[\sum_n \log\left((2\pi n)^2+1\right) - \left(2\log(1-e^{-1})+1\right)\right]\Bigg] \\
&\sim 2\sum_\ell \mathrm{Deg}_D(\ell)\left[\beta\omega_\ell + \log(1-e^{-\beta(\omega_\ell+\mu_i)}) + \log(1-e^{-\beta(\omega_\ell-\mu_i)})\right].
\end{aligned}
\tag{C.18}
$$

where we have ignored all constant terms independent of temperature and volume. Finally, the grand potential $\Omega \overset{N\to\infty}{=} -\log(Z_{gc})/((2N)\beta V)$ reads

$$
\Omega(\mu_1,\dots) = \frac{1}{2N}\sum_i \zeta_i^2\left(r+\langle\sigma\rangle-\mu_i^2\right) + \frac{1}{(2N)V}\sum_i\sum_\ell \mathrm{Deg}_D(\ell)\left[\omega_\ell + \frac{1}{\beta}\log(1-e^{-\beta(\omega_\ell+\mu_i)})(1-e^{-\beta(\omega_\ell-\mu_i)})\right],
\tag{C.19}
$$

which is convergent as long as $|\mu_i| \le \sqrt{r+\langle\sigma\rangle}$. We remark that this result can be generalized from the sphere to any other spatial manifold $\mathcal{M}$ by simply replacing the eigenvalues $\omega_\ell \to \omega(\mathbf{p}) = \sqrt{E(\mathbf{p})^2 + r + \langle\sigma\rangle}$ and the corresponding multiplicities $\mathrm{Deg}_D(\ell) \to \mathrm{Deg}_D(\mathbf{p})$.

[2] For any other spatial manifold we can simply replace the eigenvalue/energy $\omega_\ell=\sqrt{\ell(\ell+D-2)}/r_0$ by the appropriate eigenvalue $E(p)$.

## C.3 Resurgence of the four-sphere

### C.3.1 Borel re-summation

In resurgence, Borel re-summation is a tool aimed at transforming an asymptotic series into a resurgent function (for an introduction see [243]). In resurgence, if we have a factorially divergent asymptotic series,

$$
\Phi_0(t) = \sum_n a_n t^n, \qquad a_n \sim \sum_k \frac{S_k}{2\pi i}\frac{\beta_k}{A_k^{n\beta_k+b_k}}\sum_{l\ge 0} a_{l;k} A_k^l \Gamma\left(n\beta_k + b_k - l\right),
\tag{C.20}
$$

the general assumption is that said asymptotic expansion can be completed into a resurgent trans-series of the form

$$
\Phi(S_k,t) = \Phi_0(t) + \sum_{k\neq 0} \sigma_k e^{-\frac{A_k}{t^{1/\beta_k}}} t^{-b_k/\beta_k}\Phi_k(t), \qquad \Phi_k(t) \sim \sum_{l\ge 0} a_{l;k} t^{l/\beta_k}.
\tag{C.21}
$$

The coefficient $\sigma_k$ are ambiguities arising from the fact that the full family of resurgent trans-series associated to different values of $\sigma_k$ all correspond to the same asymptotic series $\Phi_0(t)$.
The standard ways of completing an asymptotic expansion into a trans-series is via a Borel transform followed by a Borel re-summation. The closed- form Borel transform of a not necessarily asymptotic series is defined as

$$
\mathscr{B}[\Phi_0](y) = \sum_{n\ge 0} \frac{a_n}{\Gamma\left(\max_k(\beta_k)\cdot n + \max_k(b_k)\right)} y^n .
\tag{C.22}
$$

Starting from the Borel transform, the directional Borel re-summation of the series $\Phi_0$ is given by

$$\mathscr{S}_\theta[\Phi_0](z) = \frac{1}{\max_k(\beta_k)} \int_0^{e^{i\theta}\infty} \frac{d\zeta}{\zeta} \left(\frac{\zeta}{z}\right)^{\frac{b}{\beta}} e^{-(\zeta/z)^{\frac{1}{\beta}}} \mathscr{B}[\Phi_0](\zeta). \tag{C.23}$$

This reproduces the original asymptotic series $\Phi_0(t)$ in Eq. (C.20) for $z = te^{i\theta}$, $t \to 0^+$. However, the Borel re-summation now defines a function computable for a wider range of values $t$ and $\theta$.

Ambiguities in the Borel re-summation arise in the case where the Borel transform exhibits singularities along the contour or ray of integration $e^{i\theta}[0,\infty]$.[3] In these cases we can define two lateral Borel re-summations $\mathscr{S}_\theta^\pm$ obtained by deforming the contour of integration around the singularities, either below or above the original contour. The new paths of integration created in this way we denote by $C_\theta^\pm$. This behaviour in the case of singularities along the ray of integration indicates the presence of a branch cut of the directional Borel re-summation $\mathscr{S}_\theta[\Phi_0](z)$ at $z = te^{i\theta}$, with the discontinuity across the branch cut being given by $\left(\mathscr{S}_\theta^- - \mathscr{S}_\theta^+\right)[\Phi_0](t)$. This discontinuity represents a purely non-perturbative effect. It includes only the exponential corrections plus the expansions $\Phi_k(t)$ around them. If we are presented with a Borel re-summable asymptotic series, then the structure of the non-perturbative corrections in the trans-series (exponential corrections plus the expansions around them) and hence the structure of the lateral Borel re-summations $\mathscr{S}_\theta^\pm[\Phi_k](t)$ is provided by and can be read off the discontinuity along the branch cut. This procedure, however, will not remove the ambiguities in the choice of parameters $\sigma_k^\pm$ for the exponential corrections. Hence, in order to promote the asymptotic expansion into a full resurgent trans-series we generally need to impose extra conditions motivated by the underlying physics.

### C.3.2 Zeta function on the four sphere

For the interacting fixed point of the $\varphi^4$ model in $D = 5$ the large-charge expansion is an asymptotic expansion related to the heat kernel trace on the four-sphere which can be computed starting from the result on the two-sphere, see the discussion around Eq. (3.149) in Section 3.2.5. Therefore, we can apply the exact same resurgent techniques used in [191] in order to study the non-perturbative corrections and find a closed-form resurgent function for the trace of the heat kernel on $S^4_{r_0}$. Using the results from [191], the asymptotic expansion of the heat kernel in the two-sphere,

$$\mathrm{Tr}\left[e^{t\left(\Delta_{S^2_{r_0}} - 1/(4r_0^2)\right)}\right] \sim \frac{r_0^2}{t} \Phi_0^{S^2_{r_0}}(t), \tag{C.24}$$

is given by

$$\Phi_0^{S^2_{r_0}}(t) = \sum_{n\geq 0} a_n^{S^2} \frac{t^n}{r_0^{2n}}, \qquad a_n^{S^2} = \frac{(1-2^{1-2n})B_{2n}}{(-1)^{n+1} n!} = \frac{\Gamma\left(n+\frac{1}{2}\right)}{\sqrt{\pi}} \sum_{k\neq 0} \frac{(-1)^{k+1}}{(\pi k)^{2n}}. \tag{C.25}$$

Using the relationship in Eq. (3.152) the asymptotic expansion of the heat kernel trace on the four-sphere,

$$\mathrm{Tr}\left[e^{t\left(\Delta_{S^4_{r_0}} - 9/(4r_0^2)\right)}\right] \sim \frac{r_0^4}{t^2} \Phi_0^{S^4_{r_0}}(t), \tag{C.26}$$

[3] Additionally, it can be the case that the behaviour at the endpoints $\{0, e^{i\theta}\infty\}$ is of importance as well.

is computed to be

$$\Phi_0^{S^4_{r_0}}(t) = \frac{1}{6} + \sum_{\substack{n\geq 1\\ k\neq 0}} \frac{(-1)^k}{(\pi k)^{2n}} \frac{1}{6\sqrt{\pi}} \left[\Gamma\big(n+\frac{3}{2}\big) - \frac{3}{2}\Gamma\big(n+\frac{1}{2}\big) + \frac{(\pi k)^2}{4}\Gamma\big(n-\frac{1}{2}\big)\right] \frac{t^n}{r_0^{2n}}. \tag{C.27}$$

From this expression we can extract the heat kernel coefficients (see also [319]). Comparing this expression with the general form in Eq. (C.20) we see that

$$A_k = (\pi k)^2\,, \qquad b_k = \frac{3}{2}\,, \qquad \beta_k = 1\,, \qquad \begin{cases} \frac{S_k}{2\pi i}\, a_{0;k} = (-1)^k |k|^3 \frac{\pi^{5/2}}{6} \\ \frac{S_k}{2\pi i}\, a_{1;k} = (-1)^{k+1} |k| \frac{\sqrt{\pi}}{4} \\ \frac{S_k}{2\pi i}\, a_{2;k} = (-1)^k |k| \frac{\sqrt{\pi}}{24} \end{cases}, \qquad a_{\ell>2;k} = 0\,. \tag{C.28}$$

Hence, the resurgent trans-series of the heat kernel trace on the four-sphere includes exponential corrections of the form

$$\text{exponential corrections} \sim 2i\left(\frac{r_0^2\pi}{t}\right)^{\frac{7}{2}} e^{-\frac{(\pi r_0 k)^2}{t}} (-1)^k |k| \left(\frac{k^2}{6} - \frac{t}{4r_0^2\pi^2} + \frac{t^2}{24 r_0^4 \pi^2}\right). \tag{C.29}$$

The Borel re-summation of the heat kernel trace on the four-sphere can directly be derived from the expression for the two-sphere found in [191]. On the two-sphere the Borel re-summation of the two-sphere heat kernel trace can be written in the compact form[4]

$$\mathscr{S}_0[\Phi_0^{S^2_{r_0}}](t) = \frac{2r_0}{\sqrt{\pi t}} \int_0^\infty \mathrm{d}y\, \frac{y e^{-\frac{y^2 r_0^2}{t}}}{\sin(y)}\,. \tag{C.30}$$

Using Eq. (3.152) we can directly infer the form of the Borel re-summed heat kernel trace on the four-sphere

$$\mathscr{S}_0[\Phi_0^{S^4_{r_0}}](t) = \frac{2r_0^3}{\sqrt{\pi t^3}} \int_0^\infty \mathrm{d}y\, \frac{y e^{-\frac{y^2 r_0^2}{t}}}{\sin(y)} \left[\frac{t}{4r_0^2} - \frac{y^2}{6} - \frac{t^2}{24 r_0^4}\right]. \tag{C.31}$$

This integral is ill-defined as it has simple poles at $\zeta = k\pi$, $k \in \mathbb{Z}$. The discontinuity between the lateral Borel re-summations are computable via the Residue theorem as

$$\left(\mathscr{S}_0^- - \mathscr{S}_0^+\right)[\Phi_0^{S^4_{r_0}}](t) = 2i\frac{t^2}{r_0^4}\left(\frac{r_0^2\pi}{t}\right)^{\frac{7}{2}} \sum_{k\neq 0} (-1)^k |k|\, e^{-\frac{k^2 r_0^2 \pi^2}{t}} \left[\frac{|k|^2}{6} - \frac{t}{4\pi^2 r_0^2} + \frac{t^2}{24\pi^2 r_0^4}\right]. \tag{C.32}$$

This agrees perfectly with the form of the exponential corrections derived in Eq. (C.29). Using this

[4] In order to avoid dealing with a Borel transform that includes branch cuts we perform the mapping $\zeta \to \zeta^2$ in Eq. (C.23).

result, the Borel re-summed trace of the heat kernel on the four-sphere with ambiguities is given by

$$\mathrm{Tr}\Big[e^{t\left(\Delta_{S^4_{r_0}}-9/(4r_0^2)\right)}\Big] \sim \frac{r_0^4}{t^2}\mathscr{S}_0^{\pm}[\Phi_0^{S^4_{r_0}}](t) = \frac{2r_0^7}{\sqrt{\pi t^7}}\int_{C_0^{\pm}} \mathrm{d}y\, \frac{ye^{-\frac{r_0^2y^2}{t}}}{\sin(y)}\left[\frac{t}{4r_0^2}-\frac{y^2}{6}-\frac{t^2}{24r_0^4}\right]$$
$$+\, i\left(r_0^2\frac{\pi}{t}\right)^{\frac{7}{2}}\sum_{k\neq 0}\frac{\sigma_k^{\pm}(-1)^k|k|}{\pi^2}e^{-\frac{k^2r_0^2\pi^2}{t}}\left[\frac{k^2\pi^2}{3}-\frac{t}{2r_0^2}+\frac{t^2}{12r_0^4}\right]. \quad \text{(C.33)}$$

In this particular case it is possible to fix the ambiguities in the trans-series expression for the trace of the heat kernel, which will carry over to all the other quantities that we are interested in. This can be achieved in two ways here: either we find a path-integral definition in which the structure of the trans-series arises automatically (see [191]) and the underlying physics fix the ambiguities, or we can impose the reality of the heat kernel trace.[5] To keep the discussion short, will choose the latter here. Either way, the final result can be written in the convenient form

$$\mathrm{Tr}\Big[e^{t\left(\Delta_{S^4_{r_0}}-9/(4r_0^2)\right)}\Big] = \frac{2r_0^7}{\sqrt{\pi t^7}}\mathrm{P.V.}\int_{C_0^{\pm}} \mathrm{d}y\, \frac{ye^{-r_0^2y^2/t}}{\sin(y)}\left[\frac{t}{4r_0^2}-\frac{y^2}{6}-\frac{t^2}{24r_0^4}\right]. \quad \text{(C.34)}$$

The notation P.V. signifies that the principal value of the integral has to be considered.

The trans-series expression for the zeta function can be derived by applying a Mellin transform to the Borel re-summed trace of the heat kernel (with the reality condition imposed). However, for $s = \pm 1/2$ in order to change to order integration the integral requires analytical continuation by extracting the first three terms from the asymptotic expansion. We find that

$$\mathscr{S}_0^{\pm}[\zeta(\,\cdot\,|S^4_{r_0},\mu^2)](s) = \frac{2r_0^{2s}}{\sqrt{\pi}\Gamma(s)}\int_0^{\infty}\mathrm{d}t\int_{C_0^{\pm}}\mathrm{d}y\,\frac{e^{-\mu^2r_0^2t-\frac{y^2}{t}}}{t^{\frac{9}{2}-s}}\left[\frac{y}{\sin(y)}\left(-\frac{y^2}{6}+\frac{t}{4}-\frac{t^2}{24}\right)-\frac{y^2}{3}+\frac{y^4}{18}+\frac{17y^6}{5'400}\right]$$
$$+\frac{1}{6}\frac{r_0^4\mu^{4-2s}}{(s-1)(s-2)}-\frac{1}{24}\frac{r_0^2\mu^{2-2s}}{(s-1)}-\frac{17}{2'880}\mu^{-2s}. \quad \text{(C.35)}$$

After exchanging the order of integration for $s = -1/2$ the result is identified with the (non-standard) Borel re-summation of the asymptotic series of the grand potential $V\Omega(\mu)$,

$$\mathscr{S}_0^{\pm}[V\Omega](\mu) = \frac{r_0^3m^4}{24\pi}\int_{C_0^{\pm}}\frac{\mathrm{d}y}{y^2}\left[\frac{y}{\sin(y)}\left(2\Big[K_4(2\mu r_0y)+K_2(2\mu r_0y)\Big]+\frac{K_2(2\mu r_0y)}{(\mu r_0)^2}\right)\right.$$
$$\left.+\left(8-\frac{4y^2}{3}-\frac{17y^4}{225}\right)K_4(2\mu r_0y)\right]+\frac{2}{45}\mu^5r_0^4+\frac{1}{36}\mu^3r_0^2-\frac{17}{2880}\mu, \quad \text{(C.36)}$$

where $K_n(x)$ denotes the order $n$ modified Bessel function of the second kind. The discontinuity — and

[5]Importantly, in the general case the reality condition does not completely fix the ambiguities in the non-perturbative corrections, since it is generally possible that the coefficients $2i\sigma_k^{\pm}$ possess a real part as well. In our case, however, imposing reality is sufficient due to the absence of a real part.

hence the form exponential corrections — given by

$$\left(\mathscr{S}_0^- - \mathscr{S}_0^+\right)[V\Omega](\mu) = -i\frac{r_0^3\mu^4}{12}\sum_{k\neq 0}\frac{(-1)^k}{|k|}\left(2\Big[K_4(2\pi\mu r_0 k) + K_2(2\pi\mu r_0 k)\Big] + \frac{1}{(\mu r_0)^2}K_2(2\pi\mu r_0 k)\right). \tag{C.37}$$

Finally, after imposing the reality condition for the heat kernel trace to remove the ambiguities, the grand potential can be written in the form

$$V\Omega(\mu) = \frac{r_0^3\mu^4}{24\pi}\text{P.V.}\int_0^\infty \frac{\mathrm{d}y}{y\sin(y)}\left(2\Big[K_4(2\mu r_0 y) + K_2(2\mu r_0 y)\Big] + \frac{1}{(\mu r_0)^2}K_2(2\mu r_0 y)\right). \tag{C.38}$$

For the purpose of this discussion it suffices to consider only the leading-order non-perturbative corrections that appear in the free energy. To extract these contributions we will use Hankel's asymptotic expansion for the modified Bessel functions of the second kind,

$$K_\alpha(z) \sim \sqrt{\frac{\pi}{2z}}e^{-z}\left[1 + \frac{(4\alpha^2-1)}{8z}\left(1 + \frac{1}{2}\frac{(4\alpha^2-3^2)}{8z}\left[1 + \frac{1}{3}\frac{(4\alpha^2-5^2)}{8z}\left(1 + \dots\ \right)\right]\right)\right], \qquad -3\frac{\pi}{2} < \arg(z) < 3\frac{\pi}{2}. \tag{C.39}$$

Using this result the leading-order non-perturbative exponential corrections of the grand potential are

$$V\Omega(\mu) \supset -i\frac{r_0^3\mu^4}{12}\frac{(-1)^k}{|k|}\left(4 + \frac{1}{(r_0\mu)^2}\right)\frac{1}{\sqrt{4r_0\mu|k|}}e^{-2\pi r_0\mu|k|} \sim i\frac{(r_0\mu)^{\frac{7}{2}}}{6r_0}\frac{(-1)^k}{|k|^{\frac{3}{2}}}e^{-2\pi r_0\mu|k|}. \tag{C.40}$$

Using only the first term on the result for the order-by-order Legendre transform of the perturbative part,

$$\frac{Q}{2N} = -(r_0\mu)^4/9, \qquad\qquad r_0 f_c(Q) = \mu(Q)r_0\frac{Q}{2N} + \frac{(\mu(Q)r_0)^5}{45}, \tag{C.41}$$

the leading-order exponential correction to the free energy per DoF reads

$$r_0 f_c(Q) \supset i\frac{\left(\frac{9Q}{2N}\right)^{7/8}}{6\sqrt{f_1}}\frac{(-1)^k}{|k|^{3/2}}e^{-2\pi f_1\left(\frac{9Q}{2N}\right)^{1/4}|k|}, \tag{C.42}$$

where $f_1$ is the leading complex phase in the asymptotic expansion of $f_c(Q)$ discussed in Eq. (3.162) and Eq. (3.163).

### C.3.3 Optimal truncation

As asymptotic series factorially diverge, in order to get a meaningful result out of such a series expansion we need to truncate it. A commonly used rule-of-thumb to find a truncated sum that is as close as possible to the "actual" value is to truncate it at the term which gives the smallest contribution to the overall sum. This procedure is called optimal truncation. Given an asymptotic series $\sum a_n x^n$ whose coefficients $a_n$ diverge as $\sum a_n x^n$, the optimal truncation is found at

$$N(x) \approx \frac{1}{\beta}|Ax|^{1/\beta}. \tag{C.43}$$

The error we make during optima truncation can be approximated within resurgence and is of the order

$$\epsilon(x) \sim \exp\left[-(Ax)^{1/\beta}\right]. \tag{C.44}$$

In the special case $x \sim 1$ it suffices to look at the ratio of consecutive coefficients within the asymptotic series. At the point where this ratio exceed one, we truncate it.

For the zeta function on the four-sphere $\zeta(s|S^4_{r_0},\mu)$ at $\mu^2 r_0^2 - \frac{9}{4} \sim 1$ the optimal truncation depends on $s$. It is after the third term for $s = \frac{3}{2}$, after the fourth term for $s = \frac{1}{2}$ and after the fifth term form for $s = -\frac{1}{2}$. To go beyond this result we need to rely on our resurgent analysis. For the trace of the heat kernel we can use the large-order behaviour of the Bernoulli numbers,

$$\frac{(1-2^{1-2n})B_{2n}}{(-1)^{n+1}n!} \sim \frac{2}{\sqrt{n\pi}}\frac{n!}{\pi^{2n}}, \tag{C.45}$$

we can deduce that the optimal truncation is given by

$$N(t) \approx \pi^2 r_0^{-2} t, \qquad \epsilon(t) \sim \exp(-\pi^2 r_0^{-2} t). \tag{C.46}$$

By performing a Mellin transform we find the asymptotic expansion of the grand potential $\Omega(\mu)$,

$$\zeta(s|S^4_{r_0},\mu^2) \sim r_0^{2s}\Big(\mu^2 r_0^2 - \frac{9}{4}\Big)^{2-s} \sum_{n\geq 0} a_n^{S^4_{r_0}} \frac{\Gamma(n+s-2)}{\Gamma(s)}\Big(\mu^2 r_0^2 - \frac{9}{4}\Big)^{-n}, \tag{C.47}$$

$$V\Omega(\mu) \sim r_0^{-1}\Big(\mu^2 r_0^2 - \frac{9}{4}\Big)^{\frac{5}{2}} \sum_{n\geq 0} a_n^{S^4_{r_0}} \frac{\Gamma\big(n-\frac{5}{2}\big)}{4\sqrt{\pi}}\Big(\mu^2 r_0^2 - \frac{9}{4}\Big)^{-n} = -\frac{\big(\mu^2 r_0^2 - \frac{9}{4}\big)^{\frac{5}{2}}}{45 r_0} + \frac{\big(\mu^2 r_0^2 - \frac{9}{4}\big)^{\frac{5}{2}}}{r_0}\sum_{n\geq 1}\Omega_n\Big(\mu^2 r_0^2 - \frac{9}{4}\Big)^{-n}, \tag{C.48}$$

$$\Omega_n = \sum_{k\neq 0}\frac{(-1)^k}{(\pi k)^{2n+3}}\frac{(\pi|k|)^3}{24\pi}\Gamma\Big(n-\frac{5}{2}\Big)\left[\Gamma\Big(n+\frac{3}{2}\Big) - \frac{3}{2}\Gamma\Big(n+\frac{1}{2}\Big) + \frac{(\pi k)^2}{4}\Gamma\Big(n-\frac{1}{2}\Big)\right]. \tag{C.49}$$

This already makes it clear that the coefficients grow like $(2n)!$. Once we have resolved the double factorial structure the leading term reads

$$\Omega_n = \sum_{k\neq 0}\frac{(-1)^k}{(2\pi|k|)^{2n-\frac{1}{2}}}\frac{1}{12\pi\sqrt{|k|}}\Big[\Gamma(2n-\frac{1}{2}) + \ldots\Big]. \tag{C.50}$$

From this result we can read off the optimal truncation of the grand potential,

$$N(\mu) \approx \frac{1}{2}\left|2\pi\Big(\mu^2 r_0^2 - \frac{9}{4}\Big)\right|^{1/2}, \qquad \epsilon\Big(\mu^2 r_0^2 - \frac{9}{4}\Big) \sim \exp\Big[-\sqrt{2\pi\Big(\mu^2 r_0^2 - \frac{9}{4}\Big)}\Big]. \tag{C.51}$$

For $\mu^2 r_0^2 \gtrsim 3/2$ the optimal truncation for the grand potential is at the first term, hence

$$V\Omega(\mu) \sim \frac{\left(\mu^2 r_0^2 - \frac{9}{4}\right)^{\frac{5}{2}}}{45 r_0}. \tag{C.52}$$

### C.3.4 Flex from large charge

Since we know that the grand potential exhibits a flex for a certain finite value of the chemical potential $\mu r_0$, it is also interesting to analyse the second derivative of the grand potential. Using the identity $(\partial/\partial\mu)\zeta(s|\mathcal{M},\mu^2) = -2\mu s\zeta(s+1|\mathcal{M},\mu^2)$ we find that

$$V\Omega''(\mu) = \frac{1}{2}\left[\zeta(1/2|S^4_{r_0},\mu^2) - \mu^2\zeta(3/2|S^4_{r_0},\mu^2)\right]. \tag{C.53}$$

Around $\mu r_0 \sim$, where the flex can be found, the optimal truncation is again deduced by looking at the ratio of consecutive coefficients, which start to exceed one after the sixth term. Up to this term the grand potential reads

$$V\Omega(\mu) = \frac{r_0^4\mu^5}{45} - \frac{r_0^2\mu^3}{9} - \frac{29\mu}{45} - \frac{37\mu^{-1}}{756 r_0^2} - \frac{149\mu^{-3}}{15'120 r_0^4} - \frac{179\mu^{-5}}{55'440 r_0^6}, \tag{C.54}$$

and its second derivative is

$$V\Omega''(\mu) = \frac{4r_0^4\mu^3}{9} - \frac{2r_0^2\mu}{3} - \frac{37\mu^{-3}}{378 r_0^2} - \frac{149\mu^{-5}}{1'260 r_0^4} - \frac{179\mu^{-7}}{1'848 r_0^6}. \tag{C.55}$$

Numerically we find that the optimally truncated asymptotic expansion of $\Omega''(\mu)$ exhibits a zero at $\mu r_0 \approx 1.290$, implying that the asymptotic expansion of $\Omega(\mu)$, which is close to the actual value of $\mu_{\text{fl}} r_0 \approx 1.266$. The error for the optima truncation of $\Omega(\mu)$ at $\mu^2 r_0^2 - 9/4 \sim 1$ is $\epsilon(\mu^2 r_0^2 - 9/4 \sim 1) \sim e^{-\sqrt{2\pi}} = 0.08$.

## C.4 3D Fermions

This appendix aims at collecting important background material for fermionic theories in $D = 2+1$ and $D = 3$ spacetime dimensions. This should make the present thesis as self-contained as possible.

### C.4.1 Gamma matrices in the Dirac convention in 3D

The gamma matrices in three spacetime dimensions ($D = 2+1$ and $D = 3$) are given by the Pauli matrices,

$$\sigma_1 = \begin{pmatrix} 0 & 1 \\ 1 & 0 \end{pmatrix}, \qquad \sigma_2 = \begin{pmatrix} 0 & -i \\ i & 0 \end{pmatrix}, \qquad \sigma_3 = \begin{pmatrix} 1 & 0 \\ 0 & -1 \end{pmatrix}, \tag{C.56}$$

as follows:

$$D = 2+1: \qquad \gamma_0 = i\sigma_3\,, \qquad \gamma_{1,2} = \sigma_{1,2}\,, \tag{C.57}$$

$$D = 3: \qquad \gamma_\mu = \sigma_\mu\,, \qquad \mu = 1,2,3\,. \tag{C.58}$$

Our convention for the Clifford algebra is

$$\{\gamma_\mu, \gamma_\nu\} = 2\eta_{\mu\nu}\,, \tag{C.59}$$

where we have chosen a mostly plus signature $\eta_{\mu\nu} = (-1,1,1)$ for $D = 2+1$ dimensional Minkowski spacetime. In this signature the spatial gamma matrices are Hermitian while the temporal gamma matrix $\gamma_0$ is anti-Hermitian. Further, the gamma matrices satisfy

$$(\gamma_i)^2 = -(\gamma_0)^2 = \mathbb{1}, \qquad \gamma_0\gamma_\mu\gamma_0 = (\gamma_\mu)^\dagger\,. \tag{C.60}$$

Complex two-dimensional (Dirac) spinors $\psi$ transform in the standard representation of $SO(1,2)$, $SO(3)$ generated by these gamma matrices. The Dirac conjugate $\bar\psi$ of the spinor $\psi$ in our notation is found to be

$$d = 2+1: \qquad \bar\psi = \psi^\dagger\gamma_0\,, \tag{C.61}$$

$$d = 3: \qquad \bar\psi = \psi^\dagger\gamma_3\,. \tag{C.62}$$

The analytic continuation from Minkowski spacetime to Euclidean spacetimes is obtained by a Wick rotation,

$$t \to -i\tau\,, \qquad \partial_t \to i\partial_\tau\,, \qquad \gamma_0 \to i\gamma_3\,, \qquad \gamma_i \to \gamma_i\,. \tag{C.63}$$

Under the above transformation rules the action of a massive Dirac spinor is analytically continued in the following manner:

$$\underbrace{i\int \mathrm{d}t\mathrm{d}^2x\,\left[\bar\psi(i\gamma_\mu\partial^\mu + im)\psi\right]}_{S_M} \qquad \longrightarrow \qquad \underbrace{-\int \mathrm{d}^3x\,\left[\bar\psi(\gamma_\mu\partial^\mu + m)\psi\right]}_{S_E}\,. \tag{C.64}$$

### C.4.2 Spinors on $S^1_\beta \times S^2$

For the discussion of spinors in curved spacetime we follow closely the treatment outlined in [**Borokhov_2002**].

#### Spinors in spherical coordinates

In flat space $\mathbb{R}^3$ the Hermitian Dirac operator is given by

$$i\gamma_\mu\partial^\mu = -\gamma\cdot\mathbf{p}\,, \tag{C.65}$$

where the momentum operator $\mathbf{p}$ is given by $p^\mu = -i\partial^\mu$ and the vector $\gamma$ is comprised of the gamma matrices $\gamma_\mu$ — *i.e.* the Pauli matrices $\sigma_\mu$ — in Eq. (C.56) and Eq. (C.57). We can define the generalized angular momentum and the total angular momentum as

$$\mathbf{L} = \mathbf{r} \times \mathbf{p}\,, \qquad \mathbf{J} = \mathbf{L} + \underbrace{\frac{\gamma}{2}}_{=\mathbf{S}}\,, \qquad [\mathbf{L}, \mathbf{r}] = [\mathbf{J}, \mathbf{r}] = 0\,. \tag{C.66}$$

We note that both $\mathbf{L}$ and $\mathbf{p}$ are Hermitian operators. The eigenfunctions of $\mathbf{L}^2$ are given by the ordinary spherical harmonics in $D = 3$,

$$\mathbf{L}^2 Y_{\ell m} = \ell(\ell+1) Y_{\ell m}\,, \qquad \underbrace{L_3}_{=L_z} Y_{\ell m} = m Y_{\ell m}\,, \qquad \ell = 0, 1, 2, \ldots\,, \qquad m = -\ell, \ldots, \ell\,. \tag{C.67}$$

Using the spherical harmonics we can build simultaneous eigenfunctions of all four operators $\{\mathbf{J}^2, J_3, \mathbf{L}^2, \mathbf{S}^2\}$. These eigenfunctions are the so-called spinor spherical harmonics,

$$\phi^+_{jm_j} = \begin{pmatrix} \sqrt{\frac{\ell+m+1}{2\ell+1}}\, Y_{\ell m} \\ \sqrt{\frac{\ell-m}{2\ell+1}}\, Y_{\ell m+1} \end{pmatrix}, \qquad \phi^-_{jm_j} = \begin{pmatrix} -\sqrt{\frac{\ell-m}{2\ell+1}}\, Y_{\ell m} \\ \sqrt{\frac{\ell+m+1}{2\ell+1}}\, Y_{\ell m+1} \end{pmatrix}. \tag{C.68}$$

The eigenfunctions $\phi^\pm_{jm_j}$ correspond to the eigenvalues $j = \ell \pm 1/2$ and $m_j = m \pm 1/2$, respectively, and they posses the following quantum numbers under the three operators $\mathbf{J}^2, J_3, \mathbf{L}^2$, respectively,

$$\begin{cases} \mathbf{L}^2 \phi^\pm_{jm_j} & = \ell(\ell+1)\phi^\pm_{jm_j} \\ \mathbf{J}^2 \phi^\pm_{jm_j} & = j(j+1)\phi^\pm_{jm_j} \\ J_3 \phi^\pm_{jm_j} & = m_j \phi^\pm_{jm_j} \end{cases}, \qquad \begin{cases} j = \frac{1}{2}, \frac{3}{2}, \frac{5}{2} \cdots \\ m_j = -j \ldots j \end{cases}. \tag{C.69}$$

The eigenvalue $j$ has a degeneracy of $(2j+1)$. Any spinor in $\mathbb{R}^3$ can be decomposed in the orthonormal basis spanned by $\phi^\pm_{jm_j}$. For our purposes, it is further convenient to introduce the radial gamma matrix $\gamma_r = \gamma \cdot \mathbf{n}$ (where $\mathbf{n} = \mathbf{r}/r \in S^2_1$). The Dirac operator can then simply be written as

$$i\gamma_\mu \partial^\mu = i\gamma_r \left[\frac{\partial}{\partial r} - \frac{1}{r}\left(\mathbf{J}^2 - \mathbf{L}^2 - \frac{3}{4}\right)\right], \tag{C.70}$$

and is diagonal in the basis of spinor spherical harmonics $\phi^\pm_{jm_j}$.

### Weyl map to the cylinder

We can perform the Weyl transformation from flat space to the cylinder $\mathbb{R} \times S^2_{r_0}$ as follows:

$$r = e^\tau\,, \qquad \eta_{\mu\nu} = r_0 e^{2\tau} g_{\mu\nu}\,, \qquad \psi_{\mathbb{R}^3} = e^{-\tau}\psi_{\mathbb{R}\times S^2}\,. \tag{C.71}$$

After foliating radially, the Dirac conjugate of a spinor $\psi$ is given by $\psi^\dagger = \bar{\psi}\gamma_r$. After performing the Weyl transformation, the free Dirac action on the cylinder $\mathbb{R} \times S^2_{r_0}$ reads

$$S = \int_{\mathbb{R}^3} \bar{\psi}\gamma_\mu \partial^\mu \psi = \int_{\mathbb{R}\times S^2_{r_0}} \bar{\psi}\gamma_\mu D^\mu \psi\,, \qquad \gamma_\mu D^\mu = \gamma_r \left[\frac{\partial}{\partial \tau} - \frac{1}{r_0}\left(\mathbf{J}^2 - \mathbf{L}^2 + \frac{1}{4}\right)\right]. \tag{C.72}$$

The spinor wave function on the cylinder are given by

$$\Psi^{\pm}_{njm_j}(\tau,\mathbf{n}) = e^{-i\omega_n\tau}\phi^{\pm}_{jm_j}(\mathbf{n}), \tag{C.73}$$

where $\mathbf{n} \in S^2_1$. In the computation of functional determinants on $S^1_\beta \times S^2_{r_0}$ we can make use of the following relations,

$$\int_{S^1_\beta\times S^2_{r_0}} (\Psi^{\pm}_{jm_j})^\dagger \Psi^{\pm}_{j'm'_j} = \delta_{j'j}\delta_{m'_j m_j}\begin{pmatrix}1 & 0\\ 0 & 1\end{pmatrix}, \tag{C.74}$$

$$\int_{S^1_\beta\times S^2_{r_0}} (\Psi^{\pm}_{jm_j})^\dagger \gamma_r \Psi^{\pm}_{j'm'_j} = \delta_{j'j}\delta_{m'_j m_j}\begin{pmatrix}0 & -1\\ -1 & 0\end{pmatrix}, \tag{C.75}$$

$$\int_{S^1_\beta\times S^2_{r_0}} (\Psi^{\pm}_{jm_j})^\dagger\, i\gamma_\mu D^\mu \Psi^{\pm}_{j'm'_j} = \delta_{j'j}\delta_{m'_j m_j}\begin{pmatrix}0 & \omega_n - i\omega_j\\ \omega_n + i\omega_j & 0\end{pmatrix}, \tag{C.76}$$

where we have introduced the fermionic Matsubara frequencies $\omega_n$ and the eigenvalues of the Dirac operator on the sphere $\omega_j$. They are given by

$$\omega_n = \frac{(2n+1)\pi}{\beta}, \qquad \omega_j = \frac{1}{r_0}\left(j+\frac{1}{2}\right), \tag{C.77}$$

respectively.

### C.4.3 Reducible Representation

For fermionic theories in three-dimensional spacetime with an even number $2N$ of fermionic fields — $\psi_{a=1,\ldots,2N}$ — it is sometimes convenient to introduce a reducible representation of the Clifford algebra as follows:

$$\Gamma_\mu = \sigma_3 \otimes \gamma_\mu = \begin{pmatrix}\gamma_\mu & 0\\ 0 & -\gamma_\mu\end{pmatrix}, \qquad \Psi_i := \begin{pmatrix}\psi_i\\ \psi_{i+N}\end{pmatrix}, \qquad i = 1,\ldots,N. \tag{C.78}$$

Introducing a reducible representation as defined above allows for a notion of chiral symmetry in $D = 3$ within the theory, which however is actually part of the global symmetry of the theory. For example, we can choose

$$\Gamma_5 = \sigma_1 \otimes \mathbb{1} = \begin{pmatrix} & \mathbb{1}\\ \mathbb{1} & \end{pmatrix}, \tag{C.79}$$

and an appropriate choice of charge conjugation matrix is given by

$$C_4 = \Gamma_2 = \sigma_3 \otimes C = \begin{pmatrix}\sigma_2 & \\ & -\sigma_2\end{pmatrix}, \tag{C.80}$$

The four dimensional matrix $C_4$ as defined above then satisfies

$$C_4 = C_4^{-1} = C_4^\dagger = -C_4^T = -C_4^*, \qquad C_4\Gamma_\mu C_4 = -(\Gamma_\mu)^T. \tag{C.81}$$

the matrices $\Gamma_5$ and $C_4$ further satisfy

$$\{\Gamma_5, C_4\} = 0\,. \tag{C.82}$$

Importantly, charge conjugation is independent of the signature of spacetime [**Wetterich:2011ab**]. In terms of the spinor fields $\psi_i$ the reducible four-dimensional representation is constructed of two two-dimensional irreducible spinors,

$$\Psi_i = \left(\psi_i\,, \psi_{i+N}\right)^T, \qquad \bar{\Psi}_i = \Psi_i^\dagger \Gamma_3 = \left(\psi_i^\dagger \gamma_3\,, -\psi_{i+N}^\dagger \gamma_3\right) = \left(\bar{\psi}_i\,, -\bar{\psi}_{i+N}\right), \tag{C.83}$$

for $i = 1, \ldots, N$. As a concrete example, we consider the action of the $U(1)$-NJL model of $N$ reducible spinors. Written in terms of our reducible representation the action is given by

$$S = \int \mathrm{d}^3 x \left(\bar{\Psi}_i \Gamma_\mu \partial^\mu \Psi_i - \frac{g}{N}\left((\bar{\Psi}_i \Psi_i)^2 - (\bar{\Psi}_i \Gamma_5 \Psi_i)^2\right)\right). \tag{C.84}$$

## C.5 U(1) Pauli–Gürsey transformation

In this appendix we return to the the $U(1)$-NJL model. For simplicity we take $N = 1$, but the results discussed here are trivially generalized to arbitrary $N$. For the $U(1)$-NJL model with the action

$$S = \int \mathrm{d}^3 x \left(\bar{\Psi} \Gamma_\mu \partial^\mu \Psi - \frac{g}{N}\left((\bar{\Psi}\Psi)^2 - (\bar{\Psi}\Gamma_5 \Psi)^2\right)\right) \tag{C.85}$$

we consider the following linear transformation of the fields at the level of the path integral

$$\begin{aligned} \Psi &\mapsto \frac{1}{2}\left[(1-\Gamma_5)\Psi + (1+\Gamma_5) C_4 \bar{\Psi}^T\right], \\ \bar{\Psi} &\mapsto \frac{1}{2}\left[\bar{\Psi}(1+\Gamma_5) - \Psi^T C_4 (1-\Gamma_5)\right]. \end{aligned} \tag{C.86}$$

This is a so-called Pauli–Gursey (PG) transformation. We remark that the precise form of this transformation depends on the convention depends on the convention for the gamma matrices, in particular $\Gamma_5$ and $C_4$. Further, we note that this transformation is an involution as it maps the field $\Psi$ onto itself after twice applying the transformation.

Under the PG transformation as defined above the kinetic term in the action remains invariant,

$$\int \mathrm{d}^3 x\, \bar{\Psi} \Gamma_\mu \partial^\mu \Psi \;\mapsto\; \int \mathrm{d}^3 x\, \bar{\Psi} \Gamma_\mu \partial^\mu \Psi\,, \tag{C.87}$$

The Cooper BCS quartic interaction term $\bar{\Psi} C_4 \bar{\Psi}^T \Psi^T C_4 \Psi$ is mapped to the interaction term of the $U(1)$-NJL model

$$-\bar{\Psi} C_4 \bar{\Psi}^T \Psi^T C_4 \Psi \;\mapsto\; \bar{\Psi}(1+\Gamma_5)\Psi\, \bar{\Psi}(1-\Gamma_5)\Psi\,. \tag{C.88}$$

and vice versa. The other direction can either be inferred by the fact that the PG transformation is an involution or computed directly. In the latter approach we will need the fact that

$$\bar{\Psi}\Gamma_5 C_4 \bar{\Psi}^T = \Psi^T \Gamma_5 C_4 \Psi = 0\,, \tag{C.89}$$

which holds true in our convention for the gamma matrices.

Interestingly, the PG transformation also maps the $U(1)_B$ fermion number chemical potential term into the $U(1)_A$ chiral (axial) chemical potential term within the action, and vice versa of course. Concretely, this means that

$$\bar{\Psi}\Gamma_3\mu\Psi \mapsto \bar{\Psi}(-\Gamma_3\Gamma_5\mu)\Psi\,, \qquad \bar{\Psi}\Gamma_3\Gamma_5\mu\Psi \mapsto \bar{\Psi}(-\Gamma_3\mu)\Psi\,. \tag{C.90}$$

In total the PG transformation, as defined above, yields the following map $U(1)$–NJL model to the so-called Cooper model,

$$\begin{aligned} S &= \int \mathrm{d}^3x \left[\bar{\Psi}(\Gamma_\mu\partial^\mu - \mu\Gamma_3\Gamma_5)\Psi - \frac{g}{N}\left((\bar{\Psi}\Psi)^2 - (\bar{\Psi}\Gamma_5\Psi)^2\right)\right] \\ &= \int \mathrm{d}^3x \left[\bar{\Psi}(\Gamma_\mu\partial^\mu - \mu\Gamma_3\Gamma_5)\Psi - \frac{g}{N}\bar{\Psi}(1+\Gamma_5)\Psi\,\bar{\Psi}(1-\Gamma_5)\Psi\right] \\ &\mapsto \int \mathrm{d}^3x \left[\bar{\Psi}(\Gamma_\mu\partial^\mu + \mu\Gamma_3)\Psi + \frac{g}{N}\bar{\Psi}C_4\bar{\Psi}^T\,\Psi^T C_4\Psi\right]. \end{aligned} \tag{C.91}$$

The converse statement is of course true as well. For completeness, written in terms of collective fields at large $N$, the Cooper model is given by

$$S = \int \mathrm{d}^3x \left[\bar{\Psi}_i(\Gamma_\mu\partial^\mu + \mu\Gamma_3)\Psi_i + \mathrm{i}\frac{\Phi}{2}\bar{\Psi}_i C_4\bar{\Psi}_i^T + \mathrm{i}\frac{\Phi^*}{2}\Psi_i^T C_4\Psi_i + \frac{N}{4g}\Phi^*\Phi\right]. \tag{C.92}$$

## C.6 Computation of the free fermion determinant

We want to compute the path integral with the action in Eq. (3.190). Spinors possess no zero modes because of the anti-periodic boundary conditions they obey within the path integral. At finite temperature, the Matsubara frequencies implement the anti-periodic boundary conditions,[6]

$$\omega_n = (2n+1)\frac{\pi}{\beta}\,, \qquad \Longrightarrow \qquad \Psi(0, r_0\mathbf{n}) = -\Psi(\beta, r_0\mathbf{n})\,, \quad \mathbf{n}\in S_1^2\,. \tag{C.93}$$

The action $S_\mu$ remains quadratic after the inclusion of the $U(1)_B$ chemical potential and all spinors can be integrated out analogously to how it is done in Appendix C.2. The resulting grand-canonical partition function to leading order in $N$ reads

$$-\log\left(Z_{gc}(\mu)\right) = -N\log\det\left[\Gamma_3\left(\Gamma_\mu\partial^\mu - \mu\Gamma_3\right)\right] = -N\mathrm{Tr}\log\left[\Gamma_3\left(\Gamma_\mu\partial^\mu - \mu\Gamma_3\right)\right], \tag{C.94}$$

[6]Compare the Matsubara frequencies for spinors here to the ones for bosonic DoF (scalars) in Appendix C.2.

where we used the fact that Grassmann variables satisfy

$$\int \mathrm{d}\eta_1 \mathrm{d}\eta_1^\dagger \cdots \mathrm{d}\eta_N \mathrm{d}\eta_N^\dagger e^{\eta^\dagger D \eta} = \det D\,. \tag{C.95}$$

Using the Matsubara frequencies we can write down the trace-log term of the above operator in terms of the eigenvalues on the sphere, which are given by (see Appendix C.4.2)

$$\mathrm{spec}\left[\Gamma_3\left(\Gamma_\mu \partial^\mu - \mu\Gamma_3\right)\right] = \left\{\pm i\sqrt{\left(\omega_n + i\mu\right)^2 + \omega_j^2}\right\}, \qquad \omega_j^2 = \frac{(j+1/2)^2}{r_0^2}\,, \tag{C.96}$$

where $j \geq 1/2, 3/2, \ldots$ and $n \in \mathbb{Z}$. The multiplicities of the eigenvalues $j$ are $(2j+1)$. The log-det term can now be computed to give [194]

$$\begin{aligned} S_\mu &= -N\mathrm{Tr}(\delta_{\alpha\beta})\sum_j (2j+1)\sum_{n\in\mathbb{Z}} \log\beta^2\left[(\omega_n + i\mu)^2 + \omega_j^2\right] \\ &= -4N\sum_j(2j+1)\sum_{n\geq 0}\left\{\log\left[(2n+1)^2\pi^2 + \beta^2(\omega_j+\mu)^2\right] + \log\left[(2n+1)^2\pi^2 + \beta^2(\omega_j-\mu)^2\right]\right\}. \end{aligned} \tag{C.97}$$

The determinantal operation is performed over both the frequency-momentum space and the Dirac indices, with $\mathrm{Tr}(\delta_{\alpha\beta}) = 4$ being the trace over the unit matrix in spinor space. This result is logarithmically divergent and needs to be regularized. To do so we use will use zeta function regularization (here $\zeta(s;a)$ denotes the Hurwitz zeta function),

$$\sum_{n\geq 0}\log\left[(2n+1)^2\pi^2\right] = -2\frac{\mathrm{d}}{\mathrm{d}s}\sum_{n\geq 0}\frac{1}{[(2n+1)\pi]^s}\bigg|_{s=0} = \frac{2}{[2\pi]^s}\left[\log(2\pi)\zeta(s;1/2) - \zeta'(s;1/2)\right]\bigg|_{s=0} = \log(2)\,. \tag{C.98}$$

where we have used the results

$$\zeta(0;a) = \frac{1}{2} - a\,, \qquad \zeta'(0;a) = \log(\Gamma(a)) - \frac{1}{2}\log(2\pi)\,. \tag{C.99}$$

The regularized action reads

$$S_\mu = -4N\sum_j(2j+1)\left(2\log 2 + \sum_{n\geq 0}\log\left[\frac{(2n+1)^2\pi^2 + \beta^2(\omega_j\pm\mu)^2}{(2n+1)^2\pi^2}\right]\left[\frac{(2n+1)^2\pi^2 + \beta^2(\omega_j-\mu)^2}{(2n+1)^2\pi^2}\right]\right). \tag{C.100}$$

We can apply the identities[7]

$$\log\left[\frac{(2n+1)^2\pi^2 + \beta^2(\omega_j\pm\mu)^2}{(2n+1)^2\pi^2}\right] = \int_0^{\beta^2(\omega_j\pm\mu)^2}\mathrm{d}\theta^2\,\frac{1}{\theta^2 + (2n+1)^2\pi^2}\,, \qquad \sum_{n\in\mathbb{Z}}\frac{1}{(n-x)(n-y)} = \frac{\pi(\cot\pi x - \cot\pi y)}{(y-x)}\,, \tag{C.101}$$

to derive that

$$\sum_{n\geq 0}\log\left[\frac{(2n+1)^2\pi^2 + \beta^2(\omega_j\pm\mu)^2}{(2n+1)^2\pi^2}\right] = \frac{\beta}{2}\left(\omega_j\pm\mu\right) + \log\left(1 + e^{-\beta(\omega_j\pm\mu)}\right) - \log(2)\,. \tag{C.102}$$

[7]The second identity can further be used to show that $\sum_{n\in\mathbb{Z}}\frac{1}{(2n+1)^2\pi^2+\theta^2} = \frac{1}{\theta}\left(\frac{1}{2} - \frac{1}{e^\theta+1}\right)$.

The constant factor precisely cancels the factor from the zeta function regularization and the grand potential now reads

$$\Omega(\mu) = -\frac{\log\big(Z_{gc}(\mu)\big)}{(2N)\beta V} = -\frac{1}{\beta V}\sum_{j=\frac{1}{2},\frac{3}{2},\dots} 2(2j+1)\sum_{n\geq 0}\Big[\beta\omega_\ell + \log\Big(1+e^{-\beta(\omega_\ell-\mu)}\Big) + \log\Big(1+e^{-\beta(\omega_\ell+\mu)}\Big)\Big], \tag{C.103}$$

We quickly remark that the computation of the log-det term for the GN model and the NJL-type models in Sections 3.3.2, 3.3.3 and 3.3.4 are computed analogously after adding the appropriate mass and chemical potential terms, at least in the limit $N\to\infty$.

## C.7 Finite-density loop integrals and Matsubara sums

Here, we collect some important machinery used in the computation of finite-density loop integrals, in particular for the GN model in Section 3.3.2.

### C.7.1 Fourier transforms on $S^1_\beta \times \mathbb{R}^2$ and Matsubara sums

In the following we denote a point in $S^1_\beta \times \mathbb{R}^2$ by $x = (\tau, \mathbf{x})$ and a point in Fourier space by $P = (\omega_n, \mathbf{p})$, where $\omega_n = (2n+1)\pi/\beta$ denote the fermionic Matsubara frequencies. In our normalization conventions for Fourier transforms we have

$$\delta(x-x') = \sum\!\!\!\!\!\!\int \frac{\mathrm{d}^2 p}{\beta(2\pi)^2}\, e^{-iP\cdot(x-x')}, \qquad \delta_{n'n}\delta(p-p') = \int \frac{\mathrm{d}\tau\mathrm{d}^2 x}{\beta(2\pi)^2}\, e^{-ix\cdot(P-P')}, \tag{C.104}$$

$$f(x) = \sum\!\!\!\!\!\!\int \frac{\mathrm{d}^2 p}{\sqrt{\beta(2\pi)^2}}\, e^{-iP\cdot x}\tilde{f}(P), \qquad \tilde{f}(P) = \int \frac{\mathrm{d}\tau\mathrm{d}^2 x}{\sqrt{\beta(2\pi)^2}}\, e^{iP\cdot x} f(x). \tag{C.105}$$

In computations of Dirac determinants we will encounter sums over the fermionic Matsubara frequencies. Generally speaking, these can all be performed using the following formula:

$$\sum_{n\in\mathbb{Z}} \log\left[\frac{(2n+1)^2\pi^2 + A^2}{(2n+1)^2\pi^2+1}\right] = A + 2\log\big(1+e^{-A}\big). \tag{C.106}$$

### C.7.2 GN scalar integrals at finite $\mu$, $\beta$

In the present appendix we collect scalar integrals required to obtain the results for the fluctuations within the large-$N$ GN model discussed in Section 3.3.2. One-loop integrals computed at finite temperature and chemical potential can generically be derived from the following massive scalar integral [195]:

$$\int \frac{\mathrm{d}^d k}{(2\pi)^d}\frac{1}{[k^2+m^2]^\alpha} = \frac{1}{(4\pi)^{\frac{d}{2}}}\frac{\Gamma(\alpha-d/2)}{\Gamma(\alpha)}(m^2)^{-\alpha+\frac{d}{2}}. \tag{C.107}$$

Using the notation $K^{(\mu)} = (\omega_n - I\mu, \mathbf{k})$ the first scalar integral appearing in the computations in

Section 3.3.2 can be computed as

$$
\begin{aligned}
I_1 &= \sum\!\!\!\!\!\!\int \frac{\mathrm{d}^d k}{\beta(2\pi)^d} \frac{1}{\left(K^{(\mu)}\right)^2}\Bigg|_{d=2} \\
&= \frac{\Gamma(1-d/2)}{(4\pi)^{\frac{d}{2}}} \sum_{n\in\mathbb{Z}} \frac{1}{[(\omega_n - i\mu)^2]^{1-\frac{d}{2}}}\Bigg|_{d=2} \\
&= \frac{\Gamma(1-d/2)}{(4\pi)^{\frac{d}{2}}} \left(\frac{2\pi}{\beta}\right)^{-2+d} \sum_{n\in\mathbb{Z}} \frac{1}{\left[\left(n+\frac{1}{2} - i\beta\mu/(2\pi)\right)^2\right]^{1-\frac{d}{2}}}\Bigg|_{d=2} \\
&= \frac{\Gamma(1-d/2)}{(4\pi)^{\frac{d}{2}}} \left(\frac{2\pi}{\beta}\right)^{-2+d} \left[\zeta\left(2-d, \frac{1}{2} - i\beta\mu/(2\pi)\right) + \zeta\left(2-d, \frac{1}{2} + i\beta\mu/(2\pi)\right)\right]\Bigg|_{d=2}.
\end{aligned}
\tag{C.108}
$$

At zero temperature after setting $d = 2$ this result becomes

$$
\lim_{\beta\to\infty} I_1 = -\frac{\mu}{4\pi}. \tag{C.109}
$$

The $I_2$ integral is slightly more involved as it includes three scales: $\beta$, $\mu$ and $P$, where $P = (\omega_m, \mathbf{p})$ is the external momentum. However, it can still be computed similarly to $I_1$ after a Feynman parametrization is used to merge the two propagators in the denominator. We find that

$$
\begin{aligned}
I_2 &= \sum\!\!\!\!\!\!\int \frac{\mathrm{d}^2 k\, \mathrm{d}^2 q}{\left(\beta(2\pi)^2\right)^2} \frac{\delta(K+Q-P)}{\left(K^{(\mu)}\right)^2 \left(Q^{(\mu)\dagger}\right)^2} \\
&= \int_0^1 \mathrm{d}x \sum\!\!\!\!\!\!\int \frac{\mathrm{d}^2 k}{\beta(2\pi)^2} \frac{1}{\left[\mathbf{k}^2 + \left(x(1-x)\mathbf{p}^2 + (1-x)(\omega_n - i\mu)^2 + x(\omega_m - \omega_n + i\mu)^2\right)\right]^2} \\
&= \frac{\Gamma(2-d/2)}{\beta(4\pi)^{\frac{d}{2}}} \left(\frac{2\pi}{\beta}\right)^{d-4} \int_0^1 \mathrm{d}x \sum_{n\in\mathbb{Z}} \frac{1}{\left[\left(n+\frac{1}{2} - i\beta\frac{\mu}{2\pi} - x\beta\frac{\omega_m}{2\pi}\right)^2 + x(1-x)\left((\beta\frac{\mathbf{p}}{2\pi})^2 + (\beta\frac{\omega_m}{2\pi})^2\right)\right]^{2-\frac{d}{2}}}\Bigg|_{d=2} \\
&= \frac{\Gamma(2-d/2)}{\beta(4\pi)^{\frac{d}{2}}} \left(\frac{2\pi}{\beta}\right)^{d-4} \int_0^1 \mathrm{d}x \sum_{n\in\mathbb{Z}} \frac{1}{\left[(n+A)^2 + B\right]^{2-\frac{d}{2}}}\Bigg|_{d=2} \\
&= \frac{\Gamma(2-d/2)}{\beta(4\pi)^{\frac{d}{2}}} \left(\frac{2\pi}{\beta}\right)^{d-4} \int_0^1 \mathrm{d}x \left[\frac{1}{[A^2+B]^{2-\frac{d}{2}}} + F(2-d/2; A, B) + F(2-d/2; -A, B)\right]\Bigg|_{d=2},
\end{aligned}
\tag{C.110}
$$

where we introduced the quantities

$$
A = \frac{1}{2} - i\beta\frac{\mu}{2\pi} - x\beta\frac{\omega_m}{2\pi}, \qquad B = x(1-x)\left[(\beta\frac{\mathbf{p}}{2\pi})^2 + (\beta\frac{\omega_m}{2\pi})^2\right]. \tag{C.111}
$$

The functions $F$ denote a family of special zeta function which can be found in [237]. At zero temperature and after setting $d = 2$ the $I_2$-integral reduces to

$$
\lim_{\beta\to\infty} I_2 = \frac{1}{8\sqrt{\omega_m^2 + \mathbf{p}^2}} = \frac{1}{8\sqrt{P^2}}. \tag{C.112}
$$

### C.7.3 NJL loop integrals

In the computations for the fluctuations within the NJL-type models discussed in Section 3.3.3 we have to evaluate the following integrals:

$$\tilde{G}^{-1}_{\sigma\sigma}(P) = -\int \frac{\mathrm{d}^3 k}{(2\pi)^3} \operatorname{Tr}\left[\tilde{D}^{(\mu,\langle\Phi\rangle)}(K)\,\tilde{D}^{(-\mu,-\langle\Phi\rangle)}(P-K)\right], \tag{C.113}$$

$$\tilde{G}^{-1}_{\sigma\pi}(P) = -i\int \frac{\mathrm{d}^3 k}{(2\pi)^3} \operatorname{Tr}\left[\tilde{D}^{(\mu,\langle\Phi\rangle)}(K)\,\Gamma_5\,\tilde{D}^{(-\mu,-\langle\Phi\rangle)}(P-K)\right], \tag{C.114}$$

$$\tilde{G}^{-1}_{\pi\sigma}(P) = -i\int \frac{\mathrm{d}^3 k}{(2\pi)^3} \operatorname{Tr}\left[\Gamma_5\,\tilde{D}^{(\mu,\langle\Phi\rangle)}(K)\,\tilde{D}^{(-\mu,-\langle\Phi\rangle)}(P-K)\right], \tag{C.115}$$

$$\tilde{G}^{-1}_{\pi\pi}(P) = \int \frac{\mathrm{d}^3 k}{(2\pi)^3} \operatorname{Tr}\left[\Gamma_5\,\tilde{D}^{(\mu,\langle\Phi\rangle)}(K)\,\Gamma_5\,\tilde{D}^{(-\mu,-\langle\Phi\rangle)}(P-K)\right], \tag{C.116}$$

where the fermion propagator in momentum space $\tilde{D}^{(\mu,\langle\Phi\rangle)}(P)$ is given by

$$\begin{aligned}\tilde{D}^{(\mu,\langle\Phi\rangle)}(P) &= \left(-i\Gamma_\mu P^\mu + \langle\Phi\rangle - \mu\Gamma_3\Gamma_5\right)^{-1}\\ &= \frac{\left(\omega^2+\mathbf{p}^2+\langle\Phi\rangle^2-\mu^2+2\mu(i\omega\Gamma_3+\langle\Phi\rangle)\Gamma_3\Gamma_5\right)}{\left(\omega^2+\langle\Phi\rangle^2+(\mu+|\mathbf{p}|)^2\right)\left(\omega^2+\langle\Phi\rangle^2+(\mu-|\mathbf{p}|)^2\right)}\left(i\Gamma_\mu P^\mu+\langle\Phi\rangle-\mu\Gamma_3\Gamma_5\right).\end{aligned} \tag{C.117}$$

The gamma matrices used here are the ones from the reducible representation defined in Appendix C.4.3. Since we only need to compute the quadratic order in $P$, we can evaluate these integrals in an expansion in $P/\mu$.

#### Zeroth order in $P/\mu$

To zeroth order in $P/\mu$ the above integrals read

$$\begin{aligned}\tilde{G}^{-1}_{\sigma\sigma}(P)\Big|_{\mathcal{O}(0)} = \int \frac{\mathrm{d}^2 k\,\mathrm{d}\omega_k}{(2\pi)^3}\Bigg[&4\langle\Phi\rangle^2\left(\frac{1}{\left[(|\mathbf{k}|+\mu)^2+\omega_k^2+\langle\Phi\rangle^2\right]^2}+\frac{1}{\left[(|\mathbf{k}|-\mu)^2+\omega_k^2+\langle\Phi\rangle^2\right]^2}\right)\\ &-\frac{2}{(|\mathbf{k}|-\mu)^2+\omega_k^2+\langle\Phi\rangle^2}-\frac{2}{(|\mathbf{k}|+\mu)^2+\omega_k^2+\langle\Phi\rangle^2}\Bigg],\end{aligned} \tag{C.118}$$

$$\tilde{G}^{-1}_{\sigma\pi}(P)\Big|_{\mathcal{O}(0)} = 0, \tag{C.119}$$

$$\tilde{G}^{-1}_{\pi\sigma}(P)\Big|_{\mathcal{O}(0)} = 0, \tag{C.120}$$

$$\tilde{G}^{-1}_{\pi\pi}(P)\Big|_{\mathcal{O}(0)} = -\int \frac{\mathrm{d}^2 k\,\mathrm{d}\omega_k}{(2\pi)^3}\left[\frac{2}{(|\mathbf{k}|+\mu)^2+\omega_k^2+\langle\Phi\rangle^2}+\frac{2}{(|\mathbf{k}|-\mu)^2+\omega_k^2+\langle\Phi\rangle^2}\right]. \tag{C.121}$$

First, we can perform the residue integrals over $\omega_k$,

$$\tilde{G}^{-1}_{\sigma\sigma}(P)\Big|_{\mathcal{O}(0)} = -\int \frac{\mathrm{d}^2 k}{(2\pi)^2}\left[\frac{(|\mathbf{k}|-\mu)^2}{\left[(|\mathbf{k}|-\mu)^2+\langle\Phi\rangle^2\right]^{3/2}}+\frac{(|\mathbf{k}|+\mu)^2}{\left[(|\mathbf{k}|+\mu)^2+\langle\Phi\rangle^2\right]^{3/2}}\right], \tag{C.122}$$

$$\tilde{G}^{-1}_{\pi\pi}(P)\Big|_{\mathcal{O}(0)} = \int \frac{\mathrm{d}^2 k}{(2\pi)^2}\left[\frac{1}{\sqrt{(|\mathbf{k}|+\mu)^2+\langle\Phi\rangle^2}}+\frac{1}{\sqrt{(|\mathbf{k}|-\mu)^2+\langle\Phi\rangle^2}}\right]. \tag{C.123}$$

The remaining integrals over $\mathbf{k}$ are both divergent. However, the divergence is independent of $\mu$ and hence these integrals can be regularized simply via subtraction of the expression for $\mu = 0$. After regularization, the divergent integrals to be subtracted give

$$\int \frac{\mathrm{d}^2 k}{(2\pi)^2} \frac{2\mathbf{k}^2}{\left(\mathbf{k}^2 + \langle\Phi\rangle^2\right)^{3/2}} = \frac{1}{\pi} \int \mathrm{d}k\, k \frac{k^2}{\left(k^2 + \langle\Phi\rangle^2\right)^{3/2}} := -\frac{2}{\pi}\langle\Phi\rangle \,, \qquad \int \frac{\mathrm{d}^2 k}{(2\pi)^2} \frac{2}{\sqrt{\mathbf{k}^2 + \langle\Phi\rangle^2}} := -\frac{\langle\Phi\rangle}{\pi} \,. \tag{C.124}$$

After performing the remaining spatial integrals over the momentum $\mathbf{k}$ we find that

$$\tilde{G}^{-1}_{\sigma\sigma}(P)\Big|_{\mathcal{O}(0)} = \frac{1}{\pi}\left[2\sqrt{\mu^2 + \langle\Phi\rangle^2} - \mu \,\mathrm{arctanh}\left(\frac{\mu}{\sqrt{\mu^2 + \langle\Phi\rangle^2}}\right)\right] , \tag{C.125}$$

$$\tilde{G}^{-1}_{\pi\pi}(P)\Big|_{\mathcal{O}(0)} = \frac{1}{\pi}\left[\sqrt{\mu^2 + \langle\Phi\rangle^2} - \mu \,\mathrm{arctanh}\left(\frac{\mu}{\sqrt{\mu^2 + \langle\Phi\rangle^2}}\right)\right] . \tag{C.126}$$

After applying the EoM,

$$\langle\Phi\rangle = \sqrt{\kappa_0^2 - 1}\,\mu \,, \qquad\qquad \mathrm{arctanh}\left(\frac{1}{\kappa_0}\right) = \kappa_0 \,, \tag{C.127}$$

the final result reads

$$\tilde{G}^{-1}_{\sigma\sigma}(P)\Big|_{\mathcal{O}(0)} = \frac{\kappa_0 \pi}{\mu} \,, \qquad\qquad \tilde{G}^{-1}_{\pi\pi}(P)\Big|_{\mathcal{O}(0)} = 0 \,. \tag{C.128}$$

### First order in $P/\mu$

To linear order in $P/\mu$ we find that for two elements of the inverse propagator the integrand is and odd function of the coefficients of $\mathbf{k}$. Hence, they identically vanish

$$\tilde{G}^{-1}_{\sigma\sigma}(P)\Big|_{\mathcal{O}(P/\mu)} = 0 \,, \qquad\qquad \tilde{G}^{-1}_{\pi\pi}(P)\Big|_{\mathcal{O}(P/\mu)} = 0 \,. \tag{C.129}$$

The remaining two integrals are computed as follows:

$$\tilde{G}^{-1}_{\sigma\pi}(P)\Big|_{\mathcal{O}(P/\mu)} = \int \frac{\mathrm{d}^2 k\, \mathrm{d}\omega_k}{(2\pi)^3} \frac{4\mu\omega\left[-3\mathbf{k}^4 + 2\mathbf{k}^2\left(-\omega_k^2 + \mu^2 - \langle\Phi\rangle^2\right) + \left(\omega_k^2 + \mu^2 + \langle\Phi\rangle^2\right)^2\right]}{\left[(|\mathbf{k}| - \mu)^2 + \omega_k^2 + \langle\Phi\rangle^2\right]^2 \left[(|\mathbf{k}| + \mu)^2 + \omega_k^2 + \langle\Phi\rangle^2\right]^2} \,, \tag{C.130}$$

$$\tilde{G}^{-1}_{\pi\sigma}(P)\Big|_{\mathcal{O}(P/\mu)} = -\int \frac{\mathrm{d}^2 k\, \mathrm{d}\omega_k}{(2\pi)^3} \frac{4\mu\omega\left[-3\mathbf{k}^4 + 2\mathbf{k}^2\left(-\omega_k^2 + \mu^2 - \langle\Phi\rangle^2\right) + \left(\omega_k^2 + \mu^2 + \langle\Phi\rangle^2\right)^2\right]}{\left[(|\mathbf{k}| - \mu)^2 + \omega_k^2 + \langle\Phi\rangle^2\right]^2 \left[(|\mathbf{k}| + \mu)^2 + \omega_k^2 + \langle\Phi\rangle^2\right]^2} \,. \tag{C.131}$$

We again perform the residue integrals over $\omega_k$ first. The result is not divergent, as expected, and can be directly integrated over the momenta $\mathbf{k}$,

$$\tilde{G}^{-1}_{\sigma\pi}(P)\Big|_{\mathcal{O}(P/\mu)} = \frac{\omega}{4\pi}\log\left(\frac{2\mu\left(\mu - \sqrt{\mu^2 + \langle\Phi\rangle^2}\right)}{\langle\Phi\rangle^2} + 1\right) , \quad \tilde{G}^{-1}_{\pi\sigma}(P)\Big|_{\mathcal{O}(P/\mu)} = \frac{\omega}{4\pi}\log\left(\frac{2\mu\left(\sqrt{\mu^2 + \langle\Phi\rangle^2} + \mu\right)}{\langle\Phi\rangle^2} + 1\right) . \tag{C.132}$$

After using the EoM the final result we end up with reads

$$\tilde{G}^{-1}_{\sigma\pi}(P)\Big|_{\mathcal{O}(P/\mu)} = -\frac{\kappa_0\omega}{2\pi}\,, \qquad \tilde{G}^{-1}_{\pi\sigma}(P)\Big|_{\mathcal{O}(P/\mu)} = \frac{\kappa_0\omega}{2\pi}\,. \tag{C.133}$$

**Second order in $P^2/\mu^2$**

Due to rotational invariance, the integrand for the quadratic order in $P/\mu$ must be of the form

$$A(|\mathbf{k}|)\,\omega^2 + B(|\mathbf{k}|)\,\omega\,(\mathbf{k}\cdot\mathbf{p}) + C(|\mathbf{k}|)\,\mathbf{p}^2 + D(|\mathbf{k}|)\,(\mathbf{k}\cdot\mathbf{p})^2\,. \tag{C.134}$$

The $B(|\mathbf{k}|)$-piece is identically zero due to being an odd function. In the same vein, the cross-term in $(\mathbf{k}\cdot\mathbf{p})^2$ identically vanishes. Hence, the part of the integrand that contributes is

$$A(|\mathbf{k}|)\,\omega^2 + C(|\mathbf{k}|)\,\mathbf{p}^2 + D(|\mathbf{k}|)\,(k_1^2p_1^2 + k_2^2p_2^2)\,. \tag{C.135}$$

Given the symmetry property of the integral under the exchange $k_1 \leftrightarrow k_2$, after evaluation the integral is a function of $\mathbf{p}^2 = p_1^2 + p_2^2$ (and $\omega^2$). We therefore split the computation into two parts, one proportional to $\omega^2$ and the other one proportional to $\mathbf{p}^2$. After performing the residue integral over $\omega_k$ the non-trivial parts of the integrals proportional to $\omega^2$ are given by

$$I^{\omega}_{\sigma\sigma} = \frac{\pi}{2}\int\frac{\mathrm{d}^2k}{(2\pi)^3}\,\mu^2\left(\frac{\omega}{\mu}\right)^2\left[\frac{(|\mathbf{k}|-\mu)^2}{\left[(|\mathbf{k}|-\mu)^2+\langle\Phi\rangle^2\right]^{5/2}} + \frac{(|\mathbf{k}|+\mu)^2}{\left[(|\mathbf{k}|+\mu)^2+\langle\Phi\rangle^2\right]^{5/2}}\right], \tag{C.136}$$

$$I^{\omega}_{\pi\pi} = \frac{\pi}{2}\int\frac{\mathrm{d}^2k}{(2\pi)^3}\,\mu^2\left(\frac{\omega}{\mu}\right)^2\left[\frac{1}{\left[(|\mathbf{k}|+\mu)^2+\langle\Phi\rangle^2\right]^{3/2}} + \frac{1}{\left[(|\mathbf{k}|-\mu)^2+\langle\Phi\rangle^2\right]^{3/2}}\right]. \tag{C.137}$$

We can then perform the integral over $\mathbf{k}$ and use the EoM to find that

$$I^{\omega}_{\sigma\sigma} = \frac{\omega^2 - 2\kappa_0^2\omega^2}{12\pi\kappa_0\mu - 12\pi\kappa_0^3\mu}\,, \qquad I^{\omega}_{\pi\pi} = -\frac{\kappa_0\omega^2}{4\pi\mu - 4\pi\kappa_0^2\mu}\,. \tag{C.138}$$

We simply repeat this procedure for the term proportional to $\mathbf{p}^2$. The final result reads

$$I^{p}_{\sigma\sigma} = \frac{\left(3\kappa_0^6 - 2\kappa_0^4 - 2\kappa_0^2 + 2\right)p^2}{24\pi\kappa_0^3\left(\kappa_0^2 - 1\right)\mu}\,, \qquad I^{p}_{\pi\pi} = -\frac{\kappa_0^3p^2}{8\pi\mu - 8\pi\kappa_0^2\mu}\,, \tag{C.139}$$

with the other terms vanishing. Putting both contributions together we find that

$$\tilde{G}^{-1}_{\sigma\sigma}(P)\Big|_{\mathcal{O}(P^2/\mu^2)} = \frac{\omega^2 - 2\kappa_0^2\omega^2}{12\pi\kappa_0\mu - 12\pi\kappa_0^3\mu} + \frac{\left(3\kappa_0^6 - 2\kappa_0^4 - 2\kappa_0^2 + 2\right)p^2}{24\pi\kappa_0^3\left(\kappa_0^2 - 1\right)\mu}\,, \tag{C.140}$$

$$\tilde{G}^{-1}_{\pi\pi}(P)\Big|_{\mathcal{O}(P^2/\mu^2)} = -\frac{\kappa_0\omega^2}{4\pi\mu - 4\pi\kappa_0^2\mu} - \frac{\kappa_0^3p^2}{8\pi\mu - 8\pi\kappa_0^2\mu}\,. \tag{C.141}$$